SHAHID BEHESHTI UNIVERSITY

# Development of Exotic Harmonium Model to Investigate Electron-Positively Charged Particle Correlation in Two-Component Quantum Systems

By:

Nahid Sadat Riyahi

A Thesis Submitted for the Degree of Doctor of Philosophy in Physical Chemistry

Faculty of Chemistry and Petroleum Sciences

January 2024

This thesis was originally written in Persian at Shahid Beheshti University as per university regulations. It has been translated into English to facilitate broader accessibility and academic exchange while maintaining the integrity of the original content.

## Abstract

The problem of calculating the electron- positively charged particle correlation energy poses a challenge in the field of quantum chemistry beyond the adiabatic approximation. In this study, a toy model called Exotic Harmonium is developed to enhance our understanding of this type of correlation. Since the analytical methods to solve eigenvalue equation for exotic harmonium lead to the excited rather than the ground state of the system, we employ the variation method in this study to obtain an approximate ground state wave function. By considering the asymptotic behavior of the wave function, we derive a compact but highly accurate variational wave function. Using this wave function, we are able to determine various properties of the system, including energy and its components, as well as single-particle densities. Additionally, we extend key concepts and quantities specifically tailored to study electron correlation in the harmonium model to the Exotic harmonium model. For instance, the "correlation hill" concept effectively highlights the distinct nature of electron-positively charged particle correlation compared to electron-electron correlation. Furthermore, we utilize Exotic Harmonium to assess the accuracy of five electron-positively charged particle correlation functionals developed within the context of the two-component density functional theory. Through a regression process, we obtained a compact yet precise analytical form for wave function which explicitly depends on the two crucial variables: oscillator field frequency and positively charged particle's mass. This analytical form provides valuable insights into the behavior of the system. Also, the Exotic Harmonium model enables the investigation of electron-positively charged particle correlation in a vast range of particle masses

To my parents, two angels in human guise

And to my sixteen-year-old self, who, upon discovering this path, imbued me with an unwavering passion for this remarkable journey, and whose whispers reminded me that even on this Pale Blue Dot[1], "Knowledge Is Power[2]."

---

[1] Carl Sagan

[2] Sir Francis Bacon

With heartfelt gratitude, I acknowledge the esteemed guidance and support of my supervisor, Dr. Shant Shahbazian,

and my advisor, Dr. Mohammad Goli.

# Contents

# 1 Introduction

## 1.1 Quantum Chemistry

Theoretical chemistry provides a systematic picture of the laws governing chemical phenomena and uses physics and mathematics to describe the structure and interaction of atoms and molecules. By the end of the 19th century, chemistry was still largely considered a descriptive science, based on concepts such as atomic and molecular weights, combinatorial ratios, thermodynamic quantities, and fundamental ideas of molecular stereochemistry. Although much more logical compared to its ancient roots in alchemy, it was still largely a collection of empirical facts about the behavior of matter. Immanuel Kant, in his famous book "Critique of Pure Reason," asserts: " In any special doctrine of nature there can be only as much proper science as there is mathematics therein." This assertion can be considered a rational and philosophical reason for emphasizing the use of mathematical methods in chemistry.

The developments in physics at the beginning of the 20th century made it possible to explain all of chemistry fundamentally within the framework of quantum mechanics. As Paul Dirac succinctly stated: " The underlying physical laws necessary for the mathematical theory of a large part of physics and the whole of chemistry are thus completely known, and the difficulty is only that the exact application of these laws leads to equations much too complicated to be soluble." Quantum mechanics, by its nature, is mathematical physics. However, the gap that Dirac pointed out, namely the existence of chemically complex problems from a mathematical perspective, justifies the survival of parts of chemistry as an experimental science. Semi-empirical concepts of chemical bonding and reactivity fall into this category.

The primary goal of theoretical chemistry is to provide a coherent picture of the structure and properties of atomic and molecular systems. Techniques adapted from mathematics and theoretical physics are employed in an effort to explain and correlate the structures and dynamics of chemical systems. Given the extreme complexity of chemical systems, theoretical chemistry, unlike theoretical physics, generally uses more approximate mathematical techniques, which are often supplemented by experimental or semi-empirical methods [1].

### 1.1.1 The Success of Quantum Chemistry

Modern science has struggled to explain the most significant phenomena in the universe, such as the accelerated expansion of galaxies (apparently due to "dark energy") and the fundamental nature of the universe's matter (a significant portion of which is "dark matter"). Quantum chemistry is in a much better position and may even be exceptional in the entire field of science. Chemical phenomena can be explained using contemporary theories down to the level of individual molecules (the subject of quantum chemistry). Comparing theory and experiment has shown that solving the Schrödinger equation provides a quantitatively accurate picture in most cases, and only molecules with very heavy atoms, due to the importance of relativistic effects, need to be treated using the Dirac equation. When exceptionally high precision is rarely required, we may ultimately consider quantum electrodynamics (QED) corrections, a method whose necessity is currently very unlikely for routine applications but feasible for small multi-particle systems. This success of computational quantum chemistry is based on the following conditions:

- Atoms and molecules are made up of only two types of particles: nuclei and electrons.
- The components of atoms and molecules are usually considered as point charges. Although nuclei have a nonzero size (electrons are considered point-like particles), this size is so small compared to atomic and molecular dimensions that its effect on computational results is much less than chemical accuracy[3].
- QED corrections are much smaller than energy changes in chemical phenomena (e.g., 1:10,000,000) and can therefore be ignored in most applications.
- Nuclei are a thousand times heavier than electrons and, therefore, except in some special cases, move a thousand times slower than electrons. This makes it possible to solve the Schrödinger equation for electrons under the assumption that the nuclei are stationary, i.e., their positions in space are fixed (clamped nuclei).

[3] Chemical accuracy is considered to be equivalent to 1 kcal/mol or 4 kJ/mol.

It is generally believed that an exact analytical solution to the Schrödinger equation for any atom (except for hydrogen-like atoms) or molecule is not possible. Instead, reasonable approximate solutions can be obtained, which almost always involve calculating a large number of electron integrals and performing some algebraic manipulations on matrices constructed from these integrals. The reason for this is the computational efficiency of what is known as the algebraic approximation (algebraization) of the Schrödinger equation.

## 1.2 Computational Chemistry

Fortunately, a large part of quantum chemistry relies on conditions such as negligible relativistic and electrodynamic corrections and the significant mass ratio of the nucleus to the electron. Rapid advancements in computer technology have revolutionized theoretical chemistry, leading to the birth of computational quantum chemistry. Computational quantum chemistry provides researchers with new, ready-to-use tools that offer powerful insights into the internal molecular structure and dynamics. Commercial computational chemistry programs allow us to perform calculations for molecules and examine each molecule independently of its chemical relationship with other molecules.

In chemistry, as in any other science, it is essential not to study only specific cases one by one but to extract important general laws whenever possible. However, since these general laws are initially derived for specific cases with certain approximations, they are not necessarily applicable to all similar cases and are sometimes only partially valid. Nonetheless, such laws enable chemists to predict, explain, and ultimately be efficient. If we relied exclusively on the exact solutions of the Schrödinger equation, chemistry would not exist at all, and people would lose the ability to rationalize this branch of science, especially for designing the synthesis of new molecules. Chemists rely on the spatial structure of molecules (nuclear configurations), concepts of valence electrons, chemical bonds, single-electron pairs, and so on. These concepts sometimes do not have a unique and precise quantum definition but remain important and highly effective for describing molecular models [2].

## 1.3 Electronic Structure Theory

Solving quantum mechanics problems always begins with writing the Schrödinger equation for the system under study. Quantum chemistry, which aims to study the quantum nature of chemical systems and processes, is no exception. Therefore, quantum chemistry starts with solving the Schrödinger equation for the simplest natural chemical system, the hydrogen atom (just as the historical development of quantum mechanics began with it). It then progresses from multi-electron atoms to more complex chemical systems, namely molecules, and their corresponding Schrödinger equations. The time-independent form of this equation is as follows:

$$\hat{H}\Psi(r^e,r^n) = E\Psi(r^e,r^n) \qquad 1.1$$

where, $r^e$ and $r^n$ are the sum of electronic and nuclear coordinates, respectively, and the molecular non-relativistic Hamiltonian $\hat{H}$ for a molecule with $N$ electrons and $N'$ nuclei in atomic units is as follows[4]:

[4] Throughout this thesis, the quantities related to nuclei or positively charged particles are denoted by prime, n or p depending on the conditions.

$$\hat{H} = -\sum_{i}^{N} \frac{1}{2} \nabla_i^2 - \sum_{i'}^{N'} \frac{1}{2m_{i'}} \nabla_{i'}^2 + \sum_{i}^{N} \sum_{j>i}^{N} \frac{1}{r_{ij}} - \sum_{i}^{N} \sum_{i'}^{N'} \frac{Z_{i'}}{r_{ii'}} + \sum_{i'}^{N'} \sum_{j'>i'}^{N'} \frac{Z_{i'} Z_{j'}}{r_{i'j'}} \quad 1.2$$

The first and second terms represent the kinetic energy of the electrons and nuclei, while the subsequent terms refer to the Coulomb potential energies (electron-electron repulsion, electron-nucleus attraction, and nucleus-nucleus repulsion, respectively). Here, $m_{i'}$ and $Z_{i'}$ denote the masses of the nuclei and the atomic number, respectively. Additionally, $r_{ij} = |\vec{r}_i - \vec{r}_j|$ represents the distance between two electrons, $r_{i'j'}$ represents the distance between two nuclei, and $r_{ii'}$ represents the distance between an electron and a nucleus.

If we want to describe the distribution of electrons in detail, there is no alternative to quantum mechanics. Electrons cannot even be qualitatively described by classical mechanics. Therefore, we will focus on solving the time-independent Schrödinger equation (1-1).

Methods in which solutions are generated without reference to experimental data, in contrast to semi-empirical models, are called "ab initio" methods. The first step in solving the Schrödinger equation in quantum chemistry is the Born-Oppenheimer approximation, which neglects the coupling between nuclear and electronic motions. This approximation allows us to solve the electronic part of the Schrödinger equation by considering the positions of the nuclei as parameters and using the resulting potential energy surface as the basis for solving the nuclear Schrödinger equation. A significant portion of computational efforts in quantum chemistry is dedicated to solving the electronic

Schrödinger equation for a given nuclear arrangement, which collectively forms the "electronic structure theory."

### 1.3.1 The Nature of Approximations

Unfortunately, the undeniable fact is that equation (1-1) cannot be solved analytically even for the simplest molecule, $H_2^+$ (which consists of three particles). As a result, we must resort to approximation from the very beginning to simplify as much as possible [3]. The most famous approximation in this context is the Born-Oppenheimer approximation, or BO for short, named after a perturbative study published in 1927 by Max Born and Robert Oppenheimer [4]. However, the BO approximation itself belongs to a broader school of thought called the adiabatic approximation, which has its roots in classical physics. The adiabatic approximation in quantum mechanics was first introduced by Max Born and Vladimir Fock in 1928 under the title of the adiabatic theorem [5]. Therefore, the BO approximation can be considered a special case of the adiabatic approximation.

The word "adiabatic" is derived from the Greek word αδιαβατος, which literally means a situation where "nothing passes through something else." In fact, the response of a system to a time-dependent perturbation depends on the time scale of the perturbation. Gradual changes in external conditions characterize an adiabatic process, whereas rapid movements that result in chaotic responses indicate a non-adiabatic process. In other words, the essence of the adiabatic approach arises from the coexistence of two time scales in some phenomena: fast and slow scales [6].

In thermodynamics, an adiabatic process is one in which energy exchange occurs without the exchange of heat between a system and its surroundings, meaning that heat does not pass through the system's boundary. Similarly, in quantum mechanics, an adiabatic process refers to a process in which no sudden transition from one state to another occurs relative to the continuous change of some parameters. Therefore, the adiabatic theorem in quantum mechanics is stated as follows: "If a given perturbation acts on a physical system slowly enough and if there is a gap between the eigenvalue of the state it is in and the rest of the Hamiltonian's eigenvalue spectrum, that system remains in its instantaneous eigenstate" [5].

The main strategy for analyzing an adiabatic process is to initially solve the system's behavior with fixed external parameters and then allow those parameters to change at the end of the calculations [7]. In quantum chemistry, this strategy is precisely used as a starting point, where the change in nuclear positions is considered a change in parameter.

In fact, to overcome the difficulty of solving equation (1-1) for molecules, we adopt this strategy and take advantage of the significant difference in mass between electrons and nuclei. Due to this difference, electrons can almost instantaneously respond to the displacement of nuclei, whereas nuclei cannot do so practically, resulting in the coexistence of two different time scales. Therefore, we can adopt the conventional strategy in such a situation: instead of trying to solve the Schrödinger equation simultaneously for all particles, we first consider the nuclei as fixed and solve the Schrödinger equation for the electrons in the static electric potential created by a specific arrangement of nuclei. Then, we change the arrangement of the nuclei and repeat the calculations.

In general, the eigenvalues of the electronic Hamiltonian are obtained for a specific arrangement of nuclei, and then the arrangement that yields the lowest energy is identified. In quantum chemistry and condensed matter and molecular physics, this different behavior regarding nuclear and electronic motions is famously known as the Born-Oppenheimer approximation [6]. However, if we want to speak in more precise and technical terms, we should call it the adiabatic approximation.

The set of solutions obtained from the electronic Schrödinger equation, i.e., the relationship between molecular electronic energies and molecular geometry, allows us to reach one of the most effective concepts in theoretical chemistry: the Potential Energy Surface (PES). This concept is highly practical, functioning like a white cane for a blind chemist. A minimum on the PES curve indicates the equilibrium configuration of a molecule [8]. Additionally, the PES enables the identification of reaction pathways.

In the BO approximation, when we solve the electronic Schrödinger equation for fixed nuclear configurations and vary the nuclear coordinates, we assume that the electronic state of the system remains the same. In other words, we assume that we remain on the same PES. However, this is only true when the PES for different electronic states is well-separated; thus, in nuclear configurations where the PES of different electronic states come close or even intersect, the BO approximation is insufficient [6]. The BO approximation is generally reliable for ground states but less so for excited states [8]. Nevertheless, most chemical systems can be described using the BO approximation because it provides accurate answers. For this reason, nearly all quantum chemistry textbooks begin by introducing the BO approximation.

### 1.3.2 Equations

The electronic Hamiltonian operator can be expressed as the sum of the electronic kinetic energy and the potential energies involving the nuclei and electrons:

$$\hat{H}_e = \hat{T}_e + \hat{V}_{ee} + \hat{V}_{ne} \tag{1.3}$$

Suppose a complete set of solutions of the electron Schrödinger equation is available as follows:

$$\hat{H}_e \psi_i(r^e; r^n) = E_i(r^n)\psi_i(r^e; r^n) \tag{1.4}$$

where $r^e$ and $r^n$ refer to the total electronic and nuclear variables, respectively, with the eigenfunctions and energies parametrically dependent on $r^n$. By solving the electronic Schrödinger equation for all conceivable configurations of the nuclei, the potential energy surface is obtained. Without introducing any approximations, the exact total wave function can be written as an expansion of a complete set of electronic functions, with the condition that the coefficients of this expansion are functions of the nuclear coordinates:

$$\Psi(r^e, r^n) = \sum_{i=1}^{\infty} \psi_i(r^e; r^n)\, \psi_{ni}(r^n) \tag{1.5}$$

In the adiabatic approximation, the form of the total wave function is limited to a single electronic surface, and it is assumed that the expansion (5-1) can be reduced to a single term [2], that is:

$$\Psi(r^e, r^n) \approx \psi_l(r^e; r^n)\, \psi_{nl}(r^n) \tag{1.6}$$

This assumption is the essence of the adiabatic approximation. Except for spatially degenerate wave functions, the first-order diagonal non-adiabatic coupling elements are zero. Finally, we arrive at the nuclear Schrödinger equation in the following form:

$$\left(\hat{T}_n + E_j(r^n) + U(r^n)\right)\psi_{nj}(r^n) = E_{tot}\psi_{nj}(r^n) \quad 1.7$$

The term $U(r^n)$ is known as the diagonal correction and is approximately smaller than $E_j(r^n)$ by a factor roughly equal to the ratio of the mass of the electron to the nucleus. This term usually depends on $r^n$ and varies slowly, so the shape of the potential energy surface is well determined by $E_j(r^n)$. In the adiabatic approximation, the diagonal correction is ignored, and the resulting equation takes the form of the ordinary nuclear Schrödinger equation, where the electronic energy acts as the local potential energy surface:

$$\left(\hat{T}_n + E_j(r^n)\right)\psi_{nj}(r^n) = E_{tot}\psi_{nj}(r^n) \quad 1.8$$

The diagonal Born-Oppenheimer correction (DBOC) can be easily evaluated, as it only involves the second derivative of the electronic wave function with respect to the nuclear coordinates, and thus has a close relation to the nuclear gradient and the second derivative of the energy [9]. Solving equation (1-8) for the nuclear wave function leads to energy levels for molecular vibrations and rotations, which in turn form the basis for many forms of spectroscopy, such as infrared, Raman, microwave, and others [9].

Sometimes, the BO approximation is mistaken for the clamped nucleus approximation, where the nuclear kinetic energy is entirely ignored in the molecular Hamiltonian. Similarly, the BO approximation is occasionally equated with the classical nucleus approximation, where nuclei are considered classical point particles.

However, molecular energies, even in the Born-Oppenheimer approximation, include contributions from the nuclear kinetic energy operator [10]. Thus, the clamped nucleus approximation can be regarded as the initial stage in the BO approximation. Born and Huang, or BH for short, developed a variational alternative to the BO approximation [11] and presented it in the sense we know today (equation 1-5).

## 1.4 Why Go Beyond the Born-Oppenheimer Approximation?

There are significant molecular phenomena that cannot be sufficiently analyzed using conventional methods based on the Born-Oppenheimer approximation. Generally, two major reasons compel us to go beyond the BO approximation and extend our framework to a non-adiabatic approach:

1. Achieving Spectroscopic Accuracy[5]: One of the most critical challenges in quantum chemistry is achieving spectroscopic accuracy. All the systems we study possess some degree of non-adiabaticity [12], which prevents us from achieving such accuracy. To describe these small effects, we need a suitable new formalism. The most common approach is to introduce some non-adiabatic corrections to the results obtained from separate electronic-nuclear Hamiltonian calculations.

[5] Indeed, spectroscopic accuracy varies across different types of spectroscopies. For example, in vibrational spectroscopy, spectroscopic accuracy is approximately equivalent to 1 $cm^{-1}$.

2. Inherently Non-Adiabatic Systems and Phenomena: There are systems and phenomena that are inherently non-adiabatic, where the BO approximation is ineffective from the outset. Specifically, we cannot describe systems that include strongly coupled particles (e.g., slight differences in particle masses) or coupled states (e.g., closely spaced electronic levels) within the adiabatic framework [13].

It is important to note that when methods beyond the Born-Oppenheimer approximation are employed, concepts such as molecular geometries become ambiguous, and energy levels no longer have unique definitions. In a quantum description, nuclei are delocalized, and "bond lengths" do not have unique definitions. The largest Born-Oppenheimer diagonal correction (DBOC) is expected for molecules containing hydrogen atoms since hydrogen has the lightest nucleus. The absolute magnitude of the DBOC for a water molecule is approximately 7 kcal/mol. These effects are expected to be much smaller for systems with heavier nuclei [9]. Proton-coupled electron transfer (PCET) reactions [14], [15] are examples of phenomena where there is strong coupling between electron and proton dynamics. Therefore, while we know the BO approximation is generally suitable for protons under normal conditions, it is no longer effective in situations such as PCET reactions. Another example includes atoms and molecules containing exotic particles such as positrons and muons, which have garnered increasing interest in recent decades. Given that these particles are lighter than atomic nuclei, using the BO approximation to separate the wave functions of these system components can introduce significant errors in calculations [16].

In addition to the above, there are other ancillary reasons for extending fully non-adiabatic computations: these calculations in fundamental physics represent a technological advancement for calculating the stationary states of atoms and molecules with very high accuracy. Although a non-adiabatic quantum theory is a natural starting point for quantum chemistry, it has historically been applied only to specific systems due to a lack of computational resources. However, with advancements in computational chemistry and increasing computational capabilities, there will be less emphasis on low-cost methods, and non-BO methods may be chosen for problems where high accuracy is desired [17]. Moreover, interest in non-adiabatic effects has been growing, inspired by the maturation of fields such as ultrafast spectroscopy and molecular dynamics. Experimental studies in these areas are more likely to encounter BO failure and observe non-adiabatic effects [17].

## 1.5 Exotic Atoms and Molecules

An exotic atom refers to a system where the nucleus of a hydrogen atom is replaced with another positively charged fundamental particle (such as a positive muon to form a muonium atom or a positron to form a positronium atom), or an electron in a regular atom is replaced with another negatively charged fundamental particle (such as a muon, pion, or antiproton), or even both simultaneously (such as in antihydrogen) [18].

Table 1-1 :Properties of particles. The mass mm refers to the rest mass of the particle.

| Particle | Symbol | Spin | Half-life (s) | Mass relative to e ($m/m_e$) | Energy (MeV) | Type |
|---|---|---|---|---|---|---|
| Electron | $e^-$ | 1/2 | $\infty$ | 1 | 0.511 | Lepton |
| Positive Muon | $\mu^+$ | 1/2 | $2.197 \times 10^{-6}$ | 207 | 105.7 | Lepton |
| Positive Pion | $\pi^+$ | 0 | $2.603 \times 10^{-8}$ | 273 | 139.6 | Meson |
| Positive Kaon | $K^+$ | 0 | $1.237 \times 10^{-8}$ | 966 | 493.6 | Meson |
| Proton | $p^+$ | 1/2 | $\infty$ | 1836 | 938.3 | Baryon |
| Sigma | $\Sigma^+$ | 1/2 | $1.48 \times 10^{-10}$ | 2343 | 1197.4 | Baryon |

When a negatively charged particle with a sufficient lifetime slows down and comes to rest in a material, it replaces an atomic electron and becomes bound in an atomic orbit. Exotic atoms formed in this way are named after the particles they contain, with the most common examples presented in Table 1-1. Depending on the particle type, we refer to muonic atoms, pionic atoms, kaonic atoms, antiprotonic atoms, or sometimes mesonic or hadronic atoms. Additionally, atoms formed by a positive particle and an electron (such as positronium or muonium) also belong to the family of exotic atoms. Due to their extensive applications in positron annihilation and muon spin research in solid-state physics and chemistry, they have been studied much more than exotic atoms with negatively charged particles.

In positronium, the positron and electron, each with the same mass, orbit around a common center of mass. This system is exactly similar to the hydrogen atom, except that the proton is approximately 1836 times heavier than the electron, giving the appearance that the latter orbits the former. By modifying the Bohr model of the hydrogen atom for positronium, we can show that the radius of the electron's ground state orbit in positronium is twice that of the corresponding orbit in the hydrogen atom. In muonium, a positive muon becomes the effective nucleus, and since the muon has a mass of about one-ninth that of the proton, the muonium atom has a larger radius than hydrogen.

Muons are unstable fundamental particles found in cosmic rays and can also be produced by particle accelerators. They have various applications in fundamental science and engineering. Scientific research using muons focuses on both the fundamental properties of this particle and the microscopic interactions (at the atomic level) between muons and surrounding particles such as nuclei, electrons, atoms, and molecules. Examples of research that can be conducted using muons include muon-catalyzed fusion, the use of muon spin probes to study the microscopic magnetic properties of materials, electron tagging to help understand the microscopic level of electron transfer in proteins, and non-destructive elemental analysis. Cosmic ray muons can even be used to study the internal structure of volcanoes [19].

Muonium is the bound atomic state of a positive muon and an electron, with a mass of 1/9 that of a regular hydrogen atom. This exotic atom is of significant interest in both physics and chemistry. In particle physics, it is used for testing fundamental symmetries of matter. In solid-state physics, it serves as a probe for measuring local magnetic fields. In chemistry, it is employed as a traceable lightweight hydrogen atom. From a field theory perspective, muonium is a bound state of two point-like leptons, making it an ideal system for studying quantum effects.

$\mu SR$ stands for Muon Spin Rotation, Relaxation, or Resonance. As mentioned above, positive muons are used to measure local magnetic fields. When polarized muons come to rest in a sample material with a magnetic field, they align along the field lines, and positrons from their decay are preferentially emitted along the muon's spin direction. This method uses the detection of emitted positrons to obtain information about the local magnetic fields in the system under study [20]. Muonium, also known as a lightweight radioisotope of hydrogen, can be considered the second radioisotope of hydrogen after tritium. It forms chemical bonds with unsaturated organic molecules and creates free radicals. The muon in these radicals acts as a radioactive probe of kinetic and structural properties [21]. Figure 1-1 shows how some of these radicals are formed [22].

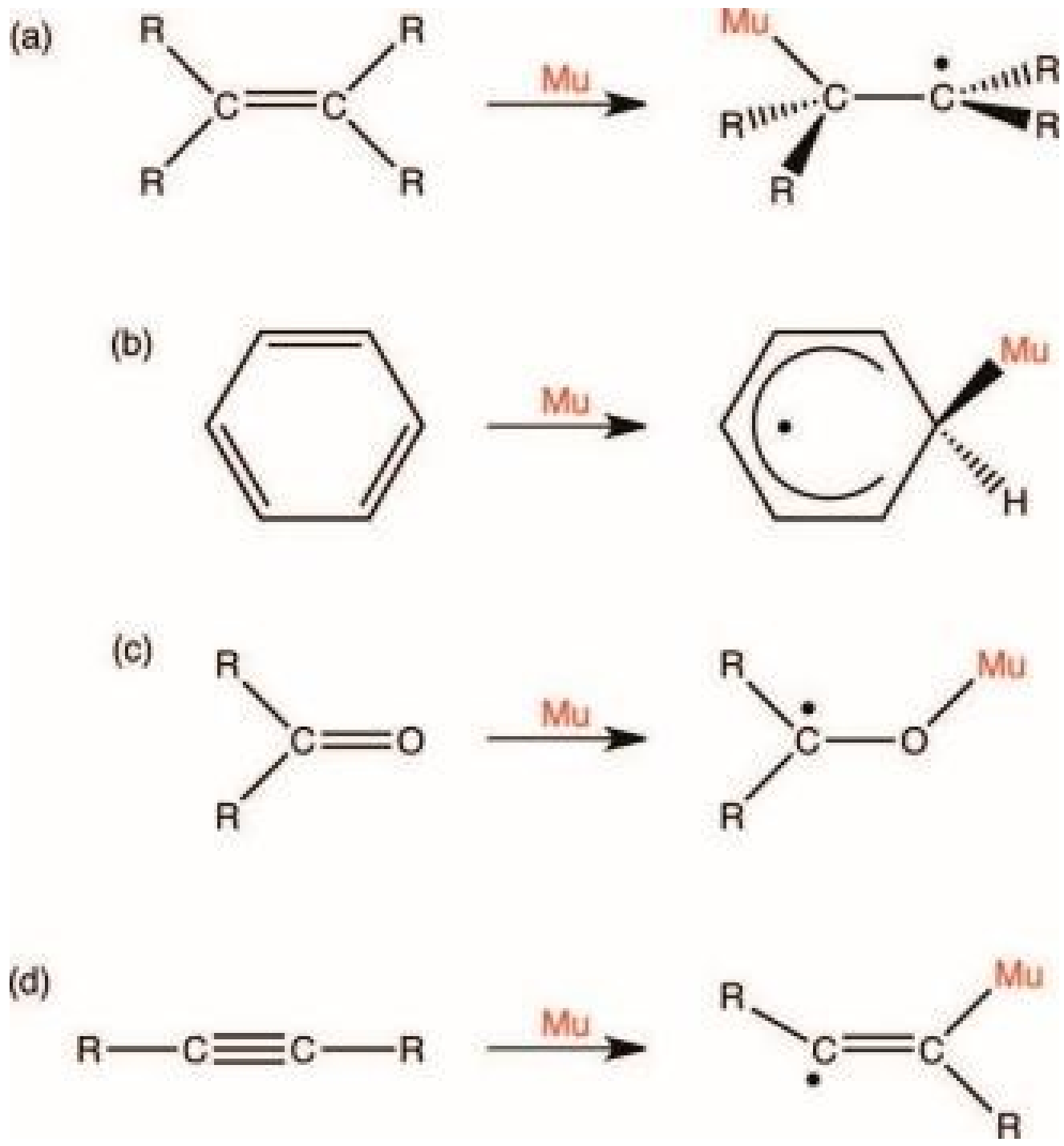

Figure 1-1: Formation of muoniated radicals; (a) alkyl radical; (b) cyclohexadienyl radical; (c) muonoxy-alkyl radical; (d) vinyl radical.

The equivalents of these radicals are also produced as a result of the reaction of hydrogen atoms with unsaturated organic molecules.

However, due to the mass difference between these two atoms, some properties of muonic bonds (such as vibrational frequency) differ from their hydrogen counterparts. This mass difference also causes the reaction rates of muonic and hydrogen radicals to differ, despite having the same mechanism.

Muoniated radicals, which can be detected by µSR\mu SR spectroscopy, provide a unique opportunity to obtain information about solids such as zeolites, solutions, or even unusual solvents like supercritical water [22]. Additionally, positive muons embedded in semiconductors with diamond and zinc sulfide structures often form paramagnetic muonium states, the characteristics of which can also be studied using the $\mu SR$ technique [23].

Positron attachment to molecules also has significant applications in material science and chemistry. When a positron comes into contact with a molecule, it annihilates with a molecular electron through a process called positron annihilation, resulting in gamma radiation. However, before this annihilation occurs, the positron can temporarily form a positronium atom, providing valuable information about the electronic structure of molecules [24]. Various experimental methods based on positron annihilation have become important tools for investigating the structure and properties of condensed matter, such as studying defects in solids [25]. Positron annihilation spectroscopy is especially suitable for studying vacancy defects in semiconductors. Combining advanced experimental and theoretical methods allows for precise identification of defects and the surrounding chemical environment. Charge states and defect levels in the energy gap can also be examined [26].

Positron Emission Tomography (PET) is a powerful metabolic imaging technique that provides high-quality images. Clinical PET

imaging is currently used in oncology, cardiology, cardiac surgery, and neurology [27].

Studies of positronium in a vacuum and its decay in the environment provide useful information about the structure of matter and the biological processes of living organisms on a nanoscale. Positronium in biological materials is sensitive to intermolecular and intramolecular structures and the metabolism of living organisms, from single cells to humans. This leads to new ideas for positronium imaging in medicine, utilizing the fact that 40% of positron annihilation through the production of positronium atoms during PET occurs inside the patient's body. A new generation of highly sensitive multi-photon PET systems opens fresh windows for clinical applications of positronium as a biomarker of tissue pathology. Figure 1-2 shows a schematic diagram of such a new PET spectroscopy [28].

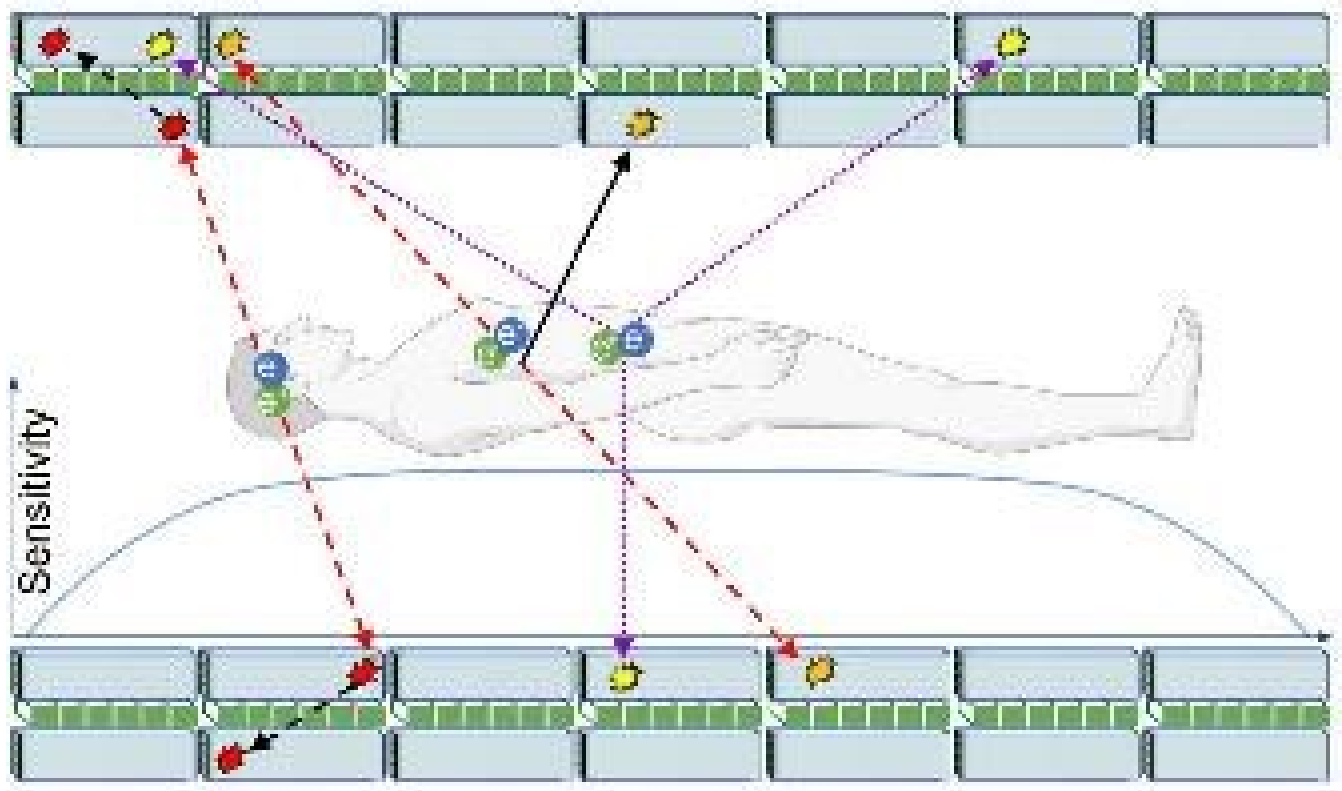

Figure 1-2: Schematic of a whole-body PET scanner for positronium imaging and quantum entanglement. This figure illustrates an axial cross-section of a tomograph design consisting of two detector layers. The red dashed arrows indicate sample responses originating from the positron annihilation process.

The spectrum of the hydrogen atom has been studied more precisely and comprehensively than any other atom. Similarly, the spectroscopy of exotic "atoms" reveals their characteristics and provides opportunities for testing physical theories. For instance, unlike the proton, which is made up of quarks, both the positron and muon are particles with no internal structure. This means that the energy state separations in these exotic atoms are determined solely by electromagnetic interactions. This makes such exotic atoms an ideal arena for testing Quantum Electrodynamics (QED), as demonstrated by recent research conducted by RIKEN Institute scientists using muonic neon [29]. All these exotic atoms, like the hydrogen atom, serve as laboratories for testing the efficiency of physical theories [30]. Research on exotic atoms is ongoing at all major and medium-energy accelerators where low-energy charged particles are produced [31]–[35].

## 1.6 Multi-Component Non-BO Ab Initio Methods

The mean-field description of electrons moving in the external electric field of clamped nuclei forms the basis of most models currently used in quantum chemistry. However, due to the presence of significant non-adiabatic systems and phenomena mentioned in sections 1-4 and 1-5, it becomes necessary to develop a more general non-BO theory that describes both electrons and nuclei in a mean-field approximation. Such a self-consistent field formulation provides a natural starting point for molecular structure theory (as opposed to electronic structure theory) and yields an initial electron-nuclear wave function reference for all hierarchies of models that include inter-particle correlation, such as Many-Body Perturbation Theory (MBPT), Configuration Interaction (CI), Coupled-Cluster (CC), and Multi-Component Self-Consistent Field (MC-SCF) [36].

A theory that describes all electrons and nuclei in a molecule simultaneously abandons the concept of the potential energy surface. This radical step enables geometry optimization in a non-iterative calculation and the calculation of dynamic processes (e.g., tunnelling of light atoms). However, hidden difficulties exist in this seemingly straightforward approach, including the invariance of the molecular Hamiltonian with respect to rotational and translational motions [37], as well as the lack of explicit dependence of wave functions (except explicitly correlated functions) on inter-particle distances [36].

In this section, we review methods that, from the start of calculations, consider at least one other component of the system in addition to electrons, simultaneously and quantum-mechanically. These methods, which can be termed Multi-Component (MC) non-BO methods, exhibit significant diversity compared to conventional methods in electronic structure theory. In these MC methods, single-particle wave functions for electrons and other quantum species are separately constructed using Gaussian-type basis sets. A notable feature of MC schemes is that nuclear quantum effects on electronic structure and molecular properties are directly obtained from single-point calculations rather than additional corrections based on previous adiabatic calculations [16].

In contrast to the uniformity of computational methods in the adiabatic approach (aiming to derive discrete eigenvalues of the electronic Hamiltonian spectrum), non-adiabatic ab initio computational strategies vary significantly [37]. Given the diversity of these methods, the following section will provide a brief overview of the most important ones that have yielded better results or have been further developed.

### 1.6.1 Overview

The first MC method was proposed by Thomas in papers from 1969 and 1970 [38], [39]. He proposed a non-adiabatic molecular orbital theory to simultaneously obtain electronic and protonic wave functions, where electrons and protons are treated quantum-mechanically while heavier nuclei are treated classically. In Thomas' method, all basis functions are placed at the origin on the clamped nucleus, creating a spherical symmetry that avoids the problem of rotational invariance. However, the main limitation of this method is the increasing computational complexity as the number of particles grows.

After Thomas, Adamowicz and colleagues [40] in 1991 proposed a non-BO computational method using explicitly correlated Gaussian (ECG) basis functions, which included the internal coordinates between nuclei and electrons. The ECG method is the most accurate and successful research program in non-BO calculations, based on initially separating the center of mass from the system and then employing the variational principle using trial wave functions that explicitly include rotational invariance [17], [41]–[43]. In this method, the n+1-particle problem is reduced to an n-pseudo particle problem, and by placing the reference particle (particle 1 with mass $M_1$ and usually the heaviest particle) at the center of the coordinate system, the Hamiltonian for internal motion is obtained as follows [43]:

$$\hat{H}_{ECG} = -\frac{1}{2}\left(\sum_{i}^{n}\frac{1}{m_i}\nabla_i^2 - \sum_{i\neq j}^{n}\frac{1}{M_1}\nabla_i'\nabla_j\right) + \sum_{i=1}^{n}\frac{Z_0 Z_i}{r_i} + \sum_{j>i}^{n}\frac{Z_i Z_j}{r_{ij}} \qquad 1.9$$

In this equation, $Z_0$ represents the charge of the reference particle, and $m_i$ is the reduced mass of the n-pseudo particle system given by $m_i = \frac{(M_1 M_{i+1})}{(M_1 + M_{i+1})}$ with charges $Z_i$. The challenge with the ECG approach is that the complexity of the integrals required to solve the problem increases rapidly with the number of particles, necessitating different coding for various n-values [44]. Additionally, since the ECG method treats system particles as pseudo-particles, previous computational strategies derived from electronic structure theory are not necessarily applicable. Therefore, other research groups have developed more practical, albeit less accurate, non-adiabatic methods to avoid these computational limitations. The goal of these methods is to develop non-adiabatic approaches that are analogous to familiar methods like Hartree-Fock, post-Hartree-Fock, and Density Functional Theory [37].

In 1998, the multi-component non-BO Hartree-Fock theory was introduced by Tachikawa, Nakai, and colleagues [45], [46]. Subsequently, in 2002, Nakai refined the initial method and introduced the Nuclear Orbital plus Molecular Orbital (NOMO) approach [47]. Nakai and his team described the Nuclear Orbital (NO) as a single-nucleus wave function, similar to the Molecular Orbital (MO) for single-electron wave functions. The total wave function is constructed by multiplying determinants made from NOs and MOs, and the coupled Hartree-Fock equations for NOs and MOs are derived within the single-particle approximation framework. The Configuration Interaction (CI) method applied to the NOMO theory yields not only electronically excited states but also vibrationally excited states of nuclei, while many-body effects in the NOMO theory are examined using Many-Body Perturbation Theory (MBPT) and Coupled-Cluster (CC) theory [48]. These studies revealed that electron-nucleus correlation energy has a larger absolute value compared to nucleus-nucleus correlation energy in the NOMO theory [49].

In the NOMO method, Gaussian-Type Functions (GTFs), which are commonly used as nuclear basis functions as well as electronic basis functions, can accurately describe a vibrational state. However, rotational and translational states are not reproduced well enough using GTFs. Therefore, accounting for rotational and translational motions is necessary to achieve high accuracy in the NOMO method [44]. Accordingly, Nakai proposed a scheme based on Sutcliffe's idea [50] to completely separate the translational motion component from the total Hamiltonian and obtain a Translation-Free Hamiltonian ($\hat{H}_{TF}$) [49]. This Hamiltonian is simply obtained by subtracting the center-of-mass kinetic energy operator (the translational Hamiltonian) from the total Hamiltonian (2-1). For this, we need the center-of-mass kinetic energy operator of nuclei, derived from a coordinate transformation [49]:

$$\hat{T}_{COM} = -\frac{1}{2M}\sum_{\mu}\nabla(x_\mu)^2 - \frac{1}{M}\sum_{v>\mu}\nabla(x_\mu)\cdot\nabla(x_v) \quad 1.10$$

where $M$ represents the total mass of the nuclei, and the first and second terms are the single-particle and two-particle operators, respectively. By subtracting this operator from the total Hamiltonian (1-2), which is contaminated by translational motion ($\hat{H}_{TRC}$), we obtain the translation-free Hamiltonian ($\hat{H}_{TF}$).

$$\hat{H}_{TF} = \hat{H}_{TRC} - \hat{T}_{CM} = \hat{T}^e + \hat{T}^p + \hat{V}^{ee} + \hat{V}^{ep} + \hat{V}^{pp} - \hat{T}_{COM} \quad 1.11$$

On the other hand, we can subtract $\hat{T}_{COM}$ separately from each of the above operators and redefine them as follows:

$$\hat{H}_{TF} = \hat{T}^e_{TF} + \hat{T}^p_{TF} + \hat{V}^{ee}_{TF} + \hat{V}^{ep}_{TF} + \hat{V}^{pp}_{TF} \quad 1.12$$

where their explicit form is as follows:

$$\hat{T}_{TF}^{e} = -\sum_{i}^{N}\frac{1}{2}\left(1-\frac{1}{M}\right)\nabla_i^2 = \sum_{i}^{N}\hat{t}_{TF}^{e} \qquad 1.13$$

$$\hat{T}_{TF}^{p} = -\sum_{i'}^{N'}\frac{1}{2}\left(\frac{1}{m_p}-\frac{1}{M}\right)\nabla_{i'}^2 = \sum_{i'}^{N'}\hat{t}_{TF}^{p} \qquad 1.14$$

$$\hat{V}_{TF}^{ee} = \sum_{i}^{N}\sum_{j>i}^{N}(\frac{1}{r_{ij}}+\frac{1}{M}\nabla_i.\nabla_j) \qquad 1.15$$

$$\hat{V}_{TF}^{ep} = \sum_{i}^{N}\sum_{i'}^{N'}(-\frac{Z_{i'}}{r_{ii'}}+\frac{1}{M}\nabla_i.\nabla_{i'}) \qquad 1.16$$

$$\hat{V}_{TF}^{pp} = \sum_{i'}^{N'}\sum_{j'>i'}^{N'}(\frac{Z_{i'}Z_{j'}}{r_{i'j'}}+\frac{1}{M}\nabla_{i'}.\nabla_{i'}) \qquad 1.17$$

where $m_p$ denotes the mass of nuclei (or positive particles). Equations (1-13) and (1-14) represent the kinetic energy of electrons and nuclei, respectively. Equations (1-15) and (1-17) indicate the Coulomb repulsion potential between electrons and between nuclei, respectively. Finally, Equation (1-16) shows the Coulomb attraction between electrons and nuclei. Using $\hat{H}_{TF}$, the accuracy of an orbital approach such as the NOMO method is significantly improved at the HF, CI, MBPT, and CC levels [51].

Nakai and his colleagues, in the next step and to remove the contribution of rotational motions, subtracted the rotational energy operator from $\hat{H}_{TF}$ and introduced a new Hamiltonian called the Translation- and Rotation-free Hamiltonian ($\hat{H}_{TRF}$) [49]. Based on these modifications, the initial total Hamiltonian, contaminated by translational motion, is referred to as $\hat{H}_{TRC}$, and the corresponding formulation is called TRC-NOMO. Similarly, the formulations related to $\hat{H}_{TF}$ and $\hat{H}_{TRF}$ are referred to as TF-NOMO and TRF-NOMO, respectively. It is noteworthy that in all versions of NOMO-HF, the total wave function is defined as the product of the Slater determinants of the electronic and nuclear wave functions.

Since the NOMO method uses Gaussian-Type Functions (GTFs) as both electronic and nuclear basis sets, efficient algorithms can be applied to evaluate the two-particle integrals. However, this method has limited accuracy, especially when used to evaluate nucleus-dependent properties such as zero-point energies. The electron-nucleus correlation in the NOMO method is not efficiently recoverable by Configuration Interaction (CI), Many-Body Perturbation Theory (MBPT), or Coupled-Cluster (CC) methods [52].

Finally, in 2002, Hammes-Schiffer and her colleagues proposed a hybrid method to simplify multi-component ab initio methods. They suggested that some nuclei in a molecule could be treated as clamped, and only the lightest nuclei (e.g., protons, deuterium) should be treated quantum-mechanically. Therefore, this method deals with three types of entities: electrons, quantum nuclei, and clamped nuclei. This approach is called the Nuclear-Electronic Orbital (NEO) method, and its Hamiltonian in atomic units is given as follows [53]:

$$\begin{aligned}\hat{H}_{NEO} = {} & \hat{T}^{e}_{NEO} + \hat{V}^{e}_{ext} + \hat{T}^{p}_{NEO} + \hat{V}^{p}_{ext} + \hat{V}^{ee}_{NEO} + \hat{V}^{ep}_{NEO} + \hat{V}^{pp}_{NEO} \\ = {} & -\sum_{i}^{N} h(i) - \sum_{i'}^{N'} h'(i') \\ & + \sum_{i}^{N}\sum_{j>i}^{N}\frac{1}{r_{ij}} - \sum_{i}^{N}\sum_{i'}^{N'}\frac{Z_{i'}}{r_{ii'}} + \sum_{i'}^{N'}\sum_{j'>i'}^{N'}\frac{Z_{i'}Z_{j'}}{r_{i'j'}}\end{aligned} \quad 1.18$$

where the first and second terms, considering the external potential that is typically interaction with clamped nuclei ($N_c$), are defined as follows:

$$h(i) = -\frac{1}{2}\nabla_i^2 - \sum_{A}^{N_c} \frac{Z_A}{r_{iA}} \qquad 1.19$$

$$h'\left(i'\right) = -\frac{1}{2M_{i'}}\nabla_{i'}^2 + \sum_{A}^{N_c} \frac{Z_A Z_{i'}}{r_{i'A}} \qquad 1.20$$

In this method, due to the presence of clamped nuclei, the issue of the total Hamiltonian's invariance with respect to translational and rotational motions is resolved. Therefore, this method is suitable for systems where only certain nuclei exhibit non-adiabatic behavior and need to be considered coupled with electrons from the outset. The mathematical details of this method will be discussed in Section 1-6-2, as this approach is directly used in this study.

Since the NEO method does not face issues related to translational and rotational invariance, the primary focus has been on improving the wave function. Initially, the Hartree-Fock wave function was extended to a multi-configurational version [54], and the following improvements were made: adding explicit electron-nucleus correlations using geminal functions to the total wave function, developing the Nuclear-Electronic Orbital Explicitly Correlated Hartree-Fock (NEO-XCHF) method [55] and extending it to multi-electron systems [56], introducing a new wave function called NEO-XCHF2 [57], developing the Nuclear-Electronic Orbital Reduced Explicitly Correlated Hartree-Fock (NEO-RXCHF) method [58], and extending it to open-shell electronic systems [59]. Reviews of the NEO method are available in [60] and [61].

Therefore, the ECG, NOMO, and NEO methods form the main approaches to multi-component non-BO ab initio methods, and other proposed methods have not been developed as extensively as these. Specifically, the NEO and NOMO methods have been developed for recovering interspecies correlations within the MC framework, starting from their own Hartree-Fock methods and extending to post-Hartree-Fock methods like Configuration Interaction, Møller–Plesset (MP) Many-Body Perturbation Theory, and Coupled-Cluster theory [61]. Additionally, Multi-Component Kohn–Sham Density Functional Theory (to be discussed in Section 1-7) has been developed as an alternative to wave function-based methods for recovering inter-particle correlations.

### 1.6.2 Multi-Component Hartree-Fock Equations

In this section, the principal equations of multi-component Hartree-Fock (MC-HF) methods are presented, with detailed derivations reported in the appendix A. In this study, the introduction of the Exotic Harmonium Model (hereafter EHM) allows for the exact variational solution of the Schrödinger equation. However, since the Hartree-Fock method is a mean-field theory, comparing it with exact solutions will reveal the correlation effects.

Since we aim to simultaneously and equally treat both electrons and nuclei as quantum particles, we need to propose a wave function that satisfies these symmetry properties of particle statistics. The most straightforward proposal for a composite two-component total wave function, using electronic and nuclear Slater determinants (assuming the nuclei are fermions), is as follows:

$$\Psi_{tot}(x^e, x^p) = \Phi_0^e(x^e)\,\Phi_0^p(x^p) \qquad 1.21$$

where

$$\Phi_0^e(x_1^e, x_2^e, \ldots, x_N^e) = (N!)^{-1/2} \begin{vmatrix} \chi_1^e(x_1^e) & \cdots & \chi_k^e(x_1^e) \\ \vdots & \ddots & \vdots \\ \chi_1^e(x_N^e) & \cdots & \chi_k^e(x_N^e) \end{vmatrix} \qquad 1.22$$

$$\Phi_0^p\left(x_1^p, x_2^p, \ldots, x_{N'}^p\right) = (N'!)^{-1/2} \begin{vmatrix} \chi_1^p(x_1^p) & \cdots & \chi_k^p(x_1^p) \\ \vdots & \ddots & \vdots \\ \chi_1^p(x_{N'}^p) & \cdots & \chi_k^p(x_{N'}^p) \end{vmatrix} \qquad 1.23$$

The notation conventions in the above equations are as follows:

$\Phi_0^e$: total electron wave function or electron Slater determinant

$\Phi_0^p$: total nuclear wave function or nuclear Slater determinant

$x^e$: set of spatial and electron spin coordinates

$x^p$: the set of spatial and nuclear spin coordinates

$k$: number of electron spin-orbitals

$k'$: number of nuclear spin-orbitals

$N$: number of electrons

$N'$: number of nuclei (or, in general, PCPs)

Therefore:

$$\chi^e(x^e) = \begin{cases} \psi^e(r^e)\alpha(\omega) \\ \text{or} \\ \psi^e(r^e)\beta(\omega) \end{cases} \qquad 1.24$$

$$\chi^p(x^p) = \begin{cases} \psi^p(r^p)\alpha(\omega) \\ \text{or} \\ \psi^p(r^p)\beta(\omega) \end{cases} \qquad 1.25$$

Similar to electronic structure theory, the total energy and spatial orbitals are obtained by simultaneously optimizing the system's energy with respect to both electronic and nuclear orbitals. To achieve this goal, we need to evaluate the expectation value of the energy, which has the general form:

$$E_0 = \langle \Psi_0 | \hat{H}_{NEO} | \Psi_0 \rangle \qquad 1.26$$

By minimizing the energy expression (1-26) and after applying certain mathematical considerations, we arrive at the final form of the coupled electron-nuclear Hartree-Fock equations for the closed-shell case:

$$f^e(1)\, \psi_i^e(1) = \varepsilon_i\, \psi_i^e(1) \qquad 1.27$$

$$f^p(1)\, \psi_{i'}^p(1) = \varepsilon_{i'}\, \psi_{i'}^p(1) \qquad 1.28$$

where $f^e$ and $f^p$ are the electronic and nuclear Fock operators, respectively, and are equal to the expressions within the brackets in the following equations:

$$\left[ h^e(1) + \sum_j^{\frac{N}{2}} \left( 2J_{NEOj}^e(1) - K_{NEOj}^e(1) \right) + \sum_{i'}^{N'} J_{NEOi'}^e \right] \psi_i^e(1) = \sum_j^{\frac{N}{2}} \varepsilon_{ij}\, \psi_j^e(1) \qquad 1.29$$

$$\left[ h^p(1) + \sum_{j'}^{N'} \left( J_{NEOj'}^p(1) - K_{TFj'}^p(1) \right) + 2 \sum_i^{\frac{N}{2}} J_{NEOi}^p(1) \right] \psi_{i'}^p(1) = \sum_{j'}^{N'} \varepsilon_{i'j'}\, \psi_{j'}^e(1) \qquad 1.30$$

where the Coulomb operators are as follows:

$$J^{e}_{NEOj}(1)\psi^{e}_{i}(1) = \left[\int dr^{e}_{2}\,\psi^{e*}_{j}(2)\,\hat{V}^{ee}_{NEO}\,\psi^{e}_{j}(2)\right]\psi^{e}_{i}(1) \quad 1.31$$

$$J^{p}_{NEOj'}(1)\psi^{p}_{i'}(1) = \left[\int dr^{p}_{2}\,\psi^{p*}_{j'}(2)\,\hat{V}^{pp}_{NEO}\,\psi^{p}_{j'}(2)\right]\psi^{p}_{i'}(1) \quad 1.32$$

$$J^{e}_{NEOi'}(1)\psi^{e}_{i}(1) = \left[\int dr^{p}_{2}\,\psi^{p*}_{i'}(2)\,\hat{V}^{pe}_{NEO}\,\psi^{p}_{i'}(2)\right]\psi^{e}_{i}(1) \quad 1.33$$

$$J^{p}_{NEOi}(1)\psi^{p}_{i'}(1) = \left[\int dr^{e}_{2}\,\psi^{e*}_{i}(2)\,\hat{V}^{pe}_{NEO}\,\psi^{e}_{i}(2)\right]\psi^{p}_{i'}(1) \quad 1.34$$

and the exchange operators are also defined as follows:

$$K^{e}_{NEOj}(1)\psi^{e}_{i}(1) = \left[\int dr^{e}_{2}\,\psi^{e*}_{j}(2)\,\hat{V}^{ee}_{NEO}\,\psi^{e}_{i}(2)\right]\psi^{e}_{j}(1) \quad 1.35$$

$$K^{p}_{NEOj'}(1)\psi^{p}_{i'}(1) = \left[\int dr^{p}_{2}\,\psi^{p*}_{j'}(2)\,\hat{V}^{pp}_{NEO}\,\psi^{p}_{i'}(2)\right]\psi^{p}_{j'}(1) \quad 1.36$$

The nuclear and electronic orbitals are expanded using Gaussian-type basis functions ($\phi^{e}_{v}$). This approach transforms the Hartree-Fock integro-differential equations into matrix equations, which are easier to solve:

$$\psi^{e}_{i}(1) = \sum_{v}^{B} c^{e}_{vi}\phi^{e}_{v}(1) \quad 1.37$$

$$\psi^{p}_{i'}(1) = \sum_{v'}^{B'} c^{p}_{v'i'}\phi^{p}_{v'}(1) \quad 1.38$$

where $B$ and $B'$ represent the number of nuclear and electronic basis functions, respectively. By substituting the above expansions into the Hartree-Fock equations (1-27) and (1-28), we arrive at the algebraic coupled Roothaan-Hall-Hartree-Fock equations:

$$\sum_{v}^{B} F_{\mu v}^{e} c_{vi}^{e} = \varepsilon_i \sum_{v}^{B} c_{vi}^{e} S_{\mu v}^{e} \qquad 1.39$$

$$\sum_{v'}^{B'} F_{\mu' v'}^{p} c_{v' i'}^{p} = \varepsilon_{i'} \sum_{v'}^{B'} c_{v' i'}^{p} S_{\mu' v'}^{p} \qquad 1.40$$

In equation (1-39), which is written for electrons, $c$ and $\varepsilon$ are $B \times B$ square matrices (similarly for nuclear equation 1-40) and overlap matrices are defined as follows:

$$S_{\mu v}^{e} = \int dr_1^e \ \phi_{\mu}^{e*}(1) \ \phi_{v}^{e}(1) \qquad 1.41$$

$$S_{\mu' v'}^{p} = \int dr_1^p \, \phi_{\mu'}^{p*}(1) \ \phi_{v'}^{p}(1) \qquad 1.42$$

And the Fock matrices are defined as follows:

$$F_{\mu v}^{e} = \int dr_1^e \ \phi_{\mu}^{e*}(1) \, f^e(1) \ \phi_{v}^{e}(1) \qquad 1.43$$

$$F_{\mu' v'}^{p} = \int dr_1^p \, \phi_{\mu'}^{p*}(1) \, f^p(1) \ \phi_{v'}^{p}(1) \qquad 1.44$$

The Fock matrix can be defined using the expansion of the basis function. For this purpose, new quantities known as density matrices need to be defined.

## 1.7 Multi-Component Non-BO Density Functional Theory

The non-BO versions of Density Functional Theory were first introduced by Parr [62] and colleagues in 1982, and later by Shigeta [63] and colleagues in 1998. Subsequently, in 2001, Gross and his collaborator published the Multi-Component Density Functional

Theory (MC-DFT) [64]. The first development of NEO-DFT dates back to 2007 when Hammes-Schiffer's group incorporated electron-electron correlation into the NEO framework, hence naming it NEO-DFT(ee) [65]. In 2009, Hammes-Schiffer's group took the initial step towards obtaining an electron-proton correlation functional by introducing exact universal functional features in a multi-component density functional theory [66], and their development continued over the following years.

In 2011, the second electron-proton correlation functional, known as EPC2, was introduced within the NEO-DFT framework [67]. In 2012, the previous functional was improved using the adiabatic connection formula in multi-component DFT and was presented as EPC2-KE [68]. A significant step was taken in 2017 with the introduction of an efficient functional called epc17 [69]. In the same year, a parameter-improved version of epc17 was introduced, named epc17-2, and the original functional was renamed epc17-1 [70]. A year later, in 2018, two more functionals, epc18-1 and epc18-2, were introduced, which performed similarly to the 2017 versions [71]. In 2019, another improved functional named epc19 was introduced, which depended on both density and density gradients [72].

A significant challenge in implementing multi-component Kohn-Sham DFT is designing accurate and practical functionals for electron-proton correlation. When electron-proton correlation is ignored, proton densities become highly localized, leading to non-physical results for dependent properties. Early attempts by Hammes-Schiffer's group to estimate electron-proton correlation functionals by analyzing the relationship between electron-proton pair density and the total wave function or specific orbitals improved proton densities, but were computationally expensive or not easily applicable to multi-particle systems. Additionally, these efforts treated functionals as post-self-consistent field energy corrections, which did not affect highly localized proton densities and, thus, were not useful for describing proton-density-based properties [69]. In 2022, the first functional for examining electron-positron correlation was proposed as part of developing electron and positive particle correlation functionals [73].

An important advantage of the NEO-DFT method is its low computational cost, making it practical for a wide range of molecular systems. However, a drawback of the NEO-DFT method is the lack of a clear path toward systematic functional improvement [74]. Unlike the NEO-DFT approach, wave function-based methods within the NEO framework are inherently improvable and free of parameters. Some of these methods include NEO-MP2 [75], [76], NEO-CI [53], and NEO-CC [74].

### 1.7.1 Equations

In this section, the final equations of the multi-component density functional theory (MC-DFT) are presented. Since this dissertation employs the inversion of two-component Kohn-Sham equations (rather than solving them self-consistently), only the main equations are provided without detailing their derivations.

MC-DFT starts with the Hohenberg-Kohn theorems, similar to its single-component counterpart. However, the form of these theorems is not obvious for different conditions and systems and must be redefined. Nevertheless, the essence of the two theorems remains the same across various DFT versions: the first theorem pertains to the existence theorem, and the second theorem proposes the variational principle in terms of single-particle densities. In a multi-component system, quantities for each component are defined separately. For example, in a two-component system consisting of electrons and positively charged particles (hereafter PCP), we deal with two single-particle densities $\rho_e(\boldsymbol{r_e})$ and $\rho_p(\boldsymbol{r_p})$ and their respective external potentials $v_e^{ext}(\boldsymbol{r_e})$ and v$v_p^{ext}(\boldsymbol{r_p})$.

The first Hohenberg-Kohn theorem for multi-component systems states that the ground state of a multi-component system with a given inter-particle interaction is a unique functional of all single-particle densities and the mass ratio of the particles. By integrating the single-particle densities, we can determine the number of particles for each component. Using the mass ratios, we can construct their kinetic energy operators, and due to the one-to-one correspondence, we can derive the external potential for each component from its single-particle density.

In a two-component system consisting of an arbitrary number of electrons and one type of PCP, the single-particle densities are defined as follows:

$$\rho_e(r_e) = N \sum_{Spin} \int \dots \int \Psi^*(\boldsymbol{r}_e^1,\boldsymbol{r}_e^2, \dots, \boldsymbol{r}_e^N,\boldsymbol{r}_p^1,\boldsymbol{r}_p^2, \dots, \boldsymbol{r}_p^{N'})\, \Psi(\boldsymbol{r}_e^1,\boldsymbol{r}_e^2, \dots, \boldsymbol{r}_e^N,\boldsymbol{r}_p^1,\boldsymbol{r}_p^2, \dots, \boldsymbol{r}_p^{N'})\, d\boldsymbol{r}_e^2\, d\boldsymbol{r}_e^3 \dots d\boldsymbol{r}_e^N\, d\boldsymbol{r}_p^1 d\boldsymbol{r}_p^2 \dots d\boldsymbol{r}_p^{N'} \quad 1.45$$

$$\rho_p(r_p) = N' \sum_{Spin} \int \dots \int \Psi^*(\boldsymbol{r}_e^1,\boldsymbol{r}_e^2, \dots, \boldsymbol{r}_e^N,\boldsymbol{r}_p^1,\boldsymbol{r}_p^2, \dots, \boldsymbol{r}_p^{N'})\, \Psi(\boldsymbol{r}_e^1,\boldsymbol{r}_e^2, \dots, \boldsymbol{r}_e^N,\boldsymbol{r}_p^1,\boldsymbol{r}_p^2, \dots, \boldsymbol{r}_p^{N'})\, d\boldsymbol{r}_e^1\, d\boldsymbol{r}_e^2 \dots d\boldsymbol{r}_e^N\, d\boldsymbol{r}_p^2 d\boldsymbol{r}_p^3 \dots d\boldsymbol{r}_p^{N'} \quad 1.46$$

The method for calculating these single-particle densities based on the system model's wave function is discussed in detail in Section 2-6.

The second Hohenberg-Kohn theorem for multi-component systems describes a constrained search process with respect to the components'

densities, leading to the determination of the wave functions, densities, and ground state energy of the system [77]:

$$E_0 = \underset{\rho_e,\rho_p}{\mathbf{Min}}\, E[\rho_e,\rho_p] = \underset{\Psi\to\rho_e,\rho_p}{\mathbf{Min}} \langle\Psi|\hat{H}|\Psi\rangle \qquad 1.47$$

The energy functional can be written as follows:

$$\begin{aligned} E[\rho_e,\rho_p] = \int \rho_e(\boldsymbol{r_e})\, v_e^{ext}(\boldsymbol{r_e})\, d\boldsymbol{r}_e \\ + \int \rho_p(\boldsymbol{r_p})\, v_p^{ext}(\boldsymbol{r_p})\, d\boldsymbol{r}_p + F_{HK}[\rho_e,\rho_p] \end{aligned} \qquad 1.48$$

where $v^{ext}$ represents the external potentials, which often but not always refer to the field caused by clamped nuclei, and $F_{HK}[\rho_e,\rho_p]$, which is equivalent to the universal functional in single-component DFT (but here, due to its implicit dependence on the masses of the components, it is not universal), is defined as follows:

$$\begin{aligned} F_{HK}[\rho_e,\rho_p] = \left(T_e[\rho_e,\rho_p] + T_p[\rho_e,\rho_p]\right) \\ + \left(V_{ep}[\rho_e,\rho_p] + V_{ee}[\rho_e,\rho_p] + V_{pp}[\rho_e,\rho_p]\right) \end{aligned} \qquad 1.49$$

where $T_e$ and $T_p$ represent the expectation values of the kinetic energies of electrons and PCPs, respectively, using the wave function (1-1). Additionally, $V_{ep}$, $V_{ee}$, and $V_{pp}$ correspond to the total interaction energies of electron-positive particle, electron-electron, and positive particle-positive particle interactions, respectively, which will be discussed in detail in Chapter 3.

The Hohenberg-Kohn theorems and the formalism of Density Functional Theory primarily apply to the ground state; however, researchers have recently extended it to ensemble states (EDFT) to explore excited states [78].

After defining the multi-component Hohenberg-Kohn theorems, the next step is to define multi-component Kohn-Sham systems, which, like their single-component counterparts, are the practical equations of computational DFT. Multi-component Kohn-Sham systems are reference, non-interacting systems whose single-particle densities for each component match those of the corresponding component in the actual interacting system. The DFT equations are then solved based on the orbitals assigned to the particles in these reference systems (Kohn-Sham orbitals).

The total wave function of the non-interacting system is also defined as the product of the Kohn-Sham determinants of electrons and PCPs:

$$\Psi^{ks}(\boldsymbol{r}_e,\boldsymbol{r}_p) = \Phi_e^{ks}(\boldsymbol{r}_e)\Phi_p^{ks}(\boldsymbol{r}_p) \qquad 1.50$$

where $\Phi_e^{ks}$ and $\Phi_p^{ks}$ are obtained using the Kohn-Sham orbitals, similar to equations (1-22) and (1-23). For a two-component two-particle system, the above equation reduces to:

$$\Psi^{ks}(\boldsymbol{r}_e,\boldsymbol{r}_p) = \phi_e^{ks}(\boldsymbol{r}_e)\phi_p^{ks}(\boldsymbol{r}_p) \qquad 1.51$$

The right-hand side of the above equation represents the electronic and positive particle Kohn-Sham orbitals, respectively, and their orthonormality condition is expressed as follows:

$$\int \phi_i^{ks^*}(1)\phi_j^{ks}(1)\, dr_1 = \left[\phi_i^{ks}\middle|\phi_j^{ks}\right] = \delta_{ij} \qquad 1.52$$

$$\int \phi_{i'}^{ks^*}(1)\phi_{j'}^{ks}(1) dr_1' = \left[\phi_{i'}^{ks}\middle|\phi_{j'}^{ks}\right] = \delta_{i'j'} \qquad 1.53$$

The electron and positive particle densities, expressed in terms of Kohn-Sham orbitals for a two-component system, are defined as follows using equations (1-45) and (1-46):

$$\rho_e(\boldsymbol{r}_e) = \sum_i n_i \left|\phi_i^{ks}(\boldsymbol{r}_e)\right|^2 \qquad 1.54$$

$$\rho_p(\boldsymbol{r}_p) = \sum_{i'} n_{i'} \left|\phi_{i'}^{ks}(\boldsymbol{r}_p)\right|^2 \qquad 1.55$$

where $n_i$ and $n_{i'}$ are the orbital occupation numbers. The kinetic energies of the Kohn-Sham system are obtained from the following equations:

$$T_e^s[\rho_e] = \left\langle \Psi^{ks*}(\boldsymbol{r}_e,\boldsymbol{r}_p) \middle| \Sigma_i^N \left(-\frac{1}{2\,m_e}\nabla_{i,e}^2\right) \middle| \Psi^{ks}(\boldsymbol{r}_e,\boldsymbol{r}_p) \right\rangle \qquad 1.56$$

$$T_p^s\left[\rho_p\right] = \left\langle \Psi^{ks*}(\boldsymbol{r}_e,\boldsymbol{r}_p) \middle| \Sigma_{i'}^{N'} \left(-\frac{1}{2\,m_p}\nabla_{i',p}^2\right) \middle| \Psi^{ks}(\boldsymbol{r}_e,\boldsymbol{r}_p) \right\rangle \qquad 1.57$$

To derive the two-component Kohn-Sham equations, we must first (similar to the single-component DFT process) write the ground state energy functional expression of a two-component system in terms of the components' densities. $F\left[\rho_e,\rho_p\right]$ in the Kohn-Sham method (Equation 1-48) can be defined as follows:

$$\begin{aligned} F\left[\rho_e,\rho_p\right] = {} & \left(T_e^s[\rho_e] + T_p^s\left[\rho_p\right]\right) \\ & + \left(J_{ee}[\rho_e] + J_{pp}\left[\rho_p\right] + J_{ep}\left[\rho_e,\rho_p\right]\right) \\ & + \left(E_{exc}[\rho_e] + E_{pxc}\left[\rho_p\right] + E_{epc}\left[\rho_e,\rho_p\right]\right) \end{aligned} \qquad 1.58$$

The third parenthesis in the above expression includes the exchange-correlation energy of electrons ($E_{exc}$), the exchange-correlation energy of PCPs ($E_{pxc}$), and the electron-PCP correlation energy ($E_{epc}$). The definitions of these three energies are as follows:

$$E_{epc}\left[\rho_e,\rho_p\right] = V_{ep}\left[\rho_e,\rho_p\right] - J_{ep}\left[\rho_e,\rho_p\right] \quad 1.59$$

$$E_{exc}[\rho_e] = \left(T_e\left[\rho_e,\rho_p\right] - T_e^s[\rho_e]\right) + \left(V_{ee}\left[\rho_e,\rho_p\right] - J_{ee}[\rho_e]\right) \quad 1.60$$

$$E_{pxc}\left[\rho_p\right] = \left(T_p\left[\rho_e,\rho_p\right] - T_p^s\left[\rho_p\right]\right) + \left(V_{pp}\left[\rho_e,\rho_p\right] - J_{pp}\left[\rho_p\right]\right) \quad 1.61$$

And their classical parts are defined as follows:

$$J_{ep}\left[\rho_e,\rho_p\right] = -\iint \frac{\rho_e(r_1^e)\rho_p(r_1^p)}{|\boldsymbol{r}_1^e - \boldsymbol{r}_1^p|} d\boldsymbol{r}_1^e d\boldsymbol{r}_1^p \quad 1.62$$

$$J_{ee}[\rho_e] = \frac{1}{2}\iint \frac{\rho_e(r_1^e)\rho_e(r_2^e)}{|\boldsymbol{r}_1^e - \boldsymbol{r}_2^e|} d\boldsymbol{r}_1^e d\boldsymbol{r}_2^e \quad 1.63$$

$$J_{pp}\left[\rho_p\right] = \frac{1}{2}\iint \frac{\rho_p(r_1^p)\rho_p(r_2^p)}{|\boldsymbol{r}_1^p - \boldsymbol{r}_2^p|} d\boldsymbol{r}_1^p d\boldsymbol{r}_2^p \quad 1.64$$

Ultimately, by minimizing the energy functional expression with respect to the Kohn-Sham orbitals, the coupled electron and PCP Kohn-Sham equations are obtained as follows:

$$\left(-\frac{1}{2m_e}\nabla_e^2 + V_{\text{eff}}^e(\boldsymbol{r}_e)\right)\phi_i^{ks} = \varepsilon_i^e\phi_i^{ks} \quad 1.65$$

$$\left(-\frac{1}{2m_p}\nabla_p^2 + V_{\text{eff}}^p(\boldsymbol{r}_p)\right)\phi_{i'}^{ks} = \varepsilon_{i'}^p\phi_{i'}^{ks} \quad 1.66$$

where $V_{\text{eff}}^e$ and $V_{\text{eff}}^p$ are called the effective Kohn-Sham potentials, and are respectively equal to the sum of the external potential, Coulomb potential, single-particle exchange-correlation, and two-particle correlation potentials as follows:

$$V_{\text{eff}}^e(\boldsymbol{r}_e) = v_e^{ext}(\boldsymbol{r}_e) + v_e^{J_{ee}}(\boldsymbol{r}_e) + v_e^{J_{ep}}(\boldsymbol{r}_e) + v_e^{xc}(\boldsymbol{r}_e) + v_e^{epc}(\boldsymbol{r}_e) \quad 1.67$$

$$V_{\text{eff}}^{p}(\boldsymbol{r}_p) = v_p^{ext}(\boldsymbol{r}_p) + v_p^{J_{pp}}(\boldsymbol{r}_p) + v_p^{J_{ep}}(\boldsymbol{r}_p) + v_p^{xc}(\boldsymbol{r}_p) + v_p^{epc}(\boldsymbol{r}_p) \quad 1.68$$

The explicit form of these potentials will be presented when calculating the correlation potential (Section 5-2-2) for the model under consideration. In general, the above equations are provided for a generic two-component system (with an arbitrary number of particles), and the equations for the EHM will be presented in the relevant sections (Sections 3-3-4 and 5-2-2).

## 1.8 Challenges and Motivation

To investigate the specific systems discussed in Section 1-4, we need computational methods beyond the Born-Oppenheimer approximation. However, the orbital-based examples of these methods introduced in Section 1-6 have fundamental issues. Research indicates that solving these problems requires a deeper understanding of a new type of correlation, namely electron-PCP (hereafter e-PCP) correlation.

Multi-component methods based on electronic and nuclear orbitals typically produce highly localized nuclear wave functions, leading to overestimated vibrational frequencies. The magnitude of errors in the stretching frequencies for hydrogen vibrations often ranges from 2000 to 3000 $cm^{-1}$, which is of the same order as the frequencies themselves. Bending frequencies involving hydrogen are often qualitatively incorrect. The excessive localization of the nuclear wave function affects not only the frequencies but also the geometries, isotopic effects, and tunneling splittings [55].

The lack of electron-PCP correlation in the product wave function leads to the unphysical localization of the nuclear wave function.

Electron-electron correlation can be included using traditional electronic structure theory methods. However, incorporating electron-PCP correlation into the wave function is more challenging due to the attractive Coulomb interaction between them. Unfortunately, for most applications, the NEO-MP2 and NEO-MCSCF methods do not provide sufficient electron-PCP correlation to overcome the localization problem. Additionally, recovering this correlation by increasing the order of perturbation theory or the orbital active space converges very slowly [56].

Providing a deeper understanding of not only electron-nucleus correlation (which is often equivalent to electron-proton correlation) but also the more general case of electron-PCP correlation with arbitrary mass is the motivation behind the present study. Based on this, the main objective of this dissertation can be summarized in one phrase: dissecting electron-PCP correlation by introducing a simple model that allows for exact solutions.

To simulate the environment of exotic molecules, we've aimed to build a theoretical laboratory using a simple model. This laboratory features an exotic hydrogen-like atom trapped in the potential well of a harmonic oscillator, with different force constants for the electron and the PCP with arbitrary mass. We were inspired by a similar model used years ago for the analytical examination of electron-electron correlation, namely the harmonium or Hooke's atom, which consists of two electrons trapped in the potential of a harmonic oscillator. Accordingly, we named our model "Exotic Harmonium," referring to a harmonium with exotic particles.

The EHM, as it only involves electron-PCP correlation, is essentially the simplest system in which this type of correlation can be studied. By adjusting the oscillator field frequencies to match real chemical environments, we can effectively simulate the effects of molecular environments within this model. Additionally, manipulating the field frequencies and the masses of the PCPs allows us to examine a wide range of systems and correlation effects within them. For instance, selecting very high frequencies can simulate environments with extremely high pressure, such as metallic hydrogen, and gradually changing the mass of the PCP can provide interesting information, such as the approximate boundary of the system's adiabatic transition.

In the case of the harmonium model, a quasi-analytical solution for the ground state wave function can be obtained [79]. However, our calculations show that using the same strategy for Exotic Harmonium only yields the wave functions of excited states. For this reason, this study employs the variational method for closed-form solutions. The initial idea involves proposing a simple yet efficient variational wave function, fitting the variational parameters, and extracting an accurate and compact analytical wave function. The parameters of this model include the mass of the PCP and the harmonic oscillator field frequency, covering a wide range of exotic particles (Section 1-5) and both weak and strong field frequencies.

Such a wave function can be used for the analytical calculation of any property concerning this system within different frameworks. In this dissertation, after developing and examining various aspects of the EHM, we also explore its more important applications, such as extracting accurate electron-PCP correlation potentials within the two-component density functional theory (TC-DFT) framework, testing the existing correlation functionals in the related literature within the TC-DFT framework, and computing the exact non-adiabatic energy of systems. It is hoped that this model and its derived results will contribute to the development of a novel first-principles method for accurately estimating electron-PCP correlation energy in the future. The simplicity of this model allows for the exact solution of the resulting mathematical expressions, embodying physics at its finest.

# 2 Development of the Exotic Harmonium Model

Since the EHM was inspired by the harmonium model, the harmonium model will first be explained. In subsequent sections, the developed model and the results obtained from it will be described.

## 2.1 Harmonium Model

The harmonium model, as we know it today, which involves two electrons repelling each other through Coulombic forces and trapped in the field or well of a harmonic oscillator, was first introduced by Kestner and Sinanoglu in 1962 [80]. At the time of its introduction, there was no exact solution for two-electron systems exhibiting electron correlation, and the best solution available was the explicitly correlated variational wavefunction proposed by Hylleraas in 1928 [81], [82]. As a result, the harmonium model, for which an exact numerical solution could be obtained, garnered significant attention. The paper begins with an introduction that highlights the significance of the proposed model.

The Hamiltonian of the simplest harmonium model is defined as follows:

$$\begin{aligned}\hat{H}(\boldsymbol{r}_1.\boldsymbol{r}_2) &= -\frac{1}{2}(\nabla_1^2 + \nabla_2^2) + \frac{1}{|\boldsymbol{r}_1 - \boldsymbol{r}_2|} + \frac{1}{2}k(\boldsymbol{r}_1^2 + \boldsymbol{r}_2^2) \\ &= -\frac{1}{2}(\nabla_1^2 + \nabla_2^2) + \frac{1}{|\boldsymbol{r}_1 - \boldsymbol{r}_2|} + \frac{1}{2}\omega^2(\boldsymbol{r}_1^2 + \boldsymbol{r}_2^2)\end{aligned} \qquad 2.1$$

where $k$ is the force constant of the oscillator and is related to the oscillator frequency ω by the following relation:

$$k = m\omega^2 \qquad 2.2$$

where $m$ is the mass of the particle. Additionally, $r_1$ and $r_2$ are defined relative to the center of the oscillator well. In general, the Hamiltonian of a bound two-electron system is given by:

$$H(\boldsymbol{r}_1.\boldsymbol{r}_2) = -\frac{1}{2}(\nabla_1^2 + \nabla_2^2) + V + W_1 + W_2 \qquad 2.3$$

where $V$ is the Coulombic repulsion potential between the two electrons $(\frac{1}{|\boldsymbol{r}_1 - \boldsymbol{r}_2|})$. $W_1$ and $W_2$ are the external potentials for the first and second electrons, respectively, with $\boldsymbol{r}_1$ and $\boldsymbol{r}_2$ defined relative to the centers of these external potentials. The most natural form of these external potentials, as they occur in atoms and molecules, is their sum as electron-nucleus Coulombic interactions, that is:

$$W_1 + W_2 = -\sum_{i=1}^{2}\sum_{A=1}^{N_c}\frac{Z_A}{|\boldsymbol{r}_i - \boldsymbol{r}_A|} \qquad 2.4$$

where $Z_A$ is the atomic number of nucleus A. However, as mentioned above, such a system (the simplest case of which is equivalent to a helium atom) cannot be solved analytically. Another possibility is to replace the electron-nucleus interaction with a different potential. One option is to substitute this potential with that of an isotropic harmonic oscillator:

$$W_1 + W_2 = \frac{1}{2}k(\boldsymbol{r}_1^2 + \boldsymbol{r}_2^2) = \frac{1}{2}\omega^2(\boldsymbol{r}_1^2 + \boldsymbol{r}_2^2) \qquad 2.5$$

Harmonium is the simplest model that not only has an exact numerical solution but also retains electron correlation. In the three decades following the proposal of the harmonium model, it was not only examined from various aspects but eventually also had a quasi-analytical solution found for it. In this regard, in 1986, Laufer and Krieger used the harmonium model to test functionals designed within the framework of density functional theory [83]. Subsequently, in 1990, Samanta and Ghosh obtained an analytical solution for the modified harmonium model by adding a linear interaction term to the Hamiltonian (2-1) [84], [85]. However, in 1993, Taut succeeded in finding an analytical solution for the harmonium model for certain frequencies of the oscillator field [79], thereby categorizing this model under quasi-exactly solvable models in quantum mechanics [86]. After separating the center of mass (with mathematical details provided in the following section), Taut started from a general power series and obtained the ground state wave function of the internal motion (without the normalization constant) for specific oscillator frequencies ($\omega = \frac{1}{2(l+1)}$) as follows:

$$u(r) = r^{l+1}e^{-r^2/8(l+1)}\left[1 + \frac{r}{2(l+1)}\right] \qquad 2.6$$

and obtained its energy as follows:

$$\varepsilon = \frac{(2l + 5)\omega}{2} \qquad 2.7$$

where $l$ is the orbital quantum number ($l = 0.1,2, \ldots$).

After presenting this quasi-analytical solution, numerous studies have been conducted by many researchers, including Karwowski and others, on the Harmonium model and its solutions [87]–[92]. The model has also been extended to 3- to 6-particle systems [93]. Additionally, this model resembles an important class of quantum systems known as confined atoms [94], [95], and thus, it can also be used as a simple model for quantum dots, which are one of the practical quantum technologies [96].

## 2.2 Analytical Solution of the EHM

We initially attempted to obtain an analytical solution for the EHM using the methods of Karwowski [87] and Taut [79] for two particles with arbitrary masses and an attractive Coulombic potential (as opposed to the repulsive Coulombic potential of electrons in the Harmonium). Trapping such a system in the potential of a harmonic oscillator leads to the quantization of the center-of-mass motion and eliminates the issues related to subtracting the center Hamiltonian, as mentioned in the NOMO method discussed in Chapter 1. In other words, the harmonic oscillator trap prevents the emergence of a continuous energy spectrum.

However, assuming variable masses and changing the type of interaction has implications that result in significant differences from the harmonium model, which will be discussed later. As mentioned in Section 1-8, using conventional approaches for the analytical solution of harmonium for this new model, which is a two-component system trapped in a dual harmonic oscillator well, does not yield an analytical solution for the ground state.

Instead, it results in the excited states of the system (the computational details of this analytical solution are provided in Section 2-2-2).

The Hamiltonian of two interacting particles trapped in the potential of a harmonic oscillator, in its most general form, is given as follows:

$$H(\boldsymbol{r}_1,\boldsymbol{r}_2) = \frac{1}{2m_1}\nabla^2 + \frac{1}{2m_2}\nabla^2 + \frac{1}{2}k_1 r_1^2 + \frac{1}{2}k_2 r_2^2 + V \qquad 2.8$$

where $m_1$ and $m_2$ represent the masses of the first and second particles, respectively. The potential $V$ specifies the type of interaction between the two particles, as it can take various forms. The most well-known type is the Coulombic potential, which is represented as follows:

$$V = \frac{\varsigma}{|\boldsymbol{r}_1 - \boldsymbol{r}_2|} \qquad \varsigma = \pm 1 \qquad 2.9$$

where the sign of $\varsigma$ indicates whether the interaction between the two particles is repulsive or attractive, respectively. The Hamiltonian (2-8) for electrons reduces to the Hamiltonian (2-1), which is the main Hamiltonian of the harmonium.

By introducing the center of mass and relative coordinates as follows:

$$\boldsymbol{R} = \frac{(m_1\boldsymbol{r_1} + m_2\boldsymbol{r_2})}{(m_1 + m_2)} \qquad 2.10$$

$$\boldsymbol{r} = |\boldsymbol{r_1} - \boldsymbol{r_2}| \qquad 2.11$$

The initial coordinates are expressed in terms of these two new coordinates as follows:

$$\boldsymbol{r}_1 = \boldsymbol{R} - \frac{m_2}{(m_1 + m_2)}\boldsymbol{r} \qquad 2.12$$

$$\boldsymbol{r}_2 = \boldsymbol{R} + \frac{m_1}{(m_1 + m_2)}\boldsymbol{r} \tag{2.13}$$

With this change of variables, the Hamiltonian is separated into two distinct parts, expressed as the sum of the center of mass (COM) and the relative motion components:

$$H(\boldsymbol{r}_1,\boldsymbol{r}_2) = H_R(\boldsymbol{R}) + H_r(\boldsymbol{r}) \tag{2.14}$$

$$H_R(\boldsymbol{R}) = -\frac{1}{2M}\nabla_R^2 + \frac{M}{2}\omega^2\boldsymbol{R}^2 \tag{2.15}$$

$$H_r(\boldsymbol{r}) = -\frac{1}{2\mu}\nabla_r^2 + \frac{\mu}{2}\omega^2\boldsymbol{r}^2 + V \tag{2.16}$$

where $M = m_1 + m_2$ and $\mu = m_1 m_2/M$. This also leads to the complete decoupling of the Hamiltonian or energy spectrum into the center-of-mass and relative motion parts. We can analyze these two parts separately:

$$H_R(\boldsymbol{R})\,\eta_{klm}(\boldsymbol{R}) = E_{kl}\,\eta_{klm}(\boldsymbol{R}) \tag{2.17}$$

$$H_r(\boldsymbol{r})\chi_{k'l'm'}(\boldsymbol{r}) = E_{k'l'}\,\chi_{k'l'm'}(\boldsymbol{r}) \tag{2.18}$$

In equation (2-17), the function $\eta_{klm}(\boldsymbol{R})$ consists of radial and angular parts as follows:

$$\eta_{klm}(\boldsymbol{R}) = N_{kl}\,\tau_{kl}(\boldsymbol{R})\,Y_{lm}(\theta,\phi) \tag{2.19}$$

where $N$ is the normalization constant, $(l,m)$ are the angular momentum quantum numbers, and $Y_{lm}(\theta,\phi)$ are the spherical harmonics. Therefore, the energy of this component is equal to that of a three-dimensional spherical harmonic oscillator:

$$E_{kl} = \omega(2k + l + \frac{3}{2}) \qquad 2.20$$

For $\chi_{k'l'm'}(\boldsymbol{r})$ in equation (2-17), we have:

$$\chi_{k'l'm'}(\boldsymbol{r}) = N_{k'l'}\ \sigma_{k'l'}(\boldsymbol{r})\ Y_{l'm'}(\theta,\phi) \qquad 2.21$$

Using spherical coordinates to separate $\boldsymbol{r}$ from the angular variables, we can write:

$$\Phi(\boldsymbol{r})_{nlm} = \frac{1}{r}\ \phi(\boldsymbol{r})_{nl}\ \mathrm{Y}_{lm}(\boldsymbol{r}) \qquad 2.22$$

Once again, the $\mathrm{Y}_{lm}$ are the spherical harmonics, and the $\boldsymbol{\phi(r)_{nl}}$ are determined by the following radial Schrödinger equation:

$$\left[-\frac{1}{2\mu}\frac{d^2}{dr^2} + \frac{l(l+1)}{2\mu r^2} + \frac{\mu\omega^2}{2}r^2 + \frac{\varsigma}{r} - E\right]\phi(\boldsymbol{r})_{nl} = 0 \qquad 2.23$$

The exact wave function of the entire system is written as the product of the wave functions of the two parts, center of mass and relative motion:

$$\Xi(\boldsymbol{R},\boldsymbol{r}) = \eta_{klm}(\boldsymbol{R})\ \chi_{k'l'm'}(\boldsymbol{r}) \qquad 2.24$$

And the total energy is equal to the sum of the energies of these two parts:

$$E_{klk'l'} = E_{kl} + E_{k'l'} \qquad 2.25$$

Equation (2-24) shows that, after defining coordinates (2-10) and (2-11), the wave function of the system will be the product of two functions, one of which depends only on the relative coordinates and the other only on the center of mass coordinates. In such a coordinate system, we will be dealing with pseudo particles (pseudo particle $\mu$ and pseudo particle *M*).

### 2.2.1 Separability of the Hamiltonian

In general, separating the Hamiltonian of a trapped or confined two-particle system, like equation (2-3), into center-of-mass and relative motion components, as in equation (2-14), is only possible for specific potentials, such as the harmonic oscillator described by equation (2-8), and with the separability condition being satisfied. To obtain this condition, we substitute equations (2-12) and (2-13) into equation (2-8), resulting in:

$$W(\boldsymbol{r}_1,\boldsymbol{r}_2) = \frac{R^2}{2}(k_1 + k_2) + \frac{r^2}{2}\left(\frac{k_1 m_2 + k_2 m_1}{M}\right) - k_1\left(\frac{m_2}{M} R\, r\right) + k_2\left(\frac{m_1}{M} R\, r\right) \qquad 2.26$$

For $W$ to be separable, the sum of the third and fourth terms (the mixed terms) must be zero. As a result, the separability condition is obtained as follows:

$$\frac{k_1}{k_2} = \frac{m_1}{m_2} \qquad 2.27$$

For the harmonium model, this ratio is equal to one, that is:

$$\frac{k_1}{k_2} = \frac{m_1}{m_2} = 1 \qquad 2.28$$

However, in a model where the masses of the two particles are different, according to equation (2-2) and to satisfy condition (2-27), it is necessary for the following equations to hold:

$$\frac{k_1}{k_2} = \frac{m_1}{m_2} \quad \& \quad \frac{\omega_1^2}{\omega_2^2} = 1 \Rightarrow \omega_1^2 = \omega_2^2 = \omega \qquad 2.29$$

$$\frac{1}{2}k_1\boldsymbol{r}_1^2 + \frac{1}{2}k_2\boldsymbol{r}_2^2 = \frac{1}{2}\omega^2(m_1\boldsymbol{r}_1^2 + m_2\boldsymbol{r}_2^2) \qquad 2.30$$

On the other hand, if we initially consider the force constant in equation (2-8) to be the same for both particles, according to equation (2-27), we conclude that the masses of the two particles must also be equal. Therefore, both methods lead us to the same result (2-29).

The above explanations clearly show that changing the masses of the particles in the harmonium model results in the creation of two harmonic oscillator traps with different force constants (proportional to the mass of each particle). In this case, although the field frequencies of the two oscillators are the same, the particles experience different harmonic potentials from the traps.

### 2.2.2 Analytical Solution

The above model, due to the variability in particle masses, can be applied to a diverse range of systems. Several specific systems are listed in the table below:

Table 2-1: Examples of Various Systems with Harmonic Oscillator Potentials

| System Type | $\mu$ | $\varsigma$ |
|---|---|---|
| Harmonium | 1/2 | 1 |
| Positronium | 1/2 | -1 |
| Hydrogen-like atom (with infinite nuclear mass) | 1 | -Z |
| Hydrogen-like atom (with finite nuclear mass) | $\frac{m_n}{1+m_n}$ | -Z |

In this model, there are two parameters that allow us to change the system's state: the strength of the external field by ω and the interaction strength between the two particles by $\varsigma$. In other words, this model has two limits: strong correlation and weak correlation. When $\omega \to \infty$, we have a system with strong oscillator confinement where the Coulomb interaction is weak in comparison, and the particle correlation is also weak in this limit. Conversely, when $\omega \to 0$, we are in the strong correlation limit.

To begin the analytical approach (based on Karwowski's approach [87]), we first substitute the following equation into equation (2-23), leading us to the next equation:

$$x = \sqrt{2\mu\omega}\, r \tag{2.31}$$

$$\left[-\frac{d^2}{dx^2} + \frac{l(l+1)}{x^2} + \frac{x^2}{4} + \frac{s}{x} - \frac{E}{\omega}\right]\phi(x) = 0 \tag{2.32}$$

where

$$s = \varsigma\sqrt{\frac{2\mu}{\omega}} \tag{2.33}$$

We then obtain the asymptotic solutions (2-32) in the two limits $r \to 0$ and $r \to \infty$. Thus, for the limit $r \to 0$ (or equivalently $x \to 0$), we have:

$$\begin{aligned}\lim_{x\to 0}\left(\left[-\frac{d^2}{dx^2} + \frac{l(l+1)}{x^2} + \frac{x^2}{4} + \frac{s}{x} - \frac{E}{\omega}\right]\phi(x) = 0\right)\\ = \left(\left[-\frac{d^2}{dx^2} + \frac{l(l+1)}{x^2} - \frac{E}{\omega}\right]\phi(x) = 0\right)\\ = \left(-\phi''(x) + \frac{l(l+1)}{x^2}\phi(x) = \frac{E}{\omega}\phi(x)\right)\end{aligned} \tag{2.34}$$

which has the following solution:

$$\phi(x) \sim x^{l+1} \tag{2.35}$$

On the other hand, for the limit $r \to \infty$ (or equivalently $x \to \infty$), we have:

$$\lim_{x\to\infty}\left(\left[-\frac{d^2}{dx^2}+\frac{l(l+1)}{x^2}+\frac{x^2}{4}+\frac{s}{x}-\frac{E}{\omega}\right]\phi(x)=0\right)$$
$$=\left(\left[-\frac{d^2}{dx^2}+\frac{x^2}{4}-\frac{E}{\omega}\right]\phi(x)=0\right) \qquad 2.36$$
$$=\left(-\phi''(x)+\frac{x^2}{4}\phi(x)=\frac{E}{\omega}\phi(x)\right)$$

which results in the one-dimensional harmonic oscillator equation, and we know its solution is given by:

$$\phi(x) \sim e^{-x^2/4} \qquad 2.37$$

As a result, the general solution can be written as the product of these two limits, with the addition of an arbitrary polynomial such as $P(x)$, as follows:

$$\phi(x) = x^{l+1}e^{-x^2/4}\ P(x) \qquad 2.38$$

The first and second derivatives of this wave function are obtained as follows:

$$\phi'(r)_{nl} = (l+1)x^l e^{-x^2/4} P(x) - \frac{1}{2}\, x^{l+2}e^{-\frac{x^2}{4}} P(x) + x^{l+1}e^{-\frac{x^2}{4}} P'(x) \qquad 2.39$$

2.40

$$\phi''(r)_{nl} = l(l+1)x^{l-1}e^{-x^2/4}\,P(x) - \frac{1}{2}\,(l+1)x^{l+1}e^{-\frac{x^2}{4}}\,P(x)$$
$$+ (l+1)x^{l}e^{-\frac{x^2}{4}}\,P'(x)$$
$$-\frac{1}{2}\,(l+2)x^{l+1}e^{-\frac{x^2}{4}}\,P(x) + \frac{1}{4}\,x^{l+3}e^{-\frac{x^2}{4}}\,P(x)$$
$$-\frac{1}{2}\,x^{l+2}e^{-\frac{x^2}{4}}\,P'(x) + (l+1)x^{l}e^{-\frac{x^2}{4}}\,P'(x)$$
$$-\frac{1}{2}x^{l+2}e^{-\frac{x^2}{4}}\,P'(x) + x^{l+1}e^{-\frac{x^2}{4}}\,P''(x)$$

By substituting the two above equations into (2-32), we obtain:

$$-l(l+1)x^{l-1}e^{-\frac{x^2}{4}}\,P(x) + \frac{1}{2}\,(l+1)x^{l+1}e^{-\frac{x^2}{4}}\,P(x)$$
$$-(l+1)x^{l}e^{-\frac{x^2}{4}}\,P'(x)$$
$$+\frac{1}{2}\,(l+2)x^{l+1}e^{-\frac{x^2}{4}}\,P(x) - \frac{1}{4}\,x^{l+3}e^{-\frac{x^2}{4}}\,P(x)$$
$$+\frac{1}{2}\,x^{l+2}e^{-\frac{x^2}{4}}\,P'(x) - (l+1)x^{l}e^{-\frac{x^2}{4}}\,P'(x) \qquad 2.41$$
$$+\frac{1}{2}x^{l+2}e^{-\frac{x^2}{4}}\,P'(x) - x^{l+1}e^{-\frac{x^2}{4}}\,P''(x)$$
$$+ l(l+1)x^{l-1}e^{-\frac{x^2}{4}}\,P(x) + s\,x^{l}e^{-\frac{x^2}{4}}\,P(x)$$
$$+\frac{1}{4}\,x^{l+3}e^{-\frac{x^2}{4}}\,P(x) - \frac{E}{\omega}\,x^{l+1}e^{-\frac{x^2}{4}}\,P(x) = 0$$

After combining the terms of equal degree $P(x)$ and then eliminating $e^{-x^2/4}$ terms, we obtain:

$$-x^{l+1}\,P''(x)+\left[\frac{1}{2}\,x^{l+2}-(l+1)x^{l}-(l+1)x^{l}\right.$$
$$\left.+\frac{1}{2}x^{l+2}\right]\,P'(x)$$
$$+\left[l(l+1)x^{l-1}+s\,x^{l}\,+\frac{1}{4}\,x^{l+3}\,-\frac{E}{\omega}\,x^{l+1}\right. \qquad 2.42$$
$$-\,l(l+1)x^{l-1}+\frac{1}{2}\,(l+1)x^{l+1}+\frac{1}{2}\,(l+2)x^{l+1}$$
$$\left.-\frac{1}{4}\,x^{l+3}\right]P(x)\;=0$$

By multiplying the above expression by $-x^{-(l+1)}$, we arrive at the following expression:

$$P''(x)+\left[\frac{2(l+1)}{x}-x\right]\,P'(x)+\left[\frac{E}{\omega}-\frac{s}{x}-l-\frac{3}{2}\right]P(x)\;=0 \qquad 2.43$$

$$\left[\frac{d^2}{dx^2}+\left(\frac{2(l+1)}{x}-x\right)\frac{d}{dx}+\left(\mathcal{E}-\frac{s}{x}\right)\right]P(x)\;=0 \qquad 2.44$$

where

$$\mathcal{E}=\frac{E}{\omega}-l-\frac{3}{2} \qquad 2.45$$

In equation (2-44), we can identify three terms with different degrees of homogeneity:

$$\Omega_{-2}=\frac{d^2}{dx^2}+\frac{2(l+1)}{x}\frac{d}{dx}\;\;,\;\;\Omega_{-1}=-\frac{s}{x}\;\;,\;\;\Omega_{0}=-x\,\frac{d}{dx}+\mathcal{E} \qquad 2.46$$

$P(x)$ can be defined as a polynomial:

$$P(x)=\sum_{m=0}^{\infty}a_m x^m \qquad 2.47$$

Since we are interested in solutions with the expansion (2-47) and truncated at $m = p$, we can write the following conditions for $a_m$:

$$a_m \begin{cases} \neq 0 & 0 \leq m \leq p \\ = 0 & m > p \;\; or\; m < 0 \end{cases} \qquad 2.48$$

Using equations (2-46) and (2-47), we have:

$$P'(x) = \sum_{m=0}^{\infty} m\, a_m x^{m-1} \qquad 2.49$$

$$P''(x) = \sum_{m=0}^{\infty} m(m-1)\, a_m x^{m-2} \qquad 2.50$$

To transform equation (2-44) into a single sum with the same power of $x$, we use the following expressions:

$$\Omega_{-2} x^n = n(n+2l+1)\, x^{n-2} \;,\;\; \Omega_{-1} x^{n-1} = -s\, x^{n-2} \;,\;\; \Omega_0 x^{n-2} = (\mathcal{E} - n + 2) \qquad 2.51$$

$$[\Omega_{-2} + \Omega_{-1} + \Omega_0] P(x) = 0 \qquad 2.52$$

After substituting and combining the terms, we arrive at a single sum as follows:

$$\sum_{m=0}^{\infty} [m(m+2l+1)a_m - s a_{m-1} + (\mathcal{E} - m + 2) a_{m-2}] x^{m-2} = 0 \qquad 2.53$$

The coefficients of each power of $x$ must be equal to zero, therefore we have:

$$\begin{aligned} m &= 0 \qquad 0 \\ m &= 1 \qquad (2l+2)a_1 - s a_0 = 0 \\ m &= 2 \qquad 2(2l+3)a_2 - s a_1 + \mathcal{E} a_0 = 0 \end{aligned} \qquad 2.54$$

$m = 3 \qquad 3(2l + 4)a_3 - sa_2 + (\mathcal{E} - 1)a_1 = 0$

...

$m = p \qquad p(2l + p + 1)a_p - sa_{p-1} + (\mathcal{E} - p + 2)a_{p-2} = 0$

$m = p + 1 \quad - sa_p + (\mathcal{E} - p + 1)a_{p-1} = 0$

$m = p + 2 \quad (\mathcal{E} - p)a_p = 0$

Since $a_p \neq 0$, the last term of the above equations shows that $\mathcal{E} = p$, which, as a result, gives us the following according to equation (2-45):

$$E = \omega\left(p + l + \frac{3}{2}\right) \qquad 2.55$$

Therefore, the determinant of the coefficients is given by:

$$\begin{vmatrix} -s & (2l+2) & 0 & 0 & \dots & 0 \\ p & -s & 2(2l+3) & 0 & \dots & 0 \\ 0 & (p-1) & -s & 3(2l+4) & \dots & 0 \\ \vdots & \vdots & \vdots & \vdots & \ddots & p(2l+p+1) \\ 0 & 0 & 0 & 0 & 2 & -s \end{vmatrix} \qquad 2.56$$

A non-trivial solution to equation (2-53) exists if the above determinant equals zero. We can check this condition for different values of $p$ and determine the dependence of energy on mass. For $p = 1$, we have:

$$\mathcal{E} = 1 \ , \quad a_1 = \frac{s\,a_0}{2l + 2} \ , \ a_2 = a_3 = \dots = 0 \qquad 2.57$$

$$E = \omega\left(l + \frac{5}{2}\right) \qquad 2.58$$

$$\begin{vmatrix} -s & (2l+2) \\ 1 & -s \end{vmatrix} = 0$$

$$s^2 = 2l + 2 \qquad 2.59$$

Using equations (2-33) and (2-59), we obtain:

$$\omega = \frac{\mu\varsigma^2}{l+1} \quad \Rightarrow E = \mu\varsigma^2 \frac{2l+5}{2l+2} \qquad 2.60$$

$$P(x) = a_0 + a_1 x \qquad 2.61$$

In conclusion, considering equations (2-31), (2-33), and (2-57), we arrive at the following result:

$$P(x) = a_0 + \frac{a_0\sqrt{2\mu\omega}\, r}{2(l+1)} \varsigma \sqrt{\frac{2\mu}{\omega}} = a_0\left(1 + \frac{\varsigma\mu}{l+1} r\right) \qquad 2.62$$

For $p = 2$, we have:

$$\mathcal{E} = 2 \text{ و } a_3 = a_4 = \dots = 0 \qquad 2.63$$

$$\begin{vmatrix} -s & (2l+2) & 0 \\ 2 & -s & 2(2l+3) \\ 0 & (\mathcal{E}-1) & -s \end{vmatrix} = 0 \qquad 2.64$$

$$s^2 = 8l + 10 \quad \Rightarrow \quad \omega = \frac{\mu\varsigma^2}{4l+5} \quad \Rightarrow \quad E = \mu\varsigma^2 \frac{2l+7}{8l+10} \qquad 2.65$$

Again, according to equation (2-47), we have:

$$P(x) = a_0 + a_1 x + a_2 x^2 \qquad 2.66$$

And for $P = 2$:

$$a_1 = \frac{s\, a_0}{2(l+1)} \ , \ a_2 = \frac{a_1}{s} = \frac{a_0}{2(l+1)} \qquad 2.67$$

Thus, considering equations (2-31) and (2-33), we have:

$$P(r) = a_0\left(1 + \frac{sx}{2(l+1)} + \frac{x^2}{2(l+1)}\right)$$
$$= a_0\left(1 + \frac{\mu\varsigma}{l+1}r + \frac{\mu^2\varsigma^2}{(l+1)(4l+5)}r^2\right) \qquad 2.68$$

The two obtained solutions, namely equations (2-62) and (2-68), are not the ground state for the attractive system ($\varsigma = -1$), and there are two reasons to prove this, one of which is mathematical and the other, which is more important, is of a physical nature.

The mathematical reason is that when $\varsigma$ is negative, the right side of these equations has positive roots, and as a result, $P(r)$ will have nodes. Since the creation of nodes depends on having positive roots, equations (2-62) and (2-68) with negative $\varsigma$ will have nodes and therefore cannot describe the ground state.

The physical reason is that the Karwowski approach creates a fundamental symmetry in the problem. Equations (2-60) and (2-65), as well as equation (30) in Karwowski's original paper for Harmonium [87], show that the energy is proportional to $\varsigma^2$. This result means that regardless of the type of interaction (repulsive or attractive), the energy value is the same, whereas we expect the ground state energies of systems with these two types of interactions to differ. Using a perturbative approach, it is easy to prove the validity of such a claim: If we consider a harmonic oscillator as the unperturbed state of the system and the Coulomb interaction as the perturbation, we can easily prove that the first-order energy of the perturbed system with attractive interaction is lower than that of the same system with repulsive interaction (in both attractive and repulsive cases, the first-order perturbation value is the same, but its sign is positive and negative, respectively, and as a result, the same value is added to and subtracted from the unperturbed system's energy, respectively). Given that we know the energy of the exotic harmonium for any given level is lower than that of the harmonium, we conclude that using the Karwowski approach, we obtain two isoenergetic states, namely the ground state of the harmonium and the first excited state of the exotic harmonium (equation 2-63).

## 2.3 EHM

As shown in the previous section, the analytical approach used for harmonium, with changing particle mass, no longer results in the ground state wave function; therefore, we need to explore other methods. Ultimately, the variational method was used to obtain the ground state energy of the system.

In this section, the developed model in this thesis, i.e., a two-component system trapped in a double-well of harmonic oscillators, is examined using the variational method. It is worth mentioning that, unlike equations (2-8) and (2-9) which include both types of interactions, this model only considers attractive interactions, which aligns with the primary goal of this thesis, namely developing a model for investigating non-BO molecular systems. Since all numerical computations in this thesis consider one electron and one PCP (with an arbitrary mass), subscripts (superscripts) $e$ for the negatively charged particle and $p$ for the PCP are used in the notation of equations and relations.

The total Hamiltonian (in atomic units) for an attractive two-component system trapped in a double-well of harmonic oscillators is as follows:

$$H(\boldsymbol{r}_e,\boldsymbol{r}_p) = -\frac{1}{2m_e}\nabla_e^2 - \frac{1}{2m_p}\nabla_p^2 - \frac{1}{|\boldsymbol{r}_e - \boldsymbol{r}_p|} + \frac{1}{2}\omega^2(m_e\boldsymbol{r}_e^2 + m_p\boldsymbol{r}_p^2) \quad 2.69$$

where ω represents the common frequency of the oscillators, and the vectors $\boldsymbol{r}_e$ and $\boldsymbol{r}_p$ are defined with respect to the centers of the potential wells. The initial coordinates are expressed based on these two new coordinates as follows:

$$\boldsymbol{r}_e = \boldsymbol{R} - \frac{m_p}{(m_e + m_p)}\boldsymbol{r}$$
$$\boldsymbol{r}_p = \boldsymbol{R} + \frac{m_e}{(m_e + m_p)}\boldsymbol{r} \qquad 2.70$$

Just like equation (2-14), using this new coordinate system, the Hamiltonian (2-69) can be written as the sum of two separate parts: the center of mass and relative motion. The difference is that here, the total mass is $M = m_e + m_p$ and the reduced mass is $\mu = m_e m_p / M$.

### 2.3.1 Finding the Relative Motion Wave Function Using the Variational Method

Given that the wave function for the center of mass motion is known, we need to obtain the wave function for the relative motion in order to find the total wave function. The trial variational function was chosen to be as simple as possible while still accurately capturing the behavior of the system's exact ground state wave function. For this purpose, the following trial function was considered:

$$\chi'_{trial} \sim e^{-\mu(\alpha' r + \beta' r^2)} \qquad 2.71$$

where the first exponential term (Slater term) represents the hydrogen-like atom function and the second exponential term (Gaussian term) represents the harmonic oscillator function, with $\alpha'$ and $\beta'$ as the variational parameters. Although the presence of $\mu$ linearly in the exponential terms is necessary, due to the zeroing of the variational parameters at certain field frequencies and the resulting numerical problems, $\mu$ is absorbed into the variational parameters $\alpha'$ and $\beta'$, and the following wave function is used as the trial variational wave function in the subsequent calculations:

$$\chi_{trial} \sim e^{-\alpha r-\beta r^2} \qquad 2.72$$

## 2.3.2 Examining the Asymptotic Behavior of the Trial Wave Function

Before applying the variational theorem to wave function (2-72), we need to ensure the correctness of its asymptotic behavior in both the strong and weak field limits of the harmonic oscillator. By substituting the Coulomb potential instead of V in Hamiltonian (2-16), it is evident that in the strong field limit ($\omega \longrightarrow \infty$), the harmonic oscillator potential energy term dominates the Coulomb interaction term, and it can be approximated by the Hamiltonian of a harmonic oscillator. In this case, the wave function should also reduce to the ground state eigenfunction of a three-dimensional harmonic oscillator (HO):

$$\phi_{HO}(r) = \left(\frac{\mu\omega}{\pi}\right)^{3/4} \exp\left(-\frac{1}{2}\mu\omega r^2\right) \qquad 2.73$$

Conversely, in the weak field limit ($\omega \longrightarrow 0$), the Coulomb interaction term dominates, and Hamiltonian (2-16) can be approximated by the Hamiltonian of a hydrogen-like atom (HL) as follows:

$$\phi_{HL}(r) = \left(\frac{\mu^3}{\pi}\right)^{1/2} \exp(-\mu r) \qquad 2.74$$

Therefore, at different field frequencies, we can observe the asymptotic behavior of the variational wave function and its similarity to the two aforementioned systems.

### 2.3.3 Other Possibilities for Variational Trial Functions

To increase accuracy, another term can be added to the exponent of wave function (2-72), which becomes zero under the limiting conditions discussed (thus ensuring the correct asymptotic behavior of the wave function) but remains non-zero at intermediate frequencies, thereby enhancing the accuracy of the wave function. As we will see, the accuracy of wave function (2-72) is very high even without introducing such a term, so adding an extra term to the wave function will not only provide minimal improvement in accuracy but will also increase computational burden and complexity of the equations. For example, when parameterizing the function with four parameters ($-\alpha r - \beta r^2 - \gamma r^3 - \delta r^4$), the fourth parameter tends to zero during the optimization process, and the third parameter does not provide a significant improvement in energy.

Another proposal for the variational trial wave function could be as follows:

$$\chi''_{trial} \sim e^{-\mu \alpha'' r^{\beta''}} \qquad 2.75$$

The asymptotic behavior of this trial wave function also converges to the correct limits. In the strong oscillator field limit ($\omega \longrightarrow \infty$), the parameter $\beta''$ approaches 2 and the parameter $\alpha''$ approaches $\frac{\omega}{2}$, resulting in the shape of the wave function being exactly like the ground state wave function of a harmonic oscillator ($e^{-\frac{\omega}{2}\mu r^2}$). In the weak oscillator field limit ($\omega \longrightarrow 0$), both the parameter $\beta''$ and the parameter $\alpha''$ approach 1, resulting in the shape of the wave function being exactly like the ground state wave function of a hydrogen-like atom ($e^{-\mu r}$). The results obtained from the optimization process of the wave function parameters (2-75) confirm this asymptotic behavior. Nevertheless, when the energy obtained from this wave function was compared with the energy obtained from wave function (2-72), it was found that the latter gives better results than the former.

### 2.3.4 Variational Theorem

According to the variational theorem, we have:

$$\frac{\int \chi^* \hat{H} \chi \, d\tau}{\int \chi^* \chi \, d\tau} = E \geq E_0 \qquad 2.76$$

where $E_0$ is the ground state energy of the system. The variational energy $E_0$ of the two-component system is given by:

$$E = \frac{\int \chi^* \hat{T} \chi \, d\tau + \int \chi^* \hat{V} \chi \, d\tau}{\int \chi^* \chi \, d\tau} \qquad 2.77$$

For the trial wave function (2-72), we have:

$$\int \chi^* \hat{T} \chi \, d\tau = \int_0^{2\pi} \int_0^{\pi} \int_0^{\infty} \chi^* \hat{T} \chi \, r^2 \sin\theta \;\, dr \, d\theta \, d\phi$$

$$= 4\pi \int_0^{\infty} \chi^* \hat{T} \chi \, r^2 dr$$

$$= -\frac{2\pi}{\mu} \Bigg( \alpha^2 \int_0^{\infty} r^2 e^{-2\alpha r - 2\beta r^2} \, dr$$

$$- 2\beta \int_0^{\infty} r^2 e^{-2\alpha r - 2\beta r^2} \, dr$$

$$+ 4\alpha\beta \int_0^{\infty} r^3 e^{-2\alpha r - 2\beta r^2} \, dr \qquad 2.78$$

$$+ 4\beta^2 \int_0^{\infty} r^4 e^{-2\alpha r - 2\beta r^2} \, dr$$

$$- 2\alpha \int_0^{\infty} r e^{-2\alpha r - 2\beta r^2} \, dr$$

$$- 4\beta \int_0^{\infty} r^2 e^{-2\alpha r - 2\beta r^2} \, dr \Bigg)$$

$$\int \chi^* \hat{V} \chi \, d\tau = 4\pi \left( - \int_0^{\infty} r e^{-2\alpha r - 2\beta r^2} \, dr \right.$$

$$\left. + \frac{1}{2} \mu \omega^2 \int_0^{\infty} r^4 e^{-2\alpha r - 2\beta r^2} \, dr \right) \qquad 2.79$$

$$\int \chi^* \chi \, d\tau = 4\pi \int_0^{\infty} r^2 e^{-2\alpha r - 2\beta r^2} \, dr \qquad 2.80$$

The values of the parameters $\alpha$ and $\beta$ are obtained using an optimization algorithm, in which the energy is optimized using the gradient and Hessian. For this purpose, the next section is dedicated to calculating the gradient and Hessian of the energy.

## 2.3.5 Calculating the Gradient and Hessian of the Energy

The gradient of the energy with respect to the variational parameters is defined as follows:

$$E_\alpha = \frac{\partial E}{\partial \alpha} \qquad E_\beta = \frac{\partial E}{\partial \beta} \tag{2.81}$$

Considering equation (2-76) for the variational energy, we have:

$$\begin{aligned} E_\alpha &= \frac{\left(\int \chi^* \hat{T} \chi \, d\tau\right)_\alpha + \left(\int \chi^* \hat{V} \chi \, d\tau\right)_\alpha - E\left(\int \chi^* \chi \, d\tau\right)_\alpha}{\int \chi^* \chi \, d\tau} \\ E_\beta &= \frac{\left(\int \chi^* \hat{T} \chi \, d\tau\right)_\beta + \left(\int \chi^* \hat{V} \chi \, d\tau\right)_\beta - E\left(\int \chi^* \chi \, d\tau\right)_\beta}{\int \chi^* \chi \, d\tau} \end{aligned} \tag{2.82}$$

where the inner terms are obtained as follows:

$$\begin{aligned} \left(\int \chi^* \hat{T} \chi \, d\tau\right)_\alpha &= \frac{\partial}{\partial \alpha} \int \chi^* \hat{T} \chi \, d\tau \\ &= -\frac{2\pi}{\mu}\left[-2\int_0^\infty r e^{-2\alpha r - 2\beta r^2}\, dr \right. \\ &\quad + 6\alpha \int_0^\infty r^2 e^{-2\alpha r - 2\beta r^2}\, dr \\ &\quad - 2\alpha^2 \int_0^\infty r^3 e^{-2\alpha r - 2\beta r^2}\, dr \\ &\quad + 16\beta \int_0^\infty r^3 e^{-2\alpha r - 2\beta r^2}\, dr \\ &\quad - 8\alpha\beta \int_0^\infty r^4 e^{-2\alpha r - 2\beta r^2}\, dr \\ &\quad \left. - 8\beta^2 \int_0^\infty r^5 e^{-2\alpha r - 2\beta r^2}\, dr \right] \end{aligned} \tag{2.83}$$

$$\left(\int \chi^* \hat{V} \chi \, d\tau\right)_\alpha = \frac{\partial}{\partial \alpha} \int \chi^* \hat{V} \chi \, d\tau$$

$$= 8\pi \left( \int_0^\infty r^2 e^{-2\alpha r - 2\beta r^2} \, dr \right. \qquad 2.84$$

$$\left. - \mu\omega^2 \int_0^\infty r^5 e^{-2\alpha r - 2\beta r^2} \, dr \right)$$

$$\left(\int \chi^* \chi \, d\tau\right)_\alpha = \frac{\partial}{\partial \alpha} \int \chi^* \chi \, d\tau = -8\pi \int_0^\infty r^3 e^{-2\alpha r - 2\beta r^2} \, dr \qquad 2.85$$

$$\left(\int \chi^* \hat{T} \chi \, d\tau\right)_\beta = \frac{\partial}{\partial \beta} \int \chi^* \hat{T} \chi \, d\tau$$

$$= -\frac{2\pi}{\mu} \left[ -6 \int_0^\infty r^2 e^{-2\alpha r - 2\beta r^2} \, dr \right.$$

$$+ 8\alpha \int_0^\infty r^3 e^{-2\alpha r - 2\beta r^2} \, dr$$

$$- 2\alpha^2 \int_0^\infty r^4 e^{-2\alpha r - 2\beta r^2} \, dr \qquad 2.86$$

$$+ 20\beta \int_0^\infty r^4 e^{-2\alpha r - 2\beta r^2} \, dr$$

$$- 8\alpha\beta \int_0^\infty r^5 e^{-2\alpha r - 2\beta r^2} \, dr$$

$$\left. - 8\beta^2 \int_0^\infty r^6 e^{-2\alpha r - 2\beta r^2} \, dr \right]$$

$$\left(\int \chi^* \hat{V} \chi \, d\tau\right)_\beta = \frac{\partial}{\partial \beta} \int \chi^* \hat{V} \chi \, d\tau$$

$$= 8\pi \left( \int_0^\infty r^3 e^{-2\alpha r - 2\beta r^2} \, dr \right. \qquad 2.87$$

$$\left. - \mu\omega^2 \int_0^\infty r^6 e^{-2\alpha r - 2\beta r^2} \, dr \right)$$

$$\left(\int \chi^* \chi \, d\tau\right)_\beta = \frac{\partial}{\partial \beta} \int \chi^* \chi \, d\tau = -8\pi \int_0^\infty r^4 e^{-2\alpha r - 2\beta r^2} \, dr \qquad 2.88$$

The Hessian matrix of the energy is also defined as follows:

$$\begin{bmatrix} E_{\alpha^2} & E_{\alpha\beta} \\ E_{\beta\alpha} & E_{\beta^2} \end{bmatrix} \quad 2.89$$

where the elements of this matrix, according to the variational energy relation, are as follows:

$$E_{\alpha^2} = \frac{\partial^2 E}{\partial \alpha^2}, \qquad E_{\beta^2} = \frac{\partial^2 E}{\partial \beta^2}, \qquad E_{\alpha\beta} = \frac{\partial^2 E}{\partial \alpha \partial \beta}, \qquad E_{\beta\alpha} = \frac{\partial^2 E}{\partial \beta \partial \alpha} \quad 2.90$$

$$E_{\alpha^2} = \frac{\left(\int \chi^* \hat{T} \chi \, d\tau\right)_{\alpha^2} + \left(\int \chi^* \hat{V} \chi \, d\tau\right)_{\alpha^2} - E_\alpha \left(\int \chi^* \chi \, d\tau\right)_\alpha - E_\alpha \left(\int \chi^* \chi \, d\tau\right)_\alpha}{\int \chi^* \chi \, d\tau}$$

$$E_{\beta^2} = \frac{\left(\int \chi^* \hat{T} \chi \, d\tau\right)_{\beta^2} + \left(\int \chi^* \hat{V} \chi \, d\tau\right)_{\beta^2} - E_\beta \left(\int \chi^* \chi \, d\tau\right)_\beta - E_\beta \left(\int \chi^* \chi \, d\tau\right)_\beta}{\int \chi^* \chi \, d\tau}$$

$$E_{\alpha\beta} = \frac{\left(\int \chi^* \hat{T} \chi \, d\tau\right)_{\alpha\beta} + \left(\int \chi^* \hat{V} \chi \, d\tau\right)_{\alpha\beta} - E_\alpha \left(\int \chi^* \chi \, d\tau\right)_\beta - E_\beta \left(\int \chi^* \chi \, d\tau\right)_\alpha}{\int \chi^* \chi \, d\tau} \quad 2.91$$

$$E_{\beta\alpha} = \frac{\left(\int \chi^* \hat{T} \chi \, d\tau\right)_{\beta\alpha} + \left(\int \chi^* \hat{V} \chi \, d\tau\right)_{\beta\alpha} - E_\beta \left(\int \chi^* \chi \, d\tau\right)_\alpha - E_\alpha \left(\int \chi^* \chi \, d\tau\right)_\beta}{\int \chi^* \chi \, d\tau}$$

The inner terms of equation (2-91) are obtained as follows:

$$\left(\int \chi^* \hat{T} \chi \, d\tau\right)_{\alpha^2} = \frac{\partial}{\partial \alpha}\left(\int \chi^* \hat{T} \chi \, d\tau\right)_{\alpha}$$

$$= -\frac{2\pi}{\mu}\left[10 \int_0^\infty r^2 e^{-2\alpha r - 2\beta r^2} \, dr\right.$$

$$- 16\alpha \int_0^\infty r^3 e^{-2\alpha r - 2\beta r^2} \, dr$$

$$+ 4\alpha^2 \int_0^\infty r^4 e^{-2\alpha r - 2\beta r^2} \, dr \qquad 2.92$$

$$- 40\beta \int_0^\infty r^4 e^{-2\alpha r - 2\beta r^2} \, dr$$

$$+ 16\alpha\beta \int_0^\infty r^5 e^{-2\alpha r - 2\beta r^2} \, dr$$

$$\left. + 16\beta^2 \int_0^\infty r^6 e^{-2\alpha r - 2\beta r^2} \, dr\right]$$

$$\left(\int \chi^* \hat{V} \chi \, d\tau\right)_{\alpha^2} = \frac{\partial}{\partial \alpha}\left(\int \chi^* \hat{V} \chi \, d\tau\right)_{\alpha}$$

$$= -16\pi\left(\int_0^\infty r^3 e^{-2\alpha r - 2\beta r^2} \, dr\right. \qquad 2.93$$

$$\left. + 2\mu\omega^2 \int_0^\infty r^6 e^{-2\alpha r - 2\beta r^2} \, dr\right)$$

$$\left(\int \chi^* \chi \, d\tau\right)_{\alpha^2} = \frac{\partial}{\partial \alpha}\left(\int \chi^* \chi \, d\tau\right)_{\alpha} = 16\pi \int_0^\infty r^4 e^{-2\alpha r - 2\beta r^2} \, dr \qquad 2.94$$

$$\left(\int \chi^* \hat{T} \chi \, d\tau\right)_{\beta^2} = \frac{\partial}{\partial \beta}\left(\int \chi^* \hat{T} \chi \, d\tau\right)_\beta$$

$$= -\frac{2\pi}{\mu}\left[32\int_0^\infty r^4 e^{-2\alpha r-2\beta r^2}\, dr\right.$$

$$-\, 24\alpha \int_0^\infty r^5 e^{-2\alpha r-2\beta r^2}\, dr$$

$$+\, 4\alpha^2 \int_0^\infty r^6 e^{-2\alpha r-2\beta r^2}\, dr \qquad 2.95$$

$$-\, 56\beta \int_0^\infty r^6 e^{-2\alpha r-2\beta r^2}\, dr$$

$$+\, 16\alpha\beta \int_0^\infty r^7 e^{-2\alpha r-2\beta r^2}\, dr$$

$$\left.+\, 16\beta^2 \int_0^\infty r^8 e^{-2\alpha r-2\beta r^2}\, dr\right]$$

$$\left(\int \chi^* \hat{V} \chi \, d\tau\right)_{\beta^2} = \frac{\partial}{\partial \beta}\left(\int \chi^* \hat{V} \chi \, d\tau\right)_\beta$$

$$= -16\pi\left(\int_0^\infty r^5 e^{-2\alpha r-2\beta r^2}\, dr\right. \qquad 2.96$$

$$\left.+\, 2\mu\omega^2 \int_0^\infty r^8 e^{-2\alpha r-2\beta r^2}\, dr\right)$$

$$\left(\int \chi^* \chi \, d\tau\right)_{\beta^2} = \frac{\partial}{\partial \beta}\left(\int \chi^* \chi \, d\tau\right)_\beta = 16\pi \int_0^\infty r^6 e^{-2\alpha r-2\beta r^2}\, dr \qquad 2.97$$

$$\left(\int \chi^* \hat{T} \chi \, d\tau\right)_{\alpha\beta} = \left(\int \chi^* \hat{T} \chi \, d\tau\right)_{\beta\alpha} = \frac{\partial}{\partial\beta}\left(\int \chi^* \hat{T} \chi \, d\tau\right)_{\alpha}$$

$$= \frac{\partial}{\partial\alpha}\left(\int \chi^* \hat{T} \chi \, d\tau\right)_{\beta}$$

$$= -\frac{2\pi}{\mu}\left[20\int_0^\infty r^3 e^{-2\alpha r - 2\beta r^2} \, dr\right.$$

$$- 20\alpha \int_0^\infty r^4 e^{-2\alpha r - 2\beta r^2} \, dr$$

$$+ 4\alpha^2 \int_0^\infty r^5 e^{-2\alpha r - 2\beta r^2} \, dr \qquad 2.98$$

$$- 48\beta \int_0^\infty r^5 e^{-2\alpha r - 2\beta r^2} \, dr$$

$$+ 16\alpha\beta \int_0^\infty r^6 e^{-2\alpha r - 2\beta r^2} \, dr$$

$$\left. + 16\beta^2 \int_0^\infty r^7 e^{-2\alpha r - 2\beta r^2} \, dr\right]$$

$$\left(\int \chi^* \hat{V} \chi \, d\tau\right)_{\alpha\beta} = \left(\int \chi^* \hat{V} \chi \, d\tau\right)_{\beta\alpha} = \frac{\partial}{\partial\beta}\left(\int \chi^* \hat{V} \chi \, d\tau\right)_{\alpha}$$

$$= \frac{\partial}{\partial\alpha}\left(\int \chi^* \hat{V} \chi \, d\tau\right)_{\beta}$$

$$= -16\pi\left(\int_0^\infty r^4 e^{-2\alpha r - 2\beta r^2} \, dr\right. \qquad 2.99$$

$$\left. - 2\mu\omega^2 \int_0^\infty r^7 e^{-2\alpha r - 2\beta r^2} \, dr\right)$$

$$\left(\int \chi^* \chi \, d\tau\right)_{\alpha\beta} = \left(\int \chi^* \chi \, d\tau\right)_{\beta\alpha} = \frac{\partial}{\partial\beta}\left(\int \chi^* \chi \, d\tau\right)_{\alpha}$$

$$= \frac{\partial}{\partial\alpha}\left(\int \chi^* \chi \, d\tau\right)_{\beta} = 16\pi \int_0^\infty r^5 e^{-2\alpha r - 2\beta r^2} \, dr \qquad 2.100$$

## 2.4 Coordinate System Transformations

### 2.4.1 Single-Particle Wave Function

The definition of the internal and center of mass coordinates according to equations (2-10) and (2-11), followed by the division of the Hamiltonian and the wave function into relative coordinate and center of mass parts, significantly simplifies the problem and enables us to construct an accurate variational wave function for the relative coordinate section. By substituting the variational wave function (2-72) for the relative motion part and the wave function (2-19) for the center of mass part in equation (2-24), the approximate total ground state wave function is given as follows:

$$\Psi(\boldsymbol{R},\boldsymbol{r}) = N\exp(-\alpha r - \beta r^2 - \gamma R^2) \qquad 2.101$$

where $\gamma = \frac{1}{2}M\omega$ and $N$ is the normalization factor of the wave function. In such a system, the volume elements for integration over the entire space will be in the form of spherical polar coordinates:

$$d\boldsymbol{R}\,d\boldsymbol{r} = R^2 sin\theta_R\; dR\; d\theta_R\; d\phi_R\;\; r^2 sin\theta_r\;\; dr\; d\theta_r\; d\phi_r \qquad 2.102$$

which, due to the lack of explicit angular dependence in wave function (2-101), can be simplified as follows:

$$d\boldsymbol{R}\,d\boldsymbol{r} = 16\,\pi^2 R^2\; dR\;\; r^2 dr \qquad 2.103$$

However, it is evident that for calculating single-particle quantities, we need a wave function explicitly dependent on single-particle coordinates. Therefore, we can rewrite wave function (2-101) in terms of single-particle coordinates using equations (2-10) and (2-11). By applying this substitution, we have:

$$-\gamma\left(\frac{m_e r_e + m_p r_p}{M}\right)^2 = -\frac{\gamma m_e^2 r_e^2}{M^2} - \frac{\gamma m_p^2 r_p^2}{M^2} - \frac{2\gamma m_e m_p r_e r_p cos\theta_{ep}}{M^2} \quad 2.104$$

$$-\beta\left|\boldsymbol{r}_e - \boldsymbol{r}_p\right|^2 = -\beta r_e^2 - \beta r_p^2 + 2\beta r_e r_p cos\theta_{ep} \quad 2.105$$

For simplicity and by defining three arbitrary constants as follows:

$$c_e = \frac{m_e^2\gamma}{M^2} + \beta \quad 2.106$$

$$c_p = \frac{m_p^2\gamma}{M^2} + \beta \quad 2.107$$

$$c_{ep} = 2\left(\beta - \frac{m_e m_p \gamma}{M^2}\right) \quad 2.108$$

The final form of the wave function will be as follows:

$$\Psi(\boldsymbol{r}_e, \boldsymbol{r}_p) = N\, exp\left[-c_e r_e^2 - c_p r_p^2 + c_{ep} r_e r_p \cos\theta_{ep} - \alpha\left(r_e^2 + r_p^2 - 2 r_e r_p \cos\theta_{ep}\right)^{1/2}\right] \quad 2.109$$

The volume elements for the above wave function are also in the form of the usual polar coordinates, as follows:

$$d\boldsymbol{r}_e\, d\boldsymbol{r}_p = r_e^2 sin\theta_e\, dr_e\, d\theta_e\, d\phi_e\, r_p^2 sin\theta_p\, dr_p\, d\theta_p\, d\phi_p \quad 2.110$$

where, due to the explicit presence of the angular part $\theta$ in wave function (2-109), simplification can only be applied to the angular part $\phi$ as follows:

$$d\boldsymbol{r}_e\, d\boldsymbol{r}_p = 4\,\pi^2 r_e^2 sin\theta_e\, dr_e\, d\theta_e\, r_p^2 sin\theta_p\, dr_p\, d\theta_p \quad 2.111$$

### 2.4.2 Intermediate Wave Function

The two wave functions (2-101) and (2-109) have variables corresponding to the center of mass and single-particle coordinates, respectively, along with their associated differentials. However, it is possible to obtain an intermediate wave function and the associated differentials that simultaneously include both single-particle and interparticle coordinates. For this purpose, we replace $R$ in the wave function (2-101) with the corresponding values. From equation (2-11), we have:

$$r^2 = \left|\boldsymbol{r}_e - \boldsymbol{r}_p\right|^2 = r_e^2 + r_p^2 - 2r_e r_p cos\theta_{ep} \qquad 2.112$$

As a result:

$$2r_e r_p cos\theta_{ep} = -r^2 + r_e^2 + r_p^2 \qquad 2.113$$

Therefore:

$$\begin{aligned} R^2 &= \left(\frac{m_e r_e + m_p r_p}{M}\right)^2 \\ &= \frac{1}{M^2}\left(m_e^2 r_e^2 + m_p^2 r_p^2 + 2m_e m_p r_e r_p cos\theta_{ep}\right) \\ &= \frac{1}{M^2}\left(m_e^2 r_e^2 + m_p^2 r_p^2 + m_e m_p\left[-r^2 + r_e^2 + r_p^2\right]\right) \end{aligned} \qquad 2.114$$

By substituting the above equation into the wave function (2-101), we have:

$$\begin{aligned} \Psi\left(r, r_e, r_p\right) = N \exp\Bigg(&-\alpha r - \beta r^2 \\ &- \frac{\gamma}{M^2}\left(m_e^2 r_e^2 + m_p^2 r_p^2\right. \\ &\left. + m_e m_p\left[-r^2 + r_e^2 + r_p^2\right]\right)\Bigg) \end{aligned} \qquad 2.115$$

This wave function is simultaneously dependent on both single-particle and interparticle coordinates, while the interparticle angle is not present.

Now, to define the volume elements corresponding to wave function (2-115), we first define the coordinates of the first particle (here, the electron) in the usual polar coordinates ($r_e$ ,$\theta_e$,$\phi_e$). To define the coordinates of the second particle (here, the positive particle), we rotate the coordinate system so that the z-axis lies along $r_e$. Then we define the coordinates of the second particle in spherical polar coordinates as ($r_p$ ,$\theta_p$,$\phi_p$) [97]. Since the z-axis is along $r_e$, $\theta_p$ is equivalent to $\theta_{ep}$, and we denote $\phi_p$ by $\chi$. The volume element for such coordinates is $r_p^2 sin\theta_{ep}\, dr_p\, d\theta_{ep}\, d\chi$. Essentially, we are using a relative polar coordinate system for the second particle.

Since $r_e$ and$r_p$ are allowed to take all values consistent with $r$, the integration range for the second particle's coordinates is as follows:

$$\int_0^{2\pi}\int_0^{\pi}\int_{|r_e-r|}^{r_e+r} r_p^2\, sin\theta_{ep} dr_p\, d\theta_{ep} d\chi \qquad 2.116$$

Now, to eliminate the remaining angle $\theta_{ep}$, we use the following relation:

$$r = \sqrt{r_e^2 + r_p^2 - 2r_e r_p\, cos\theta_{ep}} \Rightarrow r^2 = r_e^2 + r_p^2 - 2r_e r_p\, cos\theta_{ep} \qquad 2.117$$

which, by differentiating

$$2r\, dr = 2r_e r_p\, Sin\theta_{ep} d\theta_{ep} \qquad 2.118$$

we arrive at the following relation:

$$\frac{dr}{d\theta_{ep}} = \frac{r_e r_p\, sin\theta_{ep}}{r} \qquad 2.119$$

Therefore, the volume element for the second particle transforms as follows:

$$r_p^2 sin\theta_{ep}\, dr_p\, d\theta_{ep}\, d\chi = \frac{r_p^2 sin\theta_{ep}\, r\, dr_p\, dr\, d\chi}{r_e r_p\, sin\theta_{ep}} = \frac{r_p\, r\, dr_p\, dr\, d\chi}{r_e} \qquad 2.120$$

By multiplying the volume element of the electron by the volume element of the positive particle, the total volume element is obtained as follows:

$$r_e^2 sin\theta_e\, dr_e\, d\theta_e\, d\phi_e\, \frac{r_p\, r\, dr_p\, dr\, d\chi}{r_e} \qquad 2.121$$
$$= r_e r_p\, r\, sin\theta_e\, dr_e dr_p\, dr\, d\theta_e\, d\phi_e\, d\chi$$

Therefore, we have:

$$d\mathbf{r}_p\, d\mathbf{r}_e = r_e r_p r\, sin\theta_e\, dr_e dr_p dr\, d\theta_e\, d\phi_e\, d\chi \qquad 2.122$$

which will have the following integration ranges:

$$\int_0^{2\pi}\int_0^{2\pi}\int_0^{\pi}\int_0^{\infty}\int_0^{\infty}\int_{|r_e-r|}^{r_e+r} r_e r_p r\, sin\theta_e\, dr_p dr_e dr\, d\theta_e\, d\phi_e\, d\chi \qquad 2.123$$

Using equation (2-123) for the wave function (2-115), which does not explicitly depend on angles, simplifies the volume elements (2-122) as follows:

$$d\mathbf{r}_p\, d\mathbf{r}_e = 8\pi^2 r_e r_p r\;\, dr_e dr_p dr \qquad 2.124$$

Therefore, with the availability of three wave functions with different dependencies, each can be used according to the specific conditions of the problem. However, wave function (2-115) is more interesting both conceptually and computationally due to its simultaneous dependency on single-particle and interparticle coordinates, as well as the absence of angular parts.

## 2.5 Energy Components

Using the wave functions introduced in the previous section, the expectation values of energy components can be calculated in both single-particle and center-of-mass coordinate systems.

### 2.5.1 Single-Particle Coordinate System

The kinetic energy of the electron and the positive particle can be calculated using both wave functions (2-109) and (2-115) (although, computationally, wave function (2-115) is more efficient). The expected value of the kinetic energy of the electron and the positive particle, using wave function (2-109), is defined as follows:

$$\begin{aligned}\langle \hat{T}_e \rangle &= \int\int \Psi^*(\boldsymbol{r}_e,\boldsymbol{r}_p)\hat{T}_e\Psi(\boldsymbol{r}_e,\boldsymbol{r}_p)\, d\boldsymbol{r}_e\, d\boldsymbol{r}_p \\ &= -\frac{1}{2\,m_e}\int\int \Psi^*(\boldsymbol{r}_e,\boldsymbol{r}_p)\nabla_e^2\Psi(\boldsymbol{r}_e,\boldsymbol{r}_p)\, d\boldsymbol{r}_e\, d\boldsymbol{r}_p\end{aligned} \qquad 2.125$$

$$\begin{aligned}\langle \hat{T}_p \rangle &= \int\int \Psi^*(\boldsymbol{r}_e,\boldsymbol{r}_p)\hat{T}_p\Psi(\boldsymbol{r}_e,\boldsymbol{r}_p)\, d\boldsymbol{r}_e\, d\boldsymbol{r}_p \\ &= -\frac{1}{2\,m_p}\int\int \Psi^*(\boldsymbol{r}_e,\boldsymbol{r}_p)\nabla_p^2\Psi(\boldsymbol{r}_e,\boldsymbol{r}_p)\, d\boldsymbol{r}_e\, d\boldsymbol{r}_p\end{aligned} \qquad 2.126$$

where

$$\begin{aligned}\nabla_e^2 = &\frac{\partial^2}{\partial r_e^2} + \frac{2}{r_e}\frac{\partial}{\partial r_e} + \frac{1}{r_e^2}\frac{\partial^2}{\partial \theta_{ep}^2} + \frac{1}{r_e^2}\cot(\theta_{ep})\frac{\partial}{\partial \theta_{ep}} \\ &+ \frac{1}{r_e^2\sin^2\theta_{ep}}\frac{\partial^2}{\partial \phi^2}\end{aligned} \qquad 2.127$$

$$\nabla_p^2 = \frac{\partial^2}{\partial r_p^2} + \frac{2}{r_p}\frac{\partial}{\partial r_p} + \frac{1}{r_p^2}\frac{\partial^2}{\partial \theta_{ep}^2} + \frac{1}{r_p^2}\cot(\theta_{ep})\frac{\partial}{\partial \theta_{ep}} + \frac{1}{r_p^2 \sin^2\theta_{ep}}\frac{\partial^2}{\partial \phi^2} \qquad 2.128$$

Due to the wave function (2-109) not depending on the angular part $\boldsymbol{\phi}$, the last term in equations (2-127) and (2-128) is eliminated. By applying these Laplacians to wave function (2-109), we obtain the following expressions:

$$\begin{aligned}
&\nabla_e^2 \Psi(\boldsymbol{r}_e, \boldsymbol{r}_p) \\
&= \frac{1}{\sqrt{-2r_e r_p \cos(\theta_{\mathrm{ep}}) + r_e^2 + r_p^2}}(-2r_e r_p \cos(\theta_{\mathrm{ep}})(2c_e(\alpha \\
&+ c_{\mathrm{ep}}\sqrt{-2r_e r_p \cos(\theta_{\mathrm{ep}}) + r_e^2 + r_p^2}) + \alpha c_{\mathrm{ep}}) + 2\alpha(2c_e r_e^2 \\
&+ c_{\mathrm{ep}} r_p^2 - 1) + (4c_e^2 r_e^2 - 6c_e \\
&+ c_{\mathrm{ep}}^2 r_p^2)\sqrt{-2r_e r_p \cos(\theta_{\mathrm{ep}}) + r_e^2 + r_p^2} \\
&+ \alpha^2\sqrt{-2r_e r_p \cos(\theta_{\mathrm{ep}}) + r_e^2 + r_p^2})\exp(c_{\mathrm{ep}} r_e r_p \cos(\theta_{\mathrm{ep}}) - c_e r_e^2 \\
&- c_p r_p^2 - \alpha\sqrt{-2r_e r_p \cos(\theta_{\mathrm{ep}}) + r_e^2 + r_p^2})
\end{aligned} \qquad 2.129$$

$$
\begin{aligned}
&\nabla_p^2 \Psi(\boldsymbol{r}_e,\boldsymbol{r}_p) \\
&= \frac{1}{\sqrt{-2r_e r_p \cos(\theta_{\text{ep}}) + r_e^2 + r_p^2}} (-2r_e r_p \cos(\theta_{\text{ep}})(2c_p(\alpha \\
&+ c_{\text{ep}}\sqrt{-2r_e r_p \cos(\theta_{\text{ep}}) + r_e^2 + r_p^2}) + \alpha c_{\text{ep}}) + 2\alpha(c_{\text{ep}} r_e^2 \\
&+ 2c_p r_p^2 - 1) + (c_{\text{ep}}^2 r_e^2 + 4c_p^2 r_p^2 \\
&- 6c_p)\sqrt{-2r_e r_p \cos(\theta_{\text{ep}}) + r_e^2 + r_p^2} \\
&+ \alpha^2\sqrt{-2r_e r_p \cos(\theta_{\text{ep}}) + r_e^2 + r_p^2}) \exp(c_{\text{ep}} r_e r_p \cos(\theta_{\text{ep}}) - c_e r_e^2 \\
&- c_p r_p^2 - \alpha\sqrt{-2r_e r_p \cos(\theta_{\text{ep}}) + r_e^2 + r_p^2})
\end{aligned}
\tag{2.130}
$$

Then, by substituting equation (2-110) into equations (2-125) and (2-126), the expectation values of the kinetic energies of the electron and the positive particle are obtained from the following expression:

$$
\begin{aligned}
&\langle \hat{T}_e \rangle = \\
&-\frac{1}{2\,m_e}\int_0^{2\pi}\int_0^{\pi}\int_0^{\infty}\int_0^{2\pi}\int_0^{\pi}\int_0^{\infty} \Psi^*(\boldsymbol{r}_e,\boldsymbol{r}_p)\nabla_e^2\Psi(\boldsymbol{r}_e,\boldsymbol{r}_p) r_e^2 r_p^2\, sin\theta_e\, sin\theta_p\, dr_e d\theta_e d\phi_e\, dr_p d\theta_p \\
&= -\frac{8\pi^2}{2\,m_e}\int_0^{\infty}\int_0^{\pi}\int_0^{\infty} \Psi^*(\boldsymbol{r}_e,\boldsymbol{r}_p)\nabla_e^2\Psi(\boldsymbol{r}_e,\boldsymbol{r}_p) r_e^2 r_p^2\, sin\theta_e\; dr_e d\theta_e\, dr_p
\end{aligned}
\tag{2.131}
$$

$$
\begin{aligned}
\langle \hat{T}_p \rangle &= -\frac{1}{2\,m_p}\int_0^{2\pi}\int_0^{\pi}\int_0^{\infty}\int_0^{2\pi}\int_0^{\pi}\int_0^{\infty} \Psi^*(\boldsymbol{r}_e,\boldsymbol{r}_p)\nabla_p^2\Psi(\boldsymbol{r}_e,\boldsymbol{r}_p) r_e^2 r_p^2\, sin\theta_e\, sin\theta_p\, dr_e d\theta_e d\phi_e\, dr_p d\theta_p d\phi_e \\
&= -\frac{8\pi^2}{2\,m_p}\int_0^{\infty}\int_0^{\pi}\int_0^{\infty} \Psi^*(\boldsymbol{r}_e,\boldsymbol{r}_p)\nabla_p^2\Psi(\boldsymbol{r}_e,\boldsymbol{r}_p) r_e^2 r_p^2\, sin\theta_p\; dr_e d\theta_p\, dr_p
\end{aligned}
\tag{2.132}
$$

Another way to obtain the kinetic energy of the electron and the positive particle is by using wave function (2-115). For this purpose, by substituting equation (2-113) into equations (2-129) and (2-130), we transform them as follows:

$$\nabla_e^2\Psi(r,r_e,r_p) = \frac{1}{r}(-(r_e^2 + r_p^2 - r^2)(2c_e(\alpha + rc_{ep}) + \alpha c_{ep}) + r(4c_e^2 r_e^2 - 6c_e + c_{ep}^2 r_p^2) + 2\alpha(2c_e r_e^2 + c_{ep} r_p^2 - 1) + \alpha^2 r)\exp(\frac{1}{2}c_{ep}(r_e^2 + r_p^2 - r^2) - c_e r_e^2 - c_p r_p^2 - \alpha r) \quad 2.133$$

$$\nabla_p^2\Psi(r,r_e,r_p) = \frac{1}{r}(-(r_e^2 + r_p^2 - r^2)(\alpha c_{ep} + 2c_p(\alpha + rc_{ep})) + r(c_{ep}^2 r_e^2 + 4c_p^2 r_p^2 - 6c_p) + 2\alpha(c_{ep} r_e^2 + 2c_p r_p^2 - 1) + \alpha^2 r)exp(\frac{1}{2}c_{ep}(r_e^2 + r_p^2 - r^2) - c_e r_e^2 - c_p r_p^2 - \alpha r) \quad 2.134$$

Therefore, the kinetic energies of the electron and the positive particle using wave function (2-115) will be as follows:

$$\langle \hat{T}_e \rangle = \int_0^{2\pi}\int_0^{2\pi}\int_0^{\pi}\int_0^{\infty}\int_0^{\infty}\int_{|r_e-r|}^{r_e+r} \Psi^*(\boldsymbol{r}_e,\boldsymbol{r}_p)\, \nabla_e^2\Psi(r,r_e,r_p) r_e r_p r\, sin\theta_e\, dr_p dr_e dr\, d\theta_e\, d\phi_e\, d\chi = -\frac{8\pi^2}{2\, m_e}\int_0^{\infty}\int_0^{\infty}\int_{|r_e-r|}^{r_e+r} \Psi^*(\boldsymbol{r}_e,\boldsymbol{r}_p)\, \nabla_e^2\Psi(r,r_e,r_p) r_e r_p r\;\; dr_p dr_e dr \quad 2.135$$

$$\langle \hat{T}_p \rangle = \int_0^{2\pi}\int_0^{2\pi}\int_0^{\pi}\int_0^{\infty}\int_0^{\infty}\int_{|r_p-r|}^{r_p+r} \Psi^*(\boldsymbol{r}_e,\boldsymbol{r}_p)\, \nabla_p^2\Psi(r,r_e,r_p) r_e r_p r\, sin\theta_e\, dr_e dr_p dr\, d\theta_p\, d\phi_p\, d\chi = -\frac{8\pi^2}{2\, m_p}\int_0^{\infty}\int_0^{\infty}\int_{|r_p-r|}^{r_p+r} \Psi^*(\boldsymbol{r}_e,\boldsymbol{r}_p)\, \nabla_p^2\Psi(r,r_e,r_p) r_e r_p r\;\; dr_e dr_p dr \quad 2.136$$

The expectation value for the potential energy of the oscillator for the electron and the positive particle, using the two wave functions (2-109) and (2-115), is defined as follows:

$$
\begin{aligned}
\langle \widehat{HO}_e \rangle &= \int\int \Psi^*(\boldsymbol{r}_e,\boldsymbol{r}_p)\frac{1}{2}m_e\omega^2 r_e^2\,\Psi(\boldsymbol{r}_e,\boldsymbol{r}_p)\,d\boldsymbol{\tau}_e\,d\boldsymbol{\tau}_p \\
&= \frac{m_e\omega^2}{2}\int_0^{2\pi}\int_0^{\pi}\int_0^{\infty}\int_0^{2\pi}\int_0^{\pi}\int_0^{\infty}\left|\Psi(r_e,r_p)\right|^2 r_e^4\, r_p^2\, sin\theta_e\, sin\theta_p\, dr_e d\theta_e d\phi_e\, dr_p d\theta_p d\phi_e \\
&= \frac{m_e\omega^2}{2}\int_0^{2\pi}\int_0^{2\pi}\int_0^{\pi}\int_0^{\infty}\int_0^{\infty}\int_{|r_e-r|}^{r_e+r}\left|\Psi(r,r_e,r_p)\right|^2 r_e^3\, r_p r\, sin\theta_e\, dr_p dr_e dr\, d\theta_e\, d\phi_e\, d\chi \qquad 2.137 \\
&= \frac{8\pi^2 m_e\omega^2}{2}\int_0^{\infty}\int_0^{\pi}\int_0^{\infty}\left|\Psi(r_e,r_p)\right|^2 r_e^4 r_p^2\, sin\theta_e\;\, dr_e d\theta_e\, dr_p \\
&= \frac{8\pi^2 m_e\omega^2}{2}\int_0^{\infty}\int_0^{\infty}\int_{|r_e-r|}^{r_e+r}\left|\Psi(r,r_e,r_p)\right|^2 r_e^3\, r_p r\, dr_p dr_e dr
\end{aligned}
$$

$$
\begin{aligned}
\langle \widehat{HO}_p \rangle &= \int\int \Psi^*(\boldsymbol{r}_e,\boldsymbol{r}_p)\frac{1}{2}m_p\omega^2 r_p^2\,\Psi(\boldsymbol{r}_e,\boldsymbol{r}_p)\,d\boldsymbol{\tau}_e\,d\boldsymbol{\tau}_p \\
&= \frac{m_p\omega^2}{2}\int_0^{2\pi}\int_0^{\pi}\int_0^{\infty}\int_0^{2\pi}\int_0^{\pi}\int_0^{\infty}\left|\Psi(r_e,r_p)\right|^2 r_p^4\, r_e^2\, sin\theta_e\, sin\theta_p\, dr_e d\theta_e d\phi_e\, dr_p d\theta_p d\phi_e \\
&= \frac{m_p\omega^2}{2}\int_0^{2\pi}\int_0^{2\pi}\int_0^{\pi}\int_0^{\infty}\int_0^{\infty}\int_{|r_p-r|}^{r_p+r}\left|\Psi(r,r_e,r_p)\right|^2 r_p^3\, r_e r\, sin\theta_e\, dr_p dr_e dr\, d\theta_e\, d\phi_e\, d\chi \qquad 2.138 \\
&= \frac{8\pi^2 m_p\omega^2}{2}\int_0^{\infty}\int_0^{\pi}\int_0^{\infty}\left|\Psi(r_e,r_p)\right|^2 r_p^4\, r_e^2\, sin\theta_e\;\, dr_e d\theta_e\, dr_p \\
&= \frac{8\pi^2 m_p\omega^2}{2}\int_0^{\infty}\int_0^{\infty}\int_{|r_p-r|}^{r_p+r}\left|\Psi(r,r_e,r_p)\right|^2 r_p^3\, r_e r\;\, dr_e dr_p dr
\end{aligned}
$$

The expectation value of the interaction energy is also defined and calculated using the two wave functions (2-109) and (2-115) as follows:

$$
\begin{aligned}
INT &= \int\int \Psi^*(\boldsymbol{r}_e,\boldsymbol{r}_p)\left(-\frac{1}{\boldsymbol{r}}\right)\Psi(\boldsymbol{r}_e,\boldsymbol{r}_p)\,d\boldsymbol{r}_e\,d\boldsymbol{r}_p \\
&= \int_0^{2\pi}\int_0^{\pi}\int_0^{\infty}\int_0^{2\pi}\int_0^{\pi}\int_0^{\infty}\left|\Psi(r_e,r_p)\right|^2\frac{1}{(r_e^2+\,r_p^2-2r_e r_p \mathrm{Cos}[\theta_{ep}])^{1/2}} r_e^2\, r_p^2\, sin\theta_e\, sin\theta_p\, dr_e d\theta_e d\phi_e\, dr_p d\theta_p d\phi_e \qquad 2.139 \\
&= -8\pi^2\int_0^{2\pi}\int_0^{2\pi}\int_0^{\pi}\int_0^{\infty}\int_0^{\infty}\int_{|r_p-r|}^{r_p+r}\left|\Psi(r,r_e,r_p)\right|^2\frac{1}{r} r_e r_p r\, sin\theta_e\, dr_p dr_e dr\, d\theta_e\, d\phi_e\, d\chi
\end{aligned}
$$

## 2.5.2 Energy Components in Pseudo particle Coordinates

The energy components for two particles with reduced mass and the mass of the center of mass are calculated using wave function (2-101). As mentioned in the previous section, wave function (2-115) can also be used for calculations in this section due to its dependence on the interparticle distance.

For example, the kinetic energies of the pseudo particles are defined as follows:

$$
\begin{aligned}
\langle \hat{T}_r \rangle &= \int\int \Psi^*(\boldsymbol{R},\boldsymbol{r})\hat{T}_r\Psi(\boldsymbol{R},\boldsymbol{r})\, d\boldsymbol{r}\, d\boldsymbol{R} \\
&= -\frac{1}{2\,\mu}\int_0^{2\pi}\int_0^{\pi}\int_0^{\infty}\int_0^{2\pi}\int_0^{\pi}\int_0^{\infty}\Psi^*(\boldsymbol{R},\boldsymbol{r})\nabla_r^2\Psi(\boldsymbol{R},\boldsymbol{r})R^2 sin\theta_R\, dR\, d\theta_R\, d\phi_R\;\; r^2 sin\theta_r\;\, dr\, d\theta_r\, d\phi_r \\
&= -\frac{1}{2\,\mu}\int_0^{2\pi}\int_0^{2\pi}\int_0^{\pi}\int_0^{\infty}\int_0^{\infty}\int_{|r_p-r|}^{r_p+r}\Psi^*(r,r_e,r_p)\nabla_r^2\Psi(r,r_e,r_p)r_e r_p r\; sin\theta_e\, dr_e dr_p dr\, d\theta_p\, d\phi_p\, d\chi \\
&= -\frac{16\pi^2}{2\,\mu}\int_0^{\infty}\int_0^{\pi}\int_0^{\infty}\Psi^*(\boldsymbol{R},\boldsymbol{r})\nabla_r^2\Psi(\boldsymbol{R},\boldsymbol{r})R^2\, r^2 dR\, dr \\
&= -\frac{8\pi^2}{2\,\mu}\int_0^{\infty}\int_0^{\infty}\int_{|r_p-r|}^{r_p+r}\Psi^*(r,r_e,r_p)\nabla_r^2\Psi(r,r_e,r_p)r_e r_p r\;\; dr_e dr_p dr
\end{aligned}
\qquad 2.140
$$

$$
\begin{aligned}
\langle \hat{T}_R \rangle &= \int\int \Psi^*(\boldsymbol{R},\boldsymbol{r})\hat{T}_R\Psi(\boldsymbol{R},\boldsymbol{r})\, d\boldsymbol{r}\, d\boldsymbol{R} \\
&= -\frac{1}{2\,M}\int_0^{2\pi}\int_0^{\pi}\int_0^{\infty}\int_0^{2\pi}\int_0^{\pi}\int_0^{\infty}\Psi^*(\boldsymbol{R},\boldsymbol{r})\nabla_R^2\Psi(\boldsymbol{R},\boldsymbol{r})R^2 sin\theta_R\, dR\, d\theta_R\, d\phi_R\;\; r^2 sin\theta_r\;\, dr\, d\theta_r\, d\phi_r \\
&= -\frac{1}{2\,M}\int_0^{2\pi}\int_0^{2\pi}\int_0^{\pi}\int_0^{\infty}\int_0^{\infty}\int_{|r_p-r|}^{r_p+r}\Psi^*(r,r_e,r_p)\nabla_R^2\Psi(r,r_e,r_p)r_e r_p r\; sin\theta_e\, dr_e dr_p dr\, d\theta_p\, d\phi_p\, d\chi \\
&= -\frac{16\pi^2}{2\,M}\int_0^{\infty}\int_0^{\pi}\int_0^{\infty}\Psi^*(\boldsymbol{R},\boldsymbol{r})\nabla_R^2\Psi(\boldsymbol{R},\boldsymbol{r})R^2\, r^2\, dR\, dr \\
&= -\frac{8\pi^2}{2\,M}\int_0^{\infty}\int_0^{\infty}\int_{|r_p-r|}^{r_p+r}\Psi^*(r,r_e,r_p)\nabla_R^2\Psi(r,r_e,r_p)r_e r_p r\;\; dr_e dr_p dr
\end{aligned}
\qquad 2.141
$$

In these cases, the Laplacian operators of equations (2-127) and (2-128), considering the angular independence of the wave functions under investigation, transform as follows for the reduced mass and the center of mass:

$$\nabla_r^2 = \frac{\partial^2}{\partial r^2} + \frac{2}{r}\frac{\partial}{\partial r} \qquad 2.142$$

$$\nabla_R^2 = \frac{\partial^2}{\partial R^2} + \frac{2}{R}\frac{\partial}{\partial R} \qquad 2.143$$

However, due to the product form of wave function (2-101) being divided into two parts: the center of mass and the reduced mass, and given the independence of the operators of each part from the other, it is simpler to calculate the energy components using only wave functions (2-19) or (2-21), depending on the type of operator. In this case, the form of the integrals that appear in these calculations and their solutions are as follows:

$$\int_0^\infty x e^{-2ax-2bx^2} dx = \frac{2\sqrt{b} - a e^{\frac{a^2}{2b}} \sqrt{2\pi} \mathrm{Erfc}[\frac{a}{\sqrt{2}\sqrt{b}}]}{8b^{3/2}} = A \qquad 2.144$$

$$\int_0^\infty x^2 e^{-2ax-2bx^2} dx = \frac{-2a\sqrt{b} + (a^2+b) e^{\frac{a^2}{2b}} \sqrt{2\pi} \mathrm{Erfc}[\frac{a}{\sqrt{2}\sqrt{b}}]}{16b^{5/2}} \qquad 2.145$$

$$= \mathrm{B}$$

$$\int_0^\infty x^3 e^{-2ax-2bx^2} dx$$

$$= \frac{2\sqrt{b}(a^2+2b) - a(a^2+3b) e^{\frac{a^2}{2b}} \sqrt{2\pi} \mathrm{Erfc}[\frac{a}{\sqrt{2}\sqrt{b}}]}{32b^{7/2}} = \mathrm{C} \qquad 2.146$$

$$\int_0^\infty x^4 e^{-2ax-2bx^2} dx$$

$$= \frac{-2a\sqrt{b}(a^2+5b) + (a^4+6a^2b+3b^2) e^{\frac{a^2}{2b}} \sqrt{2\pi} \mathrm{Erfc}[\frac{a}{\sqrt{2}\sqrt{b}}]}{64b^{9/2}} = \mathrm{D} \qquad 2.147$$

For the center of mass, the contribution to the kinetic energy is as follows:

$$T_R = \left\langle \eta_{klm}(\boldsymbol{R}) \middle| -\frac{1}{2M} \nabla_R^2 \middle| \eta_{klm}(\boldsymbol{R}) \right\rangle$$

$$= -\frac{2\pi}{M} \int_0^\infty R^2 e^{-\frac{1}{2}M\omega R^2} \nabla_R^2 \, e^{-\frac{1}{2}M\omega R^2} dR = \frac{3\omega}{4} \qquad 2.148$$

and the potential energy of the harmonic oscillator is as follows:

$$HO_R = \left\langle \eta_{klm}(\boldsymbol{R}) \middle| \frac{M}{2}\omega^2 \mathrm{R}^2 \middle| \eta_{klm}(\boldsymbol{R}) \right\rangle = 2M\pi \int_0^\infty R^4 e^{-M\omega R^2}\, dR$$
$$= \frac{3\omega}{4} \qquad 2.149$$

and the total energy of the center of mass will be $\frac{3\omega}{2}$. Additionally, for the reduced mass, the kinetic energy is as follows:

$$T_r = \left\langle \chi_{k'l'm'}(\boldsymbol{r}) \middle| -\frac{1}{2\mu}\nabla_r^2 \middle| \chi_{k'l'm'}(\boldsymbol{r}) \right\rangle$$
$$= -\frac{1}{2\mu}(\alpha^2 - 2\beta + 4\alpha\beta\frac{\mathrm{C}}{\mathrm{B}} + 4\beta^2\frac{\mathrm{D}}{\mathrm{B}} - 2\alpha\frac{\mathrm{A}}{\mathrm{B}} - 4\beta) \qquad 2.150$$

The potential energy of the harmonic oscillator:

$$HO_r = \left\langle \chi_{k'l'm'}(\boldsymbol{r}) \middle| \frac{\mu}{2}\omega^2 r^2 \middle| \chi_{k'l'm'}(\boldsymbol{r}) \right\rangle = \frac{1}{2}\mu\omega^2\frac{\mathrm{D}}{\mathrm{B}} \qquad 2.151$$

and the interaction energy is as follows:

$$\mathrm{INT} = \left\langle \chi_{k'l'm'}(\boldsymbol{r}) \middle| -\frac{1}{r} \middle| \chi_{k'l'm'}(\boldsymbol{r}) \right\rangle = -\frac{\mathrm{A}}{\mathrm{B}} \qquad 2.152$$

It is important to note that the results of the angular part integration ($4\pi$) are not included in the above integrals, as they will cancel each other out when each of the integrals (2-144) to (2-147) is divided by the overlap integral (equation 2-145).

## 2.6 Single-Particle Densities

The normalization constant can be calculated using the normalization condition through any of the three wave functions mentioned in the previous sections, and it will yield the same numerical result. However, due to simplicity, we will use wave function (2-101) here. The normalization condition is defined as follows:

$$N^2 \int \Psi(\boldsymbol{r},\boldsymbol{R})^2\, d\tau$$
$$= N^2 \int\, exp(-2\alpha r - 2\beta r^2 - 2\gamma R^2)\, d\tau$$
$$= N^2 \int_0^{\infty}\int_0^{\infty}\int_0^{\pi}\int_0^{\pi}\int_0^{2\pi}\int_0^{2\pi} exp(-2\alpha r - 2\beta r^2 \qquad 2.153$$
$$- 2\gamma R^2)\ r^2 R^2 dr\, dR\, sin\,\theta_e d\theta_e\, sin\,\theta_p d\theta_p\, d\phi_e\, d\phi_p = 1$$

As a result, the normalization constant will be:

$$N = \frac{1}{\sqrt{\int |\Psi(\boldsymbol{r},\boldsymbol{R})|^2\, d\tau}} \qquad 2.154$$

Given that wave function (2-101) lacks angular dependence, integrating over the angular part leads to the following result:

$$\int_0^{2\pi}\int_0^{2\pi} d\phi_e\, d\phi_p = 4\pi^2 \qquad 2.155$$

$$\int_0^{\pi}\int_0^{\pi} sin\,\theta_e d\theta_e\, sin\,\theta_p d\theta_p = 4 \qquad 2.156$$

Then, by integrating over the variable r, we have:

$$\int_0^{\infty} exp(-2\alpha r - 2\beta r^2 - 2\gamma R^2)\, r^2 dr$$
$$= \frac{e^{-2R^2\gamma}(-2\alpha\sqrt{\beta} + e^{\frac{\alpha^2}{2\beta}}\sqrt{2\pi}(\alpha^2+\beta)\mathrm{erfc}[\frac{\alpha}{\sqrt{2\beta}}])}{16\beta^{5/2}} = \xi \qquad 2.157$$

where the complementary error function is defined as follows:

$$\mathrm{erfc}(x) = 1 - \mathrm{erf}\,(x) \qquad 2.158$$

Subsequently, by integrating $\xi$ over $R$, we obtain the following expression:

$$\int_0^{\infty} \xi\, R^2\, dR = \frac{-\sqrt{2\pi}\alpha\sqrt{\beta} + e^{\frac{\alpha^2}{2\beta}}\pi(\alpha^2+\beta)\text{erfc}\left[\frac{\alpha}{\sqrt{2\beta}}\right]}{128\beta^{5/2}\gamma^{3/2}}$$

$$= \frac{\sqrt{\pi}}{64\beta^{3/2}\gamma^{3/2}}\left(\frac{e^{\frac{\alpha^2}{2\beta}}\sqrt{\pi}(\alpha^2+\beta)\text{erfc}\left[\frac{\alpha}{\sqrt{2\beta}}\right]}{2\beta} - \frac{\alpha}{\sqrt{2\beta}}\right) \qquad 2.159$$

Therefore, the final expression for the normalization constant using the above equations will be:

$$N = 2\pi^{-5/4}\beta^{3/4}\gamma^{3/4}\left(\frac{e^{\frac{\alpha^2}{2\beta}}\sqrt{\pi}(\alpha^2+\beta)\text{erfc}\left[\frac{\alpha}{\sqrt{2\beta}}\right]}{2\beta} - \frac{\alpha}{\sqrt{2\beta}}\right)^{-1/2} \qquad 2.160$$

It is important to note that in the following sections, the steps for calculating single-particle quantities are only explained for the electron, and only the final results are presented for the positive particle. The calculation steps for the positive particle are exactly similar to those for the electron, with the substitution of the positive particle's coordinates.

To calculate the density, we use the intermediate wave function. Since the electron density is obtained by integrating over the positive particle's coordinates (and vice versa), we first need to obtain the appropriate differentials for the density in the volume elements of the intermediate wave function. We know that the fundamental definition of electron density is as follows:

$$\rho_e(r_e) = \int_0^{2\pi}\int_0^{\pi}\int_0^{\infty} \left|\Psi(\boldsymbol{r}_e,\boldsymbol{r}_p)\right|^2\, r_p^2\, sin\theta_p\, dr_p d\theta_p d\phi_p \qquad 2.161$$

Additionally, using the normalization condition, we have:

$$\int_0^{2\pi}\int_0^{\pi}\int_0^{\infty}\underbrace{\int_0^{2\pi}\int_0^{\pi}\int_0^{\infty}\left|\Psi(\boldsymbol{r}_e,\boldsymbol{r}_p)\right|^2 r_p^2\, sin\theta_p dr_p d\theta_p d\phi_p}_{\rho_e(r_e)}\; r_e^2\, sin\theta_e\, dr_e d\theta_e d\phi_e = 1 \qquad 2.162$$

where the inner part of the integral is the electron density. Now, we rewrite the above equation (or the normalization condition) based on the intermediate wave function and its volume elements:

$$\int_0^{2\pi}\int_0^{2\pi}\int_0^{\pi}\int_0^{\infty}\int_0^{\infty}\int_{|r_e-r|}^{r_e+r}\left|\Psi(r,r_e,r_p)\right|^2\; r_p r_e\, r\, sin\theta_e dr_p\; dr_e dr\; d\theta_e d\phi_e d\chi = 1 \qquad 2.163$$

By rearranging the above equation so that the electron differentials are in the outer part (like in equation 2-162), we have:

$$\int_0^{2\pi}\int_0^{\pi}\int_0^{\infty}\int_0^{2\pi}\int_0^{\infty}\int_{|r_e-r|}^{r_e+r}\left|\Psi(r,r_e,r_p)\right|^2\; r_p\, r\, dr_p dr d\chi\;\; r_e\, sin\theta_e dr_e\; d\theta_e d\phi_e = 1 \qquad 2.164$$

Then, by multiplying and dividing the above equation by $\mathrm{r_e}$, we arrive at an equation in the form of (2-162):

$$\int_0^{2\pi}\int_0^{\pi}\int_{\epsilon}^{\infty}\underbrace{\int_0^{2\pi}\int_0^{\infty}\int_{|r_e-r|}^{r_e+r}\left|\Psi(r,r_e,r_p)\right|^2\;\frac{r_p r}{r_e}\; dr_p dr d\chi}_{\rho_e(r_e)}\;\; r_e^2\, sin\theta_e dr_e\; d\theta_e d\phi_e = 1 \qquad 2.165$$

Due to the presence of $r_e$ in the denominator, it cannot be zero, and as a result, the lower limit of its integral starts from a very small non-zero value ($\epsilon$). The electron density in the intermediate coordinates is as follows:

$$\rho_e(r_e) = \int_0^{2\pi}\int_0^{\infty}\int_{|r_e-r|}^{r_e+r} \left|\Psi(r,r_e,r_p)\right|^2 \ \frac{r_p r}{r_e}\, dr_p dr\, d\chi \tag{2.166}$$

After integrating the angular part, we have:

$$\rho_e(r_e) = \frac{2\pi}{r_e}\int_0^{\infty}\int_{|r_e-r|}^{r_e+r} \left|\Psi(r,r_e,r_p)\right|^2 \ r_p r\, dr_p dr \tag{2.167}$$

And finally, after integrating over the radial parts, we have:

$$\begin{aligned}
\rho_e(r_e) &= \frac{1}{8\sqrt{2}\gamma r_e m_p\left(\gamma m_p^2 + \beta M^2\right)\sqrt{\beta + \frac{\gamma m_p^2}{M^2}}}\pi^{3/2}M^2N^2e^{-2\gamma r_e^2}((\alpha M \\
&- 2\gamma r_e m_p)\operatorname{erf}(\frac{\alpha M - 2\gamma r_e m_p}{\sqrt{2}M\sqrt{\beta + \frac{\gamma m_p^2}{M^2}}})\exp(\frac{(\alpha M - 2\gamma r_e m_p)^2}{2(\gamma m_p^2 + \beta M^2)}) \\
&- (2\gamma r_e m_p + \alpha M)\operatorname{erf}(\frac{2\gamma r_e m_p + \alpha M}{\sqrt{2}M\sqrt{\beta + \frac{\gamma m_p^2}{M^2}}})\exp(\frac{(2\gamma r_e m_p + \alpha M)^2}{2(\gamma m_p^2 + \beta M^2)}) \\
&+ (\alpha M(e^{\frac{4\alpha\gamma M r_e m_p}{\gamma m_p^2 + \beta M^2}} - 1) + 2\gamma r_e m_p(e^{\frac{4\alpha\gamma M r_e m_p}{\gamma m_p^2 + \beta M^2}} \\
&+ 1))\exp(\frac{(\alpha M - 2\gamma r_e m_p)^2}{2(\gamma m_p^2 + \beta M^2)})) \qquad , r_e \neq 0
\end{aligned} \tag{2.168}$$

$$\rho_p(r_p) = \frac{1}{8\sqrt{2}\gamma m_e r_p(\gamma m_e^2 + \beta M^2)\sqrt{\beta + \frac{\gamma m_e^2}{M^2}}} \pi^{3/2} M^2 N^2 e^{-2\gamma r_p^2} ((\alpha M - 2\gamma m_e r_p)\operatorname{erf}(\frac{\alpha M - 2\gamma m_e r_p}{\sqrt{2}M\sqrt{\beta + \frac{\gamma m_e^2}{M^2}}})\exp(\frac{(\alpha M - 2\gamma m_e r_p)^2}{2(\gamma m_e^2 + \beta M^2)}) - (2\gamma m_e r_p + \alpha M)\operatorname{erf}(\frac{2\gamma m_e r_p + \alpha M}{\sqrt{2}M\sqrt{\beta + \frac{\gamma m_e^2}{M^2}}})\exp(\frac{(2\gamma m_e r_p + \alpha M)^2}{2(\gamma m_e^2 + \beta M^2)}) + (\alpha M(e^{\frac{4\alpha\gamma M m_e r_p}{\gamma m_e^2 + \beta M^2}} - 1) + 2\gamma m_e r_p(e^{\frac{4\alpha\gamma M m_e r_p}{\gamma m_e^2 + \beta M^2}} + 1))\exp(\frac{(\alpha M - 2\gamma m_e r_p)^2}{2(\gamma m_e^2 + \beta M^2)})) \quad , r_p \neq 0 \tag{2.169}$$

Equations (2-168) and (2-169) represent the final analytical forms of the electron and positive particle densities for the intermediate wave function. Their shapes are illustrated in section 5-1-2.

# 3 Electron-Positively Charged Particle Correlation

## 3.1 Fundamental Definition of Correlation in Physics

The most general and, in a sense, fundamental mathematical definition of correlation between two particles in quantum mechanics can be presented using an inherently statistical quantity called the "pair density" or “two-particle density”. The two-particle density expresses the probability of finding two particles (distinguishable or indistinguishable), one in the volume element $d\boldsymbol{r}_1$ and the other in the volume element $\boldsymbol{dr}_2$. The two-particle density for two indistinguishable particles (such as two electrons) in a system consisting of $N$ particles, after integrating over spins, is defined as follows [98]:

$$\Gamma(\boldsymbol{r}_1,\boldsymbol{r}_2) = N(N-1)\int \dots \int |\Psi(\boldsymbol{r}_1, \dots, \boldsymbol{r}_N)|^2 \, \boldsymbol{dr}_3, \dots, \boldsymbol{dr}_N \qquad 3.1$$

and for two distinguishable particles (such as an electron and a PCP) in a system consisting of $N$ and $N'$ numbers of these two particles, after integrating over spins, it is defined as follows:

$$\Gamma(\boldsymbol{r}_1^e,\boldsymbol{r}_1^p) = NN'\int \dots \int \Psi^*(\boldsymbol{r}_1^e, \dots, \boldsymbol{r}_N^e, \boldsymbol{r}_1^p, \dots, \boldsymbol{r}_{N'}^p)\, \Psi(\boldsymbol{r}_1^e, \dots, \boldsymbol{r}_N^e, \boldsymbol{r}_1^p, \dots, \boldsymbol{r}_{N'}^p)\, \boldsymbol{dr}_2^e, \dots, \boldsymbol{dr}_N^e, \boldsymbol{dr}_2^p, \dots, \boldsymbol{dr}_{N'}^p \qquad 3.2$$

If the two particles are independent of each other, the two-particle density simply equals the product of the densities of each particle. Thus, whenever the two-particle density deviates from the product of the single-particle densities, we are dealing with correlated particles. Therefore, the two-particle density can be decomposed as follows:

$$\Gamma(\boldsymbol{r}_1^e,\boldsymbol{r}_1^p) = \rho_e(\boldsymbol{r}_1^e)\rho_p(\boldsymbol{r}_1^p) + \Gamma_c(\boldsymbol{r}_1^e,\boldsymbol{r}_1^p) \qquad 3.3$$

Where $\Gamma_c(\boldsymbol{r}_1^e,\boldsymbol{r}_1^p)$ represents the correlation part of the two-particle density, which is zero for uncorrelated systems. In this chapter, we will explore various aspects of e-PCP correlation, which is one of the main objectives of developing the EHM in this dissertation.

## 3.2 E-PCP Correlation

### 3.2.1 Nature

The e-PCP correlation differs fundamentally from electron-electron correlation in at least two aspects: the nature of the interaction and the mass of the particles. The e-PCP interaction, unlike the purely electronic case, is attractive, which lowers the energy level. As we saw in section 2-2-2, this attraction is one of the obstacles to achieving an analytical solution for the ground state of the EHM using the conventional power series method. Furthermore, as we will see, the attractive nature of the e-PCP correlation results in the creation of a correlation hill (as opposed to a correlation hole like in electronic correlation).

### 3.2.2 History

The PCPs examined in this study encompass a mass range from 1 (positron) to 1836 (proton) (details of the calculations will be provided in the next chapter). The e-PCP correlation has been extensively studied for particles at both ends of this mass spectrum, i.e., positrons and protons, due to the significance of positron annihilation spectroscopy in studying the electronic structure of solids and the presence of protons in atomic nuclei (hence the term electron-nucleus correlation is also used). However, other PCPs within this spectrum also have numerous applications, as mentioned in section 1-5. Below is a brief history of the research conducted on e-PCP correlation.

The electron-positron correlation gained attention early on, starting in the 1950s, due to the importance of positron annihilation. In this context, Ferrell published a review article titled "Theory of Positron Annihilation in Solids" in 1956 [99]. In 1960, Kahana introduced a correlation function for electron-positron pairs in metals, reconciling the Sommerfeld model of a metal with experimental data, as the Sommerfeld model alone could not correctly reproduce the experimental positron annihilation rates in metals [100]. In 1966, Hamann calculated the self-energy of the positron (due to electron-positron correlation) in an electron gas [101]. Following this, in 1969, Bergersen and Pajanne calculated the effective mass of the positron, emphasizing nonlinear terms in the electron-positron interaction [102]. In 1986, Boronski and Nieminen introduced a local electron-positron correlation function [103]. In 1981, Chakraborty developed a formalism for self-consistent calculation of positron annihilation characteristics in real metals, taking into account electron-positron correlation effects [104]. Subsequently, various calculations were carried out to incorporate electron-positron correlation effects in metals and solids [105], [106]. Electron-positron correlation remains an active research area, with recent publications on the topic [107], [108].

Electron-proton correlation has been studied in scattered research efforts, such as estimating electron-proton correlation in a hydrogen atom as an entangled system using density matrix theory and von Neumann entropy [109], or investigating the effect of electron-proton correlation (especially proton movement) on hydrogen bonds in charge transfer processes within halogen-bridged complexes [110]. However, with the introduction of multi-component ab initio frameworks beyond the Born-Oppenheimer approximation, allowing for the treatment of the proton as a quantum particle, electron-proton correlation has received increased attention. One of the earliest papers in this area was the calculation of electron-proton correlation in the hydrogen tunnelling process, published by Hammes-Schiffer and Pak in 2004 [54]. Since then, this correlation has been studied within perturbation theory frameworks [75], [111]–[113], and especially within multi-component density functional theory to develop a suitable functional for this correlation [70]–[72], [114].

Correlation of other particles, such as electron-muon correlation, has been primarily of interest in high-energy physics [115]. Recently, this correlation has entered the domain of multi-component density functional theory, and an initial functional for its calculation has been developed [73].

## 3.3 Mathematical Framework

### 3.3.1 Pair Correlation Factor

The pair correlation factor is a measure for determining the degree of particle correlation, which is related to the two-particle density through the following relationship:

$$\Gamma(r_e, r_p) = \rho_e(r_e)\,\rho_p(r_p)[1 + \mathcal{F}(r_e, r_p)] \quad 3.4$$

In other words, the correlation factor indicates the deviation of the two-particle density from the product of the single-particle densities. This factor is related to the pair distribution function $\mathcal{G}(r_e, r_p)$ through the following relationship:

$$\mathcal{G}(r_e, r_p) = 1 + \mathcal{F}(r_e, r_p) \quad 3.5$$

Based on equations (3-3), (3-4), and (3-5), we can write:

$$\mathcal{F}(r_e, r_p) = \frac{\Gamma(r_e, r_p)}{\rho_e(r_e)\,\rho_p(r_p)} - 1 = \frac{\Gamma_c(r_e, r_p)}{\rho_e(r_e)\,\rho_p(r_p)} \quad 3.6$$

$$\mathcal{G}(r_e, r_p) = \frac{\Gamma(r_e, r_p)}{\rho_e(r_e)\,\rho_p(r_p)} = \frac{\Gamma_c(r_e, r_p)}{\rho_e(r_e)\,\rho_p(r_p)} + 1 \quad 3.7$$

### 3.3.2 Correlation Hill

The correlation hill can be considered analogous to the correlation hole in electronic systems. In electronic systems, electron correlation or instantaneous interactions between two electrons lead to a depletion of density around one electron, creating a "correlation hole" around it [98]. In the EHM, due to the attractive interaction between two particles (in contrast to the repulsive interaction in electronic systems), instead of creating a hole around each particle, a hill is formed. In other words, the e-PCP correlation in a two-component system results in the accumulation of electron density around the positive particle or the accumulation of the positive particle density around the electron, creating a "correlation hill."

In electronic structure theory, the correlation hole is divided into two parts: the Fermi hole (or exchange hole) and the Coulomb hole (or correlation hole). The Fermi hole arises from the Pauli exclusion principle and exists for electrons with the same spin, while the Coulomb hole results from Coulomb repulsion between electrons. However, in the EHM, due to the distinguishability of the particles, only the Coulomb correlation hill is formed. Before presenting the mathematical framework of the correlation hill, a brief history of various definitions of the correlation hole is provided.

The quantities "correlation hole" [97], [116] and "exchange hole" [117] were first introduced by Coulson and Neilson in 1961. They defined these quantities inspired by the traditional definition of correlation energy in electronic structure theory as the difference between the exact and Hartree-Fock two-particle distribution functions (or inter-particle radial density). However, alternative definitions for these hole quantities were introduced in the 1960s, particularly in 1967 by McWeeny [118], [119]. In these definitions, instead of working with the correlation factor, which is a unitless quantity representing the difference between the two-particle density and the product of single-electron densities (equation 3-4), a new quantity called the "hole function" with the unit of single-electron density was introduced [120]. After the publication of the groundbreaking papers by Hohenberg, Kohn, and Sham [121], [122], and the birth of Density Functional Theory (DFT) in 1964, and given the better alignment of the recent hole definitions with this theory, notable papers examining various aspects of this hole definition within the DFT framework were published in subsequent years [122]–[134]. Valuable review articles on the nature and history of DFT and the development of the hole concept by prominent developers of this theory are available in references [135] and [136]. Coulson and Neilson's definitions were also used to some extent, especially within wavefunction-based frameworks [137], [138].

Since the primary goal of investigating e-PCP correlation in this study is to better understand it for developing an efficient functional within the DFT framework, we will use the appropriate correlation (hill) hole definition in DFT. However, two-particle distribution functions, mean, and variance of the two-particle distance will also be analytically provided to gain a deeper understanding of e-PCP correlation.

In electronic systems, due to the identical nature of the two particles, there is only one type of hole. In contrast, in the two-component e-PCP system, due to the different nature of the particles (both in terms of charge and mass), two types of hills (one belonging to the electron and the other to the PCP) are formed.

Since the correlation hole and hill indicate the deviation of density around one particle interacting with a reference particle, we need a quantity that describes such a situation to define them. This quantity is called the conditional density, which indicates the probability of finding particle 2 at coordinates 2 around particle 1 (reference particle) if particle 1 is at coordinates 1. It is defined for the electron and PCP, respectively, as follows:

$$\rho^{cond}(r_e|r_p) = \frac{\Gamma(r_e,r_p)}{\rho_p(r_p)} \qquad 3.8$$

$$\rho^{cond}(r_p|r_e) = \frac{\Gamma(r_e,r_p)}{\rho_e(r_e)} \qquad 3.9$$

Therefore, $\rho^{cond}(r_e|r_p)$ represents the probability of finding the electron at $r_e$ given that the PCP is at $r_p$ (and similarly for $\rho^{cond}(r_p|r_e)$). Since $\Gamma(r_e,r_p)$ can be decomposed as in equation (3-3), substituting (3-3) into (3-8) and (3-9) yields:

$$\rho^{cond}(r_e|r_p) = \rho_e(r_e) + \frac{\Gamma_c(r_e,r_p)}{\rho_p(r_p)} = \rho_e(r_e) + \rho_c^{hill}(r_e|r_p) \qquad 3.10$$

$$\rho^{cond}(r_p|r_e) = \rho_p(r_p) + \frac{\Gamma_c(r_e,r_p)}{\rho_e(r_p)} = \rho_p(r_p) + \rho_c^{hill}(r_p|r_e) \tag{3.11}$$

where $\rho_c^{hill}(r_e|r_p)$ and $\rho_c^{hill}(r_p|r_e)$ are the densities of the correlation hill for the electron and the PCP, respectively. Equations (3-10) and (3-11) demonstrate how the hill density deviates the conditional density from the single-particle density. Therefore, the correlation hills for the electron and PCP are defined as follows, respectively:

$$\begin{aligned}\rho_c^{hill}(r_e|r_p) &= \rho^{cond}(r_e|r_p) - \rho_e(r_e) = \frac{\Gamma(r_e,r_p)}{\rho_p(r_p)} - \rho_e(r_e) \\ &= \frac{\Gamma_c(r_e,r_p)}{\rho_p(r_p)}\end{aligned} \tag{3.12}$$

$$\begin{aligned}\rho_c^{hill}(r_p|r_e) &= \rho^{cond}(r_p|r_e) - \rho_p(r_p) = \frac{\Gamma(r_e,r_p)}{\rho_e(r_e)} - \rho_p(r_p) \\ &= \frac{\Gamma_c(r_e,r_p)}{\rho_e(r_e)}\end{aligned} \tag{3.13}$$

### 3.3.3 Correlation Hill Sum Rule

Since the two-particle density for a two-component system consisting of $N$ electrons and $N'$ PCPs is defined as in equation (3-2), integrating over it equals $NN'$. On the other hand, considering the decomposition of $\Gamma(\boldsymbol{r}_e,\boldsymbol{r}_p)$ according to equation (3-3) and the correlation hill expressions (equations 3-12 and 3-13) for the electron and the PCP, we need to examine the integration over conditional densities and single-particle densities to derive the correlation hill sum rule.

Given that the integration over single-particle densities is:

$$\int \rho_e(r_e)d\boldsymbol{r}_e = N \tag{3.14}$$

$$\int \rho_p(r_p)d\boldsymbol{r}_p = N' \tag{3.15}$$

Integration over the conditional densities of the electron and the PCP (equations 3-8 and 3-9) results in the following:

$$\int \rho^{cond}(r_e|r_p)\, d\boldsymbol{r}_e = \int \frac{\Gamma(\boldsymbol{r}_e,\boldsymbol{r}_p)}{\rho_p(r_p)}\, d\boldsymbol{r}_e = \frac{N\rho_p(r_p)}{\rho_p(r_p)} = N \tag{3.16}$$

$$\int \rho^{cond}(r_p|r_e)\, d\boldsymbol{r}_p = \int \frac{\Gamma(\boldsymbol{r}_e,\boldsymbol{r}_p)}{\rho_e(r_e)}\, d\boldsymbol{r}_p = \frac{\rho_e(r_e)N'}{\rho_e(r_e)} = N' \tag{3.17}$$

Therefore, according to the above results and the correlation hill expressions (equations 3-12 and 3-13), the sum rule for them will be as follows:

$$\int \rho_c^{hill}(r_e|r_p)d\boldsymbol{r}_e = N - N = 0 \tag{3.18}$$

$$\int \rho_c^{hill}(r_p|r_e)d\boldsymbol{r}_p = N' - N' = 0 \tag{3.19}$$

### 3.3.4 Correlation Energy

The correlation energy of a system, which measures the degree of instantaneous interaction between two particles, can be defined in two ways: independent of an external reference and dependent on an external reference. The latter is further divided into two categories based on the type of external reference (Hartree-Fock or Kohn-Sham).

Independent correlation is defined uniformly across all approaches using only the two-particle density of the system $\Gamma(r_e,r_p)$. Based on equation (3-3), we can decompose the total interaction energy of two particles as follows:

$$W = -\int \frac{\Gamma(r_e,r_p)}{r} d\boldsymbol{r}_e d\boldsymbol{r}_p = -\int \frac{\rho_e(r_e)\rho_p(r_p)}{r} d\boldsymbol{r}_e d\boldsymbol{r}_p - \int \frac{\Gamma_c(r_e,r_p)}{r} d\boldsymbol{r}_e d\boldsymbol{r}_p \qquad 3.20$$

For a two-component system consisting of electrons and PCPs, we can write it as follows:

$$W = \int \rho_e(r_e)\, V_{coul}(r_e)\, d\boldsymbol{r}_e - \int \frac{\rho_e(r_e)\rho_c^{hill}(r_p|r_e)}{r} d\boldsymbol{r}_e d\boldsymbol{r}_p = \int \rho_p(r_p)\, V_{coul}(r_p)\, d\boldsymbol{r}_p - \int \frac{\rho_p(r_p)\rho_c^{hill}(r_e|r_p)}{r} d\boldsymbol{r}_p d\boldsymbol{r}_e = W_{Coul} + W_c \qquad 3.21$$

where

$$V_{coul}(r_e) = v_e^{J_{ep}}(r_e) = -\int \frac{\rho_p(r_p)}{|\boldsymbol{r}_e - \boldsymbol{r}_p|} d\boldsymbol{r}_p \qquad 3.22$$

$$V_{coul}(r_p) = v_p^{J_{ep}}(r_p) = -\int \frac{\rho_e(r_e)}{|\boldsymbol{r}_e - \boldsymbol{r}_p|} d\boldsymbol{r}_e \qquad 3.23$$

These equations show that $\mathrm{W_{Coul}}$ is ultimately equal to $\mathrm{J_{ep}}$ (equation 1-62). Therefore, the independent correlation energy can be defined as follows:

$$W_c = W - W_{Coul} \qquad 3.24$$

The traditional or external-reference-dependent correlation energy in the Hartree-Fock method is defined as the difference between the exact non-relativistic energy and the Hartree-Fock energy of the system:

$$E_c^{HF} = E_{exact} - E_{HF} \qquad 3.25$$

This definition, which is the most common definition of correlation in electronic structure theory, is typically used to calculate the correlation energy using post-Hartree-Fock methods. If multi-component Hartree-Fock methods are employed, the above definition can be extended to multi-component systems. However, in this case, the calculated correlation energy is equal to the sum of all types of correlations present in the system. Separating these correlation contributions from one another in a unique manner is not possible and may vary depending on the reference definition [139]. Given that in the EHM we deal with e-PCP correlation, these complexities are not present here. To obtain the external-reference-dependent correlation energy in density functional theory, we start from the total energy functional of a two-component system consisting of an electron and a PCP (inspired by equation 1-48), defined as follows [140]:

$$E[\rho_e,\rho_p] = \int \rho_e(\boldsymbol{r_e})\, v_e^{ext}(\boldsymbol{r_e})\, d\boldsymbol{r}_e + \int \rho_p(\boldsymbol{r_p})\, v_p^{ext}(\boldsymbol{r_p})\, d\boldsymbol{r}_p + F'[\rho_e,\rho_p] \qquad 3.26$$

where

$$F'[\rho_e,\rho_p] = \left(T_e[\rho_e] + T_p[\rho_p]\right) + J_{ep}[\rho_e,\rho_p] + W_c[\rho_e,\rho_p] \qquad 3.27$$

On the other hand, by rewriting equation (1-59) and removing the exchange terms and interactions of similar particles, we get:

$$F[\rho_e,\rho_p] = \left(T_e^s[\rho_e] + T_p^s[\rho_p]\right) + J_{ep}[\rho_e,\rho_p] + E_{epc}[\rho_e,\rho_p] \qquad 3.28$$

By equating the two equations (3-26) and the rewritten form of equation (1-58), we arrive at the following result:

$$\begin{aligned} E_{epc}\left[\rho_e,\rho_p\right] &= E_c^{KS} \\ &= \left(T_e[\rho_e] - T_e^s[\rho_e]\right) + \left(T_p\left[\rho_p\right] - T_e^s\left[\rho_p\right]\right) \\ &+ W_c\left[\rho_e,\rho_p\right] = T_e^c + T_p^c + W_c \end{aligned} \qquad 3.29$$

where $T_c$ represents the difference between the exact kinetic energy and the kinetic energy of the Kohn-Sham orbitals for the electron and the PCP.

On the other hand, the expectation value of the energy for the two-component system consisting of the electron and the PCP using the Kohn-Sham orbitals and the total system Hamiltonian (2-69) is defined as follows:

$$\begin{aligned} E_{KS} &= \langle \Psi_{ks}|\hat{H}|\Psi_{ks}\rangle \\ &= \int \rho_e(\boldsymbol{r_e})\, v_e^{ext}(\boldsymbol{r_e})\, d\boldsymbol{r}_e \\ &+ \int \rho_p(\boldsymbol{r_p})\, v_p^{ext}(\boldsymbol{r_p})\, d\boldsymbol{r}_p + F''\left[\rho_e,\rho_p\right] \end{aligned} \qquad 3.30$$

where

$$F''[\rho_e,\rho_p] = \left(T_e^s[\rho_e] + T_p^s[\rho_p]\right) + J_{ep}[\rho_e,\rho_p] \qquad 3.31$$

By comparing equations (3-30) and (3-26) or the rewritten form of equation (1-58), we arrive at the following final result:

$$E_c^{KS} = E_{exact} - E_{KS} \tag{3.32}$$

Therefore, the external-reference-dependent correlation energy in density functional theory is also defined as the difference between the exact non-relativistic energy and the Kohn-Sham system energy. The values of the three defined correlation energies differ from one another. Since the Hartree-Fock determinant is the determinant with the lowest possible energy, $E_{KS}$ must necessarily be greater than $E_{HF}$, and consequently, the DFT correlation energy should be more negative (larger in absolute value) than the HF correlation energy.

### 3.3.5 Two-Particle Distribution Functions

To gain a deeper understanding of the effects of e-PCP correlation, two-particle distribution functions can be used, as they contain the correlation information between the two particles. To define the two-particle distribution function, we express the expectation value of the distance between particles using the intermediate wavefunction (2-115) as follows:

$$\langle r\rangle = \int_0^{2\pi}\int_0^{\pi}\int_0^{\infty}\int_0^{2\pi}\int_0^{\infty}\int_{|r_e-r|}^{r_e+r} r\left|\Psi(r,r_e,r_p)\right|^2\, r_p\, r\, r_e dr_p dr d\chi\, sin\theta_e dr_e\, d\theta_e d\phi_e \tag{3.33}$$

The above expression can be written as follows by introducing the distribution function f(r):

$$\langle r\rangle = \int_0^\infty r\, f(r)\, dr \qquad 3.34$$

This distribution function depends only on $r$ and is averaged over all positions of the two particles. After integrating over the angular parts of (3-33), we have:

$$\langle r\rangle = \int_0^\infty \underbrace{r\, 8\pi^2 \int_0^\infty \int_{|r_e-r|}^{r_e+r} \left|\Psi(r,r_e,r_p)\right|^2\, r_p r\, r_e dr_p\, dr_e}_{f(r)}\, dr \qquad 3.35$$

Using the normalization condition for $f(r)$ and equation (2-164), we have:

$$\int_0^\infty f(r)\, dr = 8\pi^2 \int_0^\infty \int_0^\infty \int_{|r_e-r|}^{r_e+r} \left|\Psi(r,r_e,r_p)\right|^2\, r_p\, r\, r_e dr_p dr dr_e = 1 \qquad 3.36$$

Comparison of the above equation with equation (3-35) shows that the expression for $f(r)$ is as follows:

$$f(r) = \int_0^\infty \underbrace{\int_{|r_e-r|}^{r_e+r} 8\pi^2 \left|\Psi(r,r_e,r_p)\right|^2\, r_p r\, r_e\, dr_p}_{g(r,r_e)}\, dr_e \qquad 3.37$$

In the above equation, by performing the inner integration over $r_p$, we obtain a local two-particle distribution function $g(r,r_e)$, which is as follows:

$$g(r,r_e) = \frac{2\pi^2 N^2 M r r_e e^{-2r(\alpha+\beta r)}}{\gamma m_p} \qquad 3.38$$

$$\left(e^{-\frac{2\gamma((r_e-r)m_p+m_e r_e)^2}{M^2}}-e^{-\frac{2\gamma((r_e+r)m_p+m_e r_e)^2}{M^2}}\right)$$

This local distribution function provides a more precise picture of the correlation effect, as it is defined for different positions of the electron at $r_e$ while considering the distance between the particles as $r$. Since this distribution function depends on both the distance between the particles and the location of the first particle, it will be different for the electron and the PCP. $g(r,r_p)$ is also obtained as follows (by following the steps above):

$$g(r,r_p)=\frac{2\pi^2N^2Mrr_pe^{-2r(\alpha+\beta r)}}{\gamma m_e}\left(e^{-\frac{2\gamma(m_e(r_p-r)+m_p r_p)^2}{M^2}}-e^{-\frac{2\gamma(m_e(r_p+r)+m_p r_p)^2}{M^2}}\right) \quad 3.39$$

Before continuing with the calculations to obtain the average two-particle distribution function, it is interesting to note that the single-particle densities of the electron and the PCP (2-168 and 2-169) can be calculated as follows by performing the following operations on equations (3-38) and (3-39), respectively:

$$\begin{aligned}\rho_e(r_e)&=\frac{1}{4\pi r_e^2}\int_0^\infty g(r,r_e)\,dr\\&=\frac{8\pi^2}{4\pi r_e^2}\int_0^\infty\int_{|r_e-r|}^{r_e+r}\left|\Psi(r,r_e,r_p)\right|^2\,r_p r\,r_e\,dr_p\,dr\\&=\frac{2\pi}{r_e}\int_0^\infty\int_{|r_e-r|}^{r_e+r}\left|\Psi(r,r_e,r_p)\right|^2\,r_p r\,dr_p\,dr\end{aligned} \quad 3.40$$

$$\rho_p(r_p) = \frac{1}{4\pi r_p^2}\int_0^\infty g(r,r_p)\,dr$$

$$= \frac{8\pi^2}{4\pi r_p^2}\int_0^\infty \int_{|r_p-r|}^{r_p+r} \left|\Psi(r,r_e,r_p)\right|^2 \; r_p r\, r_e\, dr_e\; dr \qquad 3.41$$

$$= \frac{2\pi}{r_p}\int_0^\infty \int_{|r_p-r|}^{r_p+r} \left|\Psi(r,r_e,r_p)\right|^2 \; r_e r\, dr_e\; dr$$

Returning to equations (3-37), (3-38), and (3-39) and integrating over $r_e$ and $r_p$ respectively, we have:

$$f(r) = \int_0^\infty g(r,r_e)\,dr_e = \int_0^\infty g(r,r_p)\,dr_p \qquad 3.42$$

and we arrive at the final expression for the average two-particle distribution function:

$$f(r) = \frac{N^2\sqrt{2}e^{-2r(\alpha+r\beta)}\pi^{5/2}r^2}{\gamma^{3/2}} \qquad 3.43$$

Using the definition of $\langle r\rangle$ for the wave functions (2-17) and (2-18), we can write:

$$\langle r\rangle = \int\int \Psi(\boldsymbol{R},\boldsymbol{r})^*\, r\, \Psi(\boldsymbol{R},\boldsymbol{r})\, d\boldsymbol{R}\, d\boldsymbol{r}$$

$$= \int\int r\, |\eta_{klm}(\boldsymbol{R})\,|^2 |\,\chi_{k'l'm'}(\boldsymbol{r})|^2\, d\boldsymbol{R}\, d\boldsymbol{r}$$

$$= \int r |\,\chi_{k'l'm'}(\boldsymbol{r})|^2\, d\boldsymbol{r} \qquad 3.44$$

$$= \int_0^{2\pi}\int_0^{\pi}\int_0^\infty r\, r^2\, |\chi_{k'l'm'}(\boldsymbol{r})|^2\, sin\theta_r dr\, d\theta_r d\phi_r$$

$$= \int_0^\infty r\, \underbrace{4\pi r^2\, |\chi_{k'l'm'}(\boldsymbol{r})|^2}_{f(r)}\, dr = \int_0^\infty r\, f(r)\, dr$$

where the fact that the wave function of the center-of-mass part has a spherical harmonic $Y_{lm}(\theta,\phi)$ and is normalized as follows is utilized:

$$\int_0^\infty |\eta_{klm}(\boldsymbol{R})|^2 d\boldsymbol{R}$$
$$= \int_0^{2\pi}\int_0^{\pi}\int_0^\infty R^2|\eta_{klm}(\boldsymbol{R})|^2 \, sin\theta_R dR \, d\theta_R d\phi_R \qquad 3.45$$
$$= N_{lm}^2 \int_0^{2\pi}\int_0^{\pi}\int_0^\infty R^2 \, \tau_{lm}^2(R) \, Y_{lm}^2(\theta,\phi) \, sin\theta_R dR \, d\theta_R d\phi_R = 1$$

Considering the form obtained for $f(r)$ in equation (3-44), it can be used to calculate the two-particle interaction energy (and other components of the energy of the internal coordinates, i.e., equations 2-148 to 2-152) as follows:

$$INT = \int\int \Psi(\boldsymbol{R},\boldsymbol{r})^* \, (-\frac{1}{r}) \, \Psi(\boldsymbol{R},\boldsymbol{r}) \, d\boldsymbol{R} \, d\boldsymbol{r}$$
$$= \int\int (-\frac{1}{r}) \, |\eta_{klm}(\boldsymbol{R})|^2 | \, \chi_{k'l'm'}(\boldsymbol{r})|^2 \, d\boldsymbol{R} \, d\boldsymbol{r} \qquad 3.46$$
$$= \int (-\frac{1}{r}) \, |\chi_{k'l'm'}(\boldsymbol{r})|^2 \, d\boldsymbol{r} = -\int_0^\infty \frac{f(r)}{r} dr$$

which, after integration, yields:

$$INT = -\int_0^\infty \frac{f(r)}{r} dr = \frac{N^2\pi^{5/2}(-2\sqrt{\beta} + e^{\frac{\alpha^2}{2\beta}}\sqrt{2\pi} \, \alpha \, \mathrm{Erfc}[\frac{\alpha}{\sqrt{2\beta}}])}{4\sqrt{2}(\beta\gamma)^{3/2}} \qquad 3.47$$

which is equal to the result of equation (2-152) after simplification (the results in Table 5-3 confirm this). Also, by differentiating equation (4-42) with respect to $r$ and setting it to zero, the value of $r$ at which $f(r)$ is maximized is obtained as follows:

$$\frac{df(r)}{dr} = 0 \quad \Rightarrow \quad r_{max} = \frac{-\alpha + \sqrt{\alpha^2 + 8\beta}}{4\beta} \qquad 3.48$$

### Differences in Two-Particle Distribution Functions

By subtracting the Hartree-Fock two-particle distribution functions from the exact two-particle distribution functions (obtained above), we arrive at the Coulson-Nielsen definition for the correlation hole. The difference in the local two-particle distribution function is defined as follows:

$$\Delta g(r,r_e) = g(r,r_e) - g_{HF}(r,r_e) \qquad 3.49$$

and the difference in the average two-particle distribution function is as follows:

$$\Delta f(r) = f(r) - f_{HF}(r) \qquad 3.50$$

The effects of correlation on the single-particle density can also be calculated as follows:

$$\Delta\rho(r) = \rho(r_e) - \rho_{HF}(r_e) \qquad 3.51$$

## 3.3.6 Mean Inter-Particle Distance and Its Variance

Now, having the final expression for the two-particle distribution function $f(r)$, the mean inter-particle distance $r$ can be calculated using equation (3-34). The desired integral can be solved analytically, and thus, the analytical form of the mean distance is obtained as follows:

$$\langle r\rangle = \frac{\mathrm{N}^2\pi^{5/2}(2\sqrt{\beta}(\alpha^2+2\beta) - e^{\frac{\alpha^2}{2\beta}}\sqrt{2\pi}\alpha(\alpha^2+3\beta)\mathrm{Erfc}[\frac{\alpha}{\sqrt{2}\sqrt{\beta}}])}{16\sqrt{2}\beta^{7/2}\gamma^{3/2}} \qquad 3.52$$

The variance of the inter-particle distance is also defined as follows:

$$S^2(r) = \int_0^\infty r^2 f(r)\,dr - \left(\int_0^\infty r\,f(r)\,dr\right)^2 \qquad 3.53$$

The explicit form for it will be as follows:

$$\begin{aligned} S^2(r) = \frac{1}{512\beta^7\gamma^3}N^2\pi^{5/2}(-N^2\pi^{5/2}(-2\sqrt{\beta}(\alpha^2+2\beta) \\ + e^{\frac{\alpha^2}{2\beta}}\sqrt{2\pi}\alpha(\alpha^2+3\beta)Erfc[\frac{\alpha}{\sqrt{2}\sqrt{\beta}}])^2 \\ + 8\sqrt{2}\beta^{5/2}\gamma^{3/2}(-2\alpha\sqrt{\beta}(\alpha^2+5\beta) + e^{\frac{\alpha^2}{2\beta}}\sqrt{2\pi}(\alpha^4 \\ + 6\alpha^2\beta + 3\beta^2)Erfc[\frac{\alpha}{\sqrt{2}\sqrt{\beta}}])) \end{aligned} \qquad 3.54$$

By using these two quantities, the width of the particle distribution can be compared under different conditions, and the degree of deviation from the mean inter-particle distance can be determined.

### 3.3.7 The Kato Condition

In the EHM, the Kato condition can be defined similarly to a two-electron system, as the derivative of the wave function with respect to the inter-particle distance at zero distance. This condition indicates the behavior of wave functions when the two particles coincide. To derive this condition in the EHM, after separating the center of mass and writing the wave function as a Taylor expansion [6] as follows and applying the Hamiltonian (2-16) with the Laplacian operator in the form (2-142), we arrive at equation (3-56):

$$\Psi|_{r\to 0} = c_0 + c_1 r + c_2 r^2 + \cdots \qquad 3.55$$

$$H_r(\boldsymbol{r})\Psi = \left(\frac{c_1}{\mu} + c_0\right)\frac{1}{r} + \cdots \qquad 3.56$$

Therefore, for $r \to \mathbf{0}$, we have:

$$\frac{c_1}{\mu} = -c_0 \qquad 3.57$$

and considering that:

$$c_1 = \left(\frac{\partial \Psi}{\partial r}\right)_{|r=0} \quad \text{و} \quad c_0 = \Psi(r = 0) \qquad 3.58$$

we arrive at the following result:

$$\left(\frac{\partial \Psi}{\partial r}\right)_{|r=0} = -\mu\, \Psi(r = 0) \qquad 3.59$$

By applying the Kato condition to the wave function (2-101), we have:

$$\left(\frac{\partial \Psi}{\partial r}\right)_{|r=0} = \frac{\partial[\exp(-\alpha r - \beta r^2 - \gamma R^2)]}{\partial r} = (-\alpha - 2\beta r)\Psi(r_e, r_p)|_{r=0} = -\alpha\Psi(r_e, r_p) \qquad 3.60$$

By comparing (3-59) and (3-60), we have:

$$\mu = \alpha \qquad 3.61$$

Thus, if $\alpha$ is equal to the reduced mass, the wave function (2-101) will satisfy the Kato condition. Numerical results for optimized $\alpha$ and the fitting results (sections 4-2 and 5-3) indicate that equation (3-61) is largely valid.

# 4 Numerical Data for Model Solution

## 4.1 Computational Details

All computations in this study are based on varying two parameters: the mass of the positive particle and the frequency of the oscillator field (in atomic units) over a wide range. These include ten different masses for the PCP, namely 1 (positron), 1.5, 2, 3, 10, 50, 207 (muon), 400, 900, and 1836 (proton), and nine frequencies that are arranged in the table below along with their equivalents in $cm^{-1}$.

Table 4-1: Range of frequencies used in the calculations of this study

| Field Frequency in $cm^{-1}$ | Field Frequency in Atomic Units |
|---|---|
| 21.9 | $10^{-4}$ |
| 219.4 | $10^{-3}$ |
| 2194.7 | $10^{-2}$ |
| 21947.5 | $10^{-1}$ |
| 219474.6 | 1 |
| $2.2\times10^{6}$ | $10^{1}$ |
| $2.2\times10^{7}$ | $10^{2}$ |
| $2.2\times10^{8}$ | $10^{3}$ |
| $2.2\times10^{9}$ | $10^{4}$ |

This wide range of parameters allows for a more detailed examination of the system's behavior under various conditions. Thus, all calculations were performed for 90 systems. Additionally, to better scale the graphs, the relevant quantities are displayed based on the natural length of the system, which corresponds to the expected value of the inter-particle distance (equation 3-52). In numerical integrations using Mathematica, to obtain the most accurate result, depending on the problem's conditions, the "Working Precision" was set to 100 digits, the "Precision Goal" to 8 digits, and the "Max Recursion" to 20 times. The "Local Adaptive" and "Cartesian Rule" integration strategies were employed. Throughout the thesis, the accuracy of the results is reported to 6 decimal places. In all calculations, the absolute error refers to the difference between the exact value and the calculated value, and the percent error is given by the following relation:

$$\boldsymbol{\delta} = \left|\frac{\boldsymbol{v_{computed}} - \boldsymbol{v_{exact}}}{\boldsymbol{v_{exact}}}\right| * \mathbf{100} \qquad 4.1$$

In this study, three methods were used: variational, two-component Hartree-Fock (TC-HF), and finite element (all three methods were programmed in Mathematica). The first two methods have a variational nature, while the latter is a standard method for solving differential equations. It's worth mentioning that the variational and finite element methods were applied after separating the center of mass and for the μ\mu particle (although in the variational method, by adding the center of mass part to the variational wave function, the total wave function is obtained). On the other hand, the Hartree-Fock method was computed based on single-particle wave functions and the primary particle coordinates of the problem.

In this study, four methods were used: variational, two-component Hartree-Fock (TC-HF), two-component density functional theory (TC-DFT), and finite element (all four methods were programmed in Mathematica). The first three methods have a variational nature, while the latter is a standard method for solving differential equations. It's worth mentioning that the variational and finite element methods were applied after separating the center of mass and for the $\mu$ pseudo-particles (although in the variational method, by adding the center of mass part to the variational wave function, the total wave function is obtained). On the other hand, the Hartree-Fock method was computed based on single-particle wave functions and the primary particle coordinates of the problem.

The variational method has been fully explained in sections 2-3-2 to 2-3-5, and the TC-DFT method will be thoroughly examined in section 5-2. In the TC-HF computations for both the electron and the positive particle, the Gaussian 7s basis set (with functions placed at the bottom of potential wells) was used, denoted as [7s:7s], and all exponents were optimized for all masses and frequencies. It is worth noting that in energy calculations, the 7s basis set for both particles essentially represents the limit of the Hartree-Fock method, and extending the basis set to [7s7p7d: 7s7p7d] does not significantly improve the Hartree-Fock energy. On the other hand, the [7s:1s] basis set (7s for the electron and 1s for the PCP) was also used to more accurately investigate the changes in the exponents of PCPs and for subsequent applications.

To solve differential equations using the Finite Element Method (FEM), a discrete representation of a region, i.e., a mesh, is required. With the differential equation and boundary conditions specified, this method can be used to solve the equations. In our case, the region of interest ranges from $r_{min} = 10^{-16}$ to $r_{max} = 50$ in atomic units, meshed with intervals of 0.001. However, this meshing at a frequency of $10^4$ reduced accuracy (see Figure B-1 in the appendix B), so intervals of $10^{-4}$ were used at this frequency. The differential equation is the radial part of the Schrödinger equation (2-18), which goes to zero for $r$ values less than $r_{min}$.

## 4.2 Optimized Variational Parameters

Without performing calculations, it can be easily seen that the variational parameters $\alpha$ and $\beta$ have a linear dependence on $\mu$ by examining the asymptotic behavior of the variational wave function (see equations 2-73 and 2-74). However, as mentioned in section 2-3-1, we avoided separating $\mu$ from the variational parameters for two reasons: 1) to maintain the simplicity of the wave function in subsequent calculations, and 2) to prevent $\beta$ from approaching zero at very low frequencies, which would result in poor fitting behavior (since $\beta$ tends to zero at very low frequencies, separating $\mu$ would lead to zeros with infinite fitting error percentages). The optimized variational parameters with and without $\mu$ are reported in Table 4-2.

Table 4-2: Optimized variational parameter values "with" and "without" consideration of the reduced mass, relative Hamiltonian energy, and their errors

| | | | $\boldsymbol{\mu}$ included | | | | | | | $\boldsymbol{\mu}$ excluded | | |
|---|---|---|---|---|---|---|---|---|---|---|---|---|
| ω/Quantities | $\boldsymbol{\alpha}$[1] | $\boldsymbol{\beta}$[2] | Variational Energy[3] | FEM Energy[4] | RMSD[5] | $\boldsymbol{\Delta}$E[6] | $(\beta/\alpha)$[7] | $\boldsymbol{\alpha}$[8] | $\boldsymbol{\beta}$[9] | Variational Energy[10] | RMSD[11] | $\boldsymbol{\Delta}$E[12] |
| | | | | | | m = 1 | | | | | | |
| 0.0001 | 0.500000 | 0.000000 | -0.250000 | -0.250000 | 0.000000 | 0.000000 | 0.000000 | 1.000000 | 0.000000 | -0.250000 | 0.000000 | 0.000000 |
| 0.001 | 0.499997 | 0.000001 | -0.249997 | -0.249997 | 0.000000 | 0.000000 | 0.000003 | 0.999994 | 0.000003 | -0.249997 | 0.000000 | 0.000000 |
| 0.01 | 0.499704 | 0.000149 | -0.249701 | -0.249701 | 0.000008 | 0.000000 | 0.000298 | 0.999407 | 0.000298 | -0.249701 | 0.000008 | 0.000000 |
| 0.1 | 0.484707 | 0.010559 | -0.223933 | -0.223957 | 0.000314 | -0.000024 | 0.021783 | 0.969414 | 0.021117 | -0.223933 | 0.000314 | -0.000024 |
| 1.0 | 0.438060 | 0.205107 | 0.612000 | 0.611853 | 0.000520 | -0.000148 | 0.468217 | 0.876121 | 0.410215 | 0.612000 | 0.000520 | -0.000148 |
| 10 | 0.414991 | 2.366272 | 12.395490 | 12.395303 | 0.000331 | -0.000187 | 5.701982 | 0.829982 | 4.732543 | 12.395490 | 0.000331 | -0.000187 |
| 100 | 0.407310 | 24.586618 | 141.942322 | 141.942127 | 0.000190 | -0.000195 | 60.363374 | 0.814620 | 49.173238 | 141.942322 | 0.000190 | -0.000195 |
| 1000 | 0.404950 | 248.701700 | 1474.690624 | 1474.690429 | 0.000108 | -0.000195 | 614.154093 | 0.809723 | 497.405074 | 1474.690624 | 0.000108 | -0.000195 |
| 10000 | 0.404560 | 2495.892300 | 14920.133739 | 14920.133541 | 0.000096 | -0.000198 | 6169.399595 | 0.808173 | 4991.813767 | 14920.133739 | 0.000061 | -0.000198 |
| | | | | | | m = 1.5 | | | | | | |
| 0.0001 | 0.600000 | 0.000000 | -0.300000 | -0.300000 | 0.000000 | 0.000000 | 0.000000 | 1.000000 | 0.000000 | -0.300000 | 0.000000 | 0.000000 |
| 0.001 | 0.599997 | 0.000001 | -0.299998 | -0.299998 | 0.000000 | 0.000000 | 0.000002 | 0.999996 | 0.000003 | -0.299998 | 0.000000 | 0.000000 |
| 0.01 | 0.599752 | 0.000149 | -0.299750 | -0.299750 | 0.000006 | 0.000000 | 0.000249 | 0.999587 | 0.000249 | -0.299750 | 0.000006 | 0.000000 |
| 0.1 | 0.585320 | 0.011363 | -0.277533 | -0.277553 | 0.000290 | -0.000020 | 0.019413 | 0.975533 | 0.018938 | -0.277533 | 0.000290 | -0.000020 |
| 1.0 | 0.529268 | 0.240597 | 0.516267 | 0.516098 | 0.000581 | -0.000169 | 0.454585 | 0.882113 | 0.400996 | 0.516267 | 0.000581 | -0.000169 |
| 10 | 0.499264 | 2.823673 | 12.137957 | 12.137735 | 0.000378 | -0.000223 | 5.655669 | 0.832107 | 4.706122 | 12.137957 | 0.000378 | -0.000223 |
| 100 | 0.489182 | 29.456033 | 141.164882 | 141.164648 | 0.000218 | -0.000234 | 60.214909 | 0.815302 | 49.093389 | 141.164882 | 0.000218 | -0.000234 |

| | | | $\boldsymbol{\mu}$ included | | | | | | | $\boldsymbol{\mu}$ excluded | | |
|---|---|---|---|---|---|---|---|---|---|---|---|---|
| ω/Quantities | $\boldsymbol{\alpha}$[1] | $\boldsymbol{\beta}$[2] | Variational Energy[3] | FEM Energy[4] | RMSD[5] | $\boldsymbol{\Delta}$E[6] | $\left(\boldsymbol{\beta}/\boldsymbol{\alpha}\right)$[7] | $\boldsymbol{\alpha}$[8] | $\boldsymbol{\beta}$[9] | Variational Energy[10] | RMSD[11] | $\boldsymbol{\Delta}$E[12] |
| 1000 | 0.485964 | 298.293868 | 1472.266767 | 1472.266533 | 0.000123 | -0.000234 | 613.818670 | 0.809940 | 497.156451 | 1472.266767 | 0.000123 | -0.000234 |
| 10000 | 0.484945 | 2994.618890 | 14912.502746 | 14912.502509 | 0.000110 | -0.000238 | 6175.176212 | 0.808241 | 4991.031476 | 14912.502746 | 0.000070 | -0.000238 |
| m = 2 | | | | | | | | | | | | |
| 0.0001 | 0.666667 | 0.000000 | -0.333333 | -0.333333 | 0.000000 | 0.000000 | 0.000000 | 1.000000 | 0.000000 | -0.333333 | 0.000000 | 0.000000 |
| 0.001 | 0.666664 | 0.000002 | -0.333331 | -0.333331 | 0.000000 | 0.000000 | 0.000002 | 0.999997 | 0.000002 | -0.333331 | 0.000000 | 0.000000 |
| 0.01 | 0.666443 | 0.000149 | -0.333109 | -0.333109 | 0.000005 | 0.000000 | 0.000224 | 0.999665 | 0.000224 | -0.333109 | 0.000005 | 0.000000 |
| 0.1 | 0.652472 | 0.011795 | -0.312780 | -0.312797 | 0.000272 | -0.000017 | 0.018077 | 0.978707 | 0.017692 | -0.312780 | 0.000272 | -0.000017 |
| 1.0 | 0.590509 | 0.263492 | 0.455715 | 0.455533 | 0.000619 | -0.000182 | 0.446212 | 0.885764 | 0.395239 | 0.455715 | 0.000619 | -0.000182 |
| 10 | 0.555616 | 3.126429 | 11.977247 | 11.977001 | 0.000408 | -0.000246 | 5.626964 | 0.833423 | 4.689644 | 11.977247 | 0.000408 | -0.000246 |
| 100 | 0.543817 | 32.695823 | 140.681475 | 140.681215 | 0.000236 | -0.000260 | 60.122804 | 0.815726 | 49.043735 | 140.681475 | 0.000236 | -0.000260 |
| 1000 | 0.540050 | 331.334674 | 1470.761234 | 1470.760975 | 0.000134 | -0.000259 | 613.526230 | 0.810074 | 497.002015 | 1470.761234 | 0.000134 | -0.000259 |
| 10000 | 0.538856 | 3327.030484 | 14907.764463 | 14907.764199 | 0.000119 | -0.000264 | 6174.249590 | 0.808284 | 4990.545724 | 14907.764463 | 0.000075 | -0.000264 |
| m = 3 | | | | | | | | | | | | |
| 0.0001 | 0.750000 | 0.000000 | -0.375000 | -0.375000 | 0.000000 | 0.000000 | 0.000000 | 1.000000 | 0.000000 | -0.375000 | 0.000000 | 0.000000 |
| 0.001 | 0.749998 | 0.000001 | -0.374998 | -0.374998 | 0.000000 | 0.000000 | 0.000002 | 0.999997 | 0.000002 | -0.374998 | 0.000000 | 0.000000 |
| 0.01 | 0.749801 | 0.000150 | -0.374800 | -0.374800 | 0.000004 | 0.000000 | 0.000199 | 0.999735 | 0.000199 | -0.374800 | 0.000004 | 0.000000 |
| 0.1 | 0.736443 | 0.012243 | -0.356440 | -0.356455 | 0.000252 | -0.000015 | 0.016625 | 0.981924 | 0.016325 | -0.356440 | 0.000252 | -0.000015 |
| 1.0 | 0.667504 | 0.291304 | 0.383003 | 0.382805 | 0.000662 | -0.000198 | 0.436408 | 0.890005 | 0.388406 | 0.383003 | 0.000662 | -0.000198 |
| 10 | 0.626232 | 3.502570 | 11.786450 | 11.786175 | 0.000445 | -0.000275 | 5.593084 | 0.834976 | 4.670093 | 11.786450 | 0.000445 | -0.000275 |
| 100 | 0.612170 | 36.738719 | 140.109312 | 140.109020 | 0.000258 | -0.000292 | 60.013926 | 0.816227 | 48.984959 | 140.109312 | 0.000258 | -0.000292 |
| 1000 | 0.607675 | 372.614531 | 1468.980867 | 1468.980576 | 0.000146 | -0.000291 | 613.180577 | 0.810234 | 496.819371 | 1468.980867 | 0.000146 | -0.000291 |

| | μ included | | | | | | | μ excluded | | | | |
|---|---|---|---|---|---|---|---|---|---|---|---|---|
| ω/Quantities | $\boldsymbol{\alpha}$[1] | $\boldsymbol{\beta}$[2] | Variational Energy[3] | FEM Energy[4] | RMSD[5] | $\boldsymbol{\Delta}$E[6] | $\left(\boldsymbol{\beta}/\boldsymbol{\alpha}\right)$[7] | $\boldsymbol{\alpha}$[8] | $\boldsymbol{\beta}$[9] | Variational Energy[10] | RMSD[11] | $\boldsymbol{\Delta}$E[12] |
| 10000 | 0.606251 | 3742.478585 | 14902.162752 | 14902.162455 | 0.000130 | -0.000297 | 6173.155369 | 0.808335 | 4989.971417 | 14902.162750 | 0.000082 | -0.000300 |
| m=10 | | | | | | | | | | | | |
| 0.0001 | 0.909091 | 0.000000 | -0.454545 | -0.454545 | 0.000000 | 0.000000 | 0.000000 | 1.000000 | 0.000000 | -0.454545 | 0.000000 | 0.000000 |
| 0.001 | 0.909089 | 0.000002 | -0.454544 | -0.454544 | 0.000000 | 0.000000 | 0.000002 | 0.999998 | 0.000002 | -0.454544 | 0.000000 | 0.000000 |
| 0.01 | 0.908926 | 0.000150 | -0.454381 | -0.454381 | 0.000003 | 0.000000 | 0.000165 | 0.999819 | 0.000165 | -0.454381 | 0.000003 | 0.000000 |
| 0.1 | 0.896744 | 0.012890 | -0.438912 | -0.438923 | 0.000216 | -0.000011 | 0.014375 | 0.986419 | 0.014179 | -0.438912 | 0.000216 | -0.000011 |
| 1.0 | 0.815719 | 0.342079 | 0.251539 | 0.251315 | 0.000736 | -0.000224 | 0.419359 | 0.897291 | 0.376287 | 0.251539 | 0.000736 | -0.000224 |
| 10 | 0.761549 | 4.213996 | 11.447417 | 11.447087 | 0.000511 | -0.000330 | 5.533453 | 0.837710 | 4.635388 | 11.447417 | 0.000511 | -0.000330 |
| 100 | 0.742795 | 44.437386 | 139.097263 | 139.096910 | 0.000297 | -0.000353 | 59.824535 | 0.817110 | 48.880985 | 139.097263 | 0.000297 | -0.000353 |
| 1000 | 0.736831 | 451.360667 | 1465.835951 | 1465.835599 | 0.000168 | -0.000351 | 612.570377 | 0.810515 | 496.496718 | 1465.835951 | 0.000168 | -0.000351 |
| 10000 | 0.734930 | 4535.415827 | 14892.271732 | 14892.271372 | 0.000150 | -0.000360 | 6171.223622 | 0.808423 | 4988.957410 | 14892.271732 | 0.000095 | -0.000360 |
| m=50 | | | | | | | | | | | | |
| 0.0001 | 0.980392 | 0.000000 | -0.490196 | -0.490196 | 0.000000 | 0.000000 | 0.000000 | 1.000000 | 0.000000 | -0.490196 | 0.000000 | 0.000000 |
| 0.001 | 0.980391 | 0.000001 | -0.490195 | -0.490195 | 0.000000 | 0.000000 | 0.000002 | 0.999998 | 0.000002 | -0.490195 | 0.000000 | 0.000000 |
| 0.01 | 0.980240 | 0.000150 | -0.490043 | -0.490043 | 0.000003 | 0.000000 | 0.000153 | 0.999844 | 0.000153 | -0.490043 | 0.000003 | 0.000000 |
| 0.1 | 0.968561 | 0.013113 | -0.475603 | -0.475612 | 0.000201 | -0.000009 | 0.013539 | 0.987933 | 0.013376 | -0.475603 | 0.000201 | -0.000009 |
| 1.0 | 0.882615 | 0.363909 | 0.195169 | 0.194935 | 0.000765 | -0.000234 | 0.412308 | 0.900267 | 0.371188 | 0.195169 | 0.000765 | -0.000234 |
| 10 | 0.822405 | 4.530152 | 11.304325 | 11.303970 | 0.000540 | -0.000355 | 5.508419 | 0.838853 | 4.620756 | 11.304325 | 0.000540 | -0.000355 |
| 100 | 0.801451 | 47.879691 | 138.671901 | 138.671520 | 0.000314 | -0.000380 | 59.741281 | 0.817480 | 48.837283 | 138.671901 | 0.000314 | -0.000380 |
| 1000 | 0.794737 | 486.628702 | 1464.515756 | 1464.515378 | 0.000178 | -0.000378 | 612.313953 | 0.810632 | 496.361283 | 1464.515756 | 0.000178 | -0.000378 |
| 10000 | 0.792600 | 4890.717855 | 14888.121166 | 14888.120778 | 0.000159 | -0.000388 | 6170.471104 | 0.808460 | 4988.531880 | 14888.121166 | 0.000100 | -0.000388 |

| | μ included | | | | | | | | μ excluded | | | |
|---|---|---|---|---|---|---|---|---|---|---|---|---|
| ω/Quantities | $\alpha$[1] | $\beta$[2] | Variational Energy[3] | FEM Energy[4] | RMSD[5] | ΔE[6] | $(\beta/\alpha)$[7] | $\alpha$[8] | $\beta$[9] | Variational Energy[10] | RMSD[11] | ΔE[12] |
| | | | | | | m = 207 | | | | | | |
| 0.0001 | 0.995192 | 0.000000 | -0.497596 | -0.497596 | 0.000000 | 0.000000 | 0.000000 | 1.000000 | 0.000000 | -0.497596 | 0.000000 | 0.000000 |
| 0.001 | 0.995191 | 0.000001 | -0.497595 | -0.497595 | 0.000000 | 0.000000 | 0.000002 | 0.999998 | 0.000002 | -0.497595 | 0.000000 | 0.000000 |
| 0.01 | 0.995042 | 0.000150 | -0.497446 | -0.497446 | 0.000003 | 0.000000 | 0.000150 | 0.999850 | 0.000150 | -0.498603 | 0.000003 | 0.000000 |
| 0.1 | 0.983465 | 0.013155 | -0.483202 | -0.483212 | 0.000199 | -0.000009 | 0.013377 | 0.988260 | 0.013195 | -0.484390 | 0.000198 | -0.000009 |
| 1.0 | 0.896535 | 0.368372 | 0.183638 | 0.183403 | 0.000771 | -0.000236 | 0.410884 | 0.900959 | 0.369991 | 0.181841 | 0.000772 | -0.000236 |
| 10 | 0.835051 | 4.595580 | 11.275221 | 11.274861 | 0.000546 | -0.000360 | 5.503352 | 0.839085 | 4.617781 | 11.275221 | 0.000546 | -0.000360 |
| 100 | 0.813624 | 48.593656 | 138.585513 | 138.585127 | 0.000318 | -0.000386 | 59.724929 | 0.817555 | 48.828408 | 138.585513 | 0.000318 | -0.000386 |
| 1000 | 0.806758 | 493.947566 | 1464.247752 | 1464.247368 | 0.000180 | -0.000384 | 612.262366 | 0.810655 | 496.333788 | 1464.247752 | 0.000180 | -0.000384 |
| 10000 | 0.804581 | 4964.462597 | 14887.278700 | 14887.278307 | 0.000161 | -0.000394 | 6170.248554 | 0.808469 | 4988.432061 | 14887.147551 | 0.000102 | -0.000395 |
| | | | | | | m = 400 | | | | | | |
| 0.0001 | 0.997506 | 0.000000 | -0.498753 | -0.498753 | 0.000000 | 0.000000 | 0.000000 | 1.000000 | 0.000000 | -0.498753 | 0.000000 | 0.000000 |
| 0.001 | 0.997505 | 0.000001 | -0.498752 | -0.498752 | 0.000000 | 0.000000 | 0.000001 | 0.999998 | 0.000002 | -0.498752 | 0.000000 | 0.000000 |
| 0.01 | 0.997356 | 0.000150 | -0.498603 | -0.498603 | 0.000003 | 0.000000 | 0.000150 | 0.999850 | 0.000150 | -0.498603 | 0.000003 | 0.000000 |
| 0.1 | 0.985796 | 0.013162 | -0.484390 | -0.484399 | 0.000198 | -0.000009 | 0.013351 | 0.988260 | 0.013195 | -0.484390 | 0.000198 | -0.000009 |
| 1.0 | 0.898713 | 0.369068 | 0.181841 | 0.181605 | 0.000772 | -0.000236 | 0.410663 | 0.900959 | 0.369991 | 0.181841 | 0.000772 | -0.000236 |
| 10 | 0.837029 | 4.605803 | 11.270688 | 11.270328 | 0.000546 | -0.000360 | 5.502564 | 0.839121 | 4.617317 | 11.270688 | 0.000546 | -0.000360 |
| 100 | 0.815527 | 48.705265 | 138.572063 | 138.571677 | 0.000318 | -0.000387 | 59.722424 | 0.817567 | 48.827026 | 138.572063 | 0.000318 | -0.000387 |
| 1000 | 0.808637 | 495.091782 | 1464.206030 | 1464.205645 | 0.000181 | -0.000385 | 612.254559 | 0.810659 | 496.329506 | 1464.206030 | 0.000181 | -0.000385 |
| 10000 | 0.806453 | 4975.992079 | 14887.147551 | 14887.147157 | 0.000161 | -0.000395 | 6170.223253 | 0.808469 | 4988.432061 | 14887.147551 | 0.000102 | -0.000395 |
| | | | | | | m = 900 | | | | | | |

| | | | **$\mu$** included | | | | | | | **$\mu$** excluded | | |
|---|---|---|---|---|---|---|---|---|---|---|---|---|
| ω/Quantities | $\alpha$[1] | $\beta$[2] | Variational Energy[3] | FEM Energy[4] | RMSD[5] | $\Delta$E[6] | $\left(\beta/\alpha\right)$[7] | $\alpha$[8] | $\beta$[9] | Variational Energy[10] | RMSD[11] | $\Delta$E[12] |
| 0.0001 | 0.998890 | 0.000000 | -0.499445 | -0.499445 | 0.000000 | 0.000000 | 0.000000 | 1.000000 | 0.000000 | -0.499445 | 0.000000 | 0.000000 |
| 0.001 | 0.998889 | 0.000001 | -0.499444 | -0.499444 | 0.000000 | 0.000000 | 0.000001 | 0.999998 | 0.000002 | -0.499444 | 0.000000 | 0.000000 |
| 0.01 | 0.998740 | 0.000150 | -0.499295 | -0.499295 | 0.000003 | 0.000000 | 0.000150 | 0.999850 | 0.000150 | -0.499295 | 0.000003 | 0.000000 |
| 0.1 | 0.987190 | 0.013166 | -0.485100 | -0.485109 | 0.000198 | -0.000009 | 0.013336 | 0.988286 | 0.013180 | -0.485100 | 0.000198 | -0.000009 |
| 1.0 | 0.900015 | 0.369483 | 0.180766 | 0.180530 | 0.000772 | -0.000236 | 0.410530 | 0.901014 | 0.369894 | 0.180766 | 0.000772 | -0.000236 |
| 10 | 0.838211 | 4.611916 | 11.267980 | 11.267619 | 0.000547 | -0.000361 | 5.502092 | 0.839143 | 4.617041 | 11.267980 | 0.000547 | -0.000361 |
| 100 | 0.816666 | 48.772012 | 138.564027 | 138.563640 | 0.000319 | -0.000387 | 59.720902 | 0.817574 | 48.826200 | 138.564027 | 0.000319 | -0.000387 |
| 1000 | 0.809761 | 495.776091 | 1464.181100 | 1464.180715 | 0.000181 | -0.000385 | 612.249716 | 0.810661 | 496.326955 | 1464.181100 | 0.000181 | -0.000385 |
| 10000 | 0.807572 | 4982.887482 | 14887.069188 | 14887.068792 | 0.000161 | -0.000395 | 6170.207675 | 0.808469 | 4988.424026 | 14887.069188 | 0.000102 | -0.000395 |
| | | | | | | m = 1836 | | | | | | |
| 0.0001 | 0.999456 | 0.000000 | -0.499728 | -0.499728 | 0.000000 | 0.000000 | 0.000000 | 1.000000 | 0.000000 | -0.499728 | 0.000000 | 0.000000 |
| 0.001 | 0.999454 | 0.000002 | -0.499726 | -0.499726 | 0.000000 | 0.000000 | 0.000002 | 0.999998 | 0.000002 | -0.499726 | 0.000000 | 0.000000 |
| 0.01 | 0.999306 | 0.000150 | -0.499578 | -0.499578 | 0.000003 | 0.000000 | 0.000150 | 0.999850 | 0.000150 | -0.499578 | 0.000003 | 0.000000 |
| 0.1 | 0.987759 | 0.013167 | -0.485391 | -0.485400 | 0.000198 | -0.000009 | 0.013331 | 0.988296 | 0.013175 | -0.485391 | 0.000198 | -0.000009 |
| 1.0 | 0.900546 | 0.369653 | 0.180327 | 0.180091 | 0.000772 | -0.000236 | 0.410477 | 0.901037 | 0.369855 | 0.180327 | 0.000772 | -0.000236 |
| 10 | 0.838695 | 4.614414 | 11.266873 | 11.266512 | 0.000547 | -0.000361 | 5.501900 | 0.839152 | 4.616927 | 11.266873 | 0.000547 | -0.000361 |
| 100 | 0.817131 | 48.799286 | 138.560744 | 138.560357 | 0.000319 | -0.000388 | 59.720281 | 0.817577 | 48.825863 | 138.560744 | 0.000319 | -0.000388 |
| 1000 | 0.810221 | 496.055714 | 1464.170918 | 1464.170533 | 0.000181 | -0.000385 | 612.247098 | 0.810662 | 496.325903 | 1464.170918 | 0.000181 | -0.000385 |
| 10000 | 0.808030 | 4985.705219 | 14887.037181 | 14887.036785 | 0.000161 | -0.000395 | 6170.201461 | 0.808470 | 4988.420744 | 14887.037181 | 0.000102 | -0.000395 |

1. Variational parameter $\alpha$ considering reduced mass
2. Variational parameter $\beta$ considering reduced mass

3. Internal variational energy of the system using parameters 1 and 2
4. FEM relative Hamiltonian energy
5. Root Mean Square error for the variational wave function with parameters 1 and 2 relative to the wave function obtained from fem (the error equation is $\sqrt{\frac{1}{n}(\sum_n(\psi_{var} - \psi_{FEM}))^2}$, where n indicates the number of points at which the wave functions were calculated)
6. Energy difference between 3 and 4
7. Ratio of parameter 2 to 1 ($\beta/\alpha$)
8. Variational parameter $\alpha$ without considering reduced mass
9. Variational parameter $\beta$ without considering reduced mass
10. Internal variational energy of the system using parameters 8 and 9
11. Root Mean Square error for the variational wave function with parameters 8 and 9 relative to the wave function obtained from FEM
12. Energy difference between 4 and 10

## 4.3 Energy and Its Components

Table 4-3 contains the total system energy data as well as the correlation energies. In this table, the total system energy is first presented using the variational method, FEM, and MC-HF (two basis sets [7s:7s] and [7s:1s]), followed by the correlation energies. The total system energy equals the sum of the relative Hamiltonian and the center of mass energies (equation 2-25). The absolute error and percentage error of the variational energy compared to the FEM energy, as well as the MC-HF/[7s:1s] energy compared to the MC-HF/[7s:7s], are also reported. The percentage error of the variational energy compared to the FEM-derived energy reaches about 0.03% in the worst-case scenario, indicating that the variational energy closely approaches the desired limit. The energy values of the wave function (2-75) and their errors for several masses are provided in Table B-1 in the appendix B, showing that this wave function has less accuracy compared to wave function (2-72).

Table 4-3: Total energy, correlation energy, and errors obtained from various methods

| | Total Energy | | | | Correlation Energy | | Error of variational energy | | Error of [*7s:1s*] Basis Set | |
|---|---|---|---|---|---|---|---|---|---|---|
| ω/Quantities | Variational [1] | FEM [2] | MC-HF/[*7s:7s*] [3] | MC-HF/[*7s:1s*] [4] | VAR-HF [5] | FEM-HF [6] | Delta E [7] | % error [8] | Delta E [9] | % error [10] |
| | | | | | m=1 | | | | | |
| 0.0001 | -0.249850 | -0.249850 | -0.108513 | -0.107219 | -0.141337 | -0.141337 | 0.000000 | 0.000000 | -0.001294 | 1.192040 |
| 0.001 | -0.248497 | -0.248497 | -0.108491 | -0.107198 | -0.140006 | -0.140006 | 0.000000 | 0.000000 | -0.001293 | 1.192142 |
| 0.01 | -0.234701 | -0.234701 | -0.106407 | -0.105127 | -0.128294 | -0.128294 | 0.000000 | 0.000006 | -0.001280 | 1.202684 |
| 0.1 | -0.073933 | -0.073957 | 0.012203 | 0.013286 | -0.086137 | -0.086160 | -0.000024 | 0.031779 | -0.001083 | 8.875661 |
| 1.0 | 2.112000 | 2.111853 | 2.171918 | 2.172834 | -0.059918 | -0.060066 | -0.000148 | 0.006987 | -0.000915 | 0.042140 |
| 10 | 27.395490 | 27.395303 | 27.448072 | 27.448929 | -0.052581 | -0.052768 | -0.000187 | 0.000682 | -0.000858 | 0.003125 |
| 100 | 291.942322 | 291.942127 | 291.992777 | 291.993616 | -0.050455 | -0.050650 | -0.000195 | 0.000067 | -0.000839 | 0.000287 |
| 1000 | 2974.690624 | 2974.690429 | 2974.740427 | 2974.741261 | -0.049803 | -0.049998 | -0.000195 | 0.000007 | -0.000834 | 0.000028 |
| 10000 | 29920.133739 | 29920.133541 | 29920.183344 | 29920.184180 | -0.049605 | -0.049804 | -0.000198 | 0.000001 | -0.000836 | 0.000003 |
| | | | | | m=1.5 | | | | | |
| 0.0001 | -0.299850 | -0.299850 | -0.131661 | -0.130508 | -0.168189 | -0.168189 | 0.000000 | 0.000000 | -0.001154 | 0.876152 |
| 0.001 | -0.298498 | -0.298498 | -0.131644 | -0.130490 | -0.166854 | -0.166854 | 0.000000 | 0.000000 | -0.001153 | 0.876104 |
| 0.01 | -0.284750 | -0.284750 | -0.129900 | -0.128764 | -0.154850 | -0.154850 | 0.000000 | 0.000003 | -0.001137 | 0.875116 |
| 0.1 | -0.127533 | -0.127553 | -0.021539 | -0.020706 | -0.105994 | -0.106014 | -0.000020 | 0.015364 | -0.000833 | 3.866629 |
| 1.0 | 2.016267 | 2.016098 | 2.087757 | 2.088293 | -0.071490 | -0.071659 | -0.000169 | 0.008391 | -0.000537 | 0.025698 |
| 10 | 27.137957 | 27.137735 | 27.199613 | 27.200050 | -0.061655 | -0.061878 | -0.000223 | 0.000820 | -0.000437 | 0.001607 |
| 100 | 291.164882 | 291.164648 | 291.223701 | 291.224105 | -0.058820 | -0.059054 | -0.000234 | 0.000080 | -0.000403 | 0.000139 |
| 1000 | 2972.266767 | 2972.266533 | 2972.324714 | 2972.325111 | -0.057948 | -0.058182 | -0.000234 | 0.000008 | -0.000396 | 0.000013 |

| | Total Energy | | | | Correlation Energy | | Error of variational energy | | Error of [*7s:1s*] Basis Set | |
|---|---|---|---|---|---|---|---|---|---|---|
| ω/Quantities | Variational[1] | FEM[2] | MC-HF/[*7s:7s*][3] | MC-HF/[*7s:1s*][4] | VAR-HF[5] | FEM-HF[6] | Delta E[7] | % error[8] | Delta E[9] | % error[10] |
| 10000 | 29912.502746 | 29912.502509 | 29912.560519 | 29912.560821 | -0.057773 | -0.058010 | -0.000238 | 0.000001 | -0.000302 | 0.000001 |
| | | | | | m=2 | | | | | |
| 0.0001 | -0.333183 | -0.333183 | -0.149328 | -0.148287 | -0.183855 | -0.183855 | 0.000000 | 0.000000 | -0.001041 | 0.696899 |
| 0.001 | -0.331831 | -0.331831 | -0.149312 | -0.148272 | -0.182519 | -0.182519 | 0.000000 | 0.000000 | -0.001040 | 0.696854 |
| 0.01 | -0.318109 | -0.318109 | -0.147746 | -0.146721 | -0.170363 | -0.170363 | 0.000000 | 0.000002 | -0.001024 | 0.693148 |
| 0.1 | -0.162780 | -0.162797 | -0.045426 | -0.044742 | -0.117354 | -0.117372 | -0.000017 | 0.010597 | -0.000684 | 1.505567 |
| 1.0 | 1.955715 | 1.955533 | 2.032554 | 2.032903 | -0.076839 | -0.077021 | -0.000182 | 0.009328 | -0.000349 | 0.017167 |
| 10 | 26.977247 | 26.977001 | 27.042353 | 27.042601 | -0.065106 | -0.065352 | -0.000246 | 0.000912 | -0.000248 | 0.000918 |
| 100 | 290.681475 | 290.681215 | 290.743202 | 290.743417 | -0.061727 | -0.061987 | -0.000260 | 0.000089 | -0.000215 | 0.000074 |
| 1000 | 2970.761234 | 2970.760975 | 2970.821923 | 2970.822132 | -0.060689 | -0.060948 | -0.000259 | 0.000009 | -0.000210 | 0.000007 |
| 10000 | 29907.764463 | 29907.764199 | 29907.824832 | 29907.825035 | -0.060368 | -0.060632 | -0.000264 | 0.000001 | -0.000204 | 0.000001 |
| | | | | | m=3 | | | | | |
| 0.0001 | -0.374850 | -0.374850 | -0.175573 | -0.174698 | -0.199277 | -0.199277 | 0.000000 | 0.000000 | -0.000875 | 0.498326 |
| 0.001 | -0.373498 | -0.373498 | -0.175559 | -0.174685 | -0.197939 | -0.197939 | 0.000000 | 0.000000 | -0.000875 | 0.498281 |
| 0.01 | -0.359800 | -0.359800 | -0.174173 | -0.173313 | -0.185627 | -0.185627 | 0.000000 | 0.000001 | -0.000860 | 0.493645 |
| 0.1 | -0.206440 | -0.206455 | -0.078661 | -0.078151 | -0.127779 | -0.127794 | -0.000015 | 0.007104 | -0.000510 | 0.648306 |
| 1.0 | 1.883003 | 1.882805 | 1.962394 | 1.962573 | -0.079392 | -0.079589 | -0.000198 | 0.010503 | -0.000179 | 0.009118 |
| 10 | 26.786450 | 26.786175 | 26.851598 | 26.851698 | -0.065147 | -0.065423 | -0.000275 | 0.001028 | -0.000100 | 0.000372 |
| 100 | 290.109312 | 290.109020 | 290.170376 | 290.170453 | -0.061064 | -0.061356 | -0.000292 | 0.000101 | -0.000077 | 0.000026 |
| 1000 | 2968.980867 | 2968.980576 | 2969.040682 | 2969.040755 | -0.059814 | -0.060105 | -0.000291 | 0.000010 | -0.000074 | 0.000002 |
| 10000 | 29902.162752 | 29902.162455 | 29902.222187 | 29902.222260 | -0.059435 | -0.059732 | -0.000297 | 0.000001 | -0.000073 | 0.000000 |

| | Total Energy | | | | Correlation Energy | | Error of variational energy | | Error of [*7s:1s*] Basis Set | |
|---|---|---|---|---|---|---|---|---|---|---|
| ω/Quantities | Variational[1] | FEM[2] | MC-HF/[*7s:7s*][3] | MC-HF/[*7s:1s*][4] | VAR-HF[5] | FEM-HF[6] | Delta E[7] | % error[8] | Delta E[9] | % error[10] |
| | | | | | m=10 | | | | | |
| 0.0001 | -0.454395 | -0.454395 | -0.257049 | -0.256606 | -0.197347 | -0.197347 | 0.000000 | 0.000000 | -0.000443 | 0.172147 |
| 0.001 | -0.453044 | -0.453044 | -0.257037 | -0.256595 | -0.196007 | -0.196007 | 0.000000 | 0.000000 | -0.000442 | 0.172111 |
| 0.01 | -0.439381 | -0.439381 | -0.255878 | -0.255445 | -0.183502 | -0.183502 | 0.000000 | 0.000001 | -0.000433 | 0.169154 |
| 0.1 | -0.288912 | -0.288923 | -0.170118 | -0.169934 | -0.118794 | -0.118805 | -0.000011 | 0.003727 | -0.000184 | 0.108200 |
| 1.0 | 1.751539 | 1.751315 | 1.810873 | 1.810890 | -0.059334 | -0.059557 | -0.000224 | 0.012763 | -0.000018 | 0.000971 |
| 10 | 26.447417 | 26.447087 | 26.490355 | 26.490359 | -0.042938 | -0.043268 | -0.000330 | 0.001249 | -0.000004 | 0.000014 |
| 100 | 289.097263 | 289.096910 | 289.135811 | 289.135812 | -0.038548 | -0.038901 | -0.000353 | 0.000122 | -0.000001 | 0.000000 |
| 1000 | 2965.835951 | 2965.835599 | 2965.873207 | 2965.873195 | -0.037256 | -0.037607 | -0.000351 | 0.000012 | 0.000012 | 0.000000 |
| 10000 | 29892.271732 | 29892.271372 | 29892.308661 | 29892.308574 | -0.036929 | -0.037289 | -0.000360 | 0.000001 | 0.000088 | 0.000000 |
| | | | | | m=50 | | | | | |
| 0.0001 | -0.490046 | -0.490046 | -0.353887 | -0.353752 | -0.136159 | -0.136159 | 0.000000 | 0.000000 | -0.000135 | 0.038033 |
| 0.001 | -0.488695 | -0.488695 | -0.353874 | -0.353739 | -0.134821 | -0.134821 | 0.000000 | 0.000000 | -0.000135 | 0.038020 |
| 0.01 | -0.475043 | -0.475043 | -0.352583 | -0.352453 | -0.122460 | -0.122460 | 0.000000 | 0.000000 | -0.000129 | 0.036671 |
| 0.1 | -0.325603 | -0.325612 | -0.261510 | -0.261482 | -0.064093 | -0.064103 | -0.000009 | 0.002887 | -0.000028 | 0.010724 |
| 1.0 | 1.695169 | 1.694935 | 1.717273 | 1.717273 | -0.022104 | -0.022337 | -0.000234 | 0.013787 | 0.000000 | 0.000023 |
| 10 | 26.304325 | 26.303970 | 26.317887 | 26.317887 | -0.013563 | -0.013917 | -0.000355 | 0.001348 | 0.000000 | 0.000000 |
| 100 | 288.671901 | 288.671520 | 288.683473 | 288.683473 | -0.011572 | -0.011953 | -0.000380 | 0.000132 | 0.000000 | 0.000000 |
| 1000 | 2964.515756 | 2964.515378 | 2964.526763 | 2964.526763 | -0.011007 | -0.011385 | -0.000378 | 0.000013 | 0.000000 | 0.000000 |
| 10000 | 29888.121166 | 29888.120778 | 29888.132005 | 29888.132001 | -0.010839 | -0.011227 | -0.000388 | 0.000001 | 0.000004 | 0.000000 |
| | | | | | m=207 | | | | | |

| | Total Energy | | | | Correlation Energy | | Error of variational energy | | Error of [*7s:1s*] Basis Set | |
|---|---|---|---|---|---|---|---|---|---|---|
| ω/Quantities | Variational [1] | FEM [2] | MC-HF/[*7s:7s*] [3] | MC-HF/[*7s:1s*] [4] | VAR-HF [5] | FEM-HF [6] | Delta E [7] | % error [8] | Delta E [9] | % error [10] |
| 0.0001 | -0.497446 | -0.497446 | -0.414686143 | -0.414646 | -0.082760 | -0.082760 | 0.000000 | 0.000000 | -0.000040 | 0.009689 |
| 0.001 | -0.496095 | -0.496095 | -0.414668065 | -0.414628 | -0.081427 | -0.081427 | 0.000000 | 0.000000 | -0.000040 | 0.009680 |
| 0.01 | -0.482446 | -0.482446 | -0.412886483 | -0.412850 | -0.069560 | -0.069560 | 0.000000 | 0.000000 | -0.000037 | 0.008883 |
| 0.1 | -0.333202 | -0.333212 | -0.306618117 | -0.306615 | -0.026584 | -0.026594 | -0.000010 | 0.003001 | -0.000003 | 0.000873 |
| 1.0 | 1.683638441 | 1.683402758 | 1.690371599 | 1.690372 | -0.006733 | -0.006969 | -0.000236 | 0.014000 | 0.000000 | 0.000004 |
| 10 | 26.27522077 | 26.27486117 | 26.27886233 | 26.278862 | -0.003642 | -0.004001 | -0.000360 | 0.001369 | 0.000000 | 0.000000 |
| 100 | 288.5855131 | 288.5851272 | 288.5884904 | 288.588490 | -0.002977 | -0.003363 | -0.000386 | 0.000134 | 0.000000 | 0.000000 |
| 1000 | 2964.247752 | 2964.247368 | 2964.250545 | 2964.250545 | -0.002793 | -0.003177 | -0.000384 | 0.000013 | 0.000000 | 0.000000 |
| 10000 | 29887.2787 | 29887.27831 | 29887.28145 | 29887.281444 | -0.002747 | -0.003137 | -0.000390 | 0.000001 | 0.000003 | 0.000000 |
| | | | | | m=400 | | | | | |
| 0.0001 | -0.498603 | -0.498603 | -0.434898 | -0.434876 | -0.063705 | -0.063705 | 0.000000 | 0.000000 | -0.000022 | 0.005113 |
| 0.001 | -0.497252 | -0.497252 | -0.434876 | -0.434854 | -0.062375 | -0.062375 | 0.000000 | 0.000000 | -0.000022 | 0.005106 |
| 0.01 | -0.483603 | -0.483603 | -0.432709 | -0.432689 | -0.050894 | -0.050894 | 0.000000 | 0.000000 | -0.000019 | 0.004457 |
| 0.1 | -0.334390 | -0.334399 | -0.318137 | -0.318136 | -0.016254 | -0.016263 | -0.000009 | 0.002722 | -0.000001 | 0.000185 |
| 1.0 | 1.681841 | 1.681605 | 1.685464 | 1.685464 | -0.003624 | -0.003860 | -0.000236 | 0.014034 | 0.000000 | 0.000008 |
| 10 | 26.270688 | 26.270328 | 26.272493 | 26.272492 | -0.001804 | -0.002165 | -0.000360 | 0.001372 | 0.000000 | 0.000000 |
| 100 | 288.572063 | 288.571677 | 288.573488 | 288.573487 | -0.001424 | -0.001811 | -0.000387 | 0.000134 | 0.000000 | 0.000000 |
| 1000 | 2964.206030 | 2964.205645 | 2964.207347 | 2964.207347 | -0.001317 | -0.001702 | -0.000385 | 0.000013 | 0.000000 | 0.000000 |
| 10000 | 29887.147551 | 29887.147157 | 29887.148995 | 29887.148842 | -0.001443 | -0.001838 | -0.000395 | 0.000001 | 0.000153 | 0.000001 |
| | | | | | m=900 | | | | | |
| 0.0001 | -0.499295 | -0.499295 | -0.453975 | -0.453965 | -0.045320 | -0.045320 | 0.000000 | 0.000000 | -0.000011 | 0.002314 |

| | Total Energy | | | | Correlation Energy | | Error of variational energy | | Error of [7s:1s] Basis Set | |
|---|---|---|---|---|---|---|---|---|---|---|
| ω/Quantities | Variational [1] | FEM [2] | MC-HF/[7s:7s] [3] | MC-HF/[7s:1s] [4] | VAR-HF [5] | FEM-HF [6] | Delta E [7] | % error [8] | Delta E [9] | % error [10] |
| 0.001 | -0.497944 | -0.497944 | -0.453946 | -0.453935 | -0.043998 | -0.043998 | 0.000000 | 0.000000 | -0.000010 | 0.002307 |
| 0.01 | -0.484295 | -0.484295 | -0.451115 | -0.451106 | -0.033180 | -0.033180 | 0.000000 | 0.000000 | -0.000008 | 0.001805 |
| 0.1 | -0.335100 | -0.335109 | -0.326729 | -0.326729 | -0.008372 | -0.008381 | -0.000009 | 0.002709 | 0.000000 | 0.000093 |
| 1.0 | 1.680766 | 1.680530 | 1.682345 | 1.682345 | -0.001579 | -0.001815 | -0.000236 | 0.014054 | 0.000000 | 0.000014 |
| 10 | 26.267980 | 26.267619 | 26.268617 | 26.268617 | -0.000638 | -0.000999 | -0.000361 | 0.001374 | 0.000000 | 0.000000 |
| 100 | 288.564027 | 288.563640 | 288.564475 | 288.564474 | -0.000448 | -0.000835 | -0.000387 | 0.000134 | 0.000000 | 0.000000 |
| 1000 | 2964.181100 | 2964.180715 | 2964.181492 | 2964.181491 | -0.000391 | -0.000777 | -0.000385 | 0.000013 | 0.000001 | 0.000000 |
| 10000 | 29887.069188 | 29887.068792 | 29887.069700 | 29887.069698 | -0.000512 | -0.000907 | -0.000395 | 0.000001 | 0.000001 | 0.000000 |
| | | | | | m=1836 | | | | | |
| 0.0001 | -0.499578 | -0.499578 | -0.466417 | -0.466412 | -0.033161 | -0.033161 | 0.000000 | 0.000000 | -0.000005 | 0.001139 |
| 0.001 | -0.498226 | -0.498226 | -0.466378 | -0.466373 | -0.031848 | -0.031848 | 0.000000 | 0.000000 | -0.000005 | 0.001134 |
| 0.01 | -0.484578 | -0.484578 | -0.462753 | -0.462749 | -0.021825 | -0.021825 | 0.000000 | 0.000000 | -0.000003 | 0.000743 |
| 0.1 | -0.335391 | -0.335400 | -0.330905 | -0.330906 | -0.004486 | -0.004495 | -0.000009 | 0.002704 | 0.000001 | 0.000382 |
| 1.0 | 1.680327 | 1.680091 | 1.681012 | 1.681011 | -0.000684 | -0.000921 | -0.000236 | 0.014062 | 0.000000 | 0.000013 |
| 10 | 26.266873 | 26.266512 | 26.267013 | 26.267013 | -0.000140 | -0.000501 | -0.000361 | 0.001375 | 0.000000 | 0.000000 |
| 100 | 288.560744 | 288.560357 | 288.560779 | 288.560778 | -0.000035 | -0.000422 | -0.000388 | 0.000134 | 0.000000 | 0.000000 |
| 1000 | 2964.170918 | 2964.170533 | 2964.170918 | 2964.170916 | 0.000000 | -0.000385 | -0.000385 | 0.000013 | 0.000002 | 0.000000 |
| 10000 | 29887.037181 | 29887.036785 | 29887.037312 | 29887.037312 | -0.000132 | -0.000527 | -0.000395 | 0.000001 | 0.000000 | 0.000000 |

1. Total variational energy, including the sum of relative and center of mass energies 2. FEM relative Hamiltonian energy plus center of mass Hamiltonian energy

3. Total MC-HF energy with basis Set [7s:7s]
4. Total MC-HF energy with basis Set [7s:1s]
5. Correlation energy obtained from the difference between the total energy calculated using the variational method compared to MC-HF/[7s:7s]
6. Correlation energy obtained from the difference between the total energy calculated using FEM compared to MC-HF/[7s:7s]
7. Difference between total variational and FEM Energy
8. Percentage error of variational energy compared to FEM energy
9. Difference between total MC-HF energy with [7s:1s] and [7s:7s] basis sets
10. Percentage error of MC-HF energy with [7s:1s] compared to [7s:7s]

The components of the total energy (kinetic energy, harmonic oscillator potential energy, and Coulomb interaction potential energy) for the variational and HF methods are reported in Table 4-4. These components can be presented in two ways: single-particle (section 2-5-1) and pseudo-particle (section 2-5-2), according to the following equations, respectively.

$$E = \langle T_e \rangle + \langle T_p \rangle + \langle HO_e \rangle + \langle HO_p \rangle + \langle INT_{ep} \rangle \quad 4.2$$

$$E = \langle T_R \rangle + \langle T_r \rangle + \langle HO_R \rangle + \langle HO_r \rangle + \langle INT_r \rangle \quad 4.3$$

Items 1 to 8 in the table indicate the single-particle components of the variational energy, items 9 to 15 represent the pseudo-particle components of the variational energy, and items 17 to 24 contain the single-particle components of the HF energy. The sum of the energy components must be equal in both single-particle and pseudo-particle systems. Additionally, the electron-positive particle interaction energy is the same in both systems:

$$\langle T_e \rangle + \langle T_p \rangle = \langle T_R \rangle + \langle T_r \rangle \quad 4.4$$

$$\langle HO_e \rangle + \langle HO_p \rangle = \langle HO_R \rangle + \langle HO_r \rangle \quad 4.5$$

$$\langle INT_{ep} \rangle = \langle INT_r \rangle \quad 4.6$$

The data indicate that the above equations hold true, with the maximum error being on the order of one-thousandth of a percent. Additionally, the virial ratio for the variational and MC-HF methods is presented in items 16 and 24, respectively. The virial ratio, which is a measure of the optimality of the wave function, is defined as follows:

$$vir = \frac{2\langle T\rangle}{\langle r \cdot \nabla V\rangle} \qquad 4.7$$

The virial ratio is expected to be close to 1 for an accurate wave function. For the variational wave function, this ratio is equal to 1 for all masses and frequencies. According to the virial theorem, $2\langle T\rangle = N\langle V\rangle$, where $N$ is the degree of homogeneity of the potential energy function. For a harmonic oscillator, $N=2$, and thus the kinetic energy and the potential energy of the oscillator will be exactly equal (see equations 2-148 and 2-149), each accounting for half of the total energy. The virial ratio for the HF method is reported as $\frac{-\langle V\rangle}{\langle T\rangle}$. Therefore, at high frequencies, where the oscillator potential dominates, this ratio approaches $-1$. At low frequencies, where the Coulomb interaction potential dominates with a degree of homogeneity $N = -1$, the virial ratio approaches 2.

The data in Table 4-4 help us observe the energy components and their relative significance. Generally, some observable trends in the data are as follows:

For the variational method:

- The magnitude of the kinetic energies increases with the oscillator frequency.
- However, at a specific frequency, the electron's kinetic energy increases with the positive particle's mass, while the positive particle's kinetic energy decreases with increasing mass.
- The harmonic oscillator potential energies (both for the electron and the positive particle) increase with frequency. At a specific frequency, they increase at a very slow rate with the positive particle's mass.

- The electron-positive particle interaction energy also increases with frequency and mass, due to the indirect effect of the decreasing average particle distance.

For the Hartree-Fock method, the trends in kinetic energy are similar to the variational method, except:

- At a specific frequency, the positive particle's kinetic energy initially increases with mass up to mass 10, then decreases with increasing mass. This can explain the change in correlation energy behavior with increasing mass.
- The harmonic oscillator potential energies of both the electron and positive particle, at a fixed mass, increase with frequency. At a fixed frequency, they decrease with increasing mass, contrary to the variational trends.
- The HF interaction energy increases with mass and frequency.

Table 4-4: Total energy, its components, and virial ratio obtained from the variational and MC-HF/[7s:7s] methods

| Quantities /ω | 0.0001 | 0.001 | 0.01 | 0.1 | 1 | 10 | 100 | 1000 | 10000 |
|---|---|---|---|---|---|---|---|---|---|
| | | | | | m=1 | | | | |
| Variation energy[1] | -0.249850 | -0.248497 | -0.234701 | -0.073933 | 2.112000 | 27.395490 | 291.942322 | 2974.690624 | 29920.133739 |
| VAR TE[2] | 0.125038 | 0.125378 | 0.129048 | 0.184559 | 0.899854 | 7.857506 | 76.037204 | 753.192946 | 7510.009565 |
| VAR TP[3] | 0.125038 | 0.125378 | 0.129048 | 0.184559 | 0.899854 | 7.857506 | 76.037204 | 753.192946 | 7510.009565 |
| Total VART[4] | 0.250075 | 0.250756 | 0.258097 | 0.369119 | 1.799707 | 15.715011 | 152.074408 | 1506.385891 | 15020.019129 |
| VAR HO E[5] | 0.000038 | 0.000376 | 0.003899 | 0.049198 | 0.651951 | 7.185084 | 74.002788 | 746.846265 | 7490.029408 |
| VAR HO P[6] | 0.000038 | 0.000376 | 0.003899 | 0.049198 | 0.651951 | 7.185084 | 74.002788 | 746.846265 | 7490.029408 |
| Total VAR HO[7] | 0.000075 | 0.000753 | 0.007799 | 0.098395 | 1.303902 | 14.370167 | 148.005576 | 1493.692530 | 14980.058815 |
| VAR INT[8] | -0.500000 | -0.500006 | -0.500596 | -0.541447 | -0.991609 | -2.689688 | -8.137662 | -25.387791 | -79.944206 |
| REL T[9] | 0.250000 | 0.250006 | 0.250597 | 0.294119 | 1.049707 | 8.215011 | 77.074408 | 756.385886 | 7520.019128 |
| REL INT[10] | -0.500000 | -0.500006 | -0.500596 | -0.541447 | -0.991609 | -2.689688 | -8.137662 | -25.387791 | -79.944206 |
| REL HO[11] | 0.000000 | 0.000003 | 0.000299 | 0.023395 | 0.553902 | 6.870167 | 73.005576 | 743.692530 | 7480.058817 |
| COM T[12] | 0.000075 | 0.000750 | 0.007500 | 0.075000 | 0.750000 | 7.500000 | 75.000000 | 750.000000 | 7500.000000 |
| COM HO[13] | 0.000075 | 0.000750 | 0.007500 | 0.075000 | 0.750000 | 7.500000 | 75.000000 | 750.000000 | 7500.000000 |
| Total RC T[14] | 0.250075 | 0.250756 | 0.258097 | 0.369119 | 1.799707 | 15.715011 | 152.074408 | 1506.385886 | 15020.019128 |
| Total RC HO[15] | 0.000075 | 0.000753 | 0.007799 | 0.098395 | 1.303902 | 14.370167 | 148.005576 | 1493.692530 | 14980.058817 |
| VAR Virial ratio[16] | 1.000000 | 1.000000 | 1.000000 | 1.000000 | 1.000000 | 1.000000 | 1.000000 | 0.999999 | 0.999998 |
| MC-HF/[*7s:7s*] E[17] | -0.108513 | -0.108491 | -0.106407 | 0.012203 | 2.171918 | 27.448072 | 291.992777 | 2974.740427 | 29920.183344 |
| MC-HF TE[18] | 0.054257 | 0.054278 | 0.056302 | 0.127546 | 0.865636 | 7.830022 | 76.011617 | 753.168064 | 7509.987785 |
| MC-HF TP[19] | 0.054257 | 0.054278 | 0.056302 | 0.127546 | 0.865636 | 7.830022 | 76.011614 | 753.168062 | 7509.987545 |

| Quantities /ω | 0.0001 | 0.001 | 0.01 | 0.1 | 1 | 10 | 100 | 1000 | 10000 |
|---|---|---|---|---|---|---|---|---|---|
| Total MC-HF T[20] | 0.108514 | 0.108556 | 0.112604 | 0.255092 | 1.731273 | 15.660044 | 152.023231 | 1506.336126 | 15019.975330 |
| MC-HF HO E[21] | 0.000000 | 0.000011 | 0.001033 | 0.044549 | 0.650532 | 7.184686 | 74.002667 | 746.846090 | 7490.026332 |
| MC-HF HO P[22] | 0.000000 | 0.000011 | 0.001033 | 0.044549 | 0.650532 | 7.184686 | 74.002670 | 746.846092 | 7490.026573 |
| Total MC-HF HO[23] | 0.000000 | 0.000021 | 0.002066 | 0.089098 | 1.301064 | 14.369372 | 148.005337 | 1493.692181 | 14980.052905 |
| MC-HF INT[24] | -0.217027 | -0.217069 | -0.221077 | -0.331987 | -0.860418 | -2.581344 | -8.035792 | -25.287881 | -79.844891 |
| MC-HF Virial Ratio[25] | 1.999989 | 1.999400 | 1.944965 | 0.952162 | -0.254521 | -0.752746 | -0.920712 | -0.974818 | -0.992026 |
| | | | | | m=1.5 | | | | |
| Variation energy[1] | -0.299850 | -0.298498 | -0.284750 | -0.127533 | 2.016267 | 27.137957 | 291.164882 | 2972.266767 | 29912.502746 |
| VAR TE[2] | 0.180030 | 0.180303 | 0.183299 | 0.233614 | 0.955011 | 7.975708 | 76.368579 | 754.202362 | 7513.166676 |
| VAR TP[3] | 0.120045 | 0.120452 | 0.124699 | 0.180742 | 0.886674 | 7.817139 | 75.912386 | 752.801576 | 7508.777784 |
| Total VART[4] | 0.300075 | 0.300755 | 0.307998 | 0.414356 | 1.841685 | 15.792847 | 152.280966 | 1507.003938 | 15021.944459 |
| VAR HO E[5] | 0.000030 | 0.000301 | 0.003150 | 0.042365 | 0.621590 | 7.086161 | 73.689169 | 745.854141 | 7486.889440 |
| VAR HO P[6] | 0.000045 | 0.000451 | 0.004600 | 0.053243 | 0.664394 | 7.224107 | 74.126113 | 747.236094 | 7491.259627 |
| Total VAR HO[7] | 0.000075 | 0.000752 | 0.007749 | 0.095608 | 1.285984 | 14.310268 | 147.815282 | 1493.090235 | 14978.149067 |
| VAR INT[8] | -0.600000 | -0.600005 | -0.600498 | -0.637497 | -1.111402 | -2.965158 | -8.931366 | -27.827407 | -87.590782 |
| REL T[9] | 0.300000 | 0.300005 | 0.300498 | 0.339356 | 1.091685 | 8.292847 | 77.280966 | 757.003939 | 7521.944461 |
| REL INT[10] | -0.600000 | -0.600005 | -0.600498 | -0.637497 | -1.111402 | -2.965158 | -8.931366 | -27.827407 | -87.590782 |
| REL HO[11] | 0.000000 | 0.000002 | 0.000249 | 0.020608 | 0.535984 | 6.810268 | 72.815282 | 743.090235 | 7478.149067 |
| COM T[12] | 0.000075 | 0.000750 | 0.007500 | 0.075000 | 0.750000 | 7.500000 | 75.000000 | 750.000000 | 7500.000000 |
| COM HO[13] | 0.000075 | 0.000750 | 0.007500 | 0.075000 | 0.750000 | 7.500000 | 75.000000 | 750.000000 | 7500.000000 |
| Total RC T[14] | 0.300075 | 0.300755 | 0.307998 | 0.414356 | 1.841685 | 15.792847 | 152.280966 | 1507.003939 | 15021.944461 |

| Quantities /ω | 0.0001 | 0.001 | 0.01 | 0.1 | 1 | 10 | 100 | 1000 | 10000 |
|---|---|---|---|---|---|---|---|---|---|
| Total RC HO[15] | 0.000075 | 0.000752 | 0.007749 | 0.095608 | 1.285984 | 14.310268 | 147.815282 | 1493.090235 | 14978.149067 |
| VAR Virial ratio[16] | 1.000000 | 1.000000 | 1.000000 | 1.000000 | 1.000000 | 1.000000 | 1.000000 | 1.000000 | 1.000000 |
| MC-HF/[$7s$:$7s$] E [17] | -0.131661 | -0.131644 | -0.129900 | -0.021539 | 2.087757 | 27.199613 | 291.223701 | 2972.324714 | 29912.560519 |
| MC-HF TE[18] | 0.071904 | 0.071923 | 0.073710 | 0.145309 | 0.902448 | 7.934121 | 76.330067 | 754.164780 | 7513.129370 |
| MC-HF TP[19] | 0.059756 | 0.059773 | 0.061403 | 0.127650 | 0.856317 | 7.793882 | 75.891153 | 752.780910 | 7508.754787 |
| Total MC-HF T[20] | 0.131660 | 0.131696 | 0.135112 | 0.272959 | 1.758765 | 15.728003 | 152.221220 | 1506.945690 | 15021.884157 |
| MC-HF HO E[21] | 0.000000 | 0.000008 | 0.000797 | 0.039402 | 0.624862 | 7.091596 | 73.695222 | 745.860380 | 7486.895746 |
| MC-HF HO P[22] | 0.000000 | 0.000010 | 0.000941 | 0.044406 | 0.657312 | 7.217609 | 74.119712 | 747.229757 | 7491.255910 |
| Total MC-HF HO[23] | 0.000000 | 0.000018 | 0.001737 | 0.083807 | 1.282174 | 14.309205 | 147.814934 | 1493.090137 | 14978.151656 |
| MC-HF INT[24] | -0.263321 | -0.263358 | -0.266750 | -0.378305 | -0.953182 | -2.837595 | -8.812453 | -27.711113 | -87.475293 |
| MC-HF Virial Ratio[25] | 2.000008 | 1.999600 | 1.961425 | 1.078908 | -0.187058 | -0.729375 | -0.913161 | -0.972417 | -0.991266 |
| | | | | m=2 | | | | | |
| Variation energy[1] | -0.333183 | -0.331831 | -0.318109 | -0.162780 | 1.955715 | 26.977247 | 290.681475 | 2970.761234 | 29907.764463 |
| VAR TE[2] | 0.222247 | 0.222475 | 0.225021 | 0.271652 | 0.996139 | 8.061391 | 76.606651 | 754.925466 | 7515.424558 |
| VAR TP[3] | 0.111161 | 0.111613 | 0.116261 | 0.173326 | 0.873070 | 7.780696 | 75.803326 | 752.462738 | 7507.712279 |
| Total VART[4] | 0.333408 | 0.334088 | 0.341282 | 0.444978 | 1.869209 | 15.842087 | 152.409976 | 1507.388204 | 15023.136838 |
| VAR HO E[5] | 0.000025 | 0.000251 | 0.002650 | 0.037711 | 0.599983 | 7.015407 | 73.464767 | 745.144321 | 7484.644733 |
| VAR HO P[6] | 0.000050 | 0.000501 | 0.005075 | 0.056355 | 0.674992 | 7.257704 | 74.232383 | 747.572161 | 7492.322367 |
| Total VAR HO[7] | 0.000075 | 0.000752 | 0.007724 | 0.094066 | 1.274975 | 14.273111 | 147.697150 | 1492.716482 | 14976.967100 |
| VAR INT[8] | -0.666667 | -0.666672 | -0.667115 | -0.701824 | -1.188468 | -3.137952 | -9.425652 | -29.343461 | -92.339475 |
| REL T[9] | 0.333333 | 0.333338 | 0.333782 | 0.369978 | 1.119209 | 8.342087 | 77.409976 | 757.388213 | 7523.136838 |

| Quantities /ω | 0.0001 | 0.001 | 0.01 | 0.1 | 1 | 10 | 100 | 1000 | 10000 |
|---|---|---|---|---|---|---|---|---|---|
| REL INT[10] | -0.666667 | -0.666671 | -0.667115 | -0.701824 | -1.188468 | -3.137951 | -9.425652 | -29.343461 | -92.339475 |
| REL HO[11] | 0.000000 | 0.000002 | 0.000224 | 0.019066 | 0.524975 | 6.773111 | 72.697150 | 742.716482 | 7476.967100 |
| COM T[12] | 0.000075 | 0.000750 | 0.007500 | 0.075000 | 0.750000 | 7.500000 | 75.000000 | 750.000000 | 7500.000000 |
| COM HO[13] | 0.000075 | 0.000750 | 0.007500 | 0.075000 | 0.750000 | 7.500000 | 75.000000 | 750.000000 | 7500.000000 |
| Total RC T[14] | 0.333408 | 0.334088 | 0.341282 | 0.444978 | 1.869209 | 15.842087 | 152.409976 | 1507.388213 | 15023.136838 |
| Total RC HO[15] | 0.000075 | 0.000752 | 0.007724 | 0.094066 | 1.274975 | 14.273111 | 147.697150 | 1492.716482 | 14976.967100 |
| VAR Virial ratio[16] | 1.000000 | 1.000000 | 1.000000 | 1.000000 | 1.000000 | 1.000000 | 1.000000 | 1.000000 | 1.000000 |
| MC-HF/[*7s:7s*] E [17] | -0.149328 | -0.149312 | -0.147746 | -0.045426 | 2.032554 | 27.042353 | 290.743202 | 2970.821923 | 29907.824832 |
| MC-HF TE[18] | 0.086372 | 0.086389 | 0.088052 | 0.159882 | 0.930716 | 8.010659 | 76.560034 | 754.880100 | 7515.379561 |
| MC-HF TP[19] | 0.062955 | 0.062970 | 0.064390 | 0.127220 | 0.848021 | 7.762520 | 75.787062 | 752.447064 | 7507.696782 |
| Total MC-HF T[20] | 0.149327 | 0.149359 | 0.152442 | 0.287101 | 1.778737 | 15.773179 | 152.347096 | 1507.327164 | 15023.076343 |
| MC-HF HO E[21] | 0.000000 | 0.000007 | 0.000672 | 0.036065 | 0.606838 | 7.025252 | 73.475454 | 745.155265 | 7484.655762 |
| MC-HF HO P[22] | 0.000000 | 0.000009 | 0.000893 | 0.044493 | 0.663592 | 7.246592 | 74.221331 | 747.561103 | 7492.311315 |
| Total MC-HF HO[23] | 0.000000 | 0.000016 | 0.001565 | 0.080558 | 1.270430 | 14.271844 | 147.696786 | 1492.716368 | 14976.967077 |
| MC-HF INT[24] | -0.298655 | -0.298687 | -0.301753 | -0.413086 | -1.016614 | -3.002670 | -9.300680 | -29.221608 | -92.218589 |
| MC-HF Virial Ratio[25] | 2.000006 | 1.999687 | 1.969194 | 1.158223 | -0.142695 | -0.714452 | -0.908426 | -0.970920 | -0.990792 |
| | | | | | m=3 | | | | |
| Variation energy[1] | -0.374850 | -0.373498 | -0.359800 | -0.206440 | 1.883003 | 26.786450 | 290.109312 | 2968.980867 | 29902.162752 |
| VAR TE[2] | 0.281269 | 0.281440 | 0.283424 | 0.325262 | 1.052436 | 8.175910 | 76.922433 | 755.882362 | 7518.410271 |
| VAR TP[3] | 0.093806 | 0.094313 | 0.099475 | 0.158421 | 0.850812 | 7.725303 | 75.640811 | 751.960794 | 7506.136756 |
| Total VART[4] | 0.375075 | 0.375754 | 0.382899 | 0.483683 | 1.903248 | 15.901213 | 152.563244 | 1507.843156 | 15024.547027 |

| Quantities /ω | 0.0001 | 0.001 | 0.01 | 0.1 | 1 | 10 | 100 | 1000 | 10000 |
|---|---|---|---|---|---|---|---|---|---|
| VAR HO E[5] | 0.000019 | 0.000189 | 0.002025 | 0.031811 | 0.571563 | 6.921916 | 73.168139 | 744.206011 | 7481.677445 |
| VAR HO P[6] | 0.000056 | 0.000563 | 0.005675 | 0.060604 | 0.690521 | 7.307305 | 74.389380 | 748.068670 | 7493.892482 |
| Total VAR HO[7] | 0.000075 | 0.000752 | 0.007700 | 0.092414 | 1.262084 | 14.229221 | 147.557519 | 1492.274681 | 14975.569926 |
| VAR INT[8] | -0.750000 | -0.750004 | -0.750399 | -0.782537 | -1.282329 | -3.343984 | -10.011451 | -31.136990 | -97.954202 |
| REL T[9] | 0.375000 | 0.375004 | 0.375399 | 0.408683 | 1.153248 | 8.401213 | 77.563244 | 757.843176 | 7524.547027 |
| REL INT[10] | -0.750000 | -0.750004 | -0.750399 | -0.782537 | -1.282329 | -3.343984 | -10.011451 | -31.136990 | -97.954202 |
| REL HO[11] | 0.000000 | 0.000002 | 0.000200 | 0.017414 | 0.512084 | 6.729221 | 72.557519 | 742.274681 | 7475.569926 |
| COM T[12] | 0.000075 | 0.000750 | 0.007500 | 0.075000 | 0.750000 | 7.500000 | 75.000000 | 750.000000 | 7499.999998 |
| COM HO[13] | 0.000075 | 0.000750 | 0.007500 | 0.075000 | 0.750000 | 7.500000 | 75.000000 | 750.000000 | 7500.000002 |
| Total RC T[14] | 0.375075 | 0.375754 | 0.382899 | 0.483683 | 1.903248 | 15.901213 | 152.563244 | 1507.843176 | 15024.547026 |
| Total RC HO[15] | 0.000075 | 0.000752 | 0.007700 | 0.092414 | 1.262084 | 14.229221 | 147.557519 | 1492.274681 | 14975.569928 |
| VAR Virial ratio[16] | 1.000000 | 1.000000 | 1.000000 | 1.000000 | 1.000000 | 1.000000 | 1.000000 | 1.000000 | 1.000000 |
| MC-HF/[*7s:7s*] E [17] | -0.175573 | -0.175559 | -0.174173 | -0.078661 | 1.962394 | 26.851598 | 290.170376 | 2969.040682 | 29902.222187 |
| MC-HF TE[18] | 0.109323 | 0.109339 | 0.110873 | 0.183072 | 0.972233 | 8.116130 | 76.868367 | 755.830037 | 7518.358479 |
| MC-HF TP[19] | 0.066251 | 0.066263 | 0.067464 | 0.125904 | 0.835049 | 7.715334 | 75.632449 | 751.952935 | 7506.129002 |
| Total MC-HF T[20] | 0.175574 | 0.175602 | 0.178338 | 0.308975 | 1.807282 | 15.831464 | 152.500816 | 1507.782972 | 15024.487481 |
| MC-HF HO E[21] | 0.000000 | 0.000005 | 0.000540 | 0.031885 | 0.582797 | 6.936918 | 73.184129 | 744.222289 | 7481.693785 |
| MC-HF HO P[22] | 0.000000 | 0.000009 | 0.000848 | 0.044886 | 0.673762 | 7.290771 | 74.372922 | 748.052211 | 7493.876075 |
| Total MC-HF HO[23] | 0.000000 | 0.000014 | 0.001389 | 0.076771 | 1.256559 | 14.227689 | 147.557051 | 1492.274500 | 14975.569861 |
| MC-HF INT[24] | -0.351147 | -0.351175 | -0.353899 | -0.464408 | -1.101447 | -3.207555 | -9.887490 | -31.016790 | -97.835154 |
| MC-HF Virial Ratio[25] | 1.999997 | 1.999758 | 1.976646 | 1.254586 | -0.085826 | -0.696091 | -0.902746 | -0.969143 | -0.990232 |

| Quantities /ω | 0.0001 | 0.001 | 0.01 | 0.1 | 1 | 10 | 100 | 1000 | 10000 |
|---|---|---|---|---|---|---|---|---|---|
| | | | | m=10 | | | | | |
| Variation energy[1] | -0.454395 | -0.453044 | -0.439381 | -0.288912 | 1.751539 | 26.447417 | 289.097263 | 2965.835951 | 29892.271732 |
| VAR TE[2] | 0.413230 | 0.413294 | 0.414204 | 0.446475 | 1.174993 | 8.416433 | 77.578073 | 757.862074 | 7524.580388 |
| VAR TP[3] | 0.041390 | 0.042004 | 0.048170 | 0.112147 | 0.792499 | 7.591643 | 75.257807 | 750.786207 | 7502.458035 |
| Total VART[4] | 0.454620 | 0.455299 | 0.462375 | 0.558622 | 1.967492 | 16.008077 | 152.835880 | 1508.648281 | 15027.038423 |
| VAR HO E[5] | 0.000007 | 0.000070 | 0.000832 | 0.020367 | 0.513343 | 6.728936 | 72.555477 | 742.267944 | 7475.548526 |
| VAR HO P[6] | 0.000068 | 0.000682 | 0.006833 | 0.069537 | 0.726334 | 7.422894 | 74.755548 | 749.226795 | 7497.554878 |
| Total VAR HO[7] | 0.000075 | 0.000752 | 0.007665 | 0.089903 | 1.239677 | 14.151830 | 147.311025 | 1491.494739 | 14973.103404 |
| VAR INT[8] | -0.909091 | -0.909094 | -0.909420 | -0.937438 | -1.455630 | -3.712490 | -11.049642 | -34.307070 | -107.870080 |
| REL T[9] | 0.454545 | 0.454549 | 0.454875 | 0.483622 | 1.217492 | 8.508077 | 77.835880 | 758.648277 | 7527.038427 |
| REL INT[10] | -0.909091 | -0.909094 | -0.909420 | -0.937438 | -1.455630 | -3.712490 | -11.049642 | -34.307069 | -107.870081 |
| REL HO[11] | 0.000000 | 0.000002 | 0.000165 | 0.014903 | 0.489677 | 6.651830 | 72.311025 | 741.494743 | 7473.103386 |
| COM T[12] | 0.000075 | 0.000750 | 0.007500 | 0.075000 | 0.750000 | 7.500000 | 75.000000 | 750.000000 | 7500.000000 |
| COM HO[13] | 0.000075 | 0.000750 | 0.007500 | 0.075000 | 0.750000 | 7.500000 | 75.000000 | 750.000000 | 7500.000000 |
| Total RC T[14] | 0.454620 | 0.455299 | 0.462375 | 0.558622 | 1.967492 | 16.008077 | 152.835880 | 1508.648277 | 15027.038427 |
| Total RC HO[15] | 0.000075 | 0.000752 | 0.007665 | 0.089903 | 1.239677 | 14.151830 | 147.311025 | 1491.494743 | 14973.103386 |
| VAR Virial ratio[16] | 1.000000 | 1.000000 | 1.000000 | 1.000000 | 1.000000 | 1.000000 | 1.000000 | 1.000000 | 1.000000 |
| MC-HF/[*7s:7s*] E [17] | -0.257049 | -0.257037 | -0.255878 | -0.170118 | 1.810873 | 26.490355 | 289.135811 | 2965.873207 | 29892.308661 |
| MC-HF TE[18] | 0.190221 | 0.190237 | 0.191653 | 0.265936 | 1.090860 | 8.363320 | 77.532992 | 757.819352 | 7524.538502 |
| MC-HF TP[19] | 0.066828 | 0.066837 | 0.067719 | 0.117340 | 0.797477 | 7.596780 | 75.262833 | 750.790852 | 7502.460544 |
| Total MC-HF T[20] | 0.257048 | 0.257074 | 0.259372 | 0.383277 | 1.888337 | 15.960100 | 152.795825 | 1508.610204 | 15026.999046 |

| Quantities /ω | 0.0001 | 0.001 | 0.01 | 0.1 | 1 | 10 | 100 | 1000 | 10000 |
|---|---|---|---|---|---|---|---|---|---|
| MC-HF HO E[21] | 0.000000 | 0.000003 | 0.000326 | 0.023032 | 0.527705 | 6.745695 | 72.572511 | 742.284996 | 7475.565480 |
| MC-HF HO P[22] | 0.000000 | 0.000008 | 0.000837 | 0.048020 | 0.705366 | 7.404457 | 74.738088 | 749.209996 | 7497.540355 |
| Total MC-HF HO[23] | 0.000000 | 0.000012 | 0.001164 | 0.071053 | 1.233070 | 14.150152 | 147.310599 | 1491.494992 | 14973.105835 |
| MC-HF INT[24] | -0.514097 | -0.514123 | -0.516413 | -0.624448 | -1.310535 | -3.619897 | -10.970613 | -34.231990 | -107.796219 |
| MC-HF Virial Ratio[25] | 2.000001 | 1.999854 | 1.986531 | 1.443852 | 0.041023 | -0.659786 | -0.892302 | -0.965964 | -0.989240 |
| | | | | | m=50 | | | | |
| Variation energy[1] | -0.490046 | -0.488695 | -0.475043 | -0.325603 | 1.695169 | 26.304325 | 288.671901 | 2964.515756 | 29888.121166 |
| VAR TE[2] | 0.480586 | 0.480602 | 0.481031 | 0.508885 | 1.236361 | 8.533208 | 77.893163 | 758.810580 | 7527.533827 |
| VAR TP[3] | 0.009685 | 0.010347 | 0.016971 | 0.083678 | 0.759727 | 7.520664 | 75.057863 | 750.176211 | 7500.550737 |
| Total VART[4] | 0.490271 | 0.490949 | 0.498002 | 0.592562 | 1.996089 | 16.053873 | 152.951026 | 1508.986791 | 15028.084564 |
| VAR HO E[5] | 0.000001 | 0.000016 | 0.000297 | 0.015183 | 0.485705 | 6.636666 | 72.262395 | 741.340705 | 7472.616132 |
| VAR HO P[6] | 0.000074 | 0.000735 | 0.007356 | 0.073804 | 0.744714 | 7.482733 | 74.945248 | 749.826814 | 7499.452322 |
| Total VAR HO[7] | 0.000075 | 0.000752 | 0.007653 | 0.088986 | 1.230419 | 14.119399 | 147.207643 | 1491.167520 | 14972.068454 |
| VAR INT[8] | -0.980392 | -0.980395 | -0.980698 | -1.007152 | -1.531339 | -3.868947 | -11.486768 | -35.638555 | -112.031793 |
| REL T[9] | 0.490196 | 0.490199 | 0.490502 | 0.517562 | 1.246089 | 8.553873 | 77.951026 | 758.986791 | 7528.084504 |
| REL INT[10] | -0.980392 | -0.980395 | -0.980698 | -1.007152 | -1.531339 | -3.868947 | -11.486768 | -35.638555 | -112.031793 |
| REL HO[11] | 0.000000 | 0.000002 | 0.000153 | 0.013986 | 0.480419 | 6.619399 | 72.207642 | 741.167520 | 7472.068455 |
| COM T[12] | 0.000075 | 0.000750 | 0.007500 | 0.075000 | 0.750000 | 7.500000 | 75.000000 | 750.000000 | 7500.000000 |
| COM HO[13] | 0.000075 | 0.000750 | 0.007500 | 0.075000 | 0.750000 | 7.500000 | 75.000000 | 750.000000 | 7500.000000 |
| Total RC T[14] | 0.490271 | 0.490949 | 0.498002 | 0.592562 | 1.996089 | 16.053873 | 152.951026 | 1508.986791 | 15028.084504 |
| Total RC HO[15] | 0.000075 | 0.000752 | 0.007653 | 0.088986 | 1.230419 | 14.119399 | 147.207642 | 1491.167520 | 14972.068455 |

| Quantities /ω | 0.0001 | 0.001 | 0.01 | 0.1 | 1 | 10 | 100 | 1000 | 10000 |
|---|---|---|---|---|---|---|---|---|---|
| VAR Virial ratio[16] | 1.000000 | 1.000000 | 1.000000 | 1.000000 | 1.000000 | 1.000000 | 1.000000 | 1.000000 | 1.000000 |
| MC-HF/[*7s:7s*] E [17] | -0.353887 | -0.353874 | -0.352583 | -0.261510 | 1.717273 | 26.317887 | 288.683473 | 2964.526763 | 29888.132005 |
| MC-HF TE[18] | 0.302339 | 0.302356 | 0.304056 | 0.382721 | 1.196668 | 8.513224 | 77.877443 | 758.796032 | 7527.519574 |
| MC-HF TP[19] | 0.051554 | 0.051563 | 0.052416 | 0.099684 | 0.766004 | 7.524732 | 75.061373 | 750.179559 | 7500.554606 |
| Total MC-HF T[20] | 0.353893 | 0.353919 | 0.356473 | 0.482406 | 1.962672 | 16.037956 | 152.938816 | 1508.975591 | 15028.074181 |
| MC-HF HO E[21] | 0.000000 | 0.000002 | 0.000219 | 0.017187 | 0.492317 | 6.643259 | 72.268755 | 741.346974 | 7472.622430 |
| MC-HF HO P[22] | 0.000000 | 0.000011 | 0.001076 | 0.056445 | 0.734331 | 7.475350 | 74.938677 | 749.820484 | 7499.445439 |
| Total MC-HF HO[23] | 0.000000 | 0.000013 | 0.001295 | 0.073632 | 1.226648 | 14.118608 | 147.207432 | 1491.167458 | 14972.067869 |
| MC-HF INT[24] | -0.707780 | -0.707806 | -0.710350 | -0.817548 | -1.472047 | -3.838676 | -11.462775 | -35.616286 | -112.010044 |
| MC-HF Virial Ratio[25] | 1.999983 | 1.999873 | 1.989088 | 1.542095 | 0.125033 | -0.640975 | -0.887575 | -0.964596 | -0.988820 |
| | | | | m=207 | | | | | |
| Variation energy[1] | -0.497446 | -0.496095 | -0.482446 | -0.333202 | 1.683638 | 26.275221 | 288.585513 | 2964.247752 | 29887.278700 |
| VAR TE[2] | 0.495204 | 0.495210 | 0.495540 | 0.522468 | 1.249601 | 8.558126 | 77.960160 | 759.012015 | 7528.160741 |
| VAR TP[3] | 0.002467 | 0.003139 | 0.009858 | 0.077162 | 0.752414 | 7.505112 | 75.014300 | 750.043536 | 7500.135624 |
| Total VART[4] | 0.497671 | 0.498349 | 0.505397 | 0.599630 | 2.002014 | 16.063238 | 152.974461 | 1509.055552 | 15028.296365 |
| VAR HO E[5] | 0.000000 | 0.000005 | 0.000186 | 0.014103 | 0.479856 | 6.617085 | 72.200178 | 741.143887 | 7471.993785 |
| VAR HO P[6] | 0.000075 | 0.000746 | 0.007465 | 0.074706 | 0.748695 | 7.495735 | 74.986474 | 749.957217 | 7499.864699 |
| Total VAR HO[7] | 0.000075 | 0.000752 | 0.007651 | 0.088809 | 1.228551 | 14.112820 | 147.186652 | 1491.101104 | 14971.858484 |
| VAR INT[8] | -0.995192 | -0.995195 | -0.995493 | -1.021642 | -1.546927 | -3.900836 | -11.575591 | -35.908903 | -112.876573 |
| REL T[9] | 0.497596 | 0.497599 | 0.497897 | 0.524630 | 1.252015 | 8.563238 | 77.974457 | 759.055552 | 7528.296783 |
| REL INT[10] | -0.995192 | -0.995195 | -0.995493 | -1.021642 | -1.546927 | -3.900837 | -11.575600 | -35.908903 | -112.876573 |

| Quantities /ω | 0.0001 | 0.001 | 0.01 | 0.1 | 1 | 10 | 100 | 1000 | 10000 |
|---|---|---|---|---|---|---|---|---|---|
| REL HO[11] | 0.000000 | 0.000002 | 0.000151 | 0.013809 | 0.478551 | 6.612820 | 72.186657 | 741.101104 | 7471.858489 |
| COM T[12] | 0.000075 | 0.000750 | 0.007500 | 0.075000 | 0.750000 | 7.500000 | 75.000000 | 750.000000 | 7500.000008 |
| COM HO[13] | 0.000075 | 0.000750 | 0.007500 | 0.075000 | 0.750000 | 7.500000 | 75.000000 | 750.000000 | 7499.999992 |
| Total RC T[14] | 0.497671 | 0.498349 | 0.505397 | 0.599630 | 2.002015 | 16.063238 | 152.974457 | 1509.055552 | 15028.296791 |
| Total RC HO[15] | 0.000075 | 0.000752 | 0.007651 | 0.088809 | 1.228551 | 14.112820 | 147.186657 | 1491.101104 | 14971.858481 |
| VAR Virial ratio[16] | 1.000000 | 1.000000 | 1.000000 | 1.000000 | 1.000000 | 1.000000 | 1.000000 | 1.000000 | 1.000000 |
| MC-HF/[*7s:7s*] E [17] | -0.414686 | -0.414668 | -0.412886 | -0.306618 | 1.690372 | 26.278862 | 288.588490 | 2964.250545 | 29887.281447 |
| MC-HF TE[18] | 0.380422 | 0.380447 | 0.382855 | 0.460480 | 1.236019 | 8.552262 | 77.955792 | 759.008048 | 7528.156876 |
| MC-HF TP[19] | 0.034267 | 0.034278 | 0.035356 | 0.086426 | 0.754936 | 7.506531 | 75.015478 | 750.044648 | 7500.136990 |
| Total MC-HF T[20] | 0.414689 | 0.414725 | 0.418211 | 0.546906 | 1.990955 | 16.058793 | 152.971270 | 1509.052696 | 15028.293866 |
| MC-HF HO E[21] | 0.000000 | 0.000002 | 0.000182 | 0.015011 | 0.482015 | 6.619078 | 72.202063 | 741.145729 | 7471.995624 |
| MC-HF HO P[22] | 0.000000 | 0.000016 | 0.001593 | 0.065086 | 0.745097 | 7.493475 | 74.984525 | 749.955355 | 7499.863016 |
| Total MC-HF HO[23] | 0.000000 | 0.000018 | 0.001774 | 0.080097 | 1.227112 | 14.112553 | 147.186588 | 1491.101084 | 14971.858640 |
| MC-HF INT[24] | -0.829375 | -0.829411 | -0.832872 | -0.933620 | -1.527695 | -3.892484 | -11.569367 | -35.903235 | -112.871059 |
| MC-HF Virial Ratio[25] | 1.999993 | 1.999863 | 1.987268 | 1.560641 | 0.150975 | -0.636416 | -0.886554 | -0.964312 | -0.988734 |
| | | | | | m = 400 | | | | |
| Variation energy[1] | -0.498603 | -0.497252 | -0.483603 | -0.334390 | 1.681841 | 26.270688 | 288.572063 | 2964.206030 | 29887.147551 |
| VAR TE[2] | 0.497510 | 0.497514 | 0.497828 | 0.524612 | 1.251687 | 8.562043 | 77.970678 | 759.043651 | 7528.259187 |
| VAR TP[3] | 0.001319 | 0.001992 | 0.008726 | 0.076124 | 0.751254 | 7.502655 | 75.007427 | 750.022610 | 7500.070071 |
| Total VART[4] | 0.498828 | 0.499506 | 0.506553 | 0.600736 | 2.002941 | 16.064698 | 152.978105 | 1509.066261 | 15028.329257 |
| VAR HO E[5] | 0.000000 | 0.000003 | 0.000169 | 0.013935 | 0.478938 | 6.614010 | 72.190415 | 741.112979 | 7471.896060 |

| Quantities /ω | 0.0001 | 0.001 | 0.01 | 0.1 | 1 | 10 | 100 | 1000 | 10000 |
|---|---|---|---|---|---|---|---|---|---|
| VAR HO P[6] | 0.000075 | 0.000748 | 0.007482 | 0.074847 | 0.749322 | 7.497785 | 74.992976 | 749.977782 | 7499.929081 |
| Total VAR HO[7] | 0.000075 | 0.000752 | 0.007650 | 0.088782 | 1.228261 | 14.111795 | 147.183391 | 1491.090762 | 14971.825141 |
| VAR INT[8] | -0.997506 | -0.997509 | -0.997806 | -1.023908 | -1.549360 | -3.905805 | -11.589431 | -35.950992 | -113.008083 |
| REL T[9] | 0.498753 | 0.498756 | 0.499053 | 0.525736 | 1.252941 | 8.564698 | 77.978106 | 759.066260 | 7528.329835 |
| REL INT[10] | -0.997506 | -0.997509 | -0.997806 | -1.023908 | -1.549361 | -3.905805 | -11.589432 | -35.950992 | -113.008083 |
| REL HO[11] | 0.000000 | 0.000002 | 0.000150 | 0.013782 | 0.478261 | 6.611795 | 72.183390 | 741.090762 | 7471.825800 |
| COM T[12] | 0.000075 | 0.000750 | 0.007500 | 0.075000 | 0.750000 | 7.500000 | 75.000000 | 750.000000 | 7500.000000 |
| COM HO[13] | 0.000075 | 0.000750 | 0.007500 | 0.075000 | 0.750000 | 7.500000 | 75.000000 | 750.000000 | 7500.000000 |
| Total RC T[14] | 0.498828 | 0.499506 | 0.506553 | 0.600736 | 2.002941 | 16.064698 | 152.978106 | 1509.066260 | 15028.329835 |
| Total RC HO[15] | 0.000075 | 0.000752 | 0.007650 | 0.088782 | 1.228261 | 14.111795 | 147.183390 | 1491.090762 | 14971.825800 |
| VAR Virial ratio[16] | 1.000000 | 1.000000 | 1.000000 | 1.000000 | 1.000000 | 1.000000 | 1.000000 | 1.000000 | 1.000000 |
| MC-HF/[$7s$:$7s$] E [17] | -0.434898 | -0.434876 | -0.432709 | -0.318137 | 1.685464 | 26.272493 | 288.573488 | 2964.207347 | 29887.148995 |
| MC-HF TE[18] | 0.407640 | 0.407671 | 0.410598 | 0.484326 | 1.244006 | 8.558972 | 77.968471 | 759.041682 | 7528.264288 |
| MC-HF TP[19] | 0.027261 | 0.027274 | 0.028544 | 0.082313 | 0.752732 | 7.503458 | 75.008085 | 750.023229 | 7500.072894 |
| Total MC-HF T[20] | 0.434901 | 0.434945 | 0.439142 | 0.566639 | 1.996738 | 16.062429 | 152.976556 | 1509.064911 | 15028.337182 |
| MC-HF HO E[21] | 0.000000 | 0.000002 | 0.000172 | 0.014501 | 0.480129 | 6.615097 | 72.191440 | 741.113975 | 7471.890187 |
| MC-HF HO P[22] | 0.000000 | 0.000021 | 0.001972 | 0.068337 | 0.747277 | 7.496544 | 74.991916 | 749.976771 | 7499.927138 |
| Total MC-HF HO[23] | 0.000000 | 0.000022 | 0.002144 | 0.082838 | 1.227406 | 14.111641 | 147.183356 | 1491.090747 | 14971.817325 |
| MC-HF INT[24] | -0.869799 | -0.869844 | -0.873995 | -0.967613 | -1.538680 | -3.901577 | -11.586424 | -35.948311 | -113.005512 |
| MC-HF Virial Ratio[25] | 1.999995 | 1.999842 | 1.985350 | 1.561445 | 0.155891 | -0.635649 | -0.886390 | -0.964268 | -0.988720 |
| | | | | m=900 | | | | | |

| Quantities /ω | 0.0001 | 0.001 | 0.01 | 0.1 | 1 | 10 | 100 | 1000 | 10000 |
|---|---|---|---|---|---|---|---|---|---|
| Variation energy[1] | -0.499295 | -0.497944 | -0.484295 | -0.335100 | 1.680766 | 26.267980 | 288.564027 | 2964.181100 | 29887.069188 |
| VAR TE[2] | 0.498891 | 0.498895 | 0.499199 | 0.525896 | 1.252936 | 8.564387 | 77.976976 | 759.062587 | 7528.318124 |
| VAR TP[3] | 0.000629 | 0.001303 | 0.008046 | 0.075501 | 0.750559 | 7.501184 | 75.003308 | 750.010071 | 7500.031464 |
| Total VART[4] | 0.499520 | 0.500198 | 0.507245 | 0.601397 | 2.003495 | 16.065571 | 152.980284 | 1509.072658 | 15028.349588 |
| VAR HO E[5] | 0.000000 | 0.000002 | 0.000158 | 0.013834 | 0.478389 | 6.612170 | 72.184564 | 741.094479 | 7471.837554 |
| VAR HO P[6] | 0.000075 | 0.000749 | 0.007492 | 0.074932 | 0.749698 | 7.499014 | 74.996872 | 749.990101 | 7499.968708 |
| Total VAR HO[7] | 0.000075 | 0.000752 | 0.007650 | 0.088766 | 1.228087 | 14.111184 | 147.181435 | 1491.084580 | 14971.806262 |
| VAR INT[8] | -0.998890 | -0.998893 | -0.999190 | -1.025263 | -1.550816 | -3.908762 | -11.597701 | -35.976141 | -113.086663 |
| REL T[9] | 0.499445 | 0.499448 | 0.499745 | 0.526397 | 1.253495 | 8.565570 | 77.980287 | 759.072657 | 7528.349589 |
| REL INT[10] | -0.998890 | -0.998893 | -0.999190 | -1.025263 | -1.550816 | -3.908774 | -11.597698 | -35.976141 | -113.086663 |
| REL HO[11] | 0.000000 | 0.000002 | 0.000150 | 0.013766 | 0.478087 | 6.611184 | 72.181438 | 741.084584 | 7471.806262 |
| COM T[12] | 0.000075 | 0.000750 | 0.007500 | 0.075000 | 0.750000 | 7.500000 | 75.000000 | 750.000000 | 7500.000000 |
| COM HO[13] | 0.000075 | 0.000750 | 0.007500 | 0.075000 | 0.750000 | 7.500000 | 75.000000 | 750.000000 | 7500.000000 |
| Total RC T[14] | 0.499520 | 0.500198 | 0.507245 | 0.601397 | 2.003495 | 16.065570 | 152.980287 | 1509.072657 | 15028.349589 |
| Total RC HO[15] | 0.000075 | 0.000752 | 0.007650 | 0.088766 | 1.228087 | 14.111184 | 147.181438 | 1491.084584 | 14971.806262 |
| VAR Virial ratio[16] | 1.000000 | 1.000000 | 1.000000 | 1.000000 | 1.000000 | 1.000000 | 1.000000 | 1.000000 | 1.000000 |
| MC-HF/[*7s*:*7s*] E[17] | -0.453975 | -0.453946 | -0.451115 | -0.326729 | 1.682345 | 26.268617 | 288.564475 | 2964.181492 | 29887.069700 |
| MC-HF TE[18] | 0.433917 | 0.433958 | 0.437715 | 0.503895 | 1.249292 | 8.563130 | 77.976156 | 759.061875 | 7528.324535 |
| MC-HF TP[19] | 0.020059 | 0.020076 | 0.021677 | 0.078944 | 0.751287 | 7.501566 | 75.003618 | 750.010409 | 7500.032322 |
| Total MC-HF T[20] | 0.453976 | 0.454034 | 0.459392 | 0.582838 | 2.000579 | 16.064696 | 152.979773 | 1509.072284 | 15028.356856 |
| MC-HF HO E[21] | 0.000000 | 0.000002 | 0.000164 | 0.014127 | 0.478928 | 6.612669 | 72.185040 | 741.094950 | 7471.831068 |

| Quantities /ω | 0.0001 | 0.001 | 0.01 | 0.1 | 1 | 10 | 100 | 1000 | 10000 |
|---|---|---|---|---|---|---|---|---|---|
| MC-HF HO P[22] | 0.000000 | 0.000028 | 0.002596 | 0.071253 | 0.748715 | 7.498434 | 74.996383 | 749.989591 | 7499.967679 |
| Total MC-HF HO[23] | 0.000000 | 0.000030 | 0.002760 | 0.085380 | 1.227642 | 14.111103 | 147.181422 | 1491.084542 | 14971.798747 |
| MC-HF INT[24] | -0.907952 | -0.908011 | -0.913266 | -0.994947 | -1.545876 | -3.907182 | -11.596721 | -35.975334 | -113.085904 |
| MC-HF Virial Ratio[25] | 1.999999 | 1.999806 | 1.981981 | 1.560582 | 0.159071 | -0.635177 | -0.886292 | -0.964241 | -0.988712 |
| | | | | m=1836 | | | | | |
| Variation energy[1] | -0.499578 | -0.498226 | -0.484578 | -0.335391 | 1.680327 | 26.266873 | 288.560744 | 2964.170918 | 29887.037181 |
| VAR TE[2] | 0.499456 | 0.499459 | 0.499759 | 0.526422 | 1.253447 | 8.565660 | 77.979554 | 759.070327 | 7528.342222 |
| VAR TP[3] | 0.000347 | 0.001022 | 0.007768 | 0.075246 | 0.750274 | 7.500580 | 75.001623 | 750.004944 | 7500.016692 |
| Total VART[4] | 0.499803 | 0.500481 | 0.507528 | 0.601667 | 2.003721 | 16.066240 | 152.981177 | 1509.075271 | 15028.358915 |
| VAR HO E[5] | 0.000000 | 0.000002 | 0.000154 | 0.013792 | 0.478164 | 6.610983 | 72.182175 | 741.086918 | 7471.813632 |
| VAR HO P[6] | 0.000075 | 0.000750 | 0.007496 | 0.074967 | 0.749852 | 7.499516 | 74.998465 | 749.995145 | 7499.984648 |
| Total VAR HO[7] | 0.000075 | 0.000752 | 0.007650 | 0.088759 | 1.228016 | 14.110499 | 147.180640 | 1491.082063 | 14971.798280 |
| VAR INT[8] | -0.999456 | -0.999459 | -0.999755 | -1.025817 | -1.551410 | -3.909978 | -11.601062 | -35.986413 | -113.118759 |
| REL T[9] | 0.499728 | 0.499731 | 0.500028 | 0.526667 | 1.253721 | 8.565927 | 77.981177 | 759.075268 | 7528.357659 |
| REL INT[10] | -0.999456 | -0.999459 | -0.999755 | -1.025817 | -1.551410 | -3.909987 | -11.601074 | -35.986413 | -113.118759 |
| REL HO[11] | 0.000000 | 0.000002 | 0.000150 | 0.013759 | 0.478016 | 6.610933 | 72.180641 | 741.082063 | 7471.798280 |
| COM T[12] | 0.000075 | 0.000750 | 0.007500 | 0.075000 | 0.750000 | 7.500000 | 75.000000 | 750.000000 | 7500.000000 |
| COM HO[13] | 0.000075 | 0.000750 | 0.007500 | 0.075000 | 0.750000 | 7.500000 | 75.000000 | 750.000000 | 7500.000000 |
| Total RC T[14] | 0.499803 | 0.500481 | 0.507528 | 0.601667 | 2.003721 | 16.065927 | 152.981177 | 1509.075268 | 15028.357659 |
| Total RC HO[15] | 0.000075 | 0.000752 | 0.007650 | 0.088759 | 1.228016 | 14.110933 | 147.180641 | 1491.082063 | 14971.798280 |
| VAR Virial ratio[16] | 1.000000 | 1.000000 | 1.000000 | 1.000000 | 1.000000 | 1.000000 | 1.000000 | 1.000000 | 1.000000 |

| Quantities /ω | 0.0001 | 0.001 | 0.01 | 0.1 | 1 | 10 | 100 | 1000 | 10000 |
|---|---|---|---|---|---|---|---|---|---|
| MC-HF/[*7s:7s*] E [17] | -0.466417 | -0.466378 | -0.462753 | -0.330905 | 1.681012 | 26.267013 | 288.560779 | 2964.170918 | 29887.037312 |
| MC-HF TE[18] | 0.451373 | 0.451428 | 0.456066 | 0.514174 | 1.251621 | 8.564874 | 77.979324 | 759.070160 | 7528.348933 |
| MC-HF TP[19] | 0.015040 | 0.015061 | 0.017054 | 0.077175 | 0.750653 | 7.500779 | 75.001780 | 750.005100 | 7500.015695 |
| Total MC-HF T[20] | 0.466413 | 0.466489 | 0.473120 | 0.591350 | 2.002274 | 16.065653 | 152.981104 | 1509.075260 | 15028.364628 |
| MC-HF HO E[21] | 0.000000 | 0.000002 | 0.000159 | 0.013946 | 0.478411 | 6.611663 | 72.182412 | 741.087157 | 7471.807132 |
| MC-HF HO P[22] | 0.000000 | 0.000037 | 0.003299 | 0.072886 | 0.749348 | 7.499221 | 74.998220 | 749.994900 | 7499.984305 |
| Total MC-HF HO[23] | 0.000000 | 0.000039 | 0.003458 | 0.086832 | 1.227758 | 14.110884 | 147.180632 | 1491.082057 | 14971.791437 |
| MC-HF INT[24] | -0.932830 | -0.932906 | -0.939331 | -1.009087 | -1.549020 | -3.909524 | -11.600958 | -35.986399 | -113.118753 |
| MC-HF Virial Ratio[25] | 2.000010 | 1.999763 | 1.978087 | 1.559575 | 0.160449 | -0.634979 | -0.886251 | -0.964230 | -0.988709 |

1. Total variational energy
2. Electron's kinetic energy component in variational method
3. PCP's kinetic energy component in variational method
4. Total kinetic energy component in variational method
5. Electron's harmonic oscillator potential energy component in variational method
6. PCP's harmonic oscillator potential energy component in variational method
7. Total harmonic oscillator potential energy component in variational method
8. E-PCP interaction energy component in variational method
9. Kinetic energy component of relative coordinates
10. Interaction energy component of relative coordinates
11. Harmonic oscillator potential energy component of relative coordinates
12. Kinetic energy component of the center of mass
13. Harmonic oscillator potential energy component of the center of mass
14. Total kinetic energy component of relative and center of mass coordinates

15. Total harmonic oscillator potential energy component of relative and center of mass coordinates
16. Variational virial ratio
17. Total MC-HF energy for basis set [7s:7s]
18. Electron's kinetic energy component in MC-HF
19. PCP's kinetic energy component in MC-HF
20. Total kinetic energy component in MC-HF
21. Electron's harmonic oscillator potential energy component in MC-HF
22. PCP's harmonic oscillator potential energy component in MC-HF
23. Total harmonic oscillator potential energy component in MC-HF
24. E-PCP interaction energy component in MC-HF
25. MC-HF virial ratio

## 4.4 Exponents and Coefficients of MC-HF Calculations

Table 4-5 is dedicated to the exponents and coefficients for the Hartree-Fock calculations using the [7s:7s] basis set, which were provided to us in an optimized form. In this table, the exponents and their corresponding coefficients are presented in order (based on the magnitude of the coefficients) for the electron and PCP orbitals (expansions 1-37 and 1-38). Similar data for the [7s:1s] basis set are reported in Table B-2 of the appendix B.

Finding a consistent and precise pattern among the coefficients and exponents is challenging, but at least a few trends can be roughly observed:

- As the frequency strength increases for a given mass, the magnitude of the exponents (for both the electron and the PCP) increases.
- The magnitude of the dominant exponent (having the largest coefficient) for the electron orbitals at each frequency does not depend on the mass of the PCP, meaning that the dominant orbital exponent at each frequency remains relatively constant regardless of the mass of the PCP.
- The dominant exponent at each frequency (having a larger linear coefficient) usually has a smaller magnitude.
- At higher frequencies, the dominance of a single exponent becomes more significant, and in some cases, the dominant

exponent can be selected as the only effective exponent in the calculations, neglecting the other exponents.

- For the PCP, the exponents are very sensitive to both increasing frequency and mass, reaching the order of millions for a mass and frequency of $10^4$.
- Additionally, for the PCP, with increasing mass, the dominant exponent shifts to a larger magnitude exponent.

Table 4-5: Optimized coefficients and exponents of the [7s:7s] basis set in the MC-HF framework. The top and bottom rows for each frequency, respectively, show the absolute values of the sorted coefficients (from largest to smallest) and their corresponding exponents.

| | Sorted Linear coefficients and corresponding exponents for Electron Orbitals | | | | | | | Sorted Linear coefficients and corresponding exponents for PC Orbitals | | | | | | |
|---|---|---|---|---|---|---|---|---|---|---|---|---|---|---|
| ω/Basis | 1s | 2s | 3s | 4s | 5s | 6s | 7s | 1s | 2s | 3s | 4s | 5s | 6s | 7s |
| m = 1 | | | | | | | | | | | | | | |
| 0.0001 | 0.3371 | 0.2945 | 0.2297 | 0.1191 | 0.0664 | 0.0036 | 0.0032 | 0.4536 | 0.2173 | 0.1781 | 0.1505 | 0.0374 | 0.0132 | 0.0030 |
| | 0.0562 | 0.0324 | 0.0241 | 0.0142 | 0.1101 | 0.0070 | 0.0095 | 0.0288 | 0.0649 | 0.0460 | 0.0146 | 0.1060 | 0.1600 | 0.1950 |
| 0.0010 | 0.3745 | 0.2937 | 0.2094 | 0.1317 | 0.0286 | 0.0086 | 0.0008 | 0.4419 | 0.2821 | 0.1792 | 0.1556 | 0.0716 | 0.0664 | 0.0064 |
| | 0.0403 | 0.0250 | 0.0703 | 0.0155 | 0.1282 | 0.0085 | 0.1646 | 0.0290 | 0.0700 | 0.0435 | 0.0148 | 0.0975 | 0.1274 | 0.1671 |
| 0.0100 | 0.3539 | 0.2825 | 0.2454 | 0.1227 | 0.0375 | 0.0041 | 0.0005 | 0.5185 | 0.3519 | 0.1494 | 0.0534 | 0.0189 | 0.0090 | 0.0003 |
| | 0.0501 | 0.0214 | 0.0329 | 0.0934 | 0.0118 | 0.2077 | 0.2700 | 0.0304 | 0.0622 | 0.0150 | 0.1529 | 0.1899 | 0.0116 | 0.5134 |
| 0.1000 | 0.5381 | 0.1703 | 0.1617 | 0.1417 | 0.0355 | 0.0289 | 0.0061 | 0.6487 | 0.3024 | 0.0544 | 0.0190 | 0.0002 | 0.0001 | 0.0001 |
| | 0.0743 | 0.0526 | 0.1208 | 0.1554 | 0.4216 | 0.5147 | 0.6643 | 0.0701 | 0.1361 | 0.0438 | 0.2970 | 8.4040 | 5.9399 | 11.9810 |
| 1 | 0.7138 | 0.1151 | 0.1028 | 0.0707 | 0.0060 | 0.0006 | 0.0002 | 0.5855 | 0.3641 | 0.0690 | 0.0139 | 0.0107 | 0.0074 | 0.0067 |
| | 0.5417 | 0.4651 | 0.7639 | 1.0044 | 1.9316 | 3.8172 | 5.4195 | 0.5000 | 0.6412 | 1.1070 | 1.4234 | 0.3872 | 1.7406 | 1.2656 |
| 10 | 0.8355 | 0.1070 | 0.0582 | 0.0024 | 0.0009 | 0.0006 | 0.0003 | 0.9270 | 0.1190 | 0.0609 | 0.0139 | 0.0023 | 0.0012 | 0.0001 |
| | 5.1246 | 4.7648 | 7.9818 | 19.7336 | 36.1043 | 59.8889 | 73.1219 | 5.0719 | 7.7344 | 8.4000 | 10.8375 | 3.8500 | 13.8906 | 3.0469 |
| 100 | 0.7978 | 0.2381 | 0.0396 | 0.0068 | 0.0025 | 0.0004 | 0.0002 | 0.6675 | 0.6598 | 0.5120 | 0.4888 | 0.0183 | 0.0133 | 0.0026 |
| | 51.7598 | 44.6831 | 39.3719 | 127.5706 | 165.3781 | 346.0825 | 402.5625 | 120.7188 | 122.0313 | 48.7500 | 52.4966 | 81.1172 | 167.2031 | 204.9219 |
| 1000 | 0.8148 | 0.1889 | 0.0047 | 0.0014 | 0.0007 | 0.0006 | 0.0000 | 0.5705 | 0.4971 | 0.0444 | 0.0334 | 0.0075 | 0.0030 | 0.0000 |
| | 508.4658 | 469.8193 | 372.3223 | 1205.0781 | 2717.5828 | 3008.7031 | 5038.5383 | 475.0000 | 536.3672 | 626.1328 | 369.5313 | 321.4844 | 971.2891 | 3865.0000 |
| 10000 | 0.7729 | 0.2436 | 0.0193 | 0.0029 | 0.0001 | 0.0000 | 0.0000 | 0.7670 | 0.4619 | 0.1751 | 0.0710 | 0.0234 | 0.0088 | 0.0026 |
| | 5094.9219 | 4643.1782 | 3577.8125 | 2833.8281 | 26213.7188 | 45381.3281 | 312053.9375 | 5301.9775 | 4346.2891 | 4051.4160 | 7264.3127 | 10207.7637 | 14162.3535 | 17021.8262 |
| m = 1.5 | | | | | | | | | | | | | | |
| 0.0001 | 0.3037 | 0.2373 | 0.2257 | 0.1998 | 0.0463 | 0.0404 | 0.0001 | 0.3196 | 0.2675 | 0.2301 | 0.1784 | 0.0289 | 0.0091 | 0.0081 |
| | 0.0562 | 0.0383 | 0.0251 | 0.0945 | 0.0142 | 0.1725 | 0.7570 | 0.0649 | 0.0460 | 0.1060 | 0.0288 | 0.1950 | 0.0146 | 0.1600 |
| 0.0010 | 0.3487 | 0.3218 | 0.2011 | 0.1079 | 0.0670 | 0.0070 | 0.0004 | 0.4285 | 0.3038 | 0.1920 | 0.0632 | 0.0453 | 0.0090 | 0.0000 |
| | 0.0403 | 0.0703 | 0.0260 | 0.1282 | 0.0155 | 0.2585 | 0.4146 | 0.0544 | 0.0975 | 0.0290 | 0.0435 | 0.1827 | 0.0148 | 0.6274 |

| | Sorted Linear coefficients and corresponding exponents for Electron Orbitals | | | | | | | Sorted Linear coefficients and corresponding exponents for PC Orbitals | | | | | | |
|---|---|---|---|---|---|---|---|---|---|---|---|---|---|---|
| ω/Basis | 1s | 2s | 3s | 4s | 5s | 6s | 7s | 1s | 2s | 3s | 4s | 5s | 6s | 7s |
| 0.0100 | 0.3892 | 0.2469 | 0.2063 | 0.1615 | 0.0302 | 0.0111 | 0.0069 | 0.5605 | 0.2532 | 0.1900 | 0.0325 | 0.0099 | 0.0072 | 0.0016 |
| | 0.0501 | 0.0934 | 0.0329 | 0.0214 | 0.1763 | 0.0118 | 0.2077 | 0.0617 | 0.0309 | 0.1215 | 0.1451 | 0.2634 | 0.2025 | 0.0086 |
| 0.1000 | 0.4673 | 0.2383 | 0.1430 | 0.1104 | 0.0705 | 0.0025 | 0.0006 | 0.5584 | 0.3315 | 0.1337 | 0.0024 | 0.0005 | 0.0004 | 0.0002 |
| | 0.0743 | 0.1208 | 0.1554 | 0.0526 | 0.2647 | 0.4216 | 0.6643 | 0.1361 | 0.0857 | 0.2638 | 0.0438 | 8.4040 | 5.9399 | 11.9810 |
| 1 | 0.6616 | 0.1677 | 0.0963 | 0.0731 | 0.0138 | 0.0005 | 0.0001 | 0.6217 | 0.5758 | 0.4259 | 0.3849 | 0.2414 | 0.0091 | 0.0000 |
| | 0.5417 | 0.7639 | 0.4651 | 1.1294 | 1.9316 | 2.8172 | 5.4195 | 0.9234 | 0.6412 | 1.7656 | 1.7719 | 0.5625 | 0.3872 | 3.1070 |
| 10 | 0.8334 | 0.0882 | 0.0757 | 0.0302 | 0.0240 | 0.0006 | 0.0000 | 0.7058 | 0.2167 | 0.0519 | 0.0243 | 0.0063 | 0.0042 | 0.0010 |
| | 5.1246 | 4.7648 | 7.9818 | 19.7336 | 21.1043 | 37.8889 | 73.1219 | 7.7344 | 7.8688 | 6.0719 | 13.3906 | 10.8375 | 3.8500 | 3.0469 |
| 100 | 0.8082 | 0.2139 | 0.0298 | 0.0096 | 0.0008 | 0.0004 | 0.0002 | 3.5325 | 3.4418 | 0.9834 | 0.1835 | 0.0709 | 0.0186 | 0.0138 |
| | 51.7598 | 44.6831 | 38.8719 | 114.0706 | 165.3781 | 346.0825 | 402.5625 | 120.3438 | 122.0313 | 81.1094 | 61.9966 | 167.2031 | 48.7500 | 203.4219 |
| 1000 | 0.7900 | 0.2136 | 0.0054 | 0.0024 | 0.0006 | 0.0005 | 0.0000 | 0.9881 | 0.0187 | 0.0049 | 0.0034 | 0.0028 | 0.0011 | 0.0000 |
| | 509.9658 | 469.8193 | 372.3223 | 1205.0781 | 2717.5828 | 3008.7031 | 5038.5383 | 747.8828 | 971.2891 | 533.3672 | 543.0000 | 369.5313 | 321.4844 | 3865.0000 |
| 10000 | 0.6629 | 0.3601 | 0.0261 | 0.0031 | 0.0001 | 0.0000 | 0.0000 | 2.0540 | 2.0268 | 0.9853 | 0.0888 | 0.0866 | 0.0147 | 0.0045 |
| | 5145.9219 | 4679.6782 | 3643.3125 | 2833.8281 | 26213.7188 | 45381.3281 | 312053.9375 | 4641.3672 | 4695.4775 | 7264.3127 | 4123.9160 | 9641.7637 | 14162.3535 | 17017.8262 |
| m=2 | | | | | | | | | | | | | | |
| 0.0001 | 0.4567 | 0.2583 | 0.1690 | 0.1162 | 0.0916 | 0.0598 | 0.0257 | 0.4004 | 0.3115 | 0.1559 | 0.1013 | 0.0367 | 0.0292 | 0.0031 |
| | 0.0562 | 0.1139 | 0.0265 | 0.0324 | 0.2664 | 0.2799 | 0.0144 | 0.1060 | 0.0649 | 0.0460 | 0.1950 | 0.0288 | 0.1600 | 0.3896 |
| 0.0010 | 0.3621 | 0.3233 | 0.1631 | 0.1408 | 0.0308 | 0.0196 | 0.0181 | 0.4335 | 0.2912 | 0.1649 | 0.1010 | 0.0413 | 0.0074 | 0.0011 |
| | 0.0703 | 0.0403 | 0.1282 | 0.0250 | 0.0155 | 0.2585 | 0.1646 | 0.0975 | 0.0513 | 0.1749 | 0.0700 | 0.0290 | 0.3774 | 0.5148 |
| 0.0100 | 0.3435 | 0.3029 | 0.1841 | 0.1050 | 0.0866 | 0.0322 | 0.0025 | 0.4057 | 0.3611 | 0.1511 | 0.1466 | 0.1171 | 0.0824 | 0.0072 |
| | 0.0817 | 0.0462 | 0.0329 | 0.1450 | 0.0195 | 0.2077 | 0.3868 | 0.0622 | 0.1053 | 0.2634 | 0.1587 | 0.2650 | 0.0357 | 0.0304 |
| 0.1000 | 0.4162 | 0.2664 | 0.1512 | 0.1158 | 0.0800 | 0.0114 | 0.0023 | 0.6135 | 0.4672 | 0.1521 | 0.1158 | 0.0441 | 0.0147 | 0.0043 |
| | 0.0743 | 0.1208 | 0.1554 | 0.2647 | 0.0526 | 0.5466 | 0.6643 | 0.1332 | 0.2521 | 0.3618 | 0.5076 | 0.7478 | 0.9185 | 0.0575 |
| 1 | 0.6362 | 0.1856 | 0.0865 | 0.0817 | 0.0250 | 0.0005 | 0.0001 | 0.6531 | 0.3176 | 0.1996 | 0.1378 | 0.1060 | 0.0133 | 0.0033 |
| | 0.5417 | 0.7639 | 1.1294 | 0.4651 | 1.9316 | 3.9195 | 2.8172 | 1.1070 | 1.2656 | 1.4859 | 1.8031 | 0.7975 | 0.5000 | 0.3872 |
| 10 | 0.8765 | 0.0812 | 0.0354 | 0.0151 | 0.0032 | 0.0004 | 0.0000 | 0.5732 | 0.4354 | 0.1280 | 0.1047 | 0.0163 | 0.0003 | 0.0000 |
| | 5.1246 | 8.0443 | 4.5148 | 16.7336 | 21.1043 | 37.8889 | 73.1219 | 10.8375 | 9.4000 | 6.7344 | 6.5719 | 17.3906 | 4.1000 | 3.2969 |

| | Sorted Linear coefficients and corresponding exponents for Electron Orbitals | | | | | | | Sorted Linear coefficients and corresponding exponents for PC Orbitals | | | | | | |
|---|---|---|---|---|---|---|---|---|---|---|---|---|---|---|
| ω/Basis | 1s | 2s | 3s | 4s | 5s | 6s | 7s | 1s | 2s | 3s | 4s | 5s | 6s | 7s |
| 100 | 0.8089 | 0.1913 | 0.0116 | 0.0109 | 0.0005 | 0.0005 | 0.0003 | 7.1415 | 6.5958 | 0.4206 | 0.0569 | 0.0185 | 0.0088 | 0.0046 |
| | 51.7598 | 45.1831 | 110.0706 | 36.3719 | 165.3781 | 346.0825 | 402.5625 | 119.2031 | 120.5313 | 90.6172 | 167.2031 | 59.4966 | 204.9219 | 48.7500 |
| 1000 | 0.7839 | 0.2189 | 0.0054 | 0.0032 | 0.0004 | 0.0004 | 0.0000 | 0.8366 | 0.1674 | 0.0678 | 0.0633 | 0.0013 | 0.0007 | 0.0001 |
| | 510.4658 | 469.8193 | 372.3223 | 1204.0781 | 3008.7031 | 2717.5828 | 5038.5383 | 1017.7891 | 923.6328 | 537.5000 | 533.3672 | 369.5313 | 321.4844 | 3812.0000 |
| 10000 | 1.0088 | 0.0322 | 0.0225 | 0.0010 | 0.0002 | 0.0000 | 0.0000 | 0.9958 | 0.0325 | 0.0286 | 0.0110 | 0.0092 | 0.0027 | 0.0011 |
| | 4958.1782 | 4195.8125 | 6361.4219 | 2842.3281 | 26211.2188 | 45381.3281 | 312053.9375 | 10028.2637 | 4495.2891 | 4839.9775 | 6784.8127 | 4123.4160 | 14162.3535 | 17021.8262 |
| m = 3 | | | | | | | | | | | | | | |
| 0.0001 | 0.3542 | 0.3477 | 0.1533 | 0.1085 | 0.0594 | 0.0362 | 0.0049 | 0.3118 | 0.2752 | 0.2415 | 0.1720 | 0.0255 | 0.0252 | 0.0178 |
| | 0.0886 | 0.0480 | 0.1549 | 0.0288 | 0.2507 | 0.0193 | 0.4758 | 0.1216 | 0.1600 | 0.0771 | 0.2575 | 0.0460 | 0.5649 | 0.5913 |
| 0.0010 | 0.6654 | 0.4322 | 0.3739 | 0.1984 | 0.1003 | 0.0403 | 0.0014 | 0.5210 | 0.2488 | 0.2387 | 0.0567 | 0.0496 | 0.0289 | 0.0030 |
| | 0.0481 | 0.0813 | 0.0546 | 0.1567 | 0.0228 | 0.2975 | 0.5155 | 0.1274 | 0.2225 | 0.0700 | 0.0460 | 0.0421 | 0.3560 | 0.5290 |
| 0.0100 | 0.4159 | 0.3793 | 0.1331 | 0.1252 | 0.0095 | 0.0009 | 0.0002 | 0.2926 | 0.2901 | 0.2366 | 0.1192 | 0.0586 | 0.0352 | 0.0002 |
| | 0.0501 | 0.1012 | 0.2077 | 0.0251 | 0.4433 | 0.0079 | 1.0200 | 0.1007 | 0.1366 | 0.2025 | 0.0622 | 0.3415 | 0.2154 | 1.3149 |
| 0.1000 | 0.3480 | 0.2837 | 0.1687 | 0.1625 | 0.0504 | 0.0294 | 0.0051 | 0.8058 | 0.7904 | 0.4673 | 0.4488 | 0.0850 | 0.0002 | 0.0000 |
| | 0.0743 | 0.1208 | 0.2647 | 0.1554 | 0.0526 | 0.4841 | 0.6643 | 0.1673 | 0.1688 | 0.1951 | 0.2970 | 0.5123 | 1.6899 | 0.7790 |
| 1 | 0.5811 | 0.1598 | 0.1569 | 0.0699 | 0.0504 | 0.0062 | 0.0034 | 0.6636 | 0.2675 | 0.1116 | 0.0965 | 0.0601 | 0.0030 | 0.0003 |
| | 0.5417 | 0.7014 | 1.0044 | 0.4651 | 1.9316 | 4.8172 | 5.4195 | 1.6227 | 1.4859 | 2.4438 | 1.1250 | 0.9850 | 0.5747 | 0.3281 |
| 10 | 0.8792 | 0.0890 | 0.0170 | 0.0166 | 0.0052 | 0.0005 | 0.0000 | 3.4301 | 3.2772 | 0.5906 | 0.5441 | 0.0269 | 0.0101 | 0.0022 |
| | 5.1246 | 7.9818 | 14.2336 | 4.2648 | 21.1043 | 37.8889 | 73.1219 | 9.9844 | 9.9625 | 13.2125 | 16.7969 | 6.7594 | 4.8500 | 3.7344 |
| 100 | 0.7674 | 0.2195 | 0.0155 | 0.0041 | 0.0036 | 0.0008 | 0.0003 | 2.1322 | 1.5441 | 0.3868 | 0.0429 | 0.0128 | 0.0067 | 0.0019 |
| | 51.7598 | 46.4331 | 97.5706 | 34.8719 | 165.3781 | 346.0825 | 402.5625 | 131.7188 | 126.3281 | 167.2031 | 90.6172 | 204.9219 | 59.4966 | 48.7500 |
| 1000 | 0.7687 | 0.2337 | 0.0059 | 0.0040 | 0.0002 | 0.0001 | 0.0000 | 0.9969 | 0.1664 | 0.1654 | 0.0045 | 0.0007 | 0.0002 | 0.0002 |
| | 511.4658 | 469.8193 | 372.3223 | 1204.0781 | 2717.5828 | 3008.7031 | 5038.5383 | 1505.2891 | 536.3672 | 534.5000 | 922.1328 | 369.5313 | 321.4844 | 3814.0000 |
| 10000 | 0.7601 | 0.2886 | 0.0538 | 0.0051 | 0.0004 | 0.0000 | 0.0000 | 0.9897 | 0.0109 | 0.0054 | 0.0048 | 0.0008 | 0.0005 | 0.0002 |
| | 5137.4219 | 4475.6782 | 3685.8125 | 2834.8281 | 22866.2188 | 45381.3281 | 312053.9375 | 14988.8535 | 17021.3262 | 4495.4775 | 4603.2891 | 4115.9160 | 10950.7637 | 7264.3127 |
| m = 10 | | | | | | | | | | | | | | |
| 0.0001 | 0.3863 | 0.3329 | 0.1780 | 0.1492 | 0.0269 | 0.0261 | 0.0154 | 0.5276 | 0.3857 | 0.0897 | 0.0364 | 0.0175 | 0.0008 | 0.0003 |

| | Sorted Linear coefficients and corresponding exponents for Electron Orbitals | | | | | | | Sorted Linear coefficients and corresponding exponents for PC Orbitals | | | | | | |
|---|---|---|---|---|---|---|---|---|---|---|---|---|---|---|
| ω/Basis | 1s | 2s | 3s | 4s | 5s | 6s | 7s | 1s | 2s | 3s | 4s | 5s | 6s | 7s |
| | 0.0866 | 0.1755 | 0.0435 | 0.3708 | 0.8058 | 0.0226 | 0.0249 | 0.5350 | 0.3200 | 1.0460 | 0.1947 | 1.3774 | 4.4040 | 6.3560 |
| 0.0010 | 2.4389 | 2.2629 | 0.5776 | 0.1568 | 0.1159 | 0.0417 | 0.0158 | 0.5274 | 0.3857 | 0.0892 | 0.0363 | 0.0166 | 0.0008 | 0.0003 |
| | 0.0996 | 0.1062 | 0.1589 | 0.0382 | 0.3645 | 0.4790 | 0.9237 | 0.5350 | 0.3200 | 1.0421 | 0.1947 | 1.3774 | 4.4040 | 6.3560 |
| 0.0100 | 0.3683 | 0.3377 | 0.1864 | 0.1256 | 0.1153 | 0.0702 | 0.0207 | 0.6204 | 0.3478 | 0.0830 | 0.0305 | 0.0028 | 0.0022 | 0.0004 |
| | 0.0661 | 0.1321 | 0.2271 | 0.0349 | 0.4141 | 0.0449 | 0.8678 | 0.5321 | 0.2911 | 1.2614 | 1.6400 | 5.4399 | 7.4040 | 11.8560 |
| 0.1000 | 0.4044 | 0.2797 | 0.2706 | 0.1068 | 0.0179 | 0.0152 | 0.0099 | 0.7369 | 0.4444 | 0.3919 | 0.2202 | 0.0011 | 0.0010 | 0.0003 |
| | 0.1443 | 0.0782 | 0.2893 | 0.6141 | 1.3278 | 0.0734 | 0.0459 | 0.7960 | 1.1400 | 1.0350 | 0.5700 | 5.9399 | 8.9040 | 11.9810 |
| 1 | 0.6302 | 0.2494 | 0.0825 | 0.0548 | 0.0179 | 0.0027 | 0.0004 | 0.9061 | 0.0854 | 0.0103 | 0.0007 | 0.0001 | 0.0001 | 0.0000 |
| | 0.5686 | 1.0044 | 2.1531 | 0.4407 | 4.7633 | 7.3379 | 13.3386 | 5.3750 | 4.4375 | 10.1372 | 20.0797 | 199.0375 | 136.9000 | 292.4000 |
| 10 | 0.7632 | 0.1223 | 0.0931 | 0.0296 | 0.0054 | 0.0001 | 0.0000 | 0.9260 | 0.0751 | 0.0016 | 0.0010 | 0.0004 | 0.0002 | 0.0001 |
| | 5.1246 | 8.1039 | 4.9602 | 18.2688 | 47.6278 | 251.2469 | 567.7014 | 51.0000 | 45.9000 | 37.4000 | 118.6719 | 172.6000 | 425.7750 | 467.5000 |
| 100 | 0.8843 | 0.2005 | 0.1032 | 0.0192 | 0.0034 | 0.0003 | 0.0000 | 0.8462 | 0.1938 | 0.0489 | 0.0235 | 0.0085 | 0.0072 | 0.0011 |
| | 51.7598 | 40.2886 | 37.9070 | 120.9788 | 316.2973 | 816.6250 | 1702.4875 | 486.2500 | 611.9043 | 949.0000 | 1359.0000 | 321.2500 | 1977.7500 | 2902.4063 |
| 1000 | 0.8946 | 0.1524 | 0.0521 | 0.0056 | 0.0012 | 0.0003 | 0.0001 | 0.8590 | 0.1797 | 0.0475 | 0.0227 | 0.0082 | 0.0067 | 0.0011 |
| | 508.9235 | 433.1982 | 401.6191 | 1252.7391 | 3526.2813 | 4413.5383 | 7435.5469 | 4862.5000 | 6130.8685 | 9487.5586 | 13510.0000 | 3212.5000 | 19850.0000 | 29185.7031 |
| 10000 | 0.8337 | 0.1846 | 0.0232 | 0.0041 | 0.0011 | 0.0001 | 0.0000 | 0.8362 | 0.2254 | 0.0556 | 0.0072 | 0.0033 | 0.0021 | 0.0000 |
| | 5094.8980 | 4498.8301 | 3480.1563 | 2833.8281 | 16863.1328 | 47646.9531 | 161858.6250 | 48624.0234 | 59724.7276 | 72750.0000 | 32446.8750 | 192150.0000 | 230775.0000 | 653117.5781 |
| m=50 | | | | | | | | | | | | | | |
| 0.0001 | 0.4294 | 0.3346 | 0.1617 | 0.1424 | 0.0361 | 0.0026 | 0.0000 | 0.5941 | 0.2894 | 0.1107 | 0.0164 | 0.0030 | 0.0015 | 0.0013 |
| | 0.1235 | 0.2868 | 0.0529 | 0.6763 | 1.6223 | 3.5866 | 6.8660 | 1.5700 | 2.3900 | 1.0350 | 3.2960 | 5.4399 | 11.4040 | 11.8560 |
| 0.0010 | 0.4282 | 0.3353 | 0.1610 | 0.1431 | 0.0362 | 0.0028 | 0.0001 | 0.5942 | 0.2895 | 0.1105 | 0.0165 | 0.0030 | 0.0015 | 0.0013 |
| | 0.1232 | 0.2857 | 0.0528 | 0.6738 | 1.6093 | 3.4538 | 6.8521 | 1.5700 | 2.3900 | 1.0350 | 3.2921 | 5.4399 | 11.4040 | 11.8560 |
| 0.0100 | 0.4273 | 0.3311 | 0.1625 | 0.1433 | 0.0390 | 0.0035 | 0.0000 | 0.5939 | 0.3065 | 0.0913 | 0.0187 | 0.0034 | 0.0010 | 0.0001 |
| | 0.1242 | 0.2861 | 0.0534 | 0.6624 | 1.5547 | 3.4577 | 6.9575 | 1.5700 | 2.3900 | 1.0350 | 3.2960 | 5.9399 | 8.3560 | 11.4040 |
| 0.1000 | 0.3581 | 0.3216 | 0.1824 | 0.1549 | 0.0706 | 0.0150 | 0.0001 | 0.8580 | 0.1690 | 0.0346 | 0.0306 | 0.0273 | 0.0128 | 0.0044 |
| | 0.1443 | 0.2873 | 0.6141 | 0.0779 | 1.4274 | 3.6047 | 18.7647 | 3.1400 | 5.0460 | 7.4399 | 1.0700 | 1.0350 | 11.4040 | 13.9810 |
| 1 | 0.5835 | 0.2815 | 0.1081 | 0.0406 | 0.0330 | 0.0073 | 0.0005 | 0.8123 | 0.1876 | 0.0002 | 0.0001 | 0.0001 | 0.0001 | 0.0000 |

| | Sorted Linear coefficients and corresponding exponents for Electron Orbitals | | | | | | | Sorted Linear coefficients and corresponding exponents for PC Orbitals | | | | | | |
|---|---|---|---|---|---|---|---|---|---|---|---|---|---|---|
| ω/Basis | 1s | 2s | 3s | 4s | 5s | 6s | 7s | 1s | 2s | 3s | 4s | 5s | 6s | 7s |
| | 0.5686 | 1.0044 | 2.2577 | 0.4407 | 5.8135 | 16.2643 | 44.4820 | 25.8750 | 24.0797 | 56.7500 | 103.5122 | 178.4000 | 281.0375 | 388.9000 |
| 10 | 0.9429 | 0.1333 | 0.1046 | 0.0355 | 0.0094 | 0.0020 | 0.0002 | 0.5286 | 0.4784 | 0.3280 | 0.3135 | 0.0218 | 0.0003 | 0.0001 |
| | 5.1246 | 8.1039 | 5.2531 | 18.2688 | 48.0672 | 138.0633 | 374.3421 | 245.2125 | 258.4000 | 549.9219 | 553.9000 | 433.5375 | 923.7500 | 90.0625 |
| 100 | 0.8966 | 0.2664 | 0.1840 | 0.0214 | 0.0045 | 0.0009 | 0.0001 | 0.6175 | 0.4049 | 0.0224 | 0.0001 | 0.0000 | 0.0000 | 0.0000 |
| | 51.7598 | 38.8238 | 37.4188 | 123.6644 | 341.1997 | 1027.5625 | 3014.9875 | 2484.0000 | 2540.2500 | 2699.0000 | 5086.0000 | 6035.0000 | 517.5000 | 258.7500 |
| 1000 | 0.9139 | 0.1214 | 0.0408 | 0.0060 | 0.0012 | 0.0002 | 0.0000 | 0.5731 | 0.4822 | 0.0366 | 0.0224 | 0.0039 | 0.0003 | 0.0000 |
| | 508.9235 | 420.9912 | 384.5293 | 1338.1883 | 3915.4914 | 12474.6094 | 38985.2656 | 26100.0000 | 23588.1250 | 18865.0000 | 30699.3750 | 13631.2500 | 7675.0000 | 4462.5000 |
| 10000 | 0.8300 | 0.1857 | 0.0199 | 0.0032 | 0.0014 | 0.0002 | 0.0000 | 0.9462 | 0.0647 | 0.0104 | 0.0007 | 0.0002 | 0.0001 | 0.0000 |
| | 5094.9219 | 4525.6855 | 3499.6875 | 2833.8281 | 16994.9688 | 54521.9531 | 187210.1875 | 251523.2834 | 219125.0000 | 182446.8750 | 546400.0000 | 805217.1875 | 51749.0234 | 25875.0000 |
| | | | | | | | m = 207 | | | | | | | |
| 0.0001 | 0.3848 | 0.3492 | 0.1762 | 0.1261 | 0.0640 | 0.0180 | 0.0034 | 0.5200 | 0.4006 | 0.0628 | 0.0193 | 0.0071 | 0.0018 | 0.0007 |
| | 0.1281 | 0.2908 | 0.6666 | 0.0562 | 1.5226 | 3.3366 | 6.8660 | 4.1400 | 5.7960 | 3.0700 | 8.4399 | 11.4040 | 13.8560 | 2.0350 |
| 0.0010 | 0.3843 | 0.3485 | 0.1765 | 0.1265 | 0.0645 | 0.0182 | 0.0034 | 0.5248 | 0.3870 | 0.0607 | 0.0279 | 0.0083 | 0.0010 | 0.0009 |
| | 0.1282 | 0.2903 | 0.6640 | 0.0563 | 1.5155 | 3.3288 | 6.8521 | 4.1400 | 5.7921 | 3.0700 | 7.4399 | 11.4040 | 15.8560 | 2.0350 |
| 0.0100 | 0.3834 | 0.3481 | 0.1770 | 0.1260 | 0.0653 | 0.0187 | 0.0035 | 0.5592 | 0.2775 | 0.1570 | 0.0081 | 0.0073 | 0.0006 | 0.0005 |
| | 0.1286 | 0.2908 | 0.6634 | 0.0566 | 1.5118 | 3.3327 | 6.9575 | 4.1400 | 5.7960 | 6.4399 | 11.4040 | 2.0700 | 15.8560 | 1.0350 |
| 0.1000 | 0.3321 | 0.3313 | 0.2075 | 0.1152 | 0.0957 | 0.0318 | 0.0060 | 0.8612 | 0.1680 | 0.0333 | 0.0072 | 0.0020 | 0.0001 | 0.0000 |
| | 0.2893 | 0.1443 | 0.6141 | 0.0782 | 1.4216 | 3.6984 | 10.7647 | 11.4040 | 14.4810 | 8.9399 | 5.7960 | 4.1400 | 2.0700 | 1.0350 |
| 1 | 0.5710 | 0.2880 | 0.1148 | 0.0383 | 0.0364 | 0.0106 | 0.0019 | 0.9587 | 0.0481 | 0.0071 | 0.0016 | 0.0009 | 0.0003 | 0.0001 |
| | 0.5686 | 1.0044 | 2.2547 | 5.8379 | 0.4407 | 17.3386 | 61.6695 | 103.5122 | 124.0797 | 178.4000 | 281.0375 | 56.7500 | 388.9000 | 25.8750 |
| 10 | 0.9288 | 0.1353 | 0.0942 | 0.0370 | 0.0104 | 0.0026 | 0.0004 | 0.8825 | 0.1176 | 0.0135 | 0.0103 | 0.0035 | 0.0005 | 0.0000 |
| | 5.1246 | 8.1039 | 5.2531 | 18.2688 | 48.7997 | 151.2469 | 567.7014 | 1038.2750 | 1017.5000 | 258.4000 | 268.6719 | 237.4000 | 472.6000 | 113.5000 |
| 100 | 0.8965 | 0.2637 | 0.1820 | 0.0221 | 0.0048 | 0.0011 | 0.0002 | 0.9977 | 0.0023 | 0.0000 | 0.0000 | 0.0000 | 0.0000 | 0.0000 |
| | 51.7598 | 38.8238 | 37.4188 | 123.6644 | 349.9888 | 1152.5625 | 4702.4875 | 10352.7500 | 10086.0000 | 1449.0000 | 2484.0000 | 1035.0000 | 517.5000 | 258.7500 |
| 1000 | 0.9151 | 0.1156 | 0.0363 | 0.0062 | 0.0012 | 0.0002 | 0.0000 | 0.9967 | 0.0033 | 0.0000 | 0.0000 | 0.0000 | 0.0000 | 0.0000 |
| | 508.9235 | 420.9912 | 382.0879 | 1350.3953 | 4101.0383 | 14310.5469 | 58047.7656 | 103510.0000 | 102350.0000 | 44840.0000 | 14490.0000 | 10350.0000 | 5175.0000 | 2587.5000 |
| 10000 | 0.8377 | 0.1827 | 0.0243 | 0.0029 | 0.0014 | 0.0002 | 0.0000 | 0.6910 | 0.3138 | 0.0049 | 0.0003 | 0.0002 | 0.0000 | 0.0000 |

| | Sorted Linear coefficients and corresponding exponents for Electron Orbitals | | | | | | | Sorted Linear coefficients and corresponding exponents for PC Orbitals | | | | | | |
|---|---|---|---|---|---|---|---|---|---|---|---|---|---|---|
| ω/Basis | 1s | 2s | 3s | 4s | 5s | 6s | 7s | 1s | 2s | 3s | 4s | 5s | 6s | 7s |
| | 5094.9219 | 4486.6230 | 3577.8125 | 2833.8281 | 17619.9688 | 63662.5781 | 305803.9375 | 1022600.0000 | 1058900.0000 | 798400.0000 | 209750.0000 | 182446.8750 | 51749.0234 | 25875.0000 |
| m=400 | | | | | | | | | | | | | | |
| 0.0001 | 0.3721 | 0.3551 | 0.1860 | 0.1131 | 0.0719 | 0.0238 | 0.0048 | 0.5434 | 0.4778 | 0.0484 | 0.0204 | 0.0170 | 0.0017 | 0.0013 |
| | 0.1281 | 0.2908 | 0.6666 | 0.0564 | 1.5226 | 3.4928 | 8.4910 | 8.4399 | 5.7960 | 4.1400 | 13.8560 | 3.0700 | 2.0350 | 11.4040 |
| 0.0010 | 0.3710 | 0.3542 | 0.1868 | 0.1136 | 0.0727 | 0.0239 | 0.0047 | 0.5088 | 0.2994 | 0.1862 | 0.0069 | 0.0050 | 0.0011 | 0.0002 |
| | 0.1282 | 0.2898 | 0.6630 | 0.0565 | 1.5155 | 3.5007 | 8.5083 | 7.4399 | 5.7921 | 9.9040 | 4.1400 | 15.8560 | 3.0700 | 2.0350 |
| 0.0100 | 0.3701 | 0.3534 | 0.1874 | 0.1129 | 0.0741 | 0.0246 | 0.0048 | 0.6626 | 0.2561 | 0.0514 | 0.0411 | 0.0050 | 0.0002 | 0.0000 |
| | 0.1286 | 0.2903 | 0.6614 | 0.0568 | 1.5118 | 3.5358 | 8.7700 | 7.9399 | 5.7960 | 11.8560 | 11.4040 | 4.1400 | 2.0700 | 1.0350 |
| 0.1000 | 0.3367 | 0.3118 | 0.2268 | 0.1111 | 0.1011 | 0.0313 | 0.0057 | 0.9015 | 0.1093 | 0.0172 | 0.0112 | 0.0045 | 0.0005 | 0.0001 |
| | 0.2844 | 0.1443 | 0.6219 | 0.0791 | 1.5515 | 4.4718 | 15.1865 | 22.4810 | 16.9040 | 8.9399 | 5.7960 | 4.1400 | 2.0700 | 1.0350 |
| 1 | 0.5448 | 0.2982 | 0.1158 | 0.0561 | 0.0366 | 0.0094 | 0.0016 | 1.0019 | 0.0065 | 0.0039 | 0.0012 | 0.0005 | 0.0001 | 0.0000 |
| | 0.5686 | 1.0090 | 2.3615 | 0.4712 | 6.5276 | 21.3058 | 87.8414 | 200.2750 | 149.0797 | 281.0375 | 103.5122 | 388.9000 | 56.7500 | 25.8750 |
| 10 | 0.8432 | 0.1272 | 0.0357 | 0.0110 | 0.0028 | 0.0006 | 0.0001 | 0.7957 | 0.2643 | 0.0771 | 0.0191 | 0.0024 | 0.0007 | 0.0004 |
| | 5.1222 | 8.2937 | 18.1009 | 46.7550 | 143.4344 | 492.9944 | 1731.8156 | 2038.2750 | 1770.1188 | 1385.8375 | 1017.5000 | 582.2500 | 268.6719 | 222.6000 |
| 100 | 0.9007 | 0.2241 | 0.1461 | 0.0220 | 0.0046 | 0.0010 | 0.0002 | 0.8291 | 0.1712 | 0.0003 | 0.0000 | 0.0000 | 0.0000 | 0.0000 |
| | 51.7598 | 38.5796 | 36.9305 | 126.1058 | 370.4966 | 1306.8594 | 6100.9250 | 20056.7031 | 19727.7500 | 13734.0000 | 2699.0000 | 1035.0000 | 517.5000 | 258.7500 |
| 1000 | 0.8153 | 0.1776 | 0.0073 | 0.0014 | 0.0003 | 0.0001 | 0.0000 | 1.0000 | 0.0000 | 0.0000 | 0.0000 | 0.0000 | 0.0000 | 0.0000 |
| | 508.9235 | 474.8547 | 1217.5903 | 3603.8133 | 11732.8742 | 40850.8379 | 147930.5781 | 200006.2500 | 192340.0000 | 451010.0000 | 71990.0000 | 10350.0000 | 5175.0000 | 2587.5000 |
| 10000 | 0.9829 | 0.0461 | 0.0347 | 0.0166 | 0.0106 | 0.0003 | 0.0000 | 0.7028 | 0.2977 | 0.0003 | 0.0002 | 0.0000 | 0.0000 | 0.0000 |
| | 5094.9100 | 3577.8125 | 8197.3889 | 10432.4688 | 2833.8281 | 62962.1406 | 377295.3906 | 2022600.0000 | 1948400.0000 | 4374024.5117 | 582446.8750 | 209750.0000 | 51749.0234 | 25875.0000 |
| m=900 | | | | | | | | | | | | | | |
| 0.0001 | 0.3598 | 0.3593 | 0.1960 | 0.1028 | 0.0809 | 0.0276 | 0.0055 | 0.6499 | 0.2487 | 0.1298 | 0.0410 | 0.0279 | 0.0109 | 0.0013 |
| | 0.1281 | 0.2903 | 0.6666 | 0.0567 | 1.5578 | 3.8991 | 11.0378 | 11.4040 | 15.8560 | 8.4399 | 5.7960 | 4.1400 | 3.0700 | 2.0350 |
| 0.0010 | 0.3593 | 0.3586 | 0.1962 | 0.1031 | 0.0815 | 0.0277 | 0.0055 | 0.7149 | 0.2250 | 0.1028 | 0.0532 | 0.0231 | 0.0079 | 0.0009 |
| | 0.1282 | 0.2898 | 0.6640 | 0.0567 | 1.5507 | 3.8913 | 11.0396 | 11.4040 | 16.1060 | 7.4399 | 5.7921 | 4.1400 | 3.0700 | 2.0350 |
| 0.0100 | 0.3568 | 0.3556 | 0.1972 | 0.1057 | 0.0834 | 0.0281 | 0.0055 | 0.6081 | 0.3982 | 0.0276 | 0.0187 | 0.0082 | 0.0010 | 0.0001 |
| | 0.1295 | 0.2908 | 0.6624 | 0.0578 | 1.5547 | 3.9772 | 11.5356 | 11.4040 | 15.8560 | 5.7960 | 6.4399 | 4.1400 | 2.0700 | 1.0350 |

| | Sorted Linear coefficients and corresponding exponents for Electron Orbitals | | | | | | | Sorted Linear coefficients and corresponding exponents for PC Orbitals | | | | | | |
|---|---|---|---|---|---|---|---|---|---|---|---|---|---|---|
| ω/Basis | 1s | 2s | 3s | 4s | 5s | 6s | 7s | 1s | 2s | 3s | 4s | 5s | 6s | 7s |
| 0.1000 | 0.3752 | 0.3149 | 0.2298 | 0.0914 | 0.0867 | 0.0260 | 0.0046 | 0.9939 | 0.0071 | 0.0029 | 0.0024 | 0.0015 | 0.0002 | 0.0000 |
| | 0.2920 | 0.1384 | 0.6990 | 1.8942 | 0.0762 | 5.9777 | 23.4522 | 47.5123 | 26.4040 | 5.7960 | 8.9399 | 4.1400 | 2.0700 | 1.0350 |
| 1 | 0.6105 | 0.3018 | 0.0981 | 0.0362 | 0.0131 | 0.0031 | 0.0005 | 0.9819 | 0.0209 | 0.0029 | 0.0004 | 0.0003 | 0.0000 | 0.0000 |
| | 0.5592 | 1.0576 | 2.4469 | 5.6059 | 15.3061 | 52.0276 | 207.7633 | 451.4000 | 406.0375 | 328.4000 | 124.0797 | 103.5122 | 56.7500 | 25.8750 |
| 10 | 0.8434 | 0.1270 | 0.0359 | 0.0110 | 0.0028 | 0.0006 | 0.0001 | 0.5413 | 0.4587 | 0.0042 | 0.0030 | 0.0013 | 0.0000 | 0.0000 |
| | 5.1231 | 8.3118 | 18.2230 | 47.8231 | 150.2703 | 544.2639 | 2278.6906 | 4472.6000 | 4534.3688 | 258.4000 | 268.6719 | 237.4000 | 1017.5000 | 113.5000 |
| 100 | 0.9067 | 0.2922 | 0.2193 | 0.0216 | 0.0044 | 0.0009 | 0.0001 | 0.9796 | 0.0204 | 0.0000 | 0.0000 | 0.0000 | 0.0000 | 0.0000 |
| | 51.7750 | 37.6030 | 36.4422 | 129.7679 | 396.8638 | 1480.6875 | 7671.2375 | 45007.8750 | 44727.7500 | 17484.0000 | 1035.0000 | 1449.0000 | 517.5000 | 258.7500 |
| 1000 | 0.8302 | 0.1631 | 0.0071 | 0.0013 | 0.0003 | 0.0001 | 0.0000 | 0.8370 | 0.1631 | 0.0000 | 0.0000 | 0.0000 | 0.0000 | 0.0000 |
| | 508.9235 | 472.4133 | 1260.0098 | 3845.5125 | 11810.9992 | 32100.8379 | 127930.5781 | 451010.0000 | 444850.0000 | 137340.0000 | 10350.0000 | 14490.0000 | 5175.0000 | 2587.5000 |
| 10000 | 0.9829 | 0.0462 | 0.0347 | 0.0166 | 0.0107 | 0.0003 | 0.0000 | 0.8964 | 0.1037 | 0.0001 | 0.0000 | 0.0000 | 0.0000 | 0.0000 |
| | 5094.9165 | 3577.8125 | 8197.3889 | 10432.4688 | 2833.8281 | 62649.6406 | 390850.0781 | 4508887.7930 | 4422600.0000 | 1898400.0000 | 232446.8750 | 209750.0000 | 51749.0234 | 25875.0000 |
| | | | | | | | m=1836 | | | | | | | |
| 0.0001 | 0.3580 | 0.3527 | 0.2039 | 0.0992 | 0.0880 | 0.0281 | 0.0052 | 0.6772 | 0.5561 | 0.3422 | 0.1999 | 0.1272 | 0.0458 | 0.0051 |
| | 0.2908 | 0.1291 | 0.6666 | 0.0573 | 1.6242 | 4.4850 | 14.3035 | 20.8560 | 12.4040 | 8.9399 | 5.7960 | 4.1400 | 3.0700 | 2.0350 |
| 0.0010 | 0.3547 | 0.3482 | 0.2065 | 0.1010 | 0.0902 | 0.0291 | 0.0054 | 0.6817 | 0.3666 | 0.1040 | 0.0789 | 0.0222 | 0.0034 | 0.0001 |
| | 0.2874 | 0.1291 | 0.6533 | 0.0577 | 1.5858 | 4.3835 | 14.1646 | 20.7310 | 13.6540 | 7.4399 | 5.7921 | 4.1400 | 3.0700 | 2.0350 |
| 0.0100 | 0.3609 | 0.3489 | 0.2079 | 0.0966 | 0.0886 | 0.0279 | 0.0051 | 0.8065 | 0.2572 | 0.2518 | 0.2442 | 0.0569 | 0.0048 | 0.0006 |
| | 0.2893 | 0.1286 | 0.6693 | 0.0573 | 1.6563 | 4.6647 | 15.3638 | 22.5435 | 13.6540 | 6.9399 | 5.7960 | 4.1400 | 2.0700 | 1.0350 |
| 0.1000 | 0.3997 | 0.3414 | 0.2225 | 0.0812 | 0.0605 | 0.0218 | 0.0039 | 0.9998 | 0.0039 | 0.0026 | 0.0026 | 0.0012 | 0.0001 | 0.0000 |
| | 0.3064 | 0.1345 | 0.7818 | 2.2399 | 0.0699 | 7.4875 | 32.4522 | 94.4810 | 8.9399 | 5.7960 | 11.4040 | 4.1400 | 2.0700 | 1.0350 |
| 1 | 0.6142 | 0.3032 | 0.0995 | 0.0329 | 0.0109 | 0.0026 | 0.0004 | 0.7167 | 0.2834 | 0.0002 | 0.0001 | 0.0001 | 0.0000 | 0.0000 |
| | 0.5603 | 1.0721 | 2.5865 | 6.4162 | 18.1076 | 64.2591 | 291.9430 | 913.9000 | 931.0375 | 328.4000 | 174.0797 | 103.5122 | 56.7500 | 25.8750 |
| 10 | 0.8425 | 0.1279 | 0.0361 | 0.0109 | 0.0028 | 0.0006 | 0.0001 | 0.7870 | 0.2193 | 0.0064 | 0.0003 | 0.0002 | 0.0000 | 0.0000 |
| | 5.1222 | 8.2946 | 18.2993 | 48.6776 | 157.1063 | 605.7874 | 2864.6281 | 9288.2750 | 8722.6000 | 6508.4000 | 1017.5000 | 862.4000 | 268.6719 | 113.5000 |
| 100 | 0.9103 | 0.0837 | 0.0212 | 0.0139 | 0.0042 | 0.0008 | 0.0001 | 0.5801 | 0.4301 | 0.0102 | 0.0000 | 0.0000 | 0.0000 | 0.0000 |
| | 51.7750 | 39.3719 | 133.1859 | 31.6216 | 421.2778 | 1634.0078 | 9241.5500 | 92852.7500 | 90086.0000 | 79046.5000 | 1035.0000 | 1449.0000 | 517.5000 | 258.7500 |

| | Sorted Linear coefficients and corresponding exponents for Electron Orbitals | | | | | | | Sorted Linear coefficients and corresponding exponents for PC Orbitals | | | | | | |
|---|---|---|---|---|---|---|---|---|---|---|---|---|---|---|
| ω/Basis | 1s | 2s | 3s | 4s | 5s | 6s | 7s | 1s | 2s | 3s | 4s | 5s | 6s | 7s |
| 1000 | 0.8212 | 0.1718 | 0.0073 | 0.0014 | 0.0003 | 0.0001 | 0.0000 | 0.9372 | 0.0628 | 0.0000 | 0.0000 | 0.0000 | 0.0000 | 0.0000 |
| | 508.9235 | 473.9392 | 1232.8491 | 3640.8332 | 10500.7859 | 28702.4004 | 130508.7031 | 918470.9375 | 911100.0000 | 14490.0000 | 10350.0000 | 44840.0000 | 5175.0000 | 2587.5000 |
| 10000 | 0.9832 | 0.0460 | 0.0347 | 0.0164 | 0.0106 | 0.0003 | 0.0000 | 0.9563 | 0.0437 | 0.0000 | 0.0000 | 0.0000 | 0.0000 | 0.0000 |
| | 5094.9213 | 3577.8125 | 8168.2637 | 10432.4306 | 2833.8281 | 63245.3438 | 414600.0781 | 9183802.3438 | 9097600.0000 | 209750.0000 | 182446.8750 | 1198400.0000 | 51749.0234 | 25875.0000 |

## 4.5 Asymptotic Behavior of the Variational Wave Function and System Nature

According to the discussion in section 2-3-2, we can compare the radial distribution function of the variational system ($\mathrm{RDF}_{\boldsymbol{var}}$)

$$\mathbf{RDF}_{var} = \boldsymbol{r^2 e^{-2\alpha r - 2\beta r^2}} \qquad 4.8$$

The radial distribution function of the hydrogen-like atom ($RDF_{HL}$)

$$RDF_{HL}(r) = 4\mu^3 r^2 \exp(-2\mu r) \qquad 4.9$$

and the radial distribution function of the harmonic oscillator ($RDF_{HO}$)

$$RDF_{HO}(r) = 4\frac{(\mu\omega)^{3/2}}{\pi^{1/2}} r^2 \exp(-\mu\omega r^2) \qquad 4.10$$

at any specific mass and frequency, and as a result, examine the overall behavior of the variational wave function and its similarity to the two mentioned systems, especially its asymptotic behavior. This comparison at frequencies $10^{-4}$ and $10^{4}$ can confirm the accuracy of the asymptotic behavior of the variational wave function (2-72). Figures 4-1 to 4-10 illustrate these points.

As expected, the behavior of the variational wave function at frequencies of $10^{-4}$ corresponds closely to that of a hydrogen-like atom; and at frequencies of $10^4$, it matches the behavior of a harmonic oscillator. This behavior can be evaluated using equation (4-8) by examining $\frac{\beta}{\alpha}$ ratio. This ratio is reported as item 7 in Table 4-2. It is expected that, at low frequencies, this ratio approaches zero, and at high frequencies, it tends towards infinity. The numerical data of this ratio confirms this behavior.

As mass increases, the rate of system behavior changes (or changes in nature) decreases. This means that the lower the mass of the PCP, the faster the system's behavior (at lower frequencies) transitions to oscillator-like behavior. This correctly shows that the lower the mass of the PCP, the more sensitive the system is to the presence of an oscillatory field. Given that the binding energy of hydrogen (mass 1836) is greater compared to positronium (mass 1) and its radius is smaller, it demonstrates more stability in maintaining atomic structure against external fields.

Based on the above explanations, it can be said that at low frequencies, the system has a nature similar to hydrogen-like atoms, and the two particles are bound to each other, making the concept of an atom and, consequently, the correlation energy meaningful. However, at high frequencies, the system's behavior diverges from that of a hydrogen-like atom and aligns with that of a harmonic oscillator. Therefore, in these regions, the two particles (even with a very small distance) are not bound to each other and are only held together by an oscillatory potential. As a result, the concept of atomic structure is meaningless, and one can only speak of two particles trapped in a powerful oscillatory potential.

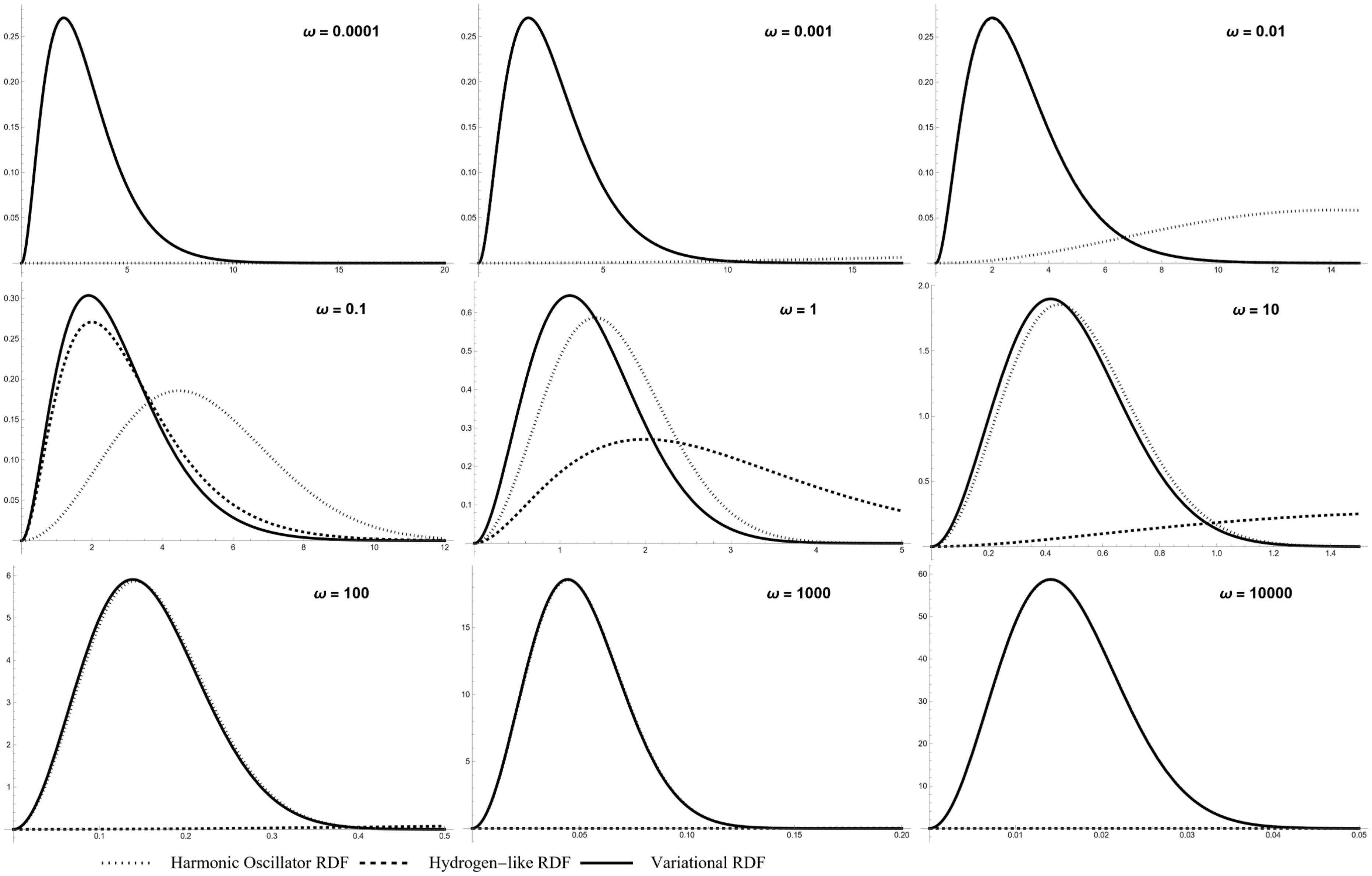

Figure 4-1: Comparison of the radial distribution functions with respect to inter-particle distance for variational, hydrogen-like, and harmonic oscillator systems at mass 1.

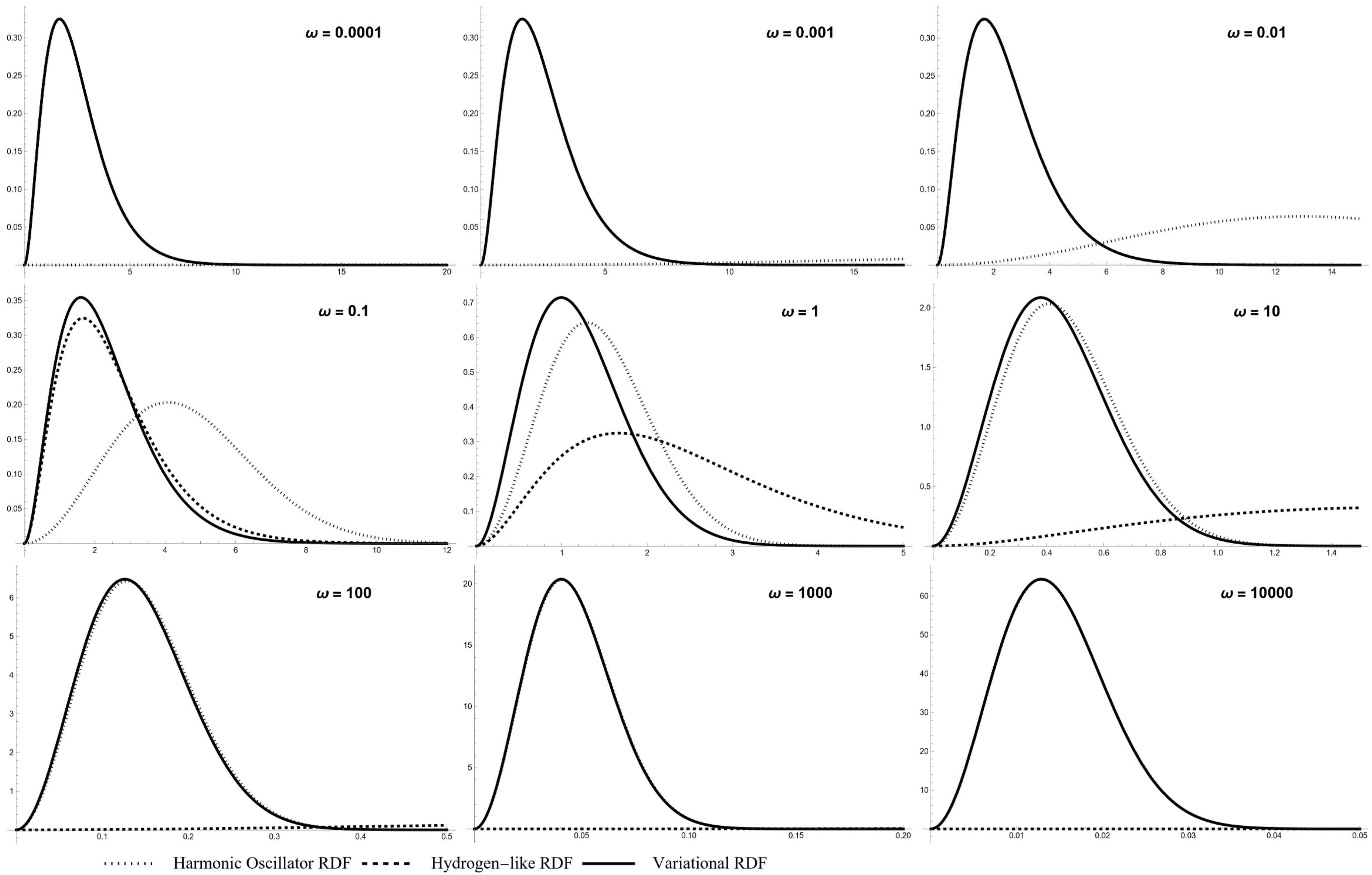

Figure 4-2: Comparison of the radial distribution functions with respect to inter-particle distance for variational, hydrogen-like, and harmonic oscillator systems at mass 1.5.

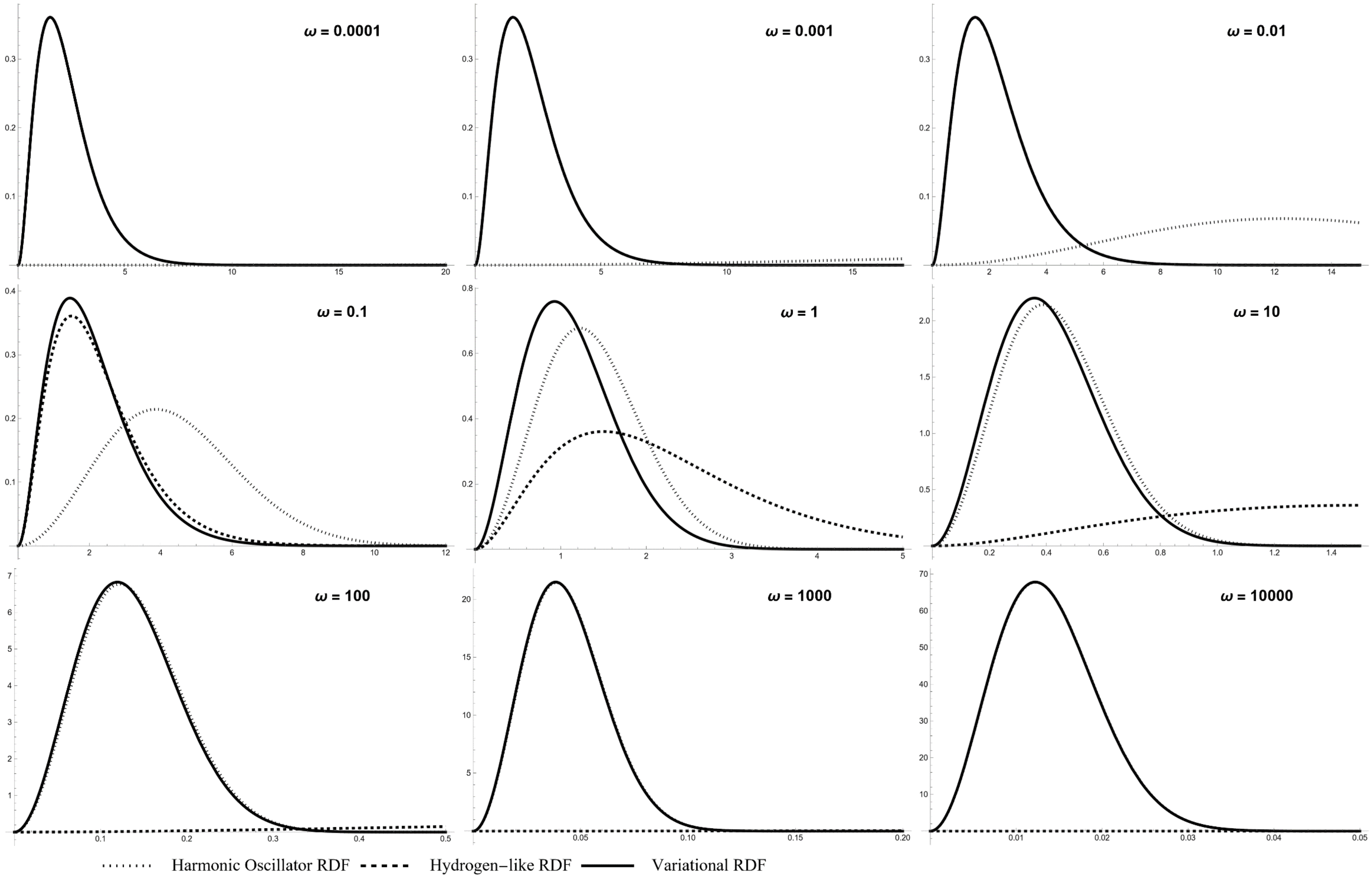

Figure 4-3: Comparison of the radial distribution functions with respect to inter-particle distance for variational, hydrogen-like, and harmonic oscillator systems at mass 2.

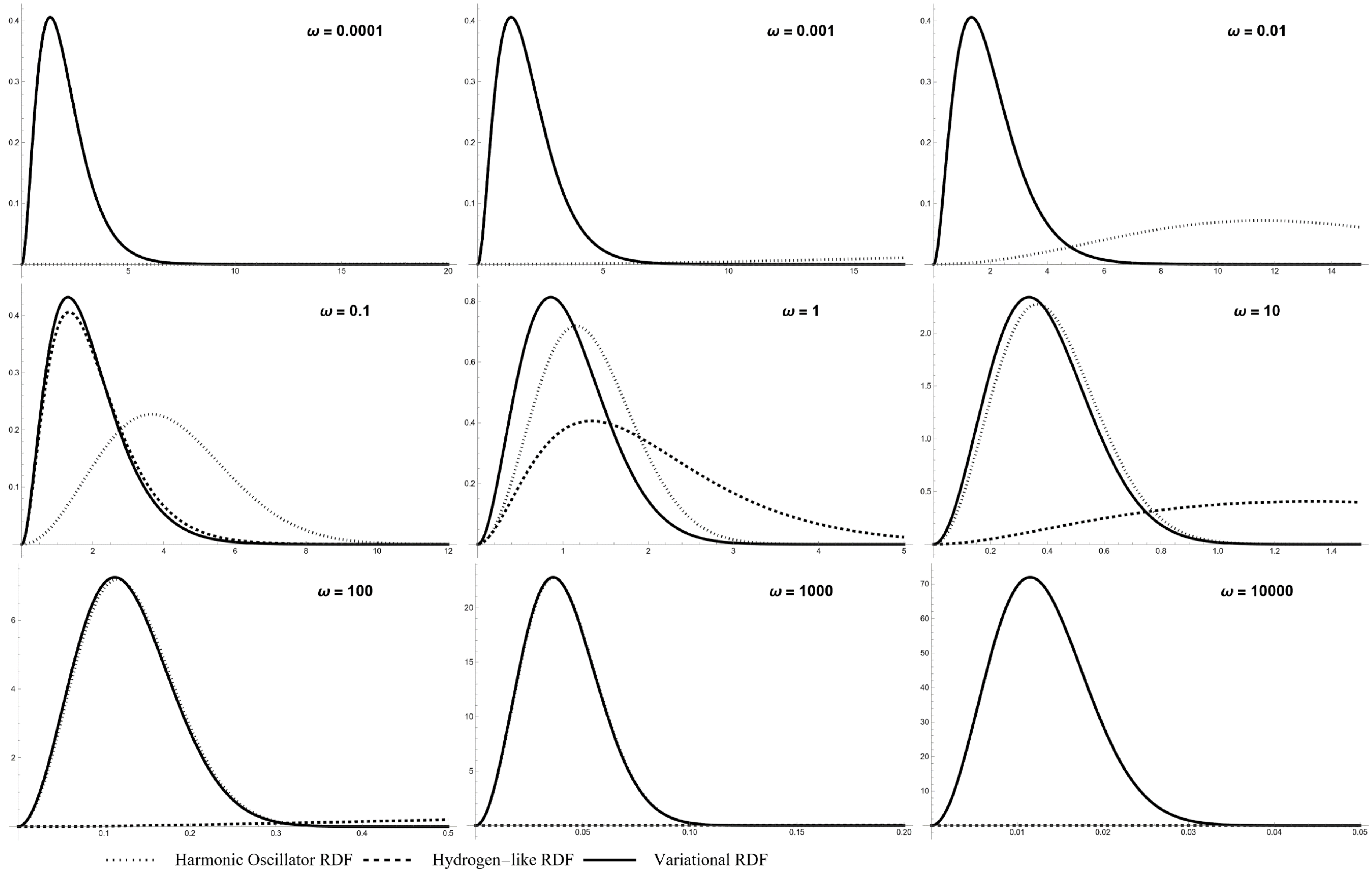

Figure 4-4: Comparison of the radial distribution functions with respect to inter-particle distance for variational, hydrogen-like, and harmonic oscillator systems at mass 3.

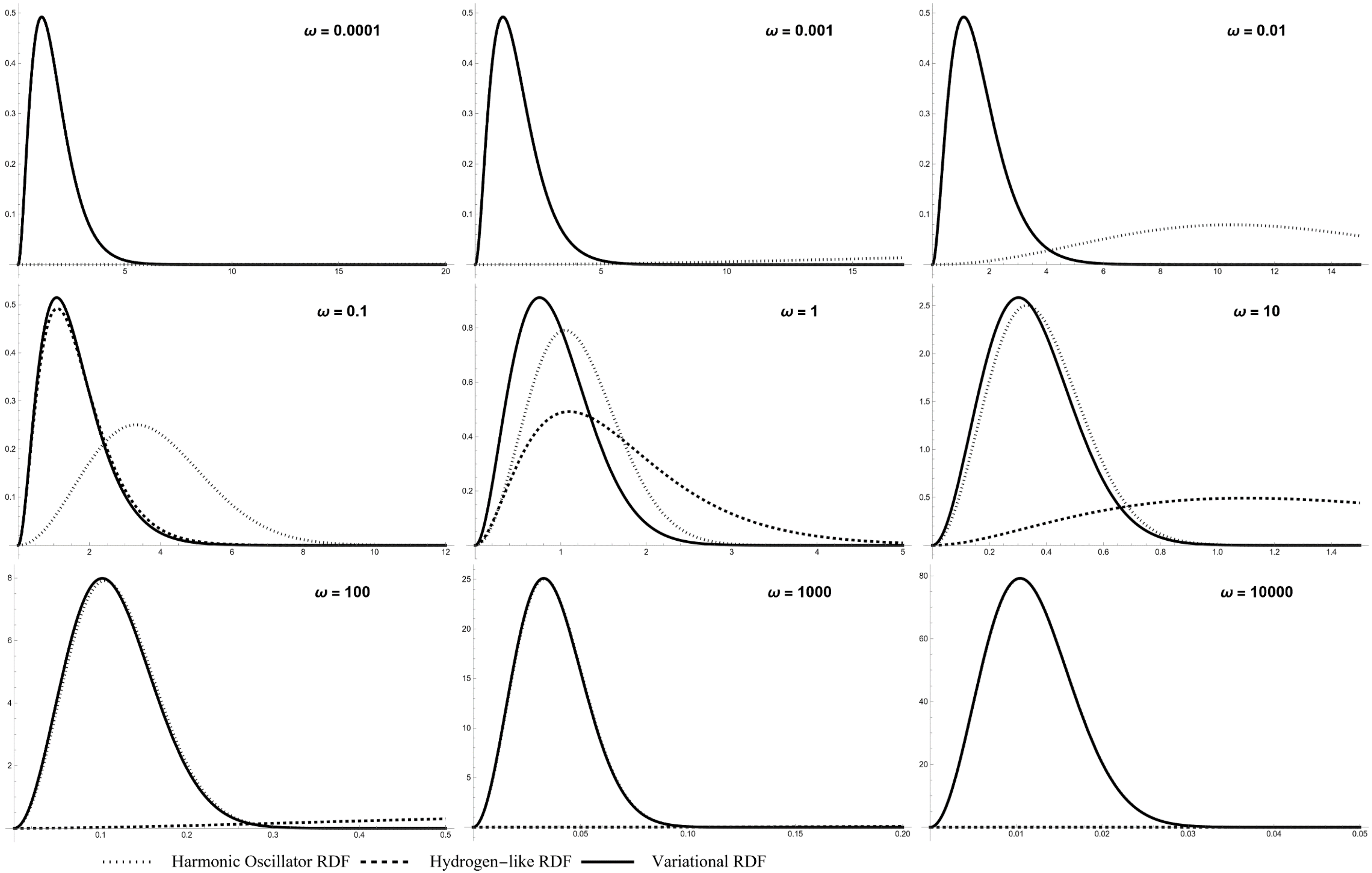

Figure 4-5: Comparison of the radial distribution functions with respect to inter-particle distance for variational, hydrogen-like, and harmonic oscillator systems at mass 10.

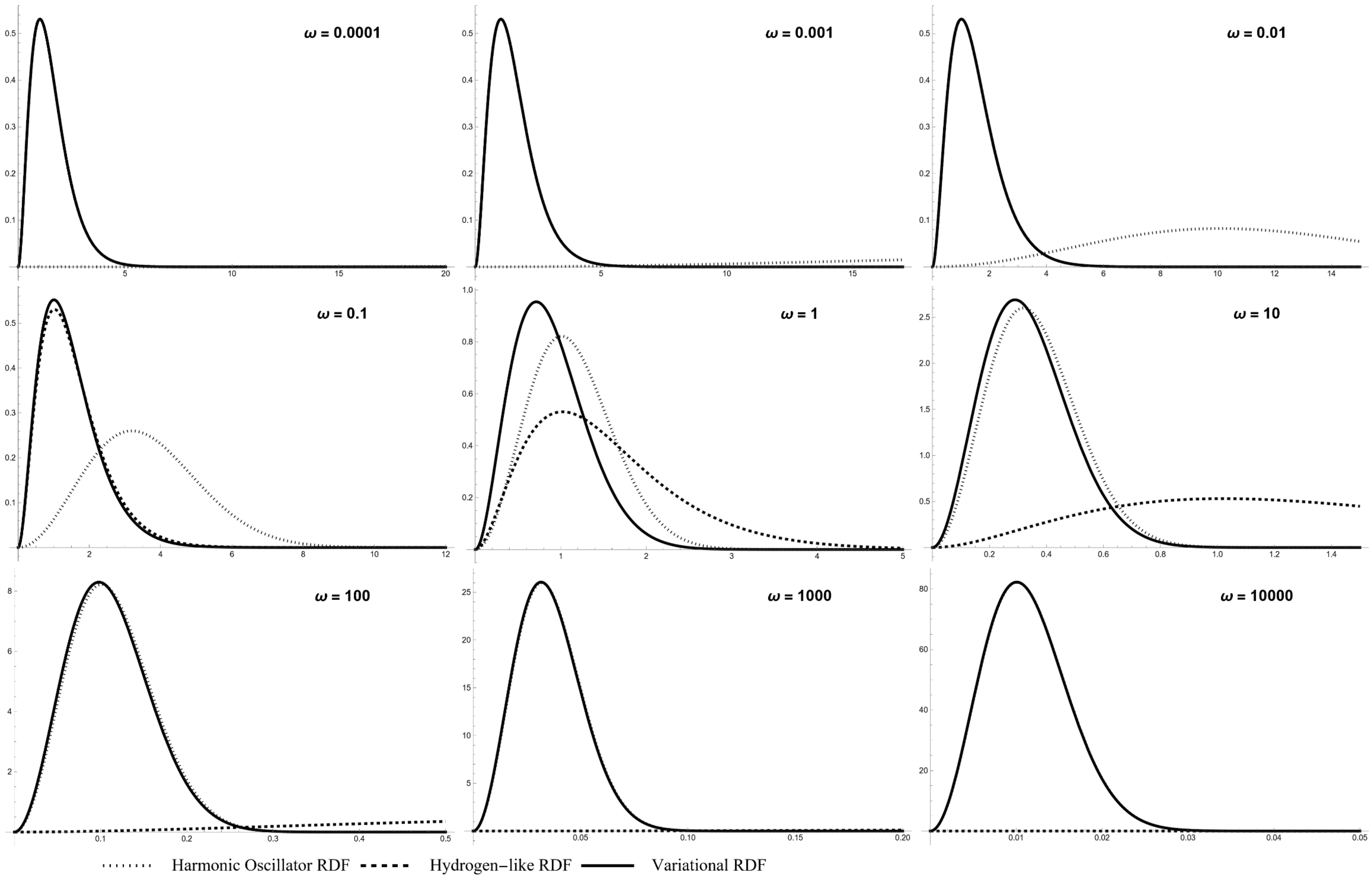

Figure 4-6: Comparison of the radial distribution functions with respect to inter-particle distance for variational, hydrogen-like, and harmonic oscillator systems at mass 50.

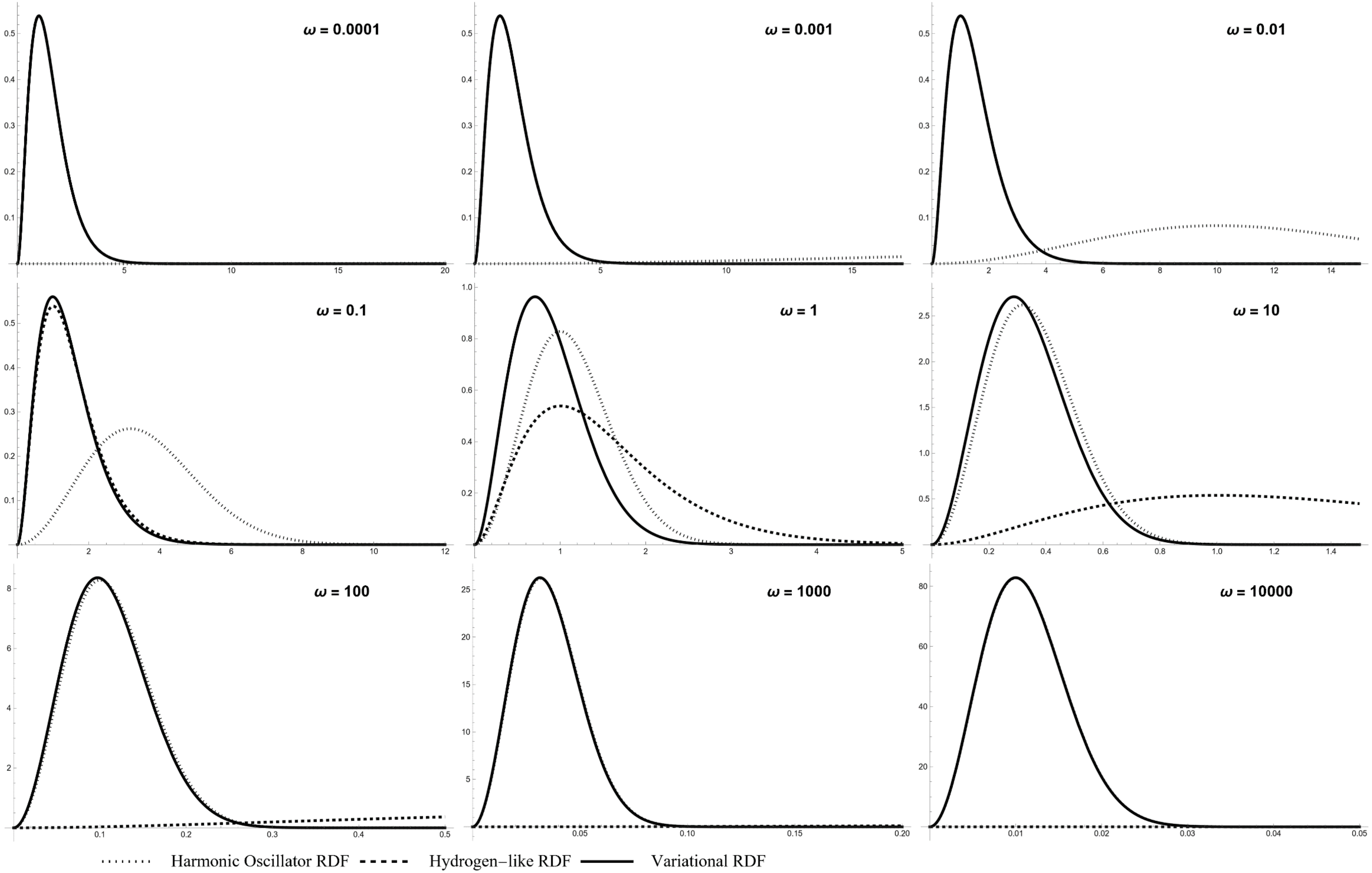

Figure 4-7: Comparison of the radial distribution functions with respect to inter-particle distance for variational, hydrogen-like, and harmonic oscillator systems at mass 207.

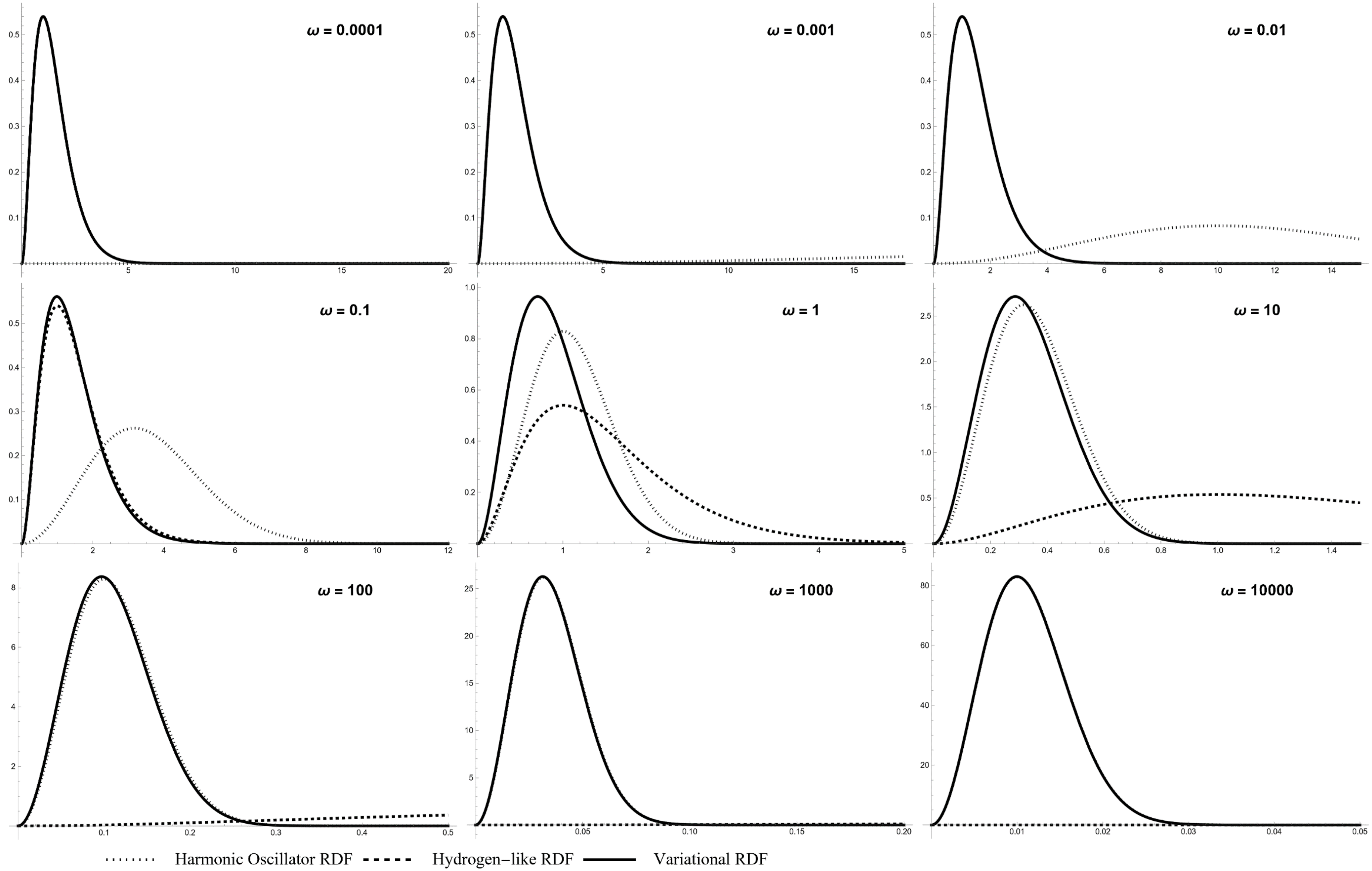

Figure 4-8: Comparison of the radial distribution functions with respect to inter-particle distance for variational, hydrogen-like, and harmonic oscillator systems at mass 400.

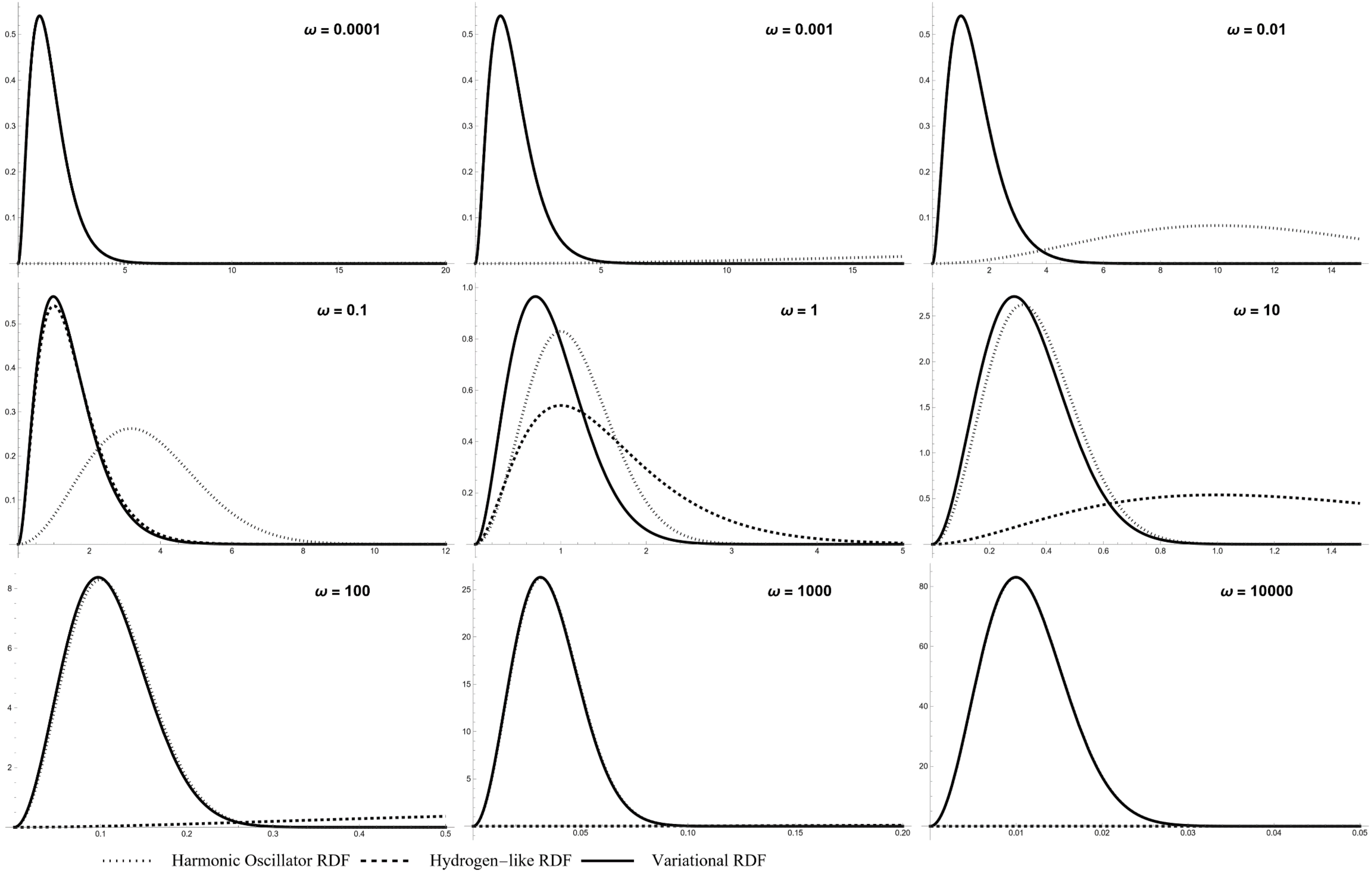

Figure 4-9: Comparison of the radial distribution functions with respect to inter-particle distance for variational, hydrogen-like, and harmonic oscillator systems at mass 900.

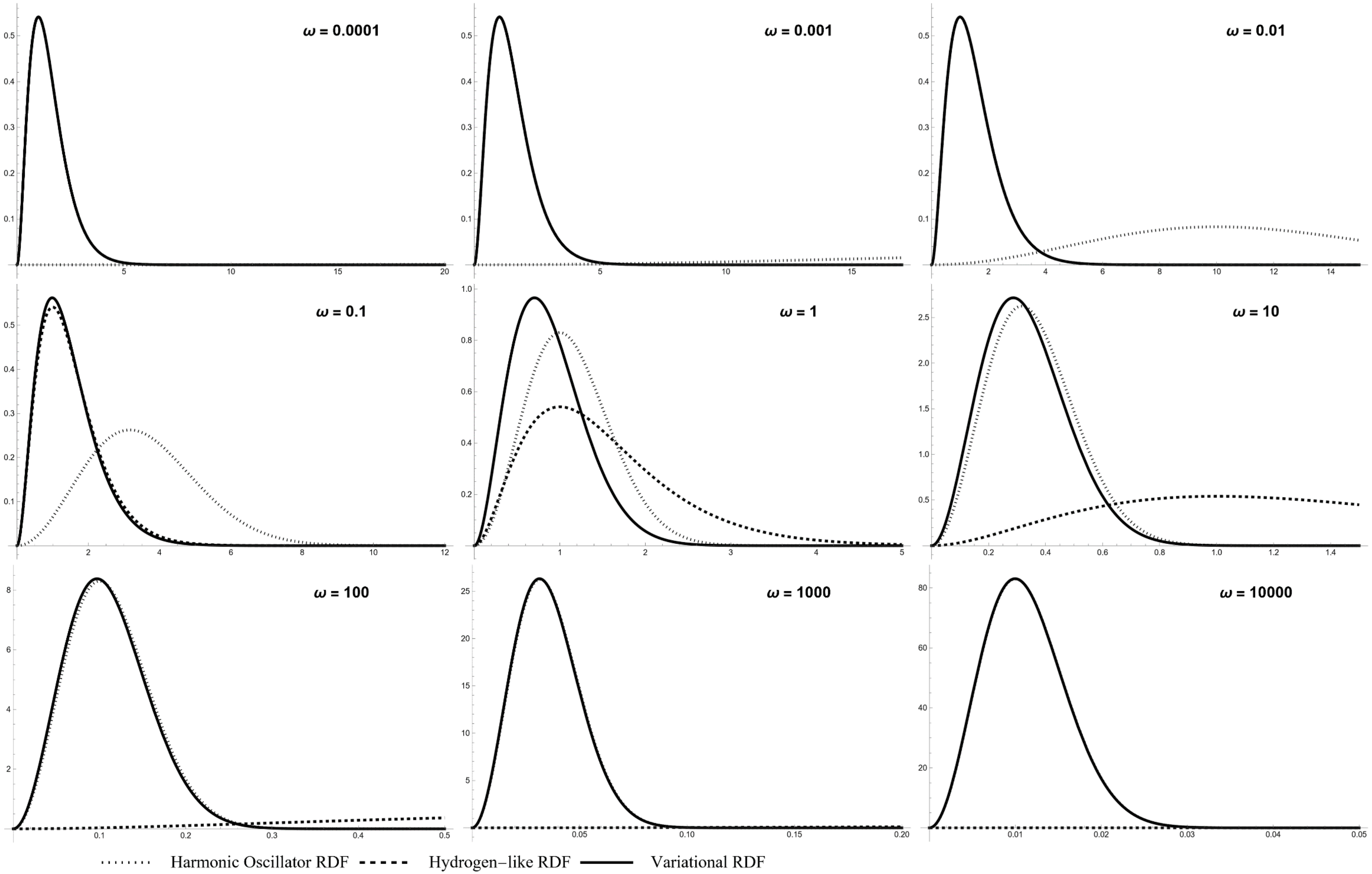

Figure 4-10: Comparison of the radial distribution functions with respect to inter-particle distance for variational, hydrogen-like, and harmonic oscillator systems at mass 1836

# 5 Electron-Positively Charged Particle Correlation in the Exotic Harmonium Model

## 5.1 Correlation in Wavefunction-Based Approaches

### 5.1.1 Correlation Energy

In this study, the correlation energies (3-24) and (3-25) for the entire range of mass and HO frequencies are calculated and reported in Table 5-1. However, DFT calculations using analytical single-particle densities were not feasible due to complexity. Therefore, the density forms were simplified through fitting only for selected masses and HO frequencies. Subsequently, the DFT calculations, including the correlation energy (3-32), were also performed only for these selected parameters.

The Hartree-Fock correlation energy, defined by equation (3-25) as the difference in energy between correlated methods and the MC-HF method, is reported for both the variational and FEM methods in items 5 and 6 of Table 4-3, respectively. These numerical data, along with Figures 5-1 and 5-2, show that the magnitude of correlation energy decreases with increasing HO frequency for a given mass. Conversely, as mass increases from 1 to 3, the magnitude of correlation energy initially increases and then decreases for masses above 10. Therefore, the magnitude of correlation energy inversely correlates with oscillator frequency and PCP mass, except for a small range between masses 1 and 3 where this relationship is direct.

Table 5-1 shows the correlation energy components relative to the MC-HF reference, obtained from the differences in variational and MC-HF energy components (Table 4-4). The difference in electron kinetic energy decreases with increasing frequency for a given mass and initially increases (up to mass 10) and then decreases with increasing mass for a given frequency. The difference in the kinetic energy of the PCP at a specific frequency decreases with increasing mass and changes sign, decreasing in magnitude for masses above 10.

The difference in the harmonic oscillator potential energy for the electron at a given mass changes sign and becomes negative with increasing frequency, with its magnitude increasing. However, its magnitude decreases with increasing mass for a given frequency. The difference in the harmonic oscillator potential energy for the PCP is always positive and increases with frequency for a given mass. At a specific frequency, it initially increases and then decreases for intermediate masses (e.g., 50 and 207). Additionally, the difference in e-PCP interaction energy is always negative, and its magnitude decreases with increasing frequency for a given mass. At a specific frequency, it initially increases and then decreases for intermediate masses.

The correlation energy has two major components: kinetic energy, which depends on mass changes, and potential energy, which depends on frequency changes.

Table 5-1: Correlation energy values relative to MC-HF reference and its components

| ω/Quantities | Correlation E[1] | ΔTE[2] | ΔTP[3] | ΔHO E[4] | ΔHO P[5] | ΔINT[6] |
|---|---|---|---|---|---|---|
| | | | m=1 | | | |
| 0.0001 | -0.141337 | 0.070781 | 0.070781 | 0.000037 | 0.000037 | -0.282973 |
| 0.001 | -0.140006 | 0.071100 | 0.071100 | 0.000366 | 0.000366 | -0.282937 |
| 0.01 | -0.128294 | 0.072746 | 0.072746 | 0.002867 | 0.002867 | -0.279520 |
| 0.1 | -0.086137 | 0.057013 | 0.057013 | 0.004648 | 0.004648 | -0.209460 |
| 1 | -0.059918 | 0.034217 | 0.034217 | 0.001419 | 0.001419 | -0.131191 |
| 10 | -0.052581 | 0.027484 | 0.027484 | 0.000398 | 0.000398 | -0.108344 |
| 100 | -0.050455 | 0.025587 | 0.025590 | 0.000121 | 0.000118 | -0.101870 |
| 1000 | -0.049803 | 0.024882 | 0.024883 | 0.000175 | 0.000173 | -0.099911 |
| 10000 | -0.049605 | 0.021779 | 0.022020 | 0.003076 | 0.002835 | -0.099315 |
| | | | m=1.5 | | | |
| 0.0001 | -0.168189 | 0.108126 | 0.060289 | 0.000030 | 0.000045 | -0.336679 |
| 0.001 | -0.166854 | 0.108380 | 0.060679 | 0.000293 | 0.000441 | -0.336647 |
| 0.01 | -0.154850 | 0.109589 | 0.063297 | 0.002353 | 0.003659 | -0.333748 |
| 0.1 | -0.105994 | 0.088305 | 0.053092 | 0.002963 | 0.008837 | -0.259192 |
| 1 | -0.071490 | 0.052563 | 0.030357 | -0.003271 | 0.007081 | -0.158220 |
| 10 | -0.061655 | 0.041587 | 0.023257 | -0.005435 | 0.006498 | -0.127562 |
| 100 | -0.058820 | 0.038512 | 0.021233 | -0.006052 | 0.006401 | -0.118913 |
| 1000 | -0.057948 | 0.037582 | 0.020665 | -0.006239 | 0.006337 | -0.116294 |
| 10000 | -0.057773 | 0.037306 | 0.022997 | -0.006305 | 0.003717 | -0.115488 |
| | | | m=2 | | | |
| 0.0001 | -0.183855 | 0.135875 | 0.048206 | 0.000025 | 0.000050 | -0.368011 |
| 0.001 | -0.182519 | 0.136086 | 0.048643 | 0.000245 | 0.000492 | -0.367985 |
| 0.01 | -0.170363 | 0.136970 | 0.051871 | 0.001978 | 0.004182 | -0.365363 |
| 0.1 | -0.117354 | 0.111770 | 0.046106 | 0.001646 | 0.011862 | -0.288738 |
| 1 | -0.076839 | 0.065423 | 0.025049 | -0.006855 | 0.011399 | -0.171855 |
| 10 | -0.065106 | 0.050733 | 0.018175 | -0.009845 | 0.011112 | -0.135281 |
| 100 | -0.061727 | 0.046616 | 0.016264 | -0.010688 | 0.011052 | -0.124972 |
| 1000 | -0.060689 | 0.045367 | 0.015674 | -0.010944 | 0.011058 | -0.121852 |
| 10000 | -0.060368 | 0.044998 | 0.015497 | -0.011029 | 0.011051 | -0.120886 |
| | | | m=3 | | | |
| 0.0001 | -0.199277 | 0.171946 | 0.027555 | 0.000019 | 0.000056 | -0.398853 |
| 0.001 | -0.197939 | 0.172102 | 0.028050 | 0.000184 | 0.000554 | -0.398829 |
| 0.01 | -0.185627 | 0.172551 | 0.032010 | 0.001484 | 0.004827 | -0.396500 |
| 0.1 | -0.127779 | 0.142191 | 0.032517 | -0.000074 | 0.015718 | -0.318130 |
| 1 | -0.079392 | 0.080203 | 0.015763 | -0.011235 | 0.016759 | -0.180882 |

| ω/Quantities | Correlation E[1] | ΔTE[2] | ΔTP[3] | ΔHO E[4] | ΔHO P[5] | ΔINT[6] |
|---|---|---|---|---|---|---|
| 10 | -0.065147 | 0.059780 | 0.009969 | -0.015002 | 0.016534 | -0.136429 |
| 100 | -0.061064 | 0.054066 | 0.008362 | -0.015990 | 0.016457 | -0.123961 |
| 1000 | -0.059814 | 0.052325 | 0.007859 | -0.016278 | 0.016460 | -0.120200 |
| 10000 | -0.059435 | 0.051791 | 0.007754 | -0.016340 | 0.016406 | -0.119048 |
| m=10 | | | | | | |
| 0.0001 | -0.197347 | 0.223009 | -0.025437 | 0.000007 | 0.000068 | -0.394994 |
| 0.001 | -0.196007 | 0.223057 | -0.024833 | 0.000066 | 0.000673 | -0.394971 |
| 0.01 | -0.183502 | 0.222552 | -0.019549 | 0.000505 | 0.005996 | -0.393007 |
| 0.1 | -0.118794 | 0.180539 | -0.005193 | -0.002666 | 0.021516 | -0.312990 |
| 1 | -0.059334 | 0.084132 | -0.004977 | -0.014362 | 0.020969 | -0.145095 |
| 10 | -0.042938 | 0.053113 | -0.005136 | -0.016759 | 0.018437 | -0.092593 |
| 100 | -0.038548 | 0.045080 | -0.005026 | -0.017034 | 0.017459 | -0.079028 |
| 1000 | -0.037256 | 0.042722 | -0.004645 | -0.017052 | 0.016799 | -0.075081 |
| 10000 | -0.036929 | 0.041886 | -0.002508 | -0.016954 | 0.014523 | -0.073860 |
| m=50 | | | | | | |
| 0.0001 | -0.136159 | 0.178247 | -0.041869 | 0.000001 | 0.000073 | -0.272612 |
| 0.001 | -0.134821 | 0.178246 | -0.041216 | 0.000014 | 0.000724 | -0.272589 |
| 0.01 | -0.122460 | 0.176975 | -0.035446 | 0.000078 | 0.006280 | -0.270348 |
| 0.1 | -0.064093 | 0.126163 | -0.016007 | -0.002004 | 0.017359 | -0.189603 |
| 1 | -0.022104 | 0.039693 | -0.006276 | -0.006612 | 0.010383 | -0.059291 |
| 10 | -0.013563 | 0.019984 | -0.004068 | -0.006593 | 0.007384 | -0.030271 |
| 100 | -0.011572 | 0.015720 | -0.003510 | -0.006360 | 0.006571 | -0.023993 |
| 1000 | -0.011007 | 0.014548 | -0.003348 | -0.006269 | 0.006330 | -0.022269 |
| 10000 | -0.010839 | 0.014253 | -0.003870 | -0.006298 | 0.006883 | -0.021748 |
| m=207 | | | | | | |
| 0.0001 | -0.082760 | 0.114782 | -0.031800 | 0.000000 | 0.000074 | -0.165817 |
| 0.001 | -0.081427 | 0.114763 | -0.031139 | 0.000003 | 0.000730 | -0.165784 |
| 0.01 | -0.069560 | 0.112685 | -0.025499 | 0.000004 | 0.005872 | -0.162622 |
| 0.1 | -0.026584 | 0.061989 | -0.009265 | -0.000908 | 0.009619 | -0.088022 |
| 1 | -0.006733 | 0.013582 | -0.002522 | -0.002159 | 0.003598 | -0.019232 |
| 10 | -0.003642 | 0.005864 | -0.001419 | -0.001994 | 0.002260 | -0.008353 |
| 100 | -0.002977 | 0.004369 | -0.001178 | -0.001885 | 0.001949 | -0.006223 |
| 1000 | -0.002793 | 0.003967 | -0.001112 | -0.001843 | 0.001862 | -0.005668 |
| 10000 | -0.002747 | 0.003865 | -0.001366 | -0.001839 | 0.001684 | -0.005513 |
| m=400 | | | | | | |
| 0.0001 | -0.063705 | 0.089870 | -0.025943 | 0.000000 | 0.000075 | -0.127707 |
| 0.001 | -0.062375 | 0.089844 | -0.025282 | 0.000002 | 0.000727 | -0.127666 |

| ω/Quantities | Correlation E[1] | ΔTE[2] | ΔTP[3] | ΔHO E[4] | ΔHO P[5] | ΔINT[6] |
|---|---|---|---|---|---|---|
| 0.01 | -0.050894 | 0.087230 | -0.019818 | -0.000004 | 0.005510 | -0.123812 |
| 0.1 | -0.016254 | 0.040286 | -0.006189 | -0.000566 | 0.006510 | -0.056295 |
| 1 | -0.003624 | 0.007681 | -0.001478 | -0.001190 | 0.002045 | -0.010680 |
| 10 | -0.001804 | 0.003071 | -0.000803 | -0.001086 | 0.001241 | -0.004228 |
| 100 | -0.001424 | 0.002208 | -0.000658 | -0.001025 | 0.001060 | -0.003007 |
| 1000 | -0.001317 | 0.001969 | -0.000620 | -0.000996 | 0.001011 | -0.002681 |
| 10000 | -0.001443 | -0.005101 | -0.002824 | 0.005873 | 0.001943 | -0.002571 |
| m=900 | | | | | | |
| 0.0001 | -0.045320 | 0.064974 | -0.019430 | 0.000000 | 0.000075 | -0.090938 |
| 0.001 | -0.043998 | 0.064936 | -0.018772 | 0.000001 | 0.000721 | -0.090882 |
| 0.01 | -0.033180 | 0.061483 | -0.013630 | -0.000006 | 0.004896 | -0.085924 |
| 0.1 | -0.008372 | 0.022002 | -0.003443 | -0.000293 | 0.003679 | -0.030317 |
| 1 | -0.001579 | 0.003644 | -0.000729 | -0.000539 | 0.000983 | -0.004939 |
| 10 | -0.000638 | 0.001257 | -0.000382 | -0.000498 | 0.000579 | -0.001580 |
| 100 | -0.000448 | 0.000821 | -0.000310 | -0.000476 | 0.000489 | -0.000980 |
| 1000 | -0.000391 | 0.000712 | -0.000338 | -0.000471 | 0.000509 | -0.000807 |
| 10000 | -0.000512 | -0.006410 | -0.000858 | 0.006486 | 0.001029 | -0.000759 |
| m=1836 | | | | | | |
| 0.0001 | -0.033161 | 0.048083 | -0.014693 | 0.000000 | 0.000075 | -0.066626 |
| 0.001 | -0.031848 | 0.048031 | -0.014040 | 0.000000 | 0.000712 | -0.066552 |
| 0.01 | -0.021825 | 0.043694 | -0.009286 | -0.000006 | 0.004197 | -0.060424 |
| 0.1 | -0.004486 | 0.012247 | -0.001929 | -0.000154 | 0.002081 | -0.016730 |
| 1 | -0.000684 | 0.001826 | -0.000378 | -0.000246 | 0.000504 | -0.002389 |
| 10 | -0.000140 | 0.000786 | -0.000199 | -0.000680 | 0.000295 | -0.000454 |
| 100 | -0.000035 | 0.000230 | -0.000157 | -0.000238 | 0.000245 | -0.000104 |
| 1000 | 0.000000 | 0.000167 | -0.000156 | -0.000239 | 0.000245 | -0.000013 |
| 10000 | -0.000132 | -0.006711 | -0.000997 | 0.006500 | 0.000343 | -0.000006 |

1. Correlation energy relative to MC-HF/[7s:7s] reference (equation 3-25)
2. Electron's kinetic energy contribution to correlation energy
3. PCP's kinetic energy contribution to correlation energy
4. Electron's harmonic oscillator potential energy contribution to correlation energy
5. PCP's harmonic oscillator potential energy contribution to correlation energy
6. E-PCP interaction energy contribution to correlation energy

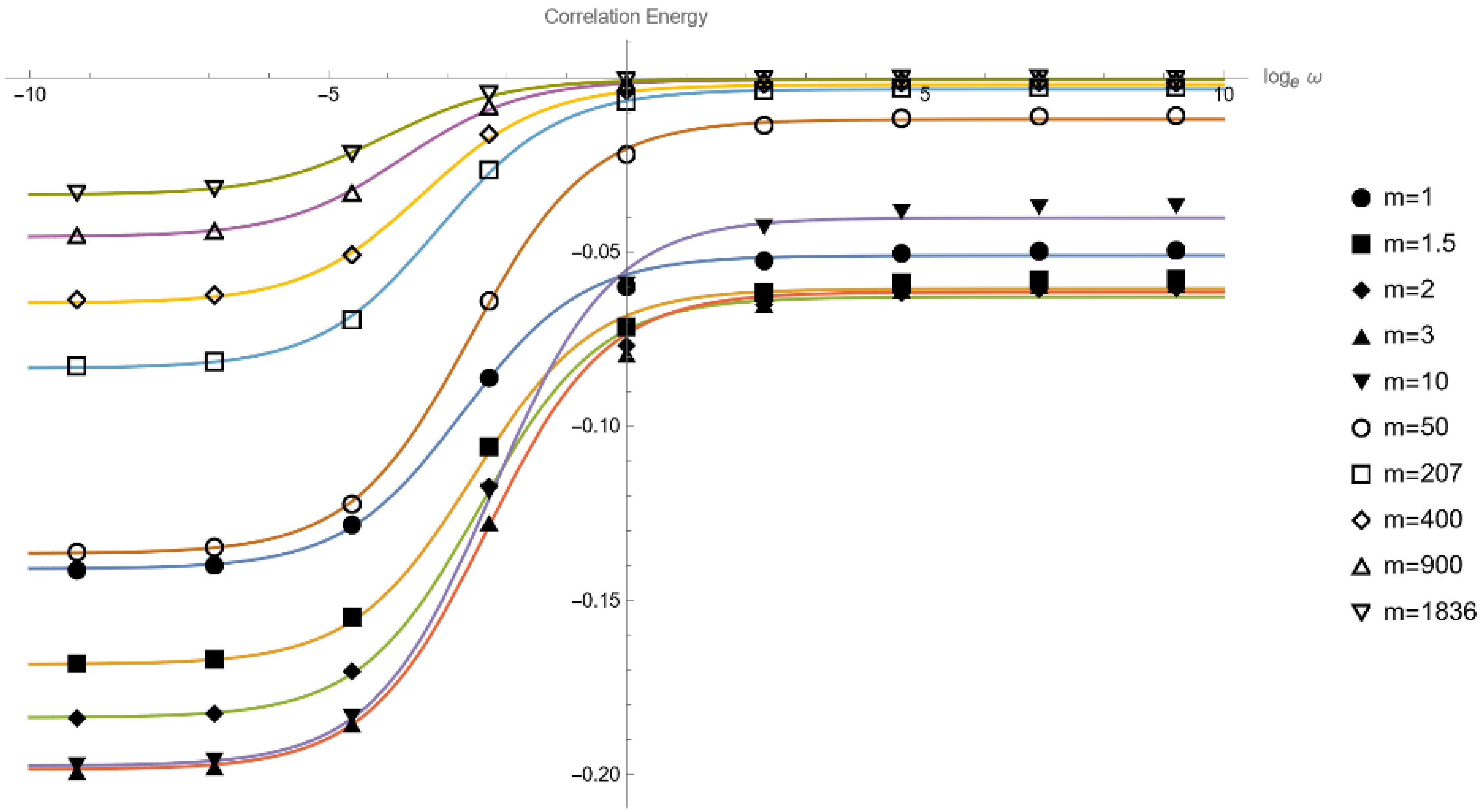

Figure 5-1: Correlation energies plotted (shapes) against the natural logarithm of HO frequency for different masses, along with lines obtained from fitting functions.

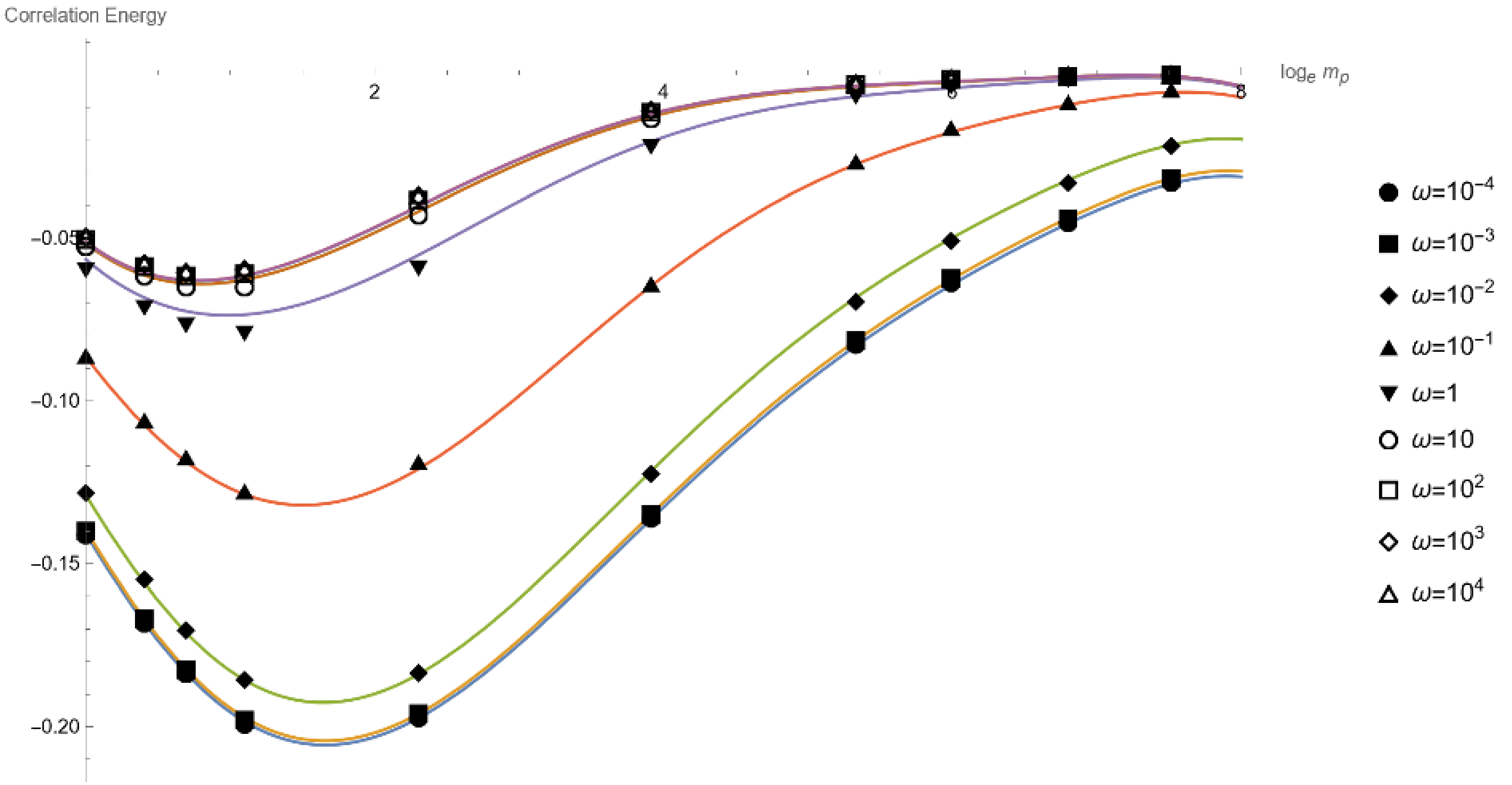

Figure 5-2: Correlation energies plotted (shapes) against the natural logarithm of the PCP's mass for different HO frequencies, along with lines obtained from fitting functions.

The reference-independent correlation energies ($w_c$), as well as the values of $w$ and $w_{col}$ in equation (3-24), are presented in Table 2-5. Since the total two-particle interaction energy ($w$) equals the interaction energy from wavefunction-based calculations (equations 2-139 and 2-152), these two values are reported for comparison in items 1 and 2, and their difference in item 6 of Table 5-2, where the zeros confirm the data's accuracy. The values of $w_c$ and $w_{col}$ are presented in items 4 and 3, respectively. To check the data's accuracy based on equation (3-21), the sum of $w_c$ and $w_{col}$ is given in item 5, which equals item 2 across all systems (while this equality seems obvious according to equation 3-21, it is noteworthy that the definition of $w_c$ in the codes was derived using the integrand of equation 3-20, and the integration was ultimately performed to obtain the numerical values. Consequently, this test indicates any potential errors in the integration process).

The data in Table 5-2 show that the trends of reference-independent correlation energy match the Hartree-Fock correlation energy trends. However, their values, especially at low masses and HO frequencies, differ significantly. At high HO frequencies, these energy values converge, reflecting the system's effective transformation into two harmonic oscillators. In such a scenario, the particles are not correlated, and the effective system's wavefunction reduces to the product of the electron and positive particle wavefunctions, resembling the two-component Hartree-Fock total wavefunction.

Table 5-2: values of e-PCP interaction energy, Hartree potential, reference-independent correlation energy, and MC-HF correlation energy

| ω/Quantities | INT[1] | $\boldsymbol{w}$ [2] | $\boldsymbol{w_{col}}$[3] | $\boldsymbol{w_c}$[4] | TEST1[5] | TEST2[6] | MC-HF corr[7] |
|---|---|---|---|---|---|---|---|
| | | | | m=1 | | | |
| 0.0001 | -0.5000 | -0.5000 | -0.0113 | -0.4887 | -0.5000 | 0.0000 | -0.141337 |
| 0.001 | -0.5000 | -0.5000 | -0.0356 | -0.4644 | -0.5000 | 0.0000 | -0.140006 |
| 0.01 | -0.5006 | -0.5006 | -0.1107 | -0.3899 | -0.5006 | 0.0000 | -0.128294 |
| 0.1 | -0.5414 | -0.5414 | -0.3127 | -0.2288 | -0.5414 | 0.0000 | -0.086137 |
| 1 | -0.9916 | -0.9916 | -0.8589 | -0.1327 | -0.9916 | 0.0000 | -0.059918 |
| 10 | -2.6897 | -2.6897 | -2.5813 | -0.1084 | -2.6897 | 0.0000 | -0.052581 |
| 100 | -8.1377 | -8.1377 | -8.0359 | -0.1017 | -8.1377 | 0.0000 | -0.050455 |
| 1000 | -25.3878 | -25.3878 | -25.2881 | -0.0997 | -25.3878 | 0.0000 | -0.049803 |
| 10000 | -79.9442 | -79.9442 | -79.8451 | -0.0991 | -79.9442 | 0.0000 | -0.049605 |
| | | | | m=1.5 | | | |
| 0.0001 | -0.6000 | -0.6000 | -0.0126 | -0.5874 | -0.6000 | 0.0000 | -0.168189 |
| 0.001 | -0.6000 | -0.6000 | -0.0398 | -0.5602 | -0.6000 | 0.0000 | -0.166854 |
| 0.01 | -0.6005 | -0.6005 | -0.1240 | -0.4765 | -0.6005 | 0.0000 | -0.154850 |
| 0.1 | -0.6375 | -0.6375 | -0.3516 | -0.2859 | -0.6375 | 0.0000 | -0.105994 |
| 1 | -1.1114 | -1.1114 | -0.9518 | -0.1596 | -1.1114 | 0.0000 | -0.071490 |
| 10 | -2.9652 | -2.9652 | -2.8378 | -0.1274 | -2.9652 | 0.0000 | -0.061655 |
| 100 | -8.9314 | -8.9314 | -8.8128 | -0.1186 | -8.9314 | 0.0000 | -0.058820 |
| 1000 | -27.8274 | -27.8274 | -27.7114 | -0.1160 | -27.8274 | 0.0000 | -0.057948 |
| 10000 | -87.5908 | -87.5908 | -87.4756 | -0.1152 | -87.5908 | 0.0000 | -0.057773 |
| | | | | m=2 | | | |
| 0.0001 | -0.6667 | -0.6667 | -0.0138 | -0.6528 | -0.6667 | 0.0000 | -0.183855 |
| 0.001 | -0.6667 | -0.6667 | -0.0436 | -0.6231 | -0.6667 | 0.0000 | -0.182519 |
| 0.01 | -0.6671 | -0.6671 | -0.1357 | -0.5314 | -0.6671 | 0.0000 | -0.170363 |
| 0.1 | -0.7018 | -0.7018 | -0.3828 | -0.3191 | -0.7018 | 0.0000 | -0.117354 |
| 1 | -1.1885 | -1.1885 | -1.0165 | -0.1720 | -1.1885 | 0.0000 | -0.076839 |
| 10 | -3.1380 | -3.1380 | -3.0035 | -0.1344 | -3.1380 | 0.0000 | -0.065106 |
| 100 | -9.4257 | -9.4257 | -9.3013 | -0.1243 | -9.4257 | 0.0000 | -0.061727 |
| 1000 | -29.3435 | -29.3435 | -29.2221 | -0.1213 | -29.3435 | 0.0000 | -0.060689 |
| 10000 | -92.3395 | -92.3395 | -92.2192 | -0.1204 | -92.3395 | 0.0000 | -0.060368 |
| | | | | m=3 | | | |
| 0.0001 | -0.7500 | -0.7500 | -0.0160 | -0.7340 | -0.7500 | 0.0000 | -0.199277 |
| 0.001 | -0.7500 | -0.7500 | -0.0504 | -0.6997 | -0.7500 | 0.0000 | -0.197939 |
| 0.01 | -0.7504 | -0.7504 | -0.1562 | -0.5942 | -0.7504 | 0.0000 | -0.185627 |
| 0.1 | -0.7825 | -0.7825 | -0.4323 | -0.3502 | -0.7825 | 0.0000 | -0.127779 |
| 1 | -1.2823 | -1.2823 | -1.1049 | -0.1775 | -1.2823 | 0.0000 | -0.079392 |

| ω/Quantities | INT[1] | $\boldsymbol{w}$[2] | $\boldsymbol{w_{col}}$[3] | $\boldsymbol{w_c}$[4] | TEST1[5] | TEST2[6] | MC-HF corr[7] |
|---|---|---|---|---|---|---|---|
| 10 | -3.3440 | -3.3440 | -3.2099 | -0.1341 | -3.3440 | 0.0000 | -0.065147 |
| 100 | -10.0115 | -10.0114 | -9.8887 | -0.1227 | -10.0114 | 0.0000 | -0.061064 |
| 1000 | -31.1370 | -31.1369 | -31.0176 | -0.1193 | -31.1369 | 0.0000 | -0.059814 |
| 10000 | -97.9542 | -97.9542 | -97.8359 | -0.1183 | -97.9542 | 0.0000 | -0.059435 |
| m=10 | | | | | | | |
| 0.0001 | -0.9091 | -0.9091 | -0.0264 | -0.8826 | -0.9091 | 0.0000 | -0.197347 |
| 0.001 | -0.9091 | -0.9091 | -0.0832 | -0.8259 | -0.9091 | 0.0000 | -0.196007 |
| 0.01 | -0.9094 | -0.9094 | -0.2515 | -0.6579 | -0.9094 | 0.0000 | -0.183502 |
| 0.1 | -0.9374 | -0.9374 | -0.6124 | -0.3250 | -0.9374 | 0.0000 | -0.118794 |
| 1 | -1.4556 | -1.4556 | -1.3269 | -0.1287 | -1.4556 | 0.0000 | -0.059334 |
| 10 | -3.7125 | -3.7125 | -3.6258 | -0.0867 | -3.7125 | 0.0000 | -0.042938 |
| 100 | -11.0496 | -11.0496 | -10.9732 | -0.0765 | -11.0496 | 0.0000 | -0.038548 |
| 1000 | -34.3071 | -34.3071 | -34.2335 | -0.0735 | -34.3071 | 0.0000 | -0.037256 |
| 10000 | -107.8701 | -107.8701 | -107.7974 | -0.0726 | -107.8701 | 0.0000 | -0.036929 |
| m=50 | | | | | | | |
| 0.0001 | -0.9804 | -0.9804 | -0.0568 | -0.9236 | -0.9804 | 0.0000 | -0.136159 |
| 0.001 | -0.9804 | -0.9804 | -0.1758 | -0.8046 | -0.9804 | 0.0000 | -0.134821 |
| 0.01 | -0.9807 | -0.9807 | -0.4723 | -0.5084 | -0.9807 | 0.0000 | -0.122460 |
| 0.1 | -1.0072 | -1.0072 | -0.8454 | -0.1617 | -1.0072 | 0.0000 | -0.064093 |
| 1 | -1.5313 | -1.5313 | -1.4860 | -0.0454 | -1.5313 | 0.0000 | -0.022104 |
| 10 | -3.8689 | -3.8689 | -3.8420 | -0.0270 | -3.8689 | 0.0000 | -0.013563 |
| 100 | -11.4868 | -11.4868 | -11.4638 | -0.0229 | -11.4868 | 0.0000 | -0.011572 |
| 1000 | -35.6386 | -35.6386 | -35.6167 | -0.0218 | -35.6386 | 0.0000 | -0.011007 |
| 10000 | -112.0318 | -112.0318 | -112.0103 | -0.0215 | -112.0318 | 0.0000 | -0.010839 |
| m=207 | | | | | | | |
| 0.0001 | -0.9952 | -0.9952 | -0.1139 | -0.8813 | -0.9952 | 0.0000 | -0.082760 |
| 0.001 | -0.9952 | -0.9952 | -0.3327 | -0.6624 | -0.9952 | 0.0000 | -0.081427 |
| 0.01 | -0.9954 | -0.9954 | -0.7068 | -0.2886 | -0.9954 | 0.0000 | -0.069560 |
| 0.1 | -1.0216 | -1.0216 | -0.9613 | -0.0603 | -1.0216 | 0.0000 | -0.026584 |
| 1 | -1.5469 | -1.5469 | -1.5232 | -0.0237 | -1.5469 | 0.0000 | -0.006733 |
| 10 | -3.9008 | -3.9008 | -3.8932 | -0.0076 | -3.9008 | 0.0000 | -0.003642 |
| 100 | -11.5756 | -11.5756 | -11.5692 | -0.0064 | -11.5756 | 0.0000 | -0.002977 |
| 1000 | -35.9089 | -35.9089 | -35.9029 | -0.0060 | -35.9089 | 0.0000 | -0.002793 |
| 10000 | -112.8766 | -112.8766 | -112.8707 | -0.0059 | -112.8766 | 0.0000 | -0.002747 |
| m=400 | | | | | | | |
| 0.0001 | -0.9975 | -0.9975 | -0.1567 | -0.8408 | -0.9975 | 0.0000 | -0.063705 |
| 0.001 | -0.9975 | -0.9975 | -0.4326 | -0.5649 | -0.9975 | 0.0000 | -0.062375 |
| 0.01 | -0.9978 | -0.9978 | -0.7984 | -0.1994 | -0.9978 | 0.0000 | -0.050894 |

| ω/Quantities | INT[1] | $\boldsymbol{w}$ [2] | $\boldsymbol{w_{col}}$[3] | $\boldsymbol{w_c}$[4] | TEST1[5] | TEST2[6] | MC-HF corr[7] |
|---|---|---|---|---|---|---|---|
| 0.1 | -1.0239 | -1.0239 | -0.9886 | -0.0354 | -1.0239 | 0.0000 | -0.016254 |
| 1 | -1.5494 | -1.5494 | -1.4583 | -0.0910 | -1.5494 | 0.0000 | -0.003624 |
| 10 | -3.9058 | -3.9058 | -3.9017 | -0.0041 | -3.9058 | 0.0000 | -0.001804 |
| 100 | -11.5894 | -11.5894 | -11.5860 | -0.0034 | -11.5894 | 0.0000 | -0.001424 |
| 1000 | -35.9510 | -35.9510 | -35.9478 | -0.0032 | -35.9510 | 0.0000 | -0.001317 |
| 10000 | -113.0081 | -113.0081 | -113.0049 | -0.0032 | -113.0081 | 0.0000 | -0.001443 |
| m=900 | | | | | | | |
| 0.0001 | -0.9989 | -0.9989 | -0.2297 | -0.7692 | -0.9989 | 0.0000 | -0.045320 |
| 0.001 | -0.9989 | -0.9989 | -0.5693 | -0.4295 | -0.9989 | 0.0000 | -0.043998 |
| 0.01 | -0.9992 | -0.9992 | -0.8826 | -0.1166 | -0.9992 | 0.0000 | -0.033180 |
| 0.1 | -1.0253 | -1.0252 | -1.0077 | -0.0175 | -1.0252 | 0.0000 | -0.008372 |
| 1 | -1.5508 | -1.5508 | -1.1976 | -0.3532 | -1.5508 | 0.0000 | -0.001579 |
| 10 | -3.9088 | -3.9088 | -3.9069 | -0.0019 | -3.9088 | 0.0000 | -0.000638 |
| 100 | -11.5977 | -11.5977 | -11.5961 | -0.0016 | -11.5977 | 0.0000 | -0.000448 |
| 1000 | -35.9761 | -35.9761 | -35.9747 | -0.0015 | -35.9761 | 0.0000 | -0.000391 |
| 10000 | -113.0867 | -113.0867 | -113.0852 | -0.0014 | -113.0867 | 0.0000 | -0.000512 |
| m=1836 | | | | | | | |
| 0.0001 | -0.9995 | -0.9995 | -0.3156 | -0.6838 | -0.9995 | 0.0000 | -0.033161 |
| 0.001 | -0.9994 | -0.9993 | -0.6875 | -0.3119 | -0.9994 | -0.0001 | -0.031848 |
| 0.01 | -0.9998 | -0.9998 | -0.9315 | -0.0683 | -0.9997 | 0.0000 | -0.021825 |
| 0.1 | -1.0258 | -1.0258 | -1.0166 | -0.0092 | -1.0258 | 0.0000 | -0.004486 |
| 1 | -1.5514 | -1.5514 | -1.5497 | -0.0018 | -1.5514 | 0.0000 | -0.000684 |
| 10 | -3.9100 | -3.9100 | -3.9090 | -0.0009 | -3.9100 | 0.0000 | -0.000140 |
| 100 | -11.6011 | -11.6011 | -11.6003 | -0.0008 | -11.6011 | 0.0000 | -0.000035 |
| 1000 | -35.9864 | -35.9864 | -35.9857 | -0.0007 | -35.9864 | 0.0000 | 0.000000 |
| 10000 | -113.1188 | -113.1188 | -113.1180 | -0.0007 | -113.1188 | 0.0000 | -0.000132 |

1. E-PCP interaction energy calculated based on wave function (equations 2-139 and 2-152)
2. E-PCP interaction energy calculated using two-particle density (equation 3-20)
3. Hartree potential energy or coulomb contribution to interaction energy (equation 3-21)
4. Reference-independent correlation energy (equation 3-24)
5. First test: sum of items 3 and 4 to ensure reproduction of item 2 according to equation (3-21)
6. Second test: difference between items 1 and 2
7. Correlation energy relative to MC-HF reference

### 5.1.2 Single-Particle Densities

The electron and PCP densities were calculated using the variational method (with equations 2-168 and 2-169) and the MC-HF method, as shown in Figures 5-3 to 5-14. These figures are grouped into three mass categories: Category 1 includes masses 1 to 3, Category 2 includes masses 10 and 50, and Category 3 includes masses 207 to 1836. This categorization is done to account for the different localization extents of various masses, ensuring image clarity and better comparison.

It is important to note that the densities given by equations (2-168) and (2-169) are not defined for $r = 0$. However, comparing the densities obtained from the numerical calculation of equation (2-161), which includes $r = 0$, shows that the densities (2-168) and (2-169) produce very accurate results at all points except $r = 0$. Figures (C-1) and (C-2) in the appendix C show the percentage error of the densities with form (2-168) compared to form (2-161).

There is a notable difference between the variational density and the HF density (illustrated in Figures C-3 to C-7 in the appendix C) at a frequency of $10^{-4}$ (especially at mass 1). This difference is not unexpected since, as shown in Table 4-3, the MC-HF method exhibits significant energy errors in this range, hence the produced density is expected to have substantial errors. A clear pattern observed in all densities is the increased localization with increasing frequency. This means that at a given mass, as the frequency increases, the potential well narrows, forcing the particles to be confined in a smaller space, which leads to increased localization of the densities with increasing frequency. Additionally, at a given frequency, increased mass of the PCP also results in more localized densities. At a frequency of $10^4$, as the mass of the PCP increases, the electron density remains constant, but the PCP density becomes more localized. This trend is consistent with the fact that heavier particles are more localized.

In general, the MC-HF densities, particularly for the PCP, clearly show over-localization. This is especially evident in the system with mass 1 and frequency $10^{-4}$. In this system, the masses of the positive and negative particles are equal, and the harmonic oscillator potential approaches zero. The variational densities demonstrate that the single-particle density in this system is significantly delocalized, nearing zero at $r_e = 200$. Meanwhile, the MC-HF single-particle density approaches zero at approximately $r_e = 10$. This over-localization is precisely the effect observed in real molecule calculations multiple times, indicating that the EHM can reproduce it accurately.

At low frequencies, the Coulomb interaction between the electron and the PCP in equation (2-16) dominates the harmonic oscillator potential, and the two particles are completely bound to each other. However, they oscillate as a single particle with mass M, according to equation (2-15). This means that at low frequencies, the lighter particle is attached to the heavier one (closer to the center of mass), and these two particles oscillate together as a total mass M. Naturally, the heavier the PCP, the smaller the oscillation amplitude of this mass M. Therefore, at mass 1, the oscillation amplitude is very large because both particles have equal and light masses, allowing either to induce oscillations in the other.

At high frequencies, the situation in equation (2-16) is reversed, and the harmonic oscillator potential now dominates the Coulomb potential. Thus, the two particles are not bound to each other. In this case, the two particles form two separate oscillators.

Finally, the normalization condition for the densities is defined as follows:

$$4\pi \int_0^\infty \rho_e(r_e)\, r_e^2\, dr_e = 1 \qquad 5.1$$

$$4\pi \int_0^\infty \rho_p(r_p)\, r_p^2\, dr_p = 1 \qquad 5.2$$

This condition was verified for the electron and PCP densities calculated at the variational, FEM, and HF levels, and all of them satisfy the above equation.

The limit of single-particle densities as $m_p \to \infty$ was examined due to the system's approach to the clamped nuclei approximation. For the EHM with $m_p \to \infty$, the electron density in the limit $\omega \to 0$ approaches the electron density of a hydrogen atom with a clamped nucleus. In the limit $\omega \to 0$, the strength of the harmonic oscillator potential and thus the perturbation caused by it becomes very small, allowing it to be approximated by a free hydrogen atom. Since $m_p \to \infty$, the electron experiences conditions identical to those of a hydrogen atom with a clamped nucleus (considering the PCP is clamped, we can assume $r_p = 0$ and therefore $r = |r_e - r_p| = r_e$), resulting in the electron density (in the limit of equation 2-168) being equal to the electron density of a hydrogen atom with a clamped nucleus:

$$\lim_{m_p \to \infty,\ \omega \to 0} \rho_e(r_e) \to \frac{e^{-2r_e}}{\pi} = \rho_H(r) \qquad 5.3$$

In this case, the density of the PCP approaches a Dirac delta function. Additionally, for sufficiently large frequencies, since the mass of the PCP is assumed to be infinite, its kinetic energy term in the Hamiltonian approaches zero. Given that in this limit, $r_p = 0$, the oscillatory term of the PCP is eliminated, and the e-PCP interaction term can be neglected. Consequently, only the electron's kinetic and oscillatory energy terms remain. Therefore, in the limit $m_p \to \infty$ and sufficiently large frequencies, the electron density approaches a Gaussian function (the ground state of the oscillator):

$$\lim_{\substack{m_p \to \infty \\ \text{large } \omega}} \rho_e(r_e) \to \left(\frac{\omega}{\pi}\right)^{3/2} e^{-\omega r_e^2} = \rho_{HO}(r) \qquad 5.4$$

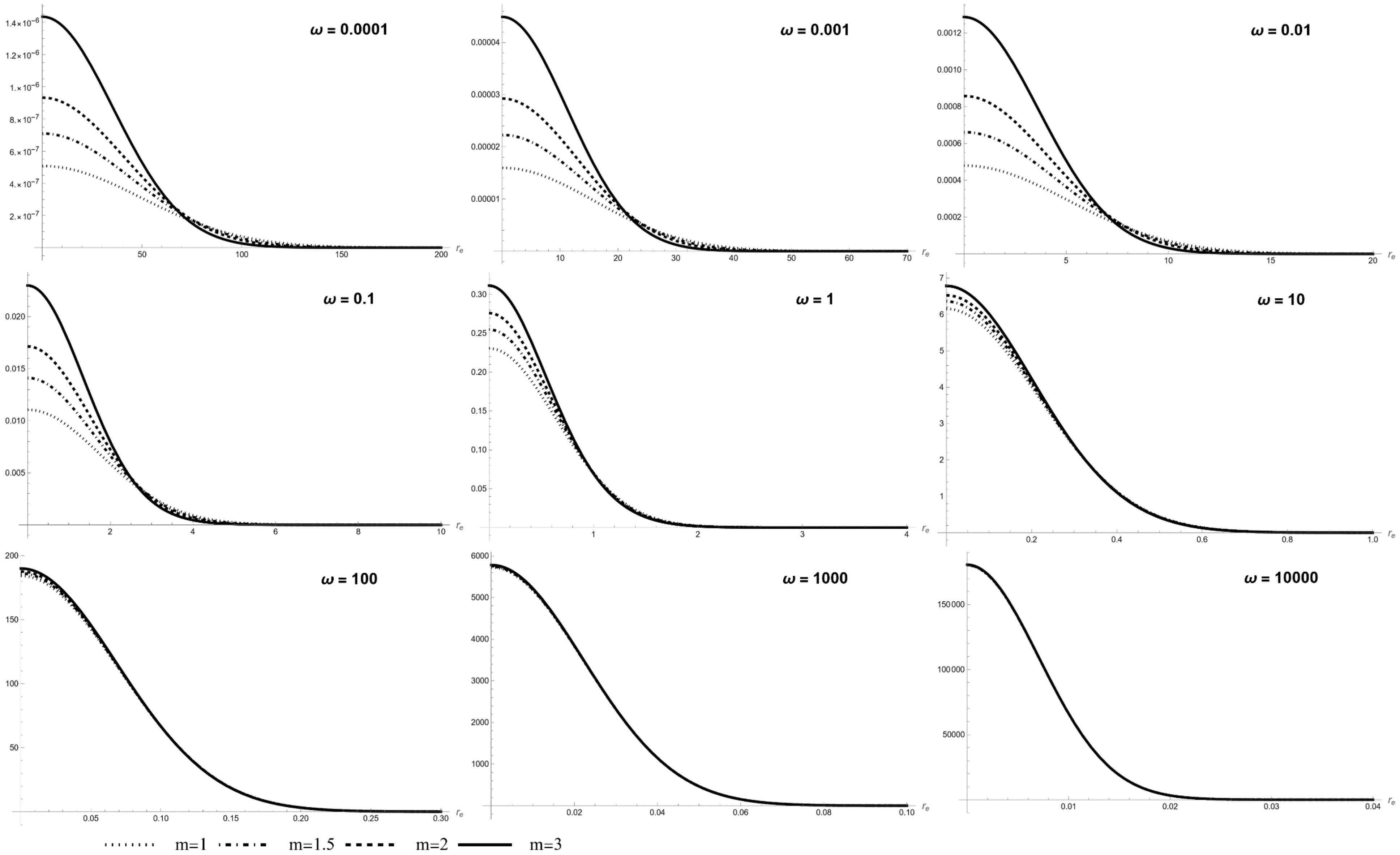

Figure 5-3: Variational electron density as a function of $r_e$ for different frequencies across the four lower masses.

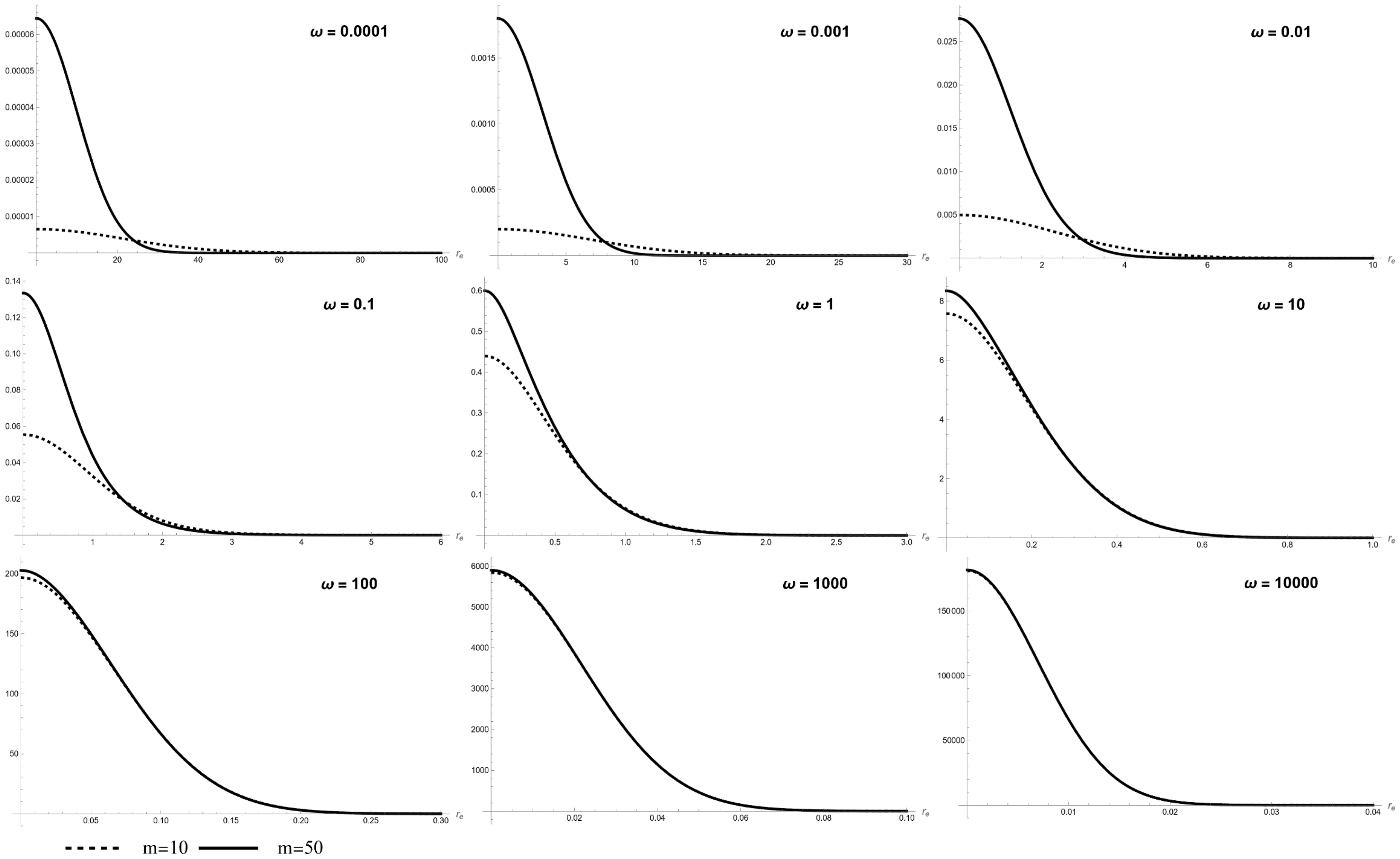

Figure 5-4: Variational electron density as a function of $r_e$ for different frequencies across the two middle masses.

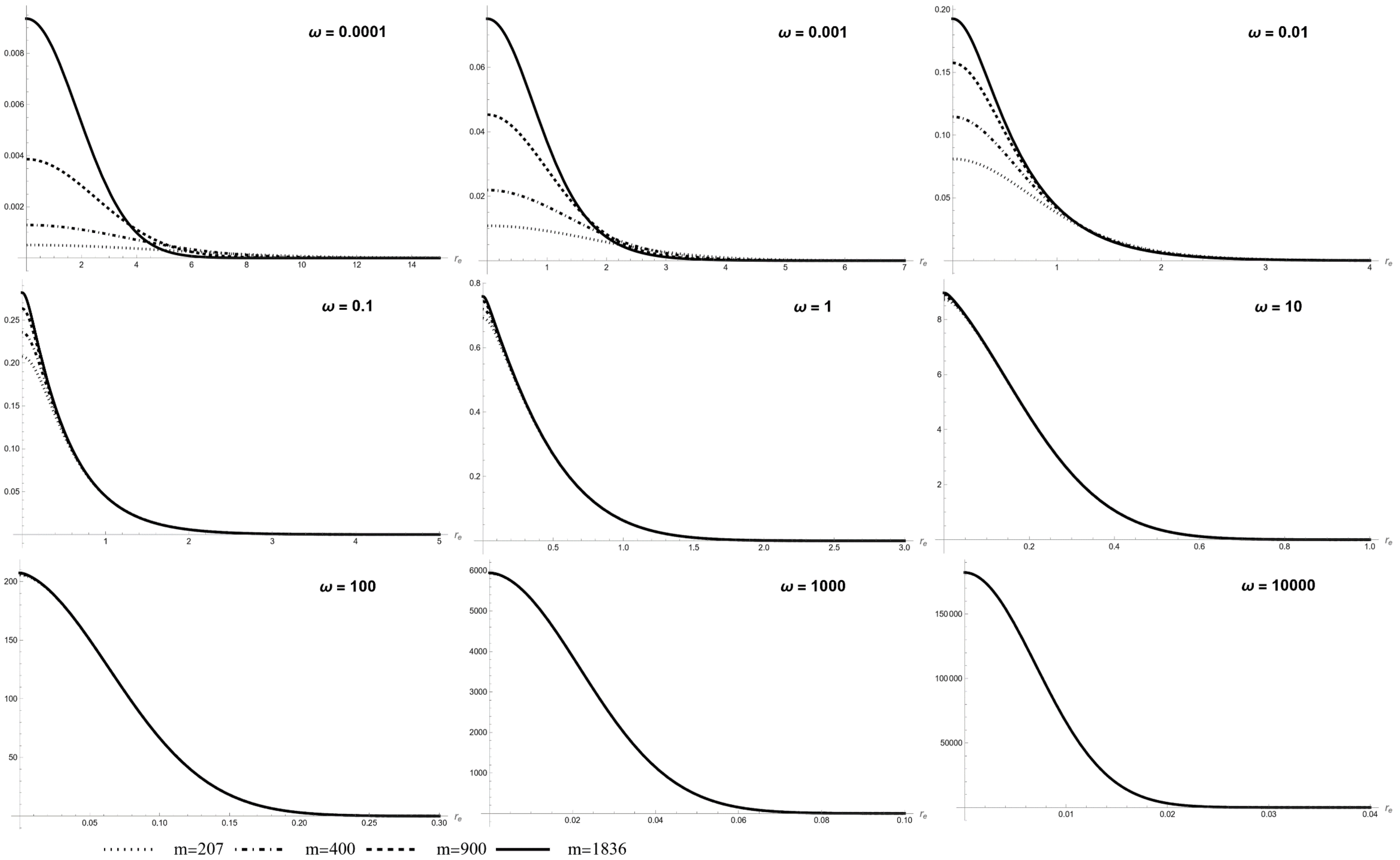

Figure 5-5: Variational electron density as a function of $r_e$ for different frequencies across the four higher masses.

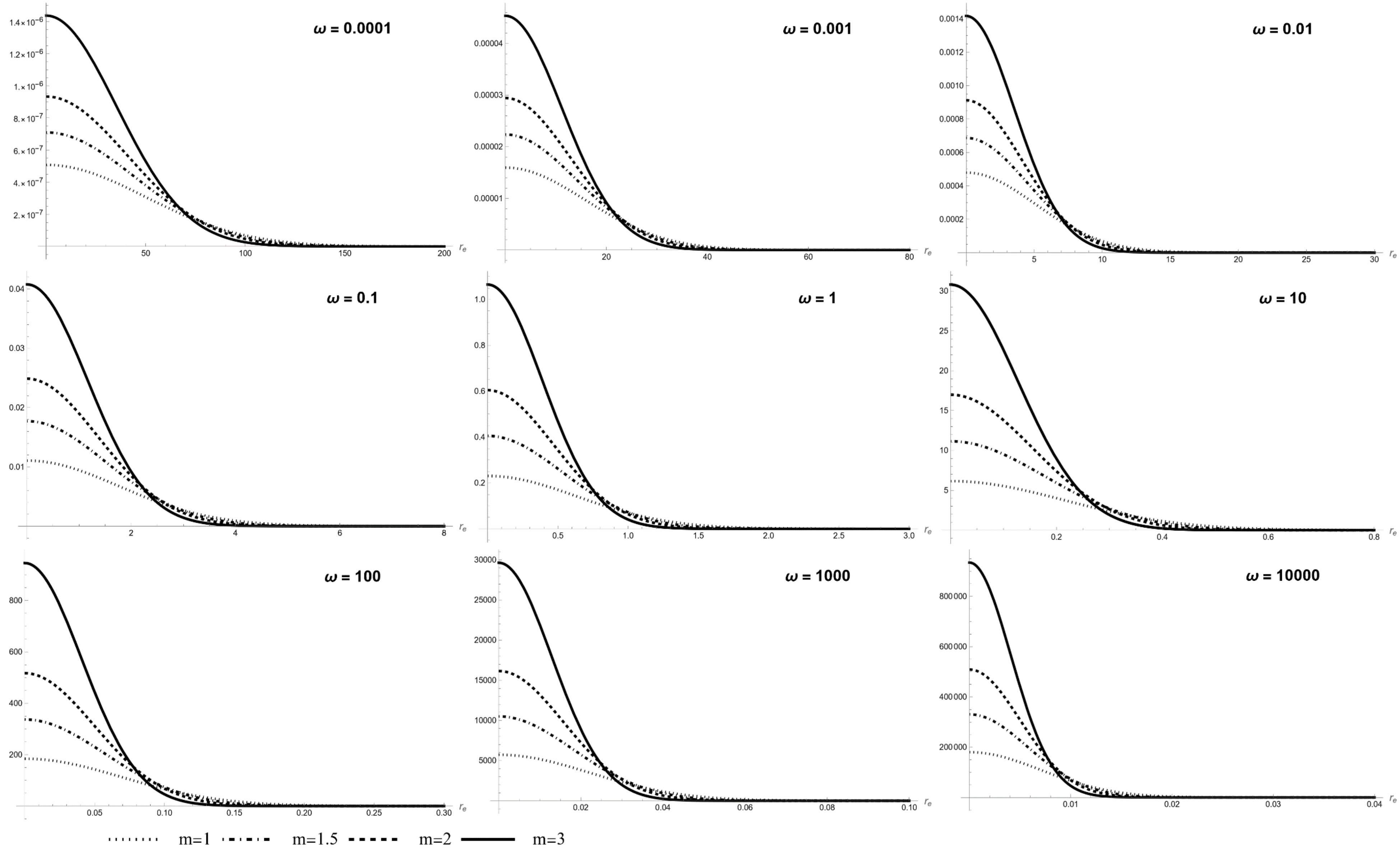

Figure 5-6: Variational PCP density as a function of $r_p$ for different frequencies across the four lower masses.

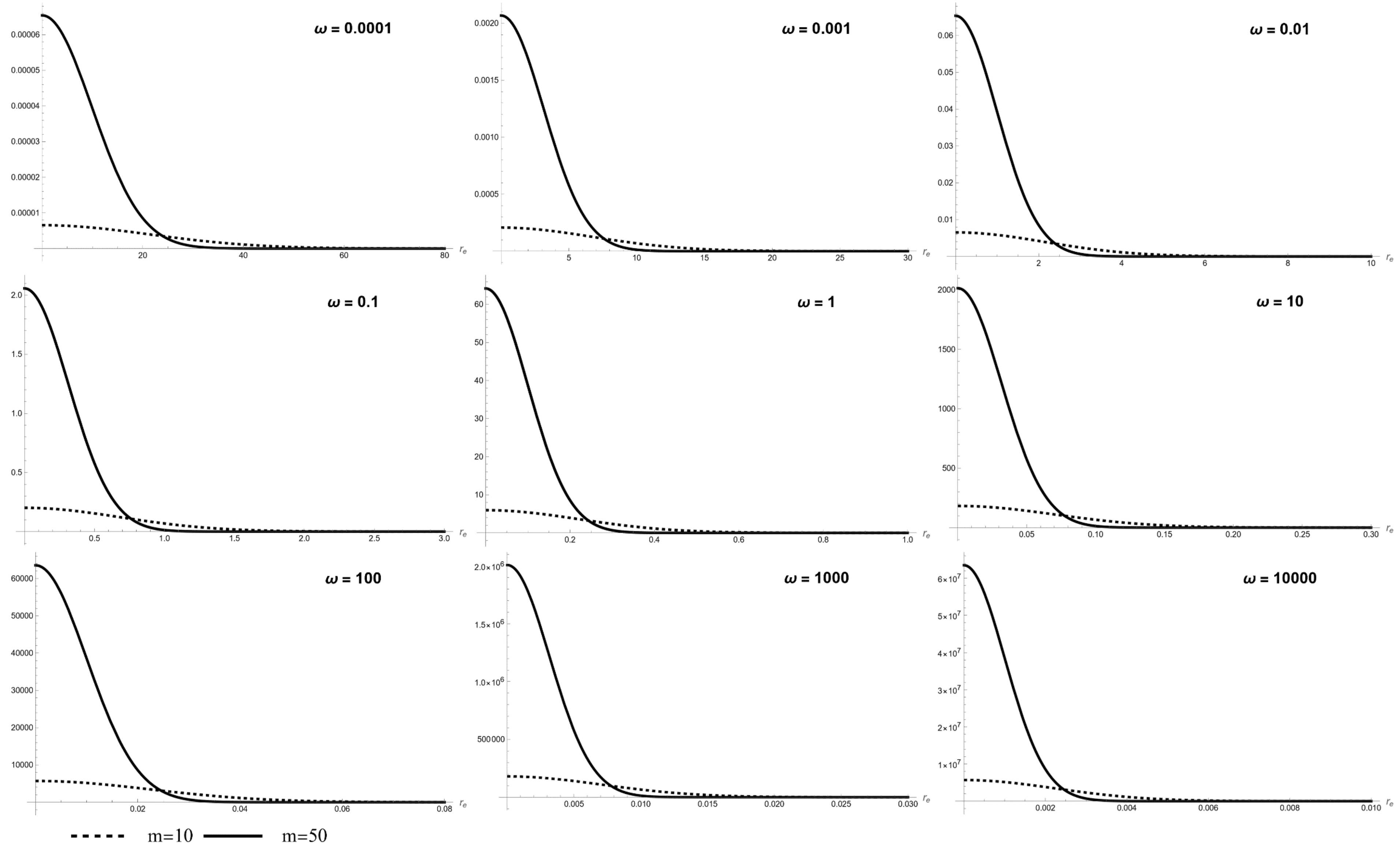

Figure 5-7: Variational PCP density as a function of $r_p$ for different frequencies across the two middle masses.

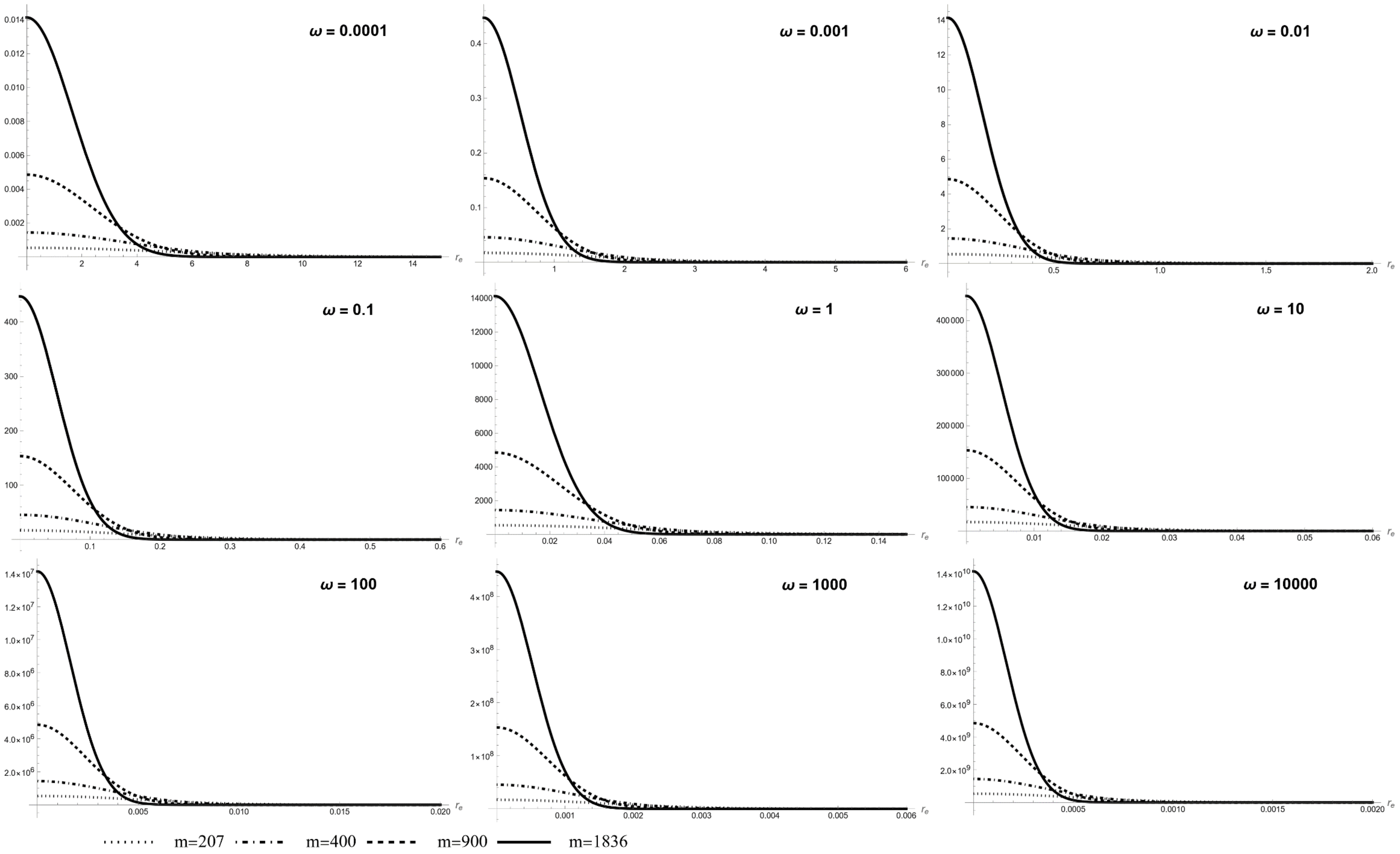

Figure 5-8: Variational PCP density as a function of $r_p$ for different frequencies across the four higher masses.

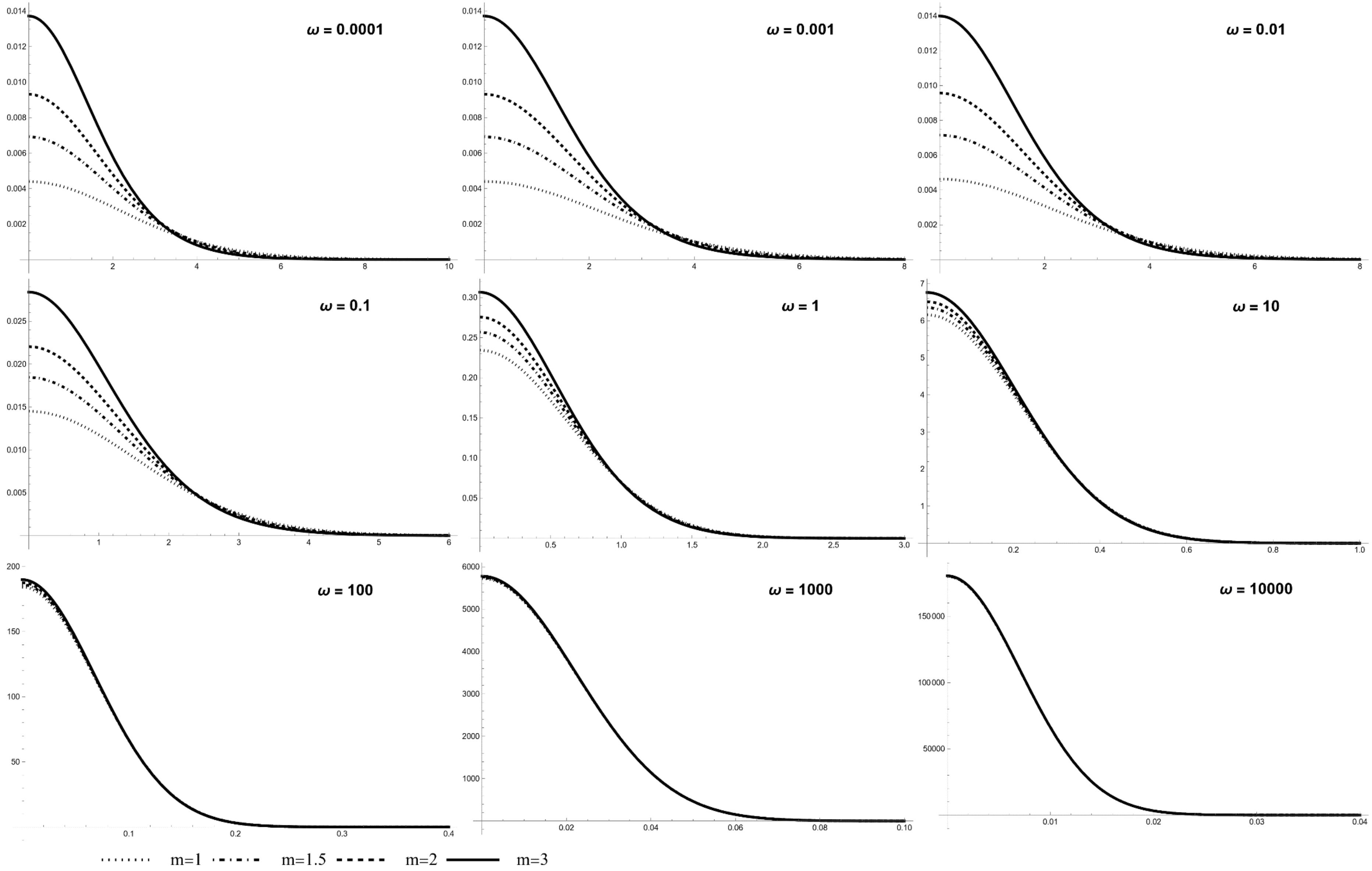

Figure 5-9: TC-HF electron density as a function of $r_e$ for different frequencies across the four lower masses.

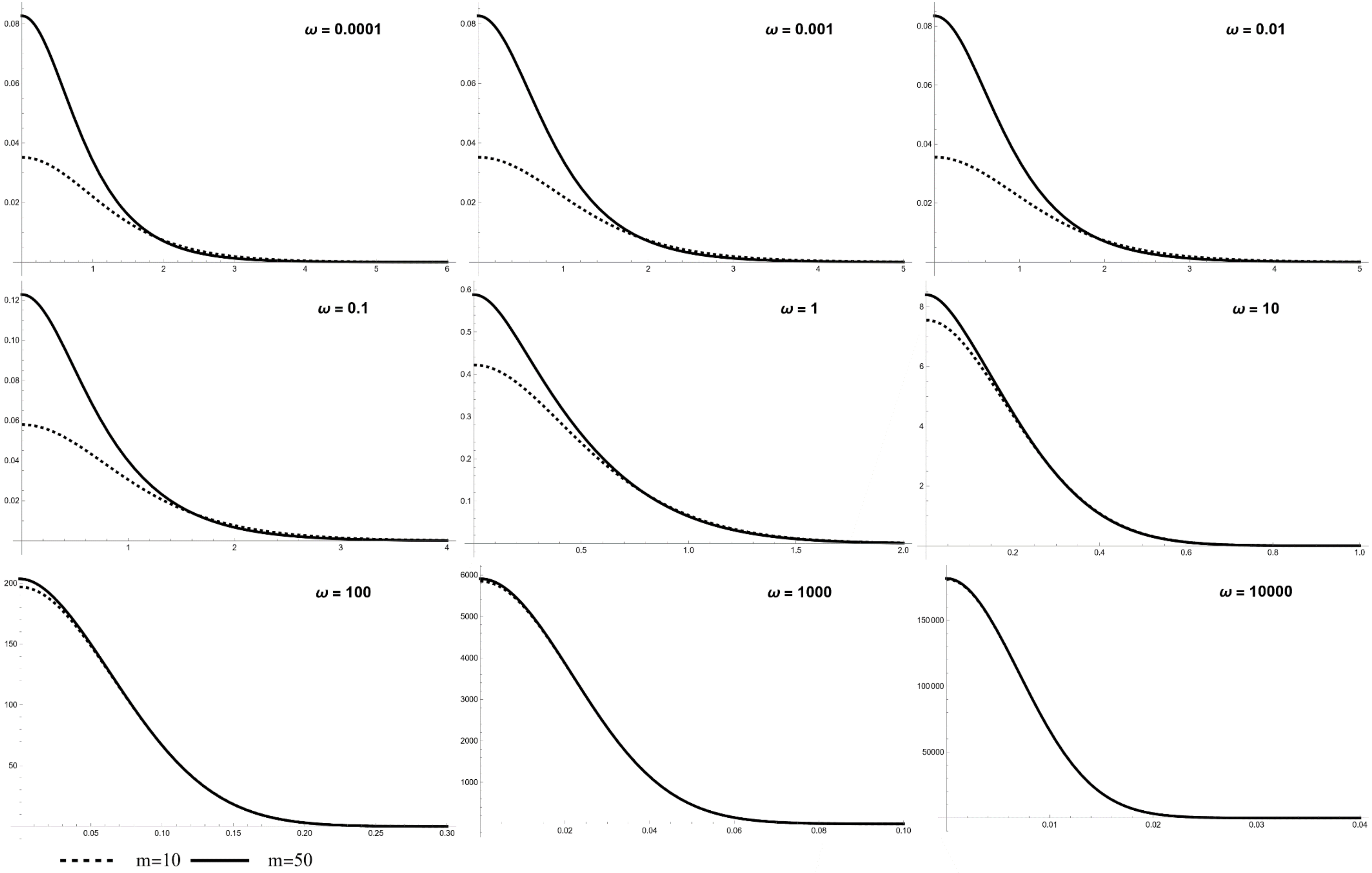

Figure 5-10: TC-HF electron density as a function of $r_s$ for different frequencies across the two middle masses.

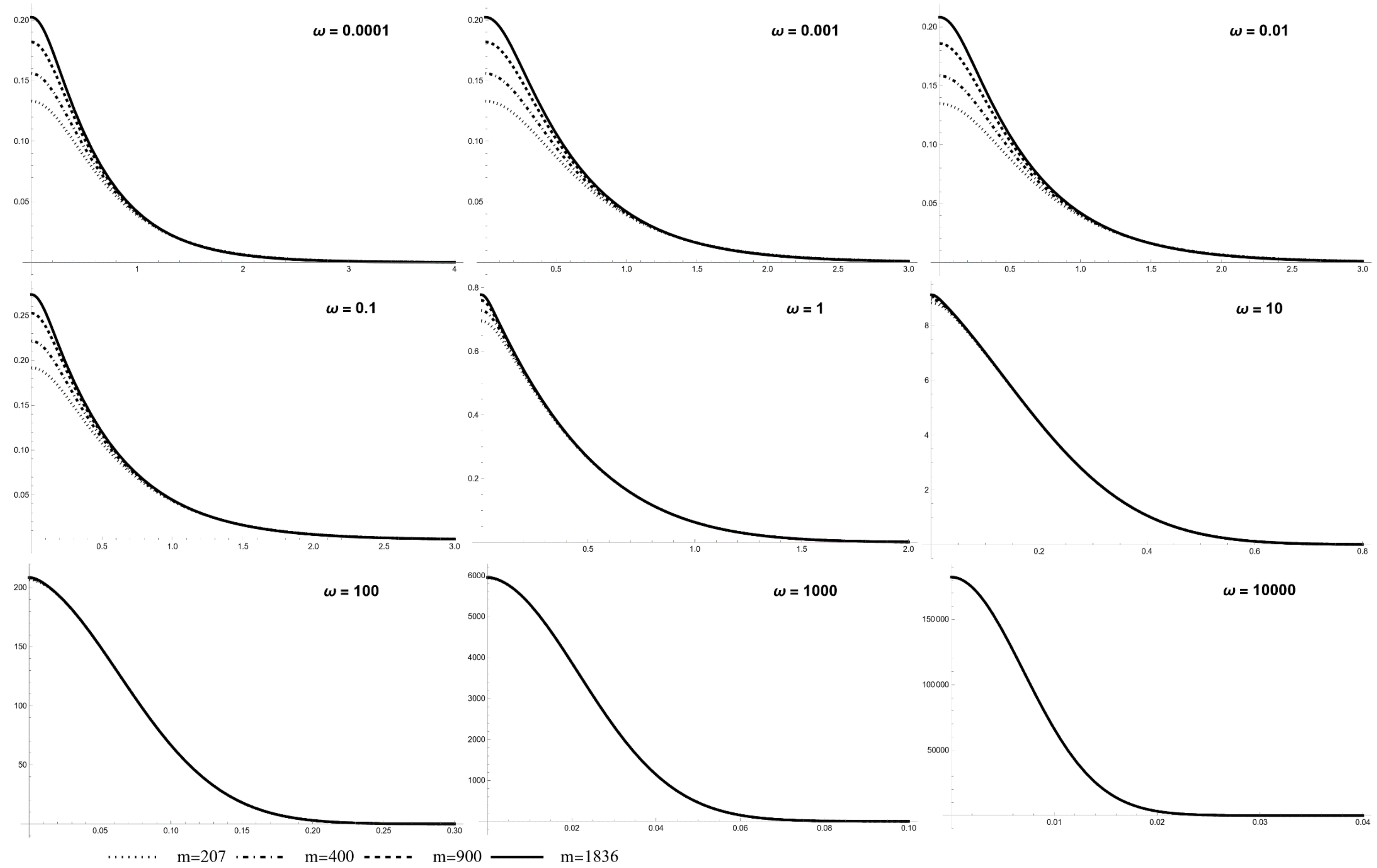

Figure 5-11: TC-HF electron density as a function of $r_e$ for different frequencies across the four higher masses.

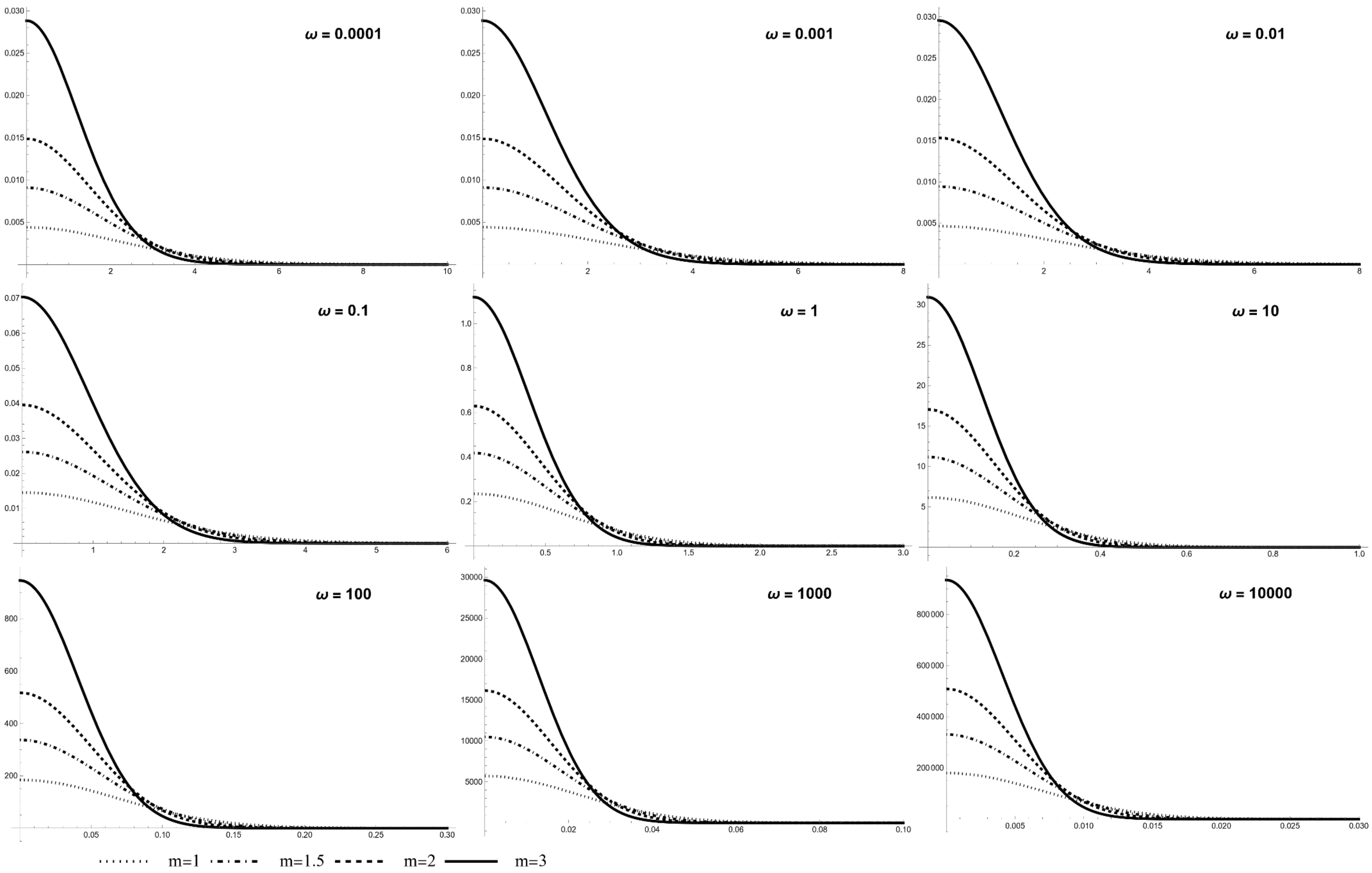

Figure 5-12: TC-HF PCP density as a function of $r_p$ for different frequencies across the four lower masses.

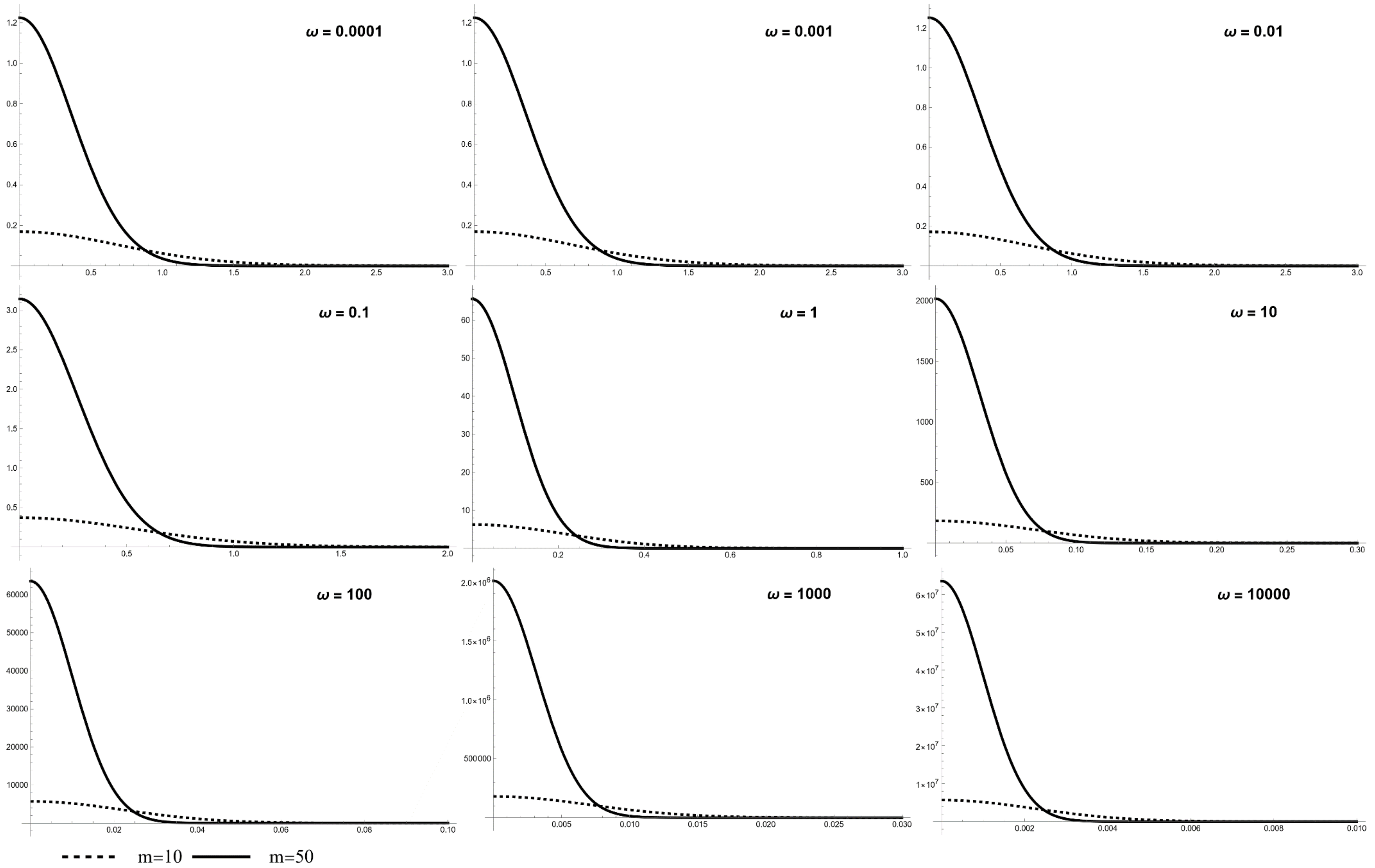

Figure 5-13: TC-HF PCP density as a function of $r_p$ for different frequencies across the two middle masses.

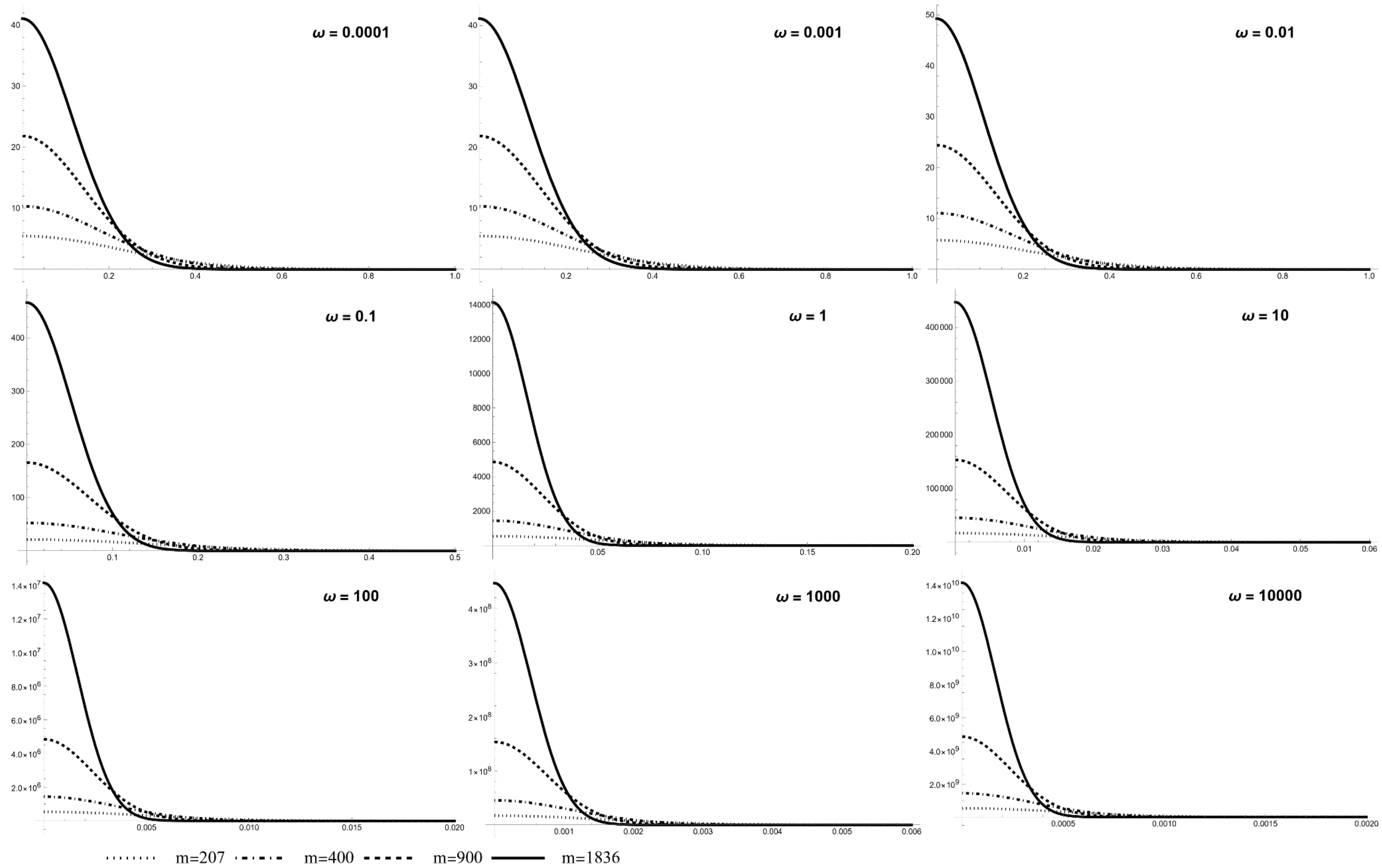

Figure 5-14: TC-HF PCP density as a function of $r_p$ for different frequencies across the four higher masses.

### 5.1.3 Correlation Hill

The correlation hills, obtained using the variational equations (3-12 and 3-13), are presented for different masses for the electron and the PCP in Figures 5-15 to 5-34. In all these figures, to simplify the computational process, the reference particle is placed at the center of the potential well (i.e., $r_e = 0$ and $r_p = 0$ for the reference electron and PCP, respectively). In this configuration, due to the isotropy of the harmonic oscillator potential, the angular dependence is eliminated, resulting in $r_p = r$ and $r_e = r$ for the reference electron and PCP, respectively.

Since, according to the aforementioned equations, the correlation hill density is derived from the difference between the conditional density and the single-particle density, both quantities are included in the correlation hill figures for better comparison. Additionally, an inset is plotted within each main graph to better compare the positive and negative parts of the graph.

The term "correlation hill" is aptly named in contrast to the "correlation hole" observed in electron-electron correlation within electronic systems. Due to the repulsive nature of the interaction, electron-electron correlation creates a hole around each electron (reference particle) and reduces the probability of another electron being nearby. Conversely, e-PCP correlation, due to the attractive nature of the interaction, creates a hill around each reference particle, increasing the probability of another particle being present. Accordingly, the correlation hill plots show high density near the reference particle. As one moves away from the reference particle, this density reaches zero and then turns negative, which is the exact opposite behavior of the electron correlation hole.

However, similar to the correlation hole, the sum rule of the correlation hill (equations 3-18 and 3-19) indicates that the sum of the positive and negative parts should be zero. Integration over the correlation hills of the electron and PCP was performed for all systems, resulting in a net value of zero, confirming the sum rule for all correlation hills.

The correlation hill figures show that as $\omega$ increases for a fixed mass, the zero-crossing boundary of the correlation hill moves to smaller distances closer to the reference particle. This distance is essentially the "effective radius" of the correlation, beyond which correlation effects do not operate, thus serving as a good measure of the correlation range in real space. This effective radius is highly sensitive to the mass of the PCP and decreases with increasing mass.

For a more detailed examination, the numerical values of the correlation radius for different masses and frequencies were calculated and presented in Table (C-1) of the appendix C. This radius indicates the distance at which the correlation hill intersects the horizontal axis and changes sign. Analysis of these values reveals that, in general, the effective radius of the PCP is smaller than that of the electron. Additionally, for both particles, increasing frequency or mass (while keeping the other parameter constant) results in a decrease in the effective radius.

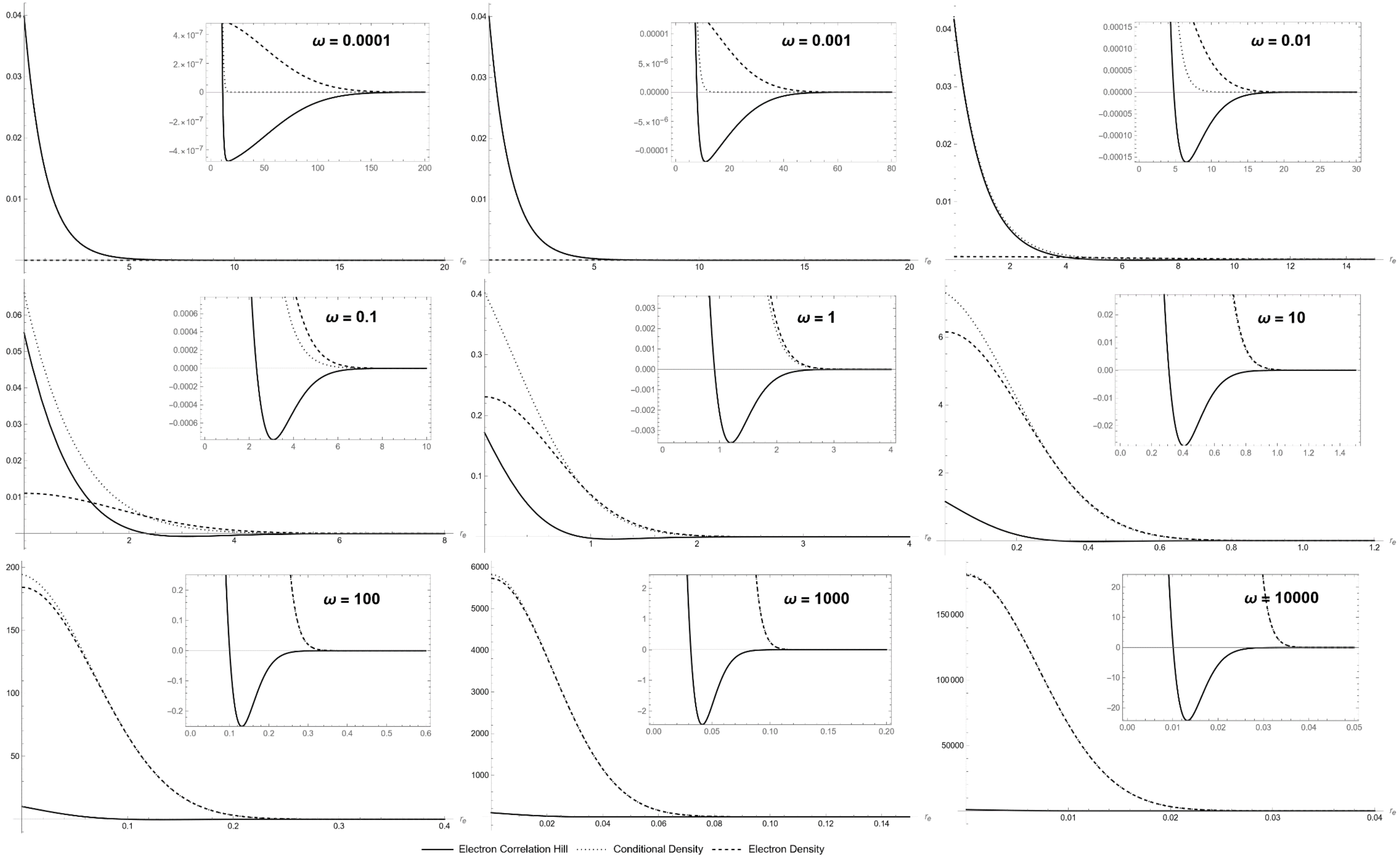

Figure 5-15: Comparison of the correlation hill, conditional density, and single-particle density of electron for mass 1.

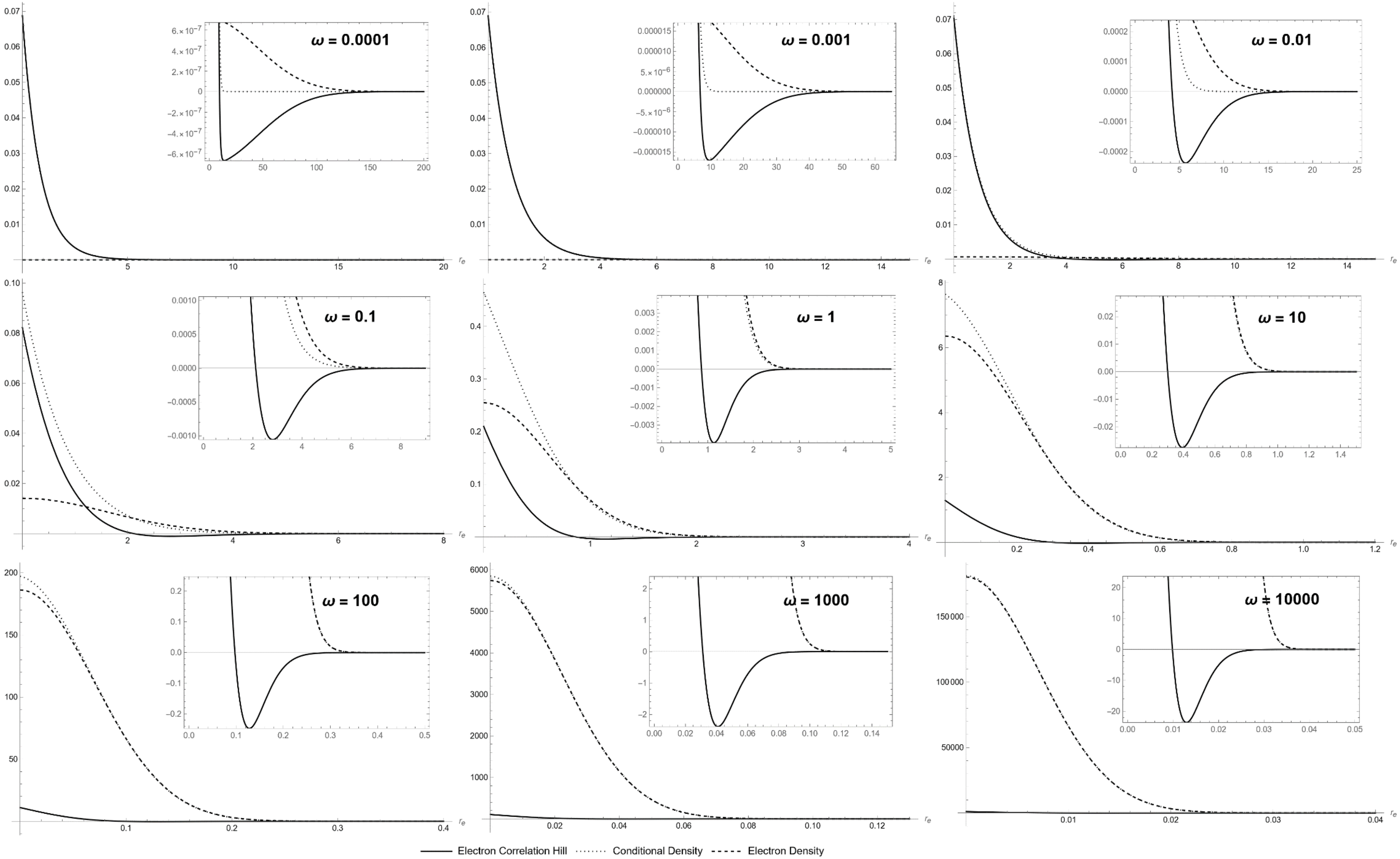

Figure 5-16: Comparison of the correlation hill, conditional density, and single-particle density of electron for mass 1.5

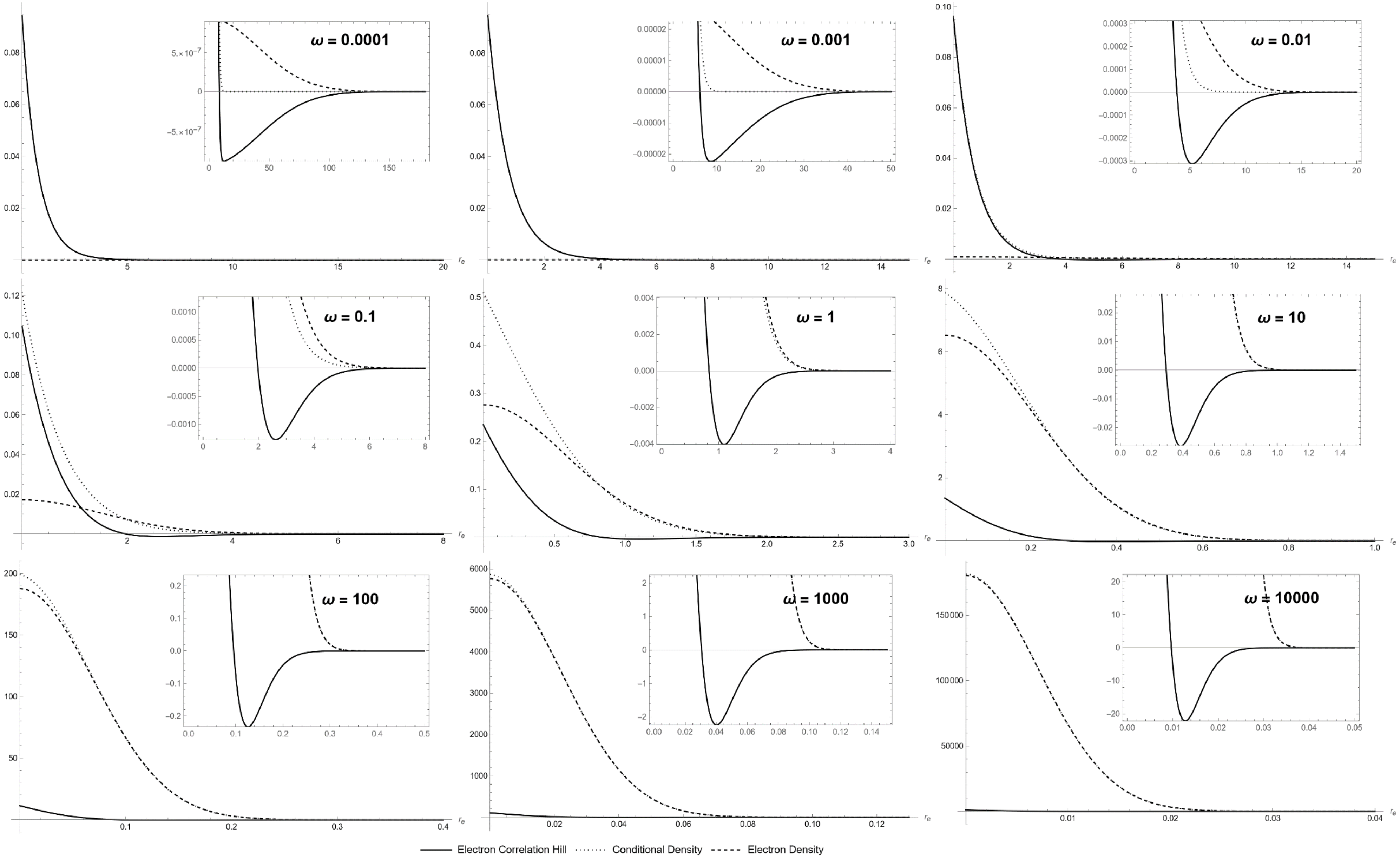

Figure 5-17: Comparison of the correlation hill, conditional density, and single-particle density of electron for mass 2

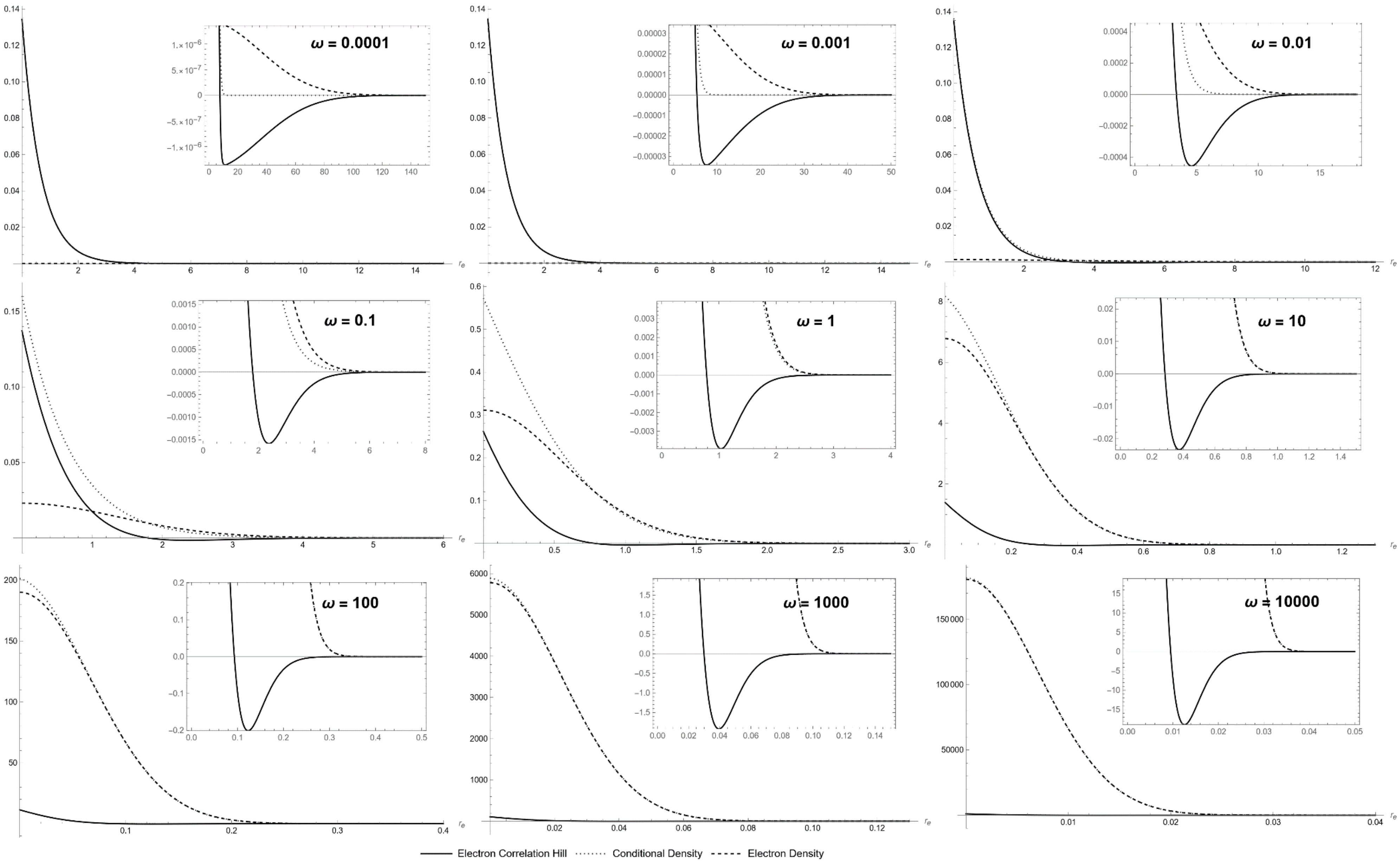

Figure 5-18: Comparison of the correlation hill, conditional density, and single-particle density of electron for mass 3

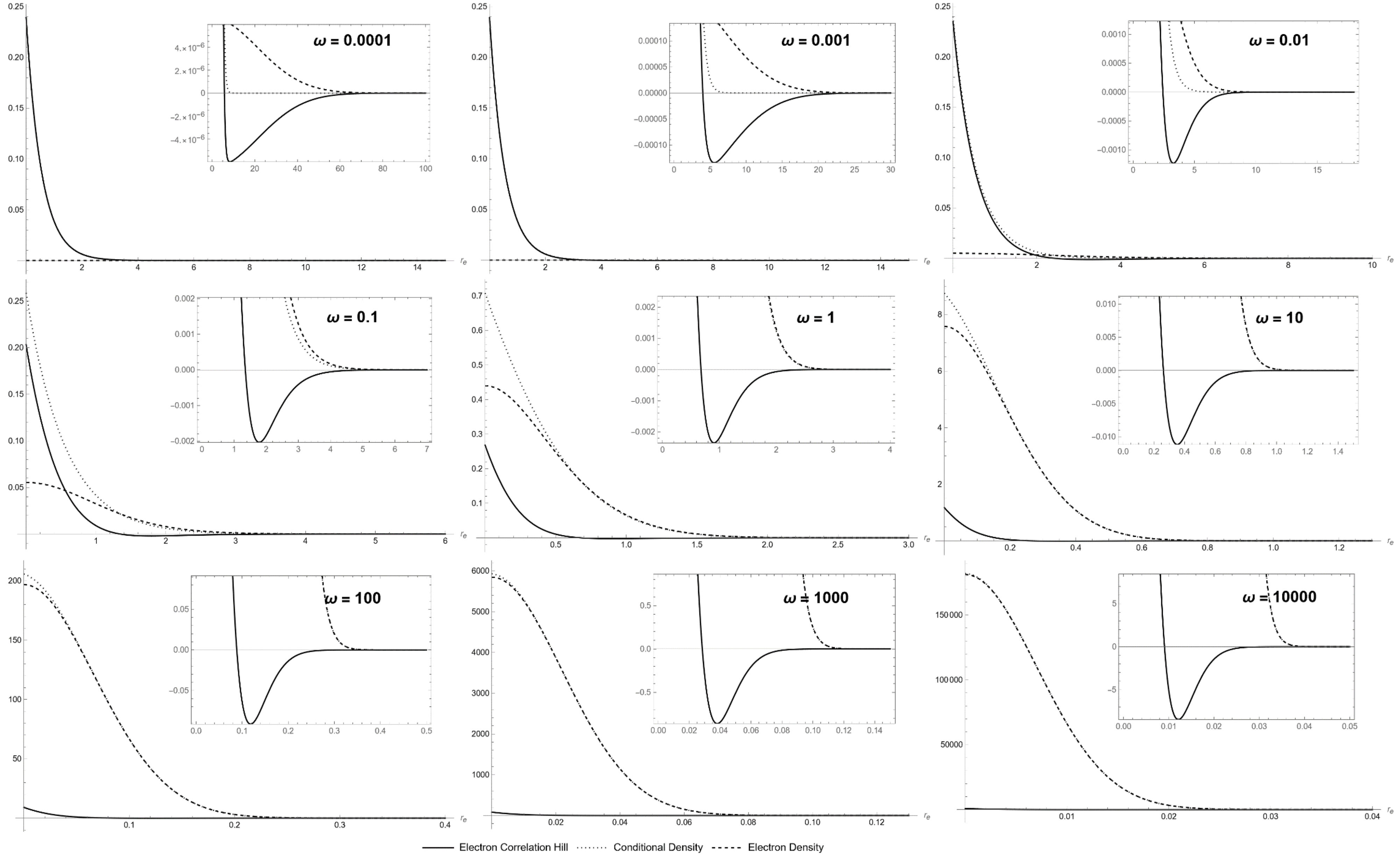

Figure 5-19: Comparison of the correlation hill, conditional density, and single-particle density of electron for mass 10

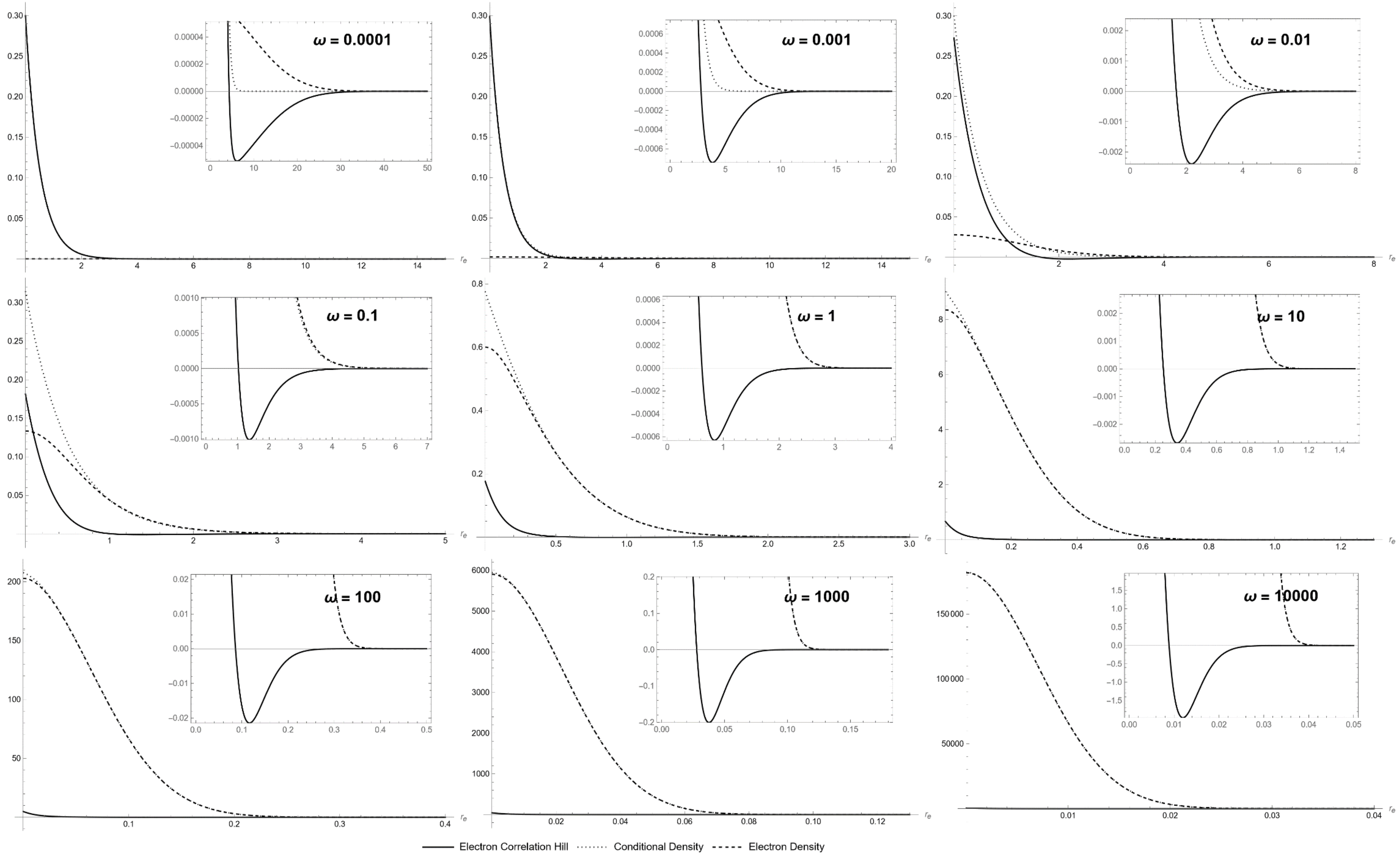

Figure 5-20: Comparison of the correlation hill, conditional density, and single-particle density of electron for mass 50

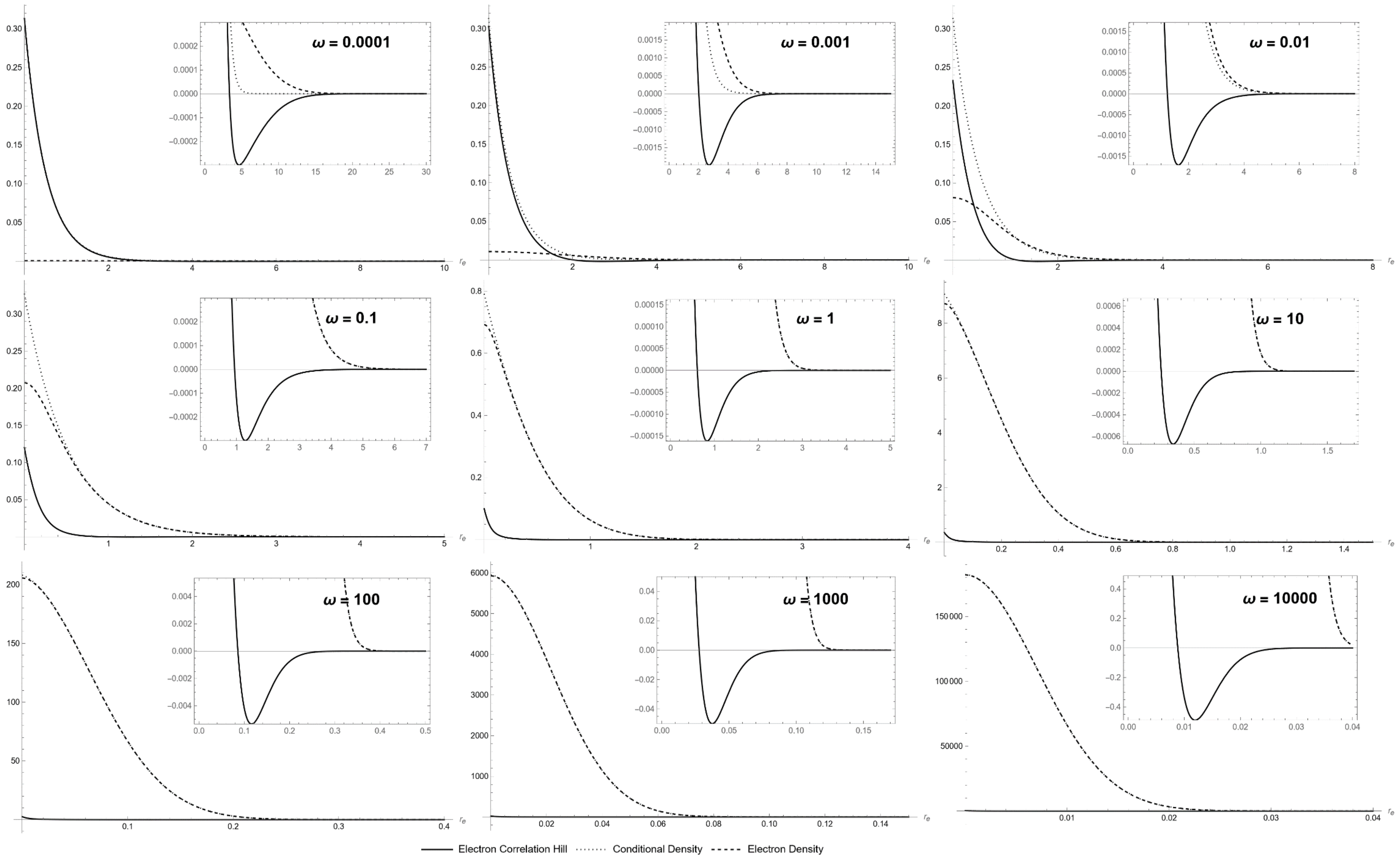

Figure 5-21: Comparison of the correlation hill, conditional density, and single-particle density of electron for mass 207

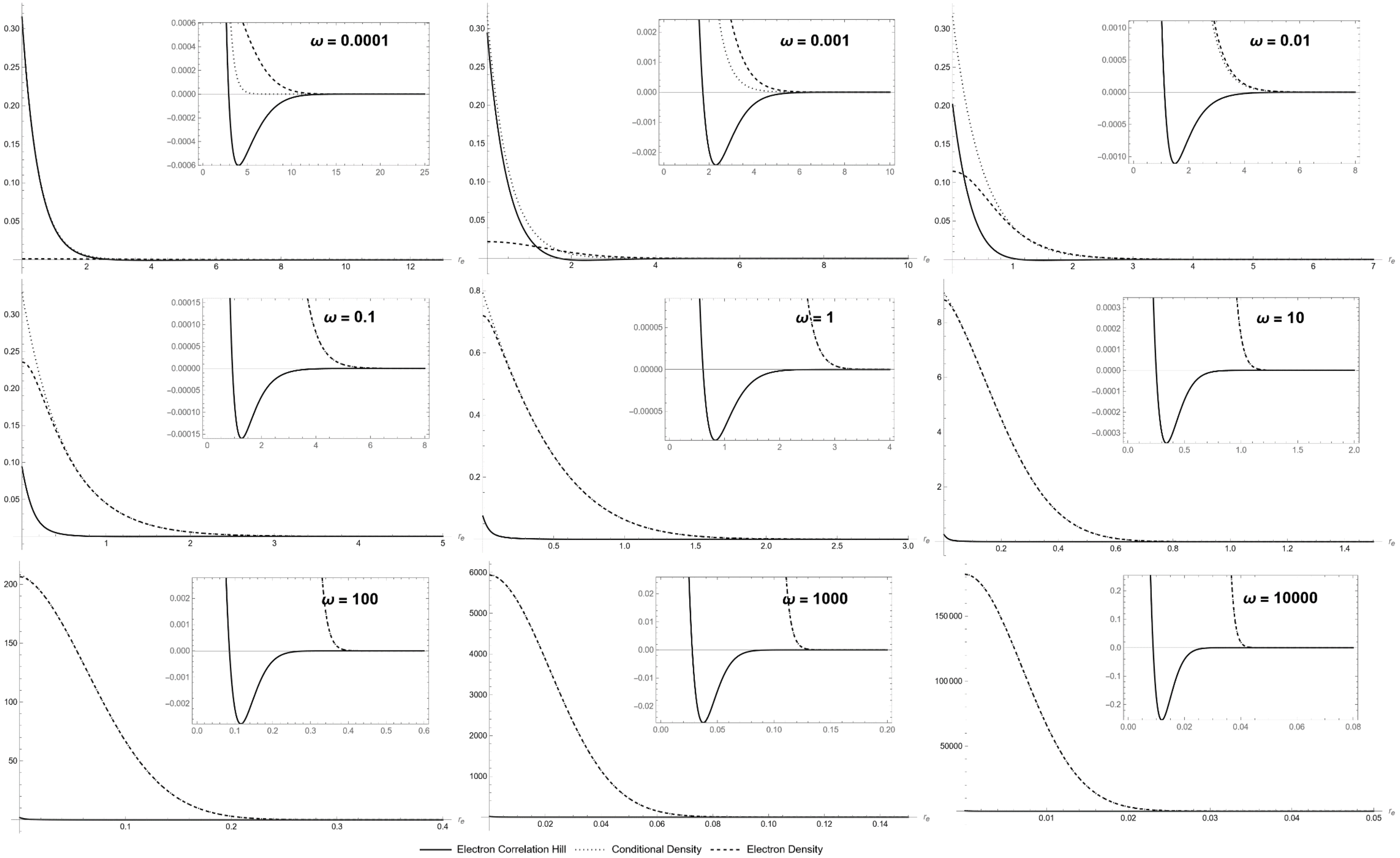

Figure 5-22: Comparison of the correlation hill, conditional density, and single-particle density of electron for mass 400

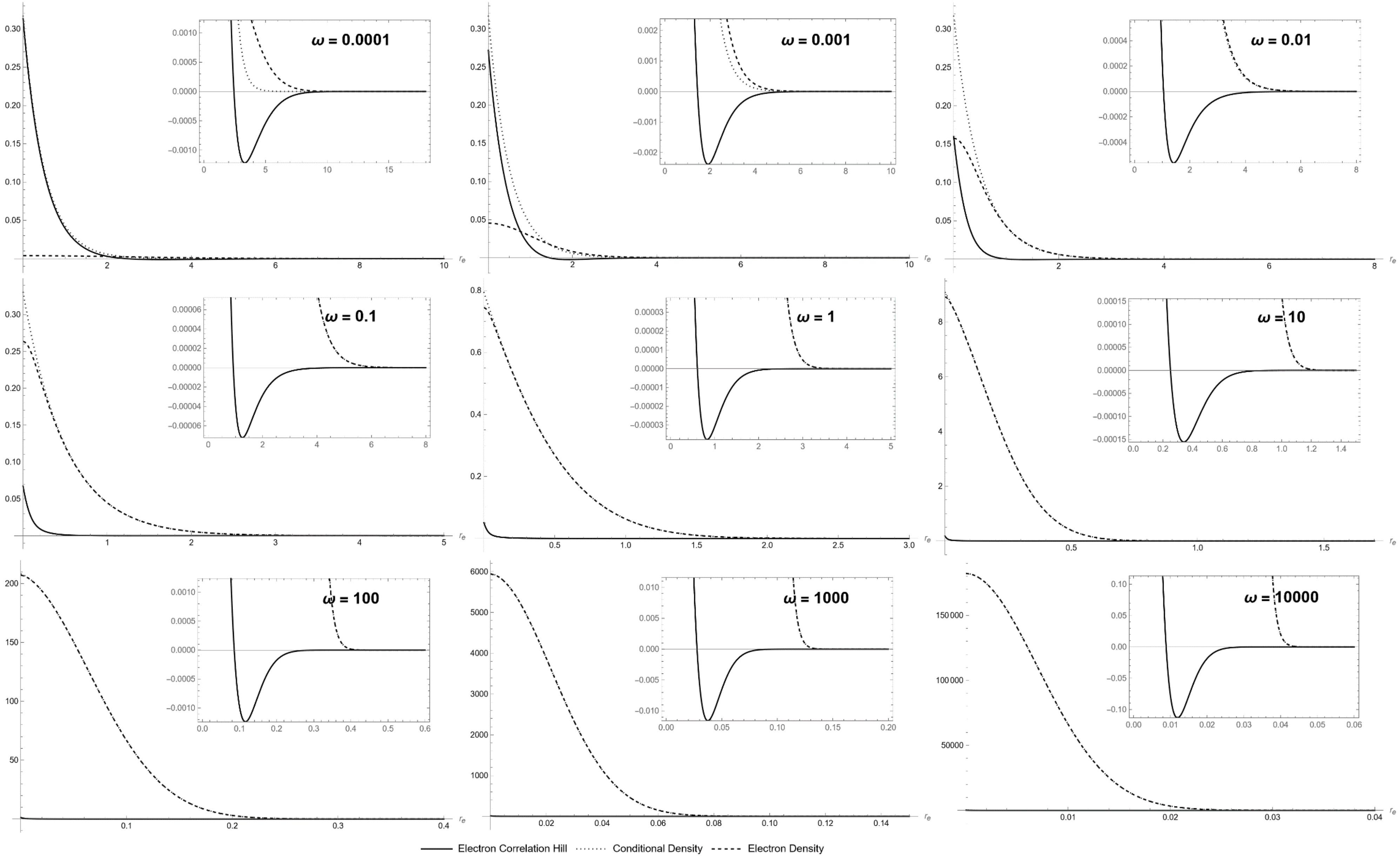

Figure 5-23: Comparison of the correlation hill, conditional density, and single-particle density of electron for mass 900

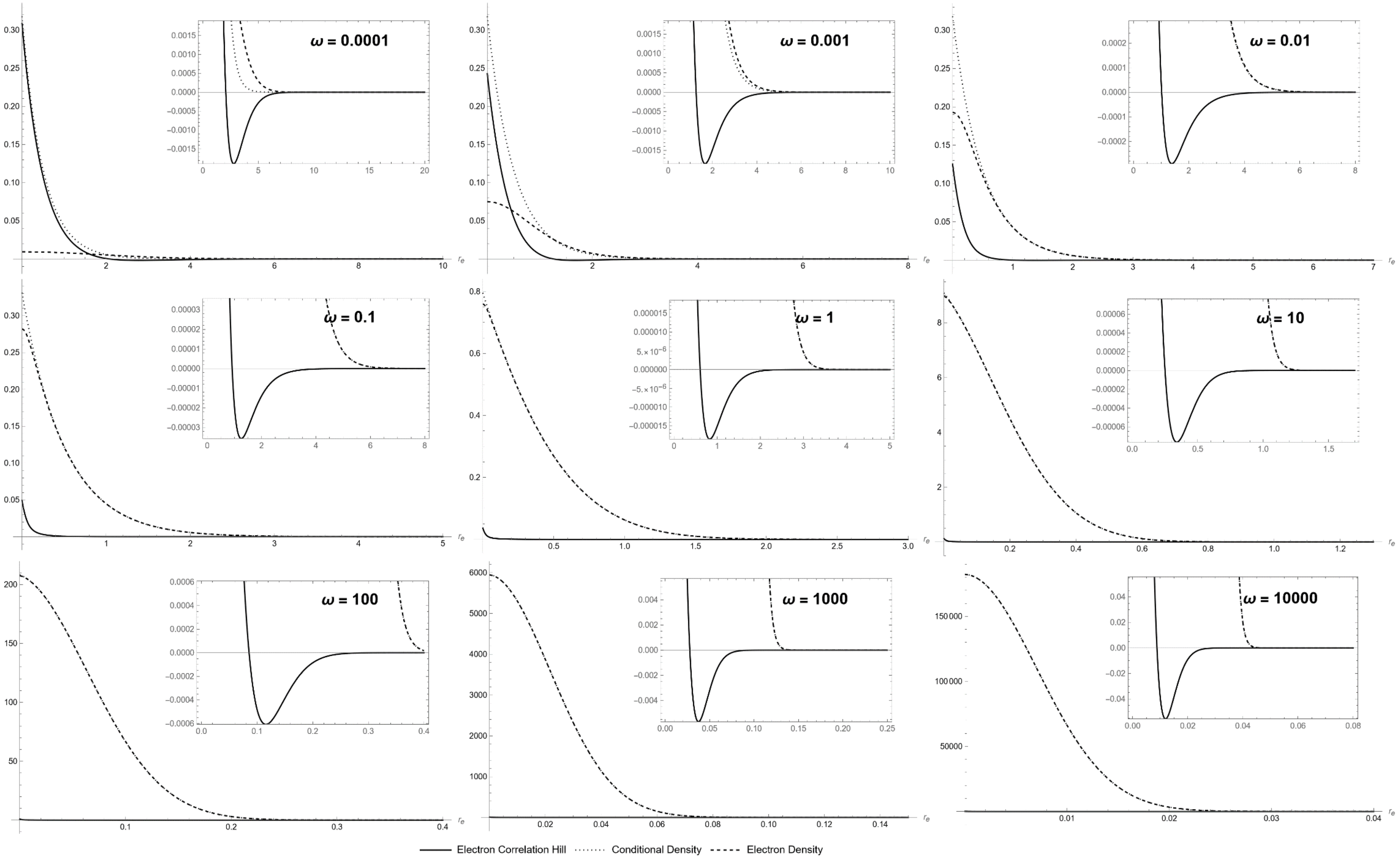

Figure 5-24: Comparison of the correlation hill, conditional density, and single-particle density of electron for mass 1836

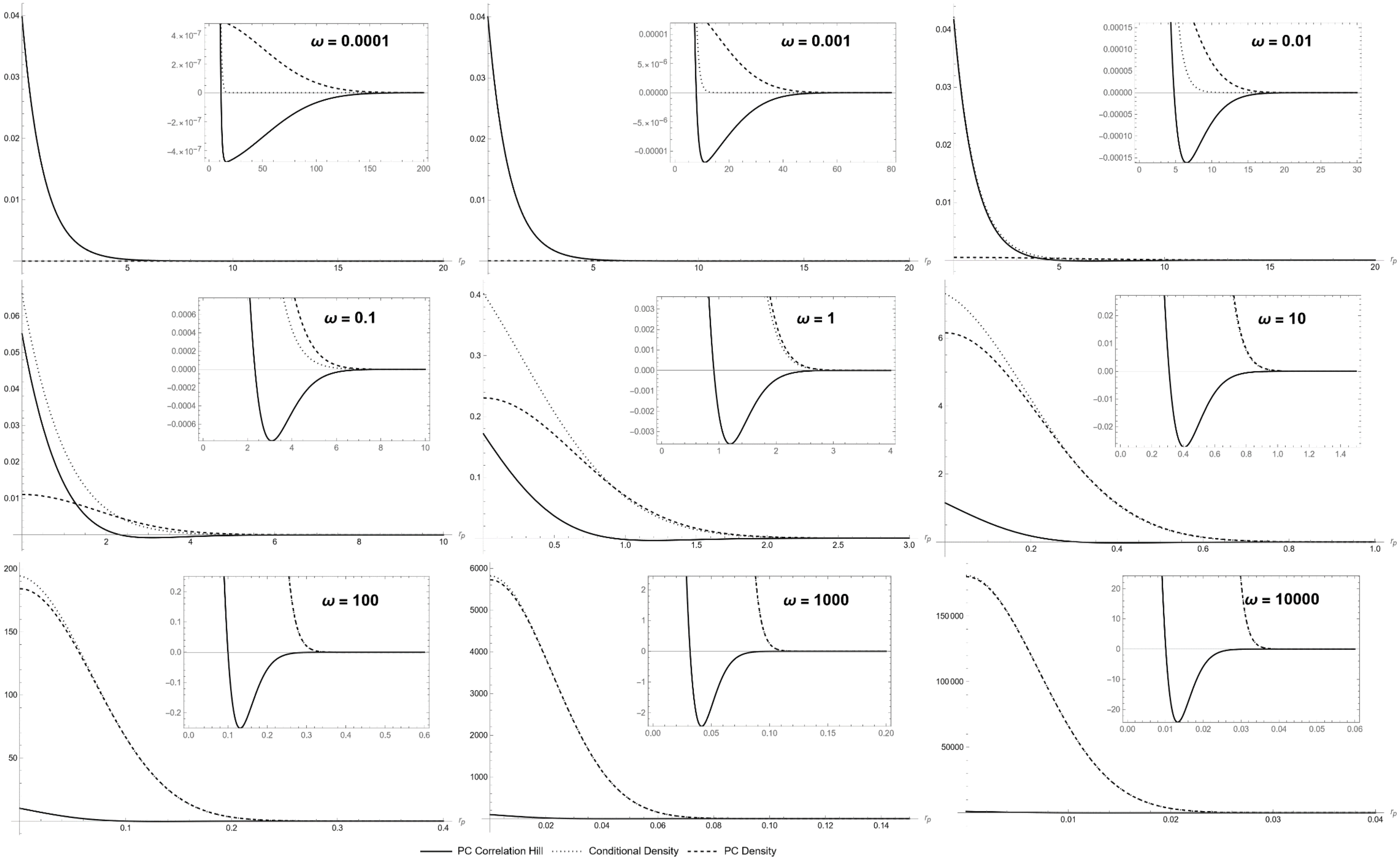

Figure 5-25: Comparison of the correlation hill, conditional density, and single-particle density of PCP for mass 1

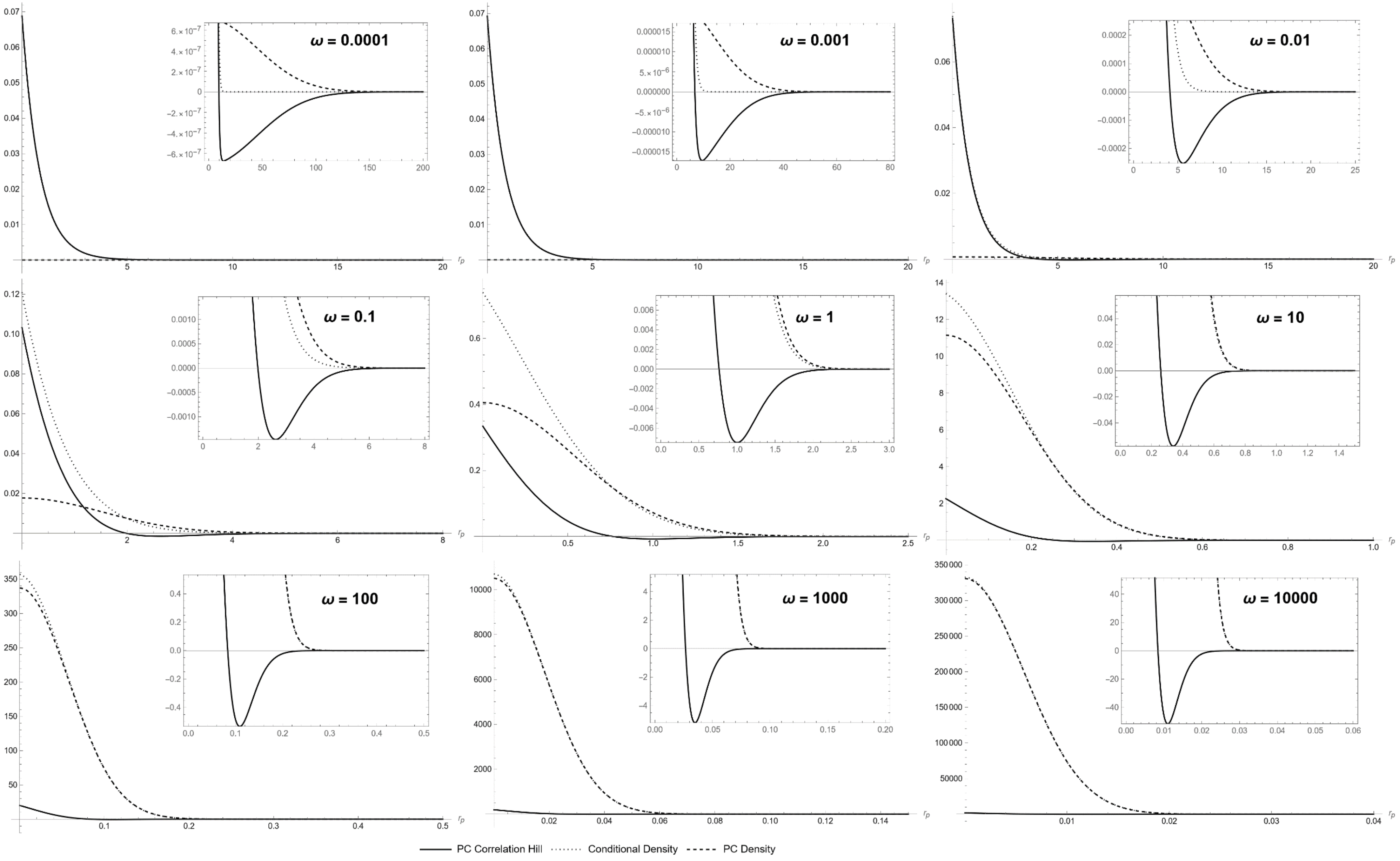

Figure 5-26: Comparison of the correlation hill, conditional density, and single-particle density of PCP for mass 1.5

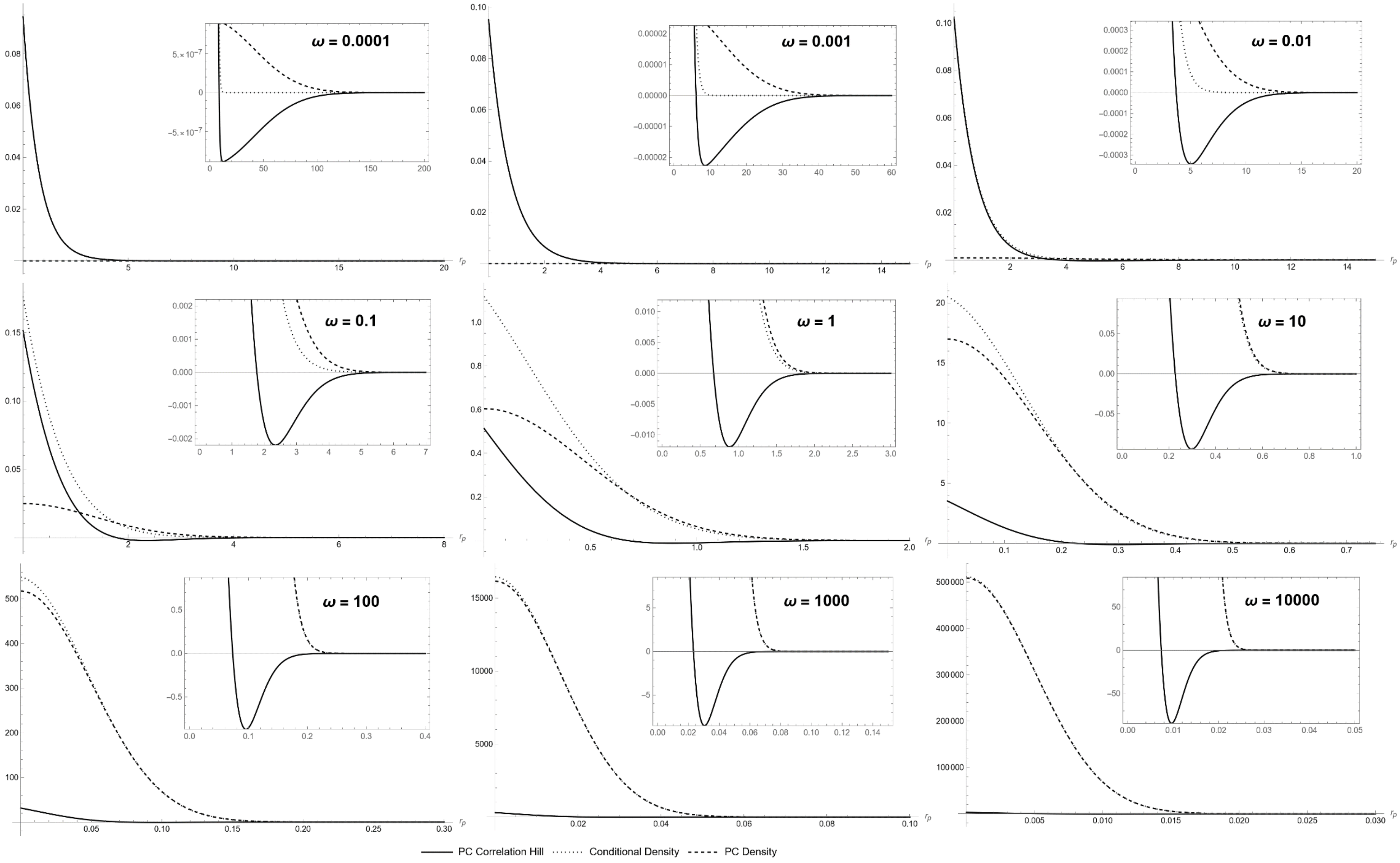

Figure 5-27: Comparison of the correlation hill, conditional density, and single-particle density of PCP for mass 2

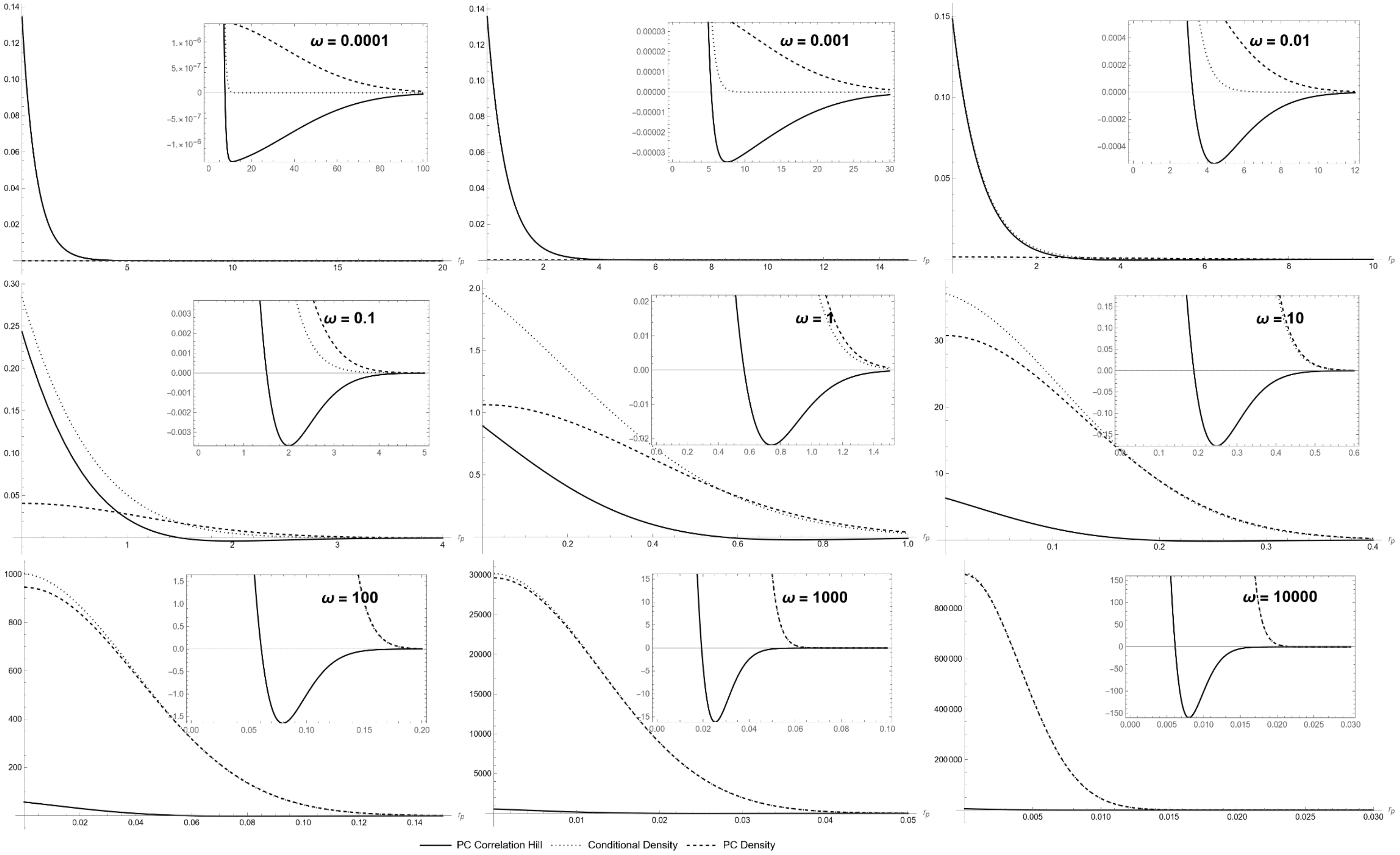

Figure 5-28: Comparison of the correlation hill, conditional density, and single-particle density of PCP for mass 3

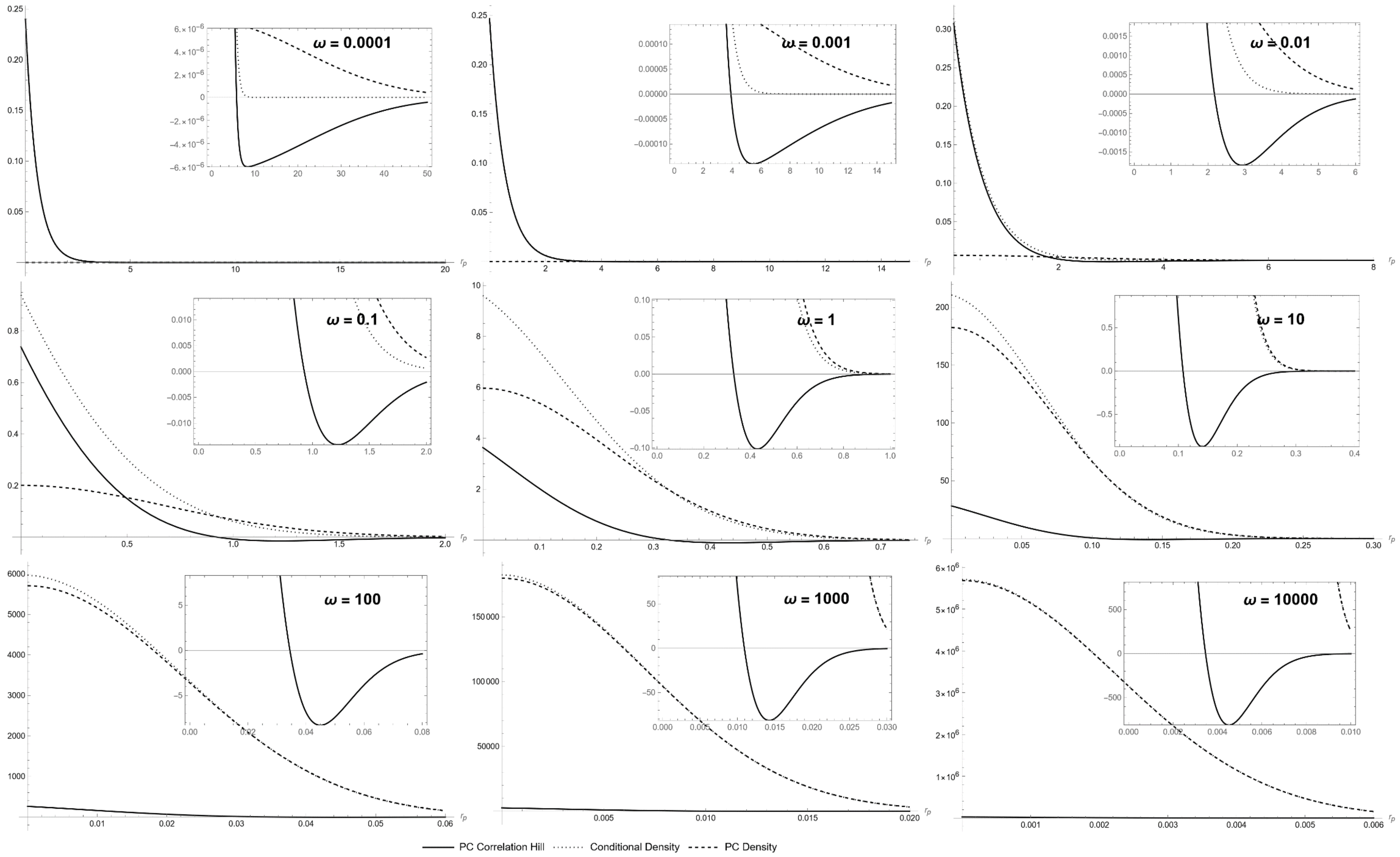

Figure 5-29: Comparison of the correlation hill, conditional density, and single-particle density of PCP for mass 10

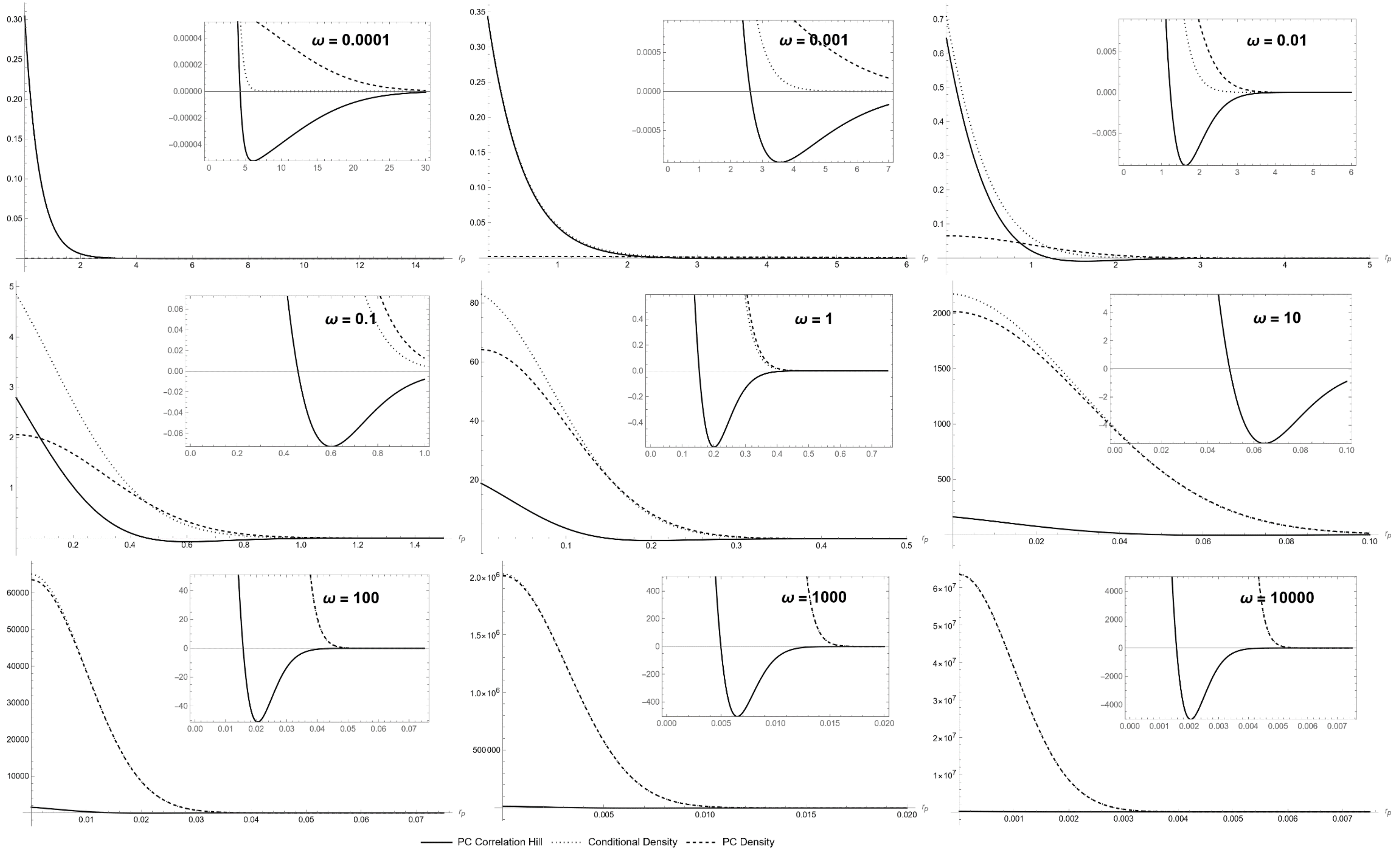

Figure 5-30: Comparison of the correlation hill, conditional density, and single-particle density of PCP for mass 50

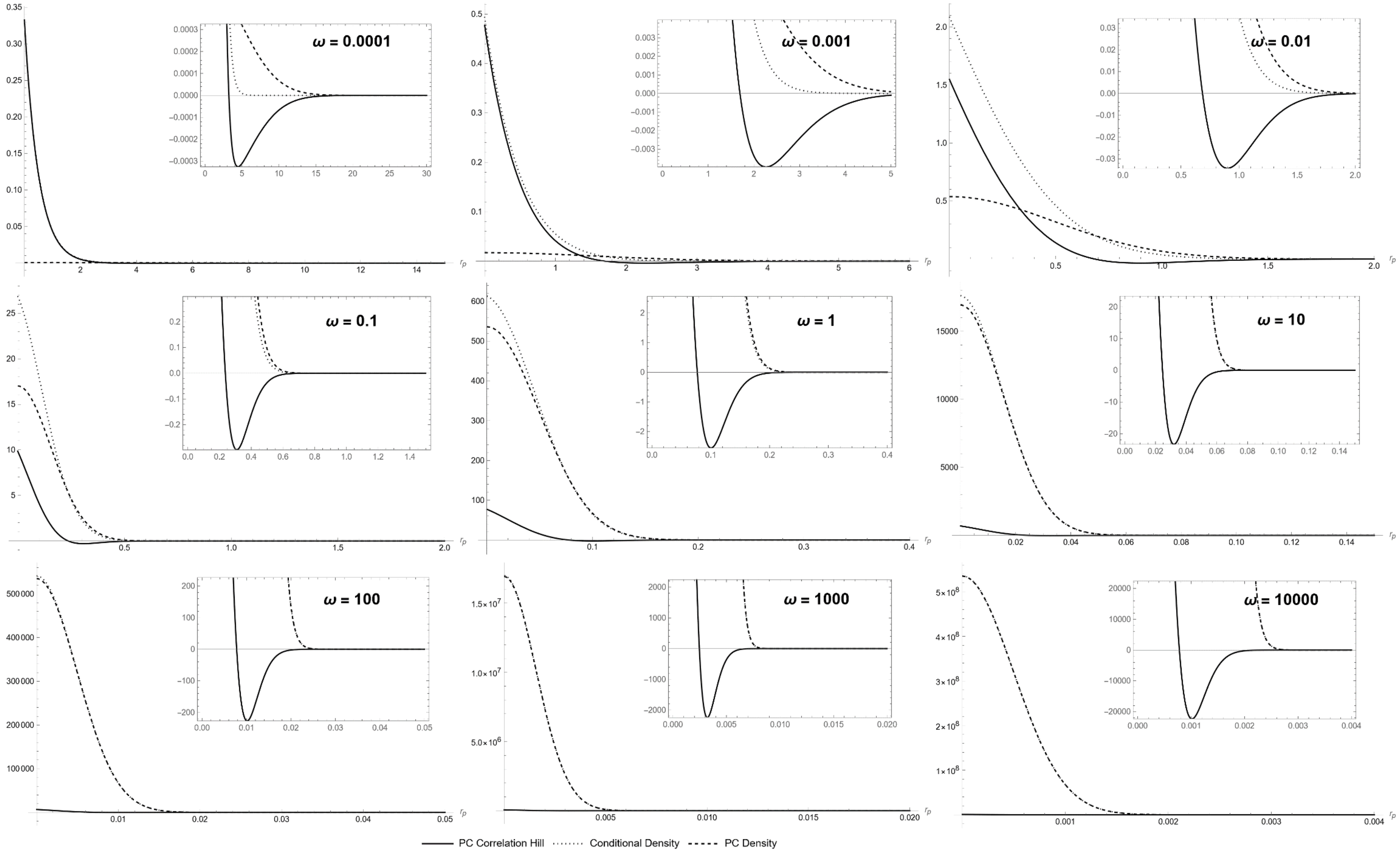

Figure 5-31: Comparison of the correlation hill, conditional density, and single-particle density of PCP for mass 207

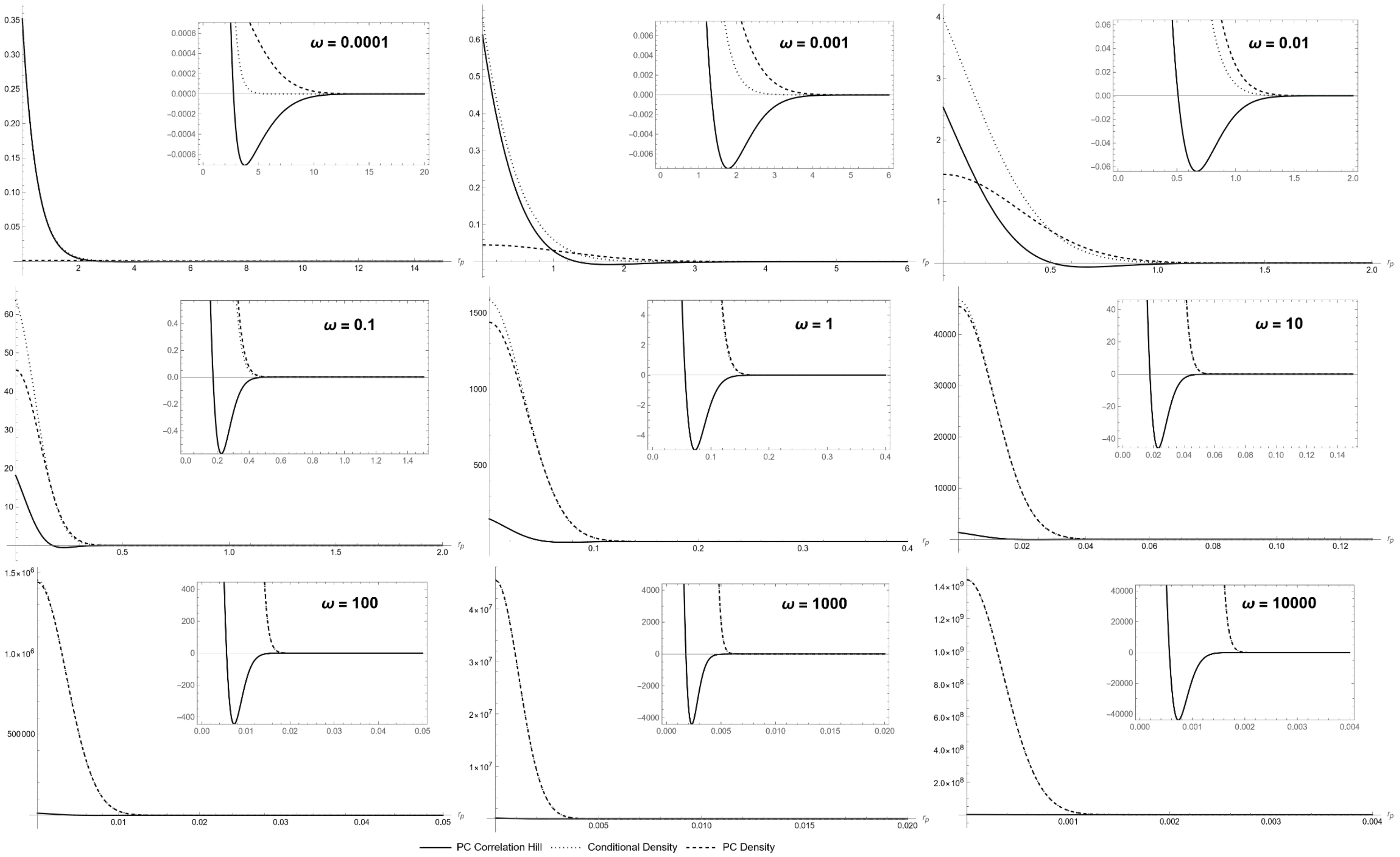

Figure 5-32: Comparison of the correlation hill, conditional density, and single-particle density of PCP for mass 400

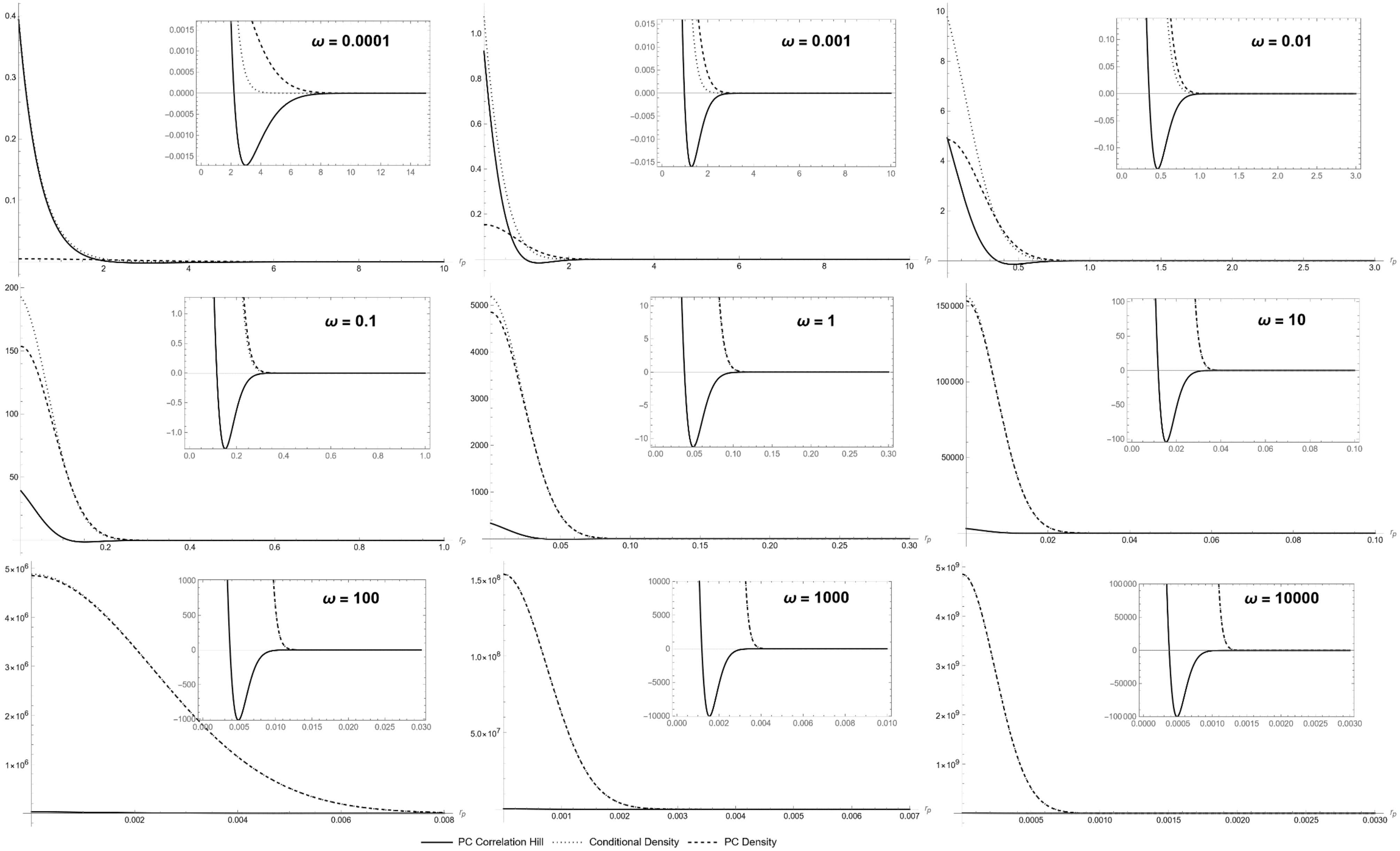

Figure 5-33: Comparison of the correlation hill, conditional density, and single-particle density of PCP for mass 900

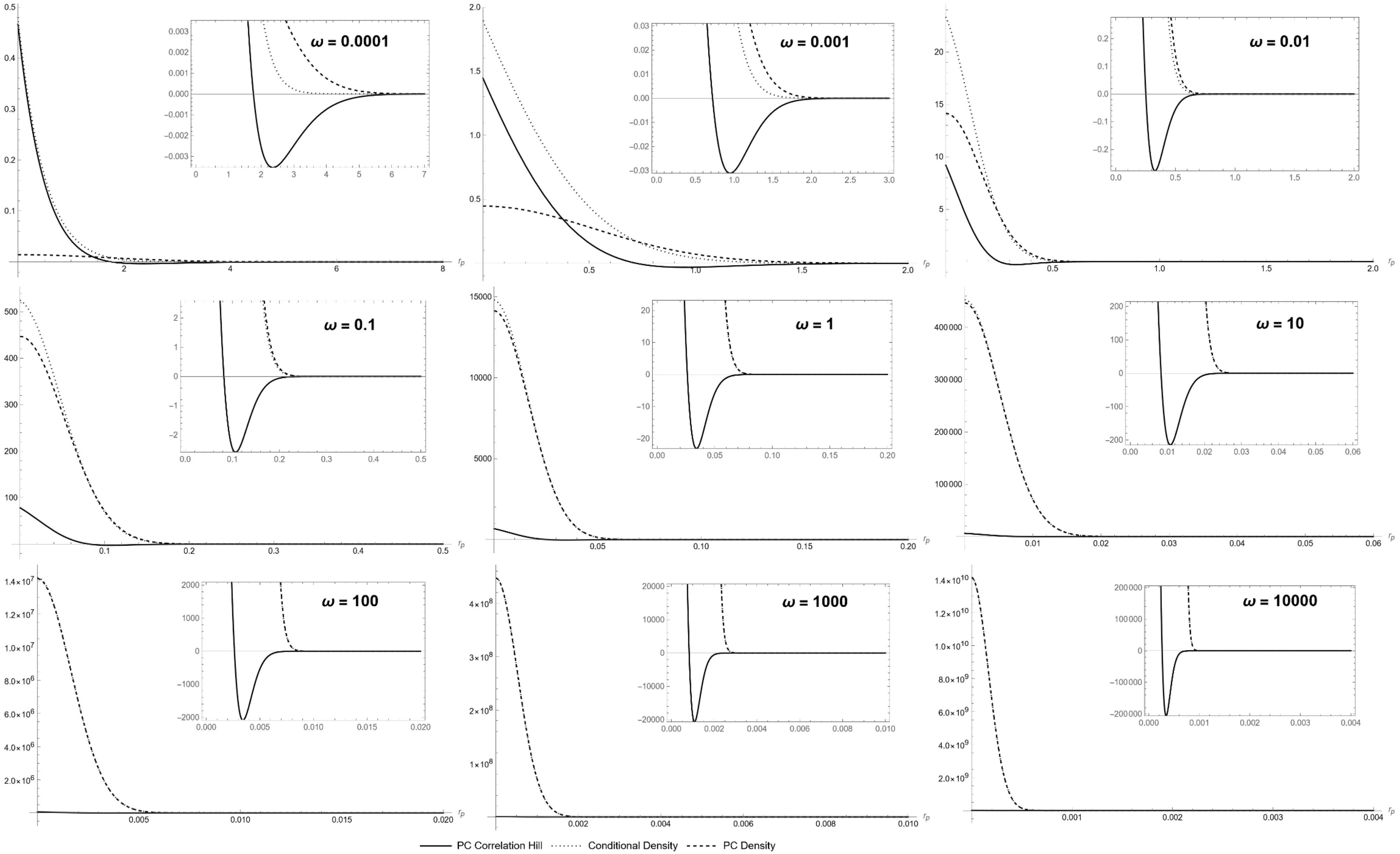

Figure 5-34: Comparison of the correlation hill, conditional density, and single-particle density of PCP for mass 1836

### 5.1.4 Two-Particle Distribution Functions

Figures 5-35, 5-36, and 5-37 respectively represent the average two-particle distribution functions (3-42) obtained using the variational, FEM, and MC-HF methods. Figure 5-38 shows the difference between the average two-particle distribution functions (equation 3-50) obtained by the variational and MC-HF methods.

The average two-particle distribution function has a Gaussian-like shape that becomes more localized with increasing frequency and mass, indicating that the two particles are closer to each other at a frequency of $10^4$ and high mass. Since the MC-HF method lacks correlation effects, the difference between the variational and MC-HF distribution functions can be interpreted as an indicator of e-PCP correlation effects. In most cases, the difference in the average two-particle distribution functions of variational and TC-HF methods shows an oscillatory pattern, being positive at short inter-particle distances and negative at long inter-particle distances. However, at high masses and frequencies, it is initially negative, then positive, then negative again, and finally positive at very short distances.

For electron correlation, the correlation effect always tends to keep electrons further apart, but conversely, for e-PCP correlation, the correlation effect, due to its attractive nature, tends to bring the two particles closer together. This is consistent with the positive correlation function differences at short inter-particle distances shown in Figure 5-38.

Integration of the average two-particle distribution function with respect to the inter-particle distance must equal 1 (equation 3-36), a condition that was verified for the average variational, numerical, and MC-HF two-particle distribution functions, all of which satisfy this requirement.

The limit of the average two-particle distribution function as $m_p \to \infty$ and $\omega \to 0$ was also examined, and the result is equal to the radial distribution function of a hydrogen atom with a clamped nucleus:

$$\lim_{m_p \to \infty,\, \omega \to 0} f(\boldsymbol{r}) = 4r^2 e^{-2r} = \mathrm{RDF_H}(\boldsymbol{r}) \qquad 5.5$$

The difference between the two-particle distribution function of the two-component system and the radial distribution function of a hydrogen atom with a clamped nucleus is due to the finite mass of the PCP and the harmonic oscillator potential, both of which disappear in the mentioned limit. This can be clearly observed in Figure 5-35 for a frequency of 0.0001 and a mass of 1836.

From Figure 5-38, it can be concluded that correlation causes the electron and PCP to be closer on average. In other words, in a fully correlated state, the electron and the PCP spend more time together, which is consistent with the concept of the correlation hill discussed in the previous section.

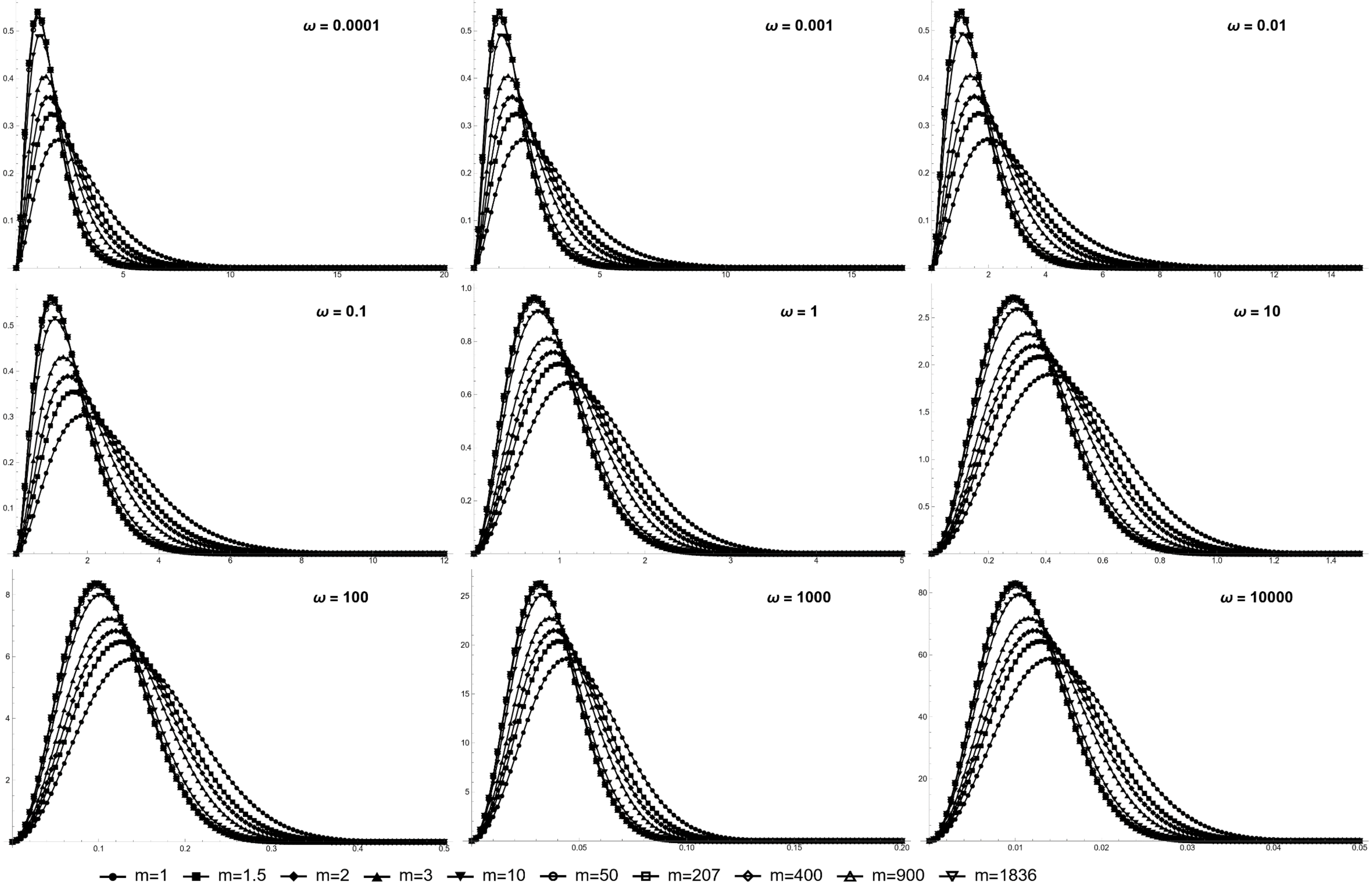

Figure 5-35: Variational average two-particle distribution function as a function of $r$ for different frequencies and masses.

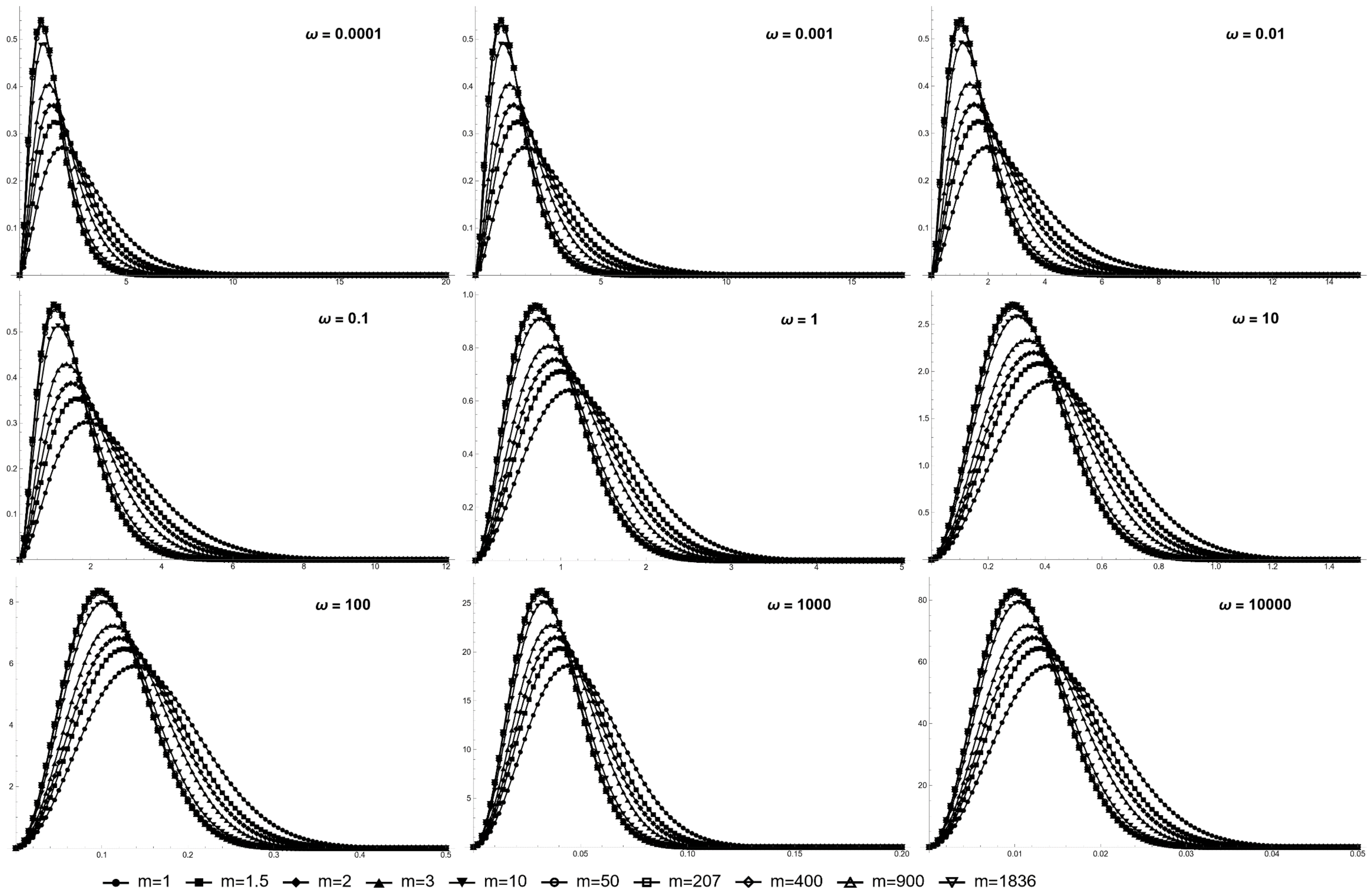

Figure 5-36: FEM average two-particle distribution function as a function of $r$ for different frequencies and masses.

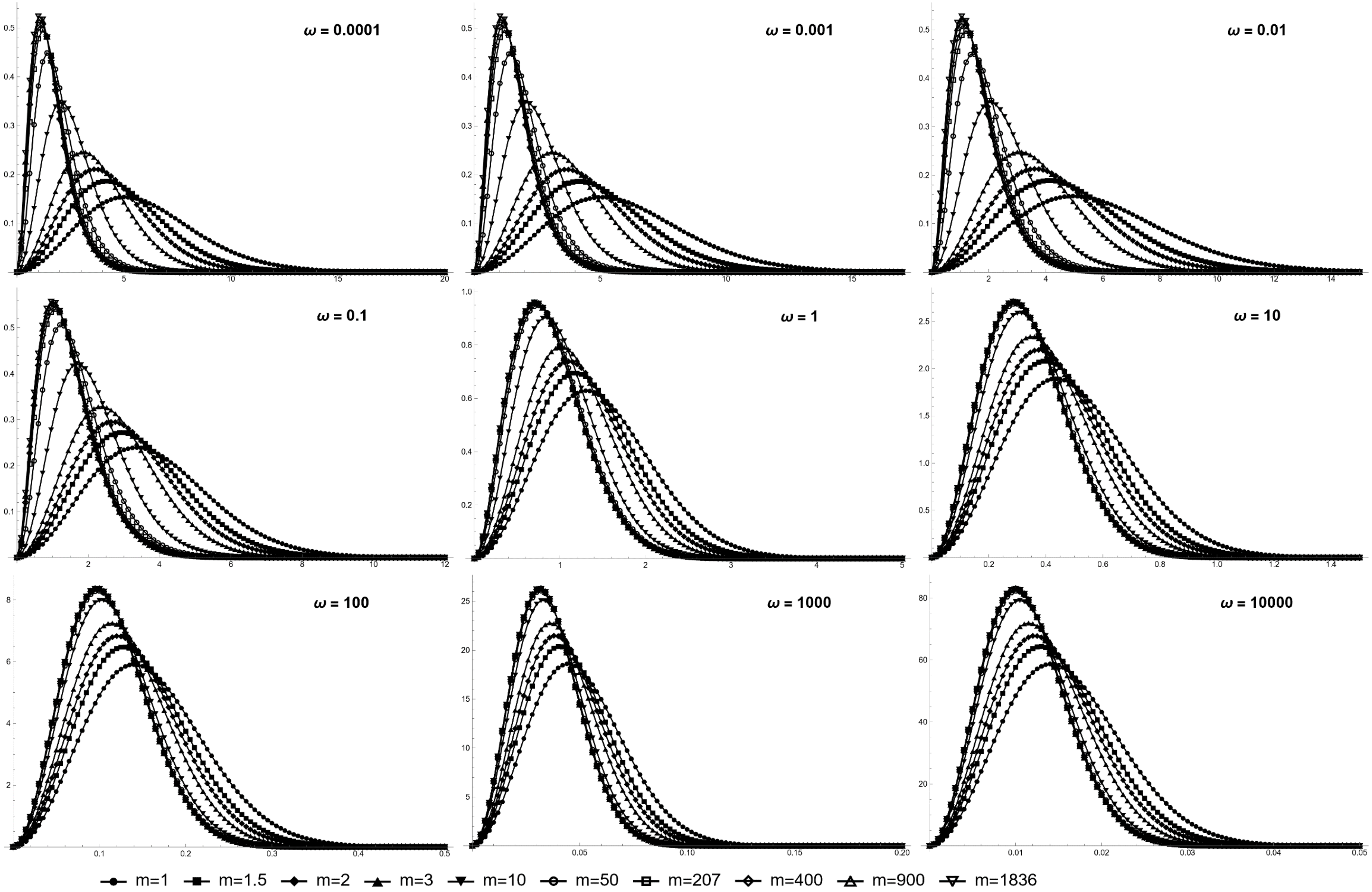

Figure 5-37: TC-HF average two-particle distribution function as a function of $r$ for different frequencies and masses.

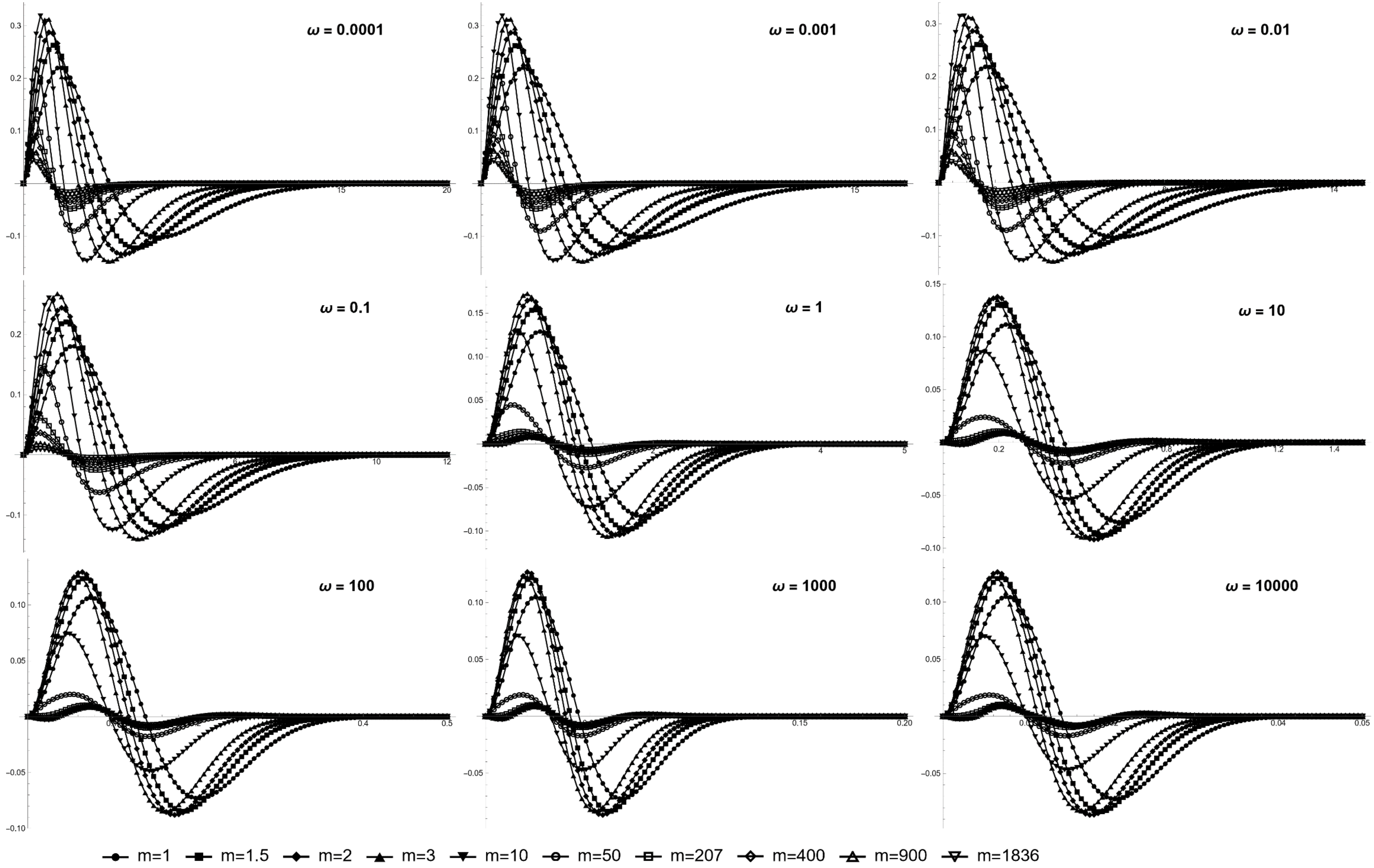

Figure 5-38: Difference between the variational and TC-HF average two-particle distribution functions

### 5.1.5 Mean Inter-particle Distance and its Variance

The results for mean inter-particle distances (equation 3-44) and their variances for the three methods—variational, FEM, and MC-HF—are reported in Table 5-3, which confirm the results mentioned in previous sections. The inter-particle distance decreases with increasing frequency and mass. The variance, which indicates the spread from the mean value, follows the same trend, which is logical given that the particles are closer to each other at higher frequencies and masses. The data show that with increasing ω, the average distance difference between the fully correlated variational method and the MC-HF method disappears. This result aligns with the idea that correlation effects decrease with increasing ω, confirming the results of previous sections. Additionally, this difference decreases with increasing mass, meaning that correlation effects are more significant for lighter PCPs. This pattern, observed for correlation energy, is exactly seen here for variance as well.

Furthermore, the results for reproducing the e-PCP interaction energy through the two-particle distribution function (equation 3-46) using variational and FEM methods are presented in items 8 and 9 of Table 5-3, which match exactly with the interaction energies presented in items 1 and 2 of Table 5-2. It is worth noting that due to computational issues, these values were not calculated for the HF method.

Table 5-3: The quantities obtained based on the average two-particle distribution function.

| | | Average of r | | | Variance of r | | | $\int$ **f/r dr** =INT | |
|---|---|---|---|---|---|---|---|---|---|
| ω/Quantities | $f_{max}$ [1] | Variational[2] | FEM[3] | MC-HF[4] | Variational[5] | FEM[6] | MC-HF[7] | Variational[8] | FEM[9] |
| m=1 | | | | | | | | | |
| 0.0001 | 2.000000 | 2.999999 | 2.999999 | 5.981602 | 2.999998 | 2.999998 | 7.190049 | -0.500000 | -0.500000 |
| 0.001 | 1.999988 | 2.999946 | 2.999946 | 5.980384 | 2.999820 | 2.999808 | 7.186729 | -0.500006 | -0.500006 |
| 0.01 | 1.998803 | 2.994651 | 2.994653 | 5.867637 | 2.982239 | 2.981093 | 6.884944 | -0.500596 | -0.500596 |
| 0.1 | 1.904998 | 2.686954 | 2.688387 | 3.871700 | 2.138349 | 2.121882 | 2.829622 | -0.541447 | -0.541407 |
| 1 | 1.116163 | 1.349412 | 1.350134 | 1.484141 | 0.394696 | 0.392679 | 0.399454 | -0.991609 | -0.991837 |
| 10 | 0.417919 | 0.480650 | 0.480729 | 0.493695 | 0.043782 | 0.043706 | 0.043653 | -2.689688 | -2.690048 |
| 100 | 0.138524 | 0.157206 | 0.157214 | 0.158491 | 0.004488 | 0.004486 | 0.004482 | -8.137662 | -8.138049 |
| 1000 | 0.044433 | 0.050227 | 0.050227 | 0.050354 | 0.000452 | 0.000452 | 0.000452 | -25.387791 | -25.388180 |
| 10000 | 0.014113 | 0.015934 | 0.015934 | 0.015947 | 0.000045 | 0.000045 | 0.000045 | -79.944206 | -79.944309 |
| m=1.5 | | | | | | | | | |
| 0.0001 | 1.666667 | 2.500000 | 2.500000 | 4.934280 | 2.083333 | 2.083332 | 4.920492 | -0.600000 | -0.600000 |
| 0.001 | 1.666660 | 2.499969 | 2.499969 | 4.933588 | 2.083247 | 2.083241 | 4.918910 | -0.600005 | -0.600005 |
| 0.01 | 1.665973 | 2.496896 | 2.496897 | 4.868394 | 2.074733 | 2.074173 | 4.773718 | -0.600498 | -0.600498 |
| 0.1 | 1.608070 | 2.296002 | 2.296978 | 3.404134 | 1.597607 | 1.586164 | 2.212973 | -0.637497 | -0.637453 |
| 1 | 0.992970 | 1.209767 | 1.210470 | 1.340977 | 0.323077 | 0.321298 | 0.327920 | -1.111402 | -1.111648 |
| 10 | 0.378914 | 0.436646 | 0.436726 | 0.449270 | 0.036349 | 0.036280 | 0.036223 | -2.965158 | -2.965584 |
| 100 | 0.126200 | 0.143301 | 0.143309 | 0.144540 | 0.003736 | 0.003734 | 0.003730 | -8.931366 | -8.931830 |
| 1000 | 0.040536 | 0.045830 | 0.045831 | 0.045953 | 0.000377 | 0.000377 | 0.000376 | -27.827407 | -27.827882 |
| 10000 | 0.012881 | 0.014544 | 0.014544 | 0.014556 | 0.000038 | 0.000038 | 0.000038 | -87.590782 | -87.591125 |
| m=2 | | | | | | | | | |

| | | Average of r | | | Variance of r | | | $\int$ **f/r dr** =INT | |
|---|---|---|---|---|---|---|---|---|---|
| ω/Quantities | $f_{max}$ [1] | Variational[2] | FEM[3] | MC-HF[4] | Variational[5] | FEM[6] | MC-HF[7] | Variational[8] | FEM[9] |
| 0.0001 | 1.500000 | 2.250000 | 2.250000 | 4.357798 | 1.687499 | 1.687499 | 3.878840 | -0.666667 | -0.666667 |
| 0.001 | 1.499995 | 2.249977 | 2.249977 | 4.357315 | 1.687443 | 1.687439 | 3.877865 | -0.666671 | -0.666671 |
| 0.01 | 1.499494 | 2.247734 | 2.247735 | 4.311394 | 1.681848 | 1.681478 | 3.786690 | -0.667115 | -0.667115 |
| 0.1 | 1.455992 | 2.092163 | 2.092922 | 3.125413 | 1.342596 | 1.333497 | 1.894119 | -0.701824 | -0.701779 |
| 1 | 0.926837 | 1.134680 | 1.135371 | 1.259215 | 0.287424 | 0.285774 | 0.291646 | -1.188468 | -1.188724 |
| 10 | 0.357940 | 0.412985 | 0.413064 | 0.424824 | 0.032637 | 0.032571 | 0.032496 | -3.137951 | -3.138421 |
| 100 | 0.119574 | 0.135825 | 0.135833 | 0.136980 | 0.003361 | 0.003358 | 0.003354 | -9.425652 | -9.426166 |
| 1000 | 0.038441 | 0.043466 | 0.043467 | 0.043580 | 0.000339 | 0.000339 | 0.000339 | -29.343461 | -29.343988 |
| 10000 | 0.012219 | 0.013796 | 0.013796 | 0.013808 | 0.000034 | 0.000034 | 0.000034 | -92.339475 | -92.339464 |
| m=3 | | | | | | | | | |
| 0.0001 | 1.333333 | 2.000000 | 2.000000 | 3.720384 | 1.333333 | 1.333333 | 2.893677 | -0.750000 | -0.750000 |
| 0.001 | 1.333330 | 1.999984 | 1.999984 | 3.720075 | 1.333298 | 1.333295 | 2.893121 | -0.750004 | -0.750004 |
| 0.01 | 1.332978 | 1.998407 | 1.998407 | 3.690477 | 1.329799 | 1.329567 | 2.842222 | -0.750399 | -0.750399 |
| 0.1 | 1.301551 | 1.882410 | 1.882969 | 2.794381 | 1.100335 | 1.093412 | 1.560897 | -0.782537 | -0.782494 |
| 1 | 0.857032 | 1.055298 | 1.055973 | 1.166177 | 0.251903 | 0.250390 | 0.254800 | -1.282329 | -1.282593 |
| 10 | 0.335763 | 0.387965 | 0.388044 | 0.398227 | 0.028929 | 0.028867 | 0.028759 | -3.343984 | -3.344507 |
| 100 | 0.112569 | 0.127921 | 0.127929 | 0.128910 | 0.002985 | 0.002983 | 0.002977 | -10.011451 | -10.012028 |
| 1000 | 0.036226 | 0.040966 | 0.040967 | 0.041064 | 0.000301 | 0.000301 | 0.000301 | -31.136990 | -31.137584 |
| 10000 | 0.011518 | 0.013006 | 0.013006 | 0.013016 | 0.000030 | 0.000030 | 0.000030 | -97.954202 | -97.954818 |
| m=10 | | | | | | | | | |
| 0.0001 | 1.100000 | 1.650000 | 1.650000 | 2.590797 | 0.907500 | 0.907500 | 1.561928 | -0.909091 | -0.909091 |
| 0.001 | 1.099998 | 1.649991 | 1.649991 | 2.590647 | 0.907483 | 0.907482 | 1.561559 | -0.909094 | -0.909094 |
| 0.01 | 1.099801 | 1.649104 | 1.649104 | 2.579139 | 0.905859 | 0.905751 | 1.547241 | -0.909420 | -0.909420 |

| | | Average of r | | | Variance of r | | | $\int$ **f/r dr** =INT | |
|---|---|---|---|---|---|---|---|---|---|
| ω/Quantities | $f_{max}$ [1] | Variational[2] | FEM[3] | MC-HF[4] | Variational[5] | FEM[6] | MC-HF[7] | Variational[8] | FEM[9] |
| 0.1 | 1.081518 | 1.578456 | 1.578774 | 2.129586 | 0.787219 | 0.782964 | 1.031733 | -0.937438 | -0.937401 |
| 1 | 0.751829 | 0.935327 | 0.935972 | 0.997221 | 0.202453 | 0.201149 | 0.202034 | -1.455630 | -1.455900 |
| 10 | 0.302230 | 0.350136 | 0.350215 | 0.355189 | 0.023745 | 0.023690 | 0.023564 | -3.712490 | -3.713113 |
| 100 | 0.101978 | 0.115970 | 0.115978 | 0.116434 | 0.002459 | 0.002457 | 0.002452 | -11.049642 | -11.050341 |
| 1000 | 0.032877 | 0.037188 | 0.037189 | 0.037233 | 0.000248 | 0.000248 | 0.000248 | -34.307069 | -34.307788 |
| 10000 | 0.010459 | 0.011811 | 0.011811 | 0.011815 | 0.000025 | 0.000025 | 0.000025 | -107.870081 | -107.870362 |
| | | | | m=50 | | | | | |
| 0.0001 | 1.020000 | 1.530000 | 1.530000 | 1.949965 | 0.780300 | 0.780300 | 1.031171 | -0.980392 | -0.980392 |
| 0.001 | 1.019998 | 1.529993 | 1.529993 | 1.949900 | 0.780288 | 0.780287 | 1.031121 | -0.980395 | -0.980395 |
| 0.01 | 1.019841 | 1.529285 | 1.529286 | 1.943441 | 0.779086 | 0.779006 | 1.024910 | -0.980698 | -0.980698 |
| 0.1 | 1.005104 | 1.471429 | 1.471680 | 1.698090 | 0.688120 | 0.684652 | 0.779651 | -1.007152 | -1.007117 |
| 1 | 0.713362 | 0.891326 | 0.891957 | 0.910909 | 0.185594 | 0.184366 | 0.184251 | -1.531339 | -1.531607 |
| 10 | 0.289923 | 0.336252 | 0.336330 | 0.337618 | 0.021971 | 0.021917 | 0.021869 | -3.868947 | -3.869613 |
| 100 | 0.098091 | 0.111585 | 0.111593 | 0.111704 | 0.002279 | 0.002277 | 0.002276 | -11.486768 | -11.487519 |
| 1000 | 0.031649 | 0.035801 | 0.035802 | 0.035813 | 0.000230 | 0.000230 | 0.000230 | -35.638555 | -35.639330 |
| 10000 | 0.010071 | 0.011372 | 0.011373 | 0.011374 | 0.000023 | 0.000023 | 0.000023 | -112.031793 | -112.031719 |
| | | | | m=207 | | | | | |
| 0.0001 | 1.004831 | 1.507246 | 1.507246 | 1.713677 | 0.757264 | 0.757264 | 0.874958 | -0.995192 | -0.995192 |
| 0.001 | 1.004829 | 1.507240 | 1.507240 | 1.713614 | 0.757252 | 0.757252 | 0.874906 | -0.995195 | -0.995195 |
| 0.01 | 1.004679 | 1.506563 | 1.506563 | 1.707620 | 0.756121 | 0.756045 | 0.870034 | -0.995493 | -0.995493 |
| 0.1 | 0.990562 | 1.450974 | 1.451212 | 1.537190 | 0.669862 | 0.666534 | 0.702137 | -1.021642 | -1.021608 |
| 1 | 0.705910 | 0.882792 | 0.883421 | 0.888875 | 0.182403 | 0.181191 | 0.181130 | -1.546927 | -1.547195 |
| 10 | 0.287535 | 0.333558 | 0.333637 | 0.333973 | 0.021634 | 0.021582 | 0.021568 | -3.900837 | -3.901511 |

| | | Average of r | | | Variance of r | | | $\int$ **f/r dr** =INT | |
|---|---|---|---|---|---|---|---|---|---|
| ω/Quantities | $f_{max}$ [1] | Variational[2] | FEM[3] | MC-HF[4] | Variational[5] | FEM[6] | MC-HF[7] | Variational[8] | FEM[9] |
| 100 | 0.097337 | 0.110734 | 0.110742 | 0.110771 | 0.002245 | 0.002243 | 0.002243 | -11.575600 | -11.576362 |
| 1000 | 0.031410 | 0.035532 | 0.035533 | 0.011263 | 0.000227 | 0.000227 | 0.001363 | -35.908903 | -35.909690 |
| 10000 | 0.009995 | 0.011287 | 0.011288 | 0.011288 | 0.000023 | 0.000023 | 0.000023 | -112.876573 | -112.876468 |
| | | | | | m=400 | | | | |
| 0.0001 | 1.002500 | 1.503750 | 1.503750 | 1.652347 | 0.753755 | 0.753755 | 0.837699 | -0.997506 | -0.997506 |
| 0.001 | 1.002498 | 1.503743 | 1.503743 | 1.652281 | 0.753743 | 0.753743 | 0.837669 | -0.997509 | -0.997509 |
| 0.01 | 1.002349 | 1.503072 | 1.503072 | 1.645959 | 0.752622 | 0.752547 | 0.832830 | -0.997806 | -0.997806 |
| 0.1 | 0.988325 | 1.447826 | 1.448062 | 1.499710 | 0.667072 | 0.663765 | 0.685196 | -1.023908 | -1.023874 |
| 1 | 0.704761 | 0.881475 | 0.882104 | 0.885056 | 0.181913 | 0.180703 | 0.180670 | -1.549361 | -1.549628 |
| 10 | 0.287167 | 0.333142 | 0.333221 | 0.333398 | 0.021583 | 0.021530 | 0.021523 | -3.905805 | -3.906481 |
| 100 | 0.097221 | 0.110603 | 0.110611 | 0.110626 | 0.002240 | 0.002238 | 0.002238 | -11.589432 | -11.590196 |
| 1000 | 0.031373 | 0.035491 | 0.035492 | 0.035493 | 0.000226 | 0.000226 | 0.000226 | -35.950992 | -35.951780 |
| 10000 | 0.009984 | 0.011274 | 0.011274 | 0.011275 | 0.000023 | 0.000023 | 0.000023 | -113.008083 | -113.007974 |
| | | | | | m=900 | | | | |
| 0.0001 | 1.001111 | 1.501667 | 1.501667 | 1.600851 | 0.751667 | 0.751667 | 0.807377 | -0.998890 | -0.998890 |
| 0.001 | 1.001110 | 1.501660 | 1.501660 | 1.600772 | 0.751656 | 0.751656 | 0.807323 | -0.998893 | -0.998893 |
| 0.01 | 1.000961 | 1.500991 | 1.500991 | 1.593711 | 0.750542 | 0.750467 | 0.802209 | -0.999190 | -0.999190 |
| 0.1 | 0.986993 | 1.445950 | 1.446185 | 1.472460 | 0.665412 | 0.662117 | 0.673072 | -1.025263 | -1.025229 |
| 1 | 0.704075 | 0.880690 | 0.881318 | 0.460130 | 0.181621 | 0.180413 | 0.292910 | -1.550816 | -1.551083 |
| 10 | 0.286947 | 0.332894 | 0.332973 | 0.333053 | 0.021552 | 0.021499 | 0.021496 | -3.908774 | -3.909451 |
| 100 | 0.097151 | 0.110525 | 0.110533 | 0.110539 | 0.002237 | 0.002235 | 0.002235 | -11.597698 | -11.598462 |
| 1000 | 0.031351 | 0.035466 | 0.035467 | 0.035467 | 0.000226 | 0.000226 | 0.000226 | -35.976141 | -35.976930 |
| 10000 | 0.009977 | 0.011266 | 0.011267 | 0.011267 | 0.000023 | 0.000023 | 0.000023 | -113.086663 | -113.086551 |

| | | Average of r | | | Variance of r | | | $\int$ **f/r dr** =INT | |
|---|---|---|---|---|---|---|---|---|---|
| ω/Quantities | $f_{max}$ [1] | Variational[2] | FEM[3] | MC-HF[4] | Variational[5] | FEM[6] | MC-HF[7] | Variational[8] | FEM[9] |
| | | | | m=1836 | | | | | |
| 0.0001 | 1.000545 | 1.500817 | 1.500817 | 1.570352 | 0.750817 | 0.750817 | 0.789786 | -0.999456 | -0.999456 |
| 0.001 | 1.000543 | 1.500810 | 1.500810 | 1.570256 | 0.750806 | 0.750805 | 0.789707 | -0.999459 | -0.999459 |
| 0.01 | 1.000394 | 1.500142 | 1.500143 | 1.562364 | 0.749694 | 0.749619 | 0.784314 | -0.999755 | -0.999755 |
| 0.1 | 0.986450 | 1.445184 | 1.445419 | 1.459412 | 0.664734 | 0.661445 | 0.667317 | -1.025817 | -1.025783 |
| 1 | 0.703795 | 0.880370 | 0.880997 | 0.454000 | 0.181502 | 0.180294 | 0.290106 | -1.551410 | -1.551677 |
| 10 | 0.286857 | 0.332793 | 0.332872 | 0.332911 | 0.021539 | 0.021487 | 0.021485 | -3.909987 | -3.910664 |
| 100 | 0.097123 | 0.110493 | 0.110501 | 0.110504 | 0.002235 | 0.002234 | 0.002234 | -11.601074 | -11.601838 |
| 1000 | 0.031343 | 0.035456 | 0.035457 | 0.035457 | 0.000226 | 0.000226 | 0.000226 | -35.986413 | -35.987203 |
| 10000 | 0.009974 | 0.011263 | 0.011263 | 0.011263 | 0.000023 | 0.000023 | 0.000023 | -113.118759 | -113.118645 |

1. The value of r at which the variational mean two-particle distribution function is maximum.
2. The mean distance $r$ calculated using the variational method.
3. The mean distance $r$ calculated using the FEM method.
4. The mean distance $r$ calculated using the MC-HF method.
5. The variance of the distance $r$ calculated using the variational method.
6. The variance of the distance $r$ calculated using the FEM method.
7. The variance of the distance $r$ calculated using the MC-HF method.
8. The e-PCP interaction energy obtained through the variational two-particle distribution function.
9. The e-PCP interaction energy obtained through the FEM two-particle distribution function.

## 5.2 E-PCP Correlation in Density Functional Theory

Since the exact wavefunction is not available for most physical systems, the conventional approach in density functional theory is to solve the Kohn-Sham equations using an approximate but suitable exchange-correlation functional to obtain the Kohn-Sham orbitals and ultimately the total energy and densities. However, fortunately, the EHM (like the harmonium model) falls into the rare category of systems for which we have the exact solution of the Schrödinger equation and the corresponding single-particle densities. Therefore, we can take a reverse approach compared to the usual method: First, using equations (1-54) and (1-55), we obtain the Kohn-Sham single-orbital for the electron and the Kohn-Sham single-orbital for the PCP. Then, by substituting them into the Kohn-Sham equations (equations 1-65 and 1-66), we calculate the e-PCP correlation potential (equations 1-67 and 1-68).

The correlation potential provides deeper insight into the nature of correlation within the DFT framework. Therefore, in this section, we first obtain the exact correlation potentials of the electron and the PCP for the EHM using the inversion of Kohn-Sham equations. Then, we compute the potentials using existing e-PCP correlation functionals and compare them with their exact counterparts. This way, in addition to analyzing the exact correlation potentials and interpreting their behavior, we can gain a better understanding of the reasons behind the success or failure of existing e-PCP correlation functionals.

### 5.2.1 Simplification of Densities

To facilitate calculations, and since the analytical form of the electron and PCP densities (equations 2-168 and 2-169, respectively) for performing calculations within the framework of two-component density functional theory for muons and protons are complex, their simplification was considered. The strategy used for simplification involves identifying the dominant terms and approximating the density with them. Accordingly, after determining the dominant term in the PCP density and calculating the normalization constant, the simplified density ($\rho_p^{sim}$) was obtained as follows:

$$\rho_p^{sim}(\boldsymbol{r}_p) = \frac{2\sqrt{2}\gamma^{3/2}}{\pi^{3/2}} e^{-2{r_p}^2\gamma} \qquad 5.6$$

However, for the analytical electron density, the constituent terms contribute almost equally to the final density, rendering simplification through identifying the dominant term ineffective. Another strategy for simplifying the electron density is to fit it with well-behaved known functions. Therefore, the electron densities and their square roots were fitted with several Gaussian functions, which were then used for working with the two-component DFT equations presented in section 1-7-1. Additionally, in all the following calculations, the system parameters were limited to the following two cases:

Table 5-4: Masses and frequencies used in DFT calculations.

| HO Frequency | PCP mass |
|---|---|
| 0.02 | 207 |
| 0.01 | 1836 |

Fitting the electron densities and Kohn-Sham orbitals (which are related through equation 1-54) were performed using 4, 5, and 6 Gaussian functions over different ranges of $r_e$. Ultimately, based on the relative errors and overlap integral values for each specific fit, the following fits were selected for Table 5-4:

Table 5-5: Characteristics of the fittings performed for the simplification of single-particle densities.

| Number of points | $r_e$ range | Number of Gaussians | Fitted quantity | Frequency | PCP mass |
|---|---|---|---|---|---|
| 200 | 0-5 | 6 | density | 0.02 | 207 |
| 200 | 0-5 | 6 | KS Orbital | 0.02 | 207 |
| 200 | 0-5 | 6 | density | 0.01 | 1836 |
| 200 | 0-5 | 6 | KS Orbital | 0.01 | 1836 |

The relative error percentages for $r_e$ for the four selected fits are shown in Figure 5-39. Additionally, the relative error percentages and overlap integral values for all fits performed are presented in Table (C-1) of the appendix C. Although the relative errors were calculated for each of the 200 points used in the fits, due to space constraints, the relative error is reported only for the first 10 points with the highest errors. The linear coefficients and Gaussian exponents obtained from the fits are also available in Table (C-3) of the appendix C. It is worth mentioning that some coefficients became zero during the fitting process, effectively resulting in four Gaussian terms instead of six.

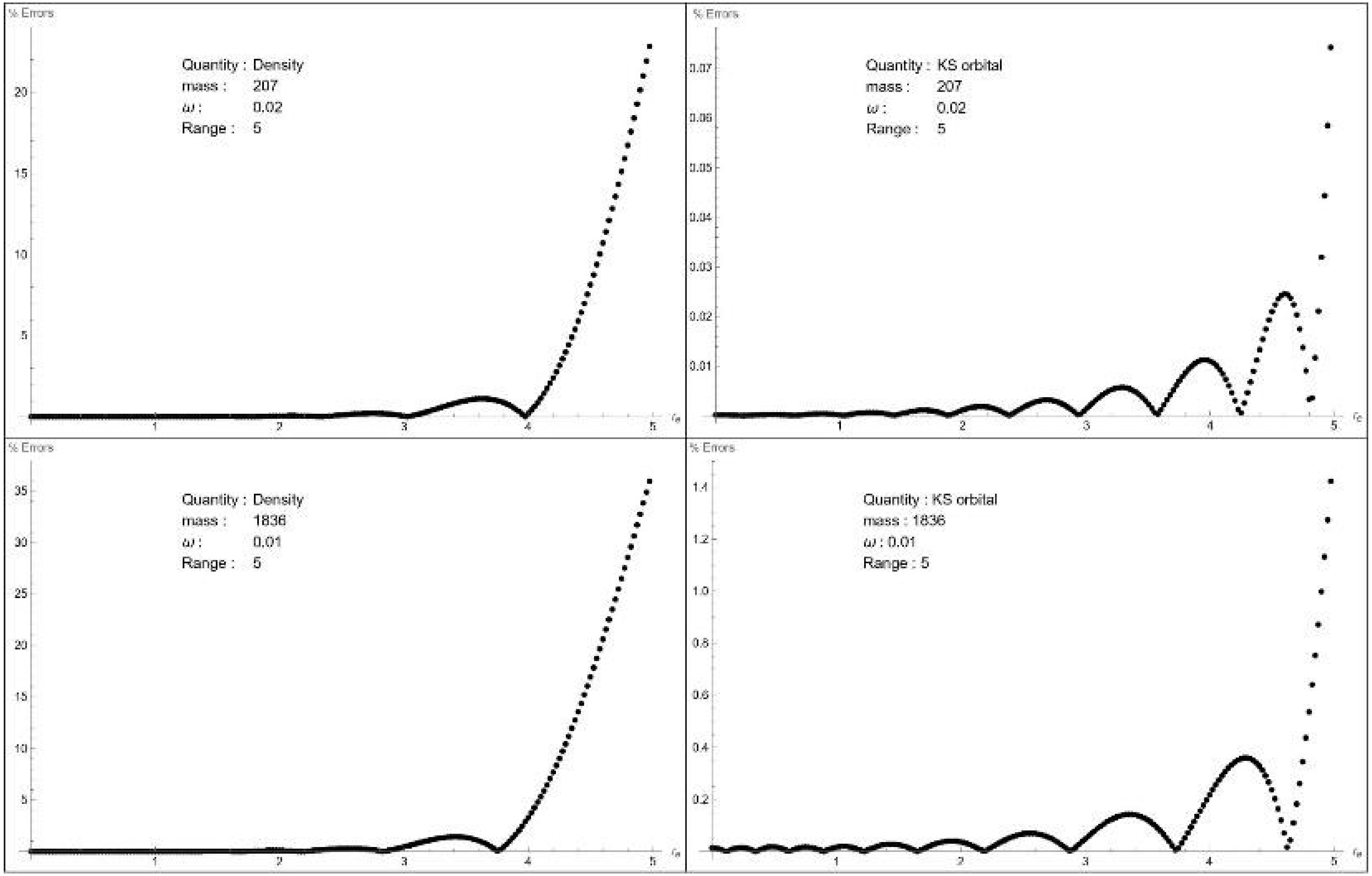

Figure 5-39: Relative error percentage versus electronic distance for the fits with four functions from Table 5-4.

The $T_e^s[\rho_e]$ and $T_p^s[\rho_p]$ (equations 1-56 and 1-57) can be calculated using different volume elements (section 2-4) as follows:

$$\langle \hat{T}_e^s \rangle = \int\int \Psi^{ks}(\boldsymbol{r}_e,\boldsymbol{r}_p)^* \hat{T}_e^s \Psi^{ks}(\boldsymbol{r}_e,\boldsymbol{r}_p)\, d\boldsymbol{r}_e\, d\boldsymbol{r}_p$$
$$= -\frac{1}{2\,m_e}\int\int \Psi^{ks}(\boldsymbol{r}_e,\boldsymbol{r}_p)^* \nabla_e^2\, \Psi^{ks}(\boldsymbol{r}_e,\boldsymbol{r}_p)\, d\boldsymbol{r}_e\, d\boldsymbol{r}_p$$
$$= -\frac{16\,\pi^2}{2\,m_e}\int\int \phi_e^{ks}(\boldsymbol{r}_e)^* \nabla_e^2\, \phi_e^{ks}(\boldsymbol{r}_e)\, \left|\phi_p^{ks}(\boldsymbol{r}_p)\right|^2 r_e^2 r_p^2\, dr_e\, dr_p \qquad 5.7$$
$$= -\frac{8\pi^2}{2\,m_e}\int_0^\infty\int_0^\infty\int_{|r_e-r|}^{r_e+r} \phi_e^{ks}(\boldsymbol{r}_e)^* \nabla_e^2\, \phi_e^{ks}(\boldsymbol{r}_e)\, \left|\phi_p^{ks}(\boldsymbol{r}_p)\right|^2 r_e r_p r\;\; dr_p dr_e dr$$
$$= -\frac{4\pi}{2\,m_e}\int_0^\infty \phi_e^{ks}(\boldsymbol{r}_e)^* \nabla_e^2\, \phi_e^{ks}(\boldsymbol{r}_e)\, r_e^2\, dr_e$$

$$\langle \hat{T}_p^s \rangle = \int\int \Psi^{ks}(\boldsymbol{r}_e,\boldsymbol{r}_p)^* \hat{T}_p^s \Psi^{ks}(\boldsymbol{r}_e,\boldsymbol{r}_p)\, d\boldsymbol{r}_e\, d\boldsymbol{r}_p$$
$$= -\frac{1}{2\,m_p}\int\int \Psi^{ks}(\boldsymbol{r}_e,\boldsymbol{r}_p)^* \nabla_p^2\, \Psi^{ks}(\boldsymbol{r}_e,\boldsymbol{r}_p)\, d\boldsymbol{r}_e\, d\boldsymbol{r}_p$$
$$= -\frac{16\,\pi^2}{2\,m_p}\int\int \phi_p^{ks}(\boldsymbol{r}_p)^* \nabla_p^2\, \phi_p^{ks}(\boldsymbol{r}_p)\, |\phi_e^{ks}(\boldsymbol{r}_e)|^2 r_e^2 r_p^2\, dr_e\, dr_p \qquad 5.8$$
$$= -\frac{8\pi^2}{2\,m_p}\int_0^\infty\int_0^\infty\int_{|r_p-r|}^{r_p+r} \phi_p^{ks}(\boldsymbol{r}_p)^* \nabla_p^2\, \phi_p^{ks}(\boldsymbol{r}_p)\, |\phi_e^{ks}(\boldsymbol{r}_e)|^2 r_e r_p r\;\; dr_e dr_p dr$$
$$= -\frac{4\pi}{2\,m_p}\int_0^\infty \phi_p^{ks}(\boldsymbol{r}_p)^* \nabla_p^2\, \phi_p^{ks}(\boldsymbol{r}_p)\, r_p^2\, dr_p$$

The external potentials in the EHM for the electron and PCP are defined as follows:

$$v_e^{ext}(\boldsymbol{r_e}) = \frac{1}{2} m_e \omega^2 r_e^2 \qquad 5.9$$

$$v_p^{ext}(\boldsymbol{r_p}) = \frac{1}{2} m_p \omega^2 r_p^2 \qquad 5.10$$

Thus, the energy components of the EHM within the DFT were calculated using the fitted densities and are presented in Table 5-6. Furthermore, to verify previous results, and since the reference system densities are the same as the actual system, the Hartree energy term $J_{ep}[\rho_e,\rho_p]$ (equation 1-62) was also calculated using the simplified densities. The data in Table 5-6 show that the Hartree energy calculated using the fitted orbitals matches the exact data more closely, and their normalization constant is nearer to one, indicating their greater accuracy compared to the data obtained from direct density fitting. The contribution of the kinetic correlation energy terms $T_e^c$ and $T_p^c$ (equation 3-29) for electrons is much larger than for muons and protons, and in both cases, they constitute the main component of the total kinetic correlation energy. Additionally, the absolute value of the KS correlation energy for the system comprising a muon and an electron is more than twice that of the system comprising a proton and an electron. Thus, as the mass of the PCP decreases, the absolute value of the KS correlation energy, like the wavefunction-based correlation energy discussed in previous sections, increases. On the other hand, the absolute value of the KS correlation energy is smaller by a factor of two compared to the reference-independent correlation energy.

Table 5-6: Quantities Calculated within the DFT

| Quantities/m | 1836 | 1836 | 207 | 207 |
|---|---|---|---|---|
| ω | 0.01 | 0.01 | 0.02 | 0.02 |
| Power[1] | 1 | 0.5 | 1 | 0.5 |
| Norm P[2] | 1 | 1 | 1 | 1 |
| Norm E[3] | 0.9970 | 0.9995 | 0.9979 | 0.9998 |
| Exact HAR[4] | -0.9315 | -0.9315 | -0.8035 | -0.8035 |
| KS HAR[5] | -0.9309 | -0.9314 | -0.8031 | -0.8035 |
| ΔHAR[6] | -0.0006 | -0.0001 | -0.0004 | 0.0000 |
| TSE1[7] | 0.4659 | 0.4649 | 0.3952 | 0.3944 |
| TSE2[8] | 0.4659 | 0.4649 | 0.3952 | 0.3944 |
| TSE3[9] | 0.4659 | 0.4649 | 0.3952 | 0.3944 |
| TSP1[10] | 0.0075 | 0.0075 | 0.0150 | 0.0151 |
| TSP2[11] | 0.0075 | 0.0075 | 0.0150 | 0.0151 |
| TSP3[12] | 0.0075 | 0.0075 | 0.0151 | 0.0151 |
| EXT E[13] | 0.0001 | 0.0002 | 0.0007 | 0.0007 |
| EXT P[14] | 0.0075 | 0.0075 | 0.0149 | 0.0149 |
| TE[15] | 0.4998 | 0.4998 | 0.4965 | 0.4965 |
| TP[16] | 0.0078 | 0.0078 | 0.0173 | 0.0173 |
| TC E[17] | 0.0339 | 0.0349 | 0.1013 | 0.1021 |
| TC P[18] | 0.0003 | 0.0003 | 0.0023 | 0.0023 |
| IND Corr[19] | -0.0683 | -0.0683 | -0.1929 | -0.1929 |
| KS Corr[20] | -0.0341 | -0.0331 | -0.0893 | -0.0885 |
| Exact Etot[21] | -0.4846 | -0.4846 | -0.4670 | -0.4670 |
| KS Etot[22] | -0.4846 | -0.4846 | -0.4670 | -0.4670 |
| ΔE[23] | 0.0000 | 0.0000 | 0.0000 | 0.0000 |

1. Density term power in the fit (1 means density, and 0.5 means KS orbital).
2. Normalization condition for electron density and KS orbital (equation 1-52).
3. Normalization condition for PCP density and KS orbital (equation 1-53).
4. Exact Hartree energy calculated using analytical densities (equation 1-62).
5. Hartree energy calculated using fitted densities (equation 1-62).
6. Difference between items 4 and 5.
7. Electron kinetic energy in the reference system using the third line of equation 5-7.
8. Electron kinetic energy in the reference system using the fourth line of equation 5-7.
9. Electron kinetic energy in the reference system using the fifth line of equation 5-7.
10. PCP kinetic energy in the reference system using the third line of equation 5-8.
11. PCP kinetic energy in the reference system using the fourth line of equation 5-8.

12. PCP kinetic energy in the reference system using the fifth line of equation 5-8.
13. Electron external potential energy (equation 5-9, comparable with item 5 of Table 4-4).
14. PCP external potential energy (equation 5-10, comparable with item 6 of Table 4-4).
15. Exact electron kinetic energy (equation 2-135 and item 2 of Table 4-4).
16. Exact PCP kinetic energy (equation 2-136 and item 3 of Table 4-4).
17. Electron kinetic correlation energy (first parenthesis in equation 3-29).
18. PCP kinetic correlation energy (second parenthesis in equation 3-29).
19. Reference-independent correlation energy (equation 3-24 and item 4 of Table 5-2).
20. Correlation energy within the DFT (equation 3-29).
21. Exact total energy (item 1 in Table 4-4).
22. Total energy of the model within the DFT (equation 3-26).
23. Difference between items 21 and 22.

### 5.2.2 Calculating Correlation Potentials via Inversion of Kohn-Sham Equations

The effective potentials in the Kohn-Sham equations (1-67) and (1-68) for the EHM reduce to the following forms:

$$V_{\text{eff}}^{e}(\boldsymbol{r}_e) = v_e^{ext}(\boldsymbol{r}_e) + v_e^{J_{ep}}(\boldsymbol{r}_e) + v_e^{epc}(\boldsymbol{r}_e) \qquad 5.11$$

$$V_{\text{eff}}^{p}(\boldsymbol{r}_p) = v_p^{ext}(\boldsymbol{r}_p) + v_p^{J_{ep}}(\boldsymbol{r}_p) + v_p^{epc}(\boldsymbol{r}_p) \qquad 5.12$$

where $v_e^{J_{ep}}$ and $v_e^{ext}$ can be calculated using equations (3-22) and (5-9) respectively, and $v_p^{J_{ep}}$ and $v_p^{ext}$ can be calculated using equations (3-23) and (5-10). By inverting the Kohn-Sham equations for the electron and the PCP (equations 1-65 and 1-66), the Kohn-Sham correlation potentials can be calculated as follows:

$$v_e^{epc}(\boldsymbol{r}_e) = \epsilon_e^{ks} - v_e^{ext}(\boldsymbol{r}_e) - v_e^{J_{ep}}(\boldsymbol{r}_e) - v_e^{KE}(\boldsymbol{r}_e) \qquad 5.13$$

$$v_p^{epc}(\boldsymbol{r}_p) = \epsilon_p^{ks} - v_p^{ext}(\boldsymbol{r}_p) - v_p^{J_{ep}}(\boldsymbol{r}_p) - v_p^{KE}(\boldsymbol{r}_p) \qquad 5.14$$

where the following terms are introduced as kinetic energy potentials:

$$v_e^{KE}(\boldsymbol{r}_e) = -\frac{1}{2m_e}\frac{\nabla_e^2\, \phi_e^{ks}(\boldsymbol{r}_e)}{\phi_e^{ks}(\boldsymbol{r}_e)} \qquad 5.15$$

$$v_p^{KE}(\boldsymbol{r}_p) = -\frac{1}{2m_p}\frac{\nabla_p^2\, \phi_p^{ks}(\boldsymbol{r}_p)}{\phi_p^{ks}(\boldsymbol{r}_p)} \qquad 5.16$$

Thus, the only unknown quantities in equations (5-13) and (5-14) are the orbital energies $\epsilon_e^{ks}$ and $\epsilon_p^{ks}$, which we can obtain by analyzing and fitting the asymptotic behavior of the correlation potential components and the condition of this potential approaching zero at infinity. It's important to note that these energies can be analytically calculated if the exact wavefunction of the system is available, by examining the asymptotic behavior of the correlation potential [83], [141]. However, since the wavefunction presented here is obtained through the variational theorem and is not the exact wavefunction of the system, this method cannot be used. Nevertheless, a similar approach can be employed to estimate the approximate values of the orbital energies.

The correlation potential's dependency is determined by the three components presented in equations (5-11) and (5-12): kinetic energy potential, external potential, and Hartree potential. The first condition regarding the dependency of the correlation potential is that it approaches zero at infinite distance. By using this condition and analyzing the asymptotic behavior of each component of the correlation potential, we can estimate the orbital energy. The simplest component is the external potential, which here is equal to the potential of a harmonic oscillator and for the electron and PCP is given by equations (5-9) and (5-10) respectively, in the form of $\lambda r^2$ where $\lambda = \frac{1}{2}m\omega^2$, thus having a $r^2$ dependency. The next component, the Hartree potential, has a $-\frac{1}{r}$ dependency at long inter-particle distances and hence should approach zero at infinite distance, as shown in Figure (5-40). Given that the density of the PCP is more localized compared to the electron density, and also considering the presence of the second particle's density in the Hartree potential of the first particle, the electron's Hartree potential reaches the $-\frac{1}{r}$ dependency faster, whereas this occurs at further distances for the PCPs.

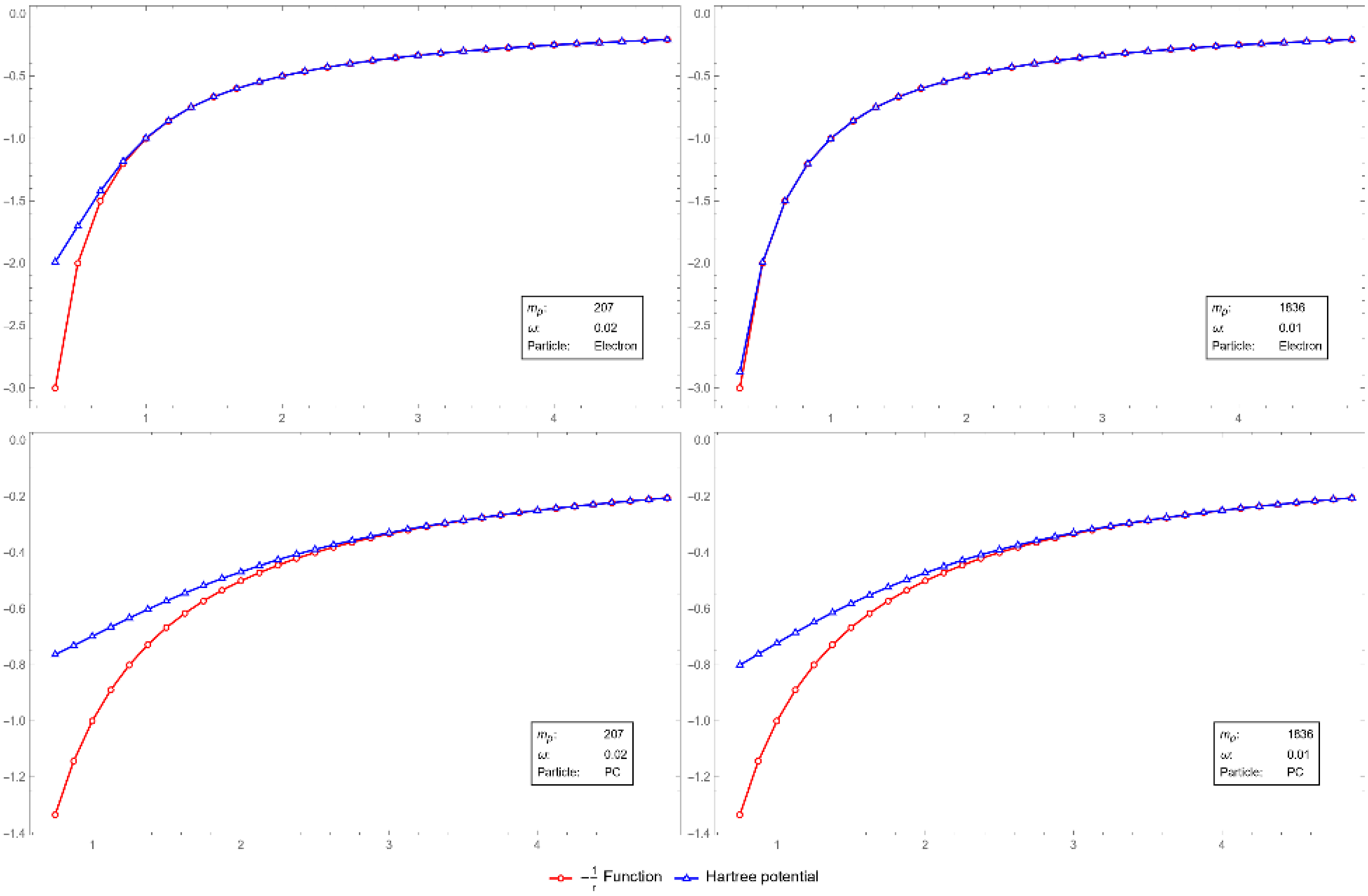

Figure 5-40: Consistency of the asymptotic behavior of the Hartree potential with the $-\frac{1}{r}$ potential.

The third component of the correlation potential is the kinetic energy potential, whose dependency on distance is not explicitly known. However, the condition that the correlation potential approaches zero at infinite distance can help determine its dependency and estimate the orbital energy. Given that both the correlation and Hartree potentials approach zero at infinite distance, we have:

$$\epsilon_e^{ks} = \lim_{r_e \to \infty} \left( v_e^{ext}(\boldsymbol{r}_e) + v_e^{asym-KE}[\rho_e] \right) \qquad 5.17$$

$$\epsilon_p^{ks} = \lim_{r_p \to \infty} \left( v_p^{ext}(\boldsymbol{r}_p) + v_p^{asymp-KE}[\rho_p] \right) \qquad 5.18$$

Given that we know the values $\epsilon_e^{ks}$ and $\epsilon_p^{ks}$ are numerical constants, the dependency of the kinetic energy potential plus a numerical constant (equivalent to the orbital energy) must be such that the sum of the harmonic oscillator potential and the orbital energy equals zero at infinite distance; that is:

$$v_e^{ext} - \epsilon_e^{ks} + (v_e^{asyKE} + \epsilon_e^{ks}) = 0 \qquad 5.19$$

$$v_p^{ext} - \epsilon_p^{ks} + (v_p^{asyKE} + \epsilon_p^{ks}) = 0 \qquad 5.20$$

Thus, the kinetic energy dependency at long distances must be in the form of $-\lambda' r^2$, where for the exact wavefunction, $\lambda' = \lambda$, and these two terms completely cancel each other out. However, for the variational wavefunction, we know that there is a slight difference between the coefficients:

$$v_e^{ext} + v_e^{asyKE} = (\lambda_e - \lambda_e')r^2 + \epsilon_e^{ks} \qquad 5.21$$

$$v_p^{ext} + v_p^{asyKE} = (\lambda_p - \lambda_p')r^2 + \epsilon_p^{ks} \qquad 5.22$$

As a result, by fitting the kinetic energy potential data to a function of the form $-ar^2 - b$, we can estimate the orbital energy, which is a numerical constant and ensures the correlation potential approaches zero at infinite distance.

In Figure 5-41, the data for the best fit of the kinetic energy potential with the function $-ar^2 - b$ in different systems and the quantities related to the fit quality are presented. As shown in the graphs, the fit quality for the PCP data is significantly better than that for the electron data, due to the oscillatory behavior of the PCP. Given that the density of the PCP can be very accurately approximated by a Gaussian function (unlike the electron), its behavior here is also similar to that of a harmonic oscillator. Therefore, the fit of the kinetic energy potential over all distances for the PCP has very high quality, whereas for the electron, the fit reaches acceptable quality only at long distances.

Considering the above explanations, the coefficients of the function ($a$ or $\lambda'$) should have values close to $\frac{1}{2}m\omega^2$, which is evident in the values presented in the tables of Figure 5-41. Furthermore, the numerical constants ($b$) or orbital energies for the PCP, which behaves similarly to a harmonic oscillator, should practically be equal to its zero-point energy ($\frac{3}{2}\omega$). The expected values of $a$ and $b$ are shown in the "Expectation" column of the tables in Figure 5-41.

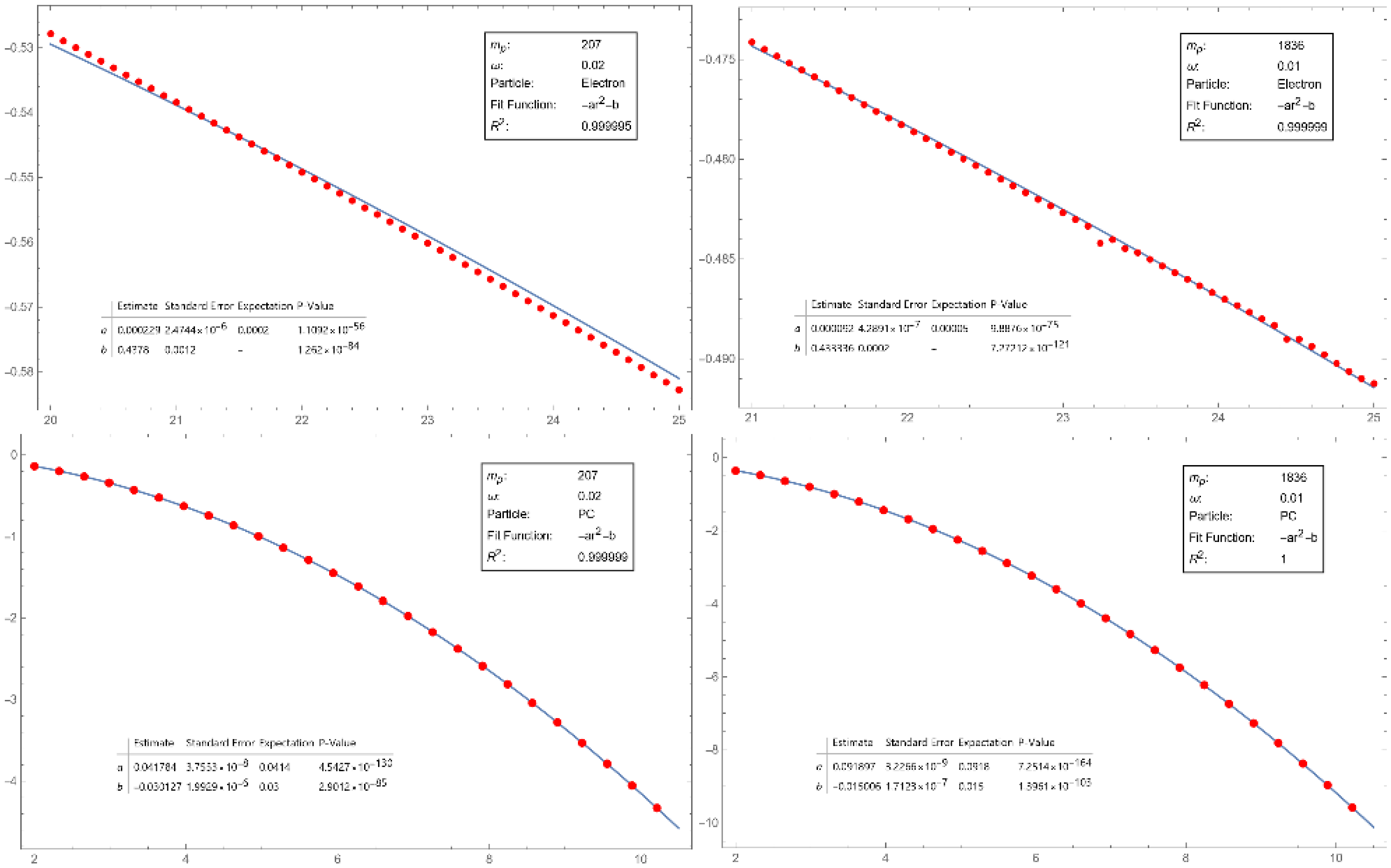

Figure 5-41: Fitting of the kinetic energy potential (dots and lines represent actual data and fitted functions, respectively).

By obtaining the orbital energies, the correlation potentials can be calculated using both FEM densities and variational (fitted) densities, as presented and compared in Figure 5-42. In all cases, there is a very high degree of consistency between them. The qualitative difference in the behavior of the correlation potentials for electrons and PCPs is quite evident in this figure.

It is noteworthy that, to ensure the correlation potential approaches zero at infinite distance, a value of $-0.06$ was added to the electron orbital energies obtained from the fit, which was $-0.43$ for both systems, resulting in a final electron orbital energy of $-0.49$ for both systems. This value is very close to the ground state energy of an electron in a free hydrogen atom.

To conduct a more precise analysis of the Kohn-Sham effective potential (the sum of Hartree, external, and correlation potentials), its components are displayed in Figure 5-43. In the Kohn-Sham effective potentials for electrons, the Hartree potential is the dominant term and controls the overall shape of the potential. This control increases with the mass of the PCP. The correlation potential, and especially the external potential, play a minor role in determining the shape of the total potential. Hence, the presence or absence of an e-PCP correlation functional has little impact on the shape of the Kohn-Sham effective potential for electrons.

In contrast, the role of the correlation potential in the Kohn-Sham potentials for the PCP is highly significant. In fact, the correlation potential neutralizes the effect of the Hartree potential, causing the final shape of the Kohn-Sham potential to align completely with the external potential (harmonic oscillator), as previous data confirm that the PCP behaves exactly like a harmonic oscillator. Therefore, the presence of an e-PCP correlation functional in the Kohn-Sham equations for the PCP is crucial.

The accuracy of the obtained orbital energies can also be verified using the equations related to the sum of orbital energy values. By extending the standard equations of DFT to their two-component versions, we get:

$$\sum_{i}^{N} \epsilon_e^{ks} = T_e^s[\rho_e] + \int \rho_e(\boldsymbol{r_e})\, V_{\text{eff}}^{e}(\boldsymbol{r}_e)\; d\boldsymbol{r}_e \qquad 5.23$$

$$\sum_{i'}^{N'} \epsilon_p^{ks} = T_p^s[\rho_p] + \int \rho_p(\boldsymbol{r_p})\, V_{\text{eff}}^{p}(\boldsymbol{r}_p)\; d\boldsymbol{r}_p \qquad 5.24$$

Considering that the functions $V_{\text{eff}}^{e}(\boldsymbol{r}_e)$ and $V_{\text{eff}}^{p}(\boldsymbol{r}_p)$ are themselves the sum of several potentials, integrating them leads to numerical difficulties. Therefore, we first fitted them using six Gaussian functions, and then used the result of this fitting in the above equations. The fitting details and results are shown in Table 5-7, which confirm the accuracy of the orbital energies obtained from the previous method. It is also worth noting that during the fitting process, some coefficients of the functions became zero, resulting in only four Gaussian functions in the final result.

Table 5-7: Details of fitting effective potentials and predicted and obtained orbital energy values

| Particle/Quantity | $m_p$ | $\boldsymbol{\omega}$ | Fitted $V_{eff}$ | $R^2$ of fitting[1] | $V_{eff}\,\boldsymbol{\rho}$ Integral[2] | $T_s$ [3] | Obtained $\boldsymbol{\epsilon}$ [4] | Expected $\boldsymbol{\epsilon}$ [5] |
|---|---|---|---|---|---|---|---|---|
| Electron | 207 | 0.02 | $-1.1828198631201177e^{-1.6958106783662794x^2}$ $-0.6030099153338793e^{-0.5137363920720693x^2}$ $-0.345016638782312e^{-0.13227251840688656x^2}$ $-0.2580958806954125e^{-0.014719066831130966x^2}$ | 0.9999996 | −0.8842 | 0.3944 | −0.4898 | -0.49 |
| PC | 207 | 0.02 | $-0.3730298834254647e^{-0.059013127132755056x^2}$ $+0.37286336846293927e^{0.05342619762675789x^2}$ | 0.9999998 | 0.0149 | 0.0151 | 0.0300 | 0.03 |
| Electron | 1836 | 0.01 | $-2.3691654995926226e^{-9.064309621783845x^2}$ $-1.4389052306037728e^{-2.175416017302567x^2}$ $-0.6836181104321617e^{-0.3893475435267434x^2}$ $-0.4079586697626854e^{-0.03108796523905667x^2}$ | 0.9999951 | −0.9595 | 0.4649 | −0.4946 | 0.-49 |
| PC | 1836 | 0.01 | $-0.905641219740176e^{-0.05317938771639458x^2}$ $+0.9055837583856249e^{0.04859194273978422 6x^2}$ | 0.9999999 | 0.0074 | 0.0075 | 0.0149 | 0.015 |

1. Coefficient of determination for fitting effective potential functions.
2. Result of the two integrals in equations (5-23) and (5-24).
3. Kohn-Sham kinetic energy values (items 7 to 13 in Table 5-6).
4. Final obtained values for orbital energies using equations (5-23) and (5-24).
5. Expected values for orbital energies (obtained from fitting the kinetic energy potentials).

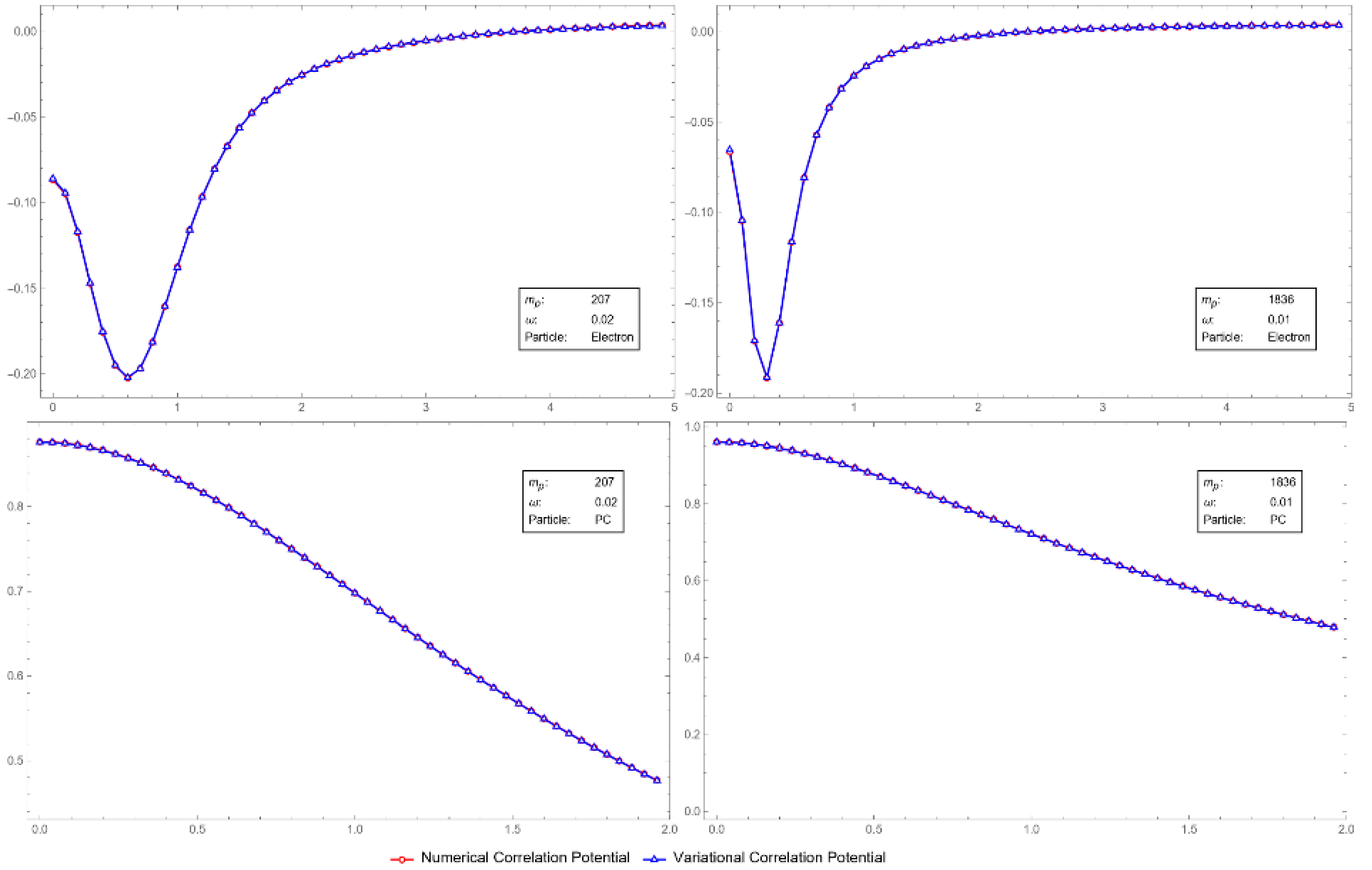

Figure 5-42: Comparison of FEM and variational correlation potentials

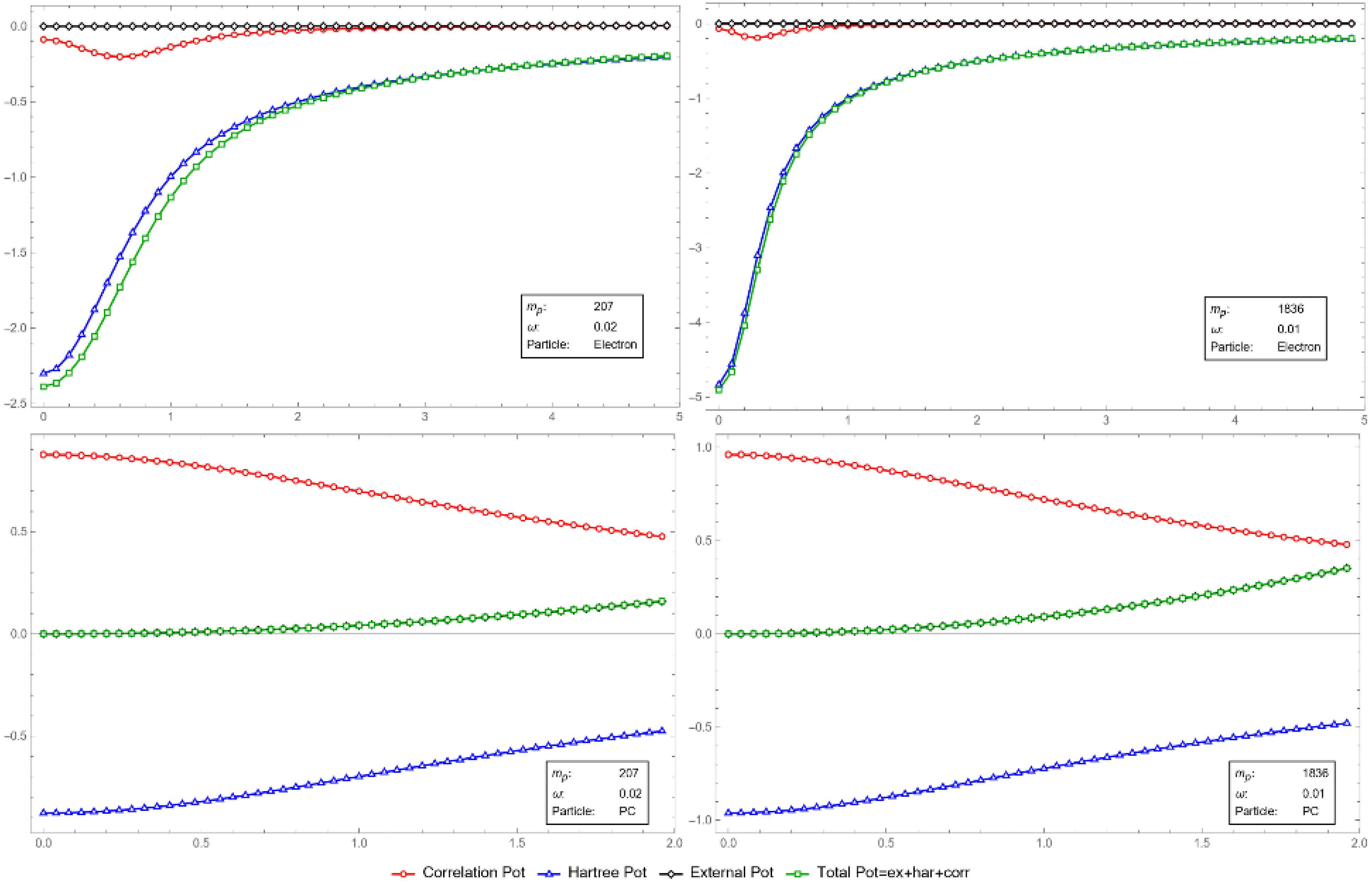

Figure 5-43: Comparison of different components of the Kohn-Sham effective potential

### 5.2.3 Evaluating the Efficiency of Existing Electron-PCP Functionals

In this section, the accuracy of five proposed local functionals for calculating the e-PCP correlation energy is examined. The parameters of four of them ($epc17\text{-}1$, $epc17\text{-}2, epc18\text{-}1, epc18\text{-}2$) have been optimized for electron-proton correlation, and one of them ($e\mu c1$) has been designed for electron-muon correlation. The explicit forms of the electron-proton correlation functionals are as follows:

$$E_{epc17}[\rho_e, \rho_p] = -\int \frac{\rho_e \rho_p}{a - b(\rho_e)^{1/2}(\rho_p)^{1/2} + c\rho_e\rho_p} d\boldsymbol{r} \qquad 5.25$$

$$E_{epc18}[\rho_e, \rho_p] = -\int \frac{\rho_e \rho_p}{\left(a - b + ({\rho_e}^{1/3} + {\rho_p}^{1/3})^3 + c\left({\rho_e}^{1/3} + {\rho_p}^{1/3}\right)^6\right)} d\boldsymbol{r} \qquad 5.26$$

where their forms are the same for both open-shell and closed-shell cases. For the two functional forms $\boldsymbol{epc}17$ and $\boldsymbol{epc}18$, two different sets of parameters are introduced, each set making the functional describe a specific observable more accurately. The parameters for these four functionals are provided in the table below:

Table 5-8: Parameters for electron-proton correlation functionals

| Functional/parameter | a | b | c |
|---|---|---|---|
| Epc17-1 | 2.35 | 2.4 | 3.2 |
| Epc17-2 | 2.35 | 2.4 | 6.6 |
| Epc18-1 | 1.8 | 0.1 | 0.03 |
| Epc18-2 | 3.9 | 0.5 | 0.06 |

The form of electron-muon correlation functional in the closed-shell case is as follows:

$$E_{e\mu c1-cs}[\rho_e, \rho_p] = -\int \frac{2\rho_e\rho_p - \rho_e(\rho_p)^{3/2}}{1 + 4\rho_e\rho_p + 2\rho_e(\rho_p)^{3/2}} d\boldsymbol{r} \qquad 5.27$$

while the open-shell case is as follows:

$$E_{e\mu c1-os}\left[\rho_e^\alpha, \rho_e^\beta, \rho_\mu\right] = -\int \frac{2\rho_e^\alpha\rho_\mu - \rho_e^\alpha(\rho_\mu)^{3/2}}{1 + 8\rho_e^\alpha\rho_\mu + 4\rho_e^\alpha(\rho_\mu)^{3/2}} + \frac{2\rho_e^\beta\rho_\mu - \rho_e^\beta(\rho_\mu)^{3/2}}{1 + 8\rho_e^\beta\rho_\mu + 4\rho_e^\beta(\rho_\mu)^{3/2}} d\boldsymbol{r} \qquad 5.28$$

where $\rho_e^\alpha$ and $\rho_e^\beta$ are the spin-up and spin-down electron densities, respectively. The following convention holds for all the above functionals:

$$\int d\boldsymbol{r} = \int d\boldsymbol{r_e} \int d\boldsymbol{r_p}\, \delta(\boldsymbol{r_e} - \boldsymbol{r_p}) \qquad 5.29$$

The e-PCP correlation energy was calculated using the five functionals introduced above, employing the exact one-particle densities. The results are presented in Table 5-9.

Table 5-9: Comparison of independent and dependent correlation energies with values obtained from 5 functionals

| Corr Energy/mp | 207 | 1836 |
|---|---|---|
| omega | 0.02 | 0.01 |
| epc17-1 | -0.0406 | -0.0564 |
| epc17-2 | -0.0367 | -0.0345 |
| epc18-1 | -0.0461 | -0.0427 |
| epc18-2 | -0.0257 | -0.0292 |
| e**μ**c1(OS) | -0.0699 | -0.0191 |
| KS | -0.0885 | -0.0331 |
| HF | -0.0592 | -0.0218 |
| IND | -0.1929 | -0.0683 |

In the case of the muon system, the *eμc-1* functional clearly has greater accuracy in reproducing the exact Kohn-Sham correlation energy compared to the other functionals. Meanwhile, for the proton system, the *epc17-2* and *epc18-2* functionals show a better match with the Kohn-Sham correlation energy. This aligns well with the idea that the parameters of these functionals have been optimized to accurately reproduce the total system energy. It appears that the designed functionals are generally successful in reproducing the correlation energy.

Additionally, the Kohn-Sham correlation energy values for a mass of 1836 at frequencies of 0.1 and 0.001 were found to be −0.0049 and −0.1380, respectively, indicating that the Kohn-Sham correlation energy is sensitive to the field frequency (ω), and the matching of correlation energies at a frequency of 0.01 is not coincidental. Therefore, the existing functionals are efficient in reproducing the correlation energy.

### 5.2.4 Calculating the Correlation Potentials of the Functionals

In this section, we extract the correlation potentials from the functionals presented in equations (5-25) to (5-28) and compare them with the analytical potentials (equations 5-13 and 5-14) to evaluate their accuracy. By differentiating the above-mentioned functionals with respect to the electron and PCP densities, we obtain the electron and PCP correlation potentials, respectively:

$$v_e^{epc}(\boldsymbol{r}_e) = \frac{\delta E_{epc}[\rho_e, \rho_p]}{\delta \rho_e} \tag{5.30}$$

$$v_p^{epc}(\boldsymbol{r}_p) = \frac{\delta E_{epc}[\rho_e, \rho_p]}{\delta \rho_p} \tag{5.31}$$

Generally, the derivative of a functional of the form below can be calculated as shown in equation (5-33) [77].

$$F[\rho] = \int f(x, \rho, \rho^{(1)}, \rho^{(2)}, \dots, \rho^{(n)})\, dx \tag{5.32}$$

$$\frac{\delta F}{\delta \rho(x)} = \frac{\partial f}{\partial \rho} - \frac{d}{dx}\left(\frac{\partial f}{\partial \rho^{(1)}}\right) + \frac{d^2}{dx^2}\left(\frac{\partial f}{\partial \rho^{(2)}}\right) - \dots + (-1)^n \frac{d^n}{dx^n}\left(\frac{\partial f}{\partial \rho^{(n)}}\right) \tag{5.33}$$

In the calculation of the electron and PCP correlation potentials from the above-mentioned local functionals, only the first term in the above expression has a non-zero value. The forms of the correlation potentials obtained from the functionals are as follows:

$$v_e^{epc171}(\boldsymbol{r}_e) = \frac{b_{171}\sqrt{\rho_e(r_e)}\rho_p(r_e)^{3/2} - 2a_{171}\rho_p(r_e)}{2(a_{171} - b_{171}\sqrt{\rho_e(r_e)}\sqrt{\rho_p(r_e)} + c_{171}\rho_e(r_e)\rho_p(r_e))^2} \quad 5.34$$

$$v_e^{epc172}(\boldsymbol{r}_e) = \frac{b_{172}\sqrt{\rho_e(r_e)}\rho_p(r_e)^{3/2} - 2a_{172}\rho_p(r_e)}{2(a_{172} - b_{172}\sqrt{\rho_e(r_e)}\sqrt{\rho_p(r_e)} + c_{172}\rho_e(r_e)\rho_p(r_e))^2} \quad 5.35$$

$$\begin{aligned} v_e^{epc181}(\boldsymbol{r}_e) = (a_{181} & + (\sqrt[3]{\rho_e(r_e)} + \sqrt[3]{\rho_p(r_e)})^3(c_{181}(\sqrt[3]{\rho_e(r_e)} \\ & + \sqrt[3]{\rho_p(r_e)})^3 - b_{181}))^{-2}\Big(\rho_p(r_e)(-a_{181} \\ & + b_{181}\sqrt[3]{\rho_p(r_e)}(\sqrt[3]{\rho_e(r_e)} + \sqrt[3]{\rho_p(r_e)})^2 \\ & + c_{181}(\sqrt[3]{\rho_e(r_e)} \\ & - \sqrt[3]{\rho_p(r_e)})(\sqrt[3]{\rho_e(r_e)} + \sqrt[3]{\rho_p(r_e)})^5)\Big) \end{aligned} \quad 5.36$$

$$\begin{aligned} v_e^{epc182}(\boldsymbol{r}_e) = (a_{182} & + (\sqrt[3]{\rho_e(r_e)} + \sqrt[3]{\rho_p(r_e)})^3(c_{182}(\sqrt[3]{\rho_e(r_e)} \\ & + \sqrt[3]{\rho_p(r_e)})^3 - b_{182}))^{-2}\Big(\rho_p(r_e)(-a_{182} \\ & + b_{182}\sqrt[3]{\rho_p(r_e)}(\sqrt[3]{\rho_e(r_e)} + \sqrt[3]{\rho_p(r_e)})^2 \\ & + c_{182}(\sqrt[3]{\rho_e(r_e)} \\ & - \sqrt[3]{\rho_p(r_e)})(\sqrt[3]{\rho_e(r_e)} + \sqrt[3]{\rho_p(r_e)})^5)\Big) \end{aligned} \quad 5.37$$

$$v_e^{e\mu c1O}(\boldsymbol{r}_e) = \frac{(\sqrt{\rho_p(r_e)} - 2)\rho_p(r_e)}{(4\rho_e(r_e)(\sqrt{\rho_p(r_e)} + 2)\rho_p(r_e) + 1)^2} \quad 5.38$$

$$v_e^{e\mu c1C}(\boldsymbol{r}_e) = \frac{(\sqrt{\rho_p(r_e)} - 2)\rho_p(r_e)}{(2\rho_e(r_e)(\sqrt{\rho_p(r_e)} + 2)\rho_p(r_e) + 1)^2} \quad 5.39$$

$$v_p^{epc171}(\boldsymbol{r}_p) = \frac{b_{17}\rho_e(r_p)^{3/2}\sqrt{\rho_p(r_p)} - 2a_{17}\rho_e(r_p)}{2(a_{17} - b_{17}\sqrt{\rho_e(r_p)}\sqrt{\rho_p(r_p)} + c_{171}\rho_e(r_p)\rho_p(r_p))^2} \quad 5.40$$

$$v_p^{epc172}(\boldsymbol{r}_p) = \frac{b_{17}\rho_e(r_p)^{3/2}\sqrt{\rho_p(r_p)} - 2a_{17}\rho_e(r_p)}{2(a_{17} - b_{17}\sqrt{\rho_e(r_p)}\sqrt{\rho_p(r_p)} + c_{172}\rho_e(r_p)\rho_p(r_p))^2} \quad 5.41$$

$$\begin{aligned} v_p^{epc181}(\boldsymbol{r}_p) = (a_{181} & \\ &+ (\sqrt[3]{\rho_e(r_p)} + \sqrt[3]{\rho_p(r_p)})^3(c_{181}(\sqrt[3]{\rho_e(r_p)} \\ &+ \sqrt[3]{\rho_p(r_p)})^3 - b_{181}))^{-2}\Big(\rho_e(r_p)(a_{181} \\ &+ (\sqrt[3]{\rho_e(r_p)} + \sqrt[3]{\rho_p(r_p)})^2(-\sqrt[3]{\rho_e(r_p)}(b_{181} \\ &+ 2c_{181}\rho_p(r_p)) + c_{181}\rho_e(r_p)^{4/3} \\ &+ 2c_{181}\rho_e(r_p)\sqrt[3]{\rho_p(r_p)} - c_{181}\rho_p(r_p)^{4/3}))\Big) \end{aligned} \quad 5.42$$

$$v_p^{epc182}(\boldsymbol{r}_p) = (a_{182} + (\sqrt[3]{\rho_e(r_p)} + \sqrt[3]{\rho_p(r_p)})^3(c_{182}(\sqrt[3]{\rho_e(r_p)} + \sqrt[3]{\rho_p(r_p)})^3 - b_{182}))^{-2}\Big(\rho_e(r_p)(a_{182} + (\sqrt[3]{\rho_e(r_p)} + \sqrt[3]{\rho_p(r_p)})^2(-\sqrt[3]{\rho_e(r_p)}(b_{182} + 2c_{182}\rho_p(r_p)) + c_{182}\rho_e(r_p)^{4/3} + 2c_{182}\rho_e(r_p)\sqrt[3]{\rho_p(r_p)} - c_{182}\rho_p(r_p)^{4/3}))\Big) \quad 5.43$$

$$v_p^{e\mu c1O}(\boldsymbol{r}_p) = -\frac{\rho_e(r_p)(-16\rho_e(r_p)\rho_p(r_p)^{3/2} - 3\sqrt{\rho_p(r_p)} + 4)}{2(4\rho_e(r_p)(\sqrt{\rho_p(r_p)} + 2)\rho_p(r_p) + 1)^2} \quad 5.44$$

$$v_p^{e\mu c1C}(\boldsymbol{r}_p) = -\frac{\rho_e(r_p)(-8\rho_e(r_p)\rho_p(r_p)^{3/2} - 3\sqrt{\rho_p(r_p)} + 4)}{2(2\rho_e(r_p)(\sqrt{\rho_p(r_p)} + 2)\rho_p(r_p) + 1)^2} \quad 5.45$$

Figures 5-44 and 5-45 show the comparison of the exact potentials with the potentials obtained from the aforementioned functionals (equations 5-34 to 5-45) for electrons and PCPs. Although these functionals reproduce the approximate shape of the exact electron correlation potential with limited accuracy, they completely fail in reproducing the exact PCP correlation potential. Consequently, it seems that the success of the functionals in reproducing the correlation energy, as mentioned in the previous section, is unrelated to their ability to accurately reproduce the exact correlation potential.

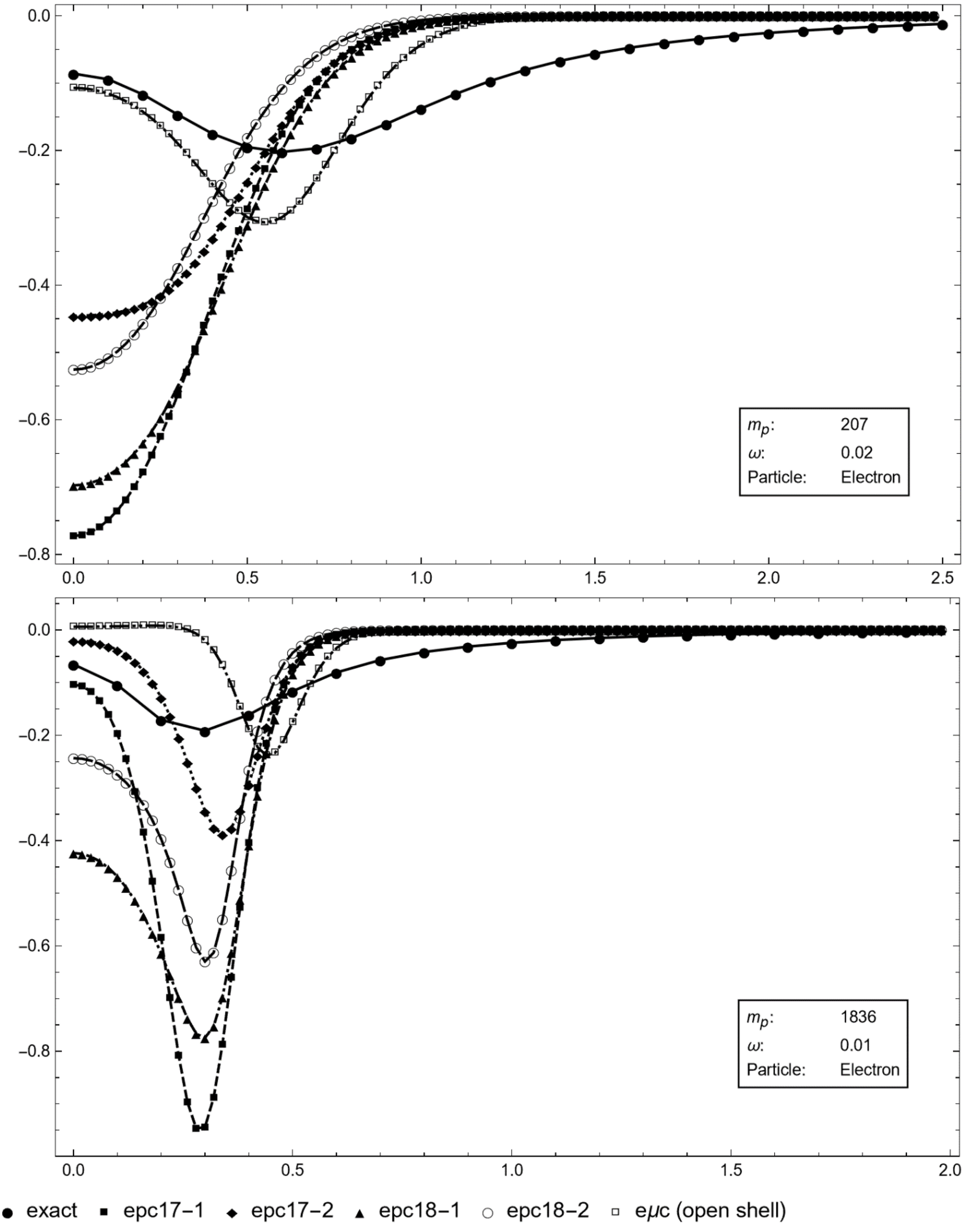

Figure 5-44: Comparison of exact correlation potentials and functional-derived potentials for electrons

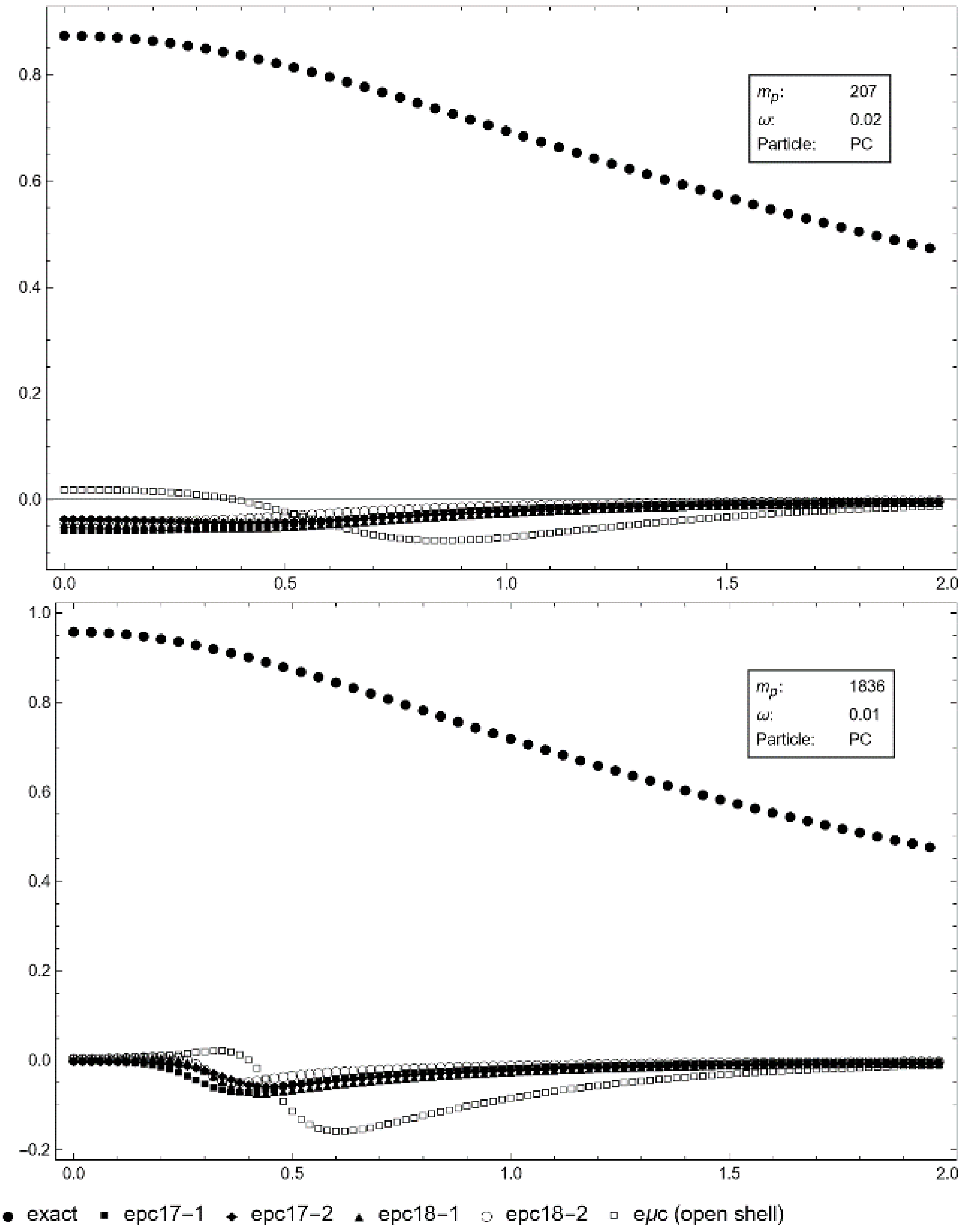

Figure 5-45: Comparison of exact correlation potentials and functional-derived potentials for PCP

## 5.3 Data Fitting

To propose an explicit functional dependence on the two main variational parameters of the problem (field frequency and the mass of the PCP), data fitting was performed in three sections: parameters of the variational wave function, nuclear exponents of Hartree-Fock wave functions, and correlation energies. In essence, data fitting allows us to propose an explicit form for the functions, thus providing a deeper insight into how the system's components behave under different conditions.

### 5.3.1 Fitting Variational Parameters

The fitting of variational parameters was performed both in two stages (i.e., first finding the dependence of the data on the frequency and then on the mass) and in a single stage (i.e., fitting the data simultaneously with respect to frequency and mass). Accordingly, the parameter $\alpha$ was fitted with a logistic function (LOG) in the form:

$$L(\kappa) = \mu\left(\frac{l_1}{1 + \exp(\kappa)} + l_2\right) \tag{5.46}$$

and a specific form of the error function (ERR) as follows:

$$\mathrm{E}(\kappa) = \mu\left(a_1 - a_2\,\mathrm{erf}(\kappa\sqrt{a_3\mu})\right) \tag{5.47}$$

where

$$\kappa = \mathrm{Log}_e[\omega] \tag{5.48}$$

Additionally, the optimized parameter $\beta$ shows good agreement with the analytical form of this parameter based on the harmonic oscillator model (SIM1):

$$\beta = \frac{1}{2}\mu\omega \tag{5.49}$$

although the data can also be fitted with the following exponential function (EXP), which can be represented in linear form as well:

$$\mathcal{Z}(\omega) = \mu b_3 \exp(\kappa) = \mu b_3\, \omega \tag{5.50}$$

However, to improve accuracy, the difference between the optimized parameter $\beta$ values and those obtained from (5-49) was also fitted to the exponential function (5-50) and added to the values obtained from (5-49). This resulted in a relatively simple yet precise form (SIM2) for this parameter, as shown below, which led to a significant increase in accuracy.

$$\mathcal{S}(\kappa) = |1/2\,\mu\omega - b_1\,\mu\,\mathrm{Exp}[b_2\,\kappa]| = |1/2\,\mu\omega - b_1\,\mu\,\omega^{b_2}| \tag{5.51}$$

With the values of the constants as shown in the table below, the expressions obtained for the fundamental parameters $\alpha$ and $\beta$ will be as shown in Table 5-11.

Table 5-10: Constants for fitting variational parameters

| | $l_1$ | $l_2$ | $a_1$ | $a_2$ | $a_3$ | $b_1$ | $b_2$ | $b_3$ |
|---|---|---|---|---|---|---|---|---|
| Variational (exact) | 0.2 | 0.8 | 0.9 | 0.096 | 0.1333 | 0.1 | 0.5 | 0.48 |

Table 5-11: Explicit forms obtained from fitting for variational parameters

| Parameter | Explicit form |
|---|---|
| $\alpha_{LOG}$ | $\frac{\mu}{5}(4+\frac{1}{1+\exp[\kappa]})$ |
| $\alpha_{ERR}$ | $\mu(a_1\text{-}a_2\ \mathrm{Erf}[\kappa\sqrt{a_3\mu}])$ |
| $\beta_{EXP}$ | $b_3\ \mu\ \mathrm{Exp}[\kappa]$ |
| $\beta_{SIM1}$ | $1/2\ \mu\omega$ |
| $\beta_{SIM2}$ | $\lvert 1/2\ \mu\omega - b_1\ \mu\ \mathrm{Exp}[b_2\ \kappa]\rvert$ |

Comparing the optimized parameter $\alpha_{OPT}$ with the two fitting parameters $\alpha_{\mathrm{LOG}}$ and $\alpha_{ERR}$ over the range of frequencies and reporting the error percentage for each case at different masses shows that, on average, the highest error percentage for both methods occurs at intermediate frequencies (particularly at $\omega = 1$). While the error percentage decreases for $\alpha_{ERR}$ at higher frequencies, it remains significant for $\alpha_{\mathrm{LOG}}$. Overall, the error percentage for $\alpha_{ERR}$ is lower (with an overall mean error of 0.82% and a variance of 0.2), making it a more suitable choice for reproducing the alpha parameter.

Comparing the optimized parameter $\beta_{OPT}$ with the three fitting parameters $\beta_{EXP}$, $\beta_{\mathrm{SIM1}}$, and $\beta_{SIM2}$ over the range of frequencies and reporting the error percentage for all three cases shows that the error percentage is very high at low frequencies (with an average of 99%). Since the beta values in these regions are very small, even a slight difference from the optimized value results in large error percentages. However, as reported in Table (C-4) of the appendix C, due to the small contribution of beta to the energy in these regions, large errors in beta do not lead to large errors in the energy (with an overall mean error of about 0.06%). The error percentage for all three forms of beta decreases with increasing frequency, and overall, $\beta_{SIM2}$ has the lowest error percentage (with an overall mean error of 35.78%).

Comparison of the energy values obtained from the combination of the two parameters $\alpha_{\text{LOG}}$ and $\alpha_{ERR}$, and the three parameters $\beta_{EXP}$, $\beta_{\text{SIM1}}$, and $\beta_{SIM2}$ over the range of frequencies, along with the error percentage report for each case relative to the energy obtained from the optimized parameters, is also presented in Table (C-4) of the appendix C. The error percentages for the six different parameter combinations show that, once again, the highest error occurs at intermediate frequencies (particularly at $\omega = 1$) and decreases at higher frequencies. This indicates that at intermediate frequencies (especially at $\omega = 1$), the alpha parameter plays a more significant role compared to the beta parameter, and thus the error pattern follows that of the alpha parameter. However, at higher frequencies, the beta parameter quickly surpasses the alpha parameter, and consequently, the energy error pattern aligns with the error pattern of the beta parameter. The data show that among the six different parameter combinations, the combination of $\alpha_{ERR}\beta_{SIM2}$ provides the best results (with the least errors). Using this combination, the total wave function is transformed into the form below:

$$\chi_{fit} = \exp\left(-\mu(a_1 - a_2\text{Erf}[\kappa\sqrt{a_3\mu}])r - \mu\left|-b_1\omega^{b_2} + \frac{\omega}{2}\right|r^2\right) \qquad 5.52$$

In this case, the variational wave function (2-72) can be simply obtained by specifying the reduced mass and the oscillator field frequency through the explicit form (5-52). Additionally, all quantities obtained in the previous sections can be derived by substituting these two parameters. For example, the average two-particle distribution function based on the fitting results will be as follows:

$$f(r) = \sqrt{2}\,\gamma^{-3/2} N^2 \pi^{5/2} r^2 e^{-2\mu r\left(r\left|\frac{\omega}{2}-\omega^{b_2} b_1\right|+(a_1-\mathrm{Erf}[\kappa\sqrt{\mu a_3}]a_2)\right)} \qquad 5.53$$

Overall, the form of the wave function obtained from the fitting shows that the hydrogen-like part, in addition to a linear dependence on $\mu$, also has a nonlinear dependence on it as well as on the field frequency. On the other hand, for the oscillator part, only a linear behavior with respect to $\mu$ is observable, but its dependence on the field frequency is both linear and nonlinear. In general, comparing (5-52) with the two equations (2-73) and (2-74) shows that the deviation of each part from its reference state (the alpha values with respect to the exponent of the hydrogen-like wave function and the beta values with respect to the exponent of the harmonic oscillator wave function) appears as a nonlinear term. Naturally, the form of this nonlinear term can vary depending on the type of function chosen for fitting and its accuracy (for example, both the error function and the logistic function can be used for alpha, with accuracy being the deciding factor for preferring one over the other).

## 5.4 Comparison of Non-Adiabatic and Correlation Energy Contributions

One of the interesting topics in multi-component quantum chemistry is the comparison of non-adiabatic energy and e-PCP correlation energy. Since accurate non-adiabatic energy of the system is required for this comparison and such capability is usually not available due to the complexity of calculations for real molecular systems, the EHM is used here to compare these two energies.

### 5.4.1 Analysis of the EHM in the Adiabatic Framework

Comparing the non-adiabatic energy and the e-PCP correlation energy helps us understand under which conditions non-adiabatic two-component methods are efficient and when they lose their effectiveness. To perform such a comparison, three energies are required: the exact non-adiabatic energy of the system, the exact adiabatic energy of the system, and the two-component Hartree-Fock limit energy of the system. The difference between the first two energies is the non-adiabatic energy; and the difference between the first and the third energies is the e-PCP correlation energy of the system:

$$E_{non-ad} = E_{exact} - E_{ad} \tag{5.54}$$

$$E_{corr} = E_{exact} - E_{TC-HF} \tag{5.55}$$

Of the three aforementioned energies, the first and third have already been calculated and are presented in Table (4-3); therefore, only the adiabatic energy of the EHM needs to be calculated. To calculate the system's energy within the adiabatic approximation framework, it is necessary, as with all adiabatic formulations, to consider the heavier particle as a clamped point charge (with infinite mass) and then place the origin of the coordinate system on this particle. After making these changes to the EHM, the Hamiltonians within the adiabatic framework are obtained as follows:

$$H(\boldsymbol{r}_e) = -\frac{1}{2}\nabla_{r_e}^2 + \frac{1}{2}\omega^2 {r_e}^2 - \frac{1}{r_e} \qquad 5.56$$

$$H(\boldsymbol{r}_p) = -\frac{1}{2m_p}\nabla_{r_p}^2 + \frac{1}{2}m_p\omega^2 {r_p}^2 \qquad 5.57$$

The energy of equation (5-67) is simply equivalent to the energy of a three-dimensional harmonic oscillator and can be calculated from equation (2-17). The energy in equation (5-56), representing the electronic energy, is similar to the relative motion part in the original EHM, except here the reduced mass equals the electron mass (1 in atomic units), and the distance between the two particles is reduced to the electron's distance from the coordinate system's center (the clamped particle). The energy of this equation can be computed variationally or using the FEM, similar to equation (2-18).

Therefore, the total ground state energy of the adiabatic system is obtained by summing the ground state energies of these two equations:

$$E_{ad} = E_{ad}^{p} + E_{ad}^{e} = E_{3D-HO} + E_{ad}^{e} = \frac{3}{2}\omega + E_{ad}^{e} \qquad 5.58$$

As a result, to calculate $E_{ad}$, it is only necessary to compute the energy $E_{ad}^{e}$, which will be addressed in the next section.

### 5.4.2 Calculation of Electronic Energy

The energy of the electronic part is calculated numerically and variationally (using the wave function 2-72), and the results are presented in Table (5-12). The electronic energy is parametrically dependent only on the frequency, and here we encounter 9 systems with different field frequencies.

Table 5-12: Electronic energies

| ω | $\boldsymbol{\alpha^{ad}}$ [1] | $\boldsymbol{\beta^{ad}}$ [2] | N [3] | FEM $\boldsymbol{E^{e}_{ad}}$ [4] | Var $\boldsymbol{E^{e}_{ad}}$ [5] | $\Delta \boldsymbol{E^{e}_{ad}}$ [6] |
|---|---|---|---|---|---|---|
| 0.0001 | 1.000000 | 0.000000 | 0.564190 | -0.500000 | -0.500000 | 0.000000 |
| 0.001 | 0.999999 | 0.000002 | 0.564191 | -0.499999 | -0.499999 | 0.000000 |
| 0.01 | 0.999850 | 0.000150 | 0.564316 | -0.499850 | -0.499850 | 0.000000 |
| 0.1 | 0.988307 | 0.013169 | 0.575925 | -0.485679 | -0.485670 | 0.000009 |
| 1.0 | 0.901059 | 0.369817 | 0.892146 | 0.179668 | 0.179905 | 0.000236 |
| 10 | 0.839160 | 4.616819 | 3.015105 | 11.265447 | 11.265809 | 0.000361 |
| 100 | 0.817579 | 48.825540 | 14.430261 | 138.557198 | 138.557585 | 0.000388 |
| 1000 | 0.810663 | 496.324903 | 77.139839 | 1464.160725 | 1464.161119 | 0.000394 |
| 10000 | 0.808470 | 4988.417578 | 426.915815991 | 14887.005984 | 14887.006379 | 0.000396 |

1. Optimized variational parameter $\alpha^{ad}$
2. Optimized variational parameter $\beta^{ad}$
3. Variational normalization coefficient
4. FEM energy
5. Variational energy
6. Difference between 4 and 5

•

To utilize the maximum accuracy of the FEM, a meshing of 0.0001 distance in atomic units was used for all frequencies. Additionally, the accuracy of the calculations was tested by comparing the results with the perturbative trends of the problem. Similar to the discussion in section 2-3-2, there are two perturbative regimes in this problem: one at low frequencies, where the interaction of the electron with the clamped positive particle dominates the Hamiltonian (equivalent to the energy of a hydrogen atom with a clamped nucleus), and the oscillator potential of the electron can be considered a perturbation. The other regime is at high frequencies, where the electron's oscillator potential becomes very large and dominant (similar to the energy of a three-dimensional harmonic oscillator), and the Coulomb interaction between the electron and the clamped positive particle can be considered a perturbation. The first-order perturbation at low frequencies is $\frac{3\pi\omega^2}{2}$ and at high frequencies is $-\frac{2\sqrt{\omega}}{\sqrt{\pi}}$. Therefore, the total adiabatic electronic energy at low and high frequencies will be as follows:

$$E_{ad}^{e-low\omega} \approx \frac{3\pi\omega^2 - 1}{2} \qquad 5.59$$

$$E_{ad}^{e-high\omega} \approx \frac{3}{2}\omega - \frac{2\sqrt{\omega}}{\sqrt{\pi}} \qquad 5.60$$

The data in Table 5-13 show that the adiabatic electronic energies are close to equation (5-59) at low frequencies and close to equation (5-60) at high frequencies. Additionally, the calculation of the components of adiabatic electronic energy indicates that the kinetic and oscillator potential energy components approach their corresponding values for a three-dimensional harmonic oscillator at high frequencies. The only factor causing deviation from these values is the expectation value of the interaction energy, which increases in absolute value with increasing frequency.

Table 5-13: Asymptotic behavior and components of electronic energy

| ω | 0.0001 | 0.001 | 0.01 | 0.1 | 1 | 10 | 100 | 1000 | 10000 |
|---|---|---|---|---|---|---|---|---|---|
| Num $E^{e}_{ad}$ [1] | -0.500000 | -0.499999 | -0.499850 | -0.485679 | 0.179668 | 11.265447 | 138.557198 | 1464.160725 | 14887.005984 |
| $E^{e}_{ad}$ low ω [2] | -0.500000 | -0.499995 | -0.499529 | -0.452876 | 4.212389 | - | - | - | - |
| ΔE low ω [3] | 0.000000 | 0.000003 | 0.000321 | 0.032803 | 4.032721 | - | - | - | - |
| % low ω [4] | 0.000011 | 0.000648 | 0.064288 | 6.754031 | 2244.534471 | - | - | - | - |
| $E^{e}_{ad}$high ω [5] | - | - | - | - | 0.371621 | 11.431752 | 138.716208 | 1464.317518 | 14887.162083 |
| ΔE high ω [6] | - | - | - | - | 0.191952 | 0.166304 | 0.159011 | 0.156793 | 0.156100 |
| % high ω [7] | - | - | - | - | 106.836989 | 1.476233 | 0.114762 | 0.010709 | 0.001049 |
| HO coms [8] | 0.000075 | 0.000750 | 0.007500 | 0.075000 | 0.750000 | 7.500000 | 75.000000 | 750.000000 | 7500.000000 |
| $\mathbf{INT}^{e}_{ad}$ [9] | -1.000000 | -1.000003 | -1.000300 | -1.026316 | -1.552250 | -3.911832 | -11.605088 | -35.997084 | -113.150437 |
| $\mathbf{HO}^{e}_{ad}$ [10] | 0.000000 | 0.000001 | 0.000150 | 0.013740 | 0.477897 | 6.610682 | 72.179871 | 741.079633 | 7471.790600 |
| $\mathbf{T}^{e}_{ad}$ [11] | 0.500001 | 0.500004 | 0.500301 | 0.526898 | 1.254023 | 8.566641 | 77.984665 | 759.139868 | 7529.547941 |

1. Numerical exact energy of the system
2. Perturbative energy of the system at low frequencies (equation 5-59)
3. Difference between 1 and 2
4. Error percentage of 2
5. Perturbative energy of the system at high frequencies (equation 5-60)
6. Difference between 1 and 5
7. Error percentage of 5
8. Components of kinetic/potential oscillator energy for a three-dimensional harmonic oscillator (equal to $\frac{3}{4}\omega$ from equations 2-148 and 2-149)
9. Expectation value of electron-clamped positive particle interaction energy
10. Expectation value of oscillator potential energy
11. Expectation value of kinetic energy

### 5.4.3 Comparison of Non-Adiabatic and Correlation Energy Contributions

After calculating the electronic energies, the total adiabatic energy can be obtained using equation (5-58). This gives us the three energies needed to compare the non-adiabatic part of the total energy and the e-PCP correlation energy (i.e., equations 5-54 and 5-55).

The numerical results of this comparison, presented in Table 5-14, show that the non-adiabatic energy contribution increases with increasing frequency for any given mass, while the correlation energy decreases at a slower rate under the same conditions. On the other hand, the non-adiabatic energy contribution decreases with increasing mass for any given frequency, showing a uniform pattern, unlike the correlation energy. It is worth mentioning that since the adiabatic energy depends only on the frequency and is therefore the same for different masses, its values for each mass are presented repeatedly in Table 5-14 for easier comparison.

Figures 5-46 and 5-47, which illustrate the differences between the non-adiabatic energy and the magnitude of the correlation energy (item 6 in Table 5-18), clearly show the aforementioned trends. To better visualize the points, the graphs are plotted on a natural logarithmic scale of frequencies. For the two masses of 1 and 1.5, the non-adiabatic energies are greater than the correlation energies across the entire frequency range. However, for larger masses, particularly at low frequencies, the non-adiabatic energy is smaller than the correlation energy, but gradually surpasses it as the frequency increases. As the mass increases, the frequency at which the sign of the difference between the non-adiabatic energy and the magnitude of the correlation energy changes shifts to higher frequencies.

Overall, the results indicate that with increasing mass, the non-adiabatic effects become smaller (as expected), eventually becoming smaller than the correlation energies.

Unfortunately, non-adiabatic energies are rarely reported due to the lack of exact solutions for obtaining the energies of systems. However, reference [12] provides these data for six hydride systems using the Monte Carlo method. Fitting the data from reference [12] and extrapolating the non-adiabatic energy for the single-electron system yields a value of 0.0003 Hartree, which matches the values obtained for the proton (mass 1836) at low frequencies in Table 5-18. The current data seem to confirm that for the physical masses of 1, 207, and 1836, only the mass of 1 (positron) has the property that the electron-positron correlation energy is smaller than the non-adiabatic energy. Consequently, it is reliable to use two-component quantum chemistry methods for calculations involving it. In the case of muons and protons, the adiabatic approximation and its perturbative corrections appear to be a better approach.

Table 5-14: Comparison of non-adiabatic and correlation energy contributions for the EHM

| omega | $\boldsymbol{E_{exact}}$[1] | $E_{TC\text{-}HF}$ [2] | $E_{ad}$ [3] | $E_{non\text{-}ad}$ [4] | $E_{Corr}$ [5] | $\Delta E$[6] |
|---|---|---|---|---|---|---|
| | | | m = 1 | | | |
| 0.0001 | -0.249850 | -0.108513 | -0.499850 | 0.250000 | -0.141337 | 0.108663 |
| 0.001 | -0.248497 | -0.108491 | -0.498499 | 0.250002 | -0.140006 | 0.109996 |
| 0.01 | -0.234701 | -0.106407 | -0.484850 | 0.250149 | -0.128294 | 0.121856 |
| 0.1 | -0.073957 | 0.012203 | -0.335679 | 0.261722 | -0.086160 | 0.175562 |
| 1.0 | 2.111853 | 2.171918 | 1.679668 | 0.432184 | -0.060066 | 0.372118 |
| 10 | 27.395303 | 27.448072 | 26.265447 | 1.129856 | -0.052768 | 1.077088 |
| 100 | 291.942127 | 291.992777 | 288.557198 | 3.384929 | -0.050650 | 3.334279 |
| 1000 | 2974.690429 | 2974.740427 | 2964.160725 | 10.529704 | -0.049998 | 10.479706 |
| 10000 | 29920.133541 | 29920.183344 | 29887.005984 | 33.127557 | -0.049804 | 33.077754 |
| | | | m = 1.5 | | | |
| 0.0001 | -0.299850 | -0.131661 | -0.499850 | 0.200000 | -0.168189 | 0.031811 |
| 0.001 | -0.298498 | -0.131644 | -0.498499 | 0.200001 | -0.166854 | 0.033147 |
| 0.01 | -0.284750 | -0.129900 | -0.484850 | 0.200100 | -0.154850 | 0.045250 |
| 0.1 | -0.127553 | -0.021539 | -0.335679 | 0.208126 | -0.106014 | 0.102112 |
| 1.0 | 2.016098 | 2.087757 | 1.679668 | 0.336429 | -0.071659 | 0.264771 |
| 10 | 27.137735 | 27.199613 | 26.265447 | 0.872287 | -0.061878 | 0.810410 |
| 100 | 291.164648 | 291.223701 | 288.557198 | 2.607450 | -0.059054 | 2.548396 |
| 1000 | 2972.266533 | 2972.324714 | 2964.160725 | 8.105808 | -0.058182 | 8.047626 |
| 10000 | 29912.502509 | 29912.560519 | 29887.005984 | 25.496525 | -0.058010 | 25.438515 |
| | | | m = 2 | | | |
| 0.0001 | -0.333183 | -0.149328 | -0.499850 | 0.166667 | -0.183855 | -0.017188 |
| 0.001 | -0.331831 | -0.149312 | -0.498499 | 0.166667 | -0.182519 | -0.015851 |
| 0.01 | -0.318109 | -0.147746 | -0.484850 | 0.166742 | -0.170363 | -0.003622 |
| 0.1 | -0.162797 | -0.045426 | -0.335679 | 0.172882 | -0.117372 | 0.055510 |
| 1.0 | 1.955533 | 2.032554 | 1.679668 | 0.275864 | -0.077021 | 0.198843 |
| 10 | 26.977001 | 27.042353 | 26.265447 | 0.711553 | -0.065352 | 0.646201 |
| 100 | 290.681215 | 290.743202 | 288.557198 | 2.124018 | -0.061987 | 2.062031 |
| 1000 | 2970.760975 | 2970.821923 | 2964.160725 | 6.600250 | -0.060948 | 6.539302 |
| 10000 | 29907.764199 | 29907.824832 | 29887.005984 | 20.758216 | -0.060632 | 20.697583 |
| | | | m = 3 | | | |
| 0.0001 | -0.374850 | -0.175573 | -0.499850 | 0.125000 | -0.199277 | -0.074277 |
| 0.001 | -0.373498 | -0.175559 | -0.498499 | 0.125001 | -0.197939 | -0.072938 |

| omega | $\boldsymbol{E_{exact}}$[1] | $E_{TC\text{-}HF}$ [2] | $E_{ad}$ [3] | $E_{non\text{-}ad}$ [4] | $E_{Corr}$ [5] | $\Delta E$[6] |
|---|---|---|---|---|---|---|
| 0.01 | -0.359800 | -0.174173 | -0.484850 | 0.125050 | -0.185627 | -0.060578 |
| 0.1 | -0.206455 | -0.078661 | -0.335679 | 0.129224 | -0.127794 | 0.001430 |
| 1.0 | 1.882805 | 1.962394 | 1.679668 | 0.203136 | -0.079589 | 0.123547 |
| 10 | 26.786175 | 26.851598 | 26.265447 | 0.520728 | -0.065423 | 0.455305 |
| 100 | 290.109020 | 290.170376 | 288.557198 | 1.551823 | -0.061356 | 1.490466 |
| 1000 | 2968.980576 | 2969.040682 | 2964.160725 | 4.819851 | -0.060105 | 4.759746 |
| 10000 | 29902.162455 | 29902.222187 | 29887.005984 | 15.156471 | -0.059732 | 15.096739 |
| | | | m=10 | | | |
| 0.0001 | -0.454395 | -0.257049 | -0.499850 | 0.045455 | -0.197347 | -0.151892 |
| 0.001 | -0.453044 | -0.257037 | -0.498499 | 0.045455 | -0.196007 | -0.150552 |
| 0.01 | -0.439381 | -0.255878 | -0.484850 | 0.045470 | -0.183502 | -0.138033 |
| 0.1 | -0.288923 | -0.170118 | -0.335679 | 0.046756 | -0.118805 | -0.072049 |
| 1.0 | 1.751315 | 1.810873 | 1.679668 | 0.071647 | -0.059557 | 0.012090 |
| 10 | 26.447087 | 26.490355 | 26.265447 | 0.181639 | -0.043268 | 0.138371 |
| 100 | 289.096910 | 289.135811 | 288.557198 | 0.539712 | -0.038901 | 0.500811 |
| 1000 | 2965.835599 | 2965.873207 | 2964.160725 | 1.674875 | -0.037607 | 1.637267 |
| 10000 | 29892.271372 | 29892.308661 | 29887.005984 | 5.265389 | -0.037289 | 5.228100 |
| | | | m=50 | | | |
| 0.0001 | -0.490046 | -0.353887 | -0.499850 | 0.009804 | -0.136159 | -0.126355 |
| 0.001 | -0.488695 | -0.353874 | -0.498499 | 0.009804 | -0.134821 | -0.125017 |
| 0.01 | -0.475043 | -0.352583 | -0.484850 | 0.009807 | -0.122460 | -0.112653 |
| 0.1 | -0.325612 | -0.261510 | -0.335679 | 0.010067 | -0.064103 | -0.054036 |
| 1.0 | 1.694935 | 1.717273 | 1.679668 | 0.015267 | -0.022337 | -0.007070 |
| 10 | 26.303970 | 26.317887 | 26.265447 | 0.038523 | -0.013917 | 0.024606 |
| 100 | 288.671520 | 288.683473 | 288.557198 | 0.114323 | -0.011953 | 0.102370 |
| 1000 | 2964.515378 | 2964.526763 | 2964.160725 | 0.354653 | -0.011385 | 0.343268 |
| 10000 | 29888.120778 | 29888.132005 | 29887.005984 | 1.114795 | -0.011227 | 1.103568 |
| | | | m=207 | | | |
| 0.0001 | -0.497446 | -0.414686 | -0.499850 | 0.002404 | -0.082760 | -0.080356 |
| 0.001 | -0.496095 | -0.414668 | -0.498499 | 0.002404 | -0.081427 | -0.079023 |
| 0.01 | -0.482446 | -0.412886 | -0.484850 | 0.002404 | -0.069560 | -0.067155 |
| 0.02 | -0.466994 | -0.407763 | -0.469401 | 0.002407 | -0.059231 | -0.056824 |
| 0.1 | -0.333212 | -0.306618 | -0.335679 | 0.002467 | -0.026594 | -0.024127 |
| 1.0 | 1.683403 | 1.690372 | 1.679668 | 0.003734 | -0.006969 | -0.003235 |
| 10 | 26.274861 | 26.278862 | 26.265447 | 0.009414 | -0.004001 | 0.005413 |

| omega | $\boldsymbol{E_{exact}}$ [1] | $E_{TC\text{-}HF}$ [2] | $E_{ad}$ [3] | $E_{non\text{-}ad}$ [4] | $E_{Corr}$ [5] | $\Delta E$ [6] |
|---|---|---|---|---|---|---|
| 100 | 288.585127 | 288.588490 | 288.557198 | 0.027930 | -0.003363 | 0.024566 |
| 1000 | 2964.247368 | 2964.250545 | 2964.160725 | 0.086643 | -0.003177 | 0.083467 |
| 10000 | 29887.278310 | 29887.281447 | 29887.005984 | 0.272326 | -0.003137 | 0.269189 |
| | | | m=400 | | | |
| 0.0001 | -0.498603 | -0.434898 | -0.499850 | 0.001247 | -0.063705 | -0.062458 |
| 0.001 | -0.497252 | -0.434876 | -0.498499 | 0.001247 | -0.062375 | -0.061128 |
| 0.01 | -0.483603 | -0.432709 | -0.484850 | 0.001247 | -0.050894 | -0.049647 |
| 0.1 | -0.334399 | -0.318137 | -0.335679 | 0.001280 | -0.016263 | -0.014983 |
| 1.0 | 1.681605 | 1.685464 | 1.679668 | 0.001936 | -0.003860 | -0.001923 |
| 10 | 26.270328 | 26.272493 | 26.265447 | 0.004880 | -0.002165 | 0.002716 |
| 100 | 288.571677 | 288.573488 | 288.557198 | 0.014479 | -0.001811 | 0.012668 |
| 1000 | 2964.205645 | 2964.207347 | 2964.160725 | 0.044921 | -0.001702 | 0.043219 |
| 10000 | 29887.147157 | 29887.148995 | 29887.005984 | 0.141173 | -0.001838 | 0.139335 |
| | | | m=900 | | | |
| 0.0001 | -0.499295 | -0.453975 | -0.499850 | 0.000555 | -0.045320 | -0.044765 |
| 0.001 | -0.497944 | -0.453946 | -0.498499 | 0.000555 | -0.043998 | -0.043443 |
| 0.01 | -0.484295 | -0.451115 | -0.484850 | 0.000555 | -0.033180 | -0.032625 |
| 0.1 | -0.335109 | -0.326729 | -0.335679 | 0.000570 | -0.008381 | -0.007811 |
| 1.0 | 1.680530 | 1.682345 | 1.679668 | 0.000862 | -0.001815 | -0.000954 |
| 10 | 26.267619 | 26.268617 | 26.265447 | 0.002171 | -0.000999 | 0.001173 |
| 100 | 288.563640 | 288.564475 | 288.557198 | 0.006442 | -0.000835 | 0.005607 |
| 1000 | 2964.180715 | 2964.181492 | 2964.160725 | 0.019991 | -0.000777 | 0.019214 |
| 10000 | 29887.068792 | 29887.069700 | 29887.005984 | 0.062809 | -0.000907 | 0.061902 |
| | | | m=1836 | | | |
| 0.0001 | -0.499578 | -0.466417 | -0.499850 | 0.000272 | -0.033161 | -0.032889 |
| 0.001 | -0.498226 | -0.466378 | -0.498499 | 0.000272 | -0.031848 | -0.031576 |
| 0.01 | -0.484578 | -0.462753 | -0.484850 | 0.000272 | -0.021825 | -0.021553 |
| 0.1 | -0.335400 | -0.330905 | -0.335679 | 0.000279 | -0.004495 | -0.004216 |
| 1.0 | 1.680091 | 1.681012 | 1.679668 | 0.000423 | -0.000921 | -0.000498 |
| 10 | 26.266512 | 26.267013 | 26.265447 | 0.001065 | -0.000501 | 0.000564 |
| 100 | 288.560357 | 288.560779 | 288.557198 | 0.003159 | -0.000422 | 0.002737 |
| 1000 | 2964.170533 | 2964.170918 | 2964.160725 | 0.009808 | -0.000385 | 0.009423 |
| 10000 | 29887.036785 | 29887.037312 | 29887.005984 | 0.030802 | -0.000527 | 0.030275 |

1. Numerical exact energy
2. Two-component Hartree-Fock energy with [7s:7s] basis sets

3. Adiabatic energy
4. Non-adiabatic energy (difference between 1 and 3)
5. Correlation energy (difference between 1 and 2)
6. Difference between 4 and the magnitude of 5 ($\Delta E = E_{non-ad} - |E_{corr}|$)

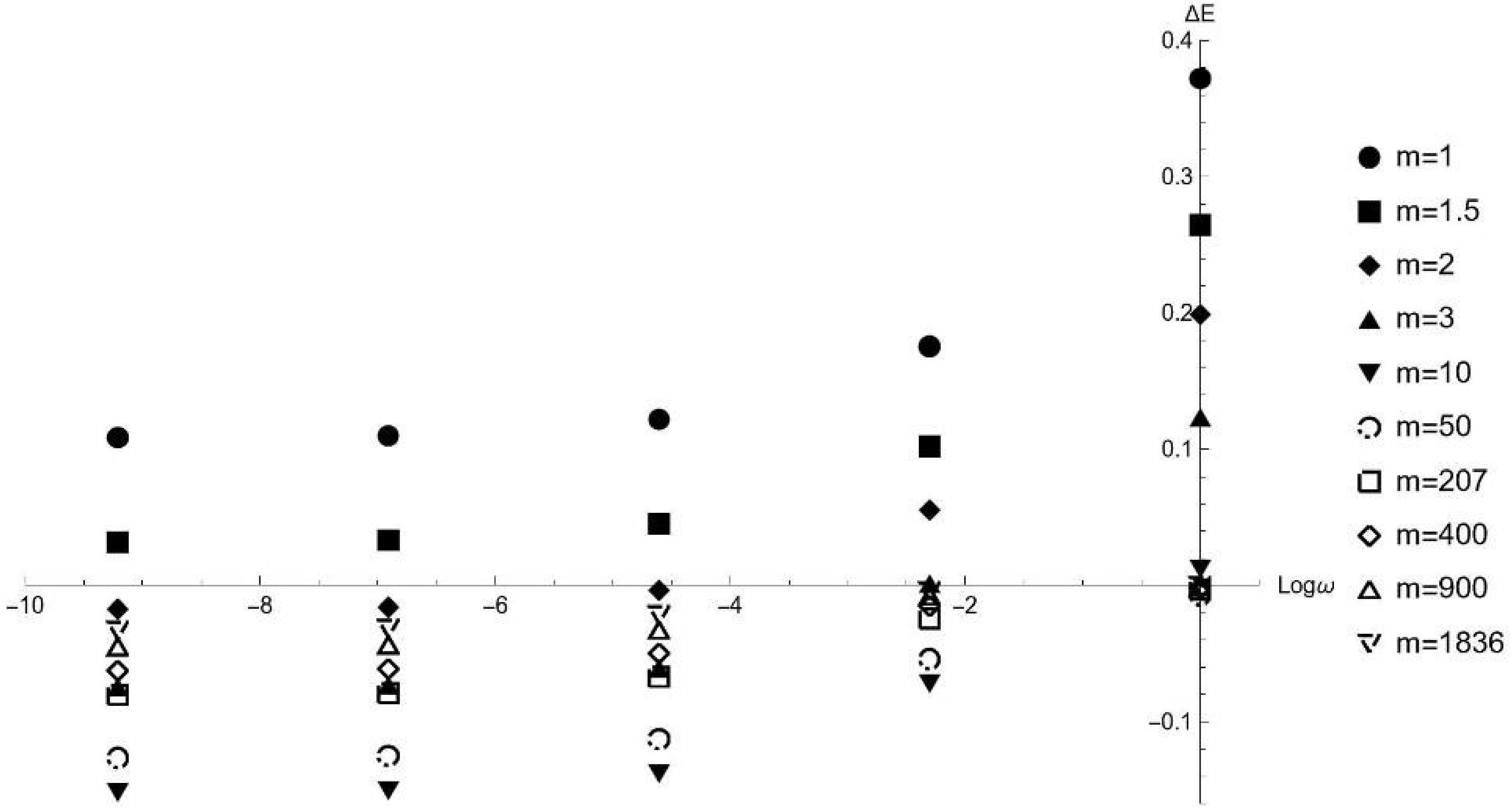

Figure 5-46: Comparison of the magnitude differences between non-adiabatic and correlation energies in the logarithmic frequency range (0.0001 to 1)

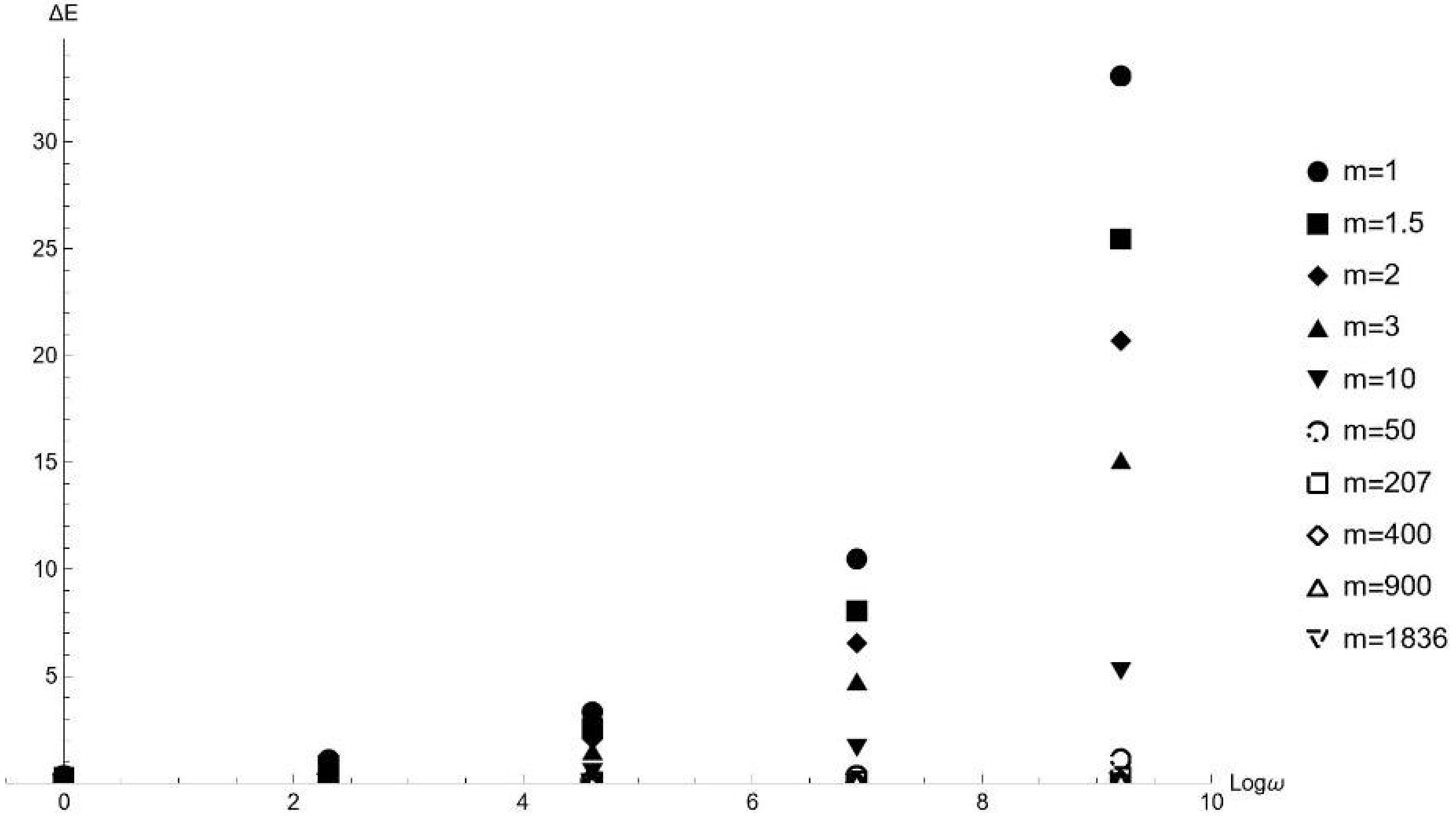

Figure 5-47: Comparison of the magnitude differences between non-adiabatic and correlation energies in the logarithmic frequency range (1 to 10000)

# 6 Conclusion

The term "model," in its most common usage, refers to a simplified version of the so-called target system, which is the part or aspect of the world that we are interested in [142]. The history of science, especially quantum physics, is replete with models that have significantly contributed to understanding real physical phenomena [143]. Models serve three important purposes: developing new theories, exploring and testing existing ones [144], and these purposes are made possible because models provide a deep understanding of the real system [145], even though they are only representations of the external world and, in this sense, are fictional systems. From the famous Solar System model to the ideal gas model, our work in this study was also inspired by a very important model in the field of electron-electron correlation: the Harmonium model.

The problem of calculating e-PCP correlation is one of the challenges of quantum chemistry beyond the adiabatic approximation. One could say that the ultimate goal of ab initio two-component methods is to better estimate this type of correlation. To this end, we developed a simple model for better understanding, inspired by its electronic counterpart, the Harmonium model, which aimed to better understand electronic correlation. However, e-PCP correlation has fundamental differences from electron-electron correlation (as discussed in Chapter 3), and since it involves particles with positive charge and variable mass, we named it the "Exotic Harmonium."

In the first step, we attempted to apply the method used to obtain analytical equations in the Harmonium model to the Exotic Harmonium, but encountered challenges that are detailed in Chapter 2. These challenges led us to the most straightforward method for finding the ground state in quantum mechanics, the variational method. Fortunately, by predicting the asymptotic behavior of the system, we were able to obtain a very simple yet accurate variational wave function. Since having the wave function essentially allows access to all the properties of the system, we obtained the main quantities such as energy and its components, as well as single-particle densities. As discussed in Chapter 2, by performing a series of transformations, we wrote the wave function in three coordinate systems, each of which could be used for a specific purpose and provided a different perspective on the system.

In Chapter 3, we developed the most important concepts and quantities for studying electronic correlation, particularly in the Harmonium model for the EHM, while also reviewing their history. The key idea introduced in Chapter 3 was the "correlation hill," which is akin to the correlation hole in the Harmonium model but has an inverse shape here, hence the name "correlation hill." This quantity clearly revealed the different nature of e-PCP correlation compared to electron-electron correlation, and precisely for this reason, e-PCP correlation cannot be considered a generalization of electron-electron correlation.

In Chapter 4, the numerical data obtained from solving the model, including the expectation values achievable with the variational wave function, were presented. Additionally, all numerical data related to the quantities and concepts introduced for extracting correlation information from the EHM were reported in Chapter 5, making this chapter the first to present applications of the EHM. Three types of correlation classifications were introduced: reference-independent, MC-HF, and MC-DFT. Although MC-DFT correlation was reported for only a few systems due to the need for simplification and density fitting, other two-particle quantities related to particle interactions were also examined.

Another application of the EHM is to benchmark existing functionals for e-PCP correlation, just as successful electron functionals like LYP have been benchmarked using the Harmonium model [83], [141], [146]. Here, we also benchmarked five existing functionals for e-PCP correlation using the EHM. Another important step reported in Chapter 5 was the fitting of the variational wave function parameters, which enabled us to derive a simple yet accurate form for the wave function in terms of the two main variables of the problem: the frequency of the oscillator field and the mass of the PCP. This form of the wave function explicitly shows its dependency on these variables, providing valuable insight into the system's behavior with respect to these two variables. Given that the range of variations in the frequency of the oscillator field and the mass of the PCP was considered very wide (from very high and very low frequencies compared to typical molecular systems to the smallest possible mass, positron, and the most significant large mass, proton), the e-PCP correlation can be examined for a wide range of real systems using the EHM.

There are many opportunities for future research in this field. The primary mission of this study was the development of the EHM. However, much like the Harmonium model, which has long been used to extract information about electron-electron correlation, the EHM will certainly have a broader scope beyond the applications presented in this dissertation.

Unlike the Harmonium model, this model deals not just with one type of particle, but with a general PCP, which can be any particle within the studied mass spectrum, such as positrons, muons, kaons, pions, protons, etc. Consequently, the EHM can be applied to each of these exotic particles (briefly discussed in Chapter 1).

Therefore, although the importance of e-PCP correlation in quantum chemistry and atomic physics calculations is not as significant as that of electron-electron correlation, the ability to investigate a wide range of PCPs makes the EHM a vital tool.

The model itself can also be considered more complex in future research. For example, to better simulate molecular environments, the harmonic oscillator can be replaced with an anharmonic oscillator, as atomic vibrations along and perpendicular to the bond axis have different frequencies. Given the capabilities of the Harmonium model in providing deep insights into electronic correlation, many studies in the field of DFT have been conducted on Harmonium, and almost all of them are applicable to Exotic Harmonium as well [147], [148], [157]–[163], [149]–[156].

The above examples show that research work in this area has just begun, and as the history of science has shown, the development of any model is equivalent to opening a window to a new world, each corner of which requires further exploration.

# 7 Appendix A

## Mathematical Framework of Ab Initio Multi-Component Methods

In this section, the mathematical details of the multi-component Hartree-Fock method are provided. Since the Hartree-Fock method is a mean-field theory, comparing it with exact methods will reveal the correlation effects. Given that the details of deriving the multi-component Hartree-Fock equations are not reported in the references, an effort has been made to present all the derivation steps in this section.

### A.1 Derivation of the Energy Expression

The general approach to obtaining MC-HF equations, similar to electronic structure theory, is to derive an expression for the system's energy and then vary it with respect to both the electronic and PCP orbitals simultaneously. To achieve this goal, we need to evaluate the expectation value of the energy, which has the following general form:

$$\begin{aligned} E_0 = \langle \Psi_0 | \hat{H}_{NEO} | \Psi_0 \rangle \\ = \left\langle \Psi_0 \middle| \hat{T}^{e}_{NEO} \middle| \Psi_0 \right\rangle + \left\langle \Psi_0 \middle| \hat{V}^{e}_{ext} \middle| \Psi_0 \right\rangle \\ + \left\langle \Psi_0 \middle| \hat{T}^{p}_{NEO} \middle| \Psi_0 \right\rangle + \left\langle \Psi_0 \middle| \hat{V}^{p}_{ext} \middle| \Psi_0 \right\rangle \\ + \left\langle \Psi_0 \middle| \hat{V}^{ee}_{NEO} \middle| \Psi_0 \right\rangle + \left\langle \Psi_0 \middle| \hat{V}^{ep}_{NEO} \middle| \Psi_0 \right\rangle \\ + \left\langle \Psi_0 \middle| \hat{V}^{pp}_{NEO} \middle| \Psi_0 \right\rangle \end{aligned} \qquad \text{A.1}$$

Kinetic Energy Operators

For the first term in Equation A-1, we have:

$$\begin{aligned} &\left\langle \Phi^e_0 \middle| \hat{T}^{e}_{NEO} + \hat{V}^{e}_{ext} \middle| \Phi^e_0 \right\rangle = \langle \Phi^e_0 | \textstyle\sum_i^N h_i \, | \Phi^e_0 \rangle \\ &= \sum_i^N \int dx^e_1 \, dx^e_2 ... \, dx^e_N \; \left[ (N!)^{-\frac{1}{2}} \left( \chi^e_1(x^e_1) ... \chi^e_k(x^e_N) \right.\right. \\ &\left.\left. - \chi^e_k(x^e_N) ... \; \chi^e_1(x^e_1) \right) \; \right]^* \\ &\times \; h_i \; \left[ (N!)^{-\frac{1}{2}} \left( \chi^e_1(x^e_1) ... \chi^e_k(x^e_N) - \chi^e_k(x^e_N) ... \; \chi^e_1(x^e_1) \right) \; \right] \\ &= \frac{1}{N!} \sum_i^N \int dx^e_1 \, dx^e_2 ... \, dx^e_N \, \{ \chi^{e*}_1(x^e_1) ... \chi^{e*}_k(x^e_N) \, h_i \, \chi^e_1(x^e_1) ... \chi^e_k(x^e_N) \\ &+ \chi^{e*}_k(x^e_N) ... \; \chi^{e*}_1(x^e_1) \, h_i \, \chi^e_k(x^e_N) ... \; \chi^e_1(x^e_1) \\ &- \chi^{e*}_1(x^e_1) ... \chi^{e*}_k(x^e_N) \, h_i \, \chi^e_k(x^e_N) ... \; \chi^e_1(x^e_1) \\ &- \chi^{e*}_k(x^e_N) ... \; \chi^e_1(x^e_1) h_i \chi^e_1(x^e_1) ... \chi^e_k(x^e_N) \} \end{aligned} \qquad \text{A.2}$$

By integrating over all variables except $x_i^e$, the above terms will be equal to one (the first two terms) or zero (the last two terms) due to the orthonormality of the electronic spin-orbitals, as follows:

$$\int dx^e\, \chi_i^{e*}\chi_j^e = \left\langle \chi_i^e \middle| \chi_j^e \right\rangle = \delta_{ij} = \begin{cases} 1 & \text{if } i = j \\ 0 & \text{if } i \neq j \end{cases} \qquad \text{A.3}$$

As a result, we will have:

$$\langle \Phi_0^e | \textstyle\sum_i^N h_i \, | \Phi_0^e \rangle = \displaystyle\sum_i^N \int dx_i^e \, \{\chi_1^{e*}(x_1^e) \ldots \chi_k^{e*}(x_N^e)\, h_i \, \chi_1^e(x_1^e) \ldots \chi_k^e(x_N^e)\} \qquad \text{A.4}$$

On the other hand, by convention, the dummy integration variables for the one-electron operators are chosen such that they correspond to the coordinates of the first electron. Therefore, we have:

$$\langle \Phi_0^e | \textstyle\sum_i^N h_i \, | \Phi_0^e \rangle = \displaystyle\sum_i^N \int dx_1^e \; \chi_i^{e*}(x_1^e)\, h_i \; \chi_i^{e*}(x_1^e) \qquad \text{A.5}$$

Following similar steps for $h_{i'}$ (Equation 1-20), we arrive at the following result:

$$\left\langle \Phi_0^p \middle| \textstyle\sum_{i'}^{N'} h_{i'} \middle| \Phi_0^p \right\rangle = \displaystyle\sum_{i'}^{N'} \int dx_1^p \; \chi_{i'}^{p*}(x_1^p)\, h_{i'}\, \chi_{i'}^p(x_1^p) \qquad \text{A.6}$$

Two-Particle Operators

Now, let's evaluate the matrix elements of the two-particle operators in Equation (A-1). Starting with $\hat{V}_{NEO}^{ee}$, we have:

$$\left\langle \Phi_0^e \middle| \sum_i^N \sum_{j>i}^N \left(\frac{1}{r_{ij}}\right) \middle| \Phi_0^e \right\rangle$$

$$= \sum_i^N \sum_{j>i}^N \left\{ \int dx_1^e\, dx_2^e \ldots\, dx_N^e \left[ (N!)^{-\frac{1}{2}} \left( \chi_1^e(x_1^e) \ldots \chi_k^e(x_N^e) - \chi_k^e(x_N^e) \ldots\, \chi_1^e(x_1^e) \right) \right]^* \times \frac{1}{r_{ij}} \left[ (N!)^{-\frac{1}{2}} \left( \chi_1^e(x_1^e) \ldots \chi_k^e(x_N^e) - \chi_k^e(x_N^e) \ldots\, \chi_1^e(x_1^e) \right) \right] \right\} \quad \text{A.7}$$

To evaluate the matrix elements of the operator $\frac{1}{r_{ij}}$, we use the general rules of electronic structure theory for evaluating two-electron integrals known as the Slater-Condon rules, resulting in the following expression [164]:

$$\left\langle \Phi_0^e \middle| \sum_i^N \sum_{j>i}^N \frac{1}{r_{ij}} \middle| \Phi_0^e \right\rangle$$

$$= \sum_i^N \sum_{j>i}^N \left[ \int dx_1^e\, dx_2^e\, \chi_i^{e*}(x_1^e) \chi_j^{e*}(x_2^e)\, \frac{1}{r_{ij}}\, \chi_i^e(x_1^e) \chi_j^e(x_2^e) - \int dx_1^e\, dx_2^e\, \chi_i^{e*}(x_1^e) \chi_j^{e*}(x_2^e)\, \frac{1}{r_{ij}}\, \chi_j^e(x_1^e)\, \chi_i^e(x_2^e) \right] \quad \text{A.8}$$

In the following, by convention, we write these expressions using standard notation. Here, we use the following notation:

$$[ij|kl] = \int dx_1 dx_2\, \chi_i^*(x_1) \chi_j(x_1) r_{12}^{-1}\, \chi_k^*(x_2) \chi_l(x_2) \quad \text{A.9}$$

Finally, we can write the two-electron integrals as follows:

$$\left\langle \Phi_0^e \middle| \sum_i^N \sum_{j>i}^N \frac{1}{r_{ij}} \middle| \Phi_0^e \right\rangle = \frac{1}{2} \sum_i^N \sum_j^N \left( \left[ \chi_i^e \chi_i^e \middle| \chi_j^e \chi_j^e \right] - \left[ \chi_i^e \chi_j^e \middle| \chi_i^e \chi_j^e \right] \right) \quad \text{A.10}$$

The coefficient $\frac{1}{2}$ appears because we have avoided imposing the condition $j > i$ for the sums. The matrix elements of the two-particle operator for the PCP also yield the following results through a similar process:

$$\left\langle \Phi_0^p \middle| \sum_{i'}^{N'} \sum_{j'>i'}^{N'} \frac{1}{r_{i'j'}} \middle| \Phi_0^p \right\rangle = \frac{1}{2} \sum_{i'}^{N'} \sum_{j'}^{N'} \left( \left[ \chi_{i'}^p \chi_{i'}^p \middle| \chi_{j'}^p \chi_{j'}^p \right] - \left[ \chi_{i'}^p \chi_{j'}^p \middle| \chi_{i'}^p \chi_{j'}^p \right] \right) \quad \text{A.11}$$

The two-particle operator for electron-PCP interaction has a form similar to other two-particle operators, but it simultaneously involves both the electronic and PCP coordinates. In other words, we should note that the exchange integrals in this case are zero because electrons and PCPs do not exhibit exchange effects with each other. Therefore, the fourth term in Equation (A-1) is calculated as follows:

$$\left\langle \Psi_0 \middle| \sum_i^N \sum_{i'}^{N'} \frac{Z_{i'}}{r_{ii'}} \middle| \Psi_0 \right\rangle = \left\langle \Phi_0^e(r^e)\, \Phi_0^p(r^p) \middle| \sum_i^N \sum_{i'}^{N'} \frac{Z_{i'}}{r_{ii'}} \middle| \Phi_0^e(r^e)\, \Phi_0^p(r^p) \right\rangle$$

$$= \frac{1}{N!} \frac{1}{N!'} \int dx_1^e\, dx_1^p \left[ \left( \chi_1^e(x_1^e)\, \chi_1^p(x_1^p) \right) \right]^* \sum_i^N \sum_{i'}^{N'} \frac{Z_{i'}}{r_{ii'}} \left[ \left( \chi_1^e(x_1^e)\, \chi_1^p(x_1^p) \right) \right] \quad \text{A.12}$$

$$= \sum_i^N \sum_{i'}^{N'} \int dx_1^e\, dx_1^p \left[ \left( \chi_1^e(x_1^e)\, \chi_1^p(x_1^p) \right) \right]^* \frac{Z_{i'}}{r_{ii'}} \left[ \left( \chi_1^e(x_1^e)\, \chi_1^p(x_1^p) \right) \right]$$

According to the rules of electronic structure theory [164], we have:

$$\left\langle \Psi_0 \middle| \Sigma_i^N \Sigma_{i'}^{N'} \frac{Z_{i'}}{r_{ii'}} \middle| \Psi_0 \right\rangle = \sum_i^N \sum_{i'}^{N'} Z_{i'} \left[ \chi_i^e \chi_i^e \middle| \chi_{i'}^p \chi_{i'}^p \right] \qquad \text{A.13}$$

In the end, the energy expression is obtained as follows:

$$\begin{aligned} E_0 = {} & \sum_i^N [\chi_i^e | h^e | \chi_i^e] + \frac{1}{2} \sum_i^N \sum_j^N \left( \left[ \chi_i^e \chi_i^e \middle| \chi_j^e \chi_j^e \right] - \left[ \chi_i^e \chi_j^e \middle| \chi_j^e \chi_i^e \right] \right) \\ & + \sum_{i'}^{N'} \left[ \chi_{i'}^p \middle| h^p \middle| \chi_{i'}^p \right] \\ & + \frac{1}{2} \sum_{i'}^{N'} \sum_{i'}^{N'} \left( \left[ \chi_{i'}^p \chi_{i'}^p \middle| \chi_{j'}^p \chi_{j'}^p \right] - \left[ \chi_{i'}^p \chi_{j'}^p \middle| \chi_{j'}^p \chi_{i'}^p \right] \right) \\ & - \sum_i^N \sum_{i'}^{N'} Z_{i'} \left[ \chi_i^e \chi_i^e \middle| \chi_{i'}^p \chi_{i'}^p \right] \end{aligned} \qquad \text{A.14}$$

### Transformation of Spin-Orbitals to Spatial Orbitals

So far, we have used spin-orbitals, but to simplify subsequent calculations, it is necessary to integrate out the spin functions from the electronic and PCP spin-orbitals. Therefore, we will use spatial orbitals instead of spin-orbitals. In the restricted Hartree-Fock approach, the $\alpha$ and $\beta$ spins have the same spatial orbitals, and each spatial orbital is occupied by two electrons with different spins. In other words, the electronic wave function consists of $N/2$ spin-orbitals with $\alpha$ spin and $N/2$ spin-orbitals with $\beta$ spin, i.e.:

$$\sum_{i}^{N} \chi_i^e = \sum_{i}^{N/2} \psi_i^e + \sum_{i}^{N/2} \bar{\psi}_i^e \qquad \text{A.15}$$

where the $\beta$ spin is indicated with a bar symbol. Using the orthonormality of the spin functions, the one-electron integrals are evaluated as follows:

$$\begin{aligned} \sum_{i}^{N} [\chi_i^e | h^e | \chi_i^e] &= \sum_{i}^{N/2} [\psi_i^e | h^e | \psi_i^e] + \sum_{\bar{i}}^{N/2} \left[\bar{\psi}_i^e \middle| h^e \middle| \bar{\psi}_i^e\right] \\ &= 2 \sum_{i}^{N/2} (\psi_i^e | h^e | \psi_i^e) \end{aligned} \qquad \text{A.16}$$

Similarly, for the double sums, we have:

$$\sum_{i}^{N}\sum_{j}^{N}\chi_i^e\chi_j^e = \sum_{i}^{N}\chi_i^e\sum_{j}^{N}\chi_j^e$$
$$= \sum_{i}^{N/2}(\psi_i^e + \bar{\psi}_i^e)\sum_{j}^{N/2}(\psi_j^e + \bar{\psi}_j^e)$$
$$= \sum_{i}^{N/2}\sum_{j}^{N/2}\psi_i^e\psi_j^e + \sum_{i}^{N/2}\sum_{j}^{N/2}\psi_i^e\bar{\psi}_j^e + \sum_{i}^{N/2}\sum_{j}^{N/2}\bar{\psi}_i^e\psi_j^e + \sum_{i}^{N/2}\sum_{j}^{N/2}\bar{\psi}_i^e\bar{\psi}_j^e \qquad \text{A.17}$$

Therefore, the integrals of the operator $\frac{1}{r_{ij}}$ can be expressed as follows:

$$\frac{1}{2}\sum_{i}^{N}\sum_{j}^{N}\left(\left[\chi_i^e\chi_i^e\middle|\chi_j^e\chi_j^e\right] - \left[\chi_i^e\chi_j^e\middle|\chi_j^e\chi_i^e\right]\right)$$
$$= \frac{1}{2}\left\{\sum_{i}^{N/2}\sum_{j}^{N/2}\left[\left(\psi_i^e\psi_i^e\middle|\psi_j^e\psi_j^e\right) - \left(\psi_i^e\psi_j^e\middle|\psi_j^e\psi_i^e\right)\right] + \sum_{i}^{N/2}\sum_{\bar{j}}^{N/2}\left[\left(\psi_i^e\psi_i^e\middle|\bar{\psi}_j^e\bar{\psi}_j^e\right) - \left(\psi_i^e\bar{\psi}_j^e\middle|\bar{\psi}_j^e\psi_i^e\right)\right] + \sum_{\bar{i}}^{N/2}\sum_{j}^{N/2}\left[\left(\bar{\psi}_i^e\bar{\psi}_i^e\middle|\psi_j^e\psi_j^e\right) - \left(\bar{\psi}_i^e\psi_j^e\middle|\psi_j^e\bar{\psi}_i^e\right)\right] + \sum_{\bar{i}}^{N/2}\sum_{\bar{j}}^{N/2}\left[\left(\bar{\psi}_i^e\bar{\psi}_i^e\middle|\bar{\psi}_j^e\bar{\psi}_j^e\right) - \left(\bar{\psi}_i^e\bar{\psi}_j^e\middle|\bar{\psi}_j^e\bar{\psi}_i^e\right)\right]\right\} \qquad \text{A.18}$$

It is clear that when the bar symbol appears on one side of the two-electron integrals, the term becomes zero due to the orthogonality of the spins.

Here, we consider the PCPs in the high-spin open-shell configuration, where all the orbitals of the PCPs have the same spin state (either all have alpha spin or beta spin). All that needs to be done here is to convert the spin-orbital notation (brackets) to spatial orbital notation (parentheses). By convention, for the integrals belonging to the operator $\frac{1}{r_{ij}}$, we have:

$$\sum_{i'}^{N'} \left[\chi_{i'}^{p}\middle| h^{p} \middle|\chi_{i'}^{p}\right] + \frac{1}{2}\sum_{i'}^{N'}\sum_{i'}^{N'} Z_{i'}Z_{j'}\left(\left[\chi_{i'}^{p}\chi_{i'}^{p}\middle|\chi_{j'}^{p}\chi_{j'}^{p}\right] - \left[\chi_{i'}^{p}\chi_{j'}^{p}\middle|\chi_{i'}^{p}\chi_{j'}^{p}\right]\right) = \sum_{i'}^{N'} h_{i'i'}^{p} + \frac{1}{2}\sum_{i'}^{N'}\sum_{i'}^{N'} Z_{i'}Z_{j'}\left[\left(\psi_{i'}^{p}\psi_{i'}^{p}\middle|\psi_{j'}^{p}\psi_{j'}^{p}\right) - \left(\psi_{i'}^{p}\psi_{j'}^{p}\middle|\psi_{i'}^{p}\psi_{j'}^{p}\right)\right] \quad \text{A.19}$$

In the final step, we need to obtain the spatial orbital notation for the electron-PCP interaction. Based on the above explanations, we can write:

$$\sum_{i}^{N}\sum_{i'}^{N'} \left[\chi_i^e\chi_i^e \middle| \chi_{i'}^p\chi_{i'}^p\right]$$

$$= \sum_{i}^{N/2}\sum_{i'}^{N'} \left[\psi_i^e\psi_i^e \middle| \psi_{i'}^p\psi_{i'}^p\right] + \sum_{\bar{i}}^{N/2}\sum_{i'}^{N'} \left[\bar{\psi}_i^e\bar{\psi}_i^e \middle| \psi_{i'}^p\psi_{i'}^p\right] = 2\sum_{i}^{N/2}\sum_{i'}^{N'} \left(\psi_i^e\psi_i^e \middle| \psi_{i'}^p\psi_{i'}^p\right) \qquad \text{A.20}$$

Finally, the energy expression in terms of spatial orbitals is obtained as follows:

$$E = 2\sum_{i}^{N/2} h_{ii}^e + \sum_{i}^{N/2}\sum_{j}^{N/2}[2(\psi_i^e\psi_i^e|\psi_j^e\psi_j^e) - (\psi_i^e\psi_j^e|\psi_j^e\psi_i^e)] + \sum_{i'}^{N'} h_{i'i'}^p + \frac{1}{2}\sum_{i'}^{N'}\sum_{i'}^{N'} Z_{i'}Z_{j'}\left[\left(\psi_{i'}^p\psi_{i'}^p \middle| \psi_{j'}^p\psi_{j'}^p\right) - \left(\psi_{i'}^p\psi_{j'}^p \middle| \psi_{i'}^p\psi_{j'}^p\right)\right] - 2\sum_{i}^{\frac{N}{2}}\sum_{i'}^{N'} Z_{i'}(\psi_i^e\psi_i^e|\psi_{i'}^p\psi_{i'}^p) \qquad \text{A.21}$$

where the integrals are defined as follows:

$$h_{ij}^{e} = \int dr_1\, \psi_i^{e*}(1)\, h^e(1)\, \psi_j^e(1) \qquad \text{A.22}$$

$$h_{i'j'}^{p} = \int dr_1\, \psi_{i'}^{p*}(1)\, h^p(1)\, \psi_{j'}^p(1) \qquad \text{A.23}$$

$$\left(\psi_i^e \psi_j^e \middle| \psi_k^e \psi_l^e\right) = \int dr_1 \int dr_2\, \psi_i^{e*}(1)\, \psi_j^e(1)\, r_{12}^{-1}\, \psi_k^{e*}(2) \psi_l^e(2) \qquad \text{A.24}$$

Minimizing the Energy Expression

At this stage, the energy expression (Equation A-21) is minimized simultaneously with respect to both the electronic and PCP orbitals under a suitable constraint using the Lagrange multiplier method. The appropriate constraints for applying the variational principle to the energy expression are the orthonormality of the electronic and PCP orbitals, as follows:

$$\int dr_1\, \psi_i^*(1) \psi_j(1) = \left[\psi_i \middle| \psi_j\right] = \delta_{ij} \qquad \text{A.25}$$

$$\int dr_1'\, \psi_{i'}^*(1) \psi_{j'}(1) = \left[\psi_{i'} \middle| \psi_{j'}\right] = \delta_{i'j'} \qquad \text{A.26}$$

$$\left[\psi_{i'} \middle| \psi_{j'}\right] - \delta_{i'j'} = 0 \qquad \text{A.27}$$

$$\left[\psi_i \middle| \psi_j\right] - \delta_{ij} = 0 \qquad \text{A.28}$$

The functional form of the Lagrangian is given by:

$$\mathcal{L}[\{\psi_i^e\},\{\psi_{i'}^p\}] = E_0[\{\psi_i^e\},\{\psi_{i'}^p\}] - \sum_i^{N/2}\sum_j^{N/2} \varepsilon_{ij}([\psi_i|\psi_j] - \delta_{ij}) - \sum_{i'}^{N'}\sum_{j'}^{N'} \varepsilon_{i'j'}([\psi_{i'}|\psi_{j'}] - \delta_{i'j'}) \quad \text{A.29}$$

where $E_0$ is the expectation value of the total wave function energy, which was finally obtained as the expression in Equation (A-21). By minimizing $\mathcal{L}[\{\psi_i^e\},\{\psi_{i'}^p\}]$, with respect to the constraints and with an infinitesimal variation of the electronic and PCP orbitals as follows:

$$\psi_i^e \longrightarrow \psi_i^e + \delta\psi_i^e \quad \text{A.30}$$

$$\psi_{i'}^p \longrightarrow \psi_{i'}^p + \delta\psi_{i'}^p \quad \text{A.31}$$

The first variation in $\mathcal{L}$ must be equal to zero:

$$\delta\mathcal{L} = \delta E_0 - 2\sum_i^{N/2}\sum_j^{N/2} \varepsilon_{ij}\delta([\psi_i|\psi_j] - \delta_{ij}) - \sum_{i'}^{N'}\sum_{j'}^{N'} \varepsilon_{i'j'}\delta([\psi_{i'}|\psi_{j'}] - \delta_{i'j'}) = 0 \quad \text{A.32}$$

Since the variation of a constant is zero, $\delta_{ij}$ and $\delta_{i'j'}$ will be zero. Now, we will evaluate each term to reach the final result. By varying the overlap integrals, we obtain:

$$\delta[\psi_i|\psi_j] = [\delta\psi_i|\psi_j] + [\psi_i|\delta\psi_j] \quad \text{A.33}$$

$$\delta[\psi_{i'}|\psi_{j'}] = [\delta\psi_{i'}|\psi_{j'}] + [\psi_{i'}|\delta\psi_{j'}] \qquad \text{A.34}$$

In the next step, by varying the energy expression, we have:

$$\begin{aligned}
\delta E_0 = \delta \Bigg\{ & 2\sum_{i}^{\frac{N}{2}} h_{ii}^{e} \\
& + \sum_{i}^{\frac{N}{2}}\sum_{j}^{\frac{N}{2}}\left[2\left(\psi_i^e\psi_i^e\middle|\psi_j^e\psi_j^e\right) - \left(\psi_i^e\psi_j^e\middle|\psi_j^e\psi_i^e\right)\right] \\
& + \sum_{i'}^{N'} h_{i'i'}^{p} \\
& + \frac{1}{2}\sum_{i'}^{N'}\sum_{j'}^{N'} Z_{i'}Z_{j'}\left[\left(\psi_{i'}^p\psi_{i'}^p\middle|\psi_{j'}^p\psi_{j'}^p\right)\right. \\
& \left. - \left(\psi_{i'}^p\psi_{j'}^p\middle|\psi_{j'}^p\psi_{i'}^p\right)\right] \\
& - 2\sum_{i}^{\frac{N}{2}}\sum_{i'}^{N'} Z_{i'}\left(\psi_i^e\psi_i^e\middle|\psi_{i'}^p\psi_{i'}^p\right)\Bigg\}
\end{aligned} \qquad \text{A.35}$$

To evaluate such a term, we need to find the complex conjugate parts and then put them all into a separate expression. Starting with $h_i^e$, we have:

$$2\sum_{i}^{\frac{N}{2}} \delta\, h_{ii}^{e} = 2\sum_{i}^{N/2} \delta(\psi_i^e|h_i^e|\psi_i^e)$$

$$= 2\sum_{i}^{N/2}[(\delta\psi_i^e|h_i^e|\psi_i^e) + (\psi_i^e|h_i^e|\delta\psi_i^e)] \qquad \text{A.36}$$

$$=2\sum_i^{N/2}(\delta\psi_i^e|h_i^e|\psi_i^e)$$

where the second term inside the brackets is the complex conjugate of the first one. For the subsequent terms, we obtain:

$$\sum_{i}^{\frac{N}{2}}\sum_{j}^{\frac{N}{2}} \delta\left[2\left(\psi_i^e\psi_i^e\middle|\psi_j^e\psi_j^e\right) - \left(\psi_i^e\psi_j^e\middle|\psi_j^e\psi_i^e\right)\right]$$

$$= 2\sum_{i}^{\frac{N}{2}}\sum_{j}^{\frac{N}{2}}\left[\left(\delta\psi_i^e\psi_i^e\middle|\psi_j^e\psi_j^e\right) + \left(\psi_i^e\delta\psi_i^e\middle|\psi_j^e\psi_j^e\right) + \left(\psi_i^e\psi_i^e\middle|\delta\psi_j^e\psi_j^e\right) + \left(\psi_i^e\psi_i^e\middle|\psi_j^e\delta\psi_j^e\right)\right]$$

$$- \sum_{i}^{\frac{N}{2}}\sum_{j}^{\frac{N}{2}}\left[\left(\delta\psi_i^e\psi_j^e\middle|\psi_j^e\psi_i^e\right) + \left(\psi_i^e\delta\psi_j^e\middle|\psi_j^e\psi_i^e\right) + \left(\psi_i^e\psi_j^e\middle|\delta\psi_j^e\psi_i^e\right) + \left(\psi_i^e\psi_j^e\middle|\psi_j^e\delta\psi_i^e\right)\right] \qquad \text{A.37}$$

In the first bracket (Coulomb terms), the first two terms are complex conjugates (CC) of each other, and the same is true for the next two terms, meaning:

$$\left(\delta\psi_i^e\psi_i^e\middle|\psi_j^e\psi_j^e\right) = \left(\psi_i^e\delta\psi_i^e\middle|\psi_j^e\psi_j^e\right)^* \qquad \text{A.38}$$

$$\left(\psi_i^e\psi_i^e\middle|\delta\psi_j^e\psi_j^e\right) = \left(\psi_i^e\psi_i^e\middle|\psi_j^e\delta\psi_j^e\right)^* \qquad \text{A.39}$$

Therefore:

$$\begin{aligned}\left(\delta\psi_i^e\psi_i^e\middle|\psi_j^e\psi_j^e\right) + \left(\psi_i^e\delta\psi_i^e\middle|\psi_j^e\psi_j^e\right) + \left(\psi_i^e\psi_i^e\middle|\delta\psi_j^e\psi_j^e\right)\\ + \left(\psi_i^e\psi_i^e\middle|\psi_j^e\delta\psi_j^e\right)\\ = \left(\delta\psi_i^e\psi_i^e\middle|\psi_j^e\psi_j^e\right) + \left(\psi_i^e\psi_i^e\middle|\delta\psi_j^e\psi_j^e\right) + \mathrm{CC}\end{aligned} \qquad \text{A.40}$$

These two terms are also equal through the change of indices in the double sum, which results in the exchange of the integral variables. In fact, since the sum over $i$ and $j$ of these quantities is equal, it is possible to exchange the variables in the sum (exchange $i$ and $j$) and the integral (exchange both sides of the parenthesis), therefore:

$$\left(\delta\psi_i^e\psi_i^e\middle|\psi_j^e\psi_j^e\right) = \left(\delta\psi_j^e\psi_j^e\middle|\psi_i^e\psi_i^e\right) = \left(\psi_i^e\psi_i^e\middle|\delta\psi_j^e\psi_j^e\right) \qquad \text{A.41}$$

We also have:

$$\begin{aligned}\left(\delta\psi_i^e\psi_i^e\middle|\psi_j^e\psi_j^e\right) + \left(\psi_i^e\delta\psi_i^e\middle|\psi_j^e\psi_j^e\right) + \left(\psi_i^e\psi_i^e\middle|\delta\psi_j^e\psi_j^e\right)\\ + \left(\psi_i^e\psi_i^e\middle|\psi_j^e\delta\psi_j^e\right) = 2\left(\delta\psi_i^e\psi_i^e\middle|\psi_j^e\psi_j^e\right) + \mathrm{CC}\end{aligned} \qquad \text{A.42}$$

We can apply these manipulations to the exchange terms as well. In this way, each of the four Coulomb term expressions is identical:

$$\left(\delta\psi_i^e\psi_j^e\middle|\psi_j^e\psi_i^e\right) = \left(\delta\psi_j^e\psi_i^e\middle|\psi_i^e\psi_j^e\right) = \left(\psi_i^e\delta\psi_j^e\middle|\psi_j^e\psi_i^e\right)^* \qquad \text{A.43}$$

$$\left(\psi_i^e\psi_j^e\middle|\delta\psi_j^e\psi_i^e\right)=\left(\psi_j^e\psi_i^e\middle|\delta\psi_i^e\psi_j^e\right)=\left(\psi_i^e\psi_j^e\middle|\psi_j^e\delta\psi_i^e\right)^* \qquad \text{A.44}$$

Therefore, we have:

$$\sum_i^{\frac{N}{2}}\sum_j^{\frac{N}{2}}\left[\left(\delta\psi_i^e\psi_j^e\middle|\psi_j^e\psi_i^e\right)+\left(\psi_i^e\delta\psi_j^e\middle|\psi_j^e\psi_i^e\right)+\left(\psi_i^e\psi_j^e\middle|\delta\psi_j^e\psi_i^e\right)+\left(\psi_i^e\psi_j^e\middle|\psi_j^e\delta\psi_i^e\right)\right]$$
$$=\left(\delta\psi_i^e\psi_j^e\middle|\psi_j^e\psi_i^e\right)+\left(\psi_i^e\psi_j^e\middle|\delta\psi_j^e\psi_i^e\right)+\mathrm{CC} \qquad \text{A.45}$$

And since:

$$\left(\delta\psi_i^e\psi_j^e\middle|\psi_j^e\psi_i^e\right)=\left(\delta\psi_j^e\psi_i^e\middle|\psi_i^e\psi_j^e\right)=\left(\psi_i^e\psi_j^e\middle|\delta\psi_j^e\psi_i^e\right) \qquad \text{A.46}$$

So, we have:

$$\sum_i^{\frac{N}{2}}\sum_j^{\frac{N}{2}}\left[\left(\delta\psi_i^e\psi_j^e\middle|\psi_j^e\psi_i^e\right)+\left(\psi_i^e\delta\psi_j^e\middle|\psi_j^e\psi_i^e\right)+\left(\psi_i^e\psi_j^e\middle|\delta\psi_j^e\psi_i^e\right)+\left(\psi_i^e\psi_j^e\middle|\psi_j^e\delta\psi_i^e\right)\right]$$
$$=\sum_i^{\frac{N}{2}}\sum_j^{\frac{N}{2}}2\left(\delta\psi_i^e\psi_j^e\middle|\psi_j^e\psi_i^e\right)+\mathrm{CC} \qquad \text{A.47}$$

Ultimately, we obtain:

$$\sum_{i}^{\frac{N}{2}}\sum_{j}^{\frac{N}{2}}\delta\left[2\left(\psi_i^e\psi_i^e\middle|\psi_j^e\psi_j^e\right)-\left(\psi_i^e\psi_j^e\middle|\psi_j^e\psi_i^e\right)\right]$$

$$=2\sum_{i}^{\frac{N}{2}}\sum_{j}^{\frac{N}{2}}\left\{2\left(\delta\psi_i^e\psi_i^e\middle|\psi_j^e\psi_j^e\right)-\left(\delta\psi_i^e\psi_j^e\middle|\psi_j^e\psi_i^e\right)+\mathrm{cc}\right\} \quad \text{A.48}$$

For the PCP terms, we also have:

$$\frac{1}{2}\sum_{i'}^{N'}\sum_{j'}^{N'}Z_{i'}Z_{j'}\delta\left[\left(\psi_{i'}^p\psi_{i'}^p\middle|\psi_{j'}^p\psi_{j'}^p\right)-\left(\psi_{i'}^p\psi_{j'}^p\middle|\psi_{j'}^p\psi_{i'}^p\right)\right]$$

$$=\sum_{i'}^{N'}\sum_{j'}^{N'}\left[\left(\delta\psi_{i'}^p\psi_{i'}^p\middle|\psi_{j'}^p\psi_{j'}^p\right)-\left(\delta\psi_{i'}^p\psi_{j'}^p\middle|\psi_{j'}^p\psi_{i'}^p\right)\right] \quad \text{A.49}$$

And the electron-PCP operator gives:

$$2\sum_{i}^{\frac{N}{2}}\sum_{i'}^{N'}Z_{i'}\left(\psi_i^e\psi_i^e\middle|\psi_{i'}^p\psi_{i'}^p\right)$$

$$=2\sum_{i}^{\frac{N}{2}}\sum_{i'}^{N'}\left(\delta\psi_i^e\psi_i^e\middle|\psi_{i'}^p\psi_{i'}^p\right)+\left(\psi_i^e\delta\psi_i^e\middle|\psi_{i'}^p\psi_{i'}^p\right)+\left(\psi_i^e\psi_i^e\middle|\delta\psi_{i'}^p\psi_{i'}^p\right)+\left(\psi_i^e\psi_i^e\middle|\psi_{i'}^p\delta\psi_{i'}^p\right) \quad \text{A.50}$$

On the other hand, we have:

$$\left(\delta\psi_i^e\psi_i^e\left|\psi_{i'}^p\psi_{i'}^p\right.\right) = \left(\psi_i^e\delta\psi_i^e\left|\psi_{i'}^p\psi_{i'}^p\right.\right)^* \quad \text{A.51}$$

$$\left(\psi_i^e\psi_i^e\left|\delta\psi_{i'}^p\psi_{i'}^p\right.\right) = \left(\psi_i^e\psi_i^e\left|\psi_{i'}^p\delta\psi_{i'}^p\right.\right)^* \quad \text{A.52}$$

Since the sum over $i$ and $i'$ for these quantities is not equal, we are not allowed to exchange variables. Therefore, ultimately, we obtain:

$$2\sum_{i}^{\frac{N}{2}}\sum_{i'}^{N'} Z_{i'}\left(\psi_i^e\psi_i^e\middle|\psi_{i'}^p\psi_{i'}^p\right) = 2\sum_{i}^{\frac{N}{2}}\sum_{i'}^{N'} Z_{i'}\left(\delta\psi_i^e\psi_i^e\middle|\psi_{i'}^p\psi_{i'}^p\right) + \left(\psi_i^e\psi_i^e\middle|\delta\psi_{i'}^p\psi_{i'}^p\right) + \mathrm{CC} \quad \text{A.53}$$

The same process applies to the integrals for PCPs. In the variation of the overlap integrals, we can exchange the dummy variables because, in these terms, we are dealing with only one electron:

$$\sum_{i}^{N/2}\sum_{j}^{N/2}\varepsilon_{ij}\left(\left[\delta\psi_i\middle|\psi_j\right] + \left[\psi_i\middle|\delta\psi_j\right]\right) = 2\sum_{i}^{N/2}\sum_{j}^{N/2}\varepsilon_{ij}\left[\delta\psi_i\middle|\psi_j\right] + \sum_{i}^{N/2}\sum_{j}^{N/2}\varepsilon_{ij}\left[\psi_i\middle|\delta\psi_j\right] \quad \text{A.54}$$

$$= 2\left(\sum_{i}^{N/2}\sum_{j}^{N/2}\varepsilon_{ij}[\delta\psi_i|\psi_j] + \sum_{i}^{N/2}\sum_{j}^{N/2}\varepsilon_{ji}[\psi_j|\delta\psi_i]\right)$$

$$= 2\left(\sum_{i}^{N/2}\sum_{j}^{N/2}\varepsilon_{ij}\left[\delta\psi_i\middle|\psi_j\right] + \sum_{i}^{N/2}\sum_{j}^{N/2}\varepsilon_{ij}^{*}\left[\delta\psi_i\middle|\psi_j\right]^{*}\right)$$

$$= 2\sum_{i}^{N/2}\sum_{j}^{N/2}\varepsilon_{ij}\left[\delta\psi_i\middle|\psi_j\right] + \mathrm{CC}$$

By applying the same process to the overlap integrals of the PCP, we obtain:

$$\sum_{i'}^{N'}\sum_{j'}^{N'}\varepsilon_{i'j'}\delta\left[\psi_{i'}\middle|\psi_{j'}\right] = \sum_{i'}^{N'}\sum_{j'}^{N'}\varepsilon_{i'j'}\left[\delta\psi_{i'}\middle|\psi_{j'}\right] + \mathrm{CC} \qquad \text{A.55}$$

We can now obtain the final form of the Hartree-Fock equation. Using the results above, the final Lagrangian expression is equal to:

$$
\begin{aligned}
\delta\mathcal{L} = 2\sum_{i}^{N/2}(\delta\psi_i^e|h^e|\psi_i^e) + \sum_{i'}^{N'}\left(\delta\psi_{i'}^p\middle|h^p\middle|\psi_{i'}^p\right) \\
+ 2\sum_{i}^{\frac{N}{2}}\sum_{j}^{\frac{N}{2}} \{2\left(\delta\psi_i^e\psi_i^e\middle|\psi_j^e\psi_j^e\right) \\
- \left(\delta\psi_i^e\psi_j^e\middle|\psi_j^e\psi_i^e\right)\} \\
+ \sum_{i'}^{N'}\sum_{i'}^{N'} Z_{i'}Z_{j'}\left[\left(\delta\psi_{i'}^p\psi_{i'}^p\middle|\psi_{j'}^p\psi_{j'}^p\right)\right. \\
\left.- \left(\delta\psi_{i'}^p\psi_{j'}^p\middle|\psi_{j'}^p\psi_{i'}^p\right)\right] \\
- 2\sum_{i}^{\frac{N}{2}}\sum_{i'}^{N'} Z_{i'}\left(\delta\psi_i^e\psi_i^e\middle|\psi_{i'}^p\psi_{i'}^p\right) \\
- 2\sum_{i}^{\frac{N}{2}}\sum_{j}^{\frac{N}{2}} \varepsilon_{ij}\left[\delta\psi_i\middle|\psi_j\right] \\
- \sum_{i'}^{N'}\sum_{j'}^{N'} \varepsilon_{i'j'}\left[\delta\psi_{i'}\middle|\psi_{j'}\right] + \mathrm{CC} = 0
\end{aligned}
\qquad \text{A.56}
$$

where the integrals are defined as follows:

$$\left(\delta\psi_i^e\psi_i^e\middle|\psi_j^e\psi_j^e\right) = \int dr_1^e\, dr_2^e\, \delta\psi_i^{e*}(1)\, \psi_i^e(1)\, r_{12}^{-1}\, \psi_j^{e*}(2)\psi_j^e(2) \qquad \text{A.57}$$

$$\left(\delta\psi_i^e\psi_i^e\middle|\psi_{i'}^p\psi_{i'}^p\right) = \int dr_1^e\, dr_2^p\, \delta\psi_i^{e*}(1)\, \psi_i^e(1)\, r_{12}^{-1}\psi_{i'}^{p*}(2)\psi_{i'}^p(2) \qquad \text{A.58}$$

Due to space limitations, we will define some operators using the operators we introduced earlier. The Coulomb operators are as follows:

$$J_j^{ee}(1)\psi_i^e(1) = \left[\iint dr_2^e\, \psi_j^{e*}(2)\, \hat{V}_{NEO}^{ee}\, \psi_j^e(2)\right] \psi_i^e(1) \qquad \text{A.59}$$

$$J_{j'}^{pp}(1)\psi_{i'}^p(1) = \left[\iint dr_2^p\, \psi_{j'}^{p*}(2)\, \hat{V}_{NEO}^{pp}\, \psi_{j'}^p(2)\right] \psi_{i'}^p(1) \qquad \text{A.60}$$

$$J_{i'}^{pe}(1)\psi_i^e(1) = \left[\iint dr_2^p\, \psi_{i'}^{p*}(2)\, \hat{V}_{NEO}^{pe}\, \psi_{i'}^p(2)\right] \psi_i^e(1) \qquad \text{A.61}$$

$$J_i^{ep}(1)\psi_{i'}^p(1) = \left[\iint dr_2^e\, \psi_i^{e*}(2)\, \hat{V}_{NEO}^{pe}\, \psi_i^e(2)\right] \psi_{i'}^p(1) \qquad \text{A.62}$$

And the exchange operators are defined as follows:

$$K_j^e(1)\psi_i^e(1) = \left[\iint dr_2^e\, \psi_j^{e*}(2)\, \hat{V}_{NEO}^{ee}\, \psi_i^e(2)\right] \psi_j^e(1) \qquad \text{A.63}$$

$$K_{j'}^p(1)\psi_{i'}^p(1) = \left[\iint dr_2^e\, \psi_{j'}^{e*}(2)\, \hat{V}_{NEO}^{pp}\, \psi_{i'}^e(2)\right] \psi_{j'}^e(1) \qquad \text{A.64}$$

Likewise, similar operators for PCPs are defined. Ultimately, we have:

$$\delta\mathcal{L} = 2\sum_{i}^{N/2}\int dr_1^e\, \delta\psi_i^{e*}(1)\Bigg[h^e(1)\psi_i^e(1) + \sum_{j}^{N/2}\left(2J_j^{ee}(1) - K_j^{ee}(1)\right)\psi_i^e(1) - \sum_{i'}^{N'} J_{i'}^{ep}\psi_i^e(1) - \sum_{j}^{N/2}\varepsilon_{ij}\,\psi_j^e(1)\Bigg] + \sum_{i'}^{N'}\int dr_1^p\, \delta\psi_{i'}^{p*}(1)\Bigg[h^p(1)\psi_{i'}^p(1) + \sum_{j'}^{N'}\left(J_{j'}^{pp}(1) - K_{j'}^{pp}(1)\right)\psi_{i'}^p(1) - 2\sum_{i}^{N/2} J_i^{pe}(1)\psi_{i'}^p(1) - \sum_{j'}^{N'}\varepsilon_{i'j'}\psi_{j'}^p(1)\Bigg] + \mathrm{CC} = 0 \qquad \text{A.65}$$

In the above expression, the quantities inside the brackets must be zero because $\delta\psi_i^{e*}(1)$ and $\delta\psi_{i'}^{p*}(1)$ are arbitrary, meaning:

$$\left[h^e(1) + \sum_{j}^{\frac{N}{2}}\left(2J_j^{ee}(1) - K_j^{ee}(1)\right) + \sum_{i'}^{N'} J_{i'}^{ep}\right]\psi_i^e(1) = \sum_{j}^{\frac{N}{2}}\varepsilon_{ij}\,\psi_j^e(1) \qquad \text{A.66}$$

$$\left[h^p(1) + \sum_{j'}^{N'} \left(J_{j'}^{pp}(1) - K_{j'}^{pp}(1)\right) + 2\sum_{i}^{\frac{N}{2}} J_i^{pe}(1)\right] \psi_{i'}^p(1) = \sum_{j'}^{N'} \varepsilon_{i'j'} \psi_{j'}^e(1) \qquad \text{A.67}$$

After performing a unitary transformation to diagonalize the matrices $\varepsilon_{ij}$ and $\varepsilon_{i'j'}$, we reach the final form of the Hartree-Fock equations:

$$f^e(1)\, \psi_i^e(1) = \varepsilon_i\, \psi_i^e(1) \qquad \text{A.68}$$

$$f^p(1)\, \psi_{i'}^p(1) = \varepsilon_i\, \psi_{i'}^p(1) \qquad \text{A.69}$$

In which $f^e$ and $f^p$ are called the electronic and PCP Fock operators, respectively, and are equal to the expressions inside the brackets in equations (A-66) and (A-67).

Roothaan Equations

We can expand the orbitals of PCPs and electrons using a series of basis functions ($\phi_v^e$), which are commonly Gaussian functions. This approach transforms an integro-differential equation into a matrix equation that is easier to solve:

$$\psi_i^e(1) = \sum_{v}^{B} c_{vi}^e \phi_v^e(1) \qquad \text{A.70}$$

$$\psi_{i'}^{p}(1) = \sum_{v'}^{B'} c_{v'i'}^{p} \phi_{v'}^{p}(1) \qquad \text{A.71}$$

where $B$ and $B'$ are the number of basis functions for the electron and the PCP, respectively. By substituting the above expansions into the Hartree-Fock equations (A-68) and (A-69), we obtain the following results:

$$\sum_{v}^{B} f^{e}(1)\, c_{vi}^{e} \phi_{v}^{e}(1) = \varepsilon_{i} \sum_{v}^{B} c_{vi}^{e} \phi_{v}^{e}(1) \qquad \text{A.72}$$

$$\sum_{v'}^{B'} f^{p}(1)\, c_{v'i'}^{p} \phi_{v'}^{p}(1) = \varepsilon_{i'} \sum_{v'}^{B'} c_{v'i'}^{p} \phi_{v'}^{p}(1) \qquad \text{A.73}$$

By multiplying $\phi_{\mu}^{e*}$ and $\phi_{\mu'}^{p*}$ on the left side of the above equations, respectively, and then integrating, we have:

$$\sum_{v}^{B} c_{vi}^{e} \int dr_{1}^{e}\, \phi_{\mu}^{e*}(1)\, f^{e}(1)\, \phi_{v}^{e}(1) = \varepsilon_{i} \sum_{v}^{B} c_{vi}^{e} \int \phi_{\mu}^{e*}(1)\, \phi_{v}^{e}(1) \qquad \text{A.74}$$

$$\begin{aligned} &\sum_{v'}^{B'} c_{v'i'}^{p} \int dr_{1}^{p}\, \phi_{\mu'}^{p*}(1)\, f^{p}(1)\, \phi_{v'}^{p}(1) \\ &\qquad = \varepsilon_{i'} \sum_{v'}^{B'} c_{v'i'}^{p} \int \phi_{\mu'}^{p*}(1)\, \phi_{v'}^{p}(1) \end{aligned} \qquad \text{A.75}$$

The above expressions can be summarized by introducing the overlap matrices and the Fock matrices as follows:

$$S^{e}_{\mu v} = \int dr^{e}_{1}\ \phi^{e*}_{\mu}(1)\ \phi^{e}_{v}(1) \qquad \text{A.76}$$

$$S^{p}_{\mu' v'} = \int dr^{p}_{1}\ \phi^{p*}_{\mu'}(1)\ \phi^{p}_{v'}(1) \qquad \text{A.77}$$

$$F^{e}_{\mu v} = \int dr^{e}_{1}\ \phi^{e*}_{\mu}(1)\ f^{e}(1)\ \phi^{e}_{v}(1) \qquad \text{A.78}$$

$$F^{p}_{\mu' v'} = \int dr^{p}_{1}\ \phi^{p*}_{\mu'}(1)\ f^{p}(1)\ \phi^{p}_{v'}(1) \qquad \text{A.79}$$

By substituting the above expressions into equations (A-68) and (A-69), we arrive at the following equations:

$$\sum_{v}^{B} F^{e}_{\mu v}\, c^{e}_{vi} = \varepsilon_i \sum_{v}^{B} c^{e}_{vi}\, S^{e}_{\mu v} \qquad \text{A.80}$$

$$\sum_{v'}^{B'} F^{p}_{\mu' v'}\, c^{p}_{v' i'} = \varepsilon_{i'} \sum_{v'}^{B'} c^{p}_{v' i'}\, S^{p}_{\mu' v'} \qquad \text{A.81}$$

These are the Roothaan equations. In equation (A-80) which pertains to electrons, $c$ and $\varepsilon$ are square matrices of size $B \times B$ (similarly for equation A-81 for PCPs). The Fock operator can be defined using the basis function expansion. In fact, the Fock matrix is a matrix representation of the Fock operator and has the following form:

$$F^{e}_{\mu v} = \int dr^{e}_{1}\ \phi^{e*}_{\mu}(1)\, f^{e}(1)\phi^{e}_{v}(1)$$

$$= \int dr^{e}_{1}\ \phi^{e*}_{\mu}(1)\, h^{e}(1)\phi^{e}_{v}(1)$$

$$+ \sum_{j}^{N/2} \int dr^{e}_{1}\ \phi^{e*}_{\mu}(1)\left[2J^{ee}_{j}(1) - K^{ee}_{j}(1)\right]\phi^{e}_{v}(1) \qquad \text{A.82}$$

$$+ \sum_{i'}^{N'} \int dr^{e}_{1}\ \phi^{e*}_{\mu}(1)\, J^{ep}_{i'}\phi^{e}_{v}(1)$$

A similar equation is obtained for PCPs by following the above process:

$$F^{p}_{\mu' v'} = \int dr^{p}_{1}\ \phi^{p*}_{\mu'}(1)\, f^{p}(1)\phi^{p}_{v'}(1)$$

$$= \int dr^{p}_{1}\ \phi^{p*}_{\mu'}(1)\, h^{p}(1)\phi^{p}_{v'}(1)$$

$$+ \sum_{j'}^{N'} \int dr^{p}_{1}\ \phi^{p*}_{\mu'}(1)\left[J^{pp}_{j'}(1) - K^{pp}_{j'}(1)\right]\phi^{p}_{v'}(1) \qquad \text{A.83}$$

$$+ \sum_{i}^{N/2} \int dr^{p}_{1}\ \phi^{p*}_{\mu'}(1)\, J^{pe}_{i}\phi^{p}_{v'}(1)$$

By defining the core Hamiltonians as follows:

$$H^{coreE}_{\mu v} = \int dr^{e}_{1}\ \phi^{e*}_{\mu}(1)\, h^{e}(1)\phi^{e}_{v}(1) \qquad \text{A.84}$$

$$H^{coreP}_{\mu' v'} = \int dr^{p}_{1}\ \phi^{p*}_{\mu'}(1)\, h^{p}(1)\phi^{p}_{v'}(1) \qquad \text{A.85}$$

We can write equation (A-82) as follows:

$$F^e_{\mu v} = H^{coreE}_{\mu v} + \sum_{j}^{N/2} \left[ 2 \left( \phi^e_\mu \, \phi^e_v \middle| \psi^e_j \psi^e_j \right) - \left( \phi^e_\mu \psi^e_j \middle| \psi^e_j \phi^e_v \right) \right] - 2 \sum_{i'}^{N'} Z_{i'} \left( \phi^e_\mu \, \phi^e_v \middle| \psi^p_{i'} \psi^p_{i'} \right) \quad \text{A.86}$$

By substituting the basis functions $\sigma$ and $\lambda$ into the above expression, we have:

$$\begin{aligned} F^e_{\mu v} &= H^{core}_{\mu v} + \sum_{j}^{N/2} \sum_{\sigma\lambda}^{B} c^e_{\lambda j} \, c^{e*}_{\sigma j} \left[ 2 \left( \phi^e_\mu \, \phi^e_v \middle| \, \phi^e_\sigma \, \phi^e_\lambda \right) \right. \\ &\quad \left. - \left( \phi^e_\mu \phi^e_\lambda \middle| \, \phi^e_\sigma \phi^e_v \right) \right] \\ &\quad - 2 \sum_{i'}^{N'} \sum_{\sigma'\lambda'}^{B'} Z_{i'} c^p_{\lambda' i'} \, c^{p*}_{\sigma' i'} \left( \phi^e_\mu \, \phi^e_v \middle| \psi^p_{\sigma'} \psi^p_{\lambda'} \right) \\ &= H^{coreE}_{\mu v} + G^e_{\mu v} - \sum_{\lambda' \sigma'}^{B'} P^p_{\lambda' \sigma'} \left( \phi^e_\mu \, \phi^e_v \middle| \psi^p_{\sigma'} \psi^p_{\lambda'} \right) \end{aligned} \quad \text{A.87}$$

By following the same procedure for the equations of PCPs, we obtain:

$$F^{p}_{\mu'\nu'} = H^{coreP}_{\mu'\nu'} + \sum_{j'}^{N'}\sum_{\sigma'\lambda'}^{B'} Z_{j'} c^{p}_{\lambda'j'} c^{p*}_{\sigma'j'} \left[\left(\phi^{p}_{\mu'}\,\phi^{p}_{\nu'}\middle|\,\phi^{p}_{\sigma'}\,\phi^{p}_{\lambda'}\right) - \left(\phi^{p}_{\mu'}\phi^{p}_{\lambda'}\middle|\,\phi^{p}_{\sigma'}\phi^{p}_{\nu'}\right)\right] - \sum_{i'}^{N'}\sum_{\sigma\lambda}^{B} Z_{i'} c^{e}_{\lambda i} c^{e*}_{\sigma i}\left(\phi^{p}_{\mu'}\,\phi^{p}_{\nu'}\middle|\psi^{e}_{\sigma}\psi^{e}_{\lambda}\right) = H^{coreP}_{\mu'\nu'} + G^{p}_{\mu'\nu'} - \sum_{\lambda\sigma}^{B} P^{e}_{\lambda\sigma}\left(\phi^{p}_{\mu'}\,\phi^{p}_{\nu'}\middle|\psi^{e}_{\sigma}\psi^{e}_{\lambda}\right) \qquad \text{A.88}$$

where

$$G^{p}_{\mu\nu} = \sum_{\sigma\lambda}^{B} P^{p}_{\lambda\sigma}\left[\left(\phi^{p}_{\mu}\,\phi^{p}_{\nu}\middle|\,\phi^{p}_{\sigma}\,\phi^{p}_{\lambda}\right) - \frac{1}{2}\left(\phi^{p}_{\mu}\phi^{p}_{\lambda}\middle|\,\phi^{p}_{\sigma}\phi^{e}_{\nu}\right)\right] \qquad \text{A.89}$$

$$G^{p}_{\mu'\nu'} = Z_{i'}\sum_{\sigma'\lambda'}^{B'} P^{p}_{\sigma'\lambda'}\left[\left(\phi^{p}_{\mu'}\,\phi^{p}_{\nu'}\middle|\,\phi^{p}_{\sigma'}\,\phi^{p}_{\lambda'}\right) - \left(\phi^{p}_{\mu'}\psi^{p}_{\sigma'}\middle|\,\phi^{p}_{\nu'}\psi^{p}_{\lambda'}\right)\right] \qquad \text{A.90}$$

In these expressions, the new quantities $P^{e}_{\nu\mu}$ and $P^{p}_{\nu'\mu'}$, called the density matrices, are defined as follows:

$$P^{e}_{\nu\mu} = 2\sum_{i}^{N/2} c^{e}_{\nu i}\, c^{e*}_{\mu i} \qquad \text{A.91}$$

$$P^{p}_{\nu'\mu'} = \sum_{i'}^{N'} Z_{i'} c^{p}_{\nu'i'}\, c^{p*}_{\mu'i'} \qquad \text{A.92}$$

The factor of 2 appearing in the electronic density matrix arises from the transition of spin-orbitals to closed-shell spatial orbitals.

Equations (A-87) and (A-88) represent the final forms of the electronic and PCP Fock operators, respectively, expanded using the basis functions. Therefore, the derivation of the multi-component Hartree-Fock equations concludes here. These equations show that the electronic (PCP) Fock matrix also includes the PCP (electronic) density matrix; hence, they are sometimes referred to as coupled Hartree-Fock equations.

The mathematical framework presented here forms the theoretical foundation of the multi-component Hartree-Fock method. However, for coding this framework, specific mathematical techniques are employed to further simplify the calculations, especially the two-particle integrals, which can be found in references [165], [166], [175], [176], [167]–[174].

# 8 Appendix B

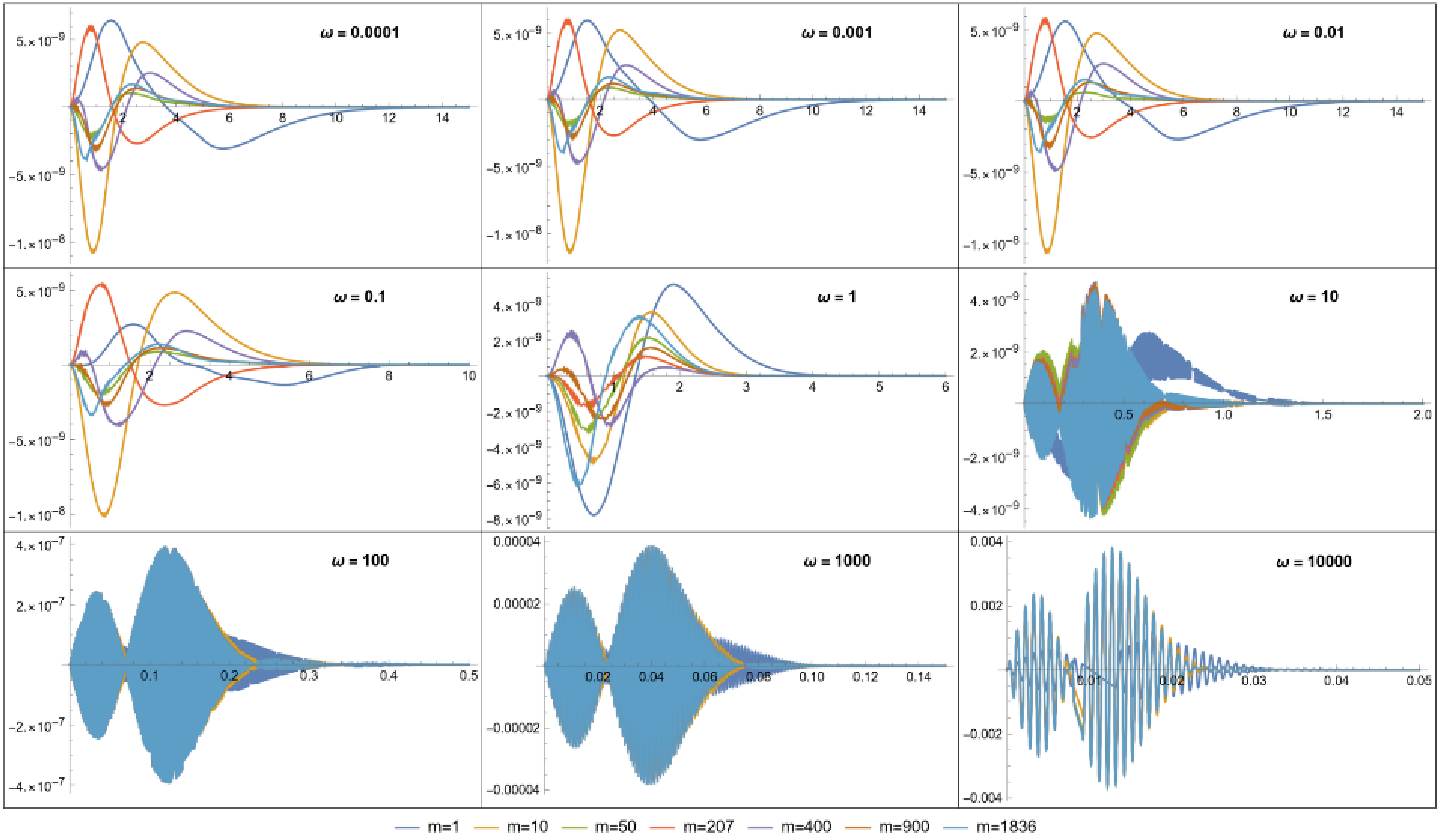

Figure B-1: Energy difference using the finite element method for two mesh sizes, 0.001 and 0.0001. This difference becomes significant only at a frequency of 10,000, therefore, a mesh size of 0.0001 was used at this frequency.

Table B-1: Variational parameters, variational energy of relative motion, and their errors for wavefunction (2-75)

| ω/Quantities | $\boldsymbol{\alpha}''$ [1] | $\boldsymbol{\beta}''$ [2] | Variational Energy[3] | FEM Energy[4] | ΔE[5] | %error[6] |
|---|---|---|---|---|---|---|
| | | | m=1 | | | |
| 0.0001 | 0.999996 | 1.000002 | -0.250000 | -0.250000 | 0.000000 | 0.000000 |
| 0.001 | 0.999989 | 1.000012 | -0.249997 | -0.249997 | 0.000000 | 0.000000 |
| 0.01 | 1.000314 | 1.000458 | -0.249700 | -0.249701 | 0.000000 | -0.000070 |
| 0.1 | 0.919436 | 1.091479 | -0.223717 | -0.223957 | -0.000240 | -0.106998 |
| 1.0 | 1.103504 | 1.559082 | 0.614439 | 0.611853 | -0.002587 | 0.422747 |
| 10 | 5.476005 | 1.851585 | 12.398910 | 12.395303 | -0.003607 | 0.029097 |
| 100 | 48.724976 | 1.952538 | 141.945987 | 141.942127 | -0.003861 | 0.002720 |
| 1000 | 487.383633 | 1.984942 | 1474.694360 | 1474.690429 | -0.003931 | 0.000267 |
| 10000 | 4932.850820 | 1.995251 | 14920.137495 | 14920.135808 | -0.001687 | 0.000011 |
| | | | m=10 | | | |
| 0.0001 | 0.999987 | 1.000010 | -0.454545 | -0.454545 | 0.000000 | 0.000000 |
| 0.001 | 0.999988 | 1.000012 | -0.454544 | -0.454544 | 0.000000 | 0.000000 |
| 0.01 | 0.999794 | 1.000429 | -0.454381 | -0.454381 | 0.000000 | -0.000003 |
| 0.1 | 0.985483 | 1.035773 | -0.438836 | -0.438923 | -0.000088 | -0.019938 |
| 1.0 | 1.183071 | 1.434114 | 0.254978 | 0.251315 | -0.003663 | 1.457446 |
| 10 | 5.325488 | 1.801301 | 11.453381 | 11.447087 | -0.006295 | 0.054991 |
| 100 | 47.379844 | 1.936102 | 139.103860 | 139.096910 | -0.006950 | 0.004997 |
| 1000 | 480.153146 | 1.979707 | 1465.842722 | 1465.835599 | -0.007123 | 0.000486 |
| 10000 | 4900.246695 | 1.993594 | 14892.278556 | 14892.271372 | -0.007183 | 0.000048 |
| | | | m=50 | | | |
| 0.0001 | 0.999985 | 1.000012 | -0.490196 | -0.490196 | 0.000000 | 0.000000 |
| 0.001 | 0.999986 | 1.000014 | -0.490195 | -0.490195 | 0.000000 | 0.000000 |
| 0.01 | 0.999847 | 1.000372 | -0.490043 | -0.490043 | 0.000000 | -0.000002 |
| 0.1 | 0.989384 | 1.031384 | -0.475538 | -0.475612 | -0.000075 | -0.015672 |
| 1.0 | 1.190500 | 1.417222 | 0.198722 | 0.194935 | -0.003787 | 1.942556 |
| 10 | 5.297276 | 1.793893 | 11.310715 | 11.303970 | -0.006745 | 0.059668 |
| 100 | 47.164546 | 1.933659 | 138.679004 | 138.671520 | -0.007484 | 0.005397 |
| 1000 | 487.383633 | 1.984942 | 1474.694360 | 1474.690429 | -0.003931 | 0.000267 |
| 10000 | 4895.212995 | 1.993348 | 14888.128524 | 14888.120778 | -0.007746 | 0.000052 |
| | | | m = 207 | | | |
| 0.0001 | 0.999985 | 1.000013 | -0.497596 | -0.497596 | 0.000000 | 0.000000 |
| 0.001 | 0.999986 | 1.000014 | -0.497595 | -0.497595 | 0.000000 | 0.000000 |
| 0.01 | 0.999857 | 1.000361 | -0.497446 | -0.497446 | 0.000000 | -0.000002 |
| 0.1 | 0.990073 | 1.030571 | -0.483139 | -0.483212 | -0.000072 | -0.014930 |
| 1.0 | 1.191882 | 1.413851 | 0.187213 | 0.183403 | -0.003810 | 2.077655 |
| 10 | 5.291400 | 1.792392 | 11.281699 | 11.274861 | -0.006838 | 0.060645 |
| 100 | 47.120482 | 1.933164 | 138.592722 | 138.585127 | -0.007595 | 0.005480 |
| 1000 | 478.795133 | 1.978769 | 1464.255161 | 1464.247368 | -0.007792 | 0.000532 |
| 10000 | 4894.205763 | 1.993299 | 14887.286169 | 14887.278307 | -0.007863 | 0.000053 |
| | | | m = 400 | | | |
| 0.0001 | 0.999985 | 1.000013 | -0.498753 | -0.498753 | 0.000000 | 0.000000 |
| 0.001 | 0.999986 | 1.000014 | -0.498752 | -0.498752 | 0.000000 | 0.000000 |
| 0.01 | 0.999856 | 1.000361 | -0.498603 | -0.498603 | 0.000000 | -0.000002 |
| 0.1 | 0.990178 | 1.030446 | -0.484327 | -0.484399 | -0.000072 | -0.014818 |
| 1.0 | 1.192093 | 1.413328 | 0.185419 | 0.181605 | -0.003814 | 2.100237 |
| 10 | 5.290481 | 1.792159 | 11.277180 | 11.270328 | -0.006852 | 0.060797 |
| 100 | 47.113599 | 1.933086 | 138.579288 | 138.571677 | -0.007612 | 0.005493 |
| 1000 | 478.759328 | 1.978744 | 1464.213456 | 1464.205645 | -0.007810 | 0.000533 |
| 10000 | 4894.031446 | 1.993290 | 14887.155037 | 14887.147157 | -0.007881 | 0.000053 |
| | | | m=900 | | | |
| 0.0001 | 0.999985 | 1.000013 | -0.499445 | -0.499445 | 0.000000 | 0.000000 |
| 0.001 | 0.999986 | 1.000014 | -0.499444 | -0.499444 | 0.000000 | 0.000000 |
| 0.01 | 0.999857 | 1.000360 | -0.499295 | -0.499295 | 0.000000 | -0.000002 |
| 0.1 | 0.990240 | 1.030372 | -0.485038 | -0.485109 | -0.000072 | -0.014752 |
| 1.0 | 1.192219 | 1.413015 | 0.184346 | 0.180530 | -0.003816 | 2.113946 |
| 10 | 5.289932 | 1.792020 | 11.274480 | 11.267619 | -0.006861 | 0.060889 |
| 100 | 47.109533 | 1.933041 | 138.571262 | 138.563640 | -0.007622 | 0.005501 |
| 1000 | 478.738007 | 1.978729 | 1464.188536 | 1464.180715 | -0.007821 | 0.000534 |
| 10000 | 4893.940977 | 1.993286 | 14887.076684 | 14887.068793 | -0.007892 | 0.000053 |
| | | | m=1836 | | | |
| 0.0001 | 0.999985 | 1.000013 | -0.499728 | -0.499728 | 0.000000 | 0.000000 |
| 0.001 | 0.999986 | 1.000014 | -0.499726 | -0.499726 | 0.000000 | 0.000000 |
| 0.01 | 0.999858 | 1.000359 | -0.499578 | -0.499578 | 0.000000 | -0.000002 |
| 0.1 | 0.990265 | 1.030342 | -0.485328 | -0.485400 | -0.000071 | -0.014725 |
| 1.0 | 1.192270 | 1.412888 | 0.183908 | 0.180091 | -0.003817 | 2.119593 |
| 10 | 5.289708 | 1.791963 | 11.273377 | 11.266512 | -0.006864 | 0.060926 |
| 100 | 47.107820 | 1.933022 | 138.567983 | 138.560357 | -0.007626 | 0.005504 |
| 1000 | 478.740821 | 1.978733 | 1464.178358 | 1464.170533 | -0.007825 | 0.000534 |
| 10000 | 4893.907688 | 1.993284 | 14887.044681 | 14887.036785 | -0.007896 | 0.000053 |

1. Variational parameter $\boldsymbol{\alpha}''$
2. Variational parameter $\beta''$
3. Variational energy of the Hamiltonian for the relative motion of the system using parameters 1 and 2
4. FEM energy of the Hamiltonian for the relative motion of the system using parameters 1 and 2
5. Energy difference between 3 and 4
6. Percentage error of the variational energy relative to the FEM energy of the Hamiltonian for relative motion

Table B-2: Coefficients and exponents of MC-HF basis set [7s:1s]: the top and bottom rows for each frequency respectively represent the absolute values of the sorted coefficients (from largest to smallest) and their corresponding exponents.

| Sorted electron and PC coefficients and corresponding exponents | | | | | | | | |
|---|---|---|---|---|---|---|---|---|
| m=1 | | | | | | | | |
| ω/basis set | 1s | 2s | 3s | 4s | 5s | 6s | 7s | 1s |
| 0.0001 | 0.319802 | 0.318093 | 0.223462 | 0.125915 | 0.051273 | 0.004274 | 0.001406 | 1 |
| | 0.056193 | 0.032409 | 0.024091 | 0.014230 | 0.100768 | 0.110112 | 0.005600 | 0.036377 |
| 0.001 | 0.480059 | 0.309318 | 0.150645 | 0.070601 | 0.023273 | 0.011134 | -0.000783 | 1 |
| | 0.037353 | 0.019479 | 0.065922 | 0.079692 | 0.009443 | 0.134079 | 0.148314 | 0.036377 |
| 0.01 | 0.751009 | 0.264841 | 0.233746 | 0.006087 | 0.000389 | -0.000569 | -0.213076 | 1 |
| | 0.036651 | 0.071272 | 0.017227 | 0.177794 | 0.914376 | 0.753855 | 0.041752 | 0.037689 |
| 0.1 | 0.426653 | 0.264646 | 0.260619 | 0.111251 | 0.005225 | -0.001106 | -0.044230 | 1 |
| | 0.074259 | 0.052490 | 0.109068 | 0.174970 | 0.351791 | 0.443185 | 0.045903 | 0.085144 |
| 1 | 0.700127 | 0.131542 | 0.121834 | 0.051798 | 0.002541 | 0.000003 | -0.000031 | 1 |
| | 0.541680 | 0.763898 | 0.465139 | 1.102035 | 1.931633 | 16.317160 | 5.419529 | 0.577278 |
| 10 | 0.830460 | 0.110435 | 0.059221 | 0.005317 | 0.001988 | -0.000579 | -0.004335 | 1 |
| | 5.124563 | 4.764844 | 7.859766 | 19.733625 | 36.104344 | 44.263938 | 26.246875 | 5.220210 |
| 100 | 0.797798 | 0.238146 | 0.006739 | 0.000421 | -0.000235 | -0.002480 | -0.039593 | 1 |
| | 51.759766 | 44.683125 | 127.570625 | 346.082500 | 402.562500 | 165.378125 | 39.371875 | 50.674434 |
| 1000 | 0.814668 | 0.189003 | 0.001467 | 0.000294 | -0.000020 | -0.000423 | -0.004737 | 1 |
| | 508.465772 | 469.819336 | 1205.078125 | 3008.703125 | 5038.538281 | 2522.270313 | 372.322266 | 502.112311 |
| 10000 | 0.797600 | 0.211149 | 0.000917 | 0.000025 | -0.000094 | -0.001887 | -0.007633 | 1 |
| | 5094.914424 | 4643.193124 | 15178.937500 | 64174.160156 | 37568.828125 | 10588.718750 | 3577.812500 | 5006.657809 |

| m = 1.5 | | | | | | | | |
|---|---|---|---|---|---|---|---|---|
| ω /basis set | 1s | 2s | 3s | 4s | 5s | 6s | 7s | 1s |
| 0.0001 | 0.521302 | 0.339283 | 0.202026 | 0.019858 | 0.012632 | -0.007539 | -0.036752 | 1 |
| | 0.053282 | 0.025131 | 0.120039 | 0.010869 | 0.368693 | 0.428893 | 0.160714 | 0.060120 |
| 0.001 | 0.324401 | 0.265691 | 0.232998 | 0.145646 | 0.048941 | 0.029086 | 0.003973 | 1 |
| | 0.059814 | 0.025981 | 0.040329 | 0.102216 | 0.014165 | 0.142455 | 0.226510 | 0.060120 |
| 0.01 | 0.356345 | 0.285918 | 0.236219 | 0.144662 | 0.026523 | 0.000002 | -0.000098 | 1 |
| | 0.035385 | 0.054391 | 0.093414 | 0.018956 | 0.177684 | 6.886778 | 0.598310 | 0.061722 |
| 0.1 | 0.472133 | 0.234186 | 0.146582 | 0.112009 | 0.065784 | 0.000671 | 0.000041 | 1 |
| | 0.074259 | 0.120849 | 0.155439 | 0.052560 | 0.264653 | 0.508041 | 1.202818 | 0.127792 |
| 1 | 0.638380 | 0.173114 | 0.116400 | 0.070425 | 0.013669 | 0.000007 | -0.000070 | 1 |
| | 0.541680 | 0.763898 | 0.473410 | 1.129379 | 1.871389 | 11.357029 | 3.806633 | 0.856476 |
| 10 | 0.890747 | 0.068393 | 0.035699 | 0.009586 | 0.000073 | -0.000006 | -0.000490 | 1 |
| | 5.124563 | 7.981836 | 4.452344 | 14.733625 | 37.888938 | 73.121875 | 21.104344 | 7.794067 |
| 100 | 0.808387 | 0.213605 | 0.009387 | 0.001201 | -0.000271 | -0.001348 | -0.029671 | 1 |
| | 51.759766 | 44.683125 | 114.070625 | 271.082500 | 352.562500 | 215.378125 | 38.871875 | 75.891419 |
| 1000 | 0.790644 | 0.212843 | 0.002496 | 0.000164 | -0.000015 | -0.000255 | -0.005466 | 1 |
| | 509.904736 | 469.819336 | 1189.453125 | 3008.703125 | 5038.538281 | 2342.582813 | 372.322266 | 752.781105 |
| 10000 | 0.662884 | 0.360082 | 0.003146 | 0.000148 | 0.000001 | -0.000040 | -0.026090 | 1 |
| | 5145.921875 | 4679.678223 | 2833.828125 | 26213.718750 | 311428.937500 | 45381.328125 | 3643.312500 | 7508.756638 |
| m = 2 | | | | | | | | |
| ω /basis set | 1s | 2s | 3s | 4s | 5s | 6s | 7s | 1s |
| 0.0001 | 0.474146 | 0.276934 | 0.248649 | 0.027532 | 0.027172 | 0.001241 | -0.000085 | 1 |
| | 0.056193 | 0.028325 | 0.113935 | 0.014342 | 0.224923 | 0.032409 | 0.795639 | 0.084515 |
| 0.001 | 0.360990 | 0.328335 | 0.172251 | 0.142398 | 0.031848 | 0.020079 | -0.000352 | 1 |
| | 0.070260 | 0.040329 | 0.128173 | 0.025004 | 0.015502 | 0.234052 | 0.359868 | 0.084515 |
| 0.01 | 0.481508 | 0.375742 | 0.165785 | 0.040314 | 0.004777 | -0.000917 | -0.012412 | 1 |
| | 0.069488 | 0.032943 | 0.153319 | 0.015706 | 0.445993 | 0.745799 | 0.227522 | 0.086327 |
| 0.1 | 0.459492 | 0.373823 | 0.080416 | 0.058739 | 0.055216 | 0.011088 | -0.001075 | 1 |
| | 0.112715 | 0.064190 | 0.261459 | 0.204772 | 0.196330 | 0.462646 | 0.357736 | 0.169869 |
| 1 | 0.636545 | 0.185479 | 0.086305 | 0.081854 | 0.024910 | 0.000411 | -0.000128 | 1 |
| | 0.541680 | 0.763898 | 1.129379 | 0.465139 | 1.931633 | 3.919529 | 2.817160 | 1.130886 |
| 10 | 0.876809 | 0.080993 | 0.035215 | 0.015527 | 0.000684 | -0.000224 | -0.003713 | 1 |
| | 5.124563 | 8.044336 | 4.514844 | 16.733625 | 37.888938 | 48.121875 | 21.104344 | 10.350143 |
| 100 | 0.802263 | 0.203540 | 0.012380 | 0.001445 | 0.000366 | -0.001273 | -0.017001 | 1 |
| | 51.759766 | 45.183125 | 108.508125 | 290.378125 | 402.562500 | 346.082500 | 37.934375 | 101.049347 |
| 1000 | 0.783587 | 0.219261 | 0.003230 | 0.000156 | -0.000018 | -0.000163 | -0.005505 | 1 |
| | 510.465772 | 469.819336 | 1200.171875 | 3008.703125 | 5038.538281 | 2217.582813 | 372.322266 | 1003.262711 |
| 10000 | 1.004939 | 0.007057 | 0.000755 | 0.000617 | 0.000000 | -0.000383 | -0.012811 | 1 |
| | 4988.695801 | 7032.808594 | 3169.476563 | 21445.593750 | 312053.937500 | 23662.578125 | 4234.875000 | 10010.263443 |

m = 3

| ω/basis set | 1s | 2s | 3s | 4s | 5s | 6s | 7s | 1s |
|---|---|---|---|---|---|---|---|---|
| 0.0001 | 0.352795 | 0.351487 | 0.158219 | 0.107902 | 0.051423 | 0.037391 | 0.003306 | 1 |
| | 0.048034 | 0.088627 | 0.154855 | 0.028849 | 0.250654 | 0.019272 | 0.426940 | 0.133343 |
| 0.001 | 0.417539 | 0.367490 | 0.109416 | 0.106654 | 0.045026 | 0.012869 | 0.003496 | 1 |
| | 0.047378 | 0.096113 | 0.205571 | 0.022998 | 0.130126 | 0.249948 | 0.466673 | 0.133343 |
| 0.01 | 0.426570 | 0.369535 | 0.132757 | 0.127499 | 0.004831 | 0.001125 | -0.000455 | 1 |
| | 0.051072 | 0.102752 | 0.025131 | 0.207685 | 0.443309 | 0.531743 | 0.789122 | 0.135727 |
| 0.1 | 0.470043 | 0.220474 | 0.174890 | 0.157130 | 0.031048 | 0.022885 | -0.029553 | 1 |
| | 0.120849 | 0.253972 | 0.063578 | 0.070871 | 0.615463 | 1.094420 | 1.034345 | 0.252075 |
| 1 | 0.562292 | 0.190519 | 0.144298 | 0.076475 | 0.044051 | 0.003650 | -0.000488 | 1 |
| | 0.541680 | 0.701398 | 1.053207 | 0.465139 | 1.980461 | 4.035910 | 5.419529 | 1.670284 |
| 10 | 0.879206 | 0.088949 | 0.016960 | 0.016570 | 0.005163 | 0.000514 | -0.000022 | 1 |
| | 5.124563 | 7.981836 | 14.233625 | 4.264844 | 21.104344 | 37.888938 | 73.121875 | 15.430786 |
| 100 | 0.766365 | 0.220560 | 0.015879 | 0.003209 | 0.000687 | -0.000018 | -0.004279 | 1 |
| | 51.759766 | 46.433125 | 97.570625 | 165.378125 | 271.082500 | 402.562500 | 34.871875 | 151.265259 |
| 1000 | 0.763792 | 0.241446 | 0.004179 | 0.000389 | 0.000150 | -0.000276 | -0.008912 | 1 |
| | 511.465772 | 469.819336 | 1188.453125 | 3258.703125 | 5717.582813 | 5038.538281 | 387.947266 | 1503.905487 |
| 10000 | 0.760091 | 0.288561 | 0.005070 | 0.000417 | 0.000001 | -0.000048 | -0.053849 | 1 |
| | 5137.421875 | 4475.678223 | 2834.828125 | 22866.218750 | 312053.937500 | 45381.328125 | 3685.812500 | 15012.256622 |

m = 10

| ω/basis set | 1s | 2s | 3s | 4s | 5s | 6s | 7s | 1s |
|---|---|---|---|---|---|---|---|---|
| 0.0001 | 0.388898 | 0.329896 | 0.176012 | 0.147766 | 0.028571 | 0.012238 | -0.000179 | 1 |
| | 0.086591 | 0.175541 | 0.043527 | 0.364731 | 0.757365 | 0.021117 | 0.610533 | 0.447388 |
| 0.001 | 2.555506 | 0.293038 | -0.060568 | -0.150463 | -0.411872 | -0.848858 | -2.459807 | 1 |
| | 0.108504 | 0.474365 | 0.686081 | 0.037615 | 0.424077 | 0.146650 | 0.099557 | 0.447578 |
| 0.01 | 0.415687 | 0.406983 | 0.167362 | 0.156054 | 0.021592 | 0.005461 | -0.090118 | 1 |
| | 0.086591 | 0.190434 | 0.041656 | 0.383605 | 0.822250 | 0.017405 | 0.242013 | 0.453301 |
| 0.1 | 0.404281 | 0.271201 | 0.267376 | 0.105556 | 0.017807 | 0.007462 | 0.000014 | 1 |
| | 0.144286 | 0.289291 | 0.078166 | 0.614083 | 1.291197 | 0.042851 | 6.323408 | 0.782700 |
| 1 | 0.630393 | 0.249179 | 0.082967 | 0.054674 | 0.024589 | 0.001483 | -0.005450 | 1 |
| | 0.568586 | 1.004379 | 2.153098 | 0.440725 | 4.763279 | 10.213555 | 4.212883 | 5.316467 |
| 10 | 0.829252 | 0.118976 | 0.029841 | 0.029229 | 0.011795 | 0.001256 | -0.006822 | 1 |
| | 5.124563 | 8.103906 | 17.672735 | 4.721738 | 48.104618 | 79.585498 | 55.578330 | 50.645339 |
| 100 | 0.882683 | 0.218492 | 0.019251 | 0.003494 | 0.000455 | -0.000157 | -0.119803 | 1 |
| | 51.759766 | 40.288594 | 119.786735 | 307.714275 | 877.660156 | 1092.135938 | 38.145450 | 501.753938 |
| 1000 | 0.887617 | 0.146405 | 0.005756 | 0.001340 | 0.000252 | -0.000491 | -0.039470 | 1 |
| | 508.923535 | 438.920288 | 1218.406787 | 3327.916992 | 6214.843750 | 4413.538281 | 401.619141 | 5005.275726 |
| 10000 | 0.833737 | 0.184573 | 0.004061 | 0.001111 | 0.000141 | 0.000001 | -0.023176 | 1 |
| | 5094.898033 | 4498.830078 | 2833.828125 | 16863.132813 | 47646.953125 | 161858.625000 | 3480.156250 | 50016.428232 |
| | | | | m = 50 | | | | |
| ω/basis set | 1s | 2s | 3s | 4s | 5s | 6s | 7s | 1s |
| 0.0001 | 0.429978 | 0.336113 | 0.160200 | 0.141464 | 0.054536 | 0.003005 | -0.018707 | 1 |
| | 0.123092 | 0.286850 | 0.052706 | 0.676339 | 1.524596 | 3.391279 | 1.397221 | 1.722260 |
| 0.001 | 0.428235 | 0.335157 | 0.161103 | 0.142172 | 0.041051 | 0.005194 | -0.006336 | 1 |
| | 0.123168 | 0.285690 | 0.052836 | 0.670724 | 1.609252 | 3.063154 | 2.164555 | 1.722260 |
| 0.01 | 0.426080 | 0.334006 | 0.161343 | 0.141930 | 0.037014 | 0.003866 | 0.002352 | 1 |
| | 0.123603 | 0.285354 | 0.053321 | 0.662423 | 1.488774 | 2.335997 | 3.457685 | 1.750488 |
| 0.1 | 0.358258 | 0.321585 | 0.181824 | 0.154887 | 0.067386 | 0.015595 | 0.003122 | 1 |
| | 0.144286 | 0.287338 | 0.614083 | 0.077922 | 1.390807 | 2.956547 | 5.167158 | 3.323288 |
| 1 | 0.584607 | 0.281076 | 0.107669 | 0.040117 | 0.032969 | 0.007394 | 0.000585 | 1 |
| | 0.568586 | 1.005905 | 2.257712 | 0.440725 | 5.776848 | 15.849297 | 40.966404 | 25.532837 |
| 10 | 0.942087 | 0.133246 | 0.035496 | 0.009349 | 0.002003 | 0.000164 | -0.103743 | 1 |
| | 5.124563 | 8.103906 | 18.268781 | 48.067234 | 137.109607 | 374.342063 | 5.253125 | 250.824745 |
| 100 | 0.896380 | 0.267331 | 0.021409 | 0.004474 | 0.000936 | 0.000114 | -0.184690 | 1 |
| | 51.759766 | 38.823750 | 123.425956 | 338.338665 | 995.137573 | 2831.882031 | 37.418750 | 2502.048564 |
| 1000 | 0.911749 | 0.135577 | 0.006065 | 0.001194 | 0.000227 | 0.000019 | -0.052943 | 1 |
| | 508.923535 | 420.991211 | 1321.022144 | 3801.050488 | 11772.705078 | 34590.734375 | 390.251343 | 25005.971909 |
| 10000 | 0.829951 | 0.185719 | 0.003163 | 0.001352 | 0.000239 | 0.000035 | -0.019860 | 1 |
| | 5094.921875 | 4525.685547 | 2833.828125 | 16994.968750 | 54521.953125 | 187210.187500 | 3499.687500 | 250018.472672 |

m = 207

| ω/basis set | 1s | 2s | 3s | 4s | 5s | 6s | 7s | 1s |
|---|---|---|---|---|---|---|---|---|
| 0.0001 | 0.384872 | 0.349120 | 0.176316 | 0.126117 | 0.063879 | 0.018202 | 0.003192 | 1 |
| | 0.128112 | 0.290756 | 0.666574 | 0.056193 | 1.522643 | 3.336591 | 6.865971 | 4.733582 |
| 0.001 | 0.384333 | 0.348393 | 0.176602 | 0.126459 | 0.064362 | 0.018336 | 0.003213 | 1 |
| | 0.128173 | 0.290329 | 0.664010 | 0.056254 | 1.515502 | 3.328779 | 6.852055 | 4.735107 |
| 0.01 | 0.383444 | 0.347966 | 0.177153 | 0.125943 | 0.065187 | 0.018876 | 0.003340 | 1 |
| | 0.128570 | 0.290756 | 0.663400 | 0.056590 | 1.511778 | 3.332685 | 6.957524 | 4.883118 |
| 0.1 | 0.332073 | 0.331309 | 0.207517 | 0.115229 | 0.095650 | 0.031774 | 0.005999 | 1 |
| | 0.289291 | 0.144286 | 0.614083 | 0.078166 | 1.421568 | 3.698408 | 10.764653 | 11.928101 |
| 1 | 0.577244 | 0.286546 | 0.115459 | 0.038367 | 0.031301 | 0.010332 | 0.001800 | 1 |
| | 0.568586 | 1.004379 | 2.254660 | 5.898918 | 0.428518 | 17.826836 | 64.110936 | 104.179688 |
| 10 | 0.928031 | 0.135150 | 0.037035 | 0.010388 | 0.002578 | 0.000436 | -0.093366 | 1 |
| | 5.124563 | 8.103906 | 18.268781 | 49.038075 | 153.392642 | 577.238181 | 5.253125 | 1035.908413 |
| 100 | 0.896466 | 0.264013 | 0.022122 | 0.004789 | 0.001058 | 0.000168 | -0.182197 | 1 |
| | 51.759766 | 38.823750 | 123.664375 | 351.896099 | 1177.358032 | 4839.816602 | 37.418750 | 10352.169991 |
| 1000 | 0.914321 | 0.116832 | 0.006144 | 0.001224 | 0.000254 | 0.000038 | -0.036783 | 1 |
| | 508.983140 | 420.991211 | 1350.395313 | 4070.520703 | 14005.371094 | 56582.921875 | 382.087891 | 103506.202698 |
| 10000 | 0.831705 | 0.190324 | 0.002972 | 0.001365 | 0.000238 | 0.000034 | -0.025996 | 1 |
| | 5098.259735 | 4486.623047 | 2833.828125 | 17886.997559 | 65585.185547 | 314348.859375 | 3593.071289 | 1035019.111633 |
| m=400 | | | | | | | | |
| ω/basis set | 1s | 2s | 3s | 4s | 5s | 6s | 7s | 1s |
| 0.0001 | 0.372058 | 0.355170 | 0.185951 | 0.113121 | 0.072028 | 0.023719 | 0.004821 | 1 |
| | 0.128112 | 0.290756 | 0.666574 | 0.056437 | 1.522643 | 3.492841 | 8.368901 | 7.275696 |
| 0.001 | 0.371008 | 0.354149 | 0.186814 | 0.113562 | 0.072696 | 0.023936 | 0.004713 | 1 |
| | 0.128173 | 0.289840 | 0.663034 | 0.056498 | 1.515502 | 3.500654 | 8.447270 | 7.279511 |
| 0.01 | 0.370070 | 0.353460 | 0.187261 | 0.112953 | 0.074224 | 0.024426 | 0.004799 | 1 |
| | 0.128570 | 0.290268 | 0.661447 | 0.056834 | 1.511778 | 3.535810 | 8.647954 | 7.615204 |
| 0.1 | 0.340196 | 0.309990 | 0.228335 | 0.110706 | 0.099598 | 0.030297 | 0.005471 | 1 |
| | 0.283931 | 0.143810 | 0.627618 | 0.079142 | 1.583876 | 4.586286 | 15.514592 | 21.950836 |
| 1 | 0.550809 | 0.298031 | 0.116308 | 0.050487 | 0.036292 | 0.009042 | 0.001493 | 1 |
| | 0.568586 | 1.008957 | 2.372916 | 0.463613 | 6.649650 | 22.038262 | 91.259373 | 200.734253 |
| 10 | 0.842814 | 0.127762 | 0.035518 | 0.010841 | 0.002920 | 0.000629 | 0.000083 | 1 |
| | 5.121582 | 8.293687 | 18.177229 | 46.449803 | 137.483447 | 462.476828 | 1631.717969 | 2000.933838 |
| 100 | 0.902013 | 0.252253 | 0.021919 | 0.004594 | 0.000974 | 0.000149 | -0.175394 | 1 |
| | 51.759766 | 38.341191 | 126.821037 | 376.695446 | 1348.821045 | 6345.065625 | 36.930469 | 20002.161980 |
| 1000 | 0.812100 | 0.180735 | 0.007367 | 0.001467 | 0.000319 | 0.000062 | 0.000008 | 1 |
| | 508.923535 | 475.331574 | 1207.815170 | 3531.334033 | 11519.251172 | 40606.697266 | 147930.578125 | 200006.122589 |
| 10000 | 1.001042 | 0.003257 | 0.001313 | 0.000648 | 0.000143 | 0.000025 | -0.005777 | 1 |
| | 5005.264568 | 9891.114471 | 3840.908203 | 27583.347656 | 93388.166016 | 433203.593750 | 4630.668945 | 2000018.310547 |
| m = 900 | | | | | | | | |
| ω / basis set | 1s | 2s | 3s | 4s | 5s | 6s | 7s | 1s |
| 0.0001 | 0.359796 | 0.359360 | 0.195998 | 0.102762 | 0.081234 | 0.027364 | 0.005330 | 1 |
| | 0.128112 | 0.290268 | 0.666574 | 0.056681 | 1.561614 | 3.944868 | 11.159916 | 12.044067 |
| 0.001 | 0.359450 | 0.358300 | 0.196746 | 0.103016 | 0.081827 | 0.027189 | 0.005315 | 1 |
| | 0.128173 | 0.289840 | 0.664010 | 0.056742 | 1.562102 | 3.959944 | 11.161625 | 12.051697 |
| 0.01 | 0.356834 | 0.355530 | 0.197377 | 0.105719 | 0.083521 | 0.027914 | 0.005407 | 1 |
| | 0.129546 | 0.290756 | 0.662423 | 0.057810 | 1.558562 | 4.007734 | 11.611943 | 13.013001 |
| 0.1 | 0.378321 | 0.321745 | 0.227120 | 0.088628 | 0.083929 | 0.024734 | 0.004307 | 1 |
| | 0.296030 | 0.138427 | 0.716211 | 1.960982 | 0.075497 | 6.237104 | 24.382939 | 47.368164 |
| 1 | 0.605674 | 0.305034 | 0.102249 | 0.034542 | 0.012201 | 0.003185 | 0.000505 | 1 |
| | 0.558167 | 1.047342 | 2.464087 | 5.880607 | 15.580742 | 51.417287 | 215.575779 | 450.780029 |
| 10 | 0.842907 | 0.127864 | 0.035649 | 0.010799 | 0.002856 | 0.000611 | 0.000087 | 1 |
| | 5.122201 | 8.311807 | 18.342214 | 47.823094 | 146.932452 | 531.866171 | 2278.690625 | 4500.905228 |
| 100 | 0.909988 | 0.288465 | 0.021206 | 0.004241 | 0.000862 | 0.000127 | -0.218327 | 1 |
| | 51.789926 | 37.126210 | 132.390495 | 412.837795 | 1565.564514 | 8312.106641 | 35.965350 | 45002.138138 |
| 1000 | 0.820272 | 0.172740 | 0.007264 | 0.001413 | 0.000309 | 0.000066 | 0.000012 | 1 |
| | 508.923535 | 474.082260 | 1229.969025 | 3650.962940 | 11772.852246 | 40889.900391 | 168457.921875 | 450006.122589 |
| 10000 | 0.982866 | 0.046181 | 0.016577 | 0.000303 | 0.000035 | -0.010655 | -0.034658 | 1 |
| | 5094.916511 | 3577.812500 | 10432.468750 | 62649.640625 | 390850.078125 | 2833.828125 | 8197.388886 | 4500018.310547 |
| m = 1836 | | | | | | | | |

| ω/basis set | 1s | 2s | 3s | 4s | 5s | 6s | 7s | 1s |
|---|---|---|---|---|---|---|---|---|
| 0.0001 | 0.357678 | 0.351737 | 0.206584 | 0.099427 | 0.087382 | 0.027318 | 0.005037 | 1 |
| | 0.289802 | 0.129089 | 0.667527 | 0.057292 | 1.649001 | 4.568952 | 14.486577 | 18.414612 |
| 0.001 | 0.354395 | 0.347732 | 0.207895 | 0.101077 | 0.090344 | 0.028465 | 0.005238 | 1 |
| | 0.286922 | 0.129150 | 0.653268 | 0.057719 | 1.601073 | 4.459760 | 14.256108 | 18.437500 |
| 0.01 | 0.361614 | 0.348298 | 0.208746 | 0.096891 | 0.087817 | 0.027372 | 0.005093 | 1 |
| | 0.289291 | 0.128570 | 0.672120 | 0.057441 | 1.675383 | 4.710493 | 15.363774 | 20.878906 |
| 0.1 | 0.397827 | 0.354195 | 0.216193 | 0.077677 | 0.061671 | 0.020193 | 0.003338 | 1 |
| | 0.316156 | 0.136905 | 0.813279 | 2.361521 | 0.069388 | 8.094007 | 35.748051 | 94.456787 |
| 1 | 0.610738 | 0.305425 | 0.102108 | 0.032507 | 0.010123 | 0.002524 | 0.000404 | 1 |
| | 0.559455 | 1.064460 | 2.591307 | 6.667961 | 18.931572 | 65.540822 | 303.661717 | 918.828125 |
| 10 | 0.841922 | 0.128848 | 0.036014 | 0.010652 | 0.002769 | 0.000578 | 0.000082 | 1 |
| | 5.121247 | 8.294641 | 18.478113 | 49.214028 | 155.437320 | 599.111655 | 2956.180859 | 9180.938721 |
| 100 | 0.911097 | 0.086886 | 0.021078 | 0.004115 | 0.000802 | 0.000114 | -0.017506 | 1 |
| | 51.789926 | 39.014247 | 134.497162 | 432.245067 | 1719.838501 | 9867.160352 | 32.098439 | 91802.234650 |
| 1000 | 0.816580 | 0.176323 | 0.007324 | 0.001452 | 0.000323 | 0.000070 | 0.000011 | 1 |
| | 508.923535 | 474.654465 | 1217.590332 | 3572.168652 | 11576.530566 | 42252.205078 | 205948.156250 | 918006.134033 |
| 10000 | 0.983239 | 0.045958 | 0.016368 | 0.000305 | 0.000035 | -0.010581 | -0.034672 | 1 |
| | 5094.920813 | 3577.812500 | 10432.430603 | 63245.343750 | 414600.078125 | 2833.828125 | 8168.263672 | 9180018.615723 |

# 9 Appendix C

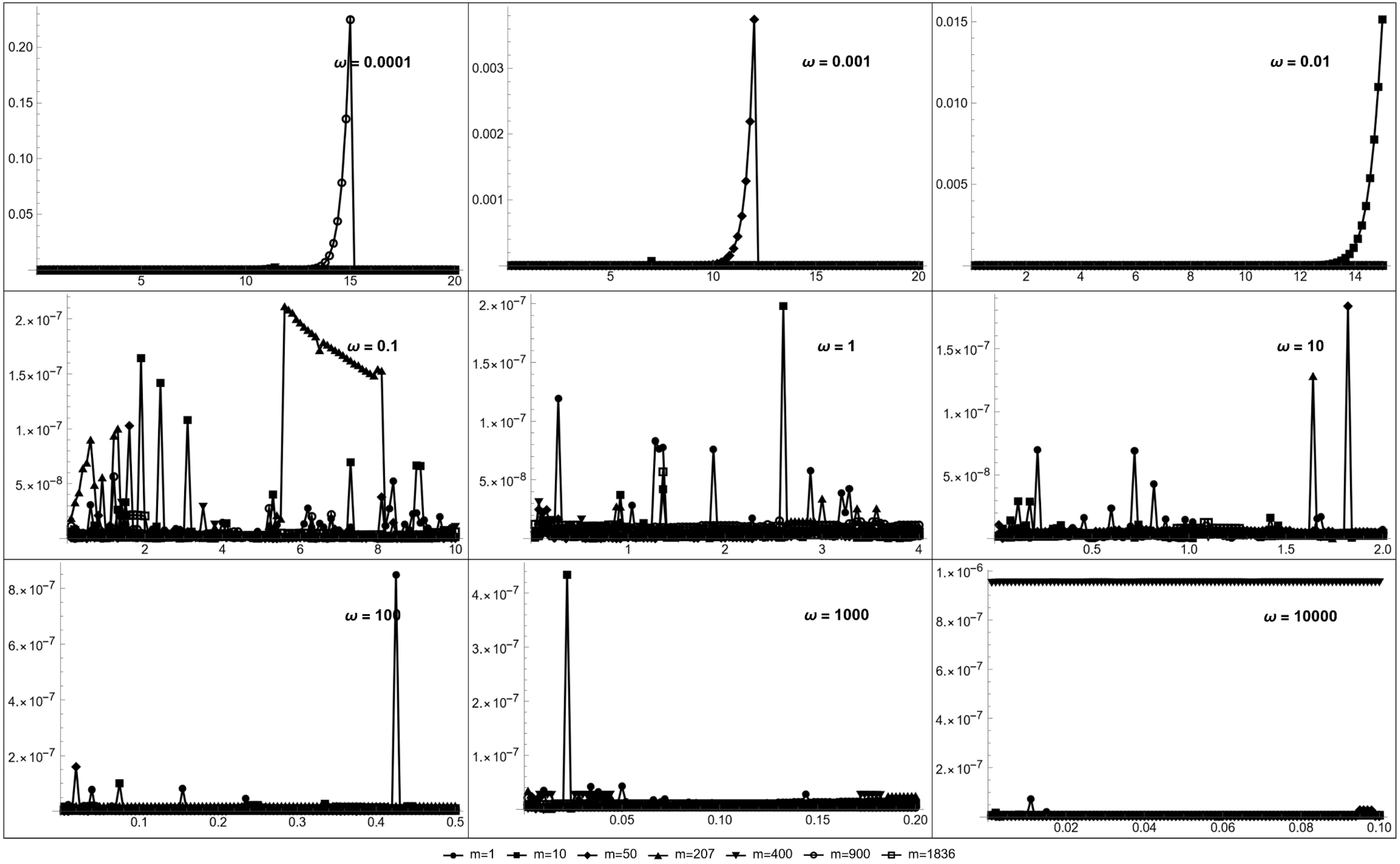

Figure C-1: Percentage error of analytical electron density from equation (2-168) relative to FEM density from equation (2-161) versus $r_e$ for different frequencies

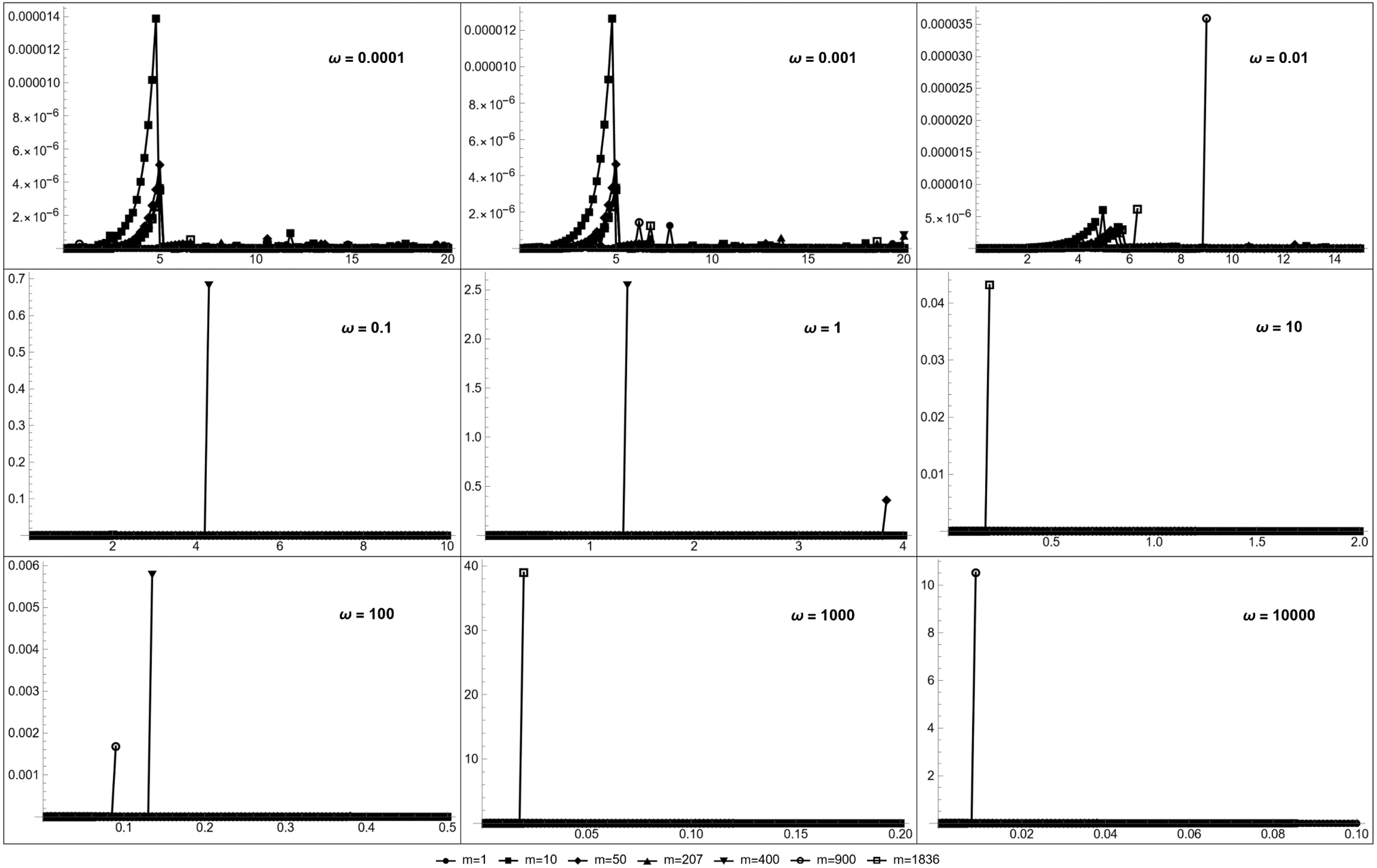

Figure C-2: Percentage error of PCP densities from equation (2-169) and FEM density from equivalent equation (2-161) versus $r_p$ for different frequencies

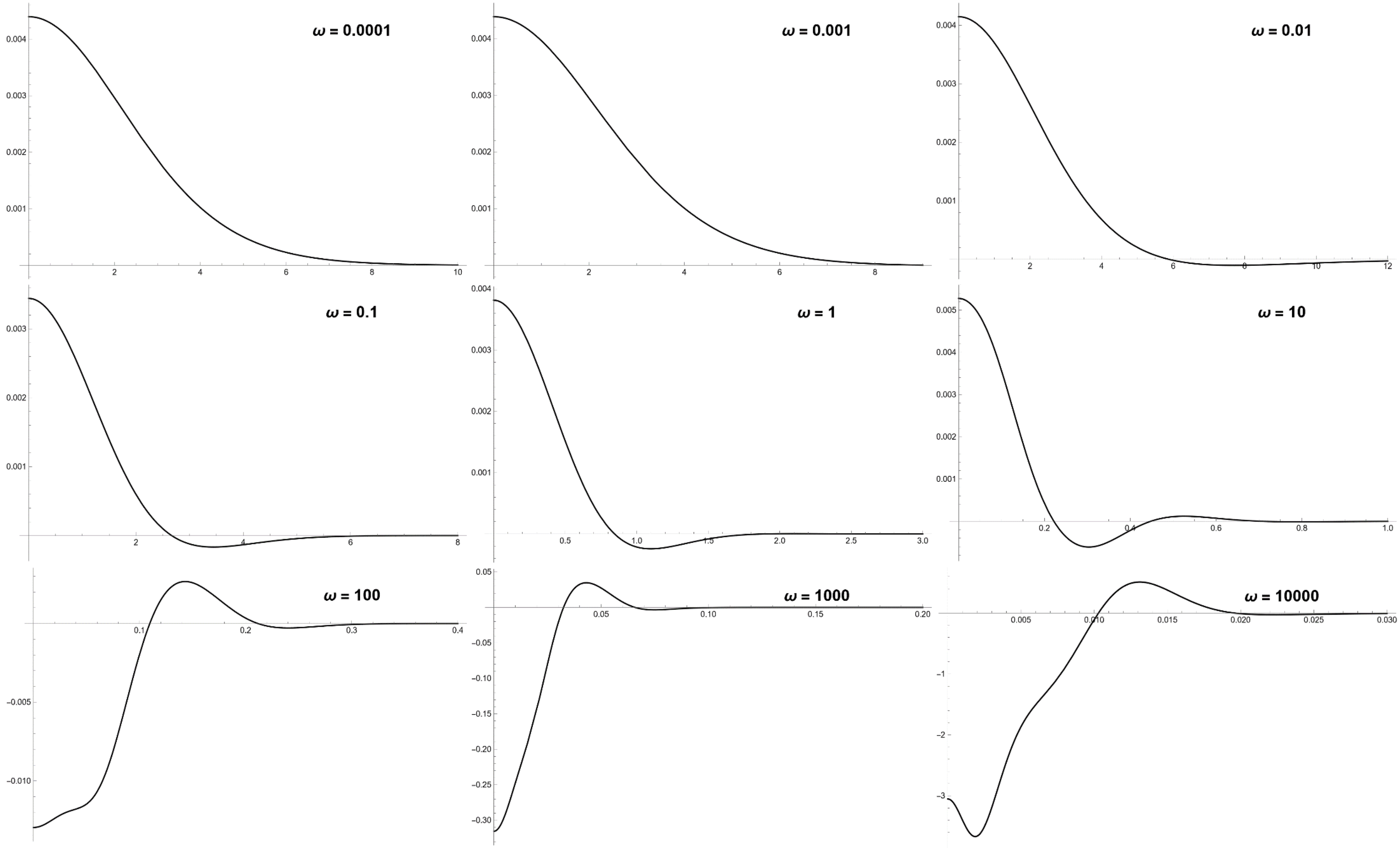

Figure C-3: Difference in electron (PCP) densities between variational and TC-HF for mass 1 at different frequencies

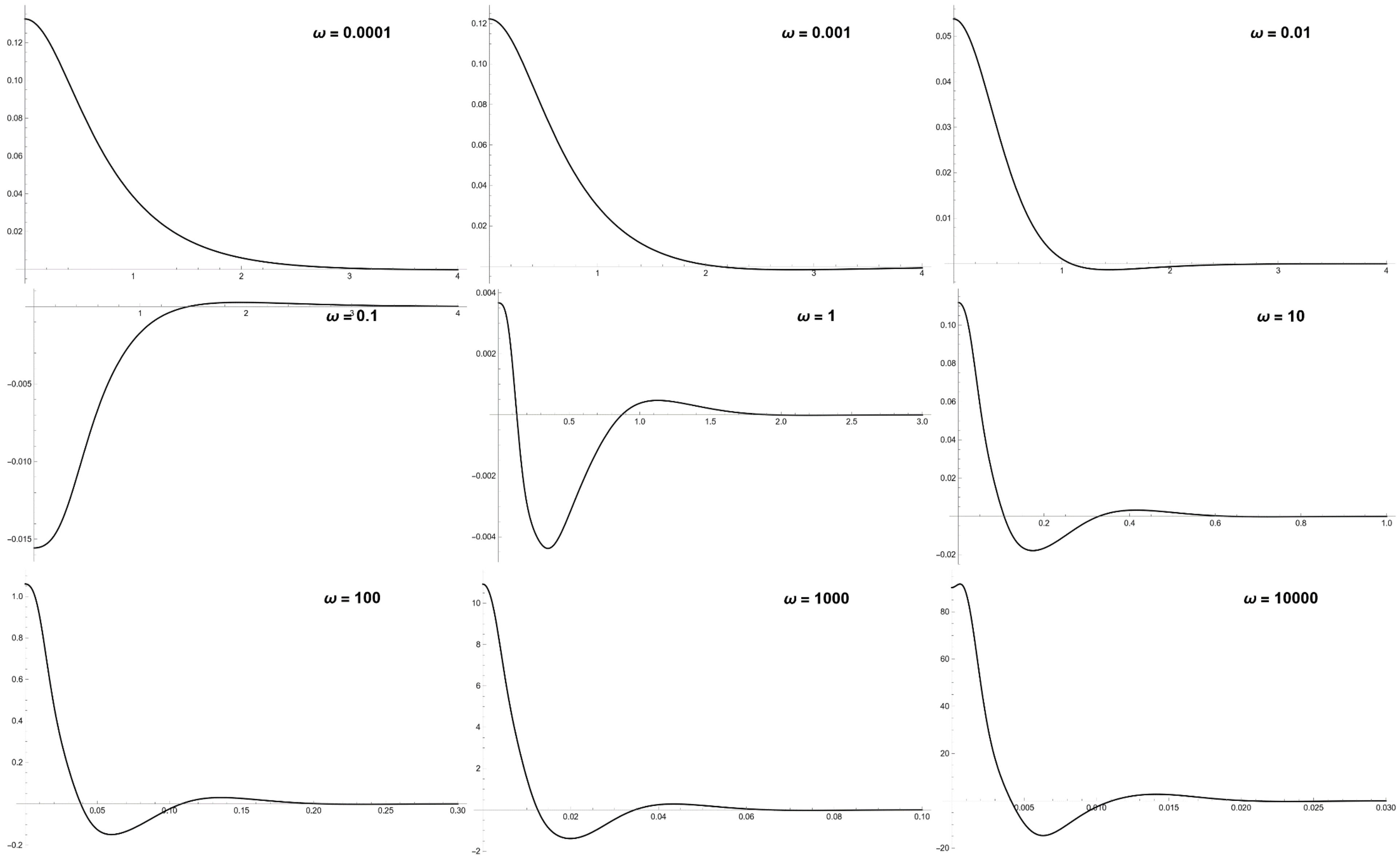

Figure C-4: Difference in electron densities between variational and TC-HF for mass 207 at different frequencies

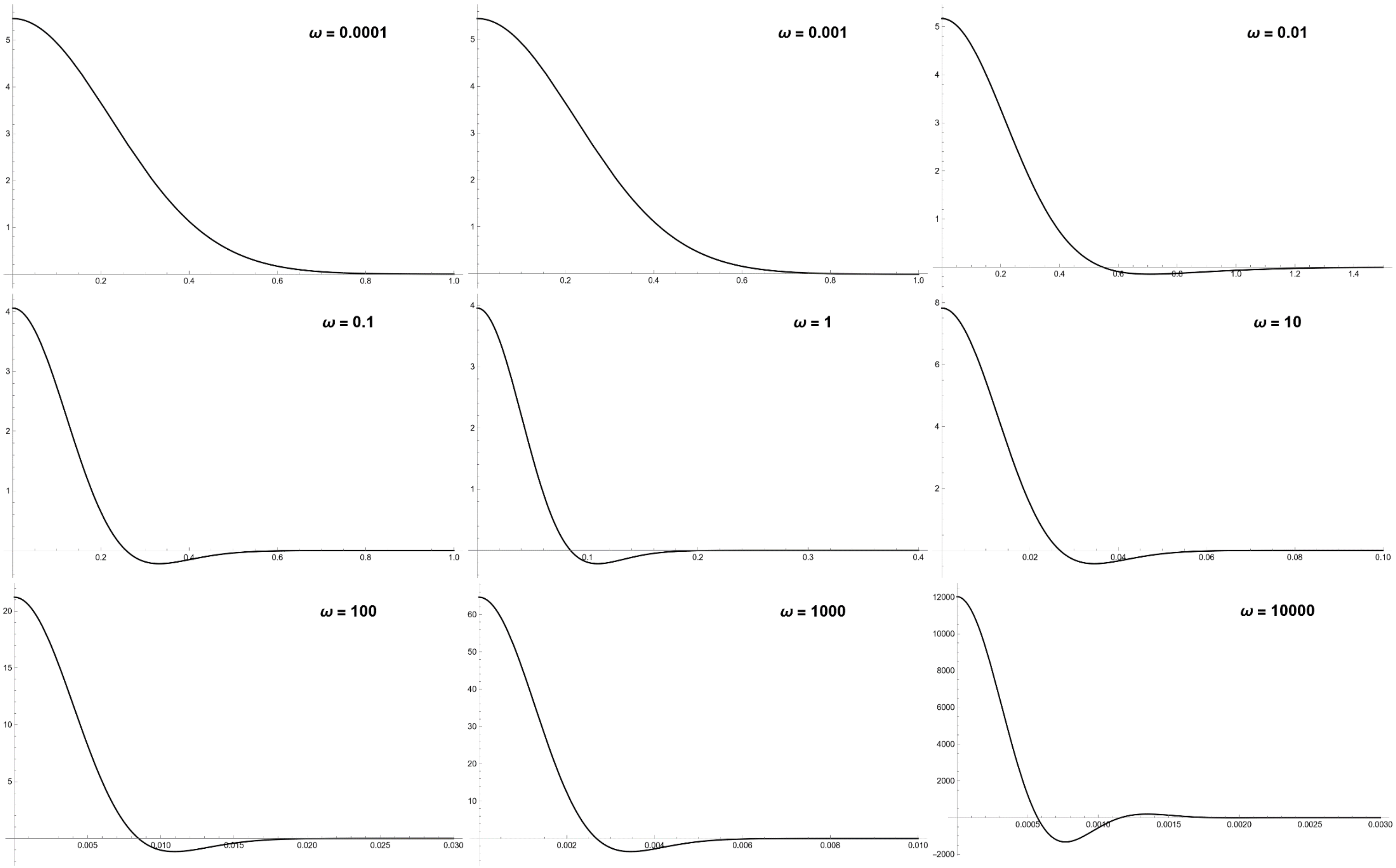

Figure C-5: Difference in PCP densities between variational and TC-HF for mass 207 at different frequencies

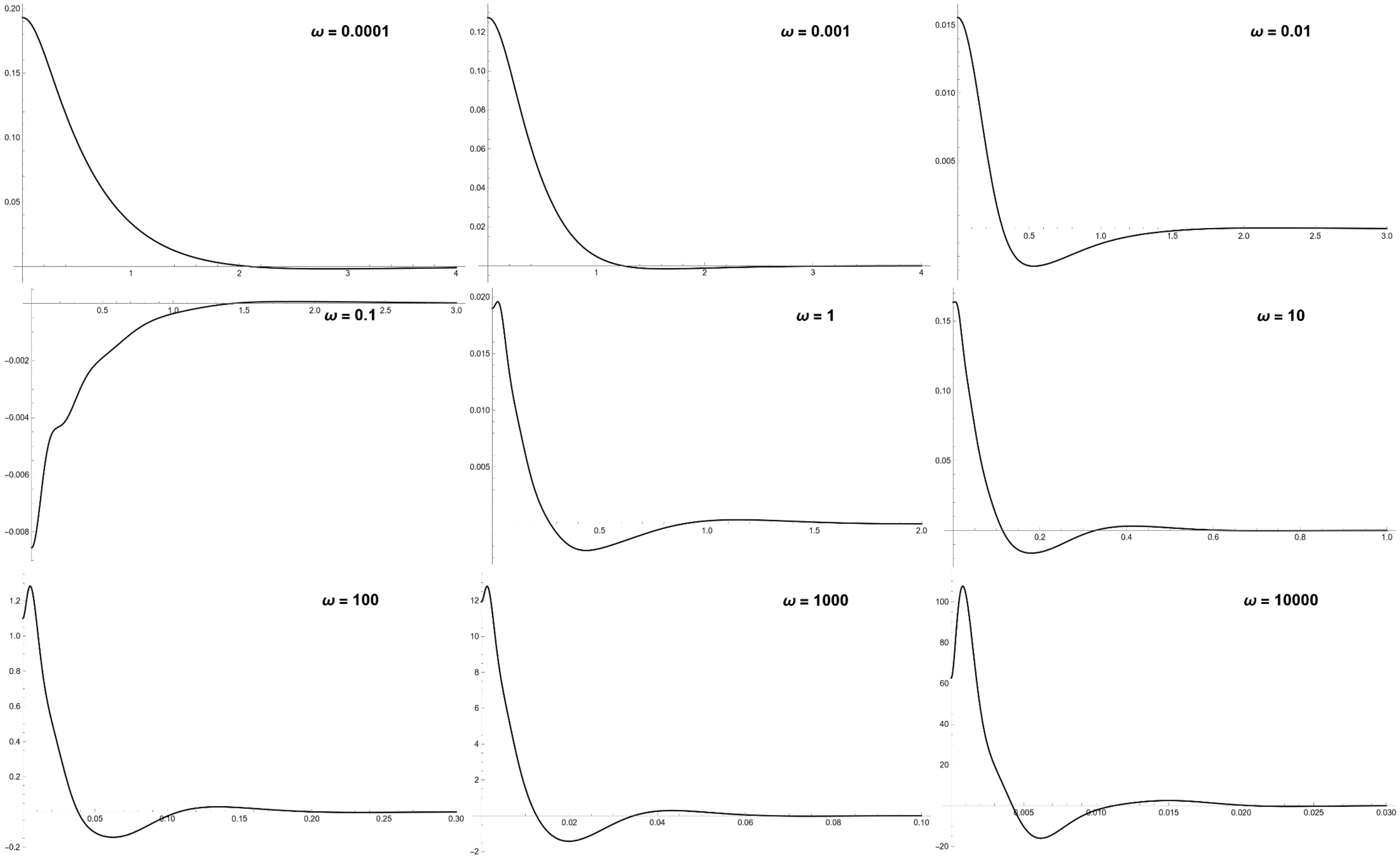

Figure C-6: Difference in electron densities between variational and TC-HF for mass 1836 at different frequencies

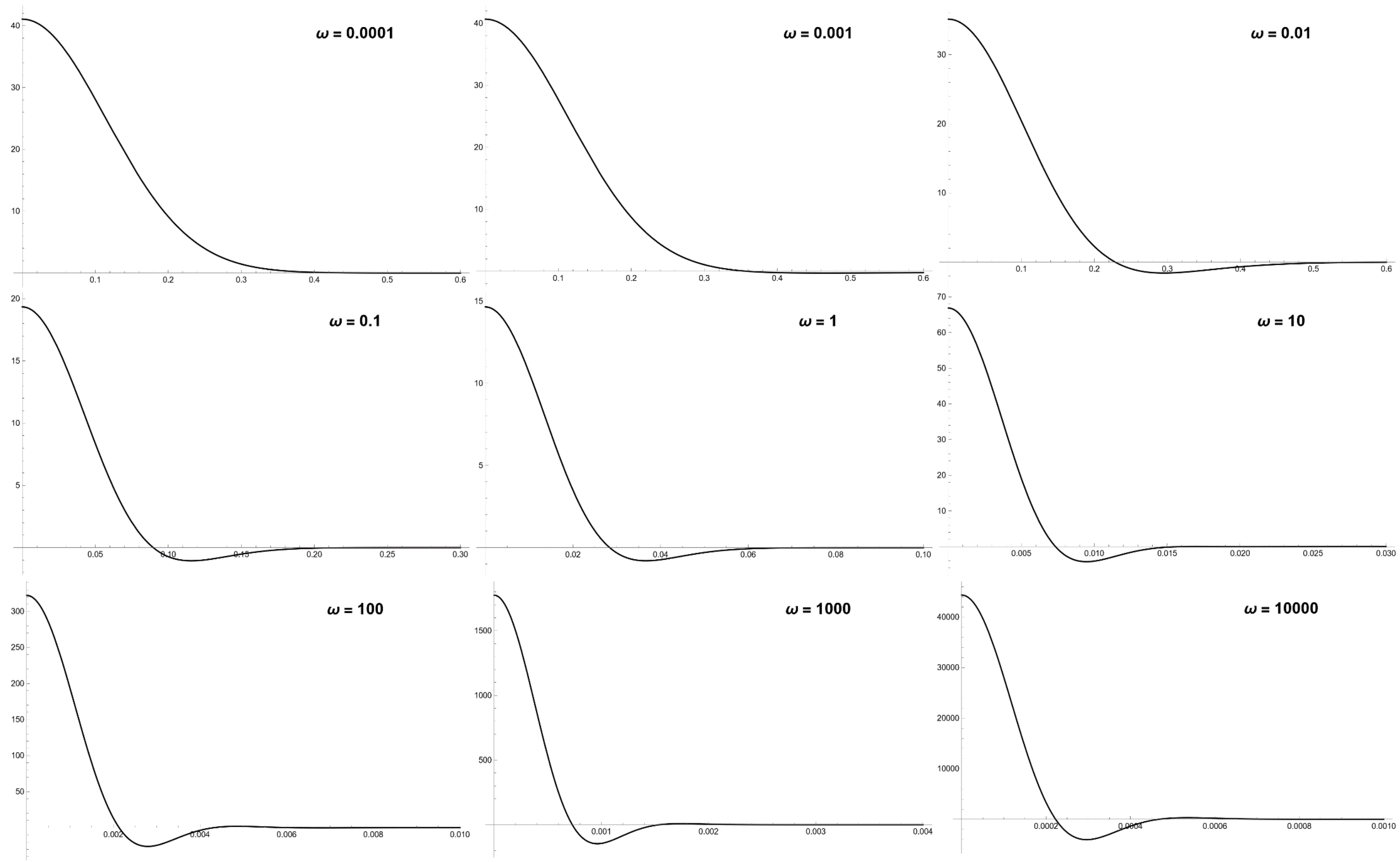

Figure C-7: Difference in PCP densities between variational and TC-HF for mass 1836 at different frequencies

Table C-1: Effective correlation hill radii for electron and PCP as a function of their position vectors ($r_e$ and $r_p$ respectively)

| ω/mp | 1 | 1.5 | 2 | 3 | 10 | 50 | 207 | 400 | 900 | 1836 |
|---|---|---|---|---|---|---|---|---|---|---|
| | | | | | Electron | | | | | |
| 0.0001 | 11.28902 | 9.584436 | 8.658497 | 7.645748 | 5.799772 | 4.356237 | 3.341128 | 2.916228 | 2.439233 | 2.070306 |
| 0.001 | 7.920257 | 6.778073 | 6.13538 | 5.408981 | 3.989535 | 2.794661 | 2.014594 | 1.729134 | 1.44638 | 1.261239 |
| 0.01 | 4.802232 | 4.185704 | 3.813969 | 3.369773 | 2.427789 | 1.641467 | 1.229876 | 1.120673 | 1.047728 | 1.018109 |
| 0.1 | 2.334938 | 2.120617 | 1.976072 | 1.787824 | 1.350063 | 1.029548 | 0.941931 | 0.928476 | 0.920483 | 0.917229 |
| 1 | 0.910885 | 0.864609 | 0.831099 | 0.785004 | 0.67705 | 0.620979 | 0.609572 | 0.6078 | 0.606742 | 0.60631 |
| 10 | 0.309557 | 0.299782 | 0.29269 | 0.283029 | 0.262315 | 0.253149 | 0.251324 | 0.251041 | 0.250872 | 0.250803 |
| 100 | 0.100111 | 0.097575 | 0.095749 | 0.093293 | 0.088289 | 0.086208 | 0.085799 | 0.085736 | 0.085699 | 0.085683 |
| 1000 | 0.031881 | 0.031136 | 0.030602 | 0.029887 | 0.028459 | 0.027879 | 0.027766 | 0.027749 | 0.027738 | 0.027734 |
| 10000 | 0.010104 | 0.009874 | 0.00971 | 0.00949 | 0.009054 | 0.008878 | 0.008844 | 0.008838 | 0.008835 | 0.008834 |
| | | | | | PCP | | | | | |
| 0.0001 | 11.28902 | 9.580604 | 8.652869 | 7.637951 | 5.783146 | 4.309265 | 3.229977 | 2.757817 | 2.207241 | 1.763979 |
| 0.001 | 7.920257 | 6.758847 | 6.107092 | 5.370028 | 3.912109 | 2.615676 | 1.695086 | 1.339603 | 0.975366 | 0.722557 |
| 0.01 | 4.802232 | 4.11391 | 3.708008 | 3.225829 | 2.178893 | 1.231347 | 0.685288 | 0.511316 | 0.351989 | 0.251319 |
| 0.1 | 2.334938 | 1.984361 | 1.770639 | 1.508874 | 0.928101 | 0.459662 | 0.236703 | 0.172503 | 0.116294 | 0.081972 |
| 1 | 0.910885 | 0.764794 | 0.67521 | 0.565573 | 0.328969 | 0.153875 | 0.077052 | 0.055708 | 0.037297 | 0.02618 |
| 10 | 0.309557 | 0.258642 | 0.227375 | 0.189177 | 0.107825 | 0.049499 | 0.024574 | 0.017725 | 0.011843 | 0.008303 |
| 100 | 0.100111 | 0.083508 | 0.073309 | 0.060859 | 0.034466 | 0.015732 | 0.007791 | 0.005615 | 0.00375 | 0.002628 |
| 1000 | 0.031881 | 0.02658 | 0.023323 | 0.019349 | 0.010935 | 0.004983 | 0.002465 | 0.001777 | 0.001186 | 0.000831 |
| 10000 | 0.010104 | 0.008422 | 0.007389 | 0.006129 | 0.003462 | 0.001576 | 0.00078 | 0.000562 | 0.000375 | 0.000263 |

Table C-2: Fit details for simplifying single-particle densities and Kohn-Sham orbitals

| mass | ω | NO[1] | Power[2] | Range[3] | Overlap[4] | Percentage errors of 10 first points sorted by absolute values[5] | | | | | | | | | |
|---|---|---|---|---|---|---|---|---|---|---|---|---|---|---|---|
| 1836 | 0.01 | fit4 | 1 | 4 | 0.969661 | 76.8 | 76.0 | 75.2 | 74.4 | 73.6 | 72.8 | 71.9 | 71.0 | 70.2 | 69.3 |
| 1836 | 0.01 | fit4 | 1 | 5 | 0.973641 | 97.1 | 96.9 | 96.7 | 96.5 | 96.3 | 96.1 | 95.8 | 95.5 | 95.3 | 95.0 |
| 1836 | 0.01 | fit5 | 1 | 4 | 0.969661 | 76.8 | 76.0 | 75.2 | 74.4 | 73.6 | 72.8 | 71.9 | 71.0 | 70.2 | 69.3 |
| 1836 | 0.01 | fit5 | 1 | 5 | 0.996404 | 36.3 | 35.2 | 34.1 | 33.1 | 32.0 | 30.9 | 29.9 | 28.8 | 27.8 | 26.8 |
| 1836 | 0.01 | fit6 | 1 | 4 | 0.985610 | 6.8 | 6.4 | 5.9 | 5.5 | 5.1 | 4.7 | 4.3 | 3.9 | 3.6 | 3.3 |
| 1836 | 0.01 | fit6 | 1 | 5 | 0.996413 | 36.0 | 34.9 | 33.8 | 32.7 | 31.7 | 30.6 | 29.6 | 28.5 | 27.5 | 26.5 |
| 1836 | 0.01 | fit4 | 0.5 | 5 | 0.989540 | 78.3 | 77.7 | 77.0 | 76.3 | 75.6 | 74.9 | 74.2 | 73.5 | 72.8 | 72.0 |
| 1836 | 0.01 | fit4 | 0.5 | 6 | 0.998900 | 55.5 | 54.6 | 53.7 | 52.7 | 51.8 | 50.8 | 49.9 | 48.9 | 47.9 | 46.9 |
| 1836 | 0.01 | fit4 | 0.5 | 7 | 0.999187 | 79.7 | 79.0 | 78.3 | 77.6 | 76.8 | 76.1 | 75.3 | 74.5 | 73.7 | 72.9 |
| 1836 | 0.01 | fit5 | 0.5 | 5 | 0.996800 | 29.0 | 28.2 | 27.3 | 26.5 | 25.7 | 24.9 | 24.1 | 23.3 | 22.5 | 21.7 |
| 1836 | 0.01 | fit5 | 0.5 | 6 | 0.999432 | 18.4 | 17.7 | 16.9 | 16.2 | 15.5 | 14.8 | 14.2 | 13.5 | 12.8 | 12.2 |
| 1836 | 0.01 | fit5 | 0.5 | 7 | 0.999859 | 38.7 | 37.7 | 36.7 | 35.7 | 34.8 | 33.8 | 32.8 | 31.8 | 30.9 | 29.9 |
| 1836 | 0.01 | fit6 | 0.5 | 5 | 0.997111 | 6.9 | 6.5 | 6.1 | 5.7 | 5.3 | 5.0 | 4.6 | 4.3 | 3.9 | 3.6 |
| 1836 | 0.01 | fit6 | 0.5 | 6 | 0.999432 | 18.4 | 17.7 | 16.9 | 16.2 | 15.5 | 14.8 | 14.2 | 13.5 | 12.8 | 12.2 |
| 1836 | 0.01 | fit6 | 0.5 | 7 | 0.999901 | 12.1 | 11.5 | 10.9 | 10.3 | 9.7 | 9.1 | 8.6 | 8.1 | 7.5 | 7.0 |
| 207 | 0.02 | fit4 | 1 | 4 | 0.982082 | 27.7 | 26.8 | 25.9 | 25.0 | 24.1 | 23.2 | 22.4 | 21.5 | 20.6 | 19.8 |
| 207 | 0.02 | fit4 | 1 | 5 | 0.992173 | 70.2 | 69.3 | 68.4 | 67.4 | 66.4 | 65.4 | 64.4 | 63.4 | 62.3 | 61.3 |
| 207 | 0.02 | fit5 | 1 | 4 | 0.982082 | 27.7 | 26.8 | 25.9 | 25.0 | 24.1 | 23.2 | 22.4 | 21.5 | 20.6 | 19.8 |
| 207 | 0.02 | fit5 | 1 | 5 | 0.992173 | 70.2 | 69.3 | 68.4 | 67.4 | 66.4 | 65.4 | 64.4 | 63.4 | 62.3 | 61.3 |
| 207 | 0.02 | fit6 | 1 | 4 | 0.983708 | 0.4 | 0.3 | 0.3 | 0.2 | 0.2 | 0.1 | 0.1 | 0.1 | 0.1 | 0.1 |
| 207 | 0.02 | fit6 | 1 | 5 | 0.996388 | 22.8 | 21.9 | 21.0 | 20.1 | 19.3 | 18.4 | 17.6 | 16.7 | 15.9 | 15.1 |
| 207 | 0.02 | fit4 | 0.5 | 5 | 0.996720 | 0.8 | 0.7 | 0.6 | 0.6 | 0.5 | 0.4 | 0.3 | 0.3 | 0.2 | 0.2 |
| 207 | 0.02 | fit4 | 0.5 | 6 | 0.999384 | 3.4 | 3.1 | 2.8 | 2.6 | 2.3 | 2.1 | 1.9 | 1.7 | 1.5 | 1.3 |
| 207 | 0.02 | fit4 | 0.5 | 7 | 0.999808 | 45.8 | 44.8 | 43.8 | 42.8 | 41.8 | 40.8 | 39.8 | 38.8 | 37.8 | 36.8 |
| 207 | 0.02 | fit5 | 0.5 | 5 | 0.996698 | 8.4 | 8.0 | 7.5 | 7.1 | 6.6 | 6.2 | 5.8 | 5.4 | 5.0 | 4.7 |
| 207 | 0.02 | fit5 | 0.5 | 6 | 0.999335 | 22.7 | 21.9 | 21.1 | 20.4 | 19.6 | 18.8 | 18.1 | 17.3 | 16.6 | 15.9 |
| 207 | 0.02 | fit5 | 0.5 | 7 | 0.999808 | 45.8 | 44.8 | 43.8 | 42.8 | 41.8 | 40.8 | 39.8 | 38.8 | 37.8 | 36.8 |
| 207 | 0.02 | fit6 | 0.5 | 5 | 0.996721 | 0.1 | 0.1 | 0.0 | 0.0 | 0.0 | 0.0 | 0.0 | 0.0 | 0.0 | 0.0 |
| 207 | 0.02 | fit6 | 0.5 | 6 | 0.999385 | 0.3 | 0.3 | 0.2 | 0.2 | 0.1 | 0.1 | 0.1 | 0.1 | 0.1 | 0.1 |
| 207 | 0.02 | fit6 | 0.5 | 7 | 0.999808 | 45.8 | 44.8 | 43.8 | 42.8 | 41.8 | 40.8 | 39.8 | 38.8 | 37.8 | 36.8 |

1. Number of Gaussian terms used in the fitting process
2. Power, where 1 means single-particle density and 0.5 means Kohn-Sham orbitals
3. Range of *r* used in the fitting process
4. Integral overlap values between the exact quantity and the fitted quantity
5. Percentage error of the fitted points relative to the original density points for the first 10 points with the highest error

Table C-3: Gaussian functions resulting from fitting electronic densities and Kohn-Sham orbitals in TC-DFT calculations

| quantity/mp | 207 | 1836 |
|---|---|---|
| One-electron density | $0.022451e^{-2.67952\mathrm{re}^2} + 0.05146e^{-1.26742\mathrm{re}^2} + 0.0363588e^{-0.563362\mathrm{re}^2} + 0.006425e^{-0.247975\mathrm{re}^2}$ | $0.0202244\,e^{-10.2106\mathrm{re}^2} + 0.262169e^{-3.94531\mathrm{re}^2} - 0.207595e^{-3.91606\mathrm{re}^2} + 0.0675395e^{-1.58444\mathrm{re}^2} + 0.0425351e^{-0.644867\mathrm{re}^2} + 0.00775279e^{-0.2684\mathrm{re}^2}$ |
| KS electronic orbital | $0.116145e^{-1.8118\mathrm{re}^2} - 0.17711e^{-1.54565\mathrm{re}^2} + 0.157897e^{-1.25662\mathrm{re}^2} + 0.112317e^{-0.503436\mathrm{re}^2} + 0.10058e^{-0.215329\mathrm{re}^2} + 0.0317891e^{-0.0863076\mathrm{re}^2}$ | $0.06247\,e^{-5.62925\mathrm{re}^2} + 0.142094e^{-1.40685\mathrm{re}^2} + 0.165614e^{-0.393182\mathrm{re}^2} + 0.0684561e^{-0.118032\mathrm{re}^2}$ |

Table C-4: Details of fitting variational parameters

| | Quantities/ ω | 0.0001 | 0.001 | 0.01 | 0.1 | 1 | 10 | 100 | 1000 | 10000 |
|---|---|---|---|---|---|---|---|---|---|---|
| m=1 | | | | | | | | | | |
| a parameter | OPT $\alpha$[1] | 0.500000 | 0.499997 | 0.499704 | 0.484707 | 0.438060 | 0.414991 | 0.407310 | 0.404861 | 0.404086 |
| | LOG $\alpha$[2] | 0.499990 | 0.499900 | 0.499010 | 0.490909 | 0.450000 | 0.409091 | 0.400990 | 0.400100 | 0.400010 |
| | ERR $\alpha$[3] | 0.497963 | 0.497440 | 0.493553 | 0.478778 | 0.450000 | 0.421222 | 0.406447 | 0.402560 | 0.402037 |
| | %LOG[4] | 0.002000 | 0.019383 | 0.139038 | 1.263419 | 2.653261 | 1.442263 | 1.576125 | 1.190062 | 1.019060 |
| | %ERR[5] | 0.409065 | 0.513939 | 1.246274 | 1.238431 | 2.653261 | 1.479361 | 0.212303 | 0.571789 | 0.509742 |
| b parameter | OPT $\beta$[6] | 0.000000 | 0.000001 | 0.000149 | 0.010559 | 0.205107 | 2.366272 | 24.586618 | 248.702541 | 2495.906884 |
| | SIM $\beta$[7] | 0.000025 | 0.000250 | 0.002500 | 0.025000 | 0.250000 | 2.500000 | 25.000000 | 250.000000 | 2500.000000 |
| | SIM2 $\beta$[8] | 0.000475 | 0.001331 | 0.002500 | 0.009189 | 0.200000 | 2.341886 | 24.500000 | 248.418861 | 2495.000000 |
| | EXP $\beta$[9] | 0.000024 | 0.000240 | 0.002400 | 0.024000 | 0.240000 | 2.400000 | 24.000000 | 240.000000 | 2400.000000 |
| | %SIM[10] | 99.952980 | 99.400089 | 94.038998 | 57.765854 | 17.957021 | 5.349138 | 1.653527 | 0.518984 | 0.163725 |
| | %SIM2[11] | 99.997525 | 99.887331 | 94.038998 | 14.908943 | 2.553724 | 1.041273 | 0.353544 | 0.114194 | 0.036348 |
| | %EXP[12] | 99.951021 | 99.375093 | 93.790623 | 56.006098 | 14.538564 | 1.405352 | 2.444243 | 3.626059 | 3.996120 |
| Energy | VAR Energy[13] | -0.250000 | -0.249997 | -0.249701 | -0.223933 | 0.612000 | 12.395490 | 141.942322 | 1474.690624 | 14920.133739 |
| | LS Energy[14] | -0.250000 | -0.249994 | -0.249474 | -0.217075 | 0.630344 | 12.412923 | 141.960439 | 1474.709359 | 14920.152718 |
| | ES Energy[15] | -0.249996 | -0.249995 | -0.249582 | -0.218556 | 0.630344 | 12.415904 | 141.961799 | 1474.709979 | 14920.153233 |
| | LE Energy[16] | -0.250000 | -0.249995 | -0.249493 | -0.217839 | 0.623950 | 12.396391 | 141.986041 | 1475.633840 | 14931.610696 |
| | EE Energy[17] | -0.249996 | -0.249995 | -0.249595 | -0.219238 | 0.623950 | 12.397175 | 141.983937 | 1475.629363 | 14931.597794 |
| | LS2 Energy[18] | -0.249989 | -0.249917 | -0.249474 | -0.223914 | 0.612047 | 12.396473 | 141.943590 | 1474.691879 | 14920.134976 |
| | ES2 Energy[19] | -0.249997 | -0.249949 | -0.249582 | -0.223730 | 0.612047 | 12.395899 | 141.943252 | 1474.691721 | 14920.134846 |

| | Quantities/ ω | 0.0001 | 0.001 | 0.01 | 0.1 | 1 | 10 | 100 | 1000 | 10000 |
|---|---|---|---|---|---|---|---|---|---|---|
| | %LS [20] | 0.000011 | 0.001064 | 0.090900 | 3.159465 | 2.910089 | 0.140437 | 0.012762 | 0.001270 | 0.000127 |
| | %LE[21] | 0.000010 | 0.000977 | 0.083307 | 2.797517 | 1.915165 | 0.007267 | 0.030791 | 0.063919 | 0.076863 |
| | %LS2[22] | 0.004237 | 0.031838 | 0.090900 | 0.008562 | 0.007715 | 0.007924 | 0.000893 | 0.000085 | 0.000008 |
| | %ES[23] | 0.001427 | 0.000749 | 0.047595 | 2.460411 | 2.910089 | 0.164414 | 0.013720 | 0.001312 | 0.000131 |
| | %EE[24] | 0.001435 | 0.000778 | 0.042332 | 2.141776 | 1.915165 | 0.013592 | 0.029309 | 0.063616 | 0.076777 |
| | %ES2[25] | 0.001325 | 0.019291 | 0.047595 | 0.091086 | 0.007715 | 0.003301 | 0.000655 | 0.000074 | 0.000007 |
| | | | | | m=1.5 | | | | | |
| a parameter | OPT a | 0.600000 | 0.599997 | 0.599752 | 0.585320 | 0.529268 | 0.499264 | 0.489182 | 0.485964 | 0.484945 |
| | LOG a | 0.599988 | 0.599880 | 0.598812 | 0.589091 | 0.540000 | 0.490909 | 0.481188 | 0.480120 | 0.480012 |
| | ERR a | 0.597587 | 0.597270 | 0.593829 | 0.577035 | 0.540000 | 0.502965 | 0.486171 | 0.482730 | 0.482413 |
| | %LOG | 0.002000 | 0.019567 | 0.157023 | 0.640109 | 1.987402 | 1.701975 | 1.661221 | 1.217252 | 1.027617 |
| | %ERR | 0.403828 | 0.456625 | 0.997417 | 1.435817 | 1.987402 | 0.735804 | 0.619313 | 0.670015 | 0.524751 |
| b parameter | OPT b | 0.000000 | 0.000001 | 0.000149 | 0.011363 | 0.240597 | 2.823673 | 29.456033 | 298.293868 | 2994.618890 |
| | SIM b | 0.000030 | 0.000300 | 0.003000 | 0.030000 | 0.300000 | 3.000000 | 30.000000 | 300.000000 | 3000.000000 |
| | SIM2 b | 0.000570 | 0.001597 | 0.003000 | 0.011026 | 0.240000 | 2.810263 | 29.400000 | 298.102633 | 2994.000000 |
| | EXP b | 0.000029 | 0.000288 | 0.002880 | 0.028800 | 0.288000 | 2.880000 | 28.800000 | 288.000000 | 2880.000000 |
| | %SIM | 99.960817 | 99.500074 | 95.022645 | 62.124846 | 19.800854 | 5.877564 | 1.813225 | 0.568711 | 0.179370 |
| | %SIM2 | 99.997938 | 99.906109 | 95.022645 | 3.049174 | 0.248932 | 0.477170 | 0.190587 | 0.064151 | 0.020671 |
| | %EXP | 99.959184 | 99.479244 | 94.815255 | 60.546715 | 16.459223 | 1.955796 | 2.277891 | 3.574260 | 3.979823 |
| Energy | VAR Energy | -0.300000 | -0.299998 | -0.299750 | -0.277533 | 0.516267 | 12.137957 | 141.164882 | 1472.266767 | 14912.502746 |
| | LS Energy | -0.300000 | -0.299995 | -0.299560 | -0.270893 | 0.536833 | 12.158441 | 141.186484 | 1472.289203 | 14912.525507 |
| | ES Energy | -0.299996 | -0.299996 | -0.299644 | -0.272214 | 0.536833 | 12.161337 | 141.187713 | 1472.289860 | 14912.526117 |

| | Quantities/ ω | 0.0001 | 0.001 | 0.01 | 0.1 | 1 | 10 | 100 | 1000 | 10000 |
|---|---|---|---|---|---|---|---|---|---|---|
| | LE Energy | -0.300000 | -0.299996 | -0.299576 | -0.271605 | 0.530186 | 12.139720 | 141.203219 | 1473.183252 | 14923.884672 |
| | EE Energy | -0.299996 | -0.299996 | -0.299656 | -0.272854 | 0.530186 | 12.140650 | 141.201573 | 1473.178978 | 14923.870779 |
| | LS2 Energy | -0.299991 | -0.299931 | -0.299560 | -0.277528 | 0.516339 | 12.138319 | 141.165405 | 1472.267268 | 14912.503233 |
| | ES2 Energy | -0.299998 | -0.299958 | -0.299644 | -0.277451 | 0.516339 | 12.138034 | 141.165225 | 1472.267172 | 14912.503146 |
| | %LS | 0.000007 | 0.000726 | 0.063548 | 2.451169 | 3.830983 | 0.168474 | 0.015300 | 0.001524 | 0.000153 |
| | %LE | 0.000007 | 0.000666 | 0.058202 | 2.182577 | 2.625281 | 0.014523 | 0.027150 | 0.062211 | 0.076267 |
| | %LS2 | 0.002948 | 0.022194 | 0.063548 | 0.001689 | 0.013945 | 0.002976 | 0.000371 | 0.000034 | 0.000003 |
| | %ES | 0.001424 | 0.000628 | 0.035430 | 1.954141 | 3.830983 | 0.192247 | 0.016171 | 0.001569 | 0.000157 |
| | %EE | 0.001432 | 0.000654 | 0.031585 | 1.714720 | 2.625281 | 0.022179 | 0.025985 | 0.061921 | 0.076173 |
| | %ES2 | 0.000795 | 0.013020 | 0.035430 | 0.029552 | 0.013945 | 0.000631 | 0.000244 | 0.000028 | 0.000003 |
| | | | | | | m = 2 | | | | |
| a parameter | OPT a | 0.666667 | 0.666664 | 0.666443 | 0.652472 | 0.590509 | 0.555616 | 0.543817 | 0.540050 | 0.538856 |
| | LOG a | 0.666653 | 0.666533 | 0.665347 | 0.654545 | 0.600000 | 0.545455 | 0.534653 | 0.533467 | 0.533347 |
| | ERR a | 0.663993 | 0.663771 | 0.660661 | 0.642776 | 0.600000 | 0.557224 | 0.539339 | 0.536229 | 0.536007 |
| | %LOG | 0.001991 | 0.019631 | 0.164833 | 0.316852 | 1.581817 | 1.862860 | 1.713983 | 1.234041 | 1.032945 |
| | %ERR | 0.402594 | 0.435954 | 0.875224 | 1.508335 | 1.581817 | 0.288593 | 0.830334 | 0.712439 | 0.531571 |
| b parameter | OPT b | 0.000000 | 0.000002 | 0.000149 | 0.011795 | 0.263492 | 3.126429 | 32.695823 | 331.334674 | 3327.030484 |
| | SIM b | 0.000033 | 0.000333 | 0.003333 | 0.033333 | 0.333333 | 3.333333 | 33.333333 | 333.333333 | 3333.333333 |
| | SIM2 b | 0.000633 | 0.001775 | 0.003333 | 0.012251 | 0.266667 | 3.122515 | 32.666667 | 331.225148 | 3326.666667 |
| | EXP b | 0.000032 | 0.000320 | 0.003200 | 0.032000 | 0.320000 | 3.200000 | 32.000000 | 320.000000 | 3200.000000 |
| | %SIM | 99.931437 | 99.544514 | 95.516568 | 64.616431 | 20.952291 | 6.207122 | 1.912531 | 0.599598 | 0.189085 |
| | %SIM2 | 99.996391 | 99.914456 | 95.516568 | 3.729827 | 1.190363 | 0.125362 | 0.089254 | 0.033067 | 0.010936 |

| | Quantities/ ω | 0.0001 | 0.001 | 0.01 | 0.1 | 1 | 10 | 100 | 1000 | 10000 |
|---|---|---|---|---|---|---|---|---|---|---|
| | %EXP | 99.928580 | 99.525536 | 95.329758 | 63.142115 | 17.658636 | 2.299086 | 2.174447 | 3.542086 | 3.969703 |
| Energy | VAR Energy | -0.333333 | -0.333331 | -0.333109 | -0.312780 | 0.455715 | 11.977247 | 140.681475 | 1470.761234 | 14907.764463 |
| | LS Energy | -0.333333 | -0.333329 | -0.332937 | -0.306322 | 0.477606 | 11.999709 | 140.705382 | 1470.786132 | 14907.789743 |
| | ES Energy | -0.333329 | -0.333329 | -0.333009 | -0.307529 | 0.477606 | 12.002497 | 140.706531 | 1470.786827 | 14907.790419 |
| | LE Energy | -0.333333 | -0.333329 | -0.332952 | -0.307002 | 0.470830 | 11.979676 | 140.716665 | 1471.661343 | 14919.087618 |
| | EE Energy | -0.333329 | -0.333329 | -0.333020 | -0.308144 | 0.470830 | 11.980658 | 140.715256 | 1471.657090 | 14919.073056 |
| | LS2 Energy | -0.333325 | -0.333271 | -0.332937 | -0.312767 | 0.455934 | 11.977383 | 140.681703 | 1470.761441 | 14907.764658 |
| | ES2 Energy | -0.333331 | -0.333297 | -0.333009 | -0.312734 | 0.455934 | 11.977250 | 140.681599 | 1470.761380 | 14907.764601 |
| | %LS | 0.000006 | 0.000581 | 0.051475 | 2.108383 | 4.583484 | 0.187193 | 0.016991 | 0.001693 | 0.000170 |
| | %LE | 0.000005 | 0.000532 | 0.047121 | 1.882285 | 3.210322 | 0.020279 | 0.025008 | 0.061163 | 0.075897 |
| | %LS2 | 0.002389 | 0.018001 | 0.051475 | 0.004119 | 0.048049 | 0.001140 | 0.000162 | 0.000014 | 0.000001 |
| | %ES | 0.001434 | 0.000607 | 0.029805 | 1.707677 | 4.583484 | 0.210376 | 0.017807 | 0.001740 | 0.000174 |
| | %EE | 0.001440 | 0.000633 | 0.026604 | 1.504499 | 3.210322 | 0.028472 | 0.024007 | 0.060874 | 0.075800 |
| | %ES2 | 0.000610 | 0.010216 | 0.029805 | 0.014645 | 0.048049 | 0.000030 | 0.000088 | 0.000010 | 0.000001 |
| | | | | | | m=3 | | | | |
| a parameter | OPT a | 0.750000 | 0.749998 | 0.749801 | 0.736443 | 0.667504 | 0.626232 | 0.612170 | 0.607675 | 0.606251 |
| | LOG a | 0.749985 | 0.749850 | 0.748515 | 0.736364 | 0.675000 | 0.613636 | 0.601485 | 0.600150 | 0.600015 |
| | ERR a | 0.746997 | 0.746856 | 0.744160 | 0.725175 | 0.675000 | 0.624825 | 0.605840 | 0.603144 | 0.603003 |
| | %LOG | 0.002000 | 0.019717 | 0.171842 | 0.010793 | 1.110486 | 2.052664 | 1.776395 | 1.253885 | 1.039226 |
| | %ERR | 0.401975 | 0.420761 | 0.758074 | 1.553884 | 1.110486 | 0.225182 | 1.044784 | 0.751156 | 0.538599 |
| b parameter | OPT b | 0.000000 | 0.000001 | 0.000150 | 0.012243 | 0.291304 | 3.502570 | 36.738719 | 372.614531 | 3742.478585 |
| | SIM b | 0.000038 | 0.000375 | 0.003750 | 0.037500 | 0.375000 | 3.750000 | 37.500000 | 375.000000 | 3750.000000 |

| | Quantities/ ω | 0.0001 | 0.001 | 0.01 | 0.1 | 1 | 10 | 100 | 1000 | 10000 |
|---|---|---|---|---|---|---|---|---|---|---|
| | SIM2 b | 0.000713 | 0.001997 | 0.003750 | 0.013783 | 0.300000 | 3.512829 | 36.750000 | 372.628292 | 3742.500000 |
| | EXP b | 0.000036 | 0.000360 | 0.003600 | 0.036000 | 0.360000 | 3.600000 | 36.000000 | 360.000000 | 3600.000000 |
| | %SIM | 99.968654 | 99.600163 | 96.011650 | 67.350968 | 22.318911 | 6.598141 | 2.030082 | 0.636125 | 0.200571 |
| | %SIM2 | 99.998350 | 99.924907 | 96.011650 | 11.169846 | 2.898639 | 0.292057 | 0.030696 | 0.003693 | 0.000572 |
| | %EXP | 99.967347 | 99.583503 | 95.845469 | 65.990592 | 19.082199 | 2.706397 | 2.051998 | 3.504036 | 3.957738 |
| Energy | VAR Energy | -0.375000 | -0.374998 | -0.374800 | -0.356440 | 0.383003 | 11.786450 | 140.109312 | 1468.980867 | 14902.162752 |
| | LS Energy | -0.375000 | -0.374996 | -0.374648 | -0.350229 | 0.406395 | 11.811329 | 140.136081 | 1469.008837 | 14902.191178 |
| | ES Energy | -0.374995 | -0.374996 | -0.374709 | -0.351292 | 0.406395 | 11.813936 | 140.137143 | 1469.009589 | 14902.191937 |
| | LE Energy | -0.375000 | -0.374996 | -0.374661 | -0.350871 | 0.399493 | 11.789789 | 140.140971 | 1469.861834 | 14913.416659 |
| | EE Energy | -0.374995 | -0.374996 | -0.374718 | -0.351878 | 0.399493 | 11.790793 | 140.139799 | 1469.857536 | 14913.401287 |
| | LS2 Energy | -0.374993 | -0.374945 | -0.374648 | -0.356411 | 0.383501 | 11.786472 | 140.109354 | 1468.980895 | 14902.162773 |
| | ES2 Energy | -0.374998 | -0.374969 | -0.374709 | -0.356414 | 0.383501 | 11.786487 | 140.109321 | 1468.980875 | 14902.162756 |
| | %LS | 0.000005 | 0.000452 | 0.040535 | 1.773442 | 5.756019 | 0.210632 | 0.019102 | 0.001904 | 0.000191 |
| | %LE | 0.000004 | 0.000414 | 0.037081 | 1.587124 | 4.127785 | 0.028322 | 0.022591 | 0.059935 | 0.075462 |
| | %LS2 | 0.001889 | 0.014233 | 0.040535 | 0.008151 | 0.129934 | 0.000188 | 0.000030 | 0.000002 | 0.000000 |
| | %ES | 0.001448 | 0.000614 | 0.024455 | 1.465585 | 5.756019 | 0.232655 | 0.019860 | 0.001955 | 0.000196 |
| | %EE | 0.001454 | 0.000639 | 0.021856 | 1.296455 | 4.127785 | 0.036835 | 0.021755 | 0.059643 | 0.075359 |
| | %ES2 | 0.000482 | 0.007681 | 0.024455 | 0.007435 | 0.129934 | 0.000315 | 0.000006 | 0.000001 | 0.000000 |
| | | | | | | m=10 | | | | |
| a parameter | OPT a | 0.909091 | 0.909089 | 0.908926 | 0.896745 | 0.815720 | 0.761554 | 0.742827 | 0.736831 | 0.734930 |
| | LOG a | 0.909073 | 0.908909 | 0.907291 | 0.892562 | 0.818182 | 0.743802 | 0.729073 | 0.727454 | 0.727291 |
| | ERR a | 0.905454 | 0.905396 | 0.903416 | 0.883033 | 0.818182 | 0.753331 | 0.732948 | 0.730968 | 0.730910 |

| | Quantities/ ω | 0.0001 | 0.001 | 0.01 | 0.1 | 1 | 10 | 100 | 1000 | 10000 |
|---|---|---|---|---|---|---|---|---|---|---|
| | %LOG | 0.001999 | 0.019796 | 0.180272 | 0.468653 | 0.300949 | 2.386751 | 1.886491 | 1.288926 | 1.050314 |
| | %ERR | 0.401661 | 0.407912 | 0.609998 | 1.552838 | 0.300949 | 1.091638 | 1.347800 | 0.802095 | 0.550021 |
| b parameter | OPT b | 0.000000 | 0.000002 | 0.000150 | 0.012890 | 0.342079 | 4.213989 | 44.437261 | 451.360667 | 4535.415827 |
| | SIM b | 0.000045 | 0.000455 | 0.004545 | 0.045455 | 0.454545 | 4.545455 | 45.454545 | 454.545455 | 4545.454545 |
| | SIM2 b | 0.000864 | 0.002420 | 0.004545 | 0.016707 | 0.363636 | 4.257975 | 44.545455 | 451.670657 | 4536.363636 |
| | EXP b | 0.000044 | 0.000436 | 0.004364 | 0.043636 | 0.436364 | 4.363636 | 43.636364 | 436.363636 | 4363.636364 |
| | %SIM | 99.974139 | 99.666836 | 96.705464 | 71.641569 | 24.742679 | 7.292248 | 2.238026 | 0.700653 | 0.220852 |
| | %SIM2 | 99.998639 | 99.937429 | 96.705464 | 22.843536 | 5.928349 | 1.033028 | 0.242884 | 0.068632 | 0.020894 |
| | %EXP | 99.973062 | 99.652954 | 96.568191 | 70.459967 | 21.606958 | 3.429426 | 1.835389 | 3.436820 | 3.936613 |
| Energy | VAR Energy | -0.454545 | -0.454544 | -0.454381 | -0.438912 | 0.251539 | 11.447417 | 139.097263 | 1465.835951 | 14892.271732 |
| | LS Energy | -0.454545 | -0.454542 | -0.454257 | -0.433193 | 0.277382 | 11.476745 | 139.129440 | 1465.869766 | 14892.306161 |
| | ES Energy | -0.454539 | -0.454541 | -0.454301 | -0.433986 | 0.277382 | 11.478902 | 139.130376 | 1465.870646 | 14892.307079 |
| | LE Energy | -0.454545 | -0.454543 | -0.454268 | -0.433772 | 0.270320 | 11.452664 | 139.123182 | 1466.683692 | 14903.403985 |
| | EE Energy | -0.454539 | -0.454541 | -0.454310 | -0.434524 | 0.270320 | 11.453602 | 139.122323 | 1466.679201 | 14903.387167 |
| | LS2 Energy | -0.454540 | -0.454500 | -0.454257 | -0.438853 | 0.252744 | 11.447614 | 139.097398 | 1465.836106 | 14892.271901 |
| | ES2 Energy | -0.454544 | -0.454523 | -0.454301 | -0.438885 | 0.252744 | 11.447801 | 139.097454 | 1465.836166 | 14892.271967 |
| | %LS | 0.000003 | 0.000298 | 0.027217 | 1.320309 | 9.316719 | 0.255541 | 0.023127 | 0.002307 | 0.000231 |
| | %LE | 0.000003 | 0.000272 | 0.024865 | 1.185062 | 6.947797 | 0.045813 | 0.018630 | 0.057800 | 0.074696 |
| | %LS2 | 0.001286 | 0.009682 | 0.027217 | 0.013473 | 0.476625 | 0.001722 | 0.000097 | 0.000011 | 0.000001 |
| | %ES | 0.001472 | 0.000674 | 0.017403 | 1.135127 | 9.316719 | 0.274290 | 0.023800 | 0.002367 | 0.000237 |
| | %EE | 0.001477 | 0.000699 | 0.015575 | 1.009911 | 6.947797 | 0.054004 | 0.018013 | 0.057494 | 0.074583 |
| | %ES2 | 0.000401 | 0.004667 | 0.017403 | 0.006157 | 0.476625 | 0.003355 | 0.000138 | 0.000015 | 0.000002 |

| | Quantities/ ω | 0.0001 | 0.001 | 0.01 | 0.1 | 1 | 10 | 100 | 1000 | 10000 |
|---|---|---|---|---|---|---|---|---|---|---|
| | | | | | m = 50 | | | | | |
| a parameter | OPT a | 0.980392 | 0.980391 | 0.980240 | 0.968561 | 0.882615 | 0.822405 | 0.801451 | 0.794737 | 0.792600 |
| | LOG a | 0.980373 | 0.980196 | 0.978451 | 0.962567 | 0.882353 | 0.802139 | 0.786255 | 0.784510 | 0.784333 |
| | ERR a | 0.976470 | 0.976432 | 0.974726 | 0.953971 | 0.882353 | 0.810735 | 0.789980 | 0.788274 | 0.788236 |
| | %LOG | 0.001995 | 0.019833 | 0.182826 | 0.622755 | 0.029713 | 2.526509 | 1.932654 | 1.303698 | 1.054025 |
| | %ERR | 0.401625 | 0.405446 | 0.565690 | 1.529464 | 0.029713 | 1.439423 | 1.452002 | 0.819909 | 0.553752 |
| b parameter | OPT b | 0.000000 | 0.000001 | 0.000150 | 0.013113 | 0.363909 | 4.530152 | 47.879691 | 486.628702 | 4890.717855 |
| | SIM b | 0.000049 | 0.000490 | 0.004902 | 0.049020 | 0.490196 | 4.901961 | 49.019608 | 490.196078 | 4901.960784 |
| | SIM2 b | 0.000931 | 0.002610 | 0.004902 | 0.018017 | 0.392157 | 4.591934 | 48.039216 | 487.095806 | 4892.156863 |
| | EXP b | 0.000047 | 0.000471 | 0.004706 | 0.047059 | 0.470588 | 4.705882 | 47.058824 | 470.588235 | 4705.882353 |
| | %SIM | 99.946254 | 99.696529 | 96.945106 | 73.248938 | 25.762490 | 7.584890 | 2.325431 | 0.727745 | 0.229356 |
| | %SIM2 | 99.997171 | 99.943005 | 96.945106 | 27.216803 | 7.203112 | 1.345427 | 0.332073 | 0.095896 | 0.029415 |
| | %EXP | 99.944015 | 99.683884 | 96.817819 | 72.134311 | 22.669260 | 3.734260 | 1.744342 | 3.408599 | 3.927754 |
| Energy | VAR Energy | -0.490196 | -0.490195 | -0.490043 | -0.475603 | 0.195169 | 11.304325 | 138.671901 | 1464.515756 | 14888.121166 |
| | LS Energy | -0.490196 | -0.490193 | -0.489929 | -0.470101 | 0.221956 | 11.335583 | 138.706479 | 1464.552184 | 14888.158282 |
| | ES Energy | -0.490189 | -0.490191 | -0.489969 | -0.470779 | 0.221956 | 11.337506 | 138.707376 | 1464.553126 | 14888.159273 |
| | LE Energy | -0.490196 | -0.490193 | -0.489939 | -0.470655 | 0.214849 | 11.310479 | 138.695595 | 1465.349772 | 14899.202603 |
| | EE Energy | -0.490189 | -0.490191 | -0.489977 | -0.471298 | 0.214849 | 11.311351 | 138.694834 | 1465.345176 | 14899.185181 |
| | LS2 Energy | -0.490191 | -0.490154 | -0.489929 | -0.475534 | 0.196730 | 11.304731 | 138.672232 | 1464.516133 | 14888.121569 |
| | ES2 Energy | -0.490194 | -0.490176 | -0.489969 | -0.475569 | 0.196730 | 11.304954 | 138.672317 | 1464.516230 | 14888.121674 |
| | %LS | 0.000003 | 0.000252 | 0.023211 | 1.170328 | 12.068611 | 0.275749 | 0.024929 | 0.002487 | 0.000249 |
| | %LE | 0.000002 | 0.000231 | 0.021191 | 1.051297 | 9.160081 | 0.054411 | 0.017084 | 0.056916 | 0.074376 |

| | Quantities/ ω | 0.0001 | 0.001 | 0.01 | 0.1 | 1 | 10 | 100 | 1000 | 10000 |
|---|---|---|---|---|---|---|---|---|---|---|
| | %LS2 | 0.001106 | 0.008318 | 0.023211 | 0.014624 | 0.793266 | 0.003588 | 0.000239 | 0.000026 | 0.000003 |
| | %ES | 0.001481 | 0.000708 | 0.015108 | 1.024639 | 12.068611 | 0.292667 | 0.025576 | 0.002552 | 0.000256 |
| | %EE | 0.001485 | 0.000733 | 0.013524 | 0.913430 | 9.160081 | 0.062117 | 0.016536 | 0.056602 | 0.074259 |
| | %ES2 | 0.000400 | 0.003797 | 0.015108 | 0.007166 | 0.793266 | 0.005568 | 0.000300 | 0.000032 | 0.000003 |
| | | | | | | m=207 | | | | |
| a parameter | OPT a | 0.995192 | 0.995191 | 0.995042 | 0.983465 | 0.896535 | 0.835051 | 0.813624 | 0.806758 | 0.804581 |
| | LOG a | 0.995172 | 0.994993 | 0.993222 | 0.977098 | 0.895673 | 0.814248 | 0.798125 | 0.796353 | 0.796174 |
| | ERR a | 0.991211 | 0.991176 | 0.989523 | 0.968708 | 0.895673 | 0.822639 | 0.801823 | 0.800170 | 0.800135 |
| | %LOG | 0.001973 | 0.019832 | 0.183285 | 0.651678 | 0.096203 | 2.554842 | 1.942031 | 1.306622 | 1.055913 |
| | %ERR | 0.401600 | 0.405063 | 0.557750 | 1.523451 | 0.096203 | 1.508864 | 1.471793 | 0.823295 | 0.555636 |
| b parameter | OPT b | 0.000000 | 0.000001 | 0.000150 | 0.013155 | 0.368372 | 4.595580 | 48.593656 | 493.947566 | 4964.462597 |
| | SIM b | 0.000050 | 0.000498 | 0.004976 | 0.049760 | 0.497596 | 4.975962 | 49.759615 | 497.596154 | 4975.961538 |
| | SIM2 b | 0.000945 | 0.002649 | 0.004976 | 0.018289 | 0.398077 | 4.661254 | 48.764423 | 494.449079 | 4966.009615 |
| | EXP b | 0.000048 | 0.000478 | 0.004777 | 0.047769 | 0.477692 | 4.776923 | 47.769231 | 477.692308 | 4776.923077 |
| | %SIM | 99.837207 | 99.698596 | 96.990537 | 73.562010 | 25.969647 | 7.644388 | 2.343183 | 0.733243 | 0.231090 |
| | %SIM2 | 99.991432 | 99.943393 | 96.990537 | 28.068596 | 7.462059 | 1.408942 | 0.350187 | 0.101429 | 0.031152 |
| | %EXP | 99.830424 | 99.686037 | 96.865143 | 72.460428 | 22.885049 | 3.796237 | 1.725851 | 3.402872 | 3.925948 |
| Energy | VAR Energy | -0.497596 | -0.497595 | -0.497446 | -0.483202 | 0.183638 | 11.275221 | 138.585513 | 1464.247752 | 14887.278700 |
| | LS Energy | -0.497596 | -0.497593 | -0.497334 | -0.477745 | 0.210611 | 11.306874 | 138.620589 | 1464.284722 | 14887.316374 |
| | ES Energy | -0.497589 | -0.497591 | -0.497372 | -0.478400 | 0.210611 | 11.308747 | 138.621479 | 1464.285678 | 14887.317380 |
| | LE Energy | -0.497596 | -0.497594 | -0.497343 | -0.478294 | 0.203496 | 11.281566 | 138.608770 | 1465.078999 | 14898.349839 |
| | EE Energy | -0.497589 | -0.497591 | -0.497380 | -0.478915 | 0.203496 | 11.282422 | 138.608027 | 1465.074380 | 14898.332295 |

| | Quantities/ ω | 0.0001 | 0.001 | 0.01 | 0.1 | 1 | 10 | 100 | 1000 | 10000 |
|---|---|---|---|---|---|---|---|---|---|---|
| | LS2 Energy | -0.497591 | -0.497554 | -0.497334 | -0.483131 | 0.185275 | 11.275678 | 138.585896 | 1464.248187 | 14887.279163 |
| | ES2 Energy | -0.497594 | -0.497577 | -0.497372 | -0.483167 | 0.185275 | 11.275907 | 138.585986 | 1464.248291 | 14887.279276 |
| | %LS | 0.000002 | 0.000244 | 0.022484 | 1.142266 | 12.806657 | 0.279951 | 0.025303 | 0.002525 | 0.000253 |
| | %LE | 0.000002 | 0.000223 | 0.020525 | 1.026232 | 9.758351 | 0.056248 | 0.016779 | 0.056737 | 0.074311 |
| | %LS2 | 0.001073 | 0.008071 | 0.022484 | 0.014785 | 0.883052 | 0.004055 | 0.000276 | 0.000030 | 0.000003 |
| | %ES | 0.001482 | 0.000715 | 0.014680 | 1.003882 | 12.806657 | 0.296465 | 0.025945 | 0.002590 | 0.000260 |
| | %EE | 0.001487 | 0.000740 | 0.013142 | 0.895267 | 9.758351 | 0.063829 | 0.016243 | 0.056422 | 0.074194 |
| | %ES2 | 0.000401 | 0.003643 | 0.014680 | 0.007405 | 0.883052 | 0.006084 | 0.000341 | 0.000037 | 0.000004 |
| | | | | | m=400 | | | | | |
| a parameter | OPT a | 0.997506 | 0.997505 | 0.997356 | 0.985796 | 0.898713 | 0.837029 | 0.815527 | 0.808637 | 0.806453 |
| | LOG a | 0.997486 | 0.997307 | 0.995531 | 0.979370 | 0.897756 | 0.816141 | 0.799980 | 0.798204 | 0.798025 |
| | ERR a | 0.993516 | 0.993481 | 0.991836 | 0.971012 | 0.897756 | 0.824499 | 0.803675 | 0.802030 | 0.801995 |
| | %LOG | 0.001997 | 0.019838 | 0.183350 | 0.656112 | 0.106591 | 2.559249 | 1.943426 | 1.307042 | 1.056055 |
| | %ERR | 0.401623 | 0.405012 | 0.556540 | 1.522480 | 0.106591 | 1.519636 | 1.474771 | 0.823783 | 0.555778 |
| b parameter | OPT b | 0.000000 | 0.000001 | 0.000150 | 0.013162 | 0.369068 | 4.605803 | 48.705265 | 495.091782 | 4975.992079 |
| | SIM b | 0.000050 | 0.000499 | 0.004988 | 0.049875 | 0.498753 | 4.987531 | 49.875312 | 498.753117 | 4987.531172 |
| | SIM2 b | 0.000948 | 0.002656 | 0.004988 | 0.018331 | 0.399002 | 4.672092 | 48.877805 | 495.598726 | 4977.556110 |
| | EXP b | 0.000048 | 0.000479 | 0.004788 | 0.047880 | 0.478803 | 4.788030 | 47.880299 | 478.802993 | 4788.029925 |
| | %SIM | 99.976431 | 99.701653 | 96.997178 | 73.610356 | 26.001949 | 7.653653 | 2.345943 | 0.734098 | 0.231359 |
| | %SIM2 | 99.998760 | 99.943968 | 96.997178 | 28.200132 | 7.502436 | 1.418833 | 0.353003 | 0.102289 | 0.031422 |
| | %EXP | 99.975449 | 99.689222 | 96.872060 | 72.510787 | 22.918696 | 3.805889 | 1.722976 | 3.401982 | 3.925668 |
| | VAR Energy | -0.498753 | -0.498752 | -0.498603 | -0.484390 | 0.181841 | 11.270688 | 138.572063 | 1464.206030 | 14887.147551 |

| | Quantities/ ω | 0.0001 | 0.001 | 0.01 | 0.1 | 1 | 10 | 100 | 1000 | 10000 |
|---|---|---|---|---|---|---|---|---|---|---|
| Energy | LS Energy | -0.498753 | -0.498750 | -0.498491 | -0.478940 | 0.208842 | 11.302404 | 138.607217 | 1464.243085 | 14887.185312 |
| | ES Energy | -0.498746 | -0.498748 | -0.498530 | -0.479591 | 0.208842 | 11.304269 | 138.608106 | 1464.244042 | 14887.186320 |
| | LE Energy | -0.498753 | -0.498751 | -0.498501 | -0.479488 | 0.201726 | 11.277064 | 138.595252 | 1465.036846 | 14898.217088 |
| | EE Energy | -0.498746 | -0.498748 | -0.498538 | -0.480105 | 0.201726 | 11.277917 | 138.594512 | 1465.032224 | 14898.199524 |
| | LS2 Energy | -0.498748 | -0.498712 | -0.498491 | -0.484318 | 0.183489 | 11.271154 | 138.572455 | 1464.206474 | 14887.148024 |
| | ES2 Energy | -0.498751 | -0.498734 | -0.498530 | -0.484354 | 0.183489 | 11.271383 | 138.572545 | 1464.206580 | 14887.148138 |
| | %LS | 0.000002 | 0.000243 | 0.022374 | 1.137965 | 12.928849 | 0.280608 | 0.025362 | 0.002531 | 0.000254 |
| | %LE | 0.000002 | 0.000222 | 0.020424 | 1.022389 | 9.857606 | 0.056537 | 0.016731 | 0.056710 | 0.074301 |
| | %LS2 | 0.001068 | 0.008033 | 0.022374 | 0.014808 | 0.898097 | 0.004130 | 0.000282 | 0.000030 | 0.000003 |
| | %ES | 0.001482 | 0.000716 | 0.014615 | 1.000699 | 12.928849 | 0.297059 | 0.026003 | 0.002596 | 0.000260 |
| | %EE | 0.001487 | 0.000741 | 0.013083 | 0.892480 | 9.857606 | 0.064097 | 0.016197 | 0.056394 | 0.074183 |
| | %ES2 | 0.000402 | 0.003619 | 0.014615 | 0.007443 | 0.898097 | 0.006166 | 0.000348 | 0.000038 | 0.000004 |
| | | | | | m=900 | | | | | |
| a parameter | OPT a | 0.998890 | 0.998889 | 0.998740 | 0.987189 | 0.900014 | 0.838211 | 0.816666 | 0.809761 | 0.807572 |
| | LOG a | 0.998870 | 0.998691 | 0.996912 | 0.980728 | 0.899001 | 0.817274 | 0.801090 | 0.799312 | 0.799132 |
| | ERR a | 0.994894 | 0.994860 | 0.993220 | 0.972390 | 0.899001 | 0.825612 | 0.804782 | 0.803143 | 0.803108 |
| | %LOG | 0.001989 | 0.019839 | 0.183392 | 0.658753 | 0.112695 | 2.561894 | 1.944300 | 1.307321 | 1.056147 |
| | %ERR | 0.401615 | 0.404980 | 0.555823 | 1.521896 | 0.112695 | 1.526080 | 1.476585 | 0.824102 | 0.555870 |
| b parameter | OPT b | 0.000000 | 0.000001 | 0.000150 | 0.013166 | 0.369484 | 4.611916 | 48.772012 | 495.776091 | 4982.887482 |
| | SIM b | 0.000050 | 0.000499 | 0.004994 | 0.049945 | 0.499445 | 4.994451 | 49.944506 | 499.445061 | 4994.450610 |
| | SIM2 b | 0.000949 | 0.002659 | 0.004994 | 0.018357 | 0.399556 | 4.678574 | 48.945616 | 496.286293 | 4984.461709 |
| | EXP b | 0.000048 | 0.000479 | 0.004795 | 0.047947 | 0.479467 | 4.794673 | 47.946726 | 479.467259 | 4794.672586 |

| | Quantities/ ω | 0.0001 | 0.001 | 0.01 | 0.1 | 1 | 10 | 100 | 1000 | 10000 |
|---|---|---|---|---|---|---|---|---|---|---|
| | %SIM | 99.918331 | 99.702440 | 97.001346 | 73.639192 | 26.021163 | 7.659192 | 2.347595 | 0.734609 | 0.231520 |
| | %SIM2 | 99.995702 | 99.944116 | 97.001346 | 28.278589 | 7.526454 | 1.424746 | 0.354688 | 0.102804 | 0.031583 |
| | %EXP | 99.914928 | 99.690042 | 96.876402 | 72.540825 | 22.938712 | 3.811659 | 1.721255 | 3.401449 | 3.925500 |
| Energy | VAR Energy | -0.499445 | -0.499444 | -0.499295 | -0.485100 | 0.180766 | 11.267980 | 138.564027 | 1464.181100 | 14887.069188 |
| | LS Energy | -0.499445 | -0.499442 | -0.499184 | -0.479654 | 0.207784 | 11.299732 | 138.599227 | 1464.218206 | 14887.107001 |
| | ES Energy | -0.499438 | -0.499440 | -0.499222 | -0.480303 | 0.207784 | 11.301592 | 138.600115 | 1464.219165 | 14887.108010 |
| | LE Energy | -0.499445 | -0.499442 | -0.499193 | -0.480202 | 0.200668 | 11.274373 | 138.587175 | 1465.011659 | 14898.137767 |
| | EE Energy | -0.499438 | -0.499440 | -0.499230 | -0.480817 | 0.200668 | 11.275225 | 138.586437 | 1465.007035 | 14898.120192 |
| | LS2 Energy | -0.499440 | -0.499404 | -0.499184 | -0.485028 | 0.182421 | 11.268450 | 138.564423 | 1464.181550 | 14887.069666 |
| | ES2 Energy | -0.499443 | -0.499426 | -0.499222 | -0.485064 | 0.182421 | 11.268680 | 138.564514 | 1464.181656 | 14887.069781 |
| | %LS | 0.000002 | 0.000242 | 0.022308 | 1.135404 | 13.002865 | 0.281001 | 0.025397 | 0.002534 | 0.000254 |
| | %LE | 0.000002 | 0.000221 | 0.020364 | 1.020101 | 9.917757 | 0.056710 | 0.016703 | 0.056693 | 0.074295 |
| | %LS2 | 0.001065 | 0.008011 | 0.022308 | 0.014822 | 0.907236 | 0.004176 | 0.000286 | 0.000031 | 0.000003 |
| | %ES | 0.001483 | 0.000717 | 0.014576 | 0.998803 | 13.002865 | 0.297413 | 0.026038 | 0.002600 | 0.000261 |
| | %EE | 0.001487 | 0.000742 | 0.013048 | 0.890820 | 9.917757 | 0.064258 | 0.016170 | 0.056378 | 0.074177 |
| | %ES2 | 0.000402 | 0.003605 | 0.014576 | 0.007465 | 0.907236 | 0.006215 | 0.000352 | 0.000038 | 0.000004 |
| | | | | | m=1836 | | | | | |
| a parameter | OPT a | 0.999456 | 0.999454 | 0.999306 | 0.987759 | 0.900546 | 0.838695 | 0.817131 | 0.810221 | 0.808030 |
| | LOG a | 0.999436 | 0.999256 | 0.997477 | 0.981284 | 0.899510 | 0.817736 | 0.801544 | 0.799764 | 0.799584 |
| | ERR a | 0.995458 | 0.995423 | 0.993785 | 0.972954 | 0.899510 | 0.826067 | 0.805235 | 0.803597 | 0.803563 |
| | %LOG | 0.001999 | 0.019831 | 0.183417 | 0.659830 | 0.115213 | 2.562955 | 1.944655 | 1.307538 | 1.056183 |
| | %ERR | 0.401625 | 0.404958 | 0.555539 | 1.521656 | 0.115213 | 1.528692 | 1.477324 | 0.824336 | 0.555905 |

| | Quantities/ ω | 0.0001 | 0.001 | 0.01 | 0.1 | 1 | 10 | 100 | 1000 | 10000 |
|---|---|---|---|---|---|---|---|---|---|---|
| b parameter | OPT b | 0.000000 | 0.000002 | 0.000150 | 0.013167 | 0.369653 | 4.614414 | 48.799286 | 496.055715 | 4985.705219 |
| | SIM b | 0.000050 | 0.000500 | 0.004997 | 0.049973 | 0.499728 | 4.997278 | 49.972782 | 499.727817 | 4997.278171 |
| | SIM2 b | 0.000949 | 0.002661 | 0.004997 | 0.018367 | 0.399782 | 4.681223 | 48.973326 | 496.567261 | 4987.283615 |
| | EXP b | 0.000048 | 0.000480 | 0.004797 | 0.047974 | 0.479739 | 4.797387 | 47.973870 | 479.738704 | 4797.387044 |
| | %SIM | 99.976477 | 99.696982 | 97.003375 | 73.650960 | 26.029034 | 7.661449 | 2.348269 | 0.734821 | 0.231585 |
| | %SIM2 | 99.998762 | 99.943091 | 97.003375 | 28.310605 | 7.536292 | 1.427155 | 0.355377 | 0.103017 | 0.031648 |
| | %EXP | 99.975497 | 99.684357 | 96.878515 | 72.553083 | 22.946910 | 3.814010 | 1.720553 | 3.401229 | 3.925432 |
| Energy | VAR Energy | -0.499728 | -0.499726 | -0.499578 | -0.485391 | 0.180327 | 11.266873 | 138.560744 | 1464.170918 | 14887.037181 |
| | LS Energy | -0.499728 | -0.499725 | -0.499467 | -0.479946 | 0.207352 | 11.298641 | 138.595963 | 1464.208044 | 14887.075015 |
| | ES Energy | -0.499720 | -0.499723 | -0.499505 | -0.480594 | 0.207352 | 11.300499 | 138.596851 | 1464.209003 | 14887.076025 |
| | LE Energy | -0.499728 | -0.499725 | -0.499476 | -0.480494 | 0.200236 | 11.273274 | 138.583876 | 1465.001371 | 14898.105369 |
| | EE Energy | -0.499720 | -0.499723 | -0.499513 | -0.481108 | 0.200236 | 11.274125 | 138.583138 | 1464.996747 | 14898.087789 |
| | LS2 Energy | -0.499722 | -0.499686 | -0.499467 | -0.485319 | 0.181985 | 11.267346 | 138.561143 | 1464.171370 | 14887.037661 |
| | ES2 Energy | -0.499726 | -0.499708 | -0.499505 | -0.485354 | 0.181985 | 11.267576 | 138.561234 | 1464.171476 | 14887.037777 |
| | %LS | 0.000002 | 0.000242 | 0.022281 | 1.134360 | 13.033316 | 0.281162 | 0.025411 | 0.002536 | 0.000254 |
| | %LE | 0.000002 | 0.000221 | 0.020339 | 1.019168 | 9.942509 | 0.056780 | 0.016692 | 0.056686 | 0.074293 |
| | %LS2 | 0.001064 | 0.008002 | 0.022281 | 0.014827 | 0.911001 | 0.004194 | 0.000287 | 0.000031 | 0.000003 |
| | %ES | 0.001483 | 0.000717 | 0.014560 | 0.998029 | 13.033316 | 0.297558 | 0.026052 | 0.002601 | 0.000261 |
| | %EE | 0.001487 | 0.000742 | 0.013034 | 0.890143 | 9.942509 | 0.064324 | 0.016159 | 0.056371 | 0.074175 |
| | %ES2 | 0.000402 | 0.003600 | 0.014560 | 0.007475 | 0.911001 | 0.006235 | 0.000353 | 0.000038 | 0.000004 |

1. Parameter α alpha using logistic function
2. Parameter α alpha using error function

3. Error percentage of parameter $\alpha$ alpha using logistic function
4. Error percentage of parameter $\alpha$ using error function
5. Optimized variational parameter $\beta$
6. Parameter $\beta$ using simple form 1
7. Parameter $\beta$ using simple form 2
8. Parameter $\beta$ using exponential function
9. Error percentage of parameter $\beta$ using simple form 1
10. Error percentage of parameter $\beta$ using simple form 2
11. Error percentage of parameter $\beta$ using exponential function
12. Variational energy of relative motion Hamiltonian using optimized parameters
13. Variational energy of relative motion Hamiltonian using parameters from items 2 and 7
14. Variational energy of relative motion Hamiltonian using parameters from items 3 and 7
15. Variational energy of relative motion Hamiltonian using parameters from items 2 and 9
16. Variational energy of relative motion Hamiltonian using parameters from items 3 and 9
17. Variational energy of relative motion Hamiltonian using parameters from items 2 and 8
18. Variational energy of relative motion Hamiltonian using parameters from items 3 and 8
19. Error percentage for item 14
20. Error percentage for item 15
21. Error percentage for item 16
22. Error percentage for item 17
23. Error percentage for item 18
24. Error percentage for item 19

# 10 References

[1] S. M. Blinder and J. E. House, "Mathematical physics in theoretical chemistry: A volume in developments in physical & theoretical chemistry," *Math. Phys. Theor. Chem.*, pp. 1–408, Jan. 2018, doi: 10.1016/C2016-0-04521-7.

[2] J. Leszczynski, *Handbook of computational chemistry*. Springer, 2012.

[3] P. W. Atkins and J. De Paula, "Atkins' Physical Chemistry, 8th Ed.; Oxford University Press," 2006.

[4] "Электронная библиотека БГУ: On the Quantum Theory of Molecules." [Online]. Available: https://elib.bsu.by/handle/123456789/154381. [Accessed: 21-May-2022].

[5] M. Born and V. Fock, "Beweis des Adiabatensatzes," *Zeitschrift für*

*Phys. 1928 513*, vol. 51, no. 3, pp. 165–180, Mar. 1928, doi: 10.1007/BF01343193.

[6] L. Piela, *Ideas of Quantum Chemistry*. Elsevier, 2020.

[7] D. J. (David J. Griffiths and D. F. Schroeter, *Introduction to quantum mechanics*, 3rd editio. Cambridge University Press, 2018.

[8] P. Atkins and R. Friedman, *Molecular Quantum Mechanics*, 5th Edition, Oxford University Press, 2011.

[9] Jensen Frank, *Introduction to Computational Chemistry*, 3rd Edition. Wiley, 2017.

[10] F. Agostini and B. F. E. Curchod, "Chemistry without the Born–Oppenheimer approximation," *Philos. Trans. R. Soc. A Math. Phys. Eng. Sci.*, vol. 380, no. 2223, May 2022, doi: 10.1098/RSTA.2020.0375.

[11] M. Born and K. Huang, *Dynamical Theory of Crystal Lattices*, New Ed. Oxford: Clarendon Press, 1954.

[12] Y. Yang, I. Kylänpää, N. M. Tubman, J. T. Krogel, S. Hammes-Schiffer, and D. M. Ceperley, "How large are nonadiabatic effects in atomic and diatomic systems?," *J. Chem. Phys.*, vol. 143, no. 12, p. 124308, Sep. 2015, doi: 10.1063/1.4931667.

[13] P. M. Kozlowski and L. Adamowicz, "Equivalent Quantum Approach to Nuclei and Electrons in Molecules," *Chem. Rev.*, vol. 93, no. 6, pp. 2007–2022, 1993, doi: 10.1021/cr00022a003.

[14] S. Hammes-Schiffer, "Current theoretical challenges in proton-coupled electron transfer: Electron-proton nonadiabaticity, proton relays, and ultrafast dynamics," *J. Phys. Chem. Lett.*, vol. 2, no. 12, pp. 1410–1416, Jun. 2011, doi: 10.1021/jz200277p.

[15] I. Navrotskaya and S. Hammes-Schiffer, "Electrochemical proton-coupled electron transfer: Beyond the golden rule," *J. Chem. Phys.*, vol. 131, no. 2, p. 024112, Jul. 2009, doi: 10.1063/1.3158828.

[16] A. Reyes, F. Moncada, and J. Charry, "The any particle molecular

orbital approach: A short review of the theory and applications," *Int. J. Quantum Chem.*, vol. 119, no. 2, p. e25705, Jan. 2019, doi: 10.1002/qua.25705.

[17] M. Cafiero, S. Bubin, and L. Adamowicz, "Non-Born-Oppenheimer calculations of atoms and molecules," *Phys. Chem. Chem. Phys.*, vol. 5, no. 8, pp. 1491–1501, Apr. 2003, doi: 10.1039/b211193d.

[18] J. Hartmann, "Exotic atoms," *Access Sci.*, 2000, doi: 10.1036/1097-8542.YB000560.

[19] K. Nagamine, *Introductory Muon Science*. Cambridge University Press, 2003.

[20] F. R. Attila Vertes, Sandor Nagy, Zoltan Klencsar, Rezso G. Lovas, Ed., *Handbook of Nuclear Chemistry*. Springer US, 2011.

[21] C. J. Rhodes, "Muonium - The second radioisotope of hydrogen - And its contribution to free radical chemistry," *J. Chem. Soc. Perkin Trans. 2*, no. 8, pp. 1379–1396, 2002, doi: 10.1039/b100699l.

[22] I. McKenzie, "The positive muon and μsR spectroscopy: Powerful tools for investigating the structure and dynamics of free radicals and spin probes in complex systems," *Annu. Reports Prog. Chem. - Sect. C*, vol. 109, pp. 65–112, 2013, doi: 10.1039/c3pc90005c.

[23] B. D. Patterson, "Muonium states in semiconductors," *Rev. Mod. Phys.*, vol. 60, no. 1, pp. 69–159, 1988, doi: 10.1103/RevModPhys.60.69.

[24] C. M. Surko, G. F. Gribakin, and S. J. Buckman, "Low-energy positron interactions with atoms and molecules," *J. Phys. B At. Mol. Opt. Phys.*, vol. 38, no. 6, 2005, doi: 10.1088/0953-4075/38/6/R01.

[25] M. J. Puska and R. M. Nieminen, "Theory of positrons," *Rev. Mod. Phys.*, vol. 66, no. 3, pp. 841–897, 1994.

[26] F. Tuomisto and I. Makkonen, "Defect identification in semiconductors with positron annihilation: Experiment and theory," *Rev. Mod. Phys.*, vol. 85, no. 4, pp. 1583–1631, 2013, doi: 10.1103/RevModPhys.85.1583.

[27] D. L. Bailey and D. W. Townsend, *Positron Emission Tomography: Basic Sciences*. Springer, 2006.

[28] S. D. Bass, S. Mariazzi, P. Moskal, and E. Stepien, "Positronium Physics and Biomedical Applications," vol. 95, no. June, pp. 1–21, 2023, doi: 10.1103/RevModPhys.95.021002.

[29] T. Okumura *et al.*, "Proof-of-Principle Experiment for Testing Strong-Field Quantum Electrodynamics with Exotic Atoms : High Precision X-Ray Spectroscopy of Muonic Neon," *Phys. Rev. Lett.*, vol. 130, no. 17, p. 173001, 2023, doi: 10.1103/PhysRevLett.130.173001.

[30] J. S. Rigden, *Hydrogen: The Essential Element*. Harvard University Press, 2003.

[31] M. Hori, H. Aghai-Khozani, A. Sótér, A. Dax, and D. Barna, "Recent results of laser spectroscopy experiments of pionic helium atoms at PSI," *SciPost Phys. Proc.*, vol. 5, no. 5, p. 026, Sep. 2021, doi: 10.21468/SCIPOSTPHYSPROC.5.026.

[32] M. Hori, A. Sótér, and V. I. Korobov, "Proposed method for laser spectroscopy of pionic helium atoms to determine the charged-pion mass," *Phys. Rev. A - At. Mol. Opt. Phys.*, vol. 89, no. 4, p. 042515, Apr. 2014, doi: 10.1103/PHYSREVA.89.042515/FIGURES/11/MEDIUM.

[33] E. D. Kena and G. B. Adera, "Solving the Dirac equation in central potential for muonic hydrogen atom with point-like nucleus," *J. Phys. Commun.*, vol. 5, no. 10, p. 105018, Oct. 2021, doi: 10.1088/2399-6528/AC2FBC.

[34] M. Hori, H. Aghai-Khozani, A. Sótér, A. Dax, and D. Barna, "Laser spectroscopy of pionic helium atoms," *Nature*, vol. 581, no. 7806, pp. 37–41, May 2020, doi: 10.1038/S41586-020-2240-X.

[35] D. Gotta, "Precision spectroscopy of light exotic atoms," *Prog. Part. Nucl. Phys.*, vol. 52, no. 1, pp. 133–195, 2004, doi: 10.1016/j.ppnp.2003.09.003.

[36] A. D. Bochevarov, E. F. Valeev, and C. D. Sherrill, "The electron and

nuclear orbitals model: Current challenges and future prospects," *Mol. Phys.*, vol. 102, no. 1, pp. 111–123, Jan. 2004, doi: 10.1080/00268970410001668525.

[37] M. Goli and S. Shahbazian, "The two-component quantum theory of atoms in molecules (TC-QTAIM): Foundations," *Theor. Chem. Acc.*, vol. 131, no. 5, pp. 1–19, Apr. 2012, doi: 10.1007/s00214-012-1208-9.

[38] I. L. Thomas and H. W. Joy, "Protonic structure of molecules. II. Methodology, center-of-mass transformation, and the structure of methane, ammonia, and water," *Phys. Rev. A*, vol. 2, no. 4, pp. 1200–1208, Oct. 1970, doi: 10.1103/PhysRevA.2.1200.

[39] I. L. Thomas, "Protonic structure of molecules. I. Ammonia molecules," *Phys. Rev.*, vol. 185, no. 1, pp. 90–94, 1969, doi: 10.1103/PhysRev.185.90.

[40] P. M. Kozlowski and L. Adamowicz, "An effective method for generating nonadiabatic many-body wave function using explicitly correlated Gaussian-type functions," *J. Chem. Phys.*, vol. 95, no. 9, pp. 6661–6668, 1991, doi: 10.1063/1.461538.

[41] S. Bubin, M. Pavanello, W. C. Tung, K. L. Sharkey, and L. Adamowicz, "Born-oppenheimer and non-born-oppenheimer, atomic and molecular calculations with explicitly correlated gaussians," *Chem. Rev.*, vol. 113, no. 1, pp. 36–79, 2013, doi: 10.1021/cr200419d.

[42] S. Bubin and L. Adamowicz, "Variational calculations of excited states with zero total angular momentum (vibrational spectrum) of H2 without use the Born-Oppenheimer approximation," *J. Chem. Phys.*, vol. 118, no. 7, pp. 3079–3082, Feb. 2003, doi: 10.1063/1.1537719.

[43] M. Cafiero and L. Adamowicz, "Non-Born-Oppenheimer molecular structure and one-particle densities for H2D+," *J. Chem. Phys.*, vol. 122, no. 18, p. 184305, 2005, doi: 10.1063/1.1891707.

[44] H. Nakai, "Nuclear orbital plus molecular orbital theory: Simultaneous determination of nuclear and electronic wave functions without Born–Oppenheimer approximation," *Int. J. Quantum Chem.*, vol. 107, no.

14, pp. 2849–2869, Nov. 2007, doi: 10.1002/qua.21379.

[45] M. Tachikawa, K. Mori, K. Suzuki, and K. Iguchi, "Full variational molecular orbital method: Application to the positron-molecule complexes," *Int. J. Quantum Chem.*, vol. 70, no. 3, pp. 491–501, 1998, doi: 10.1002/(SICI)1097-461X(1998)70:3<491::AID-QUA5>3.0.CO;2-P.

[46] M. Tachikawa, K. Mori, H. Nakai, and K. Iguchi, "An extension of ab initio molecular orbital theory to nuclear motion," *Chem. Phys. Lett.*, vol. 290, no. 4–6, pp. 437–442, Jul. 1998, doi: 10.1016/S0009-2614(98)00519-3.

[47] H. Nakai, "Simultaneous determination of nuclear and electronic wave functions without Born-Oppenheimer approximation: Ab initio NO+MO/HF theory," *Int. J. Quantum Chem.*, vol. 86, no. 6, pp. 511–517, Feb. 2002, doi: 10.1002/qua.1106.

[48] H. Nakai and K. Sodeyama, "Many-body effects in nonadiabatic molecular theory for simultaneous determination of nuclear and electronic wave functions: Ab initio NOMO/MBPT and CC methods," *J. Chem. Phys.*, vol. 118, no. 3, pp. 1119–1127, Jan. 2003, doi: 10.1063/1.1528951.

[49] H. Nakai, M. Hoshino, K. Miyamoto, and S. Hyodo, "Elimination of translational and rotational motions in nuclear orbital plus molecular orbital theory," *J. Chem. Phys.*, vol. 122, no. 16, p. 164101, Apr. 2005, doi: 10.1063/1.1871914.

[50] B. T. Sutcliffe, "The idea of a potential energy surface," *J. Mol. Struct. THEOCHEM*, vol. 341, no. 1–3, pp. 217–235, Oct. 1995, doi: 10.1016/0166-1280(95)04125-P.

[51] M. Hoshino and H. Nakai, "Elimination of translational and rotational motions in nuclear orbital plus molecular orbital theory: Application of Møller-Plesset perturbation theory," *J. Chem. Phys.*, vol. 124, no. 19, p. 194110, May 2006, doi: 10.1063/1.2193513.

[52] H. Nishizawa, M. Hoshino, Y. Imamura, and H. Nakai, "Evaluation of electron repulsion integral of the explicitly correlated Gaussian-

nuclear orbital plus molecular orbital theory," *Chem. Phys. Lett.*, vol. 521, pp. 142–149, Jan. 2012, doi: 10.1016/j.cplett.2011.11.023.

[53] S. P. Webb, T. Iordanov, and S. Hammes-Schiffer, "Multiconfigurational nuclear-electronic orbital approach: Incorporation of nuclear quantum effects in electronic structure calculations," *J. Chem. Phys.*, vol. 117, no. 9, pp. 4106–4118, Sep. 2002, doi: 10.1063/1.1494980.

[54] M. V. Pak and S. Hammes-Schiffer, "Electron-Proton Correlation for Hydrogen Tunneling Systems," *Phys. Rev. Lett.*, vol. 92, no. 10, p. 103002, Mar. 2004, doi: 10.1103/PHYSREVLETT.92.103002/FIGURES/2/MEDIUM.

[55] C. Swalina, M. V. Pak, A. Chakraborty, and S. Hammes-Schiffer, "Explicit dynamical electron-proton correlation in the nuclear-electronic orbital framework," *J. Phys. Chem. A*, vol. 110, no. 33, pp. 9983–9987, Aug. 2006, doi: 10.1021/jp0634297.

[56] A. Chakraborty, M. V. Pak, and S. Hammes-Schiffer, "Inclusion of explicit electron-proton correlation in the nuclear-electronic orbital approach using Gaussian-type geminal functions," *J. Chem. Phys.*, vol. 129, no. 1, p. 014101, Jul. 2008, doi: 10.1063/1.2943144.

[57] C. Ko, M. V. Pak, C. Swalina, and S. Hammes-Schiffer, "Alternative wavefunction ansatz for including explicit electron-proton correlation in the nuclear-electronic orbital approach," *J. Chem. Phys.*, vol. 135, no. 5, p. 054106, Aug. 2011, doi: 10.1063/1.3611054.

[58] A. Sirjoosingh, M. V. Pak, C. Swalina, and S. Hammes-Schiffer, "Reduced explicitly correlated Hartree-Fock approach within the nuclear-electronic orbital framework: Theoretical formulation," *J. Chem. Phys.*, vol. 139, no. 3, p. 034102, 2013, doi: 10.1063/1.4812257.

[59] K. R. Brorsen, A. Sirjoosingh, M. V. Pak, and S. Hammes-Schiffer, "Nuclear-electronic orbital reduced explicitly correlated Hartree-Fock approach: Restricted basis sets and open-shell systems," *J. Chem. Phys.*, vol. 142, no. 21, p. 214108, Jun. 2015, doi: 10.1063/1.4921304.

[60] F. Pavošević, T. Culpitt, and S. Hammes-Schiffer, "Multicomponent Quantum Chemistry: Integrating Electronic and Nuclear Quantum Effects via the Nuclear-Electronic Orbital Method," *Chem. Rev.*, vol. 120, no. 9, pp. 4222–4253, 2020, doi: 10.1021/acs.chemrev.9b00798.

[61] S. Hammes-Schiffer, "Nuclear-electronic orbital methods: Foundations and prospects," *J. Chem. Phys.*, vol. 155, no. 3, pp. 1–11, 2021, doi: 10.1063/5.0053576.

[62] J. F. Capitani, R. F. Nalewajski, and R. G. Parr, "NonBorn – Oppenheimer density functional theory of molecular systems Non-Born-Oppenheimer density functional theory of molecular systems," *J. Chem. Phys.*, no. 76, p. 568, 1982, doi: 10.1063/1.442703.

[63] Y. Shigeta, H. Takahashi, S. Yamanaka, M. Mitani, H. Nagao, and K. Yamaguchi, "Density functional theory without the Born–Oppenheimer approximation and its application," *Int. J. Quantum Chem.*, vol. 70, no. 45, pp. 659–669, 1998, doi: 10.1002/(sici)1097-461x(1998)70:4/5<659::aid-qua12>3.3.co;2-s.

[64] T. Kreibich and E. K. U. Gross, "Multicomponent density-functional theory for electrons and nuclei," *Phys. Rev. Lett.*, vol. 86, no. 14, pp. 2984–2987, Apr. 2001, doi: 10.1103/PhysRevLett.86.2984.

[65] M. V. Pak, A. Chakraborty, and S. Hammes-Schiffer, "Density functional theory treatment of electron correlation in the nuclear-electronic orbital approach," *J. Phys. Chem. A*, vol. 111, no. 20, pp. 4522–4526, 2007, doi: 10.1021/jp0704463.

[66] A. Chakraborty, M. V. Pak, and S. Hammes-Schiffer, "Properties of the exact universal functional in multicomponent density functional theory," *J. Chem. Phys.*, vol. 131, no. 12, p. 124115, 2009, doi: 10.1063/1.3236844.

[67] A. Sirjoosingh, M. V. Pak, and S. Hammes-Schiffer, "Derivation of an electron-proton correlation functional for multicomponent density functional theory within the nuclear-electronic orbital approach," *J. Chem. Theory Comput.*, vol. 7, no. 9, pp. 2689–2693, Sep. 2011, doi: 10.1021/ct200473r.

[68] A. Sirjoosingh, M. V. Pak, and S. Hammes-Schiffer, “Multicomponent density functional theory study of the interplay between electron-electron and electron-proton correlation,” *J. Chem. Phys.*, vol. 136, no. 17, p. 174114, May 2012, doi: 10.1063/1.4709609.

[69] Y. Yang, K. R. Brorsen, T. Culpitt, M. V. Pak, and S. Hammes-Schiffer, “Development of a practical multicomponent density functional for electron-proton correlation to produce accurate proton densities,” *J. Chem. Phys.*, vol. 147, no. 11, p. 114113, Sep. 2017, doi: 10.1063/1.4996038.

[70] K. R. Brorsen, Y. Yang, and S. Hammes-Schiffer, “Multicomponent Density Functional Theory: Impact of Nuclear Quantum Effects on Proton Affinities and Geometries,” *J. Phys. Chem. Lett.*, vol. 8, no. 15, pp. 3488–3493, 2017, doi: 10.1021/acs.jpclett.7b01442.

[71] K. R. Brorsen, P. E. Schneider, and S. Hammes-Schiffer, “Alternative forms and transferability of electron-proton correlation functionals in nuclear-electronic orbital density functional theory,” *J. Chem. Phys.*, vol. 149, no. 4, p. 044110, 2018, doi: 10.1063/1.5037945.

[72] Z. Tao, Y. Yang, and S. Hammes-Schiffer, “Multicomponent density functional theory: Including the density gradient in the electron-proton correlation functional for hydrogen and deuterium,” *J. Chem. Phys.*, vol. 151, no. 12, p. 124102, 2019, doi: 10.1063/1.5119124.

[73] M. Goli and S. Shahbazian, “Two-component density functional theory for muonic molecules: Inclusion of the electron–positive muon correlation functional,” *J. Chem. Phys.*, vol. 156, no. 4, p. 044104, Jan. 2022, doi: 10.1063/5.0077179.

[74] F. Pavošević, T. Culpitt, and S. Hammes-Schiffer, “Multicomponent Coupled Cluster Singles and Doubles Theory within the Nuclear-Electronic Orbital Framework,” *J. Chem. Theory Comput.*, vol. 15, no. 1, pp. 338–347, Jan. 2019, doi: 10.1021/acs.jctc.8b01120.

[75] C. Swalina, M. V. Pak, and S. Hammes-Schiffer, “Alternative formulation of many-body perturbation theory for electron-proton correlation,” *Chem. Phys. Lett.*, vol. 404, no. 4–6, pp. 394–399, Mar.

2005, doi: 10.1016/j.cplett.2005.01.115.

[76] F. Pavošević, B. J. G. Rousseau, and S. Hammes-Schiffer, "Multicomponent Orbital-Optimized Perturbation Theory Methods: Approaching Coupled Cluster Accuracy at Lower Cost," *J. Phys. Chem. Lett.*, vol. 11, no. 4, pp. 1578–1583, 2020, doi: 10.1021/acs.jpclett.0c00090.

[77] R. G. Parr and Y. Weitao, *Density-Functional Theory of Atoms and Molecules*. Oxford University Press, 1989.

[78] T. Gould, D. P. Kooi, P. Gori-Giorgi, and S. Pittalis, "Electronic Excited States in Extreme Limits via Ensemble Density Functionals," *Phys. Rev. Lett.*, vol. 130, no. 10, p. 106401, 2023, doi: 10.1103/physrevlett.130.106401.

[79] M. Taut, "Two electrons in an external oscillator potential: Particular analytic solutions of a Coulomb correlation problem," *Phys. Rev. A*, vol. 48, no. 5, pp. 3561–3566, Nov. 1993, doi: 10.1103/PhysRevA.48.3561.

[80] N. R. Kestner and O. Sinanoälu, "Study of Electron Correlation in Helium-Like Systems Using an Exactly Soluble Model," *Phys. Rev.*, vol. 128, no. 6, pp. 2687–2692, Dec. 1962, doi: 10.1103/PhysRev.128.2687.

[81] E. A. Hylleraas, "Über den Grundzustand des Heliumatoms," *Zeitschrift für Phys.*, vol. 48, no. 7–8, pp. 469–494, 1928, doi: 10.1007/BF01340013.

[82] E. A. Hylleraas, "Neue Berechnung der Energie des Heliums im Grundzustande, sowie des tiefsten Terms von Ortho-Helium," *Zeitschrift für Phys.*, vol. 54, no. 5–6, pp. 347–366, 1929, doi: 10.1007/BF01375457.

[83] P. M. Laufer and J. B. Krieger, "Test of density-functional approximations in an exactly soluble model," *Phys. Rev. A*, vol. 33, no. 3, pp. 1480–1491, Mar. 1986, doi: 10.1103/PhysRevA.33.1480.

[84] S. K. Ghosh and A. Samanta, "Study of correlation effects in an exactly

solvable model two-electron system," *J. Chem. Phys.*, vol. 94, no. 1, pp. 517–522, Aug. 1991, doi: 10.1063/1.460368.

[85] A. Samanta and S. K. Ghosh, "Correlation in an exactly solvable two-particle quantum system," *Phys. Rev. A*, vol. 42, no. 3, pp. 1178–1183, 1990, doi: 10.1103/PhysRevA.42.1178.

[86] A. G. Ushveridze, *Quasi-exactly solvable models in quantum mechanics*. Taylor & Francis, 1994.

[87] J. Karwowski and L. Cyrnek, "Harmonium," *Ann. Phys.*, vol. 13, no. 4, pp. 181–193, Apr. 2004, doi: 10.1002/andp.200310071.

[88] J. Karwowski, "Influence of confinement on the properties of quantum systems," *J. Mol. Struct. THEOCHEM*, vol. 727, no. 1-3 SPEC. ISS., pp. 1–7, Aug. 2005, doi: 10.1016/j.theochem.2005.02.038.

[89] J. Karwowski and H. A. Witek, "Biconfluent Heun equation in quantum chemistry: Harmonium and related systems," *Theor. Chem. Acc.*, vol. 133, no. 7, pp. 1–11, May 2014, doi: 10.1007/s00214-014-1494-5.

[90] J. Karwowski, "Inverse problems in quantum chemistry," *Int. J. Quantum Chem.*, vol. 109, no. 11, pp. 2456–2463, Jan. 2009, doi: 10.1002/qua.22048.

[91] J. Karwowski, "Few-particle systems: Quasi-exactly solvable models," *J. Phys. Conf. Ser.*, vol. 104, no. 1, p. 12033, 2008, doi: 10.1088/1742-6596/104/1/012033.

[92] J. Karwowski and H. A. Witek, "Schrödinger equations with power potentials," *Mol. Phys.*, vol. 114, no. 7–8, pp. 932–940, Apr. 2016, doi: 10.1080/00268976.2015.1115565.

[93] J. Cioslowski and K. Strasburger, "Five- and six-electron harmonium atoms: Highly accurate electronic properties and their application to benchmarking of approximate 1-matrix functionals," *J. Chem. Phys.*, vol. 148, no. 14, p. 144107, 2018, doi: 10.1063/1.5021419.

[94] K. D. Sen, *Electronic structure of quantum confined atoms and molecules*. Springer, 2014.

[95] P. C. Deshmukh, J. Jose, H. R. Varma, and S. T. Manson, “Electronic structure and dynamics of confined atoms,” *Eur. Phys. J. D*, vol. 75, no. 6, pp. 1–32, 2021, doi: 10.1140/epjd/s10053-021-00151-2.

[96] P. A. Maksym and T. Chakraborty, “Quantum dots in a magnetic field: Role of electron-electron interactions,” *Phys. Rev. Lett.*, vol. 65, no. 1, pp. 108–111, 1990, doi: 10.1103/PhysRevLett.65.108.

[97] C. A. Coulson and A. H. Neilson, “Electron correlation in the ground state of helium,” *Proc. Phys. Soc.*, vol. 78, no. 5, pp. 831–837, 1961, doi: 10.1088/0370-1328/78/5/328.

[98] W. Koch and M. C. Holthausen, *A Chemist’s Guide to Density Functional Theory*. Wiley, 2001.

[99] R. A. Ferrell, “Theory of Positron Annihilation in Solids,” *Rev. Mod. Phys.*, vol. 28, p. 308, 1956.

[100] S. Kahana, “Positron annihilation in metals,” *Phys. Rev.*, vol. 117, no. 1, pp. 123–128, 1960, doi: 10.1103/PhysRev.117.123.

[101] D. R. Hamann, “Effective mass of positrons in metals,” *Phys. Rev.*, vol. 146, no. 1, pp. 277–281, 1966, doi: 10.1103/PhysRev.146.277.

[102] B. Bergersen and E. Pajanne, “Positron-electron correlation-polarization potentials for the calculation of positron collisions with atoms and molecules,” *Phys. Rev.*, vol. 186, no. 2, pp. 375–380, Oct. 1969, doi: 10.1103/PhysRev.186.375.

[103] E. Boroński and R. M. Nieminen, “Electron-positron density-functional theory,” *Phys. Rev. B*, vol. 34, no. 6, pp. 3820–3831, 1986, doi: 10.1103/PhysRevB.34.3820.

[104] B. Chakraborty, “Effects of electron-positron correlation on positron annihilation: Self-consistent band-structure calculations in Al,” *Phys. Rev. B*, vol. 24, no. 12, pp. 7423–7426, Dec. 1981, doi: 10.1103/PhysRevB.24.7423.

[105] B. Barbiellini, M. Hakala, M. Puska, R. Nieminen, and A. Manuel, “Correlation effects for electron-positron momentum density in solids,” *Phys. Rev. B - Condens. Matter Mater. Phys.*, vol. 56, no. 12,

pp. 7136–7142, 1997, doi: 10.1103/PhysRevB.56.7136.

[106] G. Kontrym-Sznajd and J. Majsnerowski, "Electronic structure and electron-positron correlation effects in Mg," *J. Phys. Condens. Matter*, vol. 2, no. 49, pp. 9927–9939, 1990, doi: 10.1088/0953-8984/2/49/017.

[107] J. Franz, "Positron-electron correlation-polarization potentials for the calculation of positron collisions with atoms and molecules*," *Eur. Phys. J. D*, vol. 71, no. 2, 2017, doi: 10.1140/epjd/e2017-70591-2.

[108] J. A. Charry Martinez, M. Barborini, and A. Tkatchenko, "Correlated Wave Functions for Electron-Positron Interactions in Atoms and Molecules," *J. Chem. Theory Comput.*, vol. 18, no. 4, pp. 2267–2280, 2022, doi: 10.1021/acs.jctc.1c01193.

[109] P. Tommasini, E. Timmermans, and A. F. R. de Toledo Piza, "The hydrogen atom as an entangled electron–proton system," *Am. J. Phys.*, vol. 66, no. 10, pp. 881–886, 1998, doi: 10.1119/1.18977.

[110] E. Matsushita, "Model of electron-proton correlation in quasi-one-dimensional halogen-bridged mixed-valence complexes: Role of proton motion," *Phys. Rev. B. Condens. Matter*, vol. 51, no. 24, pp. 17332–17337, 1995, doi: 10.1103/PHYSREVB.51.17332.

[111] T. Ishimoto, M. Tachikawa, and U. Nagashima, "Electron-electron and electron-nucleus correlation effects on exponent values of Gaussian-type functions for quantum protons and deuterons," *J. Chem. Phys.*, vol. 125, no. 14, p. 144103, 2006, doi: 10.1063/1.2352753.

[112] O. J. Fajen and K. R. Brorsen, "Separation of electron-electron and electron-proton correlation in multicomponent orbital-optimized perturbation theory," *J. Chem. Phys.*, vol. 152, no. 19, p. 194107, 2020, doi: 10.1063/5.0006743.

[113] Z. Chen and J. Yang, "Nucleus-electron correlation revising molecular bonding fingerprints from the exact wavefunction factorization," *J. Chem. Phys.*, vol. 155, no. 10, p. 104111, 2021, doi: 10.1063/5.0056773.

[114] T. Udagawa, T. Tsuneda, and M. Tachikawa, "Electron-nucleus correlation functional for multicomponent density-functional theory," *Phys. Rev. A - At. Mol. Opt. Phys.*, vol. 89, no. 5, pp. 1–6, 2014, doi: 10.1103/PhysRevA.89.052519.

[115] Y. Akamatsu, T. Hatsuda, and T. Hirano, "Electron-muon correlation as a new probe of strongly interacting quark-gluon plasma," *Phys. Rev. C - Nucl. Phys.*, vol. 80, no. 3, pp. 33–36, 2009, doi: 10.1103/PhysRevC.80.031901.

[116] R. J. Boyd and C. A. Coulson, "Coulomb hole in some excited states of helium," *J. Phys. B At. Mol. Phys.*, vol. 6, no. 5, pp. 782–793, 1973, doi: 10.1088/0022-3700/6/5/012.

[117] C. Boyd, "On the Fermi hole in atoms," *J. Phys. B At. Mol. Phys.*, vol. 7, no. 14, pp. 1805–1816, 1975, doi: 10.1088/0022-3700/8/8/002.

[118] R. McWeeny, "Some recent advances in density matrix theory," *Rev. Mod. Phys.*, vol. 32, no. 2, pp. 335–369, 1960, doi: 10.1103/RevModPhys.32.335.

[119] R. Mcweeny, "The nature of electron correlation in molecules," *Int. J. Quantum Chem.*, vol. 1, no. 1 S, pp. 351–359, 1967, doi: 10.1002/qua.560010641.

[120] A. J. Thakkar, "Extracules, Intracules, Correlation Holes, Potentials, Coefficients and All That," in *Density Matrices and Density Functionals*, 1987, pp. 553–581, doi: 10.1007/978-94-009-3855-7_30.

[121] W. Kohn and L. J. Sham, "Self-consistent equations including exchange and correlation effects," *Phys. Rev.*, vol. 140, no. 4A, pp. A1133–A1138, Nov. 1965, doi: 10.1103/PHYSREV.140.A1133/FIGURE/1/THUMB.

[122] P. Hohenberg and W. Kohn, "Inhomogeneous electron gas," *Phys. Rev.*, vol. 136, no. 3B, pp. B864–B871, Nov. 1964, doi: 10.1103/PHYSREV.136.B864/FIGURE/1/THUMB.

[123] B. I. Gunnarsson, O. Lundqvist, "Exchange and correlation in atoms, molecules, and solids by the spin-density-functional formalism," *Phys.*

*Rev. B*, vol. 13, no. 10, pp. 4274–4298, 1976.

[124] D. C. Langreth and J. P. Perdew, "THE EXCHANGE-CORRELATION ENERGY OF A METALLIC SURFACE," *Phys. Rev. B*, vol. 15, no. 6, pp. 2884–2901, 1977.

[125] J. Harris, "The role of occupation numbers in HKS theory," *Int. J. Quantum Chem.*, vol. 16, no. 13 S, pp. 189–193, 1979, doi: 10.1002/qua.560160821.

[126] J. Harris, "Adiabatic-connection approach to Kohn-Sham theory," *Phys. Rev. A*, vol. 29, no. 4, pp. 1648–1659, 1984, doi: 10.1103/PhysRevA.29.1648.

[127] A. D. Becke, "Correlation energy of an inhomogeneous electron gas: A coordinate-space model," *J. Chem. Phys.*, vol. 88, no. 2, pp. 1053–1062, 1988, doi: 10.1063/1.454274.

[128] K. Burke, J. P. Perdew, and D. C. Langreth, "Is the local density approximation exact for short wavelength fluctuations?," *Phys. Rev. Lett.*, vol. 73, no. 9, pp. 1283–1286, 1994, doi: 10.1103/PhysRevLett.73.1283.

[129] T. Leininger, H. Stoll, H. J. Werner, and A. Savin, "Combining long-range configuration interaction with short-range density functional," *Chem. Phys. Lett.*, vol. 275, no. 3–4, pp. 151–160, 1997, doi: 10.1016/S0009-2614(97)00758-6.

[130] J. Cioslowski and G. Liu, "Electron intracule densities and Coulomb holes from energy-derivative two-electron reduced density matrices," *J. Chem. Phys.*, vol. 109, no. 19, pp. 8225–8231, 1998, doi: 10.1063/1.477484.

[131] K. Burke, J. P. Perdew, and M. Ernzerhof, "Why semilocal functionals work: Accuracy of the on-top pair density and importance of system averaging," *J. Chem. Phys.*, vol. 109, no. 10, pp. 3760–3771, 1998, doi: 10.1063/1.476976.

[132] Z. Yan, J. P. Perdew, and S. Kurth, "Density functional for short-range correlation: Accuracy of the random-phase approximation for

isoelectronic energy changes," *Phys. Rev. B - Condens. Matter Mater. Phys.*, vol. 61, no. 24, pp. 16430–16439, 2000, doi: 10.1103/PhysRevB.61.16430.

[133] T. M. Henderson and R. J. Bartlett, "Short-range corrections to the correlation hole," *Phys. Rev. A*, vol. 70, no. 2, pp. 1–12, 2004, doi: 10.1103/PhysRevA.70.022512.

[134] T. M. Henderson and R. J. Bartlett, "Theory of the short-range correlation hole model," *Mol. Phys.*, vol. 103, no. 15–16, pp. 2093–2103, 2005, doi: 10.1080/09500340500131442.

[135] A. D. Becke, "Perspective: Fifty years of density-functional theory in chemical physics," *J. Chem. Phys.*, vol. 140, no. 18, 2014, doi: 10.1063/1.4869598.

[136] A. Pribram-Jones, D. A. Gross, and K. Burke, "DFT: A theory full of holes," *Annu. Rev. Phys. Chem.*, vol. 66, pp. 283–304, 2015, doi: 10.1146/annurev-physchem-040214-121420.

[137] D. P. O'Neill and P. M. W. Gill, "Wave functions and two-electron probability distributions of the Hooke's-law atom and helium," *Phys. Rev. A - At. Mol. Opt. Phys.*, vol. 68, no. 2, p. 7, 2003, doi: 10.1103/PhysRevA.68.022505.

[138] J. Makarewicz, "Coulomb and Fermi holes in a two-electron model atom," *Am. J. Phys.*, vol. 56, no. 12, pp. 1100–1104, 1988, doi: 10.1119/1.15760.

[139] P. Cassam-Chenaï, B. Suo, and W. Liu, "A quantum chemical definition of electron–nucleus correlation," *Theor. Chem. Acc.*, vol. 136, no. 4, 2017, doi: 10.1007/s00214-017-2081-3.

[140] E. J. Baerends and O. V Gritsenko, "A Quantum Chemical View of Density Functional Theory," *J. Phys. Chem. A.*, vol. 101, no. 30, pp. 5384–5403, 1997, doi: 10.1021/jp9703768.

[141] S. Kais, D. R. Herschbach, N. C. Handy, C. W. Murray, and G. J. Laming, "Density functionals and dimensional renormalization for an exactly solvable model," *J. Chem. Phys.*, vol. 99, no. 1, pp. 417–425,

1993, doi: 10.1063/1.465765.

[142] R. Frigg, "Models in physics," *Routledge Encyclopedia of Philosophy Online*. Routledge, 2009, doi: 10.4324/9780415249126-Q135-1.

[143] B. Sutherland, *Beautiful Models*. World Scientific, 2004.

[144] M. B. Hesse, "Models in physics," *Br. J. Philos. Sci.*, vol. 4, no. 15, pp. 198–214, 1953, doi: 10.1093/bjps/IV.15.198.

[145] M. W. W. Robert S. Cohen, Ed., *Boston Studies in the Philosophy of Science*, vol. 5. springer, 1968.

[146] C. Filippi, C. J. Umrigar, and M. Taut, "Comparison of exact and approximate density functionals for an exactly soluble model," *J. Chem. Phys.*, vol. 100, no. 2, pp. 1290–1296, 1994, doi: 10.1063/1.466658.

[147] S. Majumdar and A. K. Roy, "Density functional study of atoms spatially confined inside a hard sphere," *Int. J. Quantum Chem.*, vol. 121, no. 11, pp. 1–21, 2021, doi: 10.1002/qua.26630.

[148] R. Singh, A. Kumar, M. K. Harbola, and P. Samal, "Semianalytical wavefunctions and Kohn-Sham exchange-correlation potentials for two-electron atomic systems in two-dimensions," *J. Phys. B At. Mol. Opt. Phys.*, vol. 53, no. 3, 2020, doi: 10.1088/1361-6455/ab56be.

[149] T. Chachiyo and H. Chachiyo, "Understanding electron correlation energy through density functional theory," *Comput. Theor. Chem.*, vol. 1172, no. September 2019, p. 112669, 2020, doi: 10.1016/j.comptc.2019.112669.

[150] D. P. Kooi and P. Gori-Giorgi, "Local and global interpolations along the adiabatic connection of DFT: a study at different correlation regimes," *Theor. Chem. Acc.*, vol. 137, no. 12, pp. 1–12, 2018, doi: 10.1007/s00214-018-2354-5.

[151] R. S. Chauhan and M. K. Harbola, "Study of adiabatic connection in density functional theory with an accurate wavefunction for two-electron spherical systems," *Int. J. Quantum Chem.*, vol. 117, no. 8, 2017, doi: 10.1002/qua.25344.

[152] S. F. Vyboishchikov, "A Simple Local Correlation Energy Functional for Spherically Confined Atoms from ab Initio Correlation Energy Density," *ChemPhysChem*, vol. 18, no. 23, pp. 3478–3484, 2017, doi: 10.1002/cphc.201700774.

[153] J. Cioslowski, M. Piris, and E. Matito, "Robust validation of approximate 1-matrix functionals with few-electron harmonium atoms," *J. Chem. Phys.*, vol. 143, no. 21, 2015, doi: 10.1063/1.4936583.

[154] J. P. Coe, A. Sudbery, and I. D'Amico, "Entanglement and density-functional theory: Testing approximations on Hooke's atom," *Phys. Rev. B - Condens. Matter Mater. Phys.*, vol. 77, no. 20, pp. 1–14, 2008, doi: 10.1103/PhysRevB.77.205122.

[155] M. Seidl, "Adiabatic connection in density-functional theory: Two electrons on the surface of a sphere," *Phys. Rev. A - At. Mol. Opt. Phys.*, vol. 75, no. 6, pp. 1–11, 2007, doi: 10.1103/PhysRevA.75.062506.

[156] S. Ragot, "Exact Kohn-Sham versus Hartree-Fock in momentum space: Examples of two-fermion systems," *J. Chem. Phys.*, vol. 125, no. 1, 2006, doi: 10.1063/1.2212935.

[157] D. Gómez, E. V. Ludeña, V. Karasiev, and P. Nieto, "Application of exact analytic total energy functional for Hooke's atom to He, Li+ and Be++: An examination of the universality of the energy functional in DFT," *Theor. Chem. Acc.*, vol. 116, no. 4–5, pp. 608–613, 2006, doi: 10.1007/s00214-006-0106-4.

[158] P. Capuzzi, N. H. March, and M. P. Tosi, "Differential equation for the ground-state density of artificial two-electron atoms with harmonic confinement," *J. Phys. A. Math. Gen.*, vol. 38, no. 24, 2005, doi: 10.1088/0305-4470/38/24/L01.

[159] E. V. Ludeña, D. Gómez, V. Karasiev, and P. Nieto, "Exact analytic total energy functional for Hooke's atom generated by local-scaling transformations," *Int. J. Quantum Chem.*, vol. 99, no. 4, pp. 297–307, 2004, doi: 10.1002/qua.10858.

[160] C. Amovilli and N. H. March, "Exact density matrix for a two-electron model atom and approximate proposals for realistic two-electron systems," *Phys. Rev. A - At. Mol. Opt. Phys.*, vol. 67, no. 2, p. 6, 2003, doi: 10.1103/PhysRevA.67.022509.

[161] D. Frydel, W. M. Terilla, and K. Burke, "Adiabatic connection from accurate wave-function calculations," *J. Chem. Phys.*, vol. 112, no. 12, pp. 5292–5297, 2000, doi: 10.1063/1.481099.

[162] Z. Qian and V. Sahni, "Physics of transformation from Schrödinger theory to Kohn-Sham density-functional theory: Application to an exactly solvable model," *Phys. Rev. A - At. Mol. Opt. Phys.*, vol. 57, no. 4, pp. 2527–2538, 1998, doi: 10.1103/PhysRevA.57.2527.

[163] K. Lam, F. G. Cruz, and K. Burke, "Virial Exchange-Correlation Energy Density in Hooke's Atom," *Int. J. Quantum Chem.*, vol. 69, pp. 533–540, 1998.

[164] N. S. O. Attila Szabo, *Modern Quantum Chemistry: Introduction to Advanced Electronic Structure Theory*. Dover Publications, 1982.

[165] H. Taketa, S. Huzinaga, and K. O-ohata, "Gaussian-Expansion Methods for Molecular Integrals," *J. Phys. Soc. Japan*, vol. 21, no. 11, pp. 2313–2324, 1966, doi: 10.1143/JPSJ.21.2313.

[166] I. I. Guseinov and B. A. Mamedov, "Evaluation of the Boys function using analytical relations," *J. Math. Chem.*, vol. 40, no. 2, pp. 179–183, 2006, doi: 10.1007/s10910-005-9023-3.

[167] E. W. Ng and M Geller, "Table of Integrals of the Error Functions," *U S Bur Stand. Res. Sci.*, vol. 73 B, no. 1, pp. 1–20, 1969, doi: 10.6028/jres.073b.001.

[168] E. R. D. Larry E McMurchie, "One- and two-electron integrals over cartesian gaussian functions," *J. Comput. Phys.*, vol. 26, pp. 218–231, 1978.

[169] T. Helgaker, P. Jørgensen, and J. Olsen, *Molecular electronic-structure theory*. Wiley, 2014.

[170] I. I. Guseinov, "Analytical evaluation of one- and two-center Coulomb

and two-center hybrid integrals for Slater-type orbitals," *J. Chem. Phys.*, vol. 67, no. 8, pp. 3837–3839, 1977, doi: 10.1063/1.435329.

[171] Peter M.W. Gill, "Molecular Integrals," in *Advances in Quantum Chemistry*, vol. 25, J. R. M. C. Z. Sabin, Ed. Academic Press, 1994, p. 65.

[172] J. A. Rosal Sandberg, "New efficient integral algorithms for quantum chemistry," KTH Royal Institute of Technology, 2014.

[173] G. M. J. Barca and P. F. Loos, "Recurrence Relations for Four-Electron Integrals Over Gaussian Basis Functions," *Adv. Quantum Chem.*, vol. 76, no. 3, pp. 147–165, 2018, doi: 10.1016/bs.aiq.2017.03.004.

[174] T. Helgaker and P. R. Taylor, "Gaussian Basis Sets and Molecular Integrals," in *Modern Electronic Structure Theory*, worldscientific, 1995, pp. 725–856.

[175] S. Obara and A. Saika, "Efficient recursive computation of molecular integrals over Cartesian Gaussian functions," *J. Chem. Phys.*, vol. 84, no. 7, pp. 3963–3974, 1985, doi: 10.1063/1.450106.

[176] W. H. Press, S. a Teukolsky, W. T. Vetterling, and B. P. Flannery, *Numerical Recipes 3rd Edition: The Art of Scientific Computing*, 3rd ed., vol. 1. Cambridge University Press, 2007.